%% file: main.tex
\documentclass[preprint,aps,prd,tightenlines,superscriptaddress,floatfix,eqsecnum,nofootinbib,titlepage]{revtex4-2}
\usepackage[utf8]{inputenc}

\usepackage{amsmath,amssymb,amsfonts}
\usepackage{bm}
\usepackage{graphicx}
\usepackage[usenames,dvipsnames]{xcolor}

\usepackage{makeidx}

\usepackage[sort&compress]{natbib}
\usepackage[pdfpagelabels,colorlinks=true,linkcolor=blue,filecolor=blue,urlcolor=blue,citecolor=blue,plainpages=false]{hyperref}

\newcommand{\submitdate}{July 15, 2022}
\newcommand{\revisedate}{September 30, 2022}

\renewcommand{\Re}{\ensuremath{\mathop{\rm Re}}}
\newcommand{\umd}{{u-d}}
\makeindex

\makeatletter
      \let\ps@titlepage\ps@empty 
\makeatother

\begin{document}

\preprint{FERMILAB-CONF-22-531-T}

\title{Lattice QCD and Particle Physics}

\input usqcd_ec
\input white_coord
\input white_auth
\input usqcd_members
%
\collaboration{USQCD \rule[-12pt]{0pt}{28pt}Collaboration}
\noaffiliation

\date[Submitted:~]{\submitdate \\ aa
Revised:~\revisedate}

\pagestyle{empty}
\pagenumbering{Alph}
\hypersetup{pageanchor=false}
\begin{center}
    \rule[-0.2in]{\hsize}{0.01in}\\\rule{\hsize}{0.01in}\\
    \vskip 0.1in Submitted to the  Proceedings of the US Community Study\\ 
        on the Future of Particle Physics (Snowmass 2021)\\ 
    \rule{\hsize}{0.01in}\\\rule[+0.2in]{\hsize}{0.01in}
\end{center}
\maketitle
\hypersetup{pageanchor=true}

\pagestyle{plain}
\pagenumbering{roman}
\setcounter{page}{4}

\setcounter{tocdepth}{1}
\tableofcontents

\clearpage

\pagenumbering{arabic}
\setcounter{page}{1}

\section*{Executive Summary}
\input exec

\section{Introduction}
\vspace*{-0.47pt}
\input intro_main

\index{Rare and Precision Frontier|(}
\section{Rare and Precision Frontier}
\label{sec:rare}
\input rare/intro

\index{Theory Frontier!TF06|(}\index{Theory Frontier!TF07|(}
\subsection{Weak decays of quarks}
\label{sec:weak}
\input rare/weak
\index{Theory Frontier!TF07|)}\index{Theory Frontier!TF06|)}

\subsection{Fundamental physics in small experiments}
\label{sec:small}
\index{Rare and Precision Frontier!RF3}
\input rare/small

\subsection{Baryon- and lepton-number violating processes}
\label{sec:B-L}
\index{Rare and Precision Frontier!RF4}
\input rare/B-L

\subsection{Charged lepton flavor violation}
\label{sec:clfv}
\index{Rare and Precision Frontier!RF5}
\input rare/clfv

\subsection{Hadron spectroscopy}
\label{sec:xyz}
\index{Rare and Precision Frontier!RF7}
\input rare/xyz
\index{Rare and Precision Frontier|)}

\index{Neutrino Physics Frontier!NF06|(}
\index{Neutrino Physics Frontier!NF08/TF11|(}
\index{Theory Frontier!TF11/NF08|(}
\section{Neutrino Physics Frontier}
\label{sec:nu}
\index{Neutrino Physics Frontier}
\input nu
\index{Theory Frontier!TF11/NF08|)}
\index{Neutrino Physics Frontier!NF08/TF11|)}
\index{Neutrino Physics Frontier!NF06|)}

\index{Energy Frontier|(}
\section{Energy Frontier}
\label{sec:energy}
\input energy/intro

\index{Theory Frontier!TF07|(}
\subsection{Precision QCD and Higgs boson properties}
\label{sec:asmq}
\index{Energy Frontier!EF01}
\input energy/asmq

\subsection{Parton distribution functions}
\label{sec:pdf}
\index{Energy Frontier!EF06}
\input energy/pdf

\subsection{Hot, dense QCD}
\label{sec:hot}
\index{Energy Frontier!EF07}
\input energy/hot
\index{Theory Frontier!TF07|)}

\subsection{Higgs boson as a portal to new physics}
\label{sec:composite}
\index{Energy Frontier!EF02}
\index{Theory Frontier!TF08}
\input energy/composite
\index{Energy Frontier|)}

\index{Cosmic Frontier|(}\index{Theory Frontier!TF09|(}
\section{Cosmic Frontier}
\label{sec:cosmic}
\input cosmic

\index{Theory Frontier!TF09|)}\index{Cosmic Frontier|)}

\index{Theory Frontier|(}
\section{Theory Frontier}
\label{sec:theory}
\input theory/intro

\subsection{Supersymmetry and gravity}
\label{sec:susy}
\index{Theory Frontier!TF01}
\input theory/susy

\subsection{Effective field theory techniques}
\label{sec:eft}
\index{Theory Frontier!TF02}
\input theory/eft

\subsection{Conformal field theory}
\label{sec:cft}
\index{Theory Frontier!TF03}
\input theory/cft
\index{Theory Frontier|)}

\section{Summary \& Outlook}
\label{sec:outlook}
\input outlook


\appendix
\index{USQCD|(}
\section{USQCD Collaboration}
\label{app:usqcd}
\input usqcd
\index{USQCD|)}

\index{Computational Frontier|(}
\index{Computing facilities|(}
\section{Computing Landscape}
\label{app:compute}
\input landscape
\index{Computing facilities|)}
\index{Computational Frontier|)}

\section{List of Snowmass Whitepapers}
\label{app:snow-white}

For easy reference, we list here all contributions to the Snowmass Study that mention lattice QCD and/or lattice~BSM.
\begin{itemize}
\item Properties of $B$ and $D$ mesons: lattice~\cite{Boyle:2022uba}, 
    phenomenology~\cite{Charles:2020dfl, Altmannshofer:2022hfs, Cheng:2022tog, Guadagnoli:2022oxk},
    summaries of experiments~\cite{LHCb:2022ine, BelleII:2022yoy, BESIII:2022mxl};
    new experimental tools~\cite{Sibidanov:2022gvb, Bhattacharya:2022cna}.
\item Properties of kaons and hyperons~\cite{Blum:2022wsz, Aebischer:2022vky, NA62KLEVER:2022nea, Goudzovski:2022vbt}.
\item Properties of $\eta^{(\prime)}$ mesons~\cite{REDTOP:2022slw}.
\item Standard-Model prediction of muon~$g-2$~\cite{Colangelo:2022jxc}.
\item Electric dipole moments~\cite{Alarcon:2022ero, Blinov:2022tfy, Alexander:2022rmq}
\item Baryon-number violation: proton decay and $n$-$\bar{n}$ oscillations~\cite{Dev:2022jbf}.
\item Neutrinoless double-beta decay ($0\nu\beta\beta$)~\cite{Cirigliano:2022oqy}.
\item Hadron spectroscopy~\cite{Bulava:2022ovd, Brambilla:2022ura}.
\item Neutrino-nucleus scattering~\cite{Ruso:2022qes, Ankowski:2022thw, Campbell:2022qmc}
\item Determination of the strong coupling $\alpha_s$~\cite{dEnterria:2022hzv}
\item Parton distribution functions: lattice~\cite{Constantinou:2022yye};
    phenomenology and experiment \cite{Hou:2022sdf, Amoroso:2022eow}.
\item The Electron Ion Collider (EIC)~\cite{Khalek:2022bzd}.
\item Composite Higgs models~\cite{Banerjee:2022xmu}
\item Axion dark matter \cite{Adams:2022pbo} (connection to EDMs~\cite{Blinov:2022tfy})
\item Supersymmetric lattice gauge theories~\cite{Catterall:2022qzs}.
\item Generalized symmetries~\cite{Cordova:2022ruw}.
\item Conformal field theories~\cite{Poland:2022qrs}.
\item Machine learning in lattice field theory~\cite{Boyda:2022nmh}.
\item Quantum information science and quantum computing, including quantum simulation of lattice field
    theories~\cite{Humble:2022vtm, Faulkner:2022mlp, Meurice:2022xbk, Bauer:2022hpo}.
\item Computing needs of numerical lattice gauge theory~\cite{Boyle:2022ncb}.
\end{itemize}
The USQCD Collaboration would be grateful to anyone bringing omissions to our attention via email to
\href{mailto:ask@fnal.gov?Snowmass WP}{ask@fnal.gov}.


\acknowledgments
\input acknowledgments


\bibliography{21LQCD,flag,usqcd-wp,snow-wp}


\clearpage
\twocolumngrid
\printindex

\end{document}

%% file: usqcd_ec.tex
\author{\href{mailto:ask@fnal.gov?USQCD Snowmass WP}{Andreas~S.~Kronfeld}}
\thanks{USQCD Whitepaper Coordinator (2019)}
\affiliation{Theory Division, Fermi National Accelerator Laboratory, Batavia, IL 60510, USA}
\author{Tanmoy~Bhattacharya}
\thanks{USQCD Whitepaper Author (2019)}
\affiliation{Group T-2, Los Alamos National Laboratory, Los Alamos, NM 87545, USA}

\author{Thomas~Blum}
\thanks{USQCD Whitepaper Author (2019)}
\affiliation{Department of Physics, University of Connecticut, Storrs, CT 06269, USA}
\affiliation{RIKEN BNL Research Center, Brookhaven National Laboratory, Upton, NY 11973, USA}

\author{Norman~H.~Christ}
\thanks{USQCD Whitepaper Author (2019)}
\affiliation{Department of Physics, Columbia University, New York, NY 10027, USA}

\author{Carleton~DeTar}
\affiliation{Department of Physics and Astronomy, University of Utah, \\ Salt Lake City, Utah 84112, USA}

\author{William~Detmold}
\thanks{USQCD Whitepaper Coordinator (2019)}
\affiliation{Center for Theoretical Physics, Massachusetts Institute of Technology, \\ Cambridge, MA 02139, USA}
\affiliation{The NSF Institute for Artificial Intelligence and Fundamental Interactions}

\author{Robert~Edwards}
\thanks{USQCD Whitepaper Coordinator (2019)}
\affiliation{Theory Center, Thomas Jefferson National Accelerator Facility, \\ Newport News, VA 23606, USA}

\author{Anna~Hasenfratz}
\thanks{USQCD Whitepaper Coordinator (2019)}
\affiliation{Department of Physics, University of Colorado, Boulder, CO 80309, USA}

\author{Huey-Wen~Lin}
\thanks{USQCD Whitepaper Author (2019)}
\affiliation{Department of Physics and Astronomy, Michigan State University, \\ East Lansing, MI 48824, USA}

\author{Swagato~Mukherjee}
\thanks{USQCD Whitepaper Coordinator (2019)}
\affiliation{Physics Department, Brookhaven National Laboratory, Upton, NY 11973, USA}

\author{Konstantinos~Orginos}
\thanks{USQCD Whitepaper Author (2019)}
\affiliation{Department of Physics, College of William \& Mary, Williamsburg, VA 23187, USA}
\affiliation{Theory Center, Thomas Jefferson National Accelerator Facility, \\ Newport News, VA 23606, USA}

\collaboration{USQCD Executive \rule[-12pt]{0pt}{28pt}Committee}
\noaffiliation

%% file: white_coord.tex
\author{Richard~Brower}
\thanks{USQCD Whitepaper Coordinator (2019)}
\affiliation{Department of Physics and Center for Computational Science, Boston University, \\ Boston, MA 02215, USA}

\author{Vincenzo~Cirigliano}
\thanks{USQCD Whitepaper Coordinator (2019)}
\affiliation{Institute of Nuclear Theory, University of Washington, Seattle, WA 98195 USA}

\author{Zohreh~Davoudi}
\thanks{USQCD Whitepaper Coordinator (2019)}
\affiliation{Department of Physics and Maryland Center for Fundamental Physics, University of Maryland, College Park, MD 20742, USA}

\author{Bálint~Jóo}
\thanks{USQCD Whitepaper Coordinator (2019)}
\affiliation{Oak Ridge Leadership Computing Facility, Oak Ridge National Laboratory, \\ Oak Ridge, TN 37831, USA}

\author{Chulwoo~Jung}
\thanks{USQCD Whitepaper Coordinator (2019)}
\affiliation{Physics Department, Brookhaven National Laboratory, Upton, NY 11973, USA}

\author{Christoph~Lehner}
\thanks{USQCD Whitepaper Coordinator (2019)}
\affiliation{Physics Department, Brookhaven National Laboratory, Upton, NY 11973, USA}
\affiliation{Fakultät für Physik, Universität Regensburg, D-93040, Regensburg, Germany}

\author{Stefan~Meinel}
\thanks{USQCD Whitepaper Coordinator (2019)}
\affiliation{Department of Physics, University of Arizona, Tucson, AZ 85721, USA}

\author{Ethan~T.~Neil}
\thanks{USQCD Whitepaper Coordinator (2019)}
\affiliation{Department of Physics, University of Colorado, Boulder, CO 80309, USA}

\author{Peter~Petreczky}
\thanks{USQCD Whitepaper Coordinator (2019)}
\affiliation{Physics Department, Brookhaven National Laboratory, Upton, NY 11973, USA}

\author{David~G.~Richards}
\thanks{USQCD Whitepaper Coordinator (2019)}
\affiliation{Theory Center, Thomas Jefferson National Accelerator Facility, \\ Newport News, VA 23606, USA}

%% file: white_auth.tex
\author{Alexei~Bazavov}
\thanks{USQCD Whitepaper Author (2019)}
\affiliation{Department of Physics and Astronomy, Michigan State University, \\ East Lansing, MI 48824, USA}
\affiliation{Department of Computational Mathematics, Science, and Engineering,
Michigan State University, East Lansing, MI 48824, USA}

\author{Simon~Catterall}
\thanks{USQCD Whitepaper Author (2019)}
\affiliation{Department of Physics, Syracuse University, Syracuse, NY 13244, USA}

\author{Jozef~J.~Dudek}
\thanks{USQCD Whitepaper Author (2019)}
\affiliation{Department of Physics, College of William \& Mary, Williamsburg, VA 23187, USA}

\author{Aida~X.~El-Khadra}
\thanks{USQCD Whitepaper Author (2019)}
\affiliation{Department of Physics, University of Illinois Urbana-Champaign, Urbana, IL 61801, USA}

\author{Michael~Engelhardt}
\thanks{USQCD Whitepaper Author (2019)}
\affiliation{Department of Physics, New Mexico State University, Las Cruces, NM 88003, USA}

\author{George~T.~Fleming}
\thanks{USQCD Whitepaper Author (2019)}
\affiliation{Department of Physics, Yale University, New Haven, CT 06437, USA}

\author{Joel~Giedt}
\thanks{USQCD Whitepaper Author (2019)}
\affiliation{Department of Physics, Applied Physics and Astronomy, Rensselaer Polytechnic Institute, \\ Troy, NY 12065, USA}

\author{Rajan~Gupta}
\thanks{USQCD Whitepaper Author (2019)}
\affiliation{Group T-2, Los Alamos National Laboratory, Los Alamos, NM 87545, USA}

\author{Maxwell~T.~Hansen}
\thanks{USQCD Whitepaper Author (2019)}
\affiliation{School of Physics and Astronomy, University of Edinburgh, Edinburgh EH9~3FD, United Kingdom}

\author{Taku~Izubuchi}
\thanks{USQCD Whitepaper Author (2019)}
\affiliation{Physics Department, Brookhaven National Laboratory, Upton, NY 11973, USA}

\author{Frithjof~Karsch}
\thanks{USQCD Whitepaper Author (2019)}
\affiliation{Physics Department, Brookhaven National Laboratory, Upton, NY 11973, USA}
\affiliation{Fakultät für Physik, Universität Bielefeld, D-33615 Bielefeld, Germany}

\author{Jack~Laiho}
\thanks{USQCD Whitepaper Author (2019)}
\affiliation{Department of Physics, Syracuse University, Syracuse, NY 13244, USA}

\author{Keh-Fei Liu}
\thanks{USQCD Whitepaper Author (2019)}
\affiliation{Department of Physics and Astronomy, University of Kentucky, \\ Lexington, KY 40508, USA}

\author{Aaron~S.~Meyer}
\thanks{USQCD Whitepaper Author (2019)}
\affiliation{Department of Physics, University of California, Berkeley, CA, 94720, USA}

\author{Enrico~Rinaldi}
\thanks{USQCD Whitepaper Author (2019)}
\affiliation{Department of Physics, University of Michigan, Ann Arbor, MI 48109, USA}

\author{Martin~Savage}
\thanks{USQCD Whitepaper Author (2019)}
\affiliation{Department of Physics, University of Washington, Seattle, WA 98195, USA}

\author{David~Schaich}
\thanks{USQCD Whitepaper Author (2019)}
\affiliation{Department of Mathematical Sciences, University of Liverpool, \\ Liverpool L69~7ZL, United Kingdom}

\author{Phiala~E.~Shanahan}
\thanks{USQCD Whitepaper Author (2019)}
\affiliation{Center for Theoretical Physics, Massachusetts Institute of Technology, \\ Cambridge, MA 02139, USA}
\affiliation{The NSF Institute for Artificial Intelligence and Fundamental Interactions}

\author{Stephen~R.~Sharpe}
\thanks{USQCD Whitepaper Author (2019)}
\affiliation{Department of Physics, University of Washington, Seattle, WA 98195, USA}

\author{Raza~Sufian}
\thanks{USQCD Whitepaper Author (2019)}
\affiliation{Theory Center, Thomas Jefferson National Accelerator Facility, \\ Newport News, VA 23606, USA}

\author{Sergey~Syritsyn}
\thanks{USQCD Whitepaper Author (2019)}
\affiliation{Department of Physics and Astronomy, Stony Brook University, \\ Stony Brook, NY 11794, USA}

\author{Ruth~S.~Van~de~Water}
\thanks{USQCD Whitepaper Author (2019)}
\affiliation{Theory Division, Fermi National Accelerator Laboratory, Batavia, IL 60510, USA}

\author{Michael~L.~Wagman}
\thanks{USQCD Whitepaper Author (2019)}
\affiliation{Theory Division, Fermi National Accelerator Laboratory, Batavia, IL 60510, USA}

\author{Evan~Weinberg}
\thanks{USQCD Whitepaper Author (2019)}
\affiliation{NVIDIA Corporation, Santa Clara, CA 95050, USA}

\author{Oliver~Witzel}
\thanks{USQCD Whitepaper Author (2019)}
\affiliation{Fakultät IV/Department Physik, Universität Siegen, D-57068 Siegen, Germany}

%% file: usqcd_members.tex
\author{Christopher~Aubin}
\affiliation{Department of Physics \& Engineering Physics, Fordham University, \\ Bronx, NY 10458, USA}

\author{Peter~Boyle}
\affiliation{Physics Department, Brookhaven National Laboratory, Upton, NY 11973, USA}

\author{Shailesh~Chandrasekharan}
\affiliation{Department of Physics, Duke University, Durham NC 27708, USA}

\author{Ian~C.~Clo\"et}
\affiliation{Physics Division, Argonne National Laboratory, Lemont, IL 60439, USA}

\author{Martha~Constantinou}
\affiliation{Department of Physics, Temple University, Philadelphia, PA 19122, USA}

\author{Kimmy~Cushman}
\affiliation{Department of Physics, Yale University, New Haven, CT 06437, USA}

\author{Thomas~DeGrand}
\affiliation{Department of Physics, University of Colorado, Boulder, CO 80309, USA}

\author{Zoltan~Fodor}
\affiliation{Department of Physics, Penn State University, University Park, PA 16802, USA}
\affiliation{Department of Physics, Wuppertal University, 42119 Wuppertal, Germany}
\affiliation{JSC, Forschungszentrum Jülich, 52428 Jülich, Germany}
\affiliation{Department of Physics, University of California at San Diego, La Jolla, CA 92093, USA}

\author{Sam~Foreman}
\affiliation{Computational Science Division, Argonne National Laboratory, Lemont, IL 60439, USA}
\affiliation{Leadership Computing Facility, Argonne National Laboratory, Lemont, IL 60439, USA}

\author{Steven~Gottlieb}
\affiliation{Department of Physics, Indiana University, Bloomington, IN 47405, USA}

\author{Daniel~Hoying}
\affiliation{Department of Computational Mathematics, Science, and Engineering,
Michigan State University, East Lansing, MI 48824, USA}

\author{Yong-Chull~Jang}
\affiliation{Department of Physics, Columbia University, New York, NY 10027, USA}

\author{William~I.~Jay}
\affiliation{Center for Theoretical Physics, Massachusetts Institute of Technology, \\ Cambridge, MA 02139, USA}

\author{Xiao-Yong~Jin}
\affiliation{Computational Science Division, Argonne National Laboratory, Lemont, IL 60439, USA}
\affiliation{Leadership Computing Facility, Argonne National Laboratory, Lemont, IL 60439, USA}

\author{Christopher~Kelly}
\affiliation{Computational Science Initiative, Brookhaven National Laboratory, \\ Upton, NY 11973, USA}

\author{Julius~Kuti}
\affiliation{Department of Physics, University of California at San Diego, La Jolla, CA 92093, USA}

\author{Henry~Lamm}
\affiliation{Theory Division, Fermi National Accelerator Laboratory, Batavia, IL 60510, USA}

\author{Meifeng~Lin}
\affiliation{Computational Science Initiative, Brookhaven National Laboratory, \\ Upton, NY 11973, USA}

\author{Yin~Lin}
\affiliation{Center for Theoretical Physics, Massachusetts Institute of Technology, \\ Cambridge, MA 02139, USA}

\author{Andrew~T.~Lytle}
\affiliation{Department of Physics, University of Illinois Urbana-Champaign, Urbana, IL 61801, USA}

\author{Paul~Mackenzie}
\affiliation{Theory Division, Fermi National Accelerator Laboratory, Batavia, IL 60510, USA}

\author{Jeffrey~Mandula}
\affiliation{Department of Physics, University of Washington, Seattle, WA 98195, USA}

\author{Yannick~Meurice}
\affiliation{Department of Physics and Astronomy, University of Iowa, \\ Iowa City, IA 52242, USA}

\author{Christopher~Monahan}
\affiliation{Department of Physics, College of William \& Mary, Williamsburg, VA 23187, USA}

\author{Colin~Morningstar}
\affiliation{Department of Physics, Carnegie Mellon University, Pittsburgh, PA 15213 USA}

\author{James~C.~Osborn}
\affiliation{Computational Science Division, Argonne National Laboratory, Lemont, IL 60439, USA}
\affiliation{Leadership Computing Facility, Argonne National Laboratory, Lemont, IL 60439, USA}

\author{Sungwoo~Park}
\affiliation{Theory Center, Thomas Jefferson National Accelerator Facility, \\ Newport News, VA 23606, USA}

\author{James~N.~Simone}
\affiliation{Scientific Computing Division, Fermi National Accelerator Laboratory, \\ Batavia, IL 60510, USA}
\affiliation{Theory Division, Fermi National Accelerator Laboratory, Batavia, IL 60510, USA}

\author{Alexei~Strelchenko}
\affiliation{Scientific Computing Division, Fermi National Accelerator Laboratory, \\ Batavia, IL 60510, USA}

\author{Masaaki~Tomii}
\affiliation{Department of Physics, University of Connecticut, Storrs, CT 06269, USA}

\author{Alejandro~Vaquero}
\affiliation{Department of Physics and Astronomy, University of Utah, \\ Salt Lake City, Utah 84112, USA}
\affiliation{Departamento de Física Teórica, Universidad de Zaragoza, Zaragoza 50009, Spain}

\author{Pavlos~Vranas}
\affiliation{Physical and Life Sciences, Lawrence Livermore National Laboratory, \\ Livermore, CA 94550, USA}
\affiliation{Nuclear Science Division, Lawrence Berkeley National Laboratory, \\ Berkeley, CA 94720, USA}

\author{Bigeng~Wang}
\affiliation{Department of Physics and Astronomy, University of Kentucky, \\ Lexington, KY 40508, USA}

\author{Walter~Wilcox}
\affiliation{Department of Physics, Baylor University, Waco, TX 76798, USA}

\author{Boram~Yoon}
\affiliation{Computer, Computational, and Statistical Sciences Division, Los Alamos National Laboratory, Los Alamos, NM 87545, USA}

\author{Yong~Zhao}
\affiliation{Physics Division, Argonne National Laboratory, Lemont, IL 60439, USA}


%% file: exec.tex
\vspace*{-6pt}
Lattice field theory provides a mathematically rigorous definition of quantum field theory, including gauge theories.
The rigor provides a platform for computation of strongly-coupled gauge theories, not only quantum chromodynamics (QCD) at
long-distances but also strongly-coupled sectors that might lie beyond the Standard Model (BSM) of particle physics.
Indeed, the interpretation of many experiments in particle physics, nuclear physics, and astrophysics relies, often crucially, on
results from lattice QCD or BSM.


In quark-flavor physics, lattice QCD is essential for grounding theoretical predictions.
Together with experimental measurements, lattice QCD is used to determine many of the fundamental parameters of the Standard Model's
flavor sector: five out of six quark masses and the three mixing angles and $CP$-violating phase of the Cabibbo-Kobayashi-Maskawa
(CKM) matrix.
Further Standard-Model predictions requiring lattice QCD are for processes that could reveal signatures of new physics.
This area of lattice QCD has become a precision science, as has the determination of the strong coupling, $\alpha_s$, which is
key in many areas, for example for understanding Higgs-boson decay to gluons.

Lattice-QCD calculations of the hadronic contributions to the anomalous magnetic moment of the muon are another area for which
precision is both mandatory and achievable.
While these contributions can be estimated via a combination of certain measurements and general theoretical concepts, a direct
\emph{ab initio} calculation from QCD is desirable to confirm robustly the tantalizingly large discrepancy between experiment and
the Standard Model.

Other indirect searches for new physics are complementary to quark-flavor physics and the muon anomalous magnetic moment.
Lattice QCD is an essential tool for understanding newly discovered hadrons with quark content different from the usual baryons and
mesons.
Lattice-QCD calculations are needed to interpret bounds on electric dipole moments, proton decay, neutron-antineutron oscillations,
neutrinoless double-beta decay, and charged-lepton flavor violation.
These topics all entail matrix elements of the nucleon, usually in concert with nuclear many-body theory.

The theory of the neutrino-nucleus cross section also requires these two ingredients.
To first approximation, neutrino-nucleus scattering consists of a neutrino-nucleon interaction followed by propagation of the struck
nucleon, and any produced hadrons, in a nuclear medium.
Having results with full error budgets for nucleon-level quantities will solidify theoretical treatments of this complicated
process.
Any improvement in nuclear modeling will pay dividends in extending the power of neutrino-oscillation experiments.
Lattice QCD can also be used to provide information needed in nuclear many-body theory, via calculations of multihadron
interactions.

The LHC experiments (among many others) rely on parton distribution functions for cross sections, both within and beyond the
Standard Model.
This is an area of rapid development, with precise lattice-QCD calculations foreseeable in the coming decade.
Lattice BSM is used to understand the spectrum of composite Higgs models and other strongly-coupled theories that motivate LHC
searches.
Composite models are also a popular explanation for dark matter, requiring spectroscopy from lattice gauge theory.
Lattice QCD is needed to understand the properties of the axion, a field introduced to explain the absence of observed strong $CP$
violation, and a dark matter candidate in its own right.
Lattice field theory is also increasingly pertinent to theoretical physics in general, for example in strongly-coupled
supersymmetric theories and in conformal field theories studied in several fields.

This contribution to Snowmass is drawn from seven whitepapers prepared by the
USQCD Collaboration in 2019~\cite{Detmold:2018qcd, Bazavov:2018qcd, Lehner:2018qcd, Kronfeld:2018qcd, Davoudi:2018qcd,
Brower:2018qcd, Joo:2018qcd}, with numerous updates as appropriate.

%% file: intro_main.tex
This contribution to the U.S.\ Community Study on the Future of Particle Physics (aka ``Snowmass'') outlines the physics program of
the USQCD Collaboration, as it pertains to particle physics.
The USQCD Collaboration\footnote{%
More on USQCD can be found in Appendix~\ref{app:usqcd}.} %
is a federation of scientific collaborations and individuals engaged in computational research on lattice gauge theory and other
lattice field theories, principally quantum chromodynamics (QCD).
It is a steward of infrastructure---both hardware and software---for lattice-QCD calculations.
In 2019, USQCD published seven whitepapers to build the case for funding mid-sized computing facilities.
They spell out a comprehensive program for the coming 5--10 years in particle and nuclear physics~\cite{Detmold:2018qcd,
Bazavov:2018qcd, Lehner:2018qcd, Kronfeld:2018qcd, Davoudi:2018qcd, Brower:2018qcd} and a perspective on
computing~\cite{Joo:2018qcd}.
Most of the material in this document is drawn from those works, with updates to the references and, in cases where developments
have been rapid, exposition and perspective.

The role of QCD in particle physics is easy to summarize: QCD is everywhere.
The LHC collides beams of protons; the SuperKEKB and BEPC~II $e^+e^-$ colliders are designed to produce flavored hadrons;
neutrino-oscillation experiments and searches for dark matter or charged-lepton-flavor violation use nuclear targets; collider
experiments keep discovering new (and often weird) hadrons.
Even purely leptonic experiments, if they are sensitive enough, probe virtual hadrons---the anomalous magnetic moment of the muon,
$g-2$, is the foremost example.
To interpret these experiments, it is necessary to have some level of control over hadrons and nuclei.
In some cases sub-percent total uncertainty is required and, as discussed below in several sections, increasingly feasible.

The techniques of numerical lattice gauge theory are not limited to QCD.
Models beyond the Standard Model (BSM) of particle physics often include confining gauge theories.
For example, dark matter could consist of hadron-like bound states.
The Higgs boson could well be composite, although if so, the dynamics will have to differ from QCD; in particular, it seems
necessary that the gauge coupling runs more slowly.
Various aspects of composite dark matter and composite Higgs are well suited to numerical lattice gauge theory.
A simple and general example is to examine a viable model to see whether the transition from high temperatures to low generates 
gravitational waves.
Another question is whether a confining gauge theory could yield a light scalar with properties of the  Standard-Model Higgs boson, 
with the rest of its spectrum (just) out of reach of the LHC.

Stemming from its focus on QCD, the lattice community's relevance spans funding agencies' programs in particle physics and in
nuclear physics.
In the United States, this means the DOE\index{DOE} Office of High Energy Physics\index{DOE!HEP Office} (HEP) and Office of Nuclear
Physics\index{DOE!NP Office} (NP), as well as the NSF Division of Physics\index{NSF PHY} programs in Elementary Particle Physics
(EPP) and in Nuclear Physics.
In addition to DOE and NSF funding for researchers, the DOE supports infrastructure for numerical lattice gauge theory.
DOE~HEP and NP support medium-scale computer clusters at Brookhaven National Laboratory, Fermi National Accelerator Laboratory, and
Thomas Jefferson National Accelerator Facility.
These computer clusters support large, albeit not the largest, lattice-QCD and -BSM projects.
USQCD's allocation process has an excellent record of supporting innovative projects that might not sway a multidisciplinary
allocation committee.
(Early work on muon $g-2$ started this way.) %
USQCD also uses its stewardship of these clusters to foster the careers of junior researchers in lattice gauge theory.
(Several by-now mid-career scientists established their reputations this way.) %
The largest-scale lattice-QCD calculations run on the leadership-class supercomputers available at several NSF and DOE facilities.
The high precision demanded by the Muon $g-2$ Experiment and several quark-flavor experiments is not possible without computing
campaigns spanning the clusters and the leadership-class facilities.

Computational science cannot prosper without code and algorithm development.
The DOE Office of Advanced Scientific Computing Research\index{DOE!ASCR} (ASCR) supports development of QCD software and algorithms
for the next-generation supercomputers coming on line this year via the Exascale Computing Project\index{Exascale Computing Project}
(ECP).
These machines---Frontier at Oak Ridge National Laboratory and Aurora at Argonne National Laboratory---are designed to be capable of
$10^{18}$ double-precision floating-point operations per second, or 1~exaflop/s, which is comparable to the human
brain~\cite{wiki:exascale}.
(ECP also supports application development in accelerator science, computational cosmology, and many other
subjects~\cite{Alexander:2020gty}.) %
ASCR and the other DOE offices together fund programs in Scientific Discovery through Advanced Computing\index{DOE!SciDAC}
(SciDAC): for lattice gauge theory, these are collaborations between ASCR and HEP as well as ASCR and NP.
SciDAC grants support research into algorithms for lattice gauge theory, in collaboration with applied mathematicians.

At the community level, lattice-gauge-theory research plays important roles in (at least) three units of the American Physical
Society\index{APS} (APS): the Division of Particles and Fields\index{APS!DPF} (DPF), the Division of Nuclear Physics\index{APS!DNP}
(DNP), and the Topical Group on Hadronic Physics\index{APS!GHP} (GHP).
Lattice methodology has an even broader influence.
For example, the hybrid Monte Carlo (HMC) algorithm~\cite{Duane:1987de} invented for full QCD with dynamical quarks is widely 
used\footnote{%
In other fields, HMC is often thought to stand for Hamiltonian Monte Carlo.} %
in Bayesian inference~\cite{stan:2022}, cosmological signal processing~\cite{Ensslin:2008iu}, and condensed-matter
physics~\cite{Beyl:2017kwp}.

Given this breadth of interests, it is impossible to cover everything.
We limit the material in this contribution to the intersection of the Snowmass topical groups and the physics program of the 2019
USQCD whitepapers~\cite{Detmold:2018qcd, Bazavov:2018qcd, Lehner:2018qcd, Kronfeld:2018qcd, Davoudi:2018qcd, Brower:2018qcd}.
Many contributions to Snowmass cover these topics and are cited in the main text.
Interesting topics not covered here include the following.
A large part of the USQCD physics program supports nuclear physics; a similar document emphasizing nuclear physics will be
prepared for the next long-range plan of the Nuclear Science Advisory Committee\index{NSAC} (NSAC).
\index{Computational Frontier|(}%
Several USQCD members engaged in the ECP discuss lattice simulations in a Snowmass contribution to the Computing
Frontier~\cite{Boyle:2022ncb},\index{Computational Frontier!CompF2} building on Ref.~\cite{Joo:2018qcd}, so that material is not
repeated here.
The fast-moving topics of machine learning,\index{Computational Frontier!CompF3} quantum information
science,\index{Theory Frontier!TF10} and quantum computing\index{Computational Frontier!CompF6}---discussed briefly in
Ref.~\cite{Joo:2018qcd}---are not yet central to USQCD collaboration activities.
Several USQCD members have, however, co-authored contributions to Snowmass on these topics~\cite{Boyda:2022nmh, Humble:2022vtm,
Faulkner:2022mlp, Meurice:2022xbk, Bauer:2022hpo}.
\index{Computational Frontier|)}

This contribution is organized as follows.
Section~\ref{sec:rare} covers the Rare and Precision Frontier.
It discusses how USQCD plans to continue to sharpen the search for new physics in the quark-flavor sector, to complete calculations
of the hadronic contributions to muon $g-2$ with the required precision on the timescale of Fermilab E989, and some aspects of a
program of precision nucleon matrix elements with comprehensive error budgets (i.e., electric dipole moments, proton decay,
$n$-$\bar n$ oscillations, and nucleon matrix elements involved in muon-to-electron conversion).
Following Snowmass organization, hadron spectroscopy is also in Sec.~\ref{sec:rare}.
Section~\ref{sec:nu} covers the Neutrino Physics Frontier, including the theory of neutrinos.
Here again nucleon matrix elements appear, but now they must be folded into nuclear many-body theory~\cite{Drischler:2019xuo,
Ruso:2022qes} and event generators~\cite{Ruso:2022qes, Campbell:2022qmc}.
Sections~\ref{sec:energy} and~\ref{sec:cosmic} cover the Energy Frontier and Cosmic Frontier, respectively.
In addition further nucleon matrix elements, these frontiers also profit from the exploration of gauge theories other than QCD.
Again following Snowmass organization, hot, dense QCD appears in Sec.~\ref{sec:energy}.
Section~\ref{sec:theory} is short, covering topics at the Theory Frontier with a less obvious connection to the HEP experimental
program.
Appendices give some background on the USQCD Collaboration, describe the landscape for computing (in the U.S.), and list 
contributions to Snowmass mentioning lattice QCD.

%% file: rare/intro.tex
The rare and precision frontier covers a broad range of experiments and correspondingly broad range of lattice-QCD calculations.
The U.S.\ lattice community has been influential in this area~\cite{Lehner:2018qcd, Davoudi:2018qcd, Detmold:2018qcd}.
Among the topics covered in this frontier, hadronic spectroscopy and decay and mixing properties of $B$, $D$, and $K$ mesons are
especially well developed.
Recent years have witnessed rapid development in computing the hadronic contributions to the anomalous magnetic moment of the muon,
namely the amplitude for hadronic vacuum polarization and for hadronic light-by-light scattering.
Calculations of nucleon properties related to rare processes should be mentioned~\cite{Davoudi:2018qcd, Detmold:2018qcd}.
As a rule~\cite{Parisi:1983ae, Lepage:1989hd}, nucleon correlation functions are noisier than their meson counterparts, so the
results are less precise.
Precision is less of an issue, though, because the corresponding experiments are still at the stage of setting limits.
Some examples are matrix elements for nucleon electric dipole moments, proton-decay form factors, and the first calculation of
operators that induce neutron-antineutron oscillations.

%% file: rare/weak.tex
Quark-flavor physics is, perhaps, the area of particle physics in which lattice QCD has had its greatest
impact~\cite{Lehner:2018qcd}.
In many cases, the way QCD enters the Standard-Model expression for a measurable rate is very simple.
Schematically,
\begin{equation}
    d\Gamma =
    \left(\begin{array}{c}    \text{CKM}    \\ \text{factor} \end{array}\right)
    \left(\begin{array}{c} \text{kinematic} \\ \text{factor} \end{array}\right)
    \left(\begin{array}{c}    \text{QCD}    \\ \text{factor} \end{array}\right) +
    \left[\begin{array}{c}    \text{BSM}    \\ \text{term}   \end{array}\right].
\end{equation}
If the BSM term can be assumed to be small, as in processes that proceed at the tree level of the electroweak interactions, the
combination of measurements and (lattice) QCD calculations can be used to determine the Cabibbo-Kobayashi-Maskawa (CKM) matrix.
Correspondingly, if the CKM matrix is known and the Standard-Model contribution is suppressed, the same approach can be used to put
constraints on physics beyond the Standard Model.
Indeed, in many cases a single matrix element, or set of related matrix elements, is needed to interpret a single measurement.
For example, in leptonic decays of charged mesons the QCD factor is parametrized by a single number, known as the decay constant,
and the decay vertex entails a single CKM matrix element.

An important by-product of these studies are determinations of the quark masses, which are relevant to Higgs-boson decay.
They are discussed in Sec.~\ref{sec:asmq} together with the strong coupling $\alpha_s$, which is relevant to Higgs-boson production 
and decay.
Of course, as the fundamental parameters of QCD, $\alpha_s$ and the quark masses are important in many areas, not least the 
quark-flavor physics discussed in this section.

\subsubsection{Weak decays of \texorpdfstring{$b$}{b} and \texorpdfstring{$c$}{c} quarks}
\label{sec:weak-b+c}
\index{Rare and Precision Frontier!RF1}

For many years, lattice QCD has been a vital contributor to $B$ and $D$ physics~\cite{Lehner:2018qcd}.
Here the experiments aim to (over)determine the CKM quark-mixing matrix and search for new sources of $CP$ violation (which are
needed to explain the observed baryon asymmetry of the universe) by measuring $B$- and $D$-meson decay rates, $CP$ asymmetries, and
mixing frequencies of neutral mesons.
Amplitudes of leptonic and semileptonic decays and neutral-meson mixing have been a major focus throughout the past 20 years,
and many are now at or approaching a watershed of sub-percent uncertainties.
In particular, the leptonic decay constants of $B$ and $D$ mesons (including those with strangeness) have now reached sub-percent
precision~\cite{Bazavov:2017lyh}, which is beyond the needs of the flavor factories BES~III, Belle~II, and LHCb for the foreseeable
future.

\begin{figure}[b]
    \vspace*{-8pt}
    \includegraphics[height=0.30\textheight]{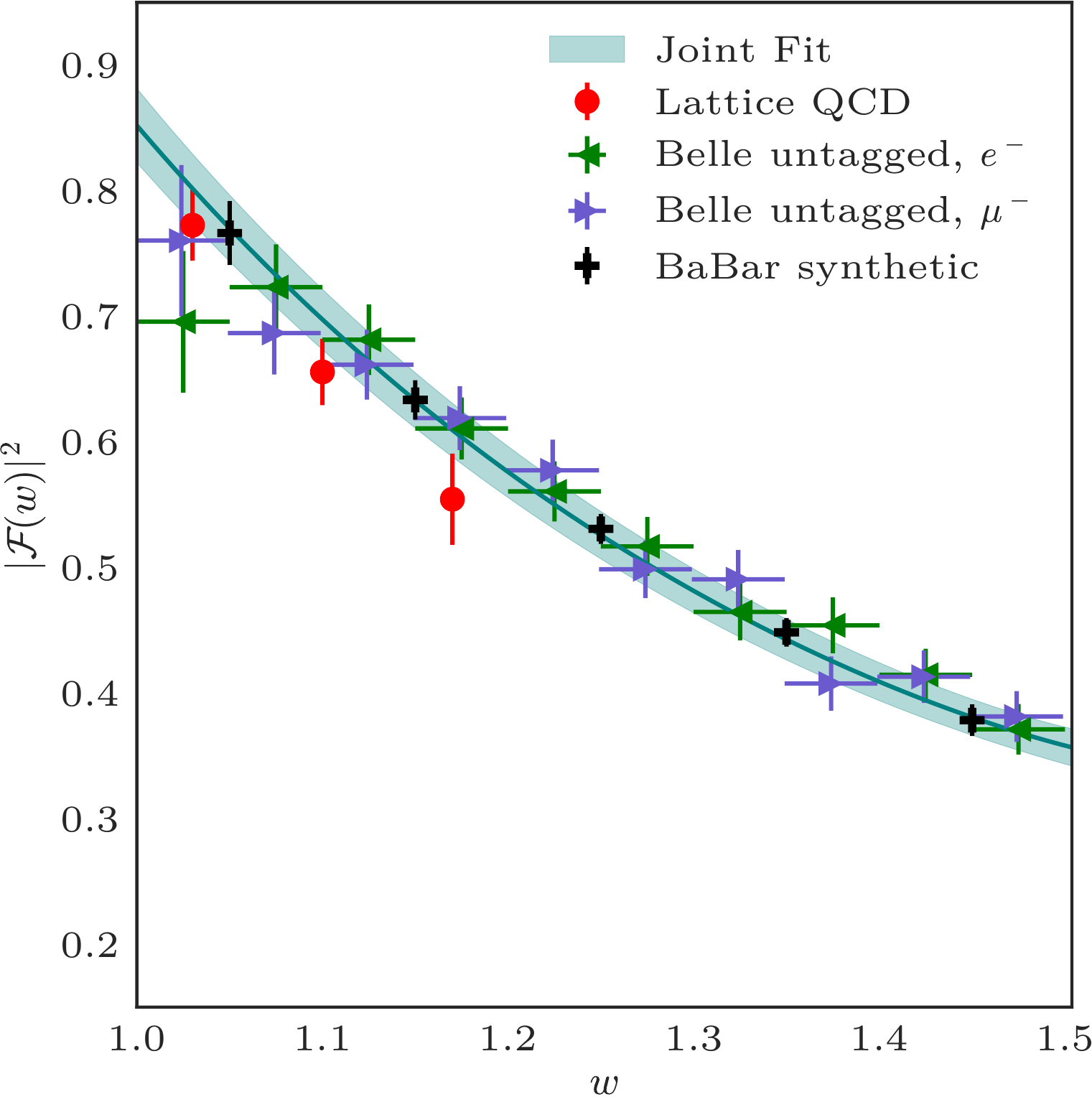} \hfill
    \includegraphics[height=0.30\textheight]{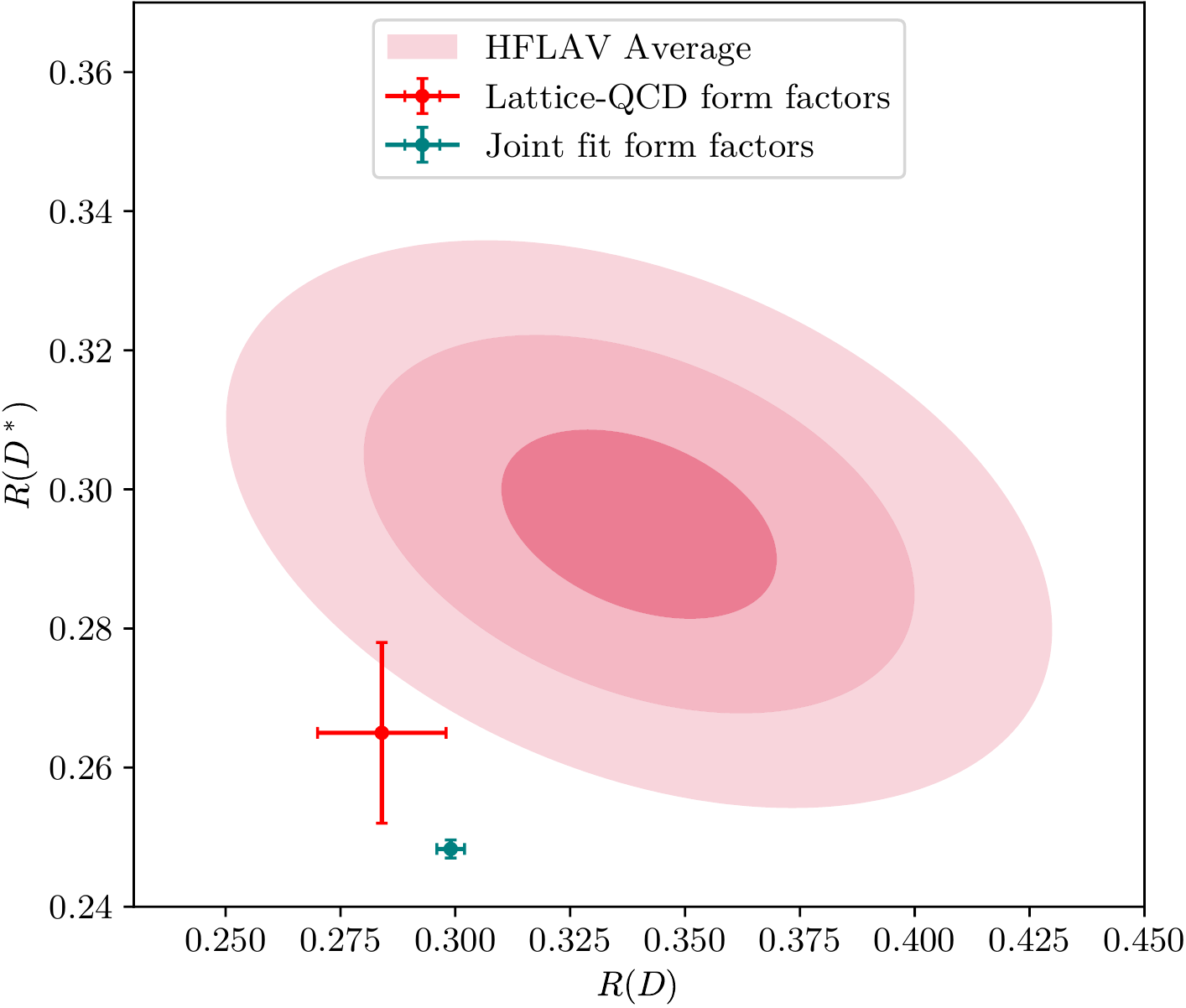}
    \caption[fig:MQ]{Semileptonic decays $B\to D^*\ell\nu$ \cite{FermilabLattice:2021cdg}.
        Left:~A joint fit to experimental and lattice data, which determines the CKM element $|V_{cb}|$.
        The variable $w$ is related to the momentum transfer, $q$, by $q^2=M_B^2+M_{D^*}^2-2wM_BM_{D^*}$.
        Right:~$R(D^*)$ vs.\ $R(D)$.
        An average of experimental measurements (red ellipses) is compared with the Standard-Model prediction either using shape
        information only from lattice QCD (red point with error bars) or using the combined fit of the shape from the $|V_{cb}|$
        determination (green point).
        The discrepancy is several~$\sigma$.}
    \label{fig:semi}
\end{figure}

Semileptonic decays are experimentally more accessible than leptonic decays, for which the rates are suppressed by the square of the
lepton mass.
Combined with lattice QCD, experimental measurements of exclusive semileptonic decays yield the most precise determinations of most
of the elements of the CKM matrix: %
$|V_{us}|$ from $K\to\pi\ell\nu$~\cite{Bazavov:2012cd, Bazavov:2013maa, Bazavov:2018kjg, Boyle:2015hfa},
$|V_{cd}|$ from $D\to\pi\ell\nu$ and $D_s\to K\ell\nu$~\cite{Na:2011mc, Li:2019phv},
$|V_{cs}|$ from $D\to K\ell\nu$~\cite{Na:2010uf, Li:2019phv, Chakraborty:2021qav},
$|V_{ub}|$ from $B\to\pi\ell\nu$~\cite{Flynn:2015mha, Lattice:2015tia, Colquhoun:2015mfa, Colquhoun:2022atw} and
$B_s\to K\ell\nu$~\cite{Bouchard:2014ypa, Flynn:2015mha, Monahan:2018lzv, FermilabLattice:2019ikx}, and
$|V_{cb}|$ from $B_{(s)}\to D_{(s)}^{(*)}\ell\nu$~\cite{Bailey:2014tva, Lattice:2015rga, Na:2015kha, Harrison:2017fmw,
Monahan:2018lzv, FermilabLattice:2021cdg, Kaneko:2021tlw}.
In all cases, not only is the normalization of the corresponding form factors computed, but also the shape, namely the dependence on
the lepton-pair invariant mass~$q^2$.
A sample joint fit to the lattice-QCD and experimental shapes is shown in Fig.~\ref{fig:semi} (left) \cite{FermilabLattice:2021cdg},
with a floating normalization that yields $|V_{cb}|$.
Further information on $|V_{ub}|/|V_{cb}|$ comes from the ratio of $b$-flavored baryon decay distributions using form factors from
lattice~QCD~\cite{Detmold:2015aaa}.
The precision of calculations of semileptonic $D$- and $B$-meson form factors is expected to reach the percent level over the 
coming few years.

Recent measurements of semileptonic decays, in both charged-current and flavor-chang\-ing-neutral-current (FCNC) processes, have
shown an abundance of deviations from the Standard Model~\cite{Altmannshofer:2022hfs, Guadagnoli:2022oxk}.
In lattice QCD, the corresponding calculations~\cite{Bailey:2015dka, Du:2015tda, Bouchard:2013pna, Horgan:2013pva} are an offshoot
of those just discussed for CKM determinations.
A sample result relying on lattice QCD is the $q^2$ distribution for the FCNC decay $B\to K\mu^+\mu^-$~\cite{Du:2015tda}, where $q$
is the four-momentum of the $\mu^+\mu^-$~pair.
The experimental results from BaBar, Belle, CDF, and LHCb lie $1.8\sigma$ ($2.2\sigma$) below the prediction, for $q^2$ below the
$J/\psi$ (above the $\psi(2S)$) resonance.
Note that the uncertainty from the form factors dominates the error in the prediction.
Beyond the differential rate of $B\to K\mu^+\mu^-$, other tensions in $b\to s$ transitions have been observed.
The differential rate and angular distributions of $B\to K^*\mu^+\mu^-$ and $B_s\to\phi\mu^+\mu^-$~\cite{Horgan:2013pva,
LHCb:2021zwz}, and ratios of the branching fractions of $K^{(*)}\mu^+\mu^-$ relative to the $K^{(*)}e^+e^-$ final state all are in
poor agreement with the Standard Model~\cite{LHCb:2017avl, LHCb:2021trn, LHCb:2021lvy}.
In addition, many measurements of the rate for the rare leptonic decay $B_s\to\mu^+\mu^-$ (i.e., $b\bar{s}$ annihilation) have been
in tension with the Standard Model~\cite{Bazavov:2017lyh, LHCb:2020zud, LHCb:2021awg}, although a recent one is
not~\cite{CMS-PAS-BPH-21-006}.
The baryon decay $\Lambda_b\to\Lambda\mu^+\mu^-$ has also been studied using lattice QCD \cite{Detmold:2016pkz, LHCb:2015tgy,
LHCb:2018jna, Blake:2019guk, Blake:2022vfl}

The $K^*$ decays to $K\pi$, so the rigorous calculation of the amplitude for $B\to K^*$ is very challenging.
An analysis with the appropriate finite-volume\footnote{%
See Sec.~\ref{sec:xyz} for more information on using finite-volume effects to understand hadronic resonances.} %
formalism~\cite{Briceno:2014uqa} is underway~\cite{Rendon:2018fem}, as is an analysis of $B\to\rho(\pi\pi)$.
These results will inform future computing projects, but the difficulty is such that the rates obtained from the lattice-QCD form
factors will be less precise than the measured rates for several years.

Neutral-meson mixing, which entails the oscillation from particle $P$ to its antiparticle $\bar{P}$ and back, is also an FCNC.
The frequency, $\Delta M_P$, is measured from the time dependence of decays, and, as is often the case with frequencies, the
measurements are very precise.
Thus, the precision of lattice-QCD calculations of these quantities have lagged experiment.
In the laboratory, this phenomenon has been observed for all stable neutral-meson systems: \linebreak $K^0$-$\bar{K}^0$,
$D^0$-$\bar{D}^0$, $B^0$-$\bar{B}^0$ and $B_s$-$\bar{B}_s$.
The most accurate theoretical results are for the $B$ systems, because these mixing processes are dominated by short-distance
virtual particles, leading to local four-quark operators.
The past few years witnessed significant improvement of all five operators that could mediate $B_{(s)}$-$\bar{B}_{(s)}$ mixing in
the Standard Model and any extension thereof~\cite{Bazavov:2016nty,Dowdall:2019bea}.
The two most precise calculations of the $B_{(s)}$-$\bar{B}_{(s)}$ mixing matrix elements are in imperfect
agreement~\cite{Bazavov:2016nty,Dowdall:2019bea}.
For this reason, and because the precision lags experiment, further work on the $B_{(s)}$ systems, $\Delta M_B$ and $\Delta
M_{B_s}$, is planned.
For remarks on the long-distance contributions that are crucial to $K^0$-$\bar{K}^0$ and $D^0$-$\bar{D}^0$ mixing, see
Sec.~\ref{sec:weak-s+d+u}.

The measured ratios $R(D^{(*)})=\text{BR}(B\to D^{(*)}\tau\nu)/\text{BR}(B\to D^{(*)}\ell\nu)$ of charged-current processes also
disagree with the Standard Model, by approximately $3\sigma$ combined~\cite{Amhis:2022mac}.
The Standard-Model prediction of these ratios requires the form factors over the full kinematically allowed range.
Results are available for $R(D)$~\cite{Lattice:2015rga,Na:2015kha} and, since 2021, for $R(D^*)$~\cite{FermilabLattice:2021cdg}.
The status after the first full lattice-QCD calculation of $R(D^*)$~\cite{FermilabLattice:2021cdg} is shown in
Fig.~\ref{fig:semi}~(right).
Lattice-QCD results and measurements are also available for the similar ratios $R(J/\psi)$~\cite{Harrison:2020nrv,LHCb:2017vlu} and
$R(\Lambda_c)$~\cite{Detmold:2015aaa,LHCb:2022piu}.
Lattice-QCD results for all ratios are of sufficient precision to meet the demands of LHCb and Belle~II for the next several years.
Even so, their precision will improve, because the relevant form factors are needed to determine the CKM element~$|V_{cb}|$.

The most persistent puzzles in $B$ physics are the discrepancies in the values $|V_{cb}|$ and~$|V_{ub}|$ determined via exclusive 
vs.\ inclusive semileptonic decays~\cite{Gambino:2020jvv}.
The theory input for exclusive decays consists of the form factors mentioned above, while the theory input for inclusive decays 
employs the operator-product expansion and heavy-quark expansion, which require coefficients calculated in perturbation theory%
\footnote{Lattice QCD for $\alpha_s$ is discussed Sec.~\ref{sec:asmq}.} %
with operator matrix elements obtained from kinematic distributions.
Lattice QCD can be used to compute (a few of) the matrix elements~\cite{Gimenez:1996av,Kronfeld:2000gk,Gambino:2017vkx,%
FermilabLattice:2018est}.
More recently, interesting ideas have been put forward to compute quantities related to the spectral function, such that a weighted
integral yields the inclusive
rate~\cite{Hashimoto:2017wqo,Hansen:2017mnd,Fukaya:2020wpp,Gambino:2020crt,Ishikawa:2021txe,DeGrand:2022lmc}.
One of these new methods~\cite{Gambino:2020crt} has recently been implemented in a full lattice-QCD study~\cite{Gambino:2022dvu}.

The possible hints of new physics in several measurements of $B$ decays lends renewed urgency to quark-flavor physics.
On the experimental side, both LHCb and Belle~II will be making more precise measurements during the coming five years, as will CMS
and ATLAS.
The $B$ program is complemented by the charm program at these experiments and BES~III.
Better precision on the fundamental parameters of the Standard Model is essential to establishing any new-physics effect in the
processes mentioned above.
For quark masses, current precision~\cite{Chakraborty:2014aca,FermilabLattice:2018est,Lytle:2018evc,Hatton:2021syc} suffices for the
time being, although further confirmation is, of course, desirable.
The magnitudes of the CKM matrix can be determined from leptonic (e.g., $B\to\tau\nu$) and charged-current semileptonic decays.
For leptonic decays, precision again suffices for the time being, although confirmation is, of course, desirable.
New calculations of the form factors for $B\to\pi\ell\nu$ and $B\to D^{(*)}\ell\nu$ are needed to match the precision of Belle~II
for $|V_{ub}|$ and $|V_{cb}|$, respectively.
Work is underway using the same general strategy that made the leptonic decays constants so precise; this work will automatically
include semileptonic $D$ decays~\cite{FermilabLattice:2021bxu}.
In addition, there is ongoing effort to support future measurements of $b$-baryon decays.

Further discussion of these and other issues related to $b$- and $c$-quark physics can be found in contributions to Snowmass;
see Ref.~\cite{Boyle:2022uba} for lattice QCD, Refs.~\cite{Charles:2020dfl, Altmannshofer:2022hfs, Cheng:2022tog} for phenomenology,
and Refs.~\cite{LHCb:2022ine, BelleII:2022yoy, BESIII:2022mxl} for experiment summaries, and Refs.~\cite{Bhattacharya:2022cna,
Sibidanov:2022gvb} for new analysis tools.

\subsubsection{Weak decays of strange and light quarks}
\label{sec:weak-s+d+u}
\index{Rare and Precision Frontier!RF2}

In addition to the short-distance processes analogous to those that dominate $B_{(s)}$-$\bar{B}_{(s)}$ mixing, $K^0$-$\bar{K}^0$
($D^0$-$\bar{D}^0$) oscillations are also mediated by processes~\cite{Golowich:2007ka, Golowich:2009ii} with two $\Delta S=1$
($\Delta C=1$) transitions separated by long, hadronic distances, such as $K^0\to\pi\pi\to\bar{K}^0$ ($D^0\to\pi
K\to\bar{D}^0$).
These long-distance effects are a challenge.
Lattice-QCD must employ a finite volume (because any computer's memory is finite), and the two-particle intermediate state is very
sensitive to finite-volume effects.
This sensitivity is, however, well understood mathematically for elastic processes such as $K\to\pi\pi$~\cite{Christ:2015pwa}.
First calculations of the long-distance contribution to $\Delta M_K$ have been carried out~\cite{Bai:2018mdv, Bai:2014cva}, although
again the precision achieved so far lags that of experiment.
These calculations are part of a broader campaign to study the $K\to\pi\pi$ reaction.
In 2015, the first lattice-QCD calculation with a complete error budget of the quantity $\Re(\epsilon'/\epsilon)$, which quantifies
direct $CP$ violation, appeared~\cite{Bai:2015nea}.
Further work~\cite{Kelly:2016mxf} led to a result
$\Re(\epsilon'/\epsilon)=21.7(8.4)\times10^{-4}$~\cite{RBC:2020kdj} 
in agreement with $\Re(\epsilon'/\epsilon)=16.6(2.3)\times10^{-4}$ from the 2002 measurements of the KTeV~\cite{AlaviHarati:2002ye}
and NA48~\cite{Batley:2002gn} experiments (at Fermilab and CERN, respectively).

In addition to $\Delta M_K$ and $\Re(\epsilon'/\epsilon)$, a few calculations in kaon physics are needed to get the most out of some
older experiments.
Even better precision than that now available (sub-percent~\cite{Bazavov:2012cd, Bazavov:2013maa, Bazavov:2018kjg, Boyle:2015hfa})
for the form factor in $K\to\pi\ell\nu$ is needed to resolve a possible $5\sigma$ tension in first row of the CKM matrix.%
\footnote{$5\sigma$ is a attained when using a new result for a certain radiative correction in nucleon $\beta$
decay~\cite{Seng:2018yzq, Seng:2022wcw, Seng:2022tjh}.} %
Such improvement requires a complete treatment of electromagnetism and isospin breaking via $m_d\neq m_u$.
This advance will require new ensembles of gauge-field configurations, which are already planned for muon $g-2$.
Now that the ingredients of a full calculation of $\Re(\epsilon'/\epsilon)$ are understood, it is time to aim for the precision of
KTeV and NA48~\cite{AlaviHarati:2002ye,Batley:2002gn}.
Last, CERN experiment NA62 is underway to improve BNL E949's measurement of the branching ratio of $K^+\to\pi^+\nu\bar\nu$.
With the recent improvement in the charm-quark mass and expected improvements in the CKM matrix, the leading theoretical uncertainty
in the Standard-Model prediction is a long-distance effect of charmed intermediate states~\cite{Bai:2018hqu}.
Technology similar to that used for $\Delta M_K$ will be used to attain a first-principles result to replace phenomenological
estimates currently in~use.
For further discussion of these and other issues related to kaon physics, see the Snowmass contributions on lattice
QCD~\cite{Blum:2022wsz}, phenomenology~\cite{Aebischer:2022vky, Goudzovski:2022vbt}, and experiment~\cite{NA62KLEVER:2022nea}.

Lattice-QCD calculations of properties of isoscalar mesons, such as $\eta$ and $\eta'$ are more difficult, because the conversion of
$q\bar{q}$ into gluons and back again is computationally more challenging, making results less precise.
Basic properties, such as the masses and mixing angle have been studied~\cite{Christ:2010dd, Michael:2013gka, Fukaya:2015ara,
Bali:2021qem, CSSMQCDSFUKQCD:2021rvs}.
The prospect of the REDTOP experiment~\cite{REDTOP:2022slw} makes further pertinent calculations compelling.

%% file: rare/small.tex
Among the many ``small'' experiments exploring fundamental physics, two kinds benefit from high-quality lattice-QCD calculations:
measurements of the (anomalous) magnetic moment of the muon~\cite{Lehner:2018qcd} and searches for permanent electric dipole moments
in the neutron, proton, and (in principle) other hadrons~\cite{Davoudi:2018qcd}.

\index{Theory Frontier!TF06|(}
\subsubsection{Muon magnetic moment (\texorpdfstring{$g-2$}{g-2})}
\label{sec:g-2}
\index{Rare and Precision Frontier!RF3!muon $g-2$}

In the Standard Model or any extension of it, the muon anomalous magnetic moment, denoted $g-2$ or $a_\mu=(g-2)/2$, is a sum of
quantum fluctuations from photons, hadrons, the top quark, $W$ and $Z$ bosons, the Higgs boson, and any as-yet undetected particles
that couple to the muon or any of the other Standard-Model particles.
Except for the hadronic contributions, perturbation theory can be used to obtain sufficient precision to match the experiments'
needs.
Thus, the hadronic contributions dominate the error budget.
The two most important hadronic contributions to $g-2$ are the leading-order HVP and the much smaller hadronic light-by-light (HLbL;
again, leading order).
To obtain these contributions, the hadron current-current correlation function (for HVP) or the four-current scattering amplitude
(for HLbL) are convolved with kernels derived in perturbative QED~\cite{Lautrup:1974ic,Blum:2002ii}.
Further hadronic contributions are the next-to-leading order (NLO) HVP and HLbL, which use the same QCD calculations but
higher-order QED kernels.

As recently as 2015, precise calculations of the HVP were beginning, and work on HLbL scattering was in an exploratory phase.
Both began to receive increasingly large computing allocations as the Fermilab experiment was being built.
By now, HVP is a fully mature subject.
Recent results for the HVP from lattice QCD come from many collaborations in the U.S., Europe, and Japan~\cite{Burger:2013jya,
Blum:2015you, Blum:2018mom, Chakraborty:2016mwy, Borsanyi:2017zdw,
Chakraborty:2017tqp, Chakraborty:2018iyb, Davies:2019efs, Davies:2022epg, DellaMorte:2017dyu, Gerardin:2019rua, Ce:2022kxy,
Giusti:2018mdh, Giusti:2019xct, Aubin:2019usy,  Aubin:2022hgm, Shintani:2019wai, Borsanyi:2020mff, Lehner:2020crt, Wang:2022lkq,
Alexandrou:2022amy}.

\begin{figure}
    \centering
    \vspace*{-12pt}
    \includegraphics[width=0.6\textwidth]{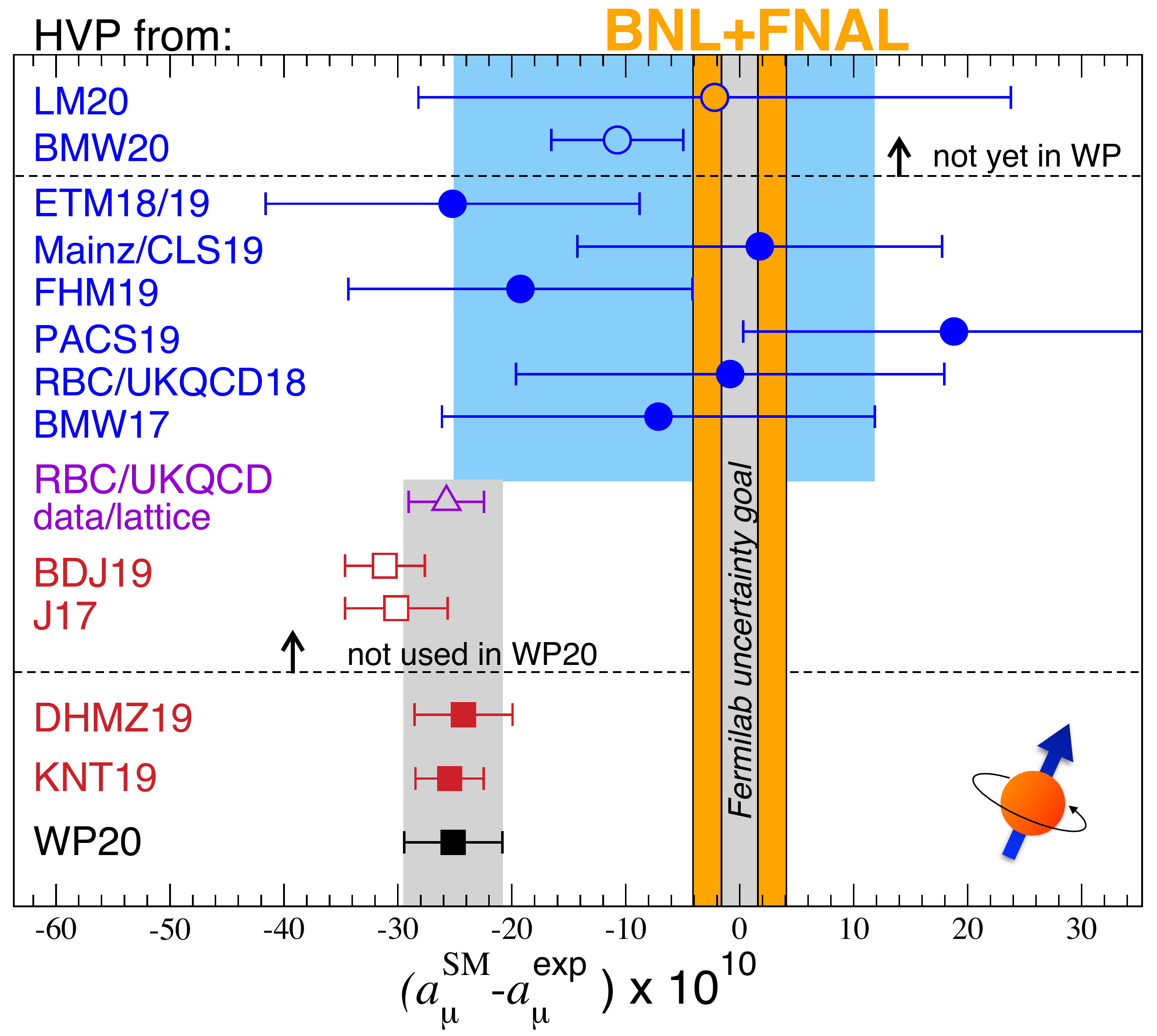}
    \caption[fig:MQ]{Comparison of $a_\mu=(g-2)/2$ calculations of the full HVP with experiment.
        To obtain $a_\mu^\text{SM}$ the other Standard-Model contributions as given in Ref.~\cite{Aoyama:2020ynm} are added to the 
        calculation of the HVP.
        Averages for dispersion-theory (gray band, lower left) and lattice-QCD (blue band) results are based on the filled symbols; 
        results shown with unfilled symbols are omitted for various reasons.
        The orange band shows the average of the BNL and Fermilab experiments, with the narrow gray band the target uncertainty of 
        Fermilab E989.
        From Ref.~\cite{Colangelo:2022jxc}.}
    \label{fig:HVP}
\end{figure}

Since the publication of the 2019 USQCD whitepaper~\cite{Lehner:2018qcd} on quark- and lepton-flavor physics, the landscape of the
$g-2$ has developed dramatically.
The Fermilab experiment reported a new result with 0.46~ppm precision~\cite{Muong-2:2021ojo, Muong-2:2021vma}, in agreement with an
earlier BNL result measured to 0.54~ppm~\cite{Muong-2:2006rrc}.
In support, the Muon $g-2$ Theory Initiative~\cite{muon:gm2:TI} produced consensus values for the hadronic
contributions~\cite{Aoyama:2020ynm}, with an update contributed to Snowmass~\cite{Colangelo:2022jxc}.
A~comparison plot~\cite{Colangelo:2022jxc} of the combination of the BNL and Fermilab experiments, $a_\mu^\text{exp}$, with
the consensus Standard-Model prediction, $a_\mu^\text{SM}$, is reproduced in Fig.~\ref{fig:HVP}.
Figure~\ref{fig:HVP} also shows individual theoretical predictions that use data for $e^+e^-\to\text{hadrons}$ and dispersion theory
to determine the HVP~\cite{Davier:2019can, Keshavarzi:2019abf}, which lead to the consensus value.
The average of the experiments~\cite{Muong-2:2021ojo, Muong-2:2021vma, Muong-2:2006rrc} and Standard-Model consensus disagree by
$4.2\sigma$.
This discrepancy is not new, and explanations of it with new physics constitute an enormous literature; see, for example,
Ref.~\cite{Athron:2021iuf} and references therein.

Meanwhile, the Budapest-Marseille-Wuppertal (BMW) collaboration~\cite{Borsanyi:2020mff} has published the first lattice-QCD result
for the HVP with a total error comparable to the latest data-driven results~\cite{Davier:2019can, Keshavarzi:2019abf}.
Using the BMW result for the HVP (labeled ``BMW20'' in Fig.~\ref{fig:HVP}) would relieve the disagreement with experiment.
Even so, it would not be the end of the drama, because the HVP with a different QED weight enters the running of $\alpha_\text{QED}$
up to the $Z$~pole~\cite{Passera:2008jk, Keshavarzi:2020bfy, Crivellin:2020zul, Miura:2022trp}.
Most of the lattice groups cited above are aiming at precision targets~\cite{Colangelo:2022jxc} of 0.5\% for the leading-order HVP
(thus surpassing the precision of the BMW calculation~\cite{Borsanyi:2020mff}) and 1\% for the NLO and NNLO~HVP.
Assuming consistency, the NLO calculations could be averaged, bringing the overall uncertainty down to the level of the Fermilab
experiment.

Even with exascale computing resources, it will be extremely difficult to attain a single such precise calculation for the
leading-order HVP ($a_\mu^\text{HVP, LO}$ in Table~\ref{tbl:milestones}).
In particular, effects of isospin breaking are needed, both from QED and $m_u\neq m_d$, and several approaches are being explored to address this challenge.
It is crucial to have independent efforts with enough differences in analysis to ameliorate correlations in the systematic errors.
These calculations will receive rigorous scrutiny from the Muon $g-2$ Theory Initiative.
Once the uncertainties are comparable to the dispersive method, they will be folded into the consensus.
The higher-order HVP, $a_\mu^\text{HVP, NLO}$, requires the same QCD calculation~\cite{Chakraborty:2018iyb}, so the same lattice-QCD
work will hit the 10\% target for this contribution too.

As a four-point function, the HLbL contribution, $a_\mu^\text{HLbL}$, is more computationally demanding (for comparable precision).
It is smaller, however, so the precision target is less demanding, again 10\%.
In the past, the HLbL contribution was suspected as the origin of the discrepancy between the BNL measurement and contemporary
Standard-Model predictions, in large part because there was no rigorous way to determine it.
The community relied on values bracketing model-based calculations~\cite{Prades:2009tw, Nyffeler:2009tw, Jegerlehner:2009ry,
Jegerlehner:2017gek}, without having a firm grasp on the uncertainties or even the robustness of those estimates.
These suspicions can now be dismissed, thanks to advances in lattice QCD and a relatively recent data-driven dispersive
method~\cite{Colangelo:2015ama}.
Building on the earlier development of viable techniques~\cite{Blum:2014oka, Blum:2015gfa, Blum:2016lnc, Blum:2017cer,
Asmussen:2016lse}, RBC/UKQCD~\cite{Blum:2019ugy} and Mainz~\cite{Chao:2021tvp, Chao:2022xzg} and both have published results with
$\sim20\%$ uncertainty, comparable to the data-driven dispersive method~\cite{Colangelo:2022jxc,Colangelo:2015ama}.
While still short of the ultimate goal, the consistency of these results makes it implausible that the HLbL contribution can be
large enough to explain the discrepancy between the consensus Standard-Model value and the experiments.
Now that the lattice-QCD methods are mature, it is expected that with exascale resources the uncertainty in HLbL can be reduced
further, again matching the needs of the final Fermilab result.

It is impossible to predict how HVP, HLbL, and the experiments will land as the work continues to unfold.
As already mentioned, $a_\mu^\text{HLbL}$ is too small to affect the outcome.
If one assumes the new data from the Fermilab experiment agree with Refs.~\cite{Muong-2:2021ojo, Muong-2:2021vma, Muong-2:2006rrc}
and reduce the uncertainty as planned, and \emph{further} assumes that lattice-QCD confirms the dispersive results for HVP, then a
very significant discrepancy would arise.
Less exciting scenarios are also possible, for example if other lattice-QCD groups confirm BMW's result for the HVP.
That would imply some sort of misunderstanding of $e^+e^-\to\text{hadrons}$, which would have repercussions
elsewhere~\cite{Crivellin:2020zul}.
\index{Theory Frontier!TF06|)}

\subsubsection{Electric dipole moments}
\label{sec:edm}
\index{Rare and Precision Frontier!RF3!nucleon electric dipole moments}

Permanent EDMs of elementary particles, nucleons, atoms and molecules in the ground state, if observed, are signals of $CP$
violation.
CKM-induced contributions are orders of magnitude tinier than experimental sensitivity~\cite{Pospelov:2005pr}.
In the Standard Model, EDMs could stem from a $CP$-violating gluonic operator in the QCD Lagrangian, commonly known as strong-$CP$
violation.
Limits on the neutron EDM lead to the strong-$CP$ problem, one of the outstanding puzzles associated with the Standard
Model~\cite{Blinov:2022tfy}.
Briefly, the current bound on the neutron EDM implies $\theta_\text{QCD}-\arg\det Y\lesssim10^{-10}$, where $\theta_\text{QCD}$ is
the coefficient of of the strong-$CP$ term, and~$Y$ is the Yukawa coupling matrix between the Higgs and quark fields.
The cancellation is baffling.
A statistically significant calculation of the required nucleon matrix element directly from (lattice) QCD is not yet available; see
Ref.~\cite{Bhattacharya:2021lol} for discussion and Ref.~\cite{Shindler:2021bcx} for a review.
These matrix elements are challenging, because of the need to sample fully the topological sectors of QCD~\cite{Shintani:2005xg}.

A popular explanation\footnote{%
The solution of the strong-$CP$ problem with $m_u=0$ is ruled out~\cite{Fodor:2016bgu, Giusti:2017dmp, FermilabLattice:2018est, 
Alexandrou:2020bkd}.} %
for the smallness of strong-$CP$ violation is the axion \cite{Blinov:2022tfy}, a field that couples to the strong-$CP$ term in a way
that dynamically cancels $\theta_\text{QCD}-\arg\det Y$.
Then a nonzero EDM would be a signal of new physics.
The axion is of further interest as a dark matter candidate; lattice-QCD input on its viability is discussed in
Sec.~\ref{sec:dm-wave}.

Several new experiments aimed at the neutron EDM are planned~\cite{Chupp:2017rkp, Alarcon:2022ero} and an experiment aimed at the
proton EDM is being developed~\cite{Anastassopoulos:2015ura, Omarov:2020kws, Alexander:2022rmq}.
In addition to the strong-$CP$ term, several higher-dimension operators induced by physics at energies at or beyond the electroweak
scale can generate EDMs; for more information, see the recent reviews~\cite{Chupp:2017rkp, Shindler:2021bcx} or a contribution to
Snowmass~\cite{Alarcon:2022ero}.
Lattice-QCD calculations of nucleon matrix elements of the Standard-Model operator and BSM operators have been carried out or are
underway~\cite{Shintani:2005xg, Guo:2015tla, Shintani:2015vsx, Abramczyk:2017oxr, Syritsyn:2018mon, Bhattacharya:2018qat,
Bhattacharya:2021lol, Dragos:2019oxn, Alexandrou:2020mds}.
Of these, the nucleon EDM induced by the quark EDM operator is a technically straightforward calculation, and results with
$\lesssim5\%$ uncertainty have been obtained~\cite{Gupta:2018lvp,Aoki:2019cca, Aoki:2021kgd}.
The calculations of the matrix elements of other leading BSM operators can be challenging because of the low statistical signal and
issues of renormalization~\cite{Bhattacharya:2015rsa, Cirigliano:2020msr, Rizik:2020naq, Mereghetti:2021nkt}.
Progress has, however, been steady and over the next five years, estimates with around $20\%$ uncertainty are expected for many EDM
matrix elements.

%% file: rare/B-L.tex
On their own, baryon number, $B$, and lepton number, $L$, are accidental symmetries of the Standard Model.
The Standard Model does allow for changes in $B-L$ via instantons or sphalerons, which are both suppressed at temperatures below the
electroweak phase transition.
Many extensions of the Standard Model break $B$ and $L$ while preserving $B-L$: an observation of baryon-number violation via proton
decay or neutron-antineutron oscillations would lend support to these ideas.
Extensions of the Standard Model that accommodate nonzero neutrino masses sometimes introduce Majorana fermions (for example,
right-handed neutrinos), leading to $\Delta L=2$ neutrinoless double-$\beta$ decay ($0\nu\beta\beta$) of nuclei.
As hadronic and nuclear transitions, proton decay, $n$-$\bar n$ oscillations, and $0\nu\beta\beta$ require a solid understanding of
the strong interactions in order to make Standard-Model predictions.

The large-scale neutrino detectors, DUNE~\cite{Kudryavtsev:2016ybl} and HyperK~\cite{Abe:2018uyc}, will set new limits on proton
decay and $n$-$\bar n$ oscillation processes~\cite{Dev:2022jbf}.
Dedicated $n$-$\bar n$ experiments are also being developed~\cite{Phillips:2014fgb} including a proposed experiment at the European
Spallation Source~\cite{Addazi:2020nlz}.
The $n$-$\bar n$ transition rates probed by these experiments can be directly connected to constraints on BSM theories using
lattice-QCD calculations of the corresponding nucleon matrix elements.
The extraction of BSM theory constraints from experimental searches for nuclear instability at DUNE and HyperK will require a
combination of lattice-QCD calculations of nucleon-level processes with nuclear effective
theories~\cite{Oosterhof:2019dlo,Haidenbauer:2019fyd} and event generators describing experimental signatures of these processes in
nuclei~\cite{Golubeva:2018mrz,Barrow:2019viz} that are under active development.
Lattice-QCD calculations of proton-decay matrix elements have been carried out for several proton decay modes~\cite{CP-PACS:2004wqk,
Aoki:2008ku, Aoki:2013yxa, Aoki:2017puj, Yoo:2021gql}, accurately enough for current limits.
During the coming decade, it will be feasible to improve them to the 10-percent level.
For $n$-$\bar n$ oscillations, the matrix elements turn out to be 5--10 times larger in lattice-QCD
calculation~\cite{Buchoff:2015qwa, Rinaldi:2018osy, Rinaldi:2019thf} than had previously been estimated using the MIT bag model and,
thus, extend the reach of current and future experiments.
A second round of calculations is needed to obtain a fuller understanding of the systematic uncertainties, but the 10-percent level
again seems feasible.

An observation of $0\nu\beta\beta$ would demonstrate that neutrinos are Majorana fermions, unlike the charged leptons and
quarks~\cite{Dolinski:2019nrj}.
Nuclear effective field theory analysis has recently demonstrated that a short-distance $nn\to pp$ interaction is required to
consistently describe $0\nu\beta\beta$ processes in nuclei~\cite{Cirigliano:2018hja}.
The corresponding low-energy constant has been estimated~\cite{Cirigliano:2021qko} and shown to lead to $\sim 30\%$ or larger
modifications of experimentally relevant nuclear matrix elements~\cite{Wirth:2021pij,Weiss:2021rig,Jokiniemi:2021qqv}.
Lattice QCD calculations of the $nn\to ppe^-e^-$ process can be used to accurately determine this low-energy constant and reduce
associated uncertainties in $0\nu\beta\beta$ nuclear matrix element predictions.
This undertaking will be challenging.
Techniques similar to those developed for $K^+\to\pi^+\nu\bar\nu$ and $a_\mu^\text{HLbL}$ are expected to be helpful.
First results are becoming available for ``warm-up'' lattice-QCD calculations: the Standard Model $2\nu\beta\beta$ process $nn\to
ppe^-e^- \overline{\nu}_e\overline{\nu}_e$~\cite{Shanahan:2017bgi,Tiburzi:2017iux}, as well as $\pi^-\to\pi^+e^-e^-$ and related
mesonic processes involving light Majorana neutrino exchange~\cite{Feng:2018pdq,Detmold:2020jqv,Tuo:2019bue}.
Lattice-QCD calculations have also been performed for four-quark operator matrix elements needed to predict $0\nu\beta\beta$ rates
from other BSM scenarios involving TeV-scale $B-L$ violation instead of long-distance Majorana neutrino
propagation~\cite{Nicholson:2018mwc}.
The push toward realistic calculations of the $nn\to ppe^-e^-$ process is expected to take at least five more
years~\cite{Davoudi:2018qcd, Cirigliano:2022oqy}.

%% file: rare/clfv.tex
With nonzero neutrino masses and mixing, the Standard Model allows charged-lepton flavor violation (CLFV), similarly to FCNCs in the quark
sector, but the rate is too small to observe, because the neutrino mass differences are so small.
An example is a muon converting to an electron, either through electromagnetic decay (i.e., $\mu\to e\gamma$) or in the field of a
nucleus ($\mu A\to eA$, often referred to as $\mu2e$).
Other possibilities include meson decays, for example $B^+\to K^+\mu^-e^+$ or $B_s^0\to\tau^\pm\mu^\mp$.
The latter require the lattice-QCD calculations discussed in Sec.~\ref{sec:weak}, while $\mu2e$ requires isoscalar nucleon
properties, as the (BSM) meditator interacts with quarks inside a nucleon inside the nucleus.
Many experiments searching for charged-lepton flavor violation are running or are on the horizon, for example the Mu2e Experiment at
Fermilab, which aims to reduce the sensitivity to $\mu A \to eA$ by four orders of magnitude.
Mediators with similar couplings to quarks are posited to couple to dark matter (DM), so the same nucleon matrix elements are needed
for limits on direct DM detection.

In the lepton conversion or DM scattering off nuclei, the energy transfers are low enough so that only $q^2=0$ nucleon matrix
elements are needed.
The interaction may be flavor singlet, in which case the current can couple to a sea quark, instead of just a valence quark as in
isovector (i.e., charged current) processes.
The sea quark can propagate from anywhere in spacetime to anywhere else and back again, and such sea-quark propagators are simply
more computationally challenging than valence-quark propagators.

\begin{figure}
    \includegraphics[height=0.275\textheight]{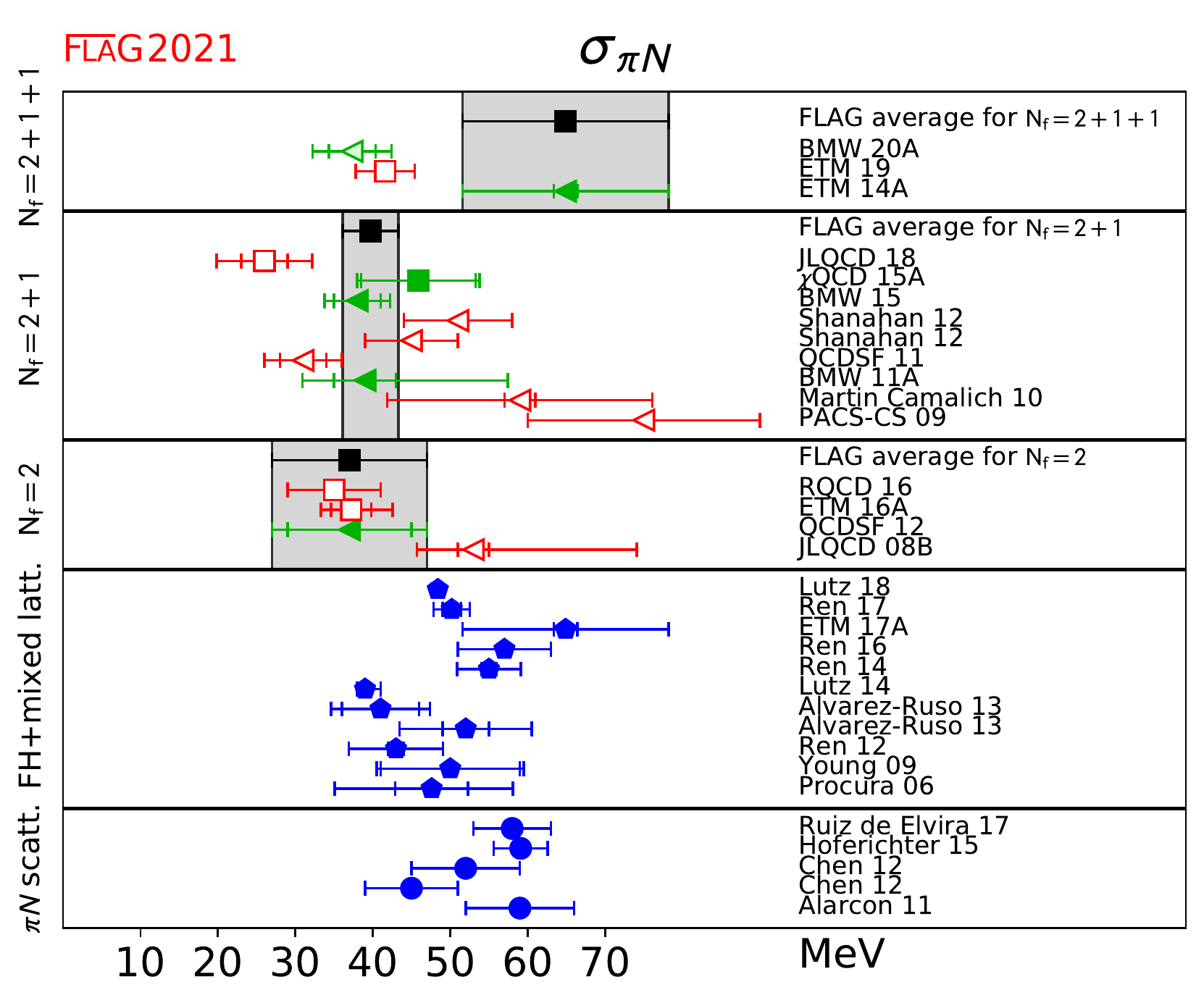} \hfill
    \includegraphics[height=0.275\textheight]{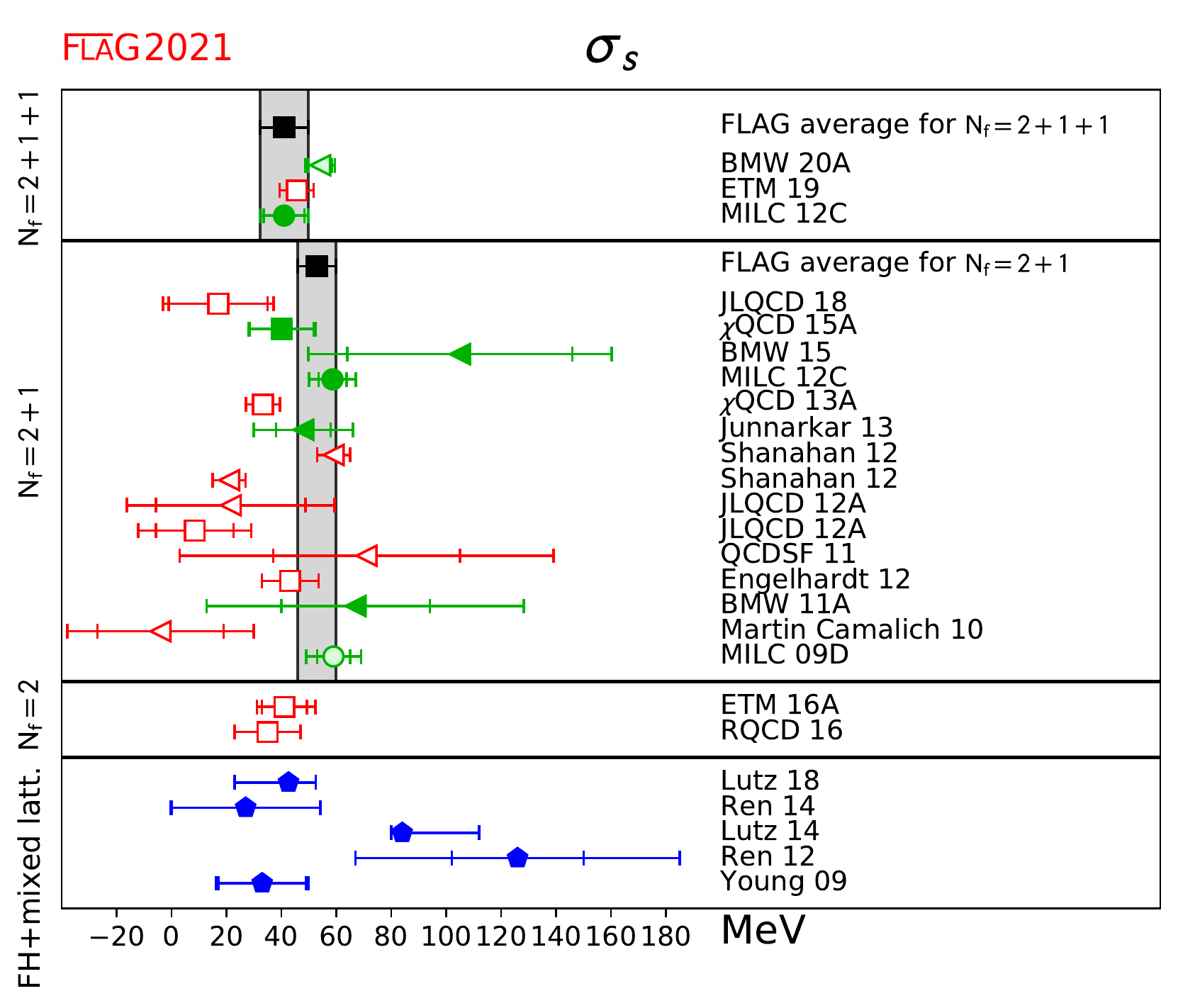}
    \caption[fig:MQ]{Comparisons of the ``nucleon sigma term'' $\sigma_{\pi N}$ (left) and the strangeness content of the nucleon 
        $\sigma_s$ (right), from Ref.~\cite{Aoki:2021kgd}.
        Green symbols~\cite{Freeman:2012ry, Junnarkar:2013ac, Alexandrou:2014sha, Yang:2015uis, Durr:2011mp, Durr:2015dna} are 
        included in the averages (gray bands); red symbols fall short of certain criteria and are omitted.
        Blue pentagons denote analyses of several data sets, often including lattice-QCD results.
        NB: the $N_f=2$ results omit the strange sea and are, thus, not recommended for phenomenology.}
    \label{fig:sigma}
\end{figure}

To interpret Mu2e, lattice-QCD calculations of the light- and strange-quark contents of the nucleon are
needed~\cite{Kosmas:1994pt,Cirigliano:2009bz}.
These are the matrix elements known as the ``sigma term'', $\sigma_{\pi N}=\frac{1}{2}(m_u+m_d)\langle N|(\bar{u}u+\bar{d}d)|N\rangle$ 
and the strangeness content $\sigma_s=m_s\langle N|\bar{s}s|N\rangle$, %
as well as the ratio $\langle N|(\bar{u}u-\bar{d}d)|N \rangle/\langle N|(\bar{u}u+\bar{d}d)|N \rangle$.
Figure~\ref{fig:sigma} shows the status for $\sigma_{\pi N}$ and $\sigma_s$.
Before lattice-QCD calculations became available~\cite{Engelhardt:2012gd, Freeman:2012ry, Junnarkar:2013ac, Alexandrou:2014sha,
Yang:2015uis, Durr:2011mp, Durr:2015dna}, estimates of $\sigma_s$ from hadronic physics were very uncertain.
The situation is much better now, thanks to lattice QCD, but further improvements are clearly needed.
The phenomenological determinations of $\sigma_{\pi N}$~\cite{Hoferichter:2015dsa, RuizdeElvira:2017stg} are much more robust
than for $\sigma_s$, providing an important benchmark for lattice-QCD calculations~\cite{Gupta:2021ahb}.

The heavy-quark content, defined in analogy with $\sigma_s$ (for charm, bottom, and top), can be related to the trace
anomaly~\cite{Shifman:1978zn}.
Because charm might not be heavy enough, lattice-QCD calculations of the charm content have been carried out~\cite{XQCD:2013odc,
Alexandrou:2019brg}.
In addition to the spin-independent matrix elements shown in Fig.~\ref{fig:sigma}, spin-dependent matrix elements are also
relevant~\cite{Cirigliano:2017azj} and computable with lattice~QCD.
First attempts to address nuclear effects on both spin-independent and spin-dependent operators are underway~\cite{Chang:2017eiq}.

The isovector versions of these matrix elements, or \emph{charges}\/ $g_A^\umd$, $g_S^\umd$, $g_T^\umd$, are important for
ultraprecise neutron decay experiments.
It is highly unlikely that lattice QCD will reach the uncertainty of the experimental average,
$g_A^\umd=1.2754(13)$~\cite{ParticleDataGroup:2022ssz} during the coming decade, but 1\% calculations should be possible and could
shed light on the
disagreement among neutron-lifetime measurements~\cite{Czarnecki:2018okw}.%
\footnote{In fact, 1\% precision is claimed already~\cite{Chang:2018uxx, Berkowitz:2017gql}, although the tension of this result 
with that of Ref.~\cite{Gupta:2018qil} (computed on the same ensembles) leads FLAG~\cite{Aoki:2021kgd} to quote an ``average'' with
2.2\% error.
See Sec,~\ref{sec:nu} for more details on~$g_A^\umd$} %
The tensor and scalar charges at similar precision also will be possible~\cite{Bhattacharya:2016zcn, Horkel:2020hpi, Liu:2021irg}.
Calculations of these charges at the 10\% level, when combined with $\beta$-decay measurements, complement the LHC search
for new quark interactions, probing effective scales of new physics close to
10~TeV~\cite{Bhattacharya:2011qm, Gupta:2018qil, Gonzalez-Alonso:2018omy}.

%% file: rare/xyz.tex
The prospect of \emph{ab initio} calculations of the hadron spectrum was one of the original attractions of numerical lattice QCD.
For the most common mesons and baryons, this task was in a sense completed about a decade ago; see Fig.~2 of
Ref.~\cite{Kronfeld:2012uk}.
In more recent years, common hadron masses are studied carefully for technical purposes such as tuning the quark masses and
converting from lattice to physical units.

At the same time, the community has moved on to more challenging and interesting questions~\cite{Detmold:2018qcd, Brambilla:2019esw,
Bulava:2022ovd, Brambilla:2022ura}, such as determining resonance widths and the masses of more exotic hadrons, such as those
discovered at BaBar, Belle, CDF, D0, and LHCb---the ``$XYZ$'' states---tetraquarks, pentaquarks, and dibaryons~\cite{NPLQCD:2011naw,
Prelovsek:2013cra, Francis:2016hui, Francis:2018qch, Francis:2018jyb, Junnarkar:2018twb, Padmanath:2021qje, Junnarkar:2019equ,
Hudspith:2020tdf, Prelovsek:2020eiw, Green:2021qol, Francis:2021vrr, Padmanath:2022cvl}.
Determining the structure of these states is a compelling and still unanswered question.

Also falling within the rubric of spectroscopy is the calculation of decay widths and scattering amplitudes, because they can be
determined from finite-volume energy levels via various universal formulas~\cite{Luscher:1986pf, Luscher:1990ux, Rummukainen:1995vs,
Lellouch:2000pv, Kim:2005gf, Lage:2009zv, Hansen:2012tf, Gockeler:2012yj, Briceno:2014oea, Briceno:2015csa, Briceno:2017tce}; for a
review, see Ref.~\cite{Briceno:2017max}.
Thus, the resonance properties of the $\rho$ and $K^*$ mesons are now well studied.
Future applications will include electromagnetic transitions such as $N\gamma\to\Delta\to N\pi$ or similarly with a weak current.
See Refs.~\cite{Briceno:2015dca,Alexandrou:2018jbt} for studies of the similar process $\pi\gamma\to\rho\to\pi\pi$.
As mentioned in Sec.~\ref{sec:weak}, weak decays to vector mesons, such as $B\to K^*l\nu$ or $B\to D^*l\nu$ play important roles
in the ``flavor anomalies'', and a completely rigorous treatment requires these finite-volume spectroscopic techniques.\footnote{%
In the case of the $D^*$, chiral perturbation theory is used to control and estimate uncertainties in $D^*\leftrightarrow D\pi$.}
Coupled-channel scattering in $D\pi$-$D\eta$-$D_s\bar{K}$ has been used to gain a QCD-based understanding of the excited $D$-meson
spectrum and its puzzling features~\cite{Moir:2016srx,Albaladejo:2016lbb,Lang:2022elg}.

Given the importance of heavy-quark physics in particle physics (cf., Sec.~\ref{sec:weak}), it is worth noting that computations of
the quarkonium~\cite{HPQCD:2011qwj, DeTar:2018uko, HadronSpectrum:2012gic, Ryan:2020iog} and heavy-light meson spectrum were
important milestones in establishing lattice QCD for heavy quarks~\cite{Kronfeld:2003sd}.
Indeed, the aim of lattice $B$ physics played a role in the invention of nonrelativistic QCD (NRQCD)~\cite{Caswell:1985ui,
Lepage:1987gg, Thacker:1990bm, Lepage:1992tx} and the leading terms of the heavy-quark effective
theory~\cite{Eichten:1987xu, Eichten:1989zv, Eichten:1990vp}.
NRQCD also played an important role in the predictions of the masses of the $B_c$~\cite{Allison:2004be, Gregory:2010gm}
(confirmed~\cite{CDF:2005yjh}), $B_c^*$~\cite{Gregory:2009hq} (not yet seen), and $B_c(2S)$~\cite{Dowdall:2012ab}
(confirmed~\cite{ATLAS:2014lga}) mesons.

%% file: nu.tex
The physics associated with neutrino mass and mixing is addressed principally through neutrino oscillation experiments, such as Daya
Bay, NOvA, T2K, DUNE, and HyperK, which compare the neutrino energy spectra in detectors at short and long baselines.
Deformations in the neutrino-energy spectrum yield the oscillation parameters of the Pontecorvo-Maki-Nakagawa-Sakata (PMNS) mixing
matrix~\cite{Pontecorvo:1957cp, Pontecorvo:1967fh, Maki:1962mu}.
The incoming neutrino energy cannot be measured directly, and it is difficult or impossible to reconstruct it without a model of the
nuclear physics of the struck nucleus~\cite{Huber:2016mki, Alvarez-Ruso:2017oui, Ruso:2022qes}, because the final-state energy of
the nuclear remnant(s) is at best measured poorly.
Scattering amplitudes at the \emph{nucleon} level are necessary ingredients to these models~\cite{Kronfeld:2018qcd}.
The uncertainties are extremely difficult to estimate because they have so many moving parts~\cite{Coloma:2012ji, Coloma:2013tba}:
lattice QCD can provide a firm anchor at the nucleon level.

At low energies, the key signal process for neutrino-nucleus scattering is quasielastic scattering off a nucleon bound in the
nucleus.
Here, the main missing ingredient is the isovector axial form factor.
In the past few years, several groups have studied this form factor's $Q^2$ dependence~\cite{Yamazaki:2009zq, Babich:2010at,
Green:2017keo, Hasan:2017wwt, Rajan:2017lxk, Jang:2019vkm, Ishikawa:2018rew, Shintani:2018ozy, Ishikawa:2021eut, RQCD:2019jai,
Alexandrou:2020okk, Park:2021ypf}.
While the precision achieved so far is considerably less than for meson form factors, discussed in Sec.~\ref{sec:weak}, the
combination of increased computer power and the increased interest (stemming from the neutrino experiments) has led to rapid
progress.

\begin{figure}[b]
    \includegraphics[width=0.48\textwidth]{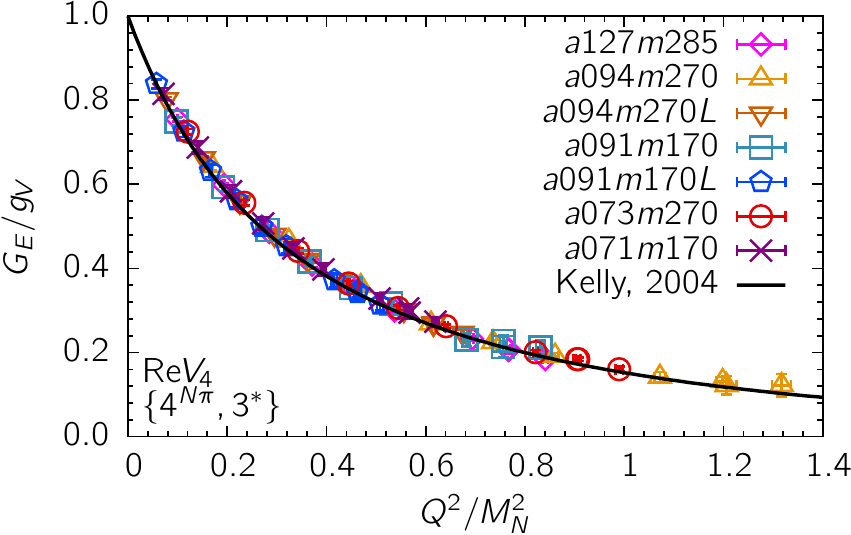} \hfill
    \includegraphics[width=0.48\textwidth]{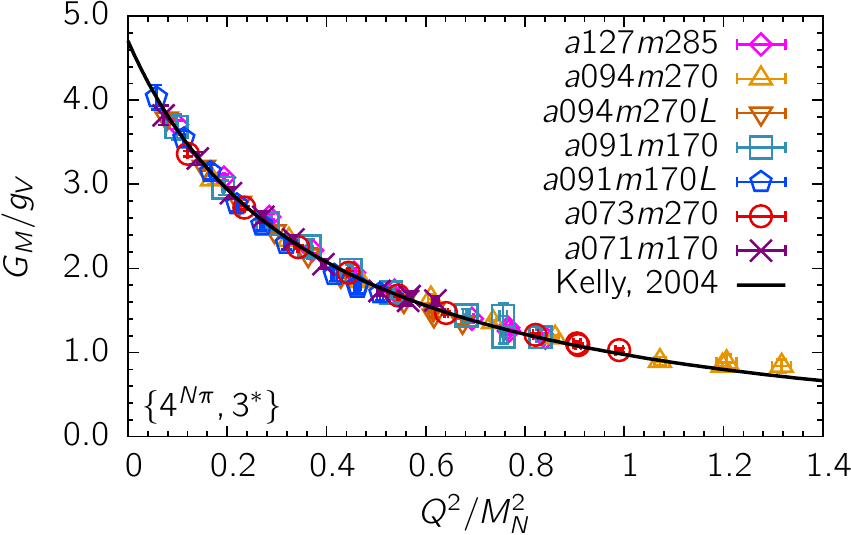}
    \vspace*{-4pt}
    \caption[fig:MQ]{Electric (left) and magnetic (right) form factors of the nucleon vs.\ squared momentum transfer $Q^2=-q^2$ in
        nucleon-mass units.
        The colored symbols denote explicit calculations at various lattice spacing ($a\approx0.13$, $0.09$, and $0.07$~fm), pion 
        mass ($M_\pi\approx285$, $270$, or $170$~MeV), volumes (``large'' or ``larger $L$'').
        The Pad\'e parametrization of Kelly~\cite{Kelly:2004hm} of experimental measurement is shown for comparison (black curve).
        From Ref.~\cite{Park:2021ypf}.}
    \label{fig:FV}
\end{figure}

Calculations of the vector form factors, which can be measured in $eN$ scattering~\cite{Ankowski:2022thw}, can provide validation.
Charge conservation makes the normalization automatic, but the radii defined by ($i$ labels form factors $G$)
\begin{equation}
    r_i^2 \equiv \frac{6}{G_i(0)} \left. \frac{dG_i}{dq^2}\right|_{q^2=0},
    \label{eq:rErMri}
\end{equation}
are of interest.
As illustrated in Fig.~\ref{fig:FV}, lattice-QCD calculations~\cite{Park:2021ypf} of the isovector electric, $G_E(Q^2)$, and
magnetic, $G_M(Q^2)$, form factors agree well with (a parametrization of) experimental measurements~\cite{Kelly:2004hm}, overall and
in particular for the radii: the Kelly parameterization gives $r_E=0.926(4)$ and 
 $r_M=0.872(7)$~\cite{Kelly:2004hm} while lattice QCD gives $r_E=0.92(12)$  and
 $r_M=0.84(18)$~fm~\cite{Park:2021ypf}.
Moreover, the shape agreement---as seen in Fig.~\ref{fig:FV}---extends well beyond $Q^2=0$.
The validation of lattice QCD is in this case is very encouraging.

\begin{figure}
    \includegraphics[width=0.48\textwidth]{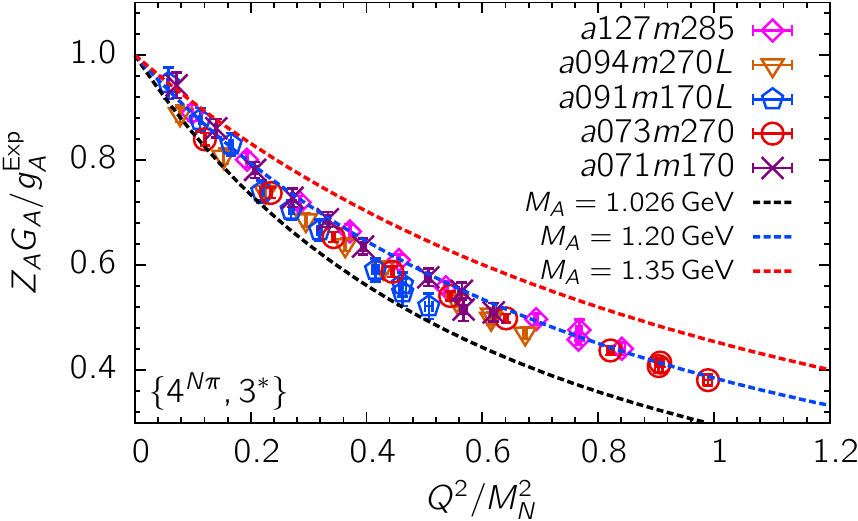} \hfill
    \includegraphics[width=0.50\textwidth,trim=16 6 48 50,clip]{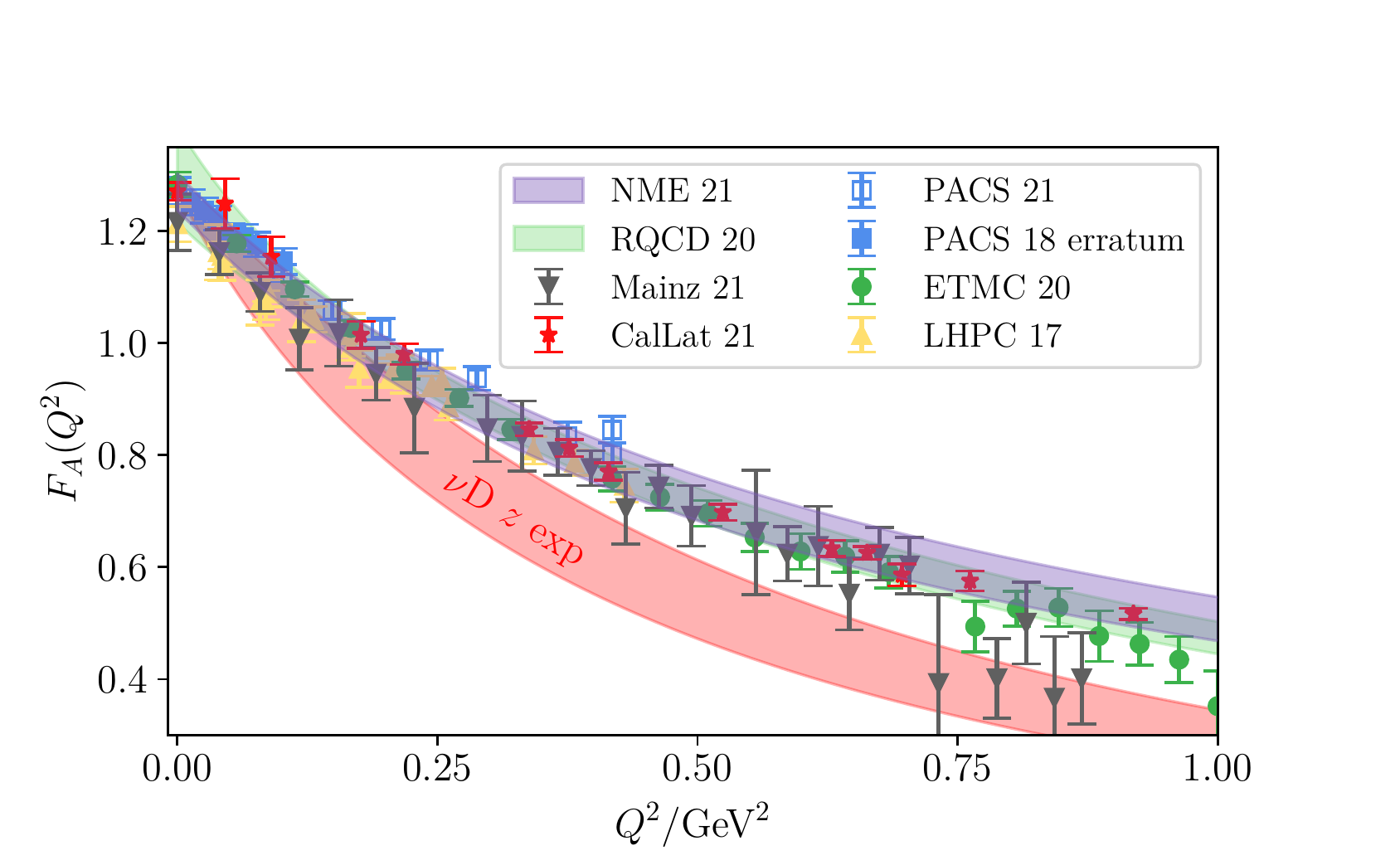}
    \caption[fig:MQ]{Isovector axial  form factors of the nucleon vs.\ squared momentum transfer $Q^2=-q^2$.
        Left: results as in Fig.~\ref{fig:FV} compared with dipole parametrization for three choices of the axial ``mass'' $M_A$;
        from Ref.~\cite{Park:2021ypf}.
        Right: compendium of results~\cite{Meyer:2022mix} compared with the $z$ expansion from $\nu\rm D$ 
        scattering~\cite{Meyer:2016oeg}; shown here are continuum limit fits from RQCD~\cite{RQCD:2019jai} and
        NME~\cite{Park:2021ypf}, and single-ensemble data points from LHPC~\cite{Green:2017keo, Hasan:2017wwt},
        PACS~\cite{Ishikawa:2018rew, Shintani:2018ozy, Ishikawa:2021eut}, ETMC~\cite{Alexandrou:2020okk},
        CalLat~\cite{Meyer:2021vfq} and Mainz~\cite{Schulz:2021kwz}; from Ref.~\cite{Meyer:2022mix}.}
    \label{fig:FA}
\end{figure}

The status of the axial form factor is less settled.
The normalization $F_A(0)=g_A=1.2754(13)$~\cite{ParticleDataGroup:2022ssz}) is known from neutron beta decay.
FLAG~\cite{Aoki:2021kgd} quotes 
averages that are  consistent but much less precise: $g_A=1.246(28)$
\{$g_A=1.248(23)$\} for 
$2+1+1$ \{$2+1$\} flavors, based on Refs.~\cite{Chang:2018uxx, Berkowitz:2017gql, Gupta:2018qil} \{Refs.~\cite{Liang:2018pis,
Harris:2019bih}\}.
The $Q^2$ dependence is shown in Fig.~\ref{fig:FA}, both from work with all sources of uncertainty under control~\cite{Park:2021ypf}
and from a compendium~\cite{Meyer:2022mix}.
Several lessons can be taken from these plots and details inferred from them.
The most striking is how the lattice data---for individual ensembles at nonzero lattice spacing and unphysical light-quark mass, but
also for continuum--physical-mass extrapolations---lie systematically above inferences from experiment.
That said, the \emph{slopes} agree at $Q^2=0$.
Two mature works obtain axial radii $r_A=0.654(47)~\text{fm}$~\cite{Park:2021ypf} 
and $r_A=0.670(31)~\text{fm}$~\cite{RQCD:2019jai} in the continuum limit.
These values are based on a model-independent parametrization of the shape founded on analyticity and unitarity, known as the
$z$~expansion.
Turning to phenomenology, the same approach to the form-factor shape and minimal assumptions on modeling the deuteron has been
used~\cite{Meyer:2016oeg} finding $r_A=0.68(16)~\text{fm}$.
Similar results with further assumptions obtain similar values with $3\%$ quoted uncertainty~\cite{Bernard:2001rs, Bodek:2007ym}.
The agreement is (now) quite good.
Note that these radii correspond closely to the black dashed line with axial ``mass'' $M_A=1.026$~GeV in the left panel of
Fig.~\ref{fig:FA}, which fails at larger values of~$Q^2$.

Earlier lattice-QCD calculations reported smaller values of the axial radius.
Such values can be extracted via the unphysical but traditional dipole form~\cite{RQCD:2019jai},
but with higher statistics the dipole leads to fits of poorer quality~\cite{Park:2021ypf, RQCD:2019jai}.
A smaller radius would increase the $\nu A$ quasielastic cross section~\cite{Hill:2017wgb} with obvious implications for neutrino
experiments.
Even with agreement for the radius, i.e., the slope, the departure of the lattice-QCD results from phenomenology for
$Q^2\gtrsim0.3~\text{GeV}^2$ influences the quasielastic cross section~\cite{Meyer:2022mix}.
An important goal, which should be achievable during the next few years, is a continuum-limit, physical-pion-mass parametrization of
the axial form factor up to, say, $Q^2=1.3~\text{GeV}^2$.

At DUNE neutrino energies, it is also necessary to understand processes in which additional pions are produced, eventually reaching
deep inelastic scattering (DIS).
In the resonance region, lattice QCD can provide transition form factors, e.g., $N\to\Delta$.
In principle, the nonzero width can be treated rigorously using the techniques described in Sec.~\ref{sec:weak} for $B$ decays to
vector mesons.
For the shallow inelastic region, which has too many pions to identify hadronic resonances but too low energy for the
operator-product expansion to hold as in DIS, there is very little information.
In this region, one can use lattice QCD to compute the hadron tensor of the nucleon~\cite{Kronfeld:2018qcd}, which, as with the
nucleon form factors, is used in nuclear many-body theory.
Work in this direction has started recently~\cite{Liang:2019frk}.
It is difficult to forecast an uncertainty at this stage---the 20\% listed in Table~\ref{tbl:milestones} is meant to suggest that a
calculation with a full error budget may be feasible on this time scale.
In the DIS region, calculations of nucleon parton distribution functions (PDFs), discussed in Sec.~\ref{sec:pdf}, will play a role.

Beyond single-baryon matrix elements lie calculations of multi-nucleon systems.
For practical reasons, these will be limited to a few nucleons, and often unphysically heavy pions, for at least a decade.
These calculations will be relevant because they can be used to constrain low-energy constants of the chiral effective theory used
to build up a systematic, theoretically based model of the nucleus.
The most prominent example is the calculation of nuclear two-body currents: in QCD language, these are matrix elements of the form
$\langle NN|J|NN\rangle$, where the $NN$ states can be bound (the deuteron) or unbound.
Exploratory calculations of axial-current matrix elements for $A=2$ and $A=3$ systems are
underway~\cite{Savage:2016kon,Chang:2017eiq,Parreno:2021ovq}.
Calculations of moments of nuclear PDFs relevant to neutrino-nucleus scattering in the DIS region have also been performed, although
only for unphysically large quark masses so far~\cite{Winter:2017bfs, Detmold:2020snb}.
See Ref.~\cite{Davoudi:2020ngi} for a recent review.

Also important for neutrino scattering is neutral-current elastic scattering.
The technical issues run parallel to those for the charged current, except that the isoscalar current can interact with sea quarks,
requiring propagators that are computationally much more demanding than those for valence quarks.
As a consequence, the precision of neutral-current matrix elements will be lower than their connected counterparts.

The PMNS matrix contains two additional $CP$-violating phases if neutrinos are Majorana particles.
This possibility can be explored via the neutrinoless double-beta ($0\nu\beta\beta$) decay of certain nuclei, which is discussed in
Sec.~\ref{sec:B-L}.

Further details can be found in a contribution to Snowmass on neutrino-nucleus scattering that contains perspectives from lattice
QCD, nuclear many-body theory, experiment, and event generators~\cite{Ruso:2022qes}.
Reference~\cite{Ankowski:2022thw} examines the connection to electron-nucleus scattering, and Ref.~\cite{Campbell:2022qmc} covers
event generators throughout high-energy physics.

%% file: energy/intro.tex
Most of the applications of computational lattice gauge theory to particle physics are QCD.
At the energy frontier, lattice QCD is just as important as elsewhere, providing determinations of $\alpha_s$ and the quark masses,
as well as calculations of the parton distribution functions that inform both tests of QCD and searches for signals of new phenomena
rising above the Standard-Model background.
To a large extent, the searches are motivated by the desire to understand the origin of electroweak symmetry: is it the
Standard-Model Higgs sector or something else?
Some ideas for ``something else'' take inspiration from QCD and posit a confining gauge theory that (to be compatible with LHC
measurements) is well-described by the Standard Model at sub-TeV energies.
Lattice QCD for the energy frontier is discussed in Secs.~\ref{sec:asmq}, \ref{sec:pdf}, and \ref{sec:hot}; lattice BSM in
Sec.~\ref{sec:composite} with related topics in Sec.~\ref{sec:cosmic}.

%% file: energy/asmq.tex
The 2012 discovery of the Higgs-like resonance at 126 GeV by the ATLAS~\cite{Aad:2012tfa} and CMS~\cite{Chatrchyan:2012xdj}
experiments at the Large Hadron Collider (LHC) provided a watershed insight into the origin of electroweak symmetry breaking.
Lattice-QCD results help turn experimental studies of this particle into a tool for further discovery, starting with the fundamental
couplings of QCD.
Of particular importance are analyses that determine from hadronic properties the strong coupling $\alpha_s$ and the quark masses,
particularly those of charm and bottom.
These quantities are needed to confront measurements of Higgs-boson branching ratios with Standard-Model predictions.

\begin{figure}[b]
    \includegraphics[height=0.275\textheight]{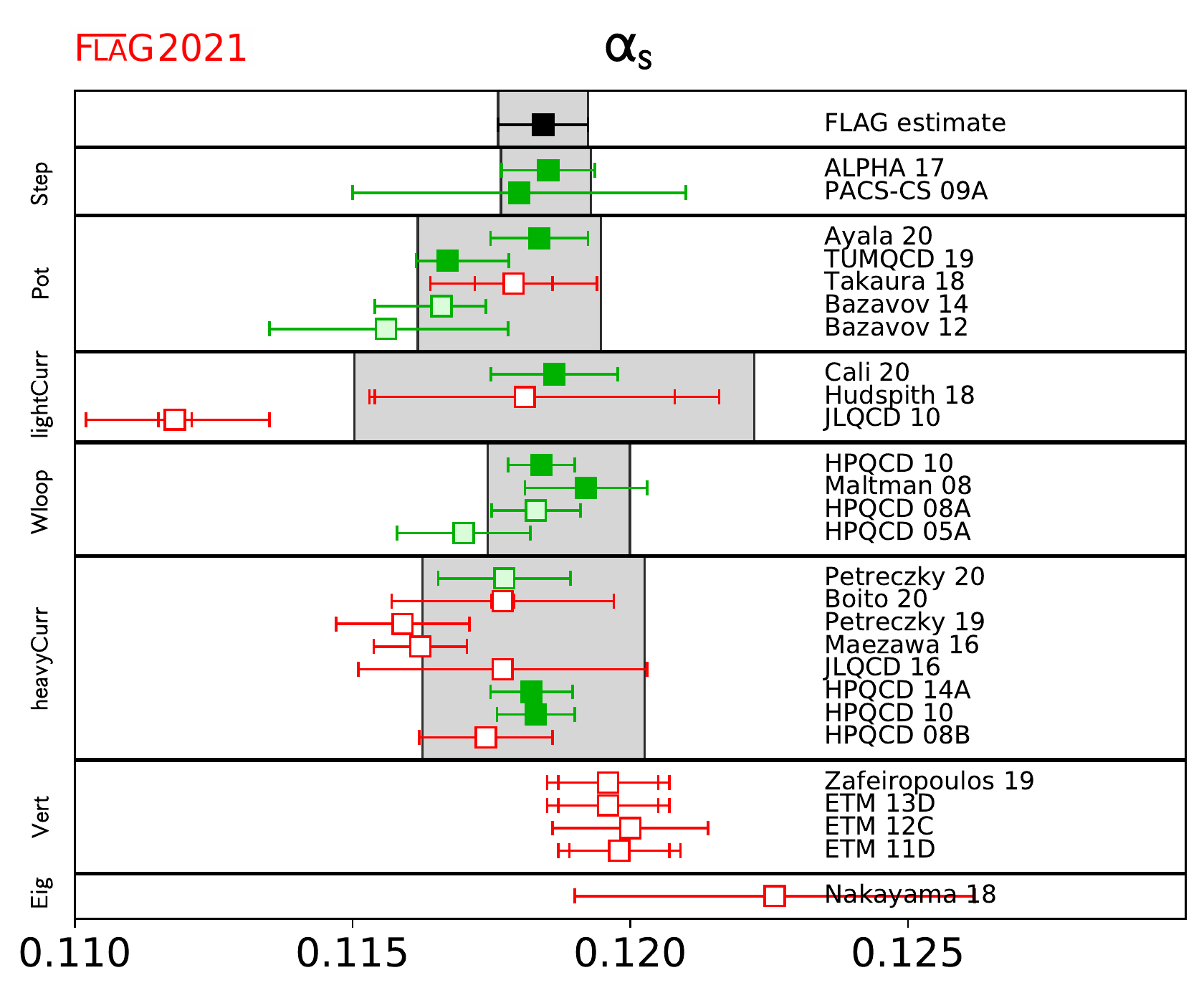} \hfill
    \includegraphics[height=0.275\textheight]{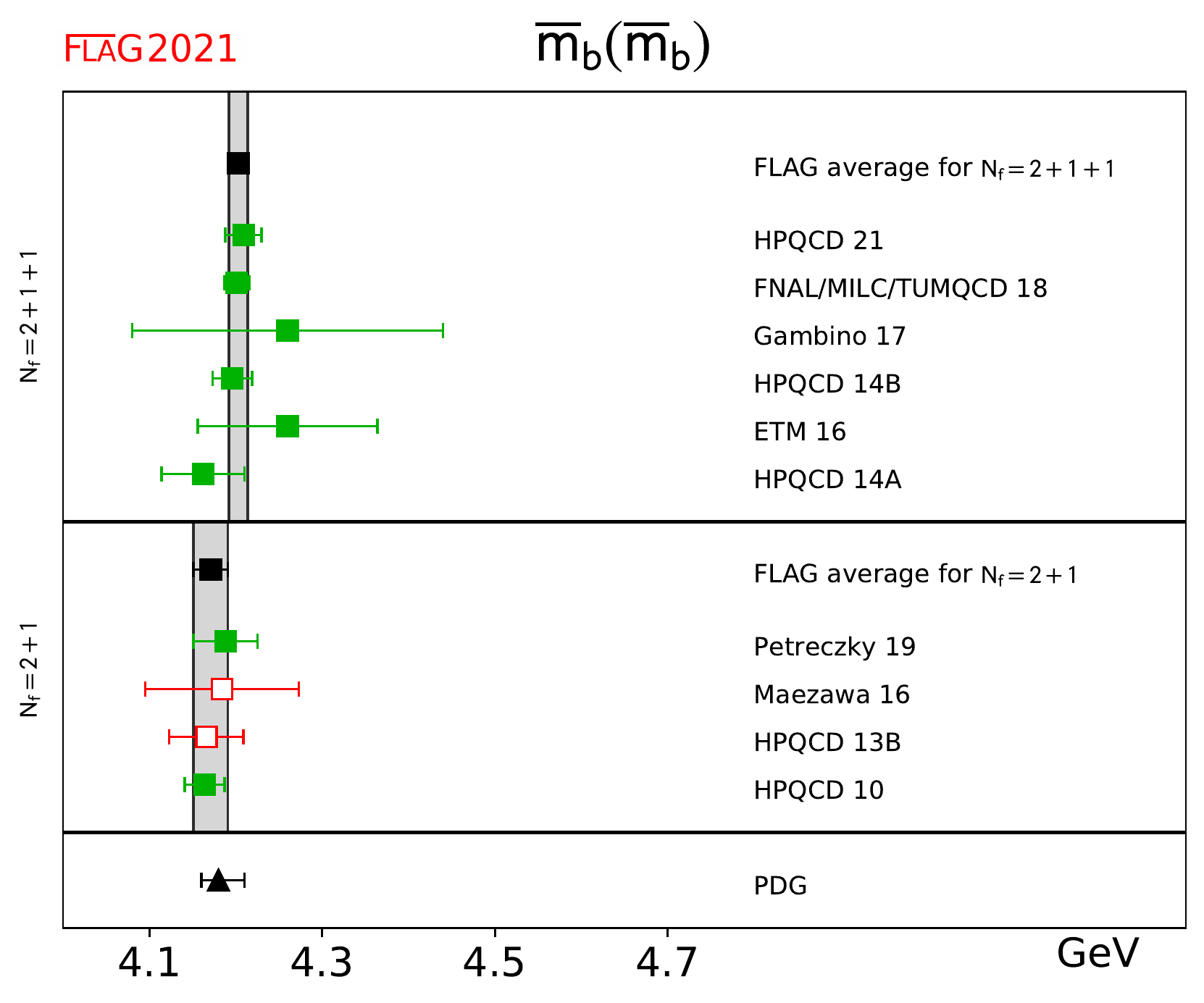}
    \caption[fig:MQ]{Comparisons of $\alpha_s(m_Z)$~\cite{Maltman:2008bx, PACS-CS:2009zxm, McNeile:2010ji, Chakraborty:2014aca,
        Bruno:2017gxd,Bazavov:2019qoo,Cali:2020hrj,Ayala:2020odx} (left) and bottom quark mass~\cite{McNeile:2010ji, 
        Colquhoun:2014ica, ETM:2016nbo, Gambino:2017vkx, FermilabLattice:2018est, Hatton:2021syc} (right).
        From FLAG 2021~\cite{Aoki:2021kgd}.
        Green symbols are included in the averages (gray bands); red symbols fall short of certain criteria and are omitted from 
        the average.}
    \label{fig:MQ}
\end{figure}

In both cases, several methods yield consistent results, with uncertainties below the percent level.
Figure~\ref{fig:MQ} shows comparisons of recent lattice-QCD results for $\alpha_s(m_Z)$ (left)~\cite{Maltman:2008bx,
PACS-CS:2009zxm, McNeile:2010ji, Chakraborty:2014aca, Maezawa:2016vgv, Bruno:2017gxd, Petreczky:2019ozv, Bazavov:2019qoo,
Cali:2020hrj, Ayala:2020odx, Petreczky:2020tky} \index{Energy Frontier!EF05} and for the bottom-quark mass
(right)~\cite{McNeile:2010ji, Chakraborty:2014aca, Colquhoun:2014ica, Yang:2014sea, ETM:2016nbo, Maezawa:2016vgv, Gambino:2017vkx,
FermilabLattice:2018est, Lytle:2018evc, Petreczky:2019ozv, Hatton:2021syc}, \index{Energy Frontier!EF01} compiled by the Flavor
Lattice Averaging Group (FLAG)~\cite{Aoki:2021kgd}.
It is worth noting that lattice QCD is the only way to determine the light-quark masses, $m_s$, $m_d$, and $m_u$ with any meaningful
precision.
Here, too, the results have become impressively precise~\cite{FermilabLattice:2018est}.

Although LHC measurements of Higgs coupling exceed expectations, the current precision of the quark masses, and probably also
$\alpha_s$, suffices for the coming decade.
For experiments at future $e^+e^-$ or $\mu^+\mu^-$ Higgs factories, some refinement in the precision of $\alpha_s$ is warranted,
while the present level of precision in the quark masses suffices~\cite{Lepage:2014fla}.
That said, many of the most precise results for $\alpha_s$ and quark masses stem from the same set of ensembles of gauge-field
configurations~\cite{MILC:2010pul, MILC:2012znn, Bazavov:2017lyh}, with staggered sea quarks.
Confirming determinations of quark masses from sets of ensembles with domain-wall or improved-Wilson sea quarks are, thus,
worthwhile; see, for example Ref.~\cite{ExtendedTwistedMass:2021gbo}.

Bottom- and charm-quark masses and $\alpha_s$ can also be extracted from high-energy decay and scattering processes, analyzed with
perturbative QCD.
A comprehensive survey of $\alpha_s$ determinations can be found in a contribution to Snowmass~\cite{dEnterria:2022hzv}.

%% file: energy/pdf.tex
At the LHC, the Higgs boson is produced in $pp$ collisions.
Therefore, like any process, the predictions of the production cross section depend on the parton distribution functions (PDFs).
Indeed, given the crucial role the PDF description of structure functions in $ep$ deep-inelastic scattering (DIS), it has been a
long-standing goal of lattice QCD to compute them.
This is a challenging problem, not least because the PDFs are functions of a kinematic variable, namely Bjorken~$x=-q^2/2p\cdot q$,
where $p$ is the target 4-momentum and $q$ the momentum transfer.

PDFs (and the related distribution amplitudes of high-energy exclusive scattering processes) are defined via operators entailing a
light-like separation, which is clearly inaccessible in the Euclidean framework of numerical lattice QCD.
One way to circumvent these problems is to focus on moments of $x^n$, which the operator-product expansion (OPE) expresses as matrix
elements of local operators, reducing the problem to that of the form factors and charges discussed in Secs.~\ref{sec:weak}
and~\ref{sec:nu}.
Unfortunately, higher moments are related to high-dimension operators, which mix (with a lattice as the ultraviolet regulator) under
renormalization with lower-dimension operators.
Thus, lattice QCD has been used to compute only the first few moments of several PDFs; for recent work on nucleon-PDF moments, see
Ref.~\cite{Mondal:2020cmt, Mondal:2020ela}.
Higher moments are accessible by introducing an intermediate ``smearing'' scale~\cite{Davoudi:2012ya, Monahan:2015lha}, taking the
continuum limit, and matching back to standard continuum renormalization schemes.

The moments can also be obtained by taking a step back to consider matrix elements of the form $\langle N(p)|J(z)J'(0)|N(p)\rangle$,
where $J^{(\prime)}$ are currents of some sort, $p$ is the momentum of the hadron (e.g., the nucleon~$N$), and $z$ is a (short)
distance.
For lattice QCD, these matrix elements are four-point functions depending on the Lorentz invariants $z^2$ and $\nu=z\cdot p$.
In the original DIS problem, the currents are electromagnetic, but here they can have different quantum numbers and even different
quark content~\cite{Detmold:2005gg, Braun:2007wv, Chambers:2017dov}.
The continuum limit of these objects can then be analyzed with the OPE to obtain expressions with the same operator matrix elements
as in DIS but different Wilson coefficients.
Such factorizable current-current matrix elements can also be used to obtain the Bjorken-$x$ dependence, either via the hadron
tensor~\cite{Liu:1993cv,Liang:2019frk} of the matrix element of two electromagnetic currents or a more general class known as ``good
lattice cross sections''~\cite{Ma:2014jla, Ma:2017pxb}.

Another way to compute the Bjorken-$x$ dependence of the PDFs directly is via the matrix element %
$\langle N(p)|\bar{q}(z)W(z,0)q(0)|N(p)\rangle$, where now $q$ ($\bar q$) is an (anti)quark field.
(For the gluon PDF, replace the quark fields with gluon field-strength tensors.) %
This idea was invigorated when Ji~\cite{Ji:2013dva} introduced the large-momentum effective field theory to show how to relate a
distribution---known as the quasi-PDF---with spacelike (i.e., Euclidean) $z^2$ to the usual Minkowski PDF~\cite{Xiong:2013bka,
Ji:2014gla, Ji:2020ect}.
Early calculations of the quasi-PDF~\cite{Lin:2014zya, Alexandrou:2015rja} stimulated a lot of theoretical
attention~\cite{Ji:2015jwa, Monahan:2016bvm, Radyushkin:2016hsy, Radyushkin:2017cyf, Radyushkin:2018nbf, Constantinou:2017sej,
Alexandrou:2017huk, Chen:2017mzz, Ji:2017oey, Ishikawa:2017faj, Green:2017xeu, Stewart:2017tvs, Izubuchi:2018srq, Li:2020xml,
Chen:2020ody, Ji:2020brr, Gao:2021hxl, LatticePartonCollaborationLPC:2021xdx}.
One of these developments is a distribution known as the pseudo-PDF, which can again be related via perturbative matching to the
Minkowski PDF~\cite{Radyushkin:2017cyf}.
Starting from the position space matrix element, the quasi-PDF is defined as a Fourier transform in $z$, while the pseudo-PDF is a
Fourier transform in $\nu$~\cite{Monahan:2018euv}.
Recovering the PDF requires taking $p_z\gg \Lambda_\text{QCD}$ at fixed $x$ for the quasi-PDF or $z^2\to0$ at fixed $\nu$ for the
pseudo-PDF.
Both quasi-PDFs~\cite{Lin:2018pvv, Alexandrou:2016jqi, Alexandrou:2018pbm, Alexandrou:2019lfo, Alexandrou:2020uyt,
Alexandrou:2020qtt, Gao:2021dbh}, and pseudo-PDFs~\cite{Orginos:2017kos, Karpie:2018zaz, Karpie:2019eiq, Joo:2019jct, Joo:2020spy,
Karpie:2021pap, DelDebbio:2020rgv, Bhat:2022zrw} are areas of active study.

These calculations are very challenging, so many studies are carried out for the simplest hadron, namely the pion.
In addition to obstacles facing all calculations, some of these methods require renormalization, and certain approaches require some
sort of inverse transform---such as the inverse Laplace transform.
On finite data sets, this problem is numerically ill-posed, so active dialog and research will be needed to attack or circumvent~it.

For more information on lattice-QCD calculations of PDFs, see Ref.~\cite{Constantinou:2022yye}.
The interplay of traditional approaches to PDFs with lattice QCD is explored in a 2017 community whitepaper~\cite{Lin:2017snn} and
two contributions to Snowmass~\cite{Amoroso:2022eow, Hou:2022sdf}.
Reference~\cite{Lin:2017snn} argues that a calculation of the isovector proton PDF at the 12\% level for $x\in[0.7, 0.9]$ will
improve our knowledge of the PDF at $x\sim1$ by more than 20\%.
This region is relevant for DUNE and for high-mass, new-physics searches at the LHC experiments ATLAS and CMS.
Given recent progress, such precision may be possible during the coming decade.
More difficult, but also under active research, are extensions of the collinear PDFs discussed here: generalized parton
distributions and transverse momentum distributions describe short-distance hadron structure in greater
detail~\cite{Constantinou:2022yye}, and their study with lattice QCD has synergy with the electron-ion
collider~\cite{Khalek:2022bzd}.

%% file: energy/hot.tex
As the universe cooled, it passed through a phase transition in which a liquid of quarks and gluons condensed into a gas of
hadrons~\cite{Muller:2013dea}.
The high-energy phase, which essentially follows from asymptotic freedom, is known as the quark-gluon plasma (QGP).
Two landmark results from lattice QCD are that the transition (at zero baryon density) is a smooth crossover~\cite{Aoki:2006we,
Bhattacharya:2014ara} at a temperature around $T_\text{c}\approx155$~MeV~\cite{Aoki:2006br, Aoki:2009sc, Borsanyi:2010bp,
Bazavov:2011nk, HotQCD:2018pds}.
In a world with massless light quarks, the transition would be second order, based on chiral symmetry.
Before definitive lattice-QCD studies were available, a first-order transition was often assumed, which would mean that bubbles of
hadronic matter would emerge from the QGP as the universe expands.
The up, down, and strange masses (i.e., the corresponding quark-Higgs Yukawa couplings) are large enough to soften the transition:
no bubbles.
The crossover temperature can be directly tested from the freeze-out of particle production in heavy-ion
collisions~\cite{Andronic:2017pug}.
It is fair to say these results from lattice QCD and experiment have changed our conception of the universe.

At zero temperature, the equation of state has been further elucidated~\cite{Borsanyi:2010cj, Borsanyi:2013bia, HotQCD:2014kol}.
The phase transition might become first order at nonzero baryon density, $\mu$, with a line in the $\mu$-$T$ plane ending in a
critical point.
A major focus of QCD thermodynamics now is to find this critical point.
The tools of this investigation are lattice QCD and the beam-energy scan of the Relativistic Heavy-Ion Collider at BNL.
A coordinated investigation of the phase transition in this region has been devised by experimentalists and theorists, including
several members of the U.S.\ lattice-QCD community~\cite{Bazavov:2018qcd}.
The key challenge for Euclidean gauge theory is that nonzero $\mu$ implies a quark determinant that is not positive definite and,
hence, a sign problem for the Monte Carlo method.
The phase diagram must therefore be explored at imaginary $\mu$ (for which the sign problem goes away) or via Taylor expansions of
thermodynamic observables around $\mu=0$, variants of multiparameter reweighting, the density of state method, or a complex Langevin
approach.

The study of hot, dense QCD is an enormous subject, and future plans will be spelled out in a future long-range plan for nuclear
science, rather than at Snowmass.
Some topics of ongoing and near-term interest include the phases and properties of baryon-rich QCD, microscopy of the QGP using
heavy-quark probes, the nature of QCD phase transitions, electromagnetic probes of QGP, and jet-energy loss in and viscosities of
the QGP~\cite{Bazavov:2018qcd}.

%% file: energy/composite.tex
If the measured Higgs-boson branching ratios deviate from the predictions of the Standard Model, speculation will ensue about the
true nature of the observed state.
One possibility is that it is a composite of more fundamental building blocks that interact via a new strong
force~\cite{Contino:2010rs, Panico:2015jxa, DelDebbio:2018szp, Banerjee:2022xmu}.
The possible composite nature of the observed Higgs boson can be studied in lattice gauge theories with fermion content that slows
the running of the gauge coupling, so that it is nearly conformal over several decades of energy scale.
Computations of the spectrum of such theories have repeatedly found a light scalar boson, a scalar almost as low in mass as the
pseudoscalars~\cite{Fodor:2016wal, Ayyar:2017qdf, Ayyar:2018zuk, Ayyar:2018glg, Brower:2015owo, Hasenfratz:2016gut,
Appelquist:2016viq, Appelquist:2018yqe, LatticeStrongDynamics:2020uwo}.
This behavior is completely different from QCD, where the scalar (the $f_0(500)$, often called the~$\sigma$) is a massive, broad
resonance, while the pseudoscalars (the pions) are the lightest particles, owing to the Nambu-Goldstone mechanism.
For this reason, near-conformal gauge theories are interesting in their own right, as well as being a part of particle-physics
phenomenology beyond the Standard Model.

The light scalar boson in these scenarios is often called the ``Higgs impostor'' because (in part, by design, to comport with LHC
measurements) its properties are close to the Standard-Model Higgs boson.
To distinguish the composite scenario from the Standard Model, it is interesting to explore the rest of the spectrum.
In analogy with QCD and chiral perturbation theory, these results can be mapped to an effective field theory framework to make
contact with phenomenology~\cite{Kilic:2009mi, Daci:2015hca, Kribs:2018oad}.
Because it is unlikely that any of the simulated models are realized in nature, it is important to uncover general 
features~\cite{Brower:2018qcd}.
If the additional states (beyond the Higgs) are seen at the LHC, the first step in identifying where to start dedicated studies is
by matching these general features.
More challenging is a study of the anomalous dimension of four-fermion operators, which are necessary to understand whether the
composite scalar boson generates mass for quarks and leptons~\cite{Panico:2015jxa}.

If the Higgs boson is the quantum of a  fundamental field, as in the Standard Model, it could couple to non-Standard-Model fields.
To the extent that such interactions are detected by the Higgs boson's effect on nucleons, the sigma terms discussed in 
Sec.~\ref{sec:clfv} are relevant; for other hadrons analogous matrix elements are also straightforward to compute.
These kinds of QCD matrix elements remain relevant  in impostor scenarios too.

%% file: cosmic.tex
Similarly to Higgs physics, lattice gauge theory can play a role in astrophysics and cosmology either through non-QCD confining 
gauge theories in a dark sector or through QCD itself to determine interaction strengths with Standard-Model matter.
Below we mention a few points of contact of lattice calculations that play a supporting role at the cosmic frontier.

In the direct detection of dark matter, the energy transfers are expected to be low enough so that only $q^2=0$ nucleon matrix
elements are needed.
As discussed in Sec.~\ref{sec:clfv}, these are the same matrix elements needed for CLFV.
Calculations of the needed quantities, such as $\sigma_{\pi N}$ and $\sigma_q$ ($q\in\{s,c,b,t\}$), at the few-percent level are
expected to be possible over the next few years, which will solidify limits set on dark matter.
As noted in Sec.~\ref{sec:clfv}, some choices made in the DM literature for $\sigma_{\pi N}$ and $\sigma_s$ suggested bounds that
were more aggressive than what the latest results (cf., Fig.~\ref{fig:sigma}) support.

Below, three BSM points of interest are discussed: dark hadrons as dark-matter candidates (Sec.~\ref{sec:dm}), the QCD axion
(Sec.~\ref{sec:dm-wave}), and the possibility of the dark-hadron thermodynamics having a first-order phase transition
(Sec.~\ref{sec:cosmo}).

\subsection{Particle-like dark matter}
\label{sec:dm}
\index{Cosmic Frontier!CF1}
\index{Theory Frontier!TF08}
\index{Energy Frontier!EF10}

Recently, models of the dark sector with QCD-like confining forces have been examined for their phenomenological viability.
To make headway, lattice-gauge-theory calculations of the spectrum of the proposed confining theories have been undertaken.
This topic is also noteworthy because it led to collaborations between dark-matter model builders and lattice experts, particularly
in the U.S\@.
This body of work is reviewed in Ref.~\cite{Kribs:2016cew}.

The underlying strong coupling in a potential composite dark sector precludes the use of perturbation theory for calculating
quantities of interest, so that lattice gauge theory is necessary to fully understand the physics of such models.
As in QCD, one is interested in the thermodynamics of the dark sector, the spectrum of dark hadrons, and their form factors.
The identity of the lightest dark hadron is also an open question: it could be a baryon (a boson for an even number of dark colors),
a meson, or a glueball.
Definitive results for dark glueballs probably lie beyond the next few years, but otherwise, exciting developments can be expected in the neare term.

\subsection{Wave-like dark matter}
\label{sec:dm-wave}
\index{Cosmic Frontier!CF2}

In QCD, the axion is a new field that couples to the strong $CP$-violating term in the Lagrangian, in such a way that it can
dynamically remove the dependence on $\theta_\text{QCD}$ and $\arg\det Y$ (cf., Sec.~\ref{sec:edm}).
With nonzero up-quark mass a firm result from lattice QCD~\cite{Fodor:2016bgu, Giusti:2017dmp, FermilabLattice:2018est, 
Alexandrou:2020bkd}, the motivation for the QCD axion is strong.
In recent years, there have been several works using lattice gauge theory to study axion phenomenology, with some emphasis on
cosmology.
The axion mass (and decay constant $f_a$) is related to the topological susceptibility $\chi_t=\int d^4x\langle q(x)q(0)\rangle$ ($q$
is the topological charge density) by $m_a^2f_a^2=\chi_t$.
The task is then to compute $\chi_t$ to temperatures well above the QCD phase transition, which has been done for pure-gauge
theory~\cite{Berkowitz:2015aua, Kitano:2015fla} and for QCD~\cite{Borsanyi:2015cka,*Borsanyi:2016ksw, Bonati:2015vqz,
Petreczky:2016vrs}.
(See also Ref.~\cite{Azcoiti:2016zbi,*Azcoiti:2017jsh} for further considerations.)
From the (steep, $\chi_t\sim T^{-8}$) fall-off, the axion relic density can be computed and a mass inferred from assuming all dark
matter consists of axions~\cite{Adams:2022pbo,Blinov:2022tfy}.

\subsection{Dark energy and cosmic acceleration}
\label{sec:cosmo}
\index{Cosmic Frontier!CF5}

As the universe evolved from the Big Bang, the phase transitions of particle physics influenced the expansion and cooling.
If a confining dark sector exists, as in the models mentioned in Sec.~\ref{sec:dm}, then it is important to understand whether the
confining transition is a smooth crossover (as it is for QCD with physical up-, down-, and strange-quark masses; cf.\
Sec.~\ref{sec:hot}) or a first-order transition (as it would be in QCD with smaller quark masses).
Thus, in addition to studying the spectrum of confined dark hadrons, it is useful to study the thermodynamics of these models as
well~\cite{Brower:2018qcd}, leveraging the extensive experience with QCD~\cite{Bazavov:2018qcd}.
An especially intriguing idea is that the violent behavior accompanying a first-order phase transition of the dark sector would
leave an imprint on gravitational waves~\cite{Schwaller:2015tja}.

%% file: theory/intro.tex
This section provides short summaries on topics not covered in detail elsewhere: the connection of lattice supersymmetry to
holography, the AdS/CFT correspondence, and string theory (TF01 in Sec.~\ref{sec:susy}); a short survey on the application of
effective field theories in numerical lattice QCD (TF02 in Sec.~\ref{sec:eft}); and computational work on conformal field theories
(TF03 in Sec.~\ref{sec:cft}).

Material for most of the other topical groups in the Theory Frontier can be found in other sections.
In applications to weak decays and the vacuum polarization for the muon $g-2$, lattice QCD is now a precision technique (TF06 in
Secs.~\ref{sec:weak} and~\ref{sec:g-2}).
Collider phenomenology is at the Energy and Rare \& Precision Frontiers (TF07 in Secs.~\ref{sec:rare} and~\ref{sec:energy}).
Lattice gauge theories beyond the Standard Model provide information on composite Higgs bosons and composite dark matter (TF08 in
Secs.~\ref{sec:composite} and~\ref{sec:dm}).
Section~\ref{sec:cosmic} contains further information related to astroparticle physics and cosmology (TF09): axion properties
(Sec.~\ref{sec:dm-wave}) and composite dark sector implications for gravitational waves (Sec.~\ref{sec:cosmo}).
Lattice QCD influences the theory of neutrino physics via inputs to neutrino cross sections (TF11/NF08 in Sec.~\ref{sec:nu}).

%% file: theory/susy.tex
Because supersymmetry is a spacetime symmetry, it is not straightforward to formulate a lattice field theory with exact
supercharges.
Recent developments with orbifolding and with topological field theory have, however, made the construction of (some) supersymmetric
lattice gauge theories possible~\cite{Catterall:2009it}.
It is now possible to address several nonperturbative questions in supersymmetric field theories.

One set of questions has to do with holography and the gauge/gravity duality, which is often used to relate a strongly coupled
(supersymmetric) gauge theory to a weakly coupled and, thus, tractable gravity problem.
Lattice supersymmetric Yang-Mills (SYM) simulations start by checking reliable analytic results and then proceed to weaker coupling
to learn about strongly coupled gravity.
For example, simulations of SYM quantum mechanics agree well with predictions for Dirichlet-0 branes~\cite{Berkowitz:2016jlq}.
Similarly, $2d$ lattice SYM with maximal supersymmetry confirms results of the black-hole--black-string phase
transition~\cite{Catterall:2010fx, Catterall:2017lub}.

The 2019 USQCD whitepaper proposes a few lines of investigation~\cite{Brower:2018qcd}.
One avenue of exploration is to test $S$~duality---the relationship in $\mathcal{N}=4$ SYM between $g^2/4\pi$ and $4\pi/g^2$---with
numerical simulations.
In the Coulomb phase of a model with spontaneous symmetry breaking, the vector boson mass is (as usual) proportional to $g^2$, while
a monopole in the model has a mass proportional to $1/g^2$.
Charged particles (electric or magnetic) can be accommodated on a torus with charge-conjugate-periodic boundary
conditions~\cite{Polley:1990tf, Kronfeld:1990qu, Wiese:1991ku}, so commonplace lattice gauge theory calculations yield the masses.
Another possibility is to monitor the free energy as a function of $g^2N$ in large-$N$, $\mathcal{N}=4$ SYM at nonzero temperature,
to test whether the known weak- and strong-coupling limits are connected by a continuous or discontinuous function of~$g^2N$.
A longer-term goal is to study supersymmetric QCD in four dimensions; work in two dimensions~\cite{Kanamori:2007yx,
Catterall:2015tta} may provide a starting point, particularly the Sugino
construction~\cite{Sugino:2004uv, Hanada:2011qx, Matsuura:2014pua}.

Further ideas can be found in Snowmass contributions on lattice $\mathcal{N}=4$ SYM~\cite{Catterall:2022qzs} and on generalized
symmetries in quantum field theory~\cite{Cordova:2022ruw}.
Researchers who would like to work on numerical lattice supersymmetry can consider a publicly available software
package~\cite{Schaich:2014pda} to get started.

%% file: theory/eft.tex
Numerical simulations generate data, which then must be combined to yield a result in the continuum limit and, in the case of QCD, 
physical quark masses.
In principle, the data all have nonzero lattice spacing and (slightly) mistuned quark masses; in practice, some data sets have 
quark masses that are considerably different from their physical values.
A~framework is needed to combine the data into final results: that framework is effective field theory~\cite{Kronfeld:2002pi}.

The guide to the continuum limit is the Symanzik effective field theory~\cite{Symanzik:1979ph, Symanzik:1983dc, Symanzik:1983gh},
which grew out of Symanzik's work on renormalization (i.e., the Callan-Symanzik equation)~\cite{Symanzik:1970rt}.
It posits a renormalized continuum field theory with a local Lagrangian, which is simply the target theory plus higher-dimension
operators multiplied by the power of the lattice spacing needed to get back to dimension~4.
Symanzik described the formalism for scalar field theories~\cite{Symanzik:1979ph, Symanzik:1983dc, Symanzik:1983gh}, while others
extended the idea to gauge theories~\cite{Weisz:1982zw, Weisz:1983bn, Curci:1983an, Luscher:1984xn} and
fermions~\cite{Sheikholeslami:1985ij, Naik:1986bn}.
The Symanzik formalism provides a framework for suppressing discretization effects order-by-order in perturbation theory (known as
Symanzik improvement)~\cite{Luscher:1985zq, Luscher:1985wf, Hart:2008sq} or even by additional powers of the lattice spacing
(nonperturbative improvement)~\cite{Jansen:1995ck}.

A large fraction of lattice-QCD data is generated with up and down quarks whose mass is larger than physical.
The tool for combining data over a range of light-quark masses is chiral perturbation theory ($\chi$PT)~\cite{Weinberg:1978kz,
Leutwyler:1993iq}, which incorporates constraints from QCD's chiral symmetries in the massless limit.
Although the original arguments for the chiral effective Lagrangian were generality, the cluster property of correlation functions,
analyticity, and unitarity, in lattice QCD unitarity is sometimes broken by choosing different sea and valence quark masses or even
different sea and valence discretizations.
The jargon for such simulations is ``partially quenched'' and ``mixed action'' with corresponding versions of
$\chi$PT~\cite{Bernard:1992mk, Bernard:1993sv, Sharpe:1997by, Sharpe:2000bc}.
It has been argued that a bounded transfer matrix can substitute for unitarity~\cite{Bernard:2013kwa} as a foundational
element~\cite{Leutwyler:1993iq} of $\chi$PT.
Numerous one-loop calculations have been worked out to support numerical computations; in the most precise cases, two-loop
calculations are necessary and available~\cite{Bijnens:2014gsa}.

Finite-volume effects for multihadron states~\cite{Luscher:1986pf, Luscher:1990ux} are discussed in Sec.~\ref{sec:xyz} in connection
with scattering and resonance properties.
That work was based on a general massive quantum field theories, which L\"uscher had used earlier to demonstrate that the
finite-volume effects on single-particle properties are exponentially suppressed~\cite{Luscher:1985dn}.
Because physical pions are so light, finite-volume effects in $\chi$PT are also considered~\cite{Gasser:1987zq}, as well as
syntheses of $\chi$PT and L\"uscher's approach~\cite{Colangelo:2005gd}.
Power-law finite-size effects can arise if the system samples topological charge incompletely, another circumstance that can be 
handled with $\chi$PT~\cite{Leutwyler:1992yt, Brower:2003yx, Aoki:2009mx, Bernard:2017npd}.
With some lattice-QCD calculations reaching a precision such that QED effects are relevant, it is also necessary to deal with
massless photons in a box; see Refs.~\cite{Duncan:1996xy, Davoudi:2018qpl, Feng:2018qpx} and a review~\cite{Patella:2017fgk} for
more information.

As mentioned in Sec.~\ref{sec:xyz}, lattice QCD was part of the motivation for the heavy-quark effective field theories
nonrelativistic QCD (NRQCD)~\cite{Lepage:1987gg, Thacker:1990bm, Lepage:1992tx} and heavy-quark effective
theory)~\cite{Eichten:1987xu, Eichten:1989zv, Eichten:1990vp}.
Discretized versions of NRQCD~\cite{Lepage:1987gg, Thacker:1990bm, Lepage:1992tx} and HQET~\cite{Hernandez:1989ch, Heitger:2003nj}
are used in heavy-quark phenomenology.
Effective field theories of heavy quarks are also used to understand and control cutoff effects of the standard fermion
formulations~\cite{El-Khadra:1996wdx, Christ:2006us, Kronfeld:2000ck, Harada:2001fi, Harada:2001fj, Oktay:2008ex}.

Effective field theories can extend the impact of lattice-QCD results in the single-, two-, and few-nucleon sectors to nuclear
many-body systems in a systematic manner.
For example, nuclear effective field theories organize the two- and few-nucleon interactions and currents at low energies within a
given power-counting scheme, assigning them given low-energy coefficients that encode the knowledge of short-distance dynamics that
are being integrated out; see Ref.~\cite{Hammer:2019poc} for a recent review.
These coefficients, in the absence of experimental data, need to be constrained by direct QCD calculations using lattice QCD.
Furthermore, direct matching of lattice QCD to in infinite volume can be bypassed by matching finite-volume effective-field-theory
calculations directly to lattice-QCD results in the same volume, with the same boundary conditions.
Such interplay between lattice QCD and effective field theories, combined with nuclear many-body calculations, is important for many
topics discussed in this document: electric dipole moments (Sec.~\ref{sec:edm}), charged-lepton-number violation
(Sec.~\ref{sec:clfv}), neutrino physics (Sec.~\ref{sec:nu}), and direct dark-matter detection (Sec.~\ref{sec:dm}).

%% file: theory/cft.tex
Conformal symmetry plays an important role in the composite Higgs models discussed in Sec.~\ref{sec:composite}, at least as a
limiting behavior.
The methods of numerical lattice field theory are widely applied to conformal systems as theoretically interesting quantum field
theories, perhaps with real applications in condensed-matter or statistical physics.
A famous example is the computation of the critical exponents of the three-dimension $\text{O}(2)$
model~\cite{Campostrini:2006ms, Hasenbusch:2019jkj}, one of which disagreed with a comparably precise experiment (on the space
shuttle)~\cite{Lipa:2000zz}, only to be confirmed by the numerical conformal bootstrap~\cite{Chester:2019ifh}.

Of course, both the lattice and the finite spacetime volume break conformal invariance explicitly.
Universality (in the sense of critical phenomena) can wash out the discretization (as in real crystals at second-order phase
transitions), and well-established relations from finite-size scaling are used to treat the finite box size; see, for example,
Ref.~\cite{Pelissetto:2000ek}.
The lattice community is developing new general purpose tools for studying conformal field theories numerically, for example the
gradient-flow renormalization group~\cite{Fodor:2017die, Kuti:2022ldb,Carosso:2018bmz, Hasenfratz:2019hpg, Peterson:2021lvb} and
radial quantization~\cite{Brower:2012vg, Brower:2018szu, Brower:2020jqj, Neuberger:2014pya}.

Further details can be found in the USQCD whitepaper~\cite{Brower:2018qcd} and (for the complementary numerical conformal bootstrap)
in a Snowmass contribution~\cite{Poland:2022qrs}.

%% file: outlook.tex
The preceding sections outline a program of lattice-QCD and -BSM calculations designed to make an impact on the experimental program
in high-energy physics.
To summarize many of the calculations needed, Table~\ref{tbl:milestones} lists several specific quantities (grouped into rough
categories) together with forecasted precision targets over the coming decade.
\input milestones
As ``forecasts'', they are contingent on many uncontrollable factors, especially funding for research and allocations of computer
time.
Historically, the least accurate USQCD forecasts have been for multiyear calculations abandoned by junior researchers taking jobs
outside the field.

The first column of Table~\ref{tbl:milestones} lists categories, with links to the sections in which they are discussed, and the
second column lists quantities of interest.
Here, $a_\mu=(g-2)_\mu/2$ is the anomalous magnetic moment of the muon, $f(q^2)$ denotes form factors for the process in the
superscript, $\Delta M$ the mass difference of a neutral-meson system, and $\epsilon^{(\prime)}$ is well-established notation for
kaon $CP$ violation.
The nucleon matrix elements are isovector axial, tensor, and scalar charges; the ``sigma'' terms (defined in Sec.~\ref{sec:clfv});
radii of nucleon form factors and axial form factor $F_A(q^2)$ itself.
The third column of Table~\ref{tbl:milestones} provides these forecasts, and the fourth column the corresponding experiments.

In many cases, the feasible precision matches that of the relevant experimental measurements.
In some cases (marked with an asterisk), the corresponding experiments require better precision than that possible in the near term.
These are simply more challenging computationally, and they represent a minimal set of topics in lattice gauge theory that will
remain relevant to particle physics beyond the coming decade.
In further cases (labeled ``NA''), precision is not the right metric; instead some aspect of the dynamics of gauge theories must be
understood via a synthesis of complementary experimental, theoretical, and numerical information.
For example, in QCD spectroscopy the structure (e.g., tetraquark vs.\ molecule) of exotic hadrons is more interesting than absolute
precision in the mass; in BSM spectroscopy, the main issues are the separation of a light scalar (the Higgs imposter) from the rest
of the spectrum and the imposter's couplings to Standard-Model particles.
Some quantities (e.g., quark masses and decay constants of pseudoscalar mesons) are not listed in Table~\ref{tbl:milestones} 
because current precision suffices for the time being.

Almost everything will rely on the mid-scale computers at BNL, Fermilab, and JLab that USQCD administers.
Many of them also require leadership-class facilities.
The underlying reason is simple: lattice gauge theory calculations proceed through a sequence of lattices with finer and finer 
spacings and, hence, larger and larger lattices.
The smaller, coarser lattices are the first steps, both in developing innovative ideas and in beginning an ``industrial-strength'' 
campaign of aimed at a specific result.
Only exploratory work (i.e., new topics not in Table~\ref{tbl:milestones}) can be completed in a few months to a year.
To make an impact on experiment require multiyear campaigns spread over many computing facilities.

Common to all numerical lattice-gauge-theory analyses are ensembles of gauge-field configurations.
Many such ensembles already exist with $2+1+1$ flavors\footnote{%
The notation $2+1$ means two equal-mass quarks, for up and down, with another tuned to the strange quark; $2+1+1$ adds
charm to the sea; $1+1+1+1$ implies the different masses for up and down.} %
of sea quark from the MIMD Lattice Computation (MILC) Collaboration~\cite{MILC:2010pul, MILC:2012znn, Bazavov:2017lyh} and the
Extended Twisted Mass (ETM) Collaboration~\cite{Baron:2010bv, Alexandrou:2018egz}, and with $2+1$ flavors of sea quark from the
RIKEN Brookhaven Columbia and United Kingdom QCD (RBC/UKQCD) Collaborations~\cite{Aoki:2010dy, RBC:2014ntl, Boyle:2017jwu}, the
Coordinated Lattice Simulations (CLS) consortium~\cite{Bruno:2014jqa, Bruno:2016plf}, and a similar effort in the
U.S.~\cite{Clover:2016csw}.
These sets of ensembles will be the starting point for the physics calculations discussed throughout this contribution.
For the most precise calculations, additional ensembles with $1+1+1+1$ flavors and/or explicit photon fields will be generated.

Table~\ref{tbl:milestones} suffers from the serious shortcoming that it does not highlight exploration and innovation.
Many lattice-QCD calculations are underway or in a stage of development that makes forecasts impossible.
For example, transverse momentum distributions are similar to but more difficult than the collinear PDFs in the table.
The transverse motion they describe may be relevant when the precision on the $W$-boson mass reaches that of the 2022 measurement
from CDF~\cite{CDF:2022hxs}.
PDFs are challenging enough: adding another layer of complexity is exciting to think through but postpones calculations of direct
relevance to experiment.
Calculations for the muon anomalous magnetic moment, $a_\mu$, are an example from recent history.
As little as ten years ago, exploration had begun but it was not clear whether useful calculations could be carried out.
Innovative methods were developed on mid-sized computer systems to bring us within reach of the precision needed by the current
Fermilab and future JPARC experiments.
It should also be noted that the enthusiasm for lattice expressed by the from $g-2$ experimental community inspired both new ideas
and hard work.

%% file: milestones.tex
\begin{table}
    \newcommand{\h}{\phantom{2}}
    \renewcommand{\o}{\phantom{1}}
    \newcommand{\s}{\phantom{*}}
    \newcommand{\fff}{\footnote{Note also the PDFs under ``Nucleon matrix elements'' in above block.}}
    \centering
    \caption{Lattice-QCD calculations supporting the U.S.\ and worldwide program in particle physics, with target precision over 
        the coming few years.
        An asterisk * indicates that the target precision \emph{falls short} of the experimental uncertainty.}
    \label{tbl:milestones} \vspace*{2pt}
    \begin{tabular*}{\textwidth}{@{\extracolsep{\fill}}ccccc}
        \hline\hline
        Category                    & Milestone                         &  Target   & Experiment(s) \\
                                    &                                   & precision & \\
        \hline
        $a_\mu=(g_\mu-2)/2$         & $a_\mu^\text{HVP, LO}$              &0.5\%\s& Muon $g-2$ (E989)\rule{0pt}{14pt} \\
        (Sec.~\ref{sec:g-2})        & $a_\mu^\text{HVP, NLO+NNLO}$        &\h1\%\s& Muon $g-2$ (E989) \\
                                    & $a_\mu^\text{HLbL}$                 & 10\%\s& Muon $g-2$ (E989) \\
        \hline
        CKM $B$ \& $D$ physics      & $f^{D \to \pi,K}(q^2)$              &\h1\%\s& Belle~II, BES~III \\     
        (Sec.~\ref{sec:weak-b+c})   & $f^{B \to D^{(*)}}(q^2)$            &\h1\%\s& Belle~II \\
                                    & $f^{B \to \pi}(q^2)$                &\h2\%\s& Belle~II \\
                                    & $f^{\Lambda_b\to p/\Lambda_c}(q^2)$ &\h2\%\s&  LHCb \\
        FCNC $B$ physics            & $f^{B \to K}(q^2)$                  &\h2\%\s& Belle~II, LHCb, ATLAS, CMS \\
        (Sec.~\ref{sec:weak-b+c})   & $f^{B \to K^*}(q^2)$                & 10\%* & Belle~II, LHCb, ATLAS, CMS \\
                                    & $f^{\Lambda_b\to\Lambda}(q^2)$      &\h2\%\s& LHCb \\
                                    & $\Delta M_{B_{(s)}}$                &\h5\%* & Belle~II, LHCb, BaBar \\
        $K$ physics                 & $f^{K \to \pi}(0)$                  &0.1\%\s& First-row CKM unitarity \\
        (Sec.~\ref{sec:weak-s+d+u}) & $\Delta M_K$                        & 20\%* & KTeV, NA48 \\
                                    & $\epsilon'/\epsilon$                & 15\%\s& KTeV, NA48 \\
                                    & $K\to\pi\nu\bar\nu$                 &\h3\%\s& NA62, K0T0 \\
        \hline
        Nucleon matrix              & Nucleon $g_A^\umd$                  &\h1\%* & Neutron lifetime puzzle \\
        elements                    & Nucleon $g_T^\umd$                  &\h1\%\s& UCNB, Nab \\
        (Secs.~\ref {sec:clfv}      & Nucleon $g_S^\umd$                  &\h3\%\s& UCNB, Nab \\
        and~\ref{sec:cosmic})       & $\sigma_{\pi N}$, $\sigma_s$        &\h5\%\s& Mu2e, LZ, CDMS \\
        (Sec.~\ref{sec:nu})         & Nucleon $r_E$, $r_M$, $r_A$         &\h5\%\s& DUNE, MicroBooNE, NOvA, T2K \\
                                    & Nucleon $F_A(q^2)$                  &\h8\%\s& DUNE, MicroBooNE, NOvA, T2K \\
        (Secs.~\ref{sec:nu}         & Nucleon tensor                      & 20\%\s& DUNE, MicroBooNE, NOvA, T2K \\
        and~\ref{sec:pdf})          & Nucleon PDFs                        & 12\%* & ATLAS, CMS, DUNE, EIC expts \\
        (Sec.~\ref{sec:B-L})        & Proton decay                        & 10\%\s& DUNE, HyperK \\
                                    & $nn\to pp$                          & 50\%* & EXO, other $0\nu\beta\beta$ experiments \\
        (Sec.~\ref{sec:edm})        & Nucleon EDM                         & 10\%* & Neutron, proton EDM experiments \\
                                    & $g_{A,T,S}$, $1<A\le4$              & 20\%* & All neutrino, DM, EDM, \ldots \\
        \hline
        Higgs + BSM                 & Light BSM spectrum                  &   NA  & ATLAS, CMS \\
        (Sec.~\ref{sec:composite})  & Anomalous dimension                 &   NA  & ATLAS, CMS \\ 
        (Sec.~\ref{sec:dm})         & Composite DM                        &   NA  & LZ, CDMS \\
        (Sec.~\ref{sec:asmq})       & $\alpha_s(m_Z)$                     &0.3\%\s& ATLAS, CMS, FCC, ILC \\               
        (Sec.~\ref{sec:susy})       & Susy                                &   NA  & ATLAS, CMS \\
        \hline
        Spectroscopy                & $XYZ$                               &   NA  & Belle (II), LHCb, BaBar, CDF, D0 \\
        (Sec.~\ref{sec:xyz})        & pentaquarks                         &   NA  & LHCb \\
                                    & exotic light hadrons                &   NA  & BES~III, CLAS, COMPASS, GlueX \\ 
        \hline
        Heavy ions (Sec.~\ref{sec:hot}) & QCD phase transition            &   NA  & (s)PHENIX, ALICE, ATLAS, CMS \\ 
        \hline\hline
    \end{tabular*}
    \vspace*{-8pt}
\end{table}

%% file: usqcd.tex
USQCD is a collaboration of almost all high-energy and nuclear physicists in the United States who are working on lattice gauge
theory.
Around 100 of USQCD's 170 members are involved in numerical projects at any given time.
The USQCD website~\cite{usqcd} covers all aspects of the USQCD collaboration and includes
the \href{https://www.usqcd.org/documents/charter.pdf}{charter},
the \href{https://www.usqcd.org/documents/code.pdf}{code of conduct}, and the current
\href{https://www.usqcd.org/members.html}{list of members}.

Overall leadership of USQCD is vested in its Executive Committee (EC)\index{USQCD!EC}.
This committee was established in 1999, with encouragement from the DOE, to organize the community, develop plans for the
infrastructure, obtain funding to carry out these plans and oversee the implementation of them.
There are nine standing members, who rotate at the rate of roughly one replacement per year.
For example, in 2018 there were three changes, in 2017, 2019, and 2021 there were none, and in 2020 there was one.
For the past six years, one member of USQCD has been an early-career scientist elected by the USQCD membership (apart from
students).
The current EC members are Robert Edwards (Chair, and USQCD Spokesperson), Thomas Blum (Deputy), Tanmoy Bhattacharya
(\emph{ex~officio}), Norman Christ, Carleton DeTar, William Detmold, Anna Hasenfratz, Andreas Kronfeld, Huey-Wen Lin (elected),
Swagato Mukherjee, and Kostas Orginos.

The principle role of the EC is to prepare proposals for mid-scale computing clusters to the DOE Offices of High Energy Physics
(HEP)\index{DOE!HEP Office} and Nuclear Physics (NP)\index{DOE!NP Office}, followed by oversight of said systems.
For many years, HEP and NP funded a single project that designed, procured, and operated dedicated clusters for the U.S.\ 
lattice-QCD community.
NP continues with this model, deploying clusters at Jefferson Lab.
HEP has moved to a different model, known as the institutional cluster, in which Fermilab and BNL design, procure, and operate the 
cluster, while the LQCD project purchases access to a certain number of nodes each year.
With an institutional cluster, there are typically several communities purchasing access.
In this way, nodes are only idle if all parts of a large, broad user base are in a lull.

The EC also coordinates proposals for software an algorithm development.
For approximately 20 years, the main funding stream has been the Scientific Discovery through Advanced Computing
(SciDAC)\index{DOE!SciDAC} program, which is a partnership between the DOE Office for Advanced Scientific Computing Research
(ASCR)\index{DOE!ASCR} and HEP and/or NP.
Over the past several years, USQCD has also been part of the Exascale Computing Project (ECP)\index{Exascale Computing Project}.
The software created under the SciDAC grants has greatly enhanced the effectiveness with which USQCD use the hardware resources,
whether leadership-class or clusters.
Software development continues under SciDAC~4 (NP only) and the ECP.
All of the software developed under the SciDAC grants is publicly available, and can be found at
\href{https://usqcd-software.github.io}{\tt https://usqcd-software.github.io/}.
At present, a NP-ASCR SciDAC~4 project is coming to a close, a proposal to the HEP-ASCR SciDAC~5 funding opportunity was recently
awarded, and a proposal to the NP-ASCR SciDAC~5 funding opportunity has been submitted.

The EC appoints the Scientific Program Committee (SPC)\index{USQCD!SPC}, which plays a major role in setting scientific priorities
and allocating USQCD resources.
Members serve terms of 3--4 years.
The current members are %
Tanmoy Bhattacharya (Los Alamos, Chair),
Alexei Bazavov (Michigan State),
Martha Constantinou (Temple),
George Fleming (Yale),
Jack Laiho (Syracuse),
Meifeng Lin (BNL),
and Sergey Syritsyn (Stony Brook). %
The SPC Chair is an \emph{ex officio} member of the~EC.
Annually, the SPC issues a call for proposals for computer time on the mid-scale computer clusters and long-term storage.
As part of its deliberations, the SPC organizes the annual USQCD All Hands' Meeting (AHM)\index{USQCD!All Hands' Meeting}, to share
its thinking and receive input from collaboration members.
The AHM also provides structured dialog between the EC and collaboration members.

The EC and the SPC solicit advice from the Scientific Advisory Board (SAB)\index{USQCD!SAB} consisting of experimenters and
phenomenologists in the various subfields of high energy and nuclear physics that depend on lattice-gauge-theory calculations.
The current members of the SAB are %
Ayana Arce (Duke, ATLAS),
Roy Briere (Carnegie Mellon, BES~III, Belle~II),
Abhay Deshpande (Stony Brook, PHENIX, EIC),
Lawrence Gibbons (Cornell, Muon~$g-2$, CMS),
Kendall Mahn (Michigan State, T2K, DUNE),
Krishna Rajagopal (MIT, theory),
Matthew Shepherd (Indiana, GlueX, BES~III), and
Jure Zupan (University of Cincinnati, theory).

In February 2020, USQCD founded a Committee on Diversity, Equity, and Inclusion (CDEI)\index{USQCD!CDEI}.
The current members of the CDEI are %
Will Detmold (MIT, chair),
Kimmy Cushman (Yale),
Joel Giedt (RPI),
Robert Edwards (\emph{ex officio}, EC Chair),
Aida El-Khadra (UIUC),
and Huey-Wen Lin (MSU).
The committee has conducted a survey on the climate in the USQCD collaboration and is currently analyzing the results.
A similar survey was conducted in 2018 at the International Symposium on Lattice Field Theory, with results reported in the 
proceedings~\cite{Aubin:2019rdf}.

%% file: landscape.tex
To supplement remarks about computing in the rest of this document, it may be helpful to survey the landscape of
lattice-gauge-theory research in the United States.
As discussed above, understanding the long-distance properties of QCD is crucial to the scientific missions of both HEP and NP.
Consequently, funding for scientists and computing comes from both Offices.
Thus, both provide funding for mid-sized computing projects aimed at lattice QCD and other gauge theories.
The HEP initiative, known as the Lattice QCD Infrastructure Research Program extension~III (LQCD), allows USQCD to purchase
resources from the institutional clusters at BNL and Fermilab.
\index{USQCD!compute clusters}
\index{Computing facilities!USQCD clusters}
The NP initiative, known as the Nuclear and Particle Physics Lattice-QCD Computing Initiative (NPPLC), designs and builds
dedicated clusters at~JLab.
The difference between ``institutional'' and ``dedicated'' has to do with the funding model, without impact on the science.
All three sites operate, as a rule, mix of CPU and GPU clusters, and they support long-range storage of large, valuable files of 
broad community interest.

The BNL, Fermilab, and Jlab clusters are a major source of computing for lattice gauge theory in the U.S\@.
Further computer time is available for lattice QCD on some university computing facilities, although the fraction of the total is
modest.
The largest source of lattice-QCD computing comes from the leadership-computing facilities (LCFs), funded by both the
DOE\index{Computing facilities!DOE LCFs} and the NSF\index{Computing facilities!NSF resources}.
The DOE facilities are the Argonne Leadership Class Facility (\href{https://www.alcf.anl.gov}{ALCF})\index{Computing facilities!DOE
LCFs!ALCF}, the Oak Ridge Leadership Class Facility (\href{https://www.olcf.ornl.gov}{OLCF})\index{Computing facilities!DOE
LCFs!OLCF}, and the National Energy Research Scientific Computing Center (\href{https://www.nersc.gov}{NERSC}).
\index{Computing facilities!DOE LCFs!NERSC}%
Some of the recent, current, and future supercomputers that have had a big impact on lattice QCD are Mira, Theta, and Aurora (ALCF);
Titan, Summit, and Frontier (OLCF); and Edison, Cori, and Perlmutter (NERSC).
The supercomputers funded and operated by the NSF are located at many universities, for example
the Texas Advanced Computing Center (\href{https://www.tacc.utexas.edu}{TACC})\index{Computing facilities!NSF resources!TACC},
which houses two computers well-suited for lattice QCD, Frontera and Stampede2.

Computing time at the LCFs is allocated via several calls for proposals\index{Computing facilities!DOE LCFs}.
Most of the supercomputers at ALCF and OLCF is allocated annually via the Innovative \& Novel Computational Impact on Theory \&
Experiment (\href{https://www.doeleadershipcomputing.org}{INCITE}) program\index{Computing facilities!DOE LCFs!INCITE},
managed by ALCF and OLCF, aiming to maximize science.
Most of the resources at NERSC are allocated annually via the Energy Research Computing Allocations Process
(\href{https://www.nersc.gov/users/accounts/allocations/2022-call-for-proposals-to-use-nersc-resources/}{ERCAP})%
\index{Computing facilities!DOE LCFs!ERCAP}in support of the
mission of the DOE Office of Science, with mid-year supplements.
Most of the rest of the ALCF and OLCF resources, as well as a small part of NERSC computing, is allocated annually via the ASCR
Leadership Computing Challenge (\href{https://science.osti.gov/ascr/Facilities/Accessing-ASCR-Facilities/ALCC}{ALCC}),%
\index{Computing facilities!DOE LCFs!ALCC} in which the other program Offices in the DOE Office of Science (e.g., HEP and NP) have
considerable influence.
NSF supercomputers are mostly allocated through the Extreme Science and Engineering Discovery Environment
\href{https://www.xsede.org}{(XSEDE)}\index{Computing facilities!NSF resources!XSEDE} on a quarterly basis.
Frontera, however, is allocated separately via three programs, the largest of which is the annual Leadership Resource Allocation
(\href{https://frontera-portal.tacc.utexas.edu/allocations/}{LRAC})\index{Computing facilities!NSF resources!LRAC} call for
proposals.
In all cases, the allocation review panels are multidisciplinary.

\index{Computing facilities!capability vs.\ capacity|(}
It is important to appreciate that leadership-class computers and the clusters fill complementary roles.
The leadership-class computers are designed for high capability; they are suited for the largest lattices, namely those with the
smallest lattice spacing at a large volume, say $(\text{6~fm})^3$, or a somewhat coarser lattice spacing with an even larger volume.
They are also best suited for mature problems with a highly automated, industrialized workflow, because the queues on
leadership-class computers are set up to accommodate such job streams.
For lattice gauge theory, high-capability computing consists of Markov chains running as long as several years, involving the 
solution of many large linear systems of equations (scaling as lattice size $\times$ lattice size) within every link of the chain.
As complicated as this sounds, the atomic operation is three-by-three matrix multiplication (and similar operations), which is the 
kind of operation any computing device can carry out.

Lattice-QCD results also rely on high-capacity computing for the statistical analysis of millions of small to medium-sized files
containing hadron correlation functions.
These analyses are the foundation of estimates of systematic uncertainties, so they require close interaction between human
researchers and the computer.
In this mode, quick turnaround is essential.
The USQCD clusters also are of moderate capability, which makes them ideal for developing new ideas into viable computing
strategies, which can entail dozens of nontrivial simulations on small or medium-sized lattices.
Such workflows would be all but impossible on leadership-class machines.
On the other hand, in USQCD's experience the queues of both dedicated and institutional clusters are not only set up to offer the
flexibility needed to foster such innovation but also can be refined with little bureaucratic effort in unforeseen circumstances.
The LQCD and NPPLC projects have also provided significant computing for excellent proposals from junior researchers (postdocs and
even advanced graduate students), who would not yet have the reputation to fare well in competition with senior scientists from all
disciplines for access to DOE or NSF leadership-class facilities.
\index{Computing facilities!capability vs.\ capacity|)}

%% file: acknowledgments.tex
This work was supported in part by grants and contracts from the U.S.\ Department of Energy, Office of Science, Offices of High Energy Physics and Nuclear Physics, from the U.S.\ National Science Foundation, and from funding agencies in Germany, Spain, and the United Kingdom.
This work was supported in part by the U.S.\ National Science Foundation under Grant No.\ NSF~PHY-1748958.
This document was prepared using the resources of the Fermi National Accelerator Laboratory (Fermilab), a U.S.\ Department of
Energy, Office of Science, HEP User Facility.
Fermilab is managed by Fermi Research Alliance, LLC (FRA), acting under Contract No.\ DE-AC02-07CH11359.

%% file: main.bbl
\begin{thebibliography}{540}%
\makeatletter
\providecommand \@ifxundefined [1]{%
 \@ifx{#1\undefined}
}%
\providecommand \@ifnum [1]{%
 \ifnum #1\expandafter \@firstoftwo
 \else \expandafter \@secondoftwo
 \fi
}%
\providecommand \@ifx [1]{%
 \ifx #1\expandafter \@firstoftwo
 \else \expandafter \@secondoftwo
 \fi
}%
\providecommand \natexlab [1]{#1}%
\providecommand \enquote  [1]{``#1''}%
\providecommand \bibnamefont  [1]{#1}%
\providecommand \bibfnamefont [1]{#1}%
\providecommand \citenamefont [1]{#1}%
\providecommand \href@noop [0]{\@secondoftwo}%
\providecommand \href [0]{\begingroup \@sanitize@url \@href}%
\providecommand \@href[1]{\@@startlink{#1}\@@href}%
\providecommand \@@href[1]{\endgroup#1\@@endlink}%
\providecommand \@sanitize@url [0]{\catcode `\\12\catcode `\$12\catcode
  `\&12\catcode `\#12\catcode `\^12\catcode `\_12\catcode `\%12\relax}%
\providecommand \@@startlink[1]{}%
\providecommand \@@endlink[0]{}%
\providecommand \url  [0]{\begingroup\@sanitize@url \@url }%
\providecommand \@url [1]{\endgroup\@href {#1}{\urlprefix }}%
\providecommand \urlprefix  [0]{URL }%
\providecommand \Eprint [0]{\href }%
\providecommand \doibase [0]{https://doi.org/}%
\providecommand \selectlanguage [0]{\@gobble}%
\providecommand \bibinfo  [0]{\@secondoftwo}%
\providecommand \bibfield  [0]{\@secondoftwo}%
\providecommand \translation [1]{[#1]}%
\providecommand \BibitemOpen [0]{}%
\providecommand \bibitemStop [0]{}%
\providecommand \bibitemNoStop [0]{.\EOS\space}%
\providecommand \EOS [0]{\spacefactor3000\relax}%
\providecommand \BibitemShut  [1]{\csname bibitem#1\endcsname}%
\let\auto@bib@innerbib\@empty
\bibitem [{\citenamefont {Detmold}\ \emph {et~al.}(2019)\citenamefont
  {Detmold}, \citenamefont {Edwards}, \citenamefont {Dudek}, \citenamefont
  {Engelhardt}, \citenamefont {Lin}, \citenamefont {Meinel}, \citenamefont
  {Orginos},\ and\ \citenamefont {Shanahan}}]{Detmold:2018qcd}%
  \BibitemOpen
  \bibfield  {author} {\bibinfo {author} {\bibfnamefont {W.}~\bibnamefont
  {Detmold}}, \bibinfo {author} {\bibfnamefont {R.~G.}\ \bibnamefont
  {Edwards}}, \bibinfo {author} {\bibfnamefont {J.~J.}\ \bibnamefont {Dudek}},
  \bibinfo {author} {\bibfnamefont {M.}~\bibnamefont {Engelhardt}}, \bibinfo
  {author} {\bibfnamefont {H.-W.}\ \bibnamefont {Lin}}, \bibinfo {author}
  {\bibfnamefont {S.}~\bibnamefont {Meinel}}, \bibinfo {author} {\bibfnamefont
  {K.}~\bibnamefont {Orginos}},\ and\ \bibinfo {author} {\bibfnamefont
  {P.}~\bibnamefont {Shanahan}} (\bibinfo {collaboration} {USQCD}),\ }\bibfield
   {title} {\bibinfo {title} {Hadrons and nuclei},\ }\href
  {https://doi.org/10.1140/epja/i2019-12902-4} {\bibfield  {journal} {\bibinfo
  {journal} {Eur. Phys. J.}\ }\textbf {\bibinfo {volume} {A55}},\ \bibinfo
  {pages} {193} (\bibinfo {year} {2019})},\ \Eprint
  {https://arxiv.org/abs/1904.09512} {arXiv:1904.09512 [hep-lat]} \BibitemShut
  {NoStop}%
\bibitem [{\citenamefont {Bazavov}\ \emph
  {et~al.}(2019{\natexlab{a}})\citenamefont {Bazavov}, \citenamefont {Karsch},
  \citenamefont {Mukherjee},\ and\ \citenamefont
  {Petreczky}}]{Bazavov:2018qcd}%
  \BibitemOpen
  \bibfield  {author} {\bibinfo {author} {\bibfnamefont {A.}~\bibnamefont
  {Bazavov}}, \bibinfo {author} {\bibfnamefont {F.}~\bibnamefont {Karsch}},
  \bibinfo {author} {\bibfnamefont {S.}~\bibnamefont {Mukherjee}},\ and\
  \bibinfo {author} {\bibfnamefont {P.}~\bibnamefont {Petreczky}} (\bibinfo
  {collaboration} {USQCD}),\ }\bibfield  {title} {\bibinfo {title} {Hot-dense
  lattice {QCD}},\ }\href {https://doi.org/10.1140/epja/i2019-12922-0}
  {\bibfield  {journal} {\bibinfo  {journal} {Eur. Phys. J.}\ }\textbf
  {\bibinfo {volume} {A55}},\ \bibinfo {pages} {194} (\bibinfo {year}
  {2019}{\natexlab{a}})},\ \Eprint {https://arxiv.org/abs/1904.09951}
  {arXiv:1904.09951 [hep-lat]} \BibitemShut {NoStop}%
\bibitem [{\citenamefont {Lehner}\ \emph {et~al.}(2019)\citenamefont {Lehner},
  \citenamefont {Meinel}, \citenamefont {Blum}, \citenamefont {Christ},
  \citenamefont {El-Khadra}, \citenamefont {Hansen}, \citenamefont {Kronfeld},
  \citenamefont {Laiho}, \citenamefont {Neil}, \citenamefont {Sharpe},\ and\
  \citenamefont {{Van de Water}}}]{Lehner:2018qcd}%
  \BibitemOpen
  \bibfield  {author} {\bibinfo {author} {\bibfnamefont {C.}~\bibnamefont
  {Lehner}}, \bibinfo {author} {\bibfnamefont {S.}~\bibnamefont {Meinel}},
  \bibinfo {author} {\bibfnamefont {T.}~\bibnamefont {Blum}}, \bibinfo {author}
  {\bibfnamefont {N.~H.}\ \bibnamefont {Christ}}, \bibinfo {author}
  {\bibfnamefont {A.~X.}\ \bibnamefont {El-Khadra}}, \bibinfo {author}
  {\bibfnamefont {M.~T.}\ \bibnamefont {Hansen}}, \bibinfo {author}
  {\bibfnamefont {A.~S.}\ \bibnamefont {Kronfeld}}, \bibinfo {author}
  {\bibfnamefont {J.}~\bibnamefont {Laiho}}, \bibinfo {author} {\bibfnamefont
  {E.~T.}\ \bibnamefont {Neil}}, \bibinfo {author} {\bibfnamefont {S.~R.}\
  \bibnamefont {Sharpe}},\ and\ \bibinfo {author} {\bibfnamefont {R.~S.}\
  \bibnamefont {{Van de Water}}} (\bibinfo {collaboration} {USQCD}),\
  }\bibfield  {title} {\bibinfo {title} {Opportunities for lattice {QCD} in
  quark and lepton flavor physics},\ }\href
  {https://doi.org/10.1140/epja/i2019-12891-2} {\bibfield  {journal} {\bibinfo
  {journal} {Eur. Phys. J.}\ }\textbf {\bibinfo {volume} {A55}},\ \bibinfo
  {pages} {195} (\bibinfo {year} {2019})},\ \Eprint
  {https://arxiv.org/abs/1904.09479} {arXiv:1904.09479 [hep-lat]} \BibitemShut
  {NoStop}%
\bibitem [{\citenamefont {Kronfeld}\ \emph {et~al.}(2019)\citenamefont
  {Kronfeld}, \citenamefont {Richards}, \citenamefont {Detmold}, \citenamefont
  {Gupta}, \citenamefont {Lin}, \citenamefont {Liu}, \citenamefont {Meyer},
  \citenamefont {Sufian},\ and\ \citenamefont {Syritsin}}]{Kronfeld:2018qcd}%
  \BibitemOpen
  \bibfield  {author} {\bibinfo {author} {\bibfnamefont {A.~S.}\ \bibnamefont
  {Kronfeld}}, \bibinfo {author} {\bibfnamefont {D.~G.}\ \bibnamefont
  {Richards}}, \bibinfo {author} {\bibfnamefont {W.}~\bibnamefont {Detmold}},
  \bibinfo {author} {\bibfnamefont {R.}~\bibnamefont {Gupta}}, \bibinfo
  {author} {\bibfnamefont {H.-W.}\ \bibnamefont {Lin}}, \bibinfo {author}
  {\bibfnamefont {K.-F.}\ \bibnamefont {Liu}}, \bibinfo {author} {\bibfnamefont
  {A.~S.}\ \bibnamefont {Meyer}}, \bibinfo {author} {\bibfnamefont
  {R.}~\bibnamefont {Sufian}},\ and\ \bibinfo {author} {\bibfnamefont
  {S.}~\bibnamefont {Syritsin}} (\bibinfo {collaboration} {USQCD}),\ }\bibfield
   {title} {\bibinfo {title} {Lattice {QCD} and neutrino-nucleus scattering},\
  }\href {https://doi.org/10.1140/epja/i2019-12916-x} {\bibfield  {journal}
  {\bibinfo  {journal} {Eur. Phys. J.}\ }\textbf {\bibinfo {volume} {A55}},\
  \bibinfo {pages} {196} (\bibinfo {year} {2019})},\ \Eprint
  {https://arxiv.org/abs/1904.09931} {arXiv:1904.09931 [hep-lat]} \BibitemShut
  {NoStop}%
\bibitem [{\citenamefont {Cirigliano}\ \emph {et~al.}(2019)\citenamefont
  {Cirigliano}, \citenamefont {Davoudi}, \citenamefont {Bhattacharya},
  \citenamefont {Izubuchi}, \citenamefont {Shanahan}, \citenamefont
  {Syritsyn},\ and\ \citenamefont {Wagman}}]{Davoudi:2018qcd}%
  \BibitemOpen
  \bibfield  {author} {\bibinfo {author} {\bibfnamefont {V.}~\bibnamefont
  {Cirigliano}}, \bibinfo {author} {\bibfnamefont {Z.}~\bibnamefont {Davoudi}},
  \bibinfo {author} {\bibfnamefont {T.}~\bibnamefont {Bhattacharya}}, \bibinfo
  {author} {\bibfnamefont {T.}~\bibnamefont {Izubuchi}}, \bibinfo {author}
  {\bibfnamefont {P.~E.}\ \bibnamefont {Shanahan}}, \bibinfo {author}
  {\bibfnamefont {S.}~\bibnamefont {Syritsyn}},\ and\ \bibinfo {author}
  {\bibfnamefont {M.~L.}\ \bibnamefont {Wagman}} (\bibinfo {collaboration}
  {USQCD}),\ }\bibfield  {title} {\bibinfo {title} {The role of lattice {QCD}
  in searches for violations of fundamental symmetries and signals for new
  physics},\ }\href {https://doi.org/10.1140/epja/i2019-12889-8} {\bibfield
  {journal} {\bibinfo  {journal} {Eur. Phys. J.}\ }\textbf {\bibinfo {volume}
  {A55}},\ \bibinfo {pages} {197} (\bibinfo {year} {2019})},\ \Eprint
  {https://arxiv.org/abs/1904.09704} {arXiv:1904.09704 [hep-lat]} \BibitemShut
  {NoStop}%
\bibitem [{\citenamefont {Brower}\ \emph {et~al.}(2019)\citenamefont {Brower},
  \citenamefont {Hasenfratz}, \citenamefont {Neil}, \citenamefont {Catterall},
  \citenamefont {Fleming}, \citenamefont {Giedt}, \citenamefont {Rinaldi},
  \citenamefont {Schaich}, \citenamefont {Weinberg},\ and\ \citenamefont
  {Witzel}}]{Brower:2018qcd}%
  \BibitemOpen
  \bibfield  {author} {\bibinfo {author} {\bibfnamefont {R.}~\bibnamefont
  {Brower}}, \bibinfo {author} {\bibfnamefont {A.}~\bibnamefont {Hasenfratz}},
  \bibinfo {author} {\bibfnamefont {E.~T.}\ \bibnamefont {Neil}}, \bibinfo
  {author} {\bibfnamefont {S.}~\bibnamefont {Catterall}}, \bibinfo {author}
  {\bibfnamefont {G.}~\bibnamefont {Fleming}}, \bibinfo {author} {\bibfnamefont
  {J.}~\bibnamefont {Giedt}}, \bibinfo {author} {\bibfnamefont
  {E.}~\bibnamefont {Rinaldi}}, \bibinfo {author} {\bibfnamefont
  {D.}~\bibnamefont {Schaich}}, \bibinfo {author} {\bibfnamefont
  {E.}~\bibnamefont {Weinberg}},\ and\ \bibinfo {author} {\bibfnamefont
  {O.}~\bibnamefont {Witzel}} (\bibinfo {collaboration} {USQCD}),\ }\bibfield
  {title} {\bibinfo {title} {Lattice gauge theory for physics beyond the
  {Standard Model}},\ }\href {https://doi.org/10.1140/epja/i2019-12901-5}
  {\bibfield  {journal} {\bibinfo  {journal} {Eur. Phys. J.}\ }\textbf
  {\bibinfo {volume} {A55}},\ \bibinfo {pages} {198} (\bibinfo {year}
  {2019})},\ \Eprint {https://arxiv.org/abs/1904.09964} {arXiv:1904.09964
  [hep-lat]} \BibitemShut {NoStop}%
\bibitem [{\citenamefont {{Jo\'o}}\ \emph {et~al.}(2019)\citenamefont
  {{Jo\'o}}, \citenamefont {Jung}, \citenamefont {Christ}, \citenamefont
  {Detmold}, \citenamefont {Edwards}, \citenamefont {Savage},\ and\
  \citenamefont {Shanahan}}]{Joo:2018qcd}%
  \BibitemOpen
  \bibfield  {author} {\bibinfo {author} {\bibfnamefont {B.}~\bibnamefont
  {{Jo\'o}}}, \bibinfo {author} {\bibfnamefont {C.}~\bibnamefont {Jung}},
  \bibinfo {author} {\bibfnamefont {N.~H.}\ \bibnamefont {Christ}}, \bibinfo
  {author} {\bibfnamefont {W.}~\bibnamefont {Detmold}}, \bibinfo {author}
  {\bibfnamefont {R.}~\bibnamefont {Edwards}}, \bibinfo {author} {\bibfnamefont
  {M.}~\bibnamefont {Savage}},\ and\ \bibinfo {author} {\bibfnamefont
  {P.}~\bibnamefont {Shanahan}} (\bibinfo {collaboration} {USQCD}),\ }\bibfield
   {title} {\bibinfo {title} {Status and future perspectives for lattice gauge
  theory calculations to the exascale and beyond},\ }\href
  {https://doi.org/10.1140/epja/i2019-12919-7} {\bibfield  {journal} {\bibinfo
  {journal} {Eur. Phys. J.}\ }\textbf {\bibinfo {volume} {A55}},\ \bibinfo
  {pages} {199} (\bibinfo {year} {2019})},\ \Eprint
  {https://arxiv.org/abs/1904.09725} {arXiv:1904.09725 [hep-lat]} \BibitemShut
  {NoStop}%
\bibitem [{wik()}]{wiki:exascale}%
  \BibitemOpen
  \href@noop {} {\bibinfo {title} {Exascale computing}},\ \bibinfo
  {howpublished}
  {\href{https://en.wikipedia.org/wiki/Exascale_computing/}{wikipedia}},\
  \bibinfo {note} {retrieved July 2022}\BibitemShut {NoStop}%
\bibitem [{\citenamefont {Kothe}\ \emph {et~al.}(2020)\citenamefont {Kothe}
  \emph {et~al.}}]{Alexander:2020gty}%
  \BibitemOpen
  \bibfield  {author} {\bibinfo {author} {\bibfnamefont {D.~B.}\ \bibnamefont
  {Kothe}} \emph {et~al.},\ }\bibfield  {title} {\bibinfo {title} {{Exascale
  applications: skin in the game}},\ }\href
  {https://doi.org/10.1098/rsta.2019.0056} {\bibfield  {journal} {\bibinfo
  {journal} {Phil. Trans. Roy. Soc. A}\ }\textbf {\bibinfo {volume} {378}},\
  \bibinfo {pages} {20190056} (\bibinfo {year} {2020})}\BibitemShut {NoStop}%
\bibitem [{\citenamefont {Duane}\ \emph {et~al.}(1987)\citenamefont {Duane},
  \citenamefont {Kennedy}, \citenamefont {Pendleton},\ and\ \citenamefont
  {Roweth}}]{Duane:1987de}%
  \BibitemOpen
  \bibfield  {author} {\bibinfo {author} {\bibfnamefont {S.}~\bibnamefont
  {Duane}}, \bibinfo {author} {\bibfnamefont {A.~D.}\ \bibnamefont {Kennedy}},
  \bibinfo {author} {\bibfnamefont {B.~J.}\ \bibnamefont {Pendleton}},\ and\
  \bibinfo {author} {\bibfnamefont {D.}~\bibnamefont {Roweth}},\ }\bibfield
  {title} {\bibinfo {title} {Hybrid {Monte Carlo}},\ }\href
  {https://doi.org/10.1016/0370-2693(87)91197-X} {\bibfield  {journal}
  {\bibinfo  {journal} {Phys. Lett. B}\ }\textbf {\bibinfo {volume} {195}},\
  \bibinfo {pages} {216} (\bibinfo {year} {1987})}\BibitemShut {NoStop}%
\bibitem [{\citenamefont {{Stan Development Team}}(2022)}]{stan:2022}%
  \BibitemOpen
  \bibfield  {author} {\bibinfo {author} {\bibnamefont {{Stan Development
  Team}}},\ }\href@noop {} {\bibinfo {title} {\href{https://mc-stan.org}{Stan
  Modeling Language Users Guide and Reference Manual}}} (\bibinfo {year}
  {2022}),\ \bibinfo {note} {version 2.30}\BibitemShut {NoStop}%
\bibitem [{\citenamefont {En{\ss}lin}\ \emph {et~al.}(2009)\citenamefont
  {En{\ss}lin}, \citenamefont {Frommert},\ and\ \citenamefont
  {Kitaura}}]{Ensslin:2008iu}%
  \BibitemOpen
  \bibfield  {author} {\bibinfo {author} {\bibfnamefont {T.~A.}\ \bibnamefont
  {En{\ss}lin}}, \bibinfo {author} {\bibfnamefont {M.}~\bibnamefont
  {Frommert}},\ and\ \bibinfo {author} {\bibfnamefont {F.~S.}\ \bibnamefont
  {Kitaura}},\ }\bibfield  {title} {\bibinfo {title} {{Information field theory
  for cosmological perturbation reconstruction and non-linear signal
  analysis}},\ }\href {https://doi.org/10.1103/PhysRevD.80.105005} {\bibfield
  {journal} {\bibinfo  {journal} {Phys. Rev. D}\ }\textbf {\bibinfo {volume}
  {80}},\ \bibinfo {pages} {105005} (\bibinfo {year} {2009})},\ \Eprint
  {https://arxiv.org/abs/0806.3474} {arXiv:0806.3474 [astro-ph]} \BibitemShut
  {NoStop}%
\bibitem [{\citenamefont {Beyl}\ \emph {et~al.}(2018)\citenamefont {Beyl},
  \citenamefont {Goth},\ and\ \citenamefont {Assaad}}]{Beyl:2017kwp}%
  \BibitemOpen
  \bibfield  {author} {\bibinfo {author} {\bibfnamefont {S.}~\bibnamefont
  {Beyl}}, \bibinfo {author} {\bibfnamefont {F.}~\bibnamefont {Goth}},\ and\
  \bibinfo {author} {\bibfnamefont {F.~F.}\ \bibnamefont {Assaad}},\ }\bibfield
   {title} {\bibinfo {title} {Revisiting the hybrid quantum {Monte Carlo}
  method for {Hubbard} and electron-phonon models},\ }\href
  {https://doi.org/10.1103/PhysRevB.97.085144} {\bibfield  {journal} {\bibinfo
  {journal} {Phys. Rev. B}\ }\textbf {\bibinfo {volume} {97}},\ \bibinfo
  {pages} {085144} (\bibinfo {year} {2018})},\ \Eprint
  {https://arxiv.org/abs/1708.03661} {arXiv:1708.03661 [cond-mat.str-el]}
  \BibitemShut {NoStop}%
\bibitem [{\citenamefont {Boyle}\ \emph
  {et~al.}(2022{\natexlab{a}})\citenamefont {Boyle} \emph
  {et~al.}}]{Boyle:2022ncb}%
  \BibitemOpen
  \bibfield  {author} {\bibinfo {author} {\bibfnamefont {P.}~\bibnamefont
  {Boyle}} \emph {et~al.},\ }\bibfield  {title} {\bibinfo {title} {Lattice
  {QCD} and the computational frontier},\ }in\ \href@noop {} {\emph {\bibinfo
  {booktitle} {{2022 Snowmass Summer Study}}}}\ (\bibinfo {year} {2022})\
  \Eprint {https://arxiv.org/abs/2204.00039} {arXiv:2204.00039 [hep-lat]}
  \BibitemShut {NoStop}%
\bibitem [{\citenamefont {Boyda}\ \emph {et~al.}(2022)\citenamefont {Boyda}
  \emph {et~al.}}]{Boyda:2022nmh}%
  \BibitemOpen
  \bibfield  {author} {\bibinfo {author} {\bibfnamefont {D.}~\bibnamefont
  {Boyda}} \emph {et~al.},\ }\bibfield  {title} {\bibinfo {title} {Applications
  of machine learning to lattice quantum field theory},\ }in\ \href@noop {}
  {\emph {\bibinfo {booktitle} {{2022 Snowmass Summer Study}}}}\ (\bibinfo
  {year} {2022})\ \Eprint {https://arxiv.org/abs/2202.05838} {arXiv:2202.05838
  [hep-lat]} \BibitemShut {NoStop}%
\bibitem [{\citenamefont {Humble}\ \emph {et~al.}(2022)\citenamefont {Humble}
  \emph {et~al.}}]{Humble:2022vtm}%
  \BibitemOpen
  \bibfield  {author} {\bibinfo {author} {\bibfnamefont {T.~S.}\ \bibnamefont
  {Humble}} \emph {et~al.},\ }\bibfield  {title} {\bibinfo {title} {Quantum
  computing systems and software for high-energy physics research},\ }in\
  \href@noop {} {\emph {\bibinfo {booktitle} {{2022 Snowmass Summer Study}}}}\
  (\bibinfo {year} {2022})\ \Eprint {https://arxiv.org/abs/2203.07091}
  {arXiv:2203.07091 [quant-ph]} \BibitemShut {NoStop}%
\bibitem [{\citenamefont {Faulkner}\ \emph {et~al.}(2022)\citenamefont
  {Faulkner}, \citenamefont {Hartman}, \citenamefont {Headrick}, \citenamefont
  {Rangamani},\ and\ \citenamefont {Swingle}}]{Faulkner:2022mlp}%
  \BibitemOpen
  \bibfield  {author} {\bibinfo {author} {\bibfnamefont {T.}~\bibnamefont
  {Faulkner}}, \bibinfo {author} {\bibfnamefont {T.}~\bibnamefont {Hartman}},
  \bibinfo {author} {\bibfnamefont {M.}~\bibnamefont {Headrick}}, \bibinfo
  {author} {\bibfnamefont {M.}~\bibnamefont {Rangamani}},\ and\ \bibinfo
  {author} {\bibfnamefont {B.}~\bibnamefont {Swingle}},\ }\bibfield  {title}
  {\bibinfo {title} {{Quantum information in quantum field theory and quantum
  gravity}},\ }in\ \href@noop {} {\emph {\bibinfo {booktitle} {{2022 Snowmass
  Summer Study}}}}\ (\bibinfo {year} {2022})\ \Eprint
  {https://arxiv.org/abs/2203.07117} {arXiv:2203.07117 [hep-th]} \BibitemShut
  {NoStop}%
\bibitem [{\citenamefont {Meurice}\ \emph {et~al.}(2022)\citenamefont
  {Meurice}, \citenamefont {Osborn}, \citenamefont {Sakai}, \citenamefont
  {Unmuth-Yockey}, \citenamefont {Catterall},\ and\ \citenamefont
  {Somma}}]{Meurice:2022xbk}%
  \BibitemOpen
  \bibfield  {author} {\bibinfo {author} {\bibfnamefont {Y.}~\bibnamefont
  {Meurice}}, \bibinfo {author} {\bibfnamefont {J.~C.}\ \bibnamefont {Osborn}},
  \bibinfo {author} {\bibfnamefont {R.}~\bibnamefont {Sakai}}, \bibinfo
  {author} {\bibfnamefont {J.}~\bibnamefont {Unmuth-Yockey}}, \bibinfo {author}
  {\bibfnamefont {S.}~\bibnamefont {Catterall}},\ and\ \bibinfo {author}
  {\bibfnamefont {R.~D.}\ \bibnamefont {Somma}},\ }\bibfield  {title} {\bibinfo
  {title} {Tensor networks for high energy physics},\ }in\ \href@noop {} {\emph
  {\bibinfo {booktitle} {{2022 Snowmass Summer Study}}}}\ (\bibinfo {year}
  {2022})\ \Eprint {https://arxiv.org/abs/2203.04902} {arXiv:2203.04902
  [hep-lat]} \BibitemShut {NoStop}%
\bibitem [{\citenamefont {Bauer}\ \emph {et~al.}(2022)\citenamefont {Bauer}
  \emph {et~al.}}]{Bauer:2022hpo}%
  \BibitemOpen
  \bibfield  {author} {\bibinfo {author} {\bibfnamefont {C.~W.}\ \bibnamefont
  {Bauer}} \emph {et~al.},\ }\bibfield  {title} {\bibinfo {title} {Quantum
  simulation for high-energy physics},\ }in\ \href@noop {} {\emph {\bibinfo
  {booktitle} {{2022 Snowmass Summer Study}}}}\ (\bibinfo {year} {2022})\
  \Eprint {https://arxiv.org/abs/2204.03381} {arXiv:2204.03381 [quant-ph]}
  \BibitemShut {NoStop}%
\bibitem [{\citenamefont {Drischler}\ \emph {et~al.}(2021)\citenamefont
  {Drischler}, \citenamefont {Haxton}, \citenamefont {McElvain}, \citenamefont
  {Mereghetti}, \citenamefont {Nicholson}, \citenamefont {Vranas},\ and\
  \citenamefont {Walker-Loud}}]{Drischler:2019xuo}%
  \BibitemOpen
  \bibfield  {author} {\bibinfo {author} {\bibfnamefont {C.}~\bibnamefont
  {Drischler}}, \bibinfo {author} {\bibfnamefont {W.}~\bibnamefont {Haxton}},
  \bibinfo {author} {\bibfnamefont {K.}~\bibnamefont {McElvain}}, \bibinfo
  {author} {\bibfnamefont {E.}~\bibnamefont {Mereghetti}}, \bibinfo {author}
  {\bibfnamefont {A.}~\bibnamefont {Nicholson}}, \bibinfo {author}
  {\bibfnamefont {P.}~\bibnamefont {Vranas}},\ and\ \bibinfo {author}
  {\bibfnamefont {A.}~\bibnamefont {Walker-Loud}},\ }\bibfield  {title}
  {\bibinfo {title} {{Towards grounding nuclear physics in QCD}},\ }\href
  {https://doi.org/10.1016/j.ppnp.2021.103888} {\bibfield  {journal} {\bibinfo
  {journal} {Prog. Part. Nucl. Phys.}\ }\textbf {\bibinfo {volume} {121}},\
  \bibinfo {pages} {103888} (\bibinfo {year} {2021})},\ \Eprint
  {https://arxiv.org/abs/1910.07961} {arXiv:1910.07961 [nucl-th]} \BibitemShut
  {NoStop}%
\bibitem [{\citenamefont {Alvarez~Ruso}\ \emph {et~al.}(2022)\citenamefont
  {Alvarez~Ruso} \emph {et~al.}}]{Ruso:2022qes}%
  \BibitemOpen
  \bibfield  {author} {\bibinfo {author} {\bibfnamefont {L.}~\bibnamefont
  {Alvarez~Ruso}} \emph {et~al.},\ }\bibfield  {title} {\bibinfo {title}
  {{Theoretical tools for neutrino scattering: interplay between lattice QCD,
  EFTs, nuclear physics, phenomenology, and neutrino event generators}},\ }in\
  \href@noop {} {\emph {\bibinfo {booktitle} {{2022 Snowmass Summer Study}}}}\
  (\bibinfo {year} {2022})\ \Eprint {https://arxiv.org/abs/2203.09030}
  {arXiv:2203.09030 [hep-ph]} \BibitemShut {NoStop}%
\bibitem [{\citenamefont {Campbell}\ \emph {et~al.}(2022)\citenamefont
  {Campbell} \emph {et~al.}}]{Campbell:2022qmc}%
  \BibitemOpen
  \bibfield  {author} {\bibinfo {author} {\bibfnamefont {J.~M.}\ \bibnamefont
  {Campbell}} \emph {et~al.},\ }\bibfield  {title} {\bibinfo {title} {Event
  generators for high-energy physics experiments},\ }in\ \href@noop {} {\emph
  {\bibinfo {booktitle} {{2022 Snowmass Summer Study}}}}\ (\bibinfo {year}
  {2022})\ \Eprint {https://arxiv.org/abs/2203.11110} {arXiv:2203.11110
  [hep-ph]} \BibitemShut {NoStop}%
\bibitem [{\citenamefont {Parisi}(1984)}]{Parisi:1983ae}%
  \BibitemOpen
  \bibfield  {author} {\bibinfo {author} {\bibfnamefont {G.}~\bibnamefont
  {Parisi}},\ }\bibfield  {title} {\bibinfo {title} {The strategy for computing
  the hadronic mass spectrum},\ }\href
  {https://doi.org/10.1016/0370-1573(84)90081-4} {\bibfield  {journal}
  {\bibinfo  {journal} {Phys. Rept.}\ }\textbf {\bibinfo {volume} {103}},\
  \bibinfo {pages} {203} (\bibinfo {year} {1984})}\BibitemShut {NoStop}%
\bibitem [{\citenamefont {Lepage}(1989)}]{Lepage:1989hd}%
  \BibitemOpen
  \bibfield  {author} {\bibinfo {author} {\bibfnamefont {G.~P.}\ \bibnamefont
  {Lepage}},\ }\bibfield  {title} {\bibinfo {title} {The analysis of algorithms
  for lattice field theory},\ }in\ \href@noop {} {\emph {\bibinfo {booktitle}
  {From Actions to Answers}}},\ \bibinfo {editor} {edited by\ \bibinfo {editor}
  {\bibfnamefont {T.}~\bibnamefont {DeGrand}}\ and\ \bibinfo {editor}
  {\bibfnamefont {W.~D.}\ \bibnamefont {Toussaint}}}\ (\bibinfo  {publisher}
  {World Scientific},\ \bibinfo {address} {Singapore},\ \bibinfo {year}
  {1989})\ pp.\ \bibinfo {pages} {97--120}\BibitemShut {NoStop}%
\bibitem [{\citenamefont {Bazavov}\ \emph
  {et~al.}(2018{\natexlab{a}})\citenamefont {Bazavov} \emph
  {et~al.}}]{Bazavov:2017lyh}%
  \BibitemOpen
  \bibfield  {author} {\bibinfo {author} {\bibfnamefont {A.}~\bibnamefont
  {Bazavov}} \emph {et~al.} (\bibinfo {collaboration} {Fermilab Lattice,
  MILC}),\ }\bibfield  {title} {\bibinfo {title} {{$B$- and $D$-meson leptonic
  decay constants from four-flavor lattice QCD}},\ }\href
  {https://doi.org/10.1103/PhysRevD.98.074512} {\bibfield  {journal} {\bibinfo
  {journal} {Phys. Rev. D}\ }\textbf {\bibinfo {volume} {98}},\ \bibinfo
  {pages} {074512} (\bibinfo {year} {2018}{\natexlab{a}})},\ \Eprint
  {https://arxiv.org/abs/1712.09262} {arXiv:1712.09262 [hep-lat]} \BibitemShut
  {NoStop}%
\bibitem [{\citenamefont {Bazavov}\ \emph
  {et~al.}(2013{\natexlab{a}})\citenamefont {Bazavov} \emph
  {et~al.}}]{Bazavov:2012cd}%
  \BibitemOpen
  \bibfield  {author} {\bibinfo {author} {\bibfnamefont {A.}~\bibnamefont
  {Bazavov}} \emph {et~al.} (\bibinfo {collaboration} {Fermilab Lattice,
  MILC}),\ }\bibfield  {title} {\bibinfo {title} {Kaon semileptonic vector form
  factor and determination of {$|V_{us}|$} using staggered fermions},\ }\href
  {https://doi.org/10.1103/PhysRevD.87.073012} {\bibfield  {journal} {\bibinfo
  {journal} {Phys. Rev. D}\ }\textbf {\bibinfo {volume} {87}},\ \bibinfo
  {pages} {073012} (\bibinfo {year} {2013}{\natexlab{a}})},\ \Eprint
  {https://arxiv.org/abs/1212.4993} {arXiv:1212.4993 [hep-lat]} \BibitemShut
  {NoStop}%
\bibitem [{\citenamefont {Bazavov}\ \emph
  {et~al.}(2014{\natexlab{a}})\citenamefont {Bazavov} \emph
  {et~al.}}]{Bazavov:2013maa}%
  \BibitemOpen
  \bibfield  {author} {\bibinfo {author} {\bibfnamefont {A.}~\bibnamefont
  {Bazavov}} \emph {et~al.} (\bibinfo {collaboration} {Fermilab Lattice,
  MILC}),\ }\bibfield  {title} {\bibinfo {title} {{Determination of $|V_{us}|$
  from a lattice-QCD calculation of the $K\to\pi\ell\nu$ semileptonic form
  factor with physical quark masses}},\ }\href
  {https://doi.org/10.1103/PhysRevLett.112.112001} {\bibfield  {journal}
  {\bibinfo  {journal} {Phys. Rev. Lett.}\ }\textbf {\bibinfo {volume} {112}},\
  \bibinfo {pages} {112001} (\bibinfo {year} {2014}{\natexlab{a}})},\ \Eprint
  {https://arxiv.org/abs/1312.1228} {arXiv:1312.1228 [hep-ph]} \BibitemShut
  {NoStop}%
\bibitem [{\citenamefont {Bazavov}\ \emph
  {et~al.}(2019{\natexlab{b}})\citenamefont {Bazavov} \emph
  {et~al.}}]{Bazavov:2018kjg}%
  \BibitemOpen
  \bibfield  {author} {\bibinfo {author} {\bibfnamefont {A.}~\bibnamefont
  {Bazavov}} \emph {et~al.} (\bibinfo {collaboration} {Fermilab Lattice,
  MILC}),\ }\bibfield  {title} {\bibinfo {title} {{$|V_{us}|$ from $K_{\ell 3}$
  decay and four-flavor lattice QCD}},\ }\href
  {https://doi.org/10.1103/PhysRevD.99.114509} {\bibfield  {journal} {\bibinfo
  {journal} {Phys. Rev. D}\ }\textbf {\bibinfo {volume} {99}},\ \bibinfo
  {pages} {114509} (\bibinfo {year} {2019}{\natexlab{b}})},\ \Eprint
  {https://arxiv.org/abs/1809.02827} {arXiv:1809.02827 [hep-lat]} \BibitemShut
  {NoStop}%
\bibitem [{\citenamefont {Boyle}\ \emph {et~al.}(2015)\citenamefont {Boyle}
  \emph {et~al.}}]{Boyle:2015hfa}%
  \BibitemOpen
  \bibfield  {author} {\bibinfo {author} {\bibfnamefont {P.~A.}\ \bibnamefont
  {Boyle}} \emph {et~al.} (\bibinfo {collaboration} {RBC, UKQCD}),\ }\bibfield
  {title} {\bibinfo {title} {The kaon semileptonic form factor in {$N_{f}=2+1$}
  domain wall lattice qcd with physical light quark masses},\ }\href
  {https://doi.org/10.1007/JHEP06(2015)164} {\bibfield  {journal} {\bibinfo
  {journal} {JHEP}\ }\textbf {\bibinfo {volume} {06}},\ \bibinfo {pages}
  {164}},\ \Eprint {https://arxiv.org/abs/1504.01692} {arXiv:1504.01692
  [hep-lat]} \BibitemShut {NoStop}%
\bibitem [{\citenamefont {Na}\ \emph {et~al.}(2011)\citenamefont {Na},
  \citenamefont {Davies}, \citenamefont {Follana}, \citenamefont {Koponen},
  \citenamefont {Lepage},\ and\ \citenamefont {Shigemitsu}}]{Na:2011mc}%
  \BibitemOpen
  \bibfield  {author} {\bibinfo {author} {\bibfnamefont {H.}~\bibnamefont
  {Na}}, \bibinfo {author} {\bibfnamefont {C.~T.~H.}\ \bibnamefont {Davies}},
  \bibinfo {author} {\bibfnamefont {E.}~\bibnamefont {Follana}}, \bibinfo
  {author} {\bibfnamefont {J.}~\bibnamefont {Koponen}}, \bibinfo {author}
  {\bibfnamefont {G.~P.}\ \bibnamefont {Lepage}},\ and\ \bibinfo {author}
  {\bibfnamefont {J.}~\bibnamefont {Shigemitsu}} (\bibinfo {collaboration}
  {HPQCD}),\ }\bibfield  {title} {\bibinfo {title} {{$D\to\pi l\nu$}
  semileptonic decays, {$|V_{cd}|$} and $2^\text{nd}$ row unitarity from
  lattice {QCD}},\ }\href {https://doi.org/10.1103/PhysRevD.84.114505}
  {\bibfield  {journal} {\bibinfo  {journal} {Phys. Rev. D}\ }\textbf {\bibinfo
  {volume} {84}},\ \bibinfo {pages} {114505} (\bibinfo {year} {2011})},\
  \Eprint {https://arxiv.org/abs/1109.1501} {arXiv:1109.1501 [hep-lat]}
  \BibitemShut {NoStop}%
\bibitem [{\citenamefont {Li}\ \emph {et~al.}(2019)\citenamefont {Li} \emph
  {et~al.}}]{Li:2019phv}%
  \BibitemOpen
  \bibfield  {author} {\bibinfo {author} {\bibfnamefont {R.}~\bibnamefont {Li}}
  \emph {et~al.} (\bibinfo {collaboration} {Fermilab Lattice, MILC}),\
  }\bibfield  {title} {\bibinfo {title} {{$D$} meson semileptonic decay form
  factors at $q^2 = 0$},\ }\href {https://doi.org/10.22323/1.334.0269}
  {\bibfield  {journal} {\bibinfo  {journal} {PoS}\ }\textbf {\bibinfo {volume}
  {LATTICE2018}},\ \bibinfo {pages} {269} (\bibinfo {year} {2019})},\ \Eprint
  {https://arxiv.org/abs/1901.08989} {arXiv:1901.08989 [hep-lat]} \BibitemShut
  {NoStop}%
\bibitem [{\citenamefont {Na}\ \emph {et~al.}(2010)\citenamefont {Na},
  \citenamefont {Davies}, \citenamefont {Follana}, \citenamefont {Lepage},\
  and\ \citenamefont {Shigemitsu}}]{Na:2010uf}%
  \BibitemOpen
  \bibfield  {author} {\bibinfo {author} {\bibfnamefont {H.}~\bibnamefont
  {Na}}, \bibinfo {author} {\bibfnamefont {C.~T.~H.}\ \bibnamefont {Davies}},
  \bibinfo {author} {\bibfnamefont {E.}~\bibnamefont {Follana}}, \bibinfo
  {author} {\bibfnamefont {G.~P.}\ \bibnamefont {Lepage}},\ and\ \bibinfo
  {author} {\bibfnamefont {J.}~\bibnamefont {Shigemitsu}} (\bibinfo
  {collaboration} {HPQCD}),\ }\bibfield  {title} {\bibinfo {title} {The {$D \to
  K l \nu$} semileptonic decay scalar form factor and {$|V_{cs}|$} from lattice
  {QCD}},\ }\href {https://doi.org/10.1103/PhysRevD.82.114506} {\bibfield
  {journal} {\bibinfo  {journal} {Phys. Rev. D}\ }\textbf {\bibinfo {volume}
  {82}},\ \bibinfo {pages} {114506} (\bibinfo {year} {2010})},\ \Eprint
  {https://arxiv.org/abs/1008.4562} {arXiv:1008.4562 [hep-lat]} \BibitemShut
  {NoStop}%
\bibitem [{\citenamefont {Chakraborty}\ \emph {et~al.}(2021)\citenamefont
  {Chakraborty}, \citenamefont {Parrott}, \citenamefont {Bouchard},
  \citenamefont {Davies}, \citenamefont {Koponen},\ and\ \citenamefont
  {Lepage}}]{Chakraborty:2021qav}%
  \BibitemOpen
  \bibfield  {author} {\bibinfo {author} {\bibfnamefont {B.}~\bibnamefont
  {Chakraborty}}, \bibinfo {author} {\bibfnamefont {W.~G.}\ \bibnamefont
  {Parrott}}, \bibinfo {author} {\bibfnamefont {C.}~\bibnamefont {Bouchard}},
  \bibinfo {author} {\bibfnamefont {C.~T.~H.}\ \bibnamefont {Davies}}, \bibinfo
  {author} {\bibfnamefont {J.}~\bibnamefont {Koponen}},\ and\ \bibinfo {author}
  {\bibfnamefont {G.~P.}\ \bibnamefont {Lepage}} (\bibinfo {collaboration}
  {HPQCD}),\ }\bibfield  {title} {\bibinfo {title} {{Improved $V_{cs}$
  determination using precise lattice QCD form factors for $D\to K\ell\nu$}},\
  }\href {https://doi.org/10.1103/PhysRevD.104.034505} {\bibfield  {journal}
  {\bibinfo  {journal} {Phys. Rev. D}\ }\textbf {\bibinfo {volume} {104}},\
  \bibinfo {pages} {034505} (\bibinfo {year} {2021})},\ \Eprint
  {https://arxiv.org/abs/2104.09883} {arXiv:2104.09883 [hep-lat]} \BibitemShut
  {NoStop}%
\bibitem [{\citenamefont {Flynn}\ \emph {et~al.}(2015)\citenamefont {Flynn},
  \citenamefont {Izubuchi}, \citenamefont {Kawanai}, \citenamefont {Lehner},
  \citenamefont {Soni}, \citenamefont {Van~de Water},\ and\ \citenamefont
  {Witzel}}]{Flynn:2015mha}%
  \BibitemOpen
  \bibfield  {author} {\bibinfo {author} {\bibfnamefont {J.~M.}\ \bibnamefont
  {Flynn}}, \bibinfo {author} {\bibfnamefont {T.}~\bibnamefont {Izubuchi}},
  \bibinfo {author} {\bibfnamefont {T.}~\bibnamefont {Kawanai}}, \bibinfo
  {author} {\bibfnamefont {C.}~\bibnamefont {Lehner}}, \bibinfo {author}
  {\bibfnamefont {A.}~\bibnamefont {Soni}}, \bibinfo {author} {\bibfnamefont
  {R.~S.}\ \bibnamefont {Van~de Water}},\ and\ \bibinfo {author} {\bibfnamefont
  {O.}~\bibnamefont {Witzel}} (\bibinfo {collaboration} {RBC, UKQCD}),\
  }\bibfield  {title} {\bibinfo {title} {{$B \to \pi \ell \nu$ and $B_s \to K
  \ell \nu$ form factors and $|V_{ub}|$ from 2+1-flavor lattice QCD with
  domain-wall light quarks and relativistic heavy quarks}},\ }\href
  {https://doi.org/10.1103/PhysRevD.91.074510} {\bibfield  {journal} {\bibinfo
  {journal} {Phys. Rev. D}\ }\textbf {\bibinfo {volume} {91}},\ \bibinfo
  {pages} {074510} (\bibinfo {year} {2015})},\ \Eprint
  {https://arxiv.org/abs/1501.05373} {arXiv:1501.05373 [hep-lat]} \BibitemShut
  {NoStop}%
\bibitem [{\citenamefont {Bailey}\ \emph
  {et~al.}(2015{\natexlab{a}})\citenamefont {Bailey} \emph
  {et~al.}}]{Lattice:2015tia}%
  \BibitemOpen
  \bibfield  {author} {\bibinfo {author} {\bibfnamefont {J.~A.}\ \bibnamefont
  {Bailey}} \emph {et~al.} (\bibinfo {collaboration} {Fermilab Lattice,
  MILC}),\ }\bibfield  {title} {\bibinfo {title} {{$|V_{ub}|$ from
  $B\to\pi\ell\nu$ decays and (2+1)-flavor lattice QCD}},\ }\href
  {https://doi.org/10.1103/PhysRevD.92.014024} {\bibfield  {journal} {\bibinfo
  {journal} {Phys. Rev. D}\ }\textbf {\bibinfo {volume} {92}},\ \bibinfo
  {pages} {014024} (\bibinfo {year} {2015}{\natexlab{a}})},\ \Eprint
  {https://arxiv.org/abs/1503.07839} {arXiv:1503.07839 [hep-lat]} \BibitemShut
  {NoStop}%
\bibitem [{\citenamefont {Colquhoun}\ \emph {et~al.}(2016)\citenamefont
  {Colquhoun}, \citenamefont {Dowdall}, \citenamefont {Koponen}, \citenamefont
  {Davies},\ and\ \citenamefont {Lepage}}]{Colquhoun:2015mfa}%
  \BibitemOpen
  \bibfield  {author} {\bibinfo {author} {\bibfnamefont {B.}~\bibnamefont
  {Colquhoun}}, \bibinfo {author} {\bibfnamefont {R.~J.}\ \bibnamefont
  {Dowdall}}, \bibinfo {author} {\bibfnamefont {J.}~\bibnamefont {Koponen}},
  \bibinfo {author} {\bibfnamefont {C.~T.~H.}\ \bibnamefont {Davies}},\ and\
  \bibinfo {author} {\bibfnamefont {G.~P.}\ \bibnamefont {Lepage}} (\bibinfo
  {collaboration} {HPQCD}),\ }\bibfield  {title} {\bibinfo {title}
  {{$B\to\pi\ell\nu$} at zero recoil from lattice {QCD} with physical $u$/$d$
  quarks},\ }\href {https://doi.org/10.1103/PhysRevD.93.034502} {\bibfield
  {journal} {\bibinfo  {journal} {Phys. Rev. D}\ }\textbf {\bibinfo {volume}
  {93}},\ \bibinfo {pages} {034502} (\bibinfo {year} {2016})},\ \Eprint
  {https://arxiv.org/abs/1510.07446} {arXiv:1510.07446 [hep-lat]} \BibitemShut
  {NoStop}%
\bibitem [{\citenamefont {Colquhoun}\ \emph {et~al.}(2022)\citenamefont
  {Colquhoun}, \citenamefont {Hashimoto}, \citenamefont {Kaneko},\ and\
  \citenamefont {Koponen}}]{Colquhoun:2022atw}%
  \BibitemOpen
  \bibfield  {author} {\bibinfo {author} {\bibfnamefont {B.}~\bibnamefont
  {Colquhoun}}, \bibinfo {author} {\bibfnamefont {S.}~\bibnamefont
  {Hashimoto}}, \bibinfo {author} {\bibfnamefont {T.}~\bibnamefont {Kaneko}},\
  and\ \bibinfo {author} {\bibfnamefont {J.}~\bibnamefont {Koponen}} (\bibinfo
  {collaboration} {JLQCD}),\ }\bibfield  {title} {\bibinfo {title} {{Form
  factors of $B\to\pi\ell\nu$ and a determination of $|V_{ub}|$ with M\"obius
  domain-wall-fermions}},\ }\href@noop {} {\  (\bibinfo {year} {2022})},\
  \Eprint {https://arxiv.org/abs/2203.04938} {arXiv:2203.04938 [hep-lat]}
  \BibitemShut {NoStop}%
\bibitem [{\citenamefont {Bouchard}\ \emph {et~al.}(2014)\citenamefont
  {Bouchard}, \citenamefont {Lepage}, \citenamefont {Monahan}, \citenamefont
  {Na},\ and\ \citenamefont {Shigemitsu}}]{Bouchard:2014ypa}%
  \BibitemOpen
  \bibfield  {author} {\bibinfo {author} {\bibfnamefont {C.~M.}\ \bibnamefont
  {Bouchard}}, \bibinfo {author} {\bibfnamefont {G.~P.}\ \bibnamefont
  {Lepage}}, \bibinfo {author} {\bibfnamefont {C.}~\bibnamefont {Monahan}},
  \bibinfo {author} {\bibfnamefont {H.}~\bibnamefont {Na}},\ and\ \bibinfo
  {author} {\bibfnamefont {J.}~\bibnamefont {Shigemitsu}} (\bibinfo
  {collaboration} {HPQCD}),\ }\bibfield  {title} {\bibinfo {title} {{$B_s \to K
  \ell \nu$ form factors from lattice QCD}},\ }\href
  {https://doi.org/10.1103/PhysRevD.90.054506} {\bibfield  {journal} {\bibinfo
  {journal} {Phys. Rev. D}\ }\textbf {\bibinfo {volume} {90}},\ \bibinfo
  {pages} {054506} (\bibinfo {year} {2014})},\ \Eprint
  {https://arxiv.org/abs/1406.2279} {arXiv:1406.2279 [hep-lat]} \BibitemShut
  {NoStop}%
\bibitem [{\citenamefont {Monahan}\ \emph {et~al.}(2018)\citenamefont
  {Monahan}, \citenamefont {Bouchard}, \citenamefont {Lepage}, \citenamefont
  {Na},\ and\ \citenamefont {Shigemitsu}}]{Monahan:2018lzv}%
  \BibitemOpen
  \bibfield  {author} {\bibinfo {author} {\bibfnamefont {C.~J.}\ \bibnamefont
  {Monahan}}, \bibinfo {author} {\bibfnamefont {C.~M.}\ \bibnamefont
  {Bouchard}}, \bibinfo {author} {\bibfnamefont {G.~P.}\ \bibnamefont
  {Lepage}}, \bibinfo {author} {\bibfnamefont {H.}~\bibnamefont {Na}},\ and\
  \bibinfo {author} {\bibfnamefont {J.}~\bibnamefont {Shigemitsu}} (\bibinfo
  {collaboration} {HPQCD}),\ }\bibfield  {title} {\bibinfo {title} {{Form
  factor ratios for $B_s\to K\ell\nu$ and $B_s\to D_s\ell\nu$ semileptonic
  decays and $|V_{ub}/V_{cb}|$}},\ }\href
  {https://doi.org/10.1103/PhysRevD.98.114509} {\bibfield  {journal} {\bibinfo
  {journal} {Phys. Rev. D}\ }\textbf {\bibinfo {volume} {98}},\ \bibinfo
  {pages} {114509} (\bibinfo {year} {2018})},\ \Eprint
  {https://arxiv.org/abs/1808.09285} {arXiv:1808.09285 [hep-lat]} \BibitemShut
  {NoStop}%
\bibitem [{\citenamefont {Bazavov}\ \emph
  {et~al.}(2019{\natexlab{c}})\citenamefont {Bazavov} \emph
  {et~al.}}]{FermilabLattice:2019ikx}%
  \BibitemOpen
  \bibfield  {author} {\bibinfo {author} {\bibfnamefont {A.}~\bibnamefont
  {Bazavov}} \emph {et~al.} (\bibinfo {collaboration} {Fermilab Lattice,
  MILC}),\ }\bibfield  {title} {\bibinfo {title} {{$B_s\to K\ell\nu$ decay from
  lattice QCD}},\ }\href {https://doi.org/10.1103/PhysRevD.100.034501}
  {\bibfield  {journal} {\bibinfo  {journal} {Phys. Rev. D}\ }\textbf {\bibinfo
  {volume} {100}},\ \bibinfo {pages} {034501} (\bibinfo {year}
  {2019}{\natexlab{c}})},\ \Eprint {https://arxiv.org/abs/1901.02561}
  {arXiv:1901.02561 [hep-lat]} \BibitemShut {NoStop}%
\bibitem [{\citenamefont {Bailey}\ \emph {et~al.}(2014)\citenamefont {Bailey}
  \emph {et~al.}}]{Bailey:2014tva}%
  \BibitemOpen
  \bibfield  {author} {\bibinfo {author} {\bibfnamefont {J.~A.}\ \bibnamefont
  {Bailey}} \emph {et~al.} (\bibinfo {collaboration} {Fermilab Lattice,
  MILC}),\ }\bibfield  {title} {\bibinfo {title} {Update of {$|V_{cb}|$} from
  the {$\bar{B}\to D^*\ell\bar{\nu}$} form factor at zero recoil with
  three-flavor lattice {QCD}},\ }\href
  {https://doi.org/10.1103/PhysRevD.89.114504} {\bibfield  {journal} {\bibinfo
  {journal} {Phys. Rev. D}\ }\textbf {\bibinfo {volume} {89}},\ \bibinfo
  {pages} {114504} (\bibinfo {year} {2014})},\ \Eprint
  {https://arxiv.org/abs/1403.0635} {arXiv:1403.0635 [hep-lat]} \BibitemShut
  {NoStop}%
\bibitem [{\citenamefont {Bailey}\ \emph
  {et~al.}(2015{\natexlab{b}})\citenamefont {Bailey} \emph
  {et~al.}}]{Lattice:2015rga}%
  \BibitemOpen
  \bibfield  {author} {\bibinfo {author} {\bibfnamefont {J.~A.}\ \bibnamefont
  {Bailey}} \emph {et~al.} (\bibinfo {collaboration} {Fermilab Lattice,
  MILC}),\ }\bibfield  {title} {\bibinfo {title} {{$B\to D\ell\nu$} form
  factors at nonzero recoil and {$|V_{cb}|$} from 2+1-flavor lattice {QCD}},\
  }\href {https://doi.org/10.1103/PhysRevD.92.034506} {\bibfield  {journal}
  {\bibinfo  {journal} {Phys. Rev. D}\ }\textbf {\bibinfo {volume} {92}},\
  \bibinfo {pages} {034506} (\bibinfo {year} {2015}{\natexlab{b}})},\ \Eprint
  {https://arxiv.org/abs/1503.07237} {arXiv:1503.07237 [hep-lat]} \BibitemShut
  {NoStop}%
\bibitem [{\citenamefont {Na}\ \emph {et~al.}(2015)\citenamefont {Na},
  \citenamefont {Bouchard}, \citenamefont {Lepage}, \citenamefont {Monahan},\
  and\ \citenamefont {Shigemitsu}}]{Na:2015kha}%
  \BibitemOpen
  \bibfield  {author} {\bibinfo {author} {\bibfnamefont {H.}~\bibnamefont
  {Na}}, \bibinfo {author} {\bibfnamefont {C.~M.}\ \bibnamefont {Bouchard}},
  \bibinfo {author} {\bibfnamefont {G.~P.}\ \bibnamefont {Lepage}}, \bibinfo
  {author} {\bibfnamefont {C.}~\bibnamefont {Monahan}},\ and\ \bibinfo {author}
  {\bibfnamefont {J.}~\bibnamefont {Shigemitsu}} (\bibinfo {collaboration}
  {HPQCD}),\ }\bibfield  {title} {\bibinfo {title} {{$B \rightarrow D l \nu$
  form factors at nonzero recoil and extraction of $|V_{cb}|$}},\ }\href
  {https://doi.org/10.1103/PhysRevD.92.054510} {\bibfield  {journal} {\bibinfo
  {journal} {Phys. Rev. D}\ }\textbf {\bibinfo {volume} {92}},\ \bibinfo
  {pages} {054510} (\bibinfo {year} {2015})},\ \bibinfo {note} {(E)
  \href{https://doi.org/10.1103/PhysRevD.93.119906}{\textbf{93}, 119906
  (2016)}},\ \Eprint {https://arxiv.org/abs/1505.03925} {arXiv:1505.03925
  [hep-lat]} \BibitemShut {NoStop}%
\bibitem [{\citenamefont {Harrison}\ \emph {et~al.}(2018)\citenamefont
  {Harrison}, \citenamefont {Davies},\ and\ \citenamefont
  {Wingate}}]{Harrison:2017fmw}%
  \BibitemOpen
  \bibfield  {author} {\bibinfo {author} {\bibfnamefont {J.}~\bibnamefont
  {Harrison}}, \bibinfo {author} {\bibfnamefont {C.}~\bibnamefont {Davies}},\
  and\ \bibinfo {author} {\bibfnamefont {M.}~\bibnamefont {Wingate}} (\bibinfo
  {collaboration} {HPQCD}),\ }\bibfield  {title} {\bibinfo {title} {{Lattice
  QCD calculation of the ${{B}_{(s)}\to D_{(s)}^{*}\ell{\nu}}$ form factors at
  zero recoil and implications for ${|V_{cb}|}$}},\ }\href
  {https://doi.org/10.1103/PhysRevD.97.054502} {\bibfield  {journal} {\bibinfo
  {journal} {Phys. Rev. D}\ }\textbf {\bibinfo {volume} {97}},\ \bibinfo
  {pages} {054502} (\bibinfo {year} {2018})},\ \Eprint
  {https://arxiv.org/abs/1711.11013} {arXiv:1711.11013 [hep-lat]} \BibitemShut
  {NoStop}%
\bibitem [{\citenamefont {Bazavov}\ \emph {et~al.}(2021)\citenamefont {Bazavov}
  \emph {et~al.}}]{FermilabLattice:2021cdg}%
  \BibitemOpen
  \bibfield  {author} {\bibinfo {author} {\bibfnamefont {A.}~\bibnamefont
  {Bazavov}} \emph {et~al.} (\bibinfo {collaboration} {Fermilab Lattice,
  MILC}),\ }\bibfield  {title} {\bibinfo {title} {{Semileptonic form factors
  for $B \to D^\ast\ell\nu$ at nonzero recoil from 2+1-flavor lattice QCD}},\
  }\href@noop {} {\  (\bibinfo {year} {2021})},\ \Eprint
  {https://arxiv.org/abs/2105.14019} {arXiv:2105.14019 [hep-lat]} \BibitemShut
  {NoStop}%
\bibitem [{\citenamefont {Kaneko}\ \emph {et~al.}(2022)\citenamefont {Kaneko},
  \citenamefont {Aoki}, \citenamefont {Colquhoun}, \citenamefont {Faur},
  \citenamefont {Fukaya}, \citenamefont {Hashimoto}, \citenamefont {Koponen},\
  and\ \citenamefont {Kou}}]{Kaneko:2021tlw}%
  \BibitemOpen
  \bibfield  {author} {\bibinfo {author} {\bibfnamefont {T.}~\bibnamefont
  {Kaneko}}, \bibinfo {author} {\bibfnamefont {Y.}~\bibnamefont {Aoki}},
  \bibinfo {author} {\bibfnamefont {B.}~\bibnamefont {Colquhoun}}, \bibinfo
  {author} {\bibfnamefont {M.}~\bibnamefont {Faur}}, \bibinfo {author}
  {\bibfnamefont {H.}~\bibnamefont {Fukaya}}, \bibinfo {author} {\bibfnamefont
  {S.}~\bibnamefont {Hashimoto}}, \bibinfo {author} {\bibfnamefont
  {J.}~\bibnamefont {Koponen}},\ and\ \bibinfo {author} {\bibfnamefont
  {E.}~\bibnamefont {Kou}},\ }\bibfield  {title} {\bibinfo {title} {{$B\to
  D^{(*)}\ell\nu$} semileptonic decays in lattice {QCD} with domain-wall heavy
  quarks},\ }\href {https://doi.org/10.22323/1.396.0561} {\bibfield  {journal}
  {\bibinfo  {journal} {PoS}\ }\textbf {\bibinfo {volume} {LATTICE2021}},\
  \bibinfo {pages} {561} (\bibinfo {year} {2022})},\ \Eprint
  {https://arxiv.org/abs/2112.13775} {arXiv:2112.13775 [hep-lat]} \BibitemShut
  {NoStop}%
\bibitem [{\citenamefont {Detmold}\ \emph {et~al.}(2015)\citenamefont
  {Detmold}, \citenamefont {Lehner},\ and\ \citenamefont
  {Meinel}}]{Detmold:2015aaa}%
  \BibitemOpen
  \bibfield  {author} {\bibinfo {author} {\bibfnamefont {W.}~\bibnamefont
  {Detmold}}, \bibinfo {author} {\bibfnamefont {C.}~\bibnamefont {Lehner}},\
  and\ \bibinfo {author} {\bibfnamefont {S.}~\bibnamefont {Meinel}},\
  }\bibfield  {title} {\bibinfo {title} {{$\Lambda_b \to p \ell^-
  \bar{\nu}_\ell$ and $\Lambda_b\to \Lambda_c \ell^- \bar{\nu}_\ell$ form
  factors from lattice QCD with relativistic heavy quarks}},\ }\href
  {https://doi.org/10.1103/PhysRevD.92.034503} {\bibfield  {journal} {\bibinfo
  {journal} {Phys. Rev. D}\ }\textbf {\bibinfo {volume} {92}},\ \bibinfo
  {pages} {034503} (\bibinfo {year} {2015})},\ \Eprint
  {https://arxiv.org/abs/1503.01421} {arXiv:1503.01421 [hep-lat]} \BibitemShut
  {NoStop}%
\bibitem [{\citenamefont {Altmannshofer}\ and\ \citenamefont
  {Archilli}(2022)}]{Altmannshofer:2022hfs}%
  \BibitemOpen
  \bibfield  {author} {\bibinfo {author} {\bibfnamefont {W.}~\bibnamefont
  {Altmannshofer}}\ and\ \bibinfo {author} {\bibfnamefont {F.}~\bibnamefont
  {Archilli}},\ }\bibfield  {title} {\bibinfo {title} {{Rare decays of $b$ and
  $c$ hadrons}},\ }in\ \href@noop {} {\emph {\bibinfo {booktitle} {{2022
  Snowmass Summer Study}}}}\ (\bibinfo {year} {2022})\ \Eprint
  {https://arxiv.org/abs/2206.11331} {arXiv:2206.11331 [hep-ph]} \BibitemShut
  {NoStop}%
\bibitem [{\citenamefont {Guadagnoli}\ and\ \citenamefont
  {Koppenburg}(2022)}]{Guadagnoli:2022oxk}%
  \BibitemOpen
  \bibfield  {author} {\bibinfo {author} {\bibfnamefont {D.}~\bibnamefont
  {Guadagnoli}}\ and\ \bibinfo {author} {\bibfnamefont {P.}~\bibnamefont
  {Koppenburg}},\ }\bibfield  {title} {\bibinfo {title} {{Lepton-flavor
  violation and lepton-flavor-universality violation in $b$ and $c$ decays}},\
  }in\ \href@noop {} {\emph {\bibinfo {booktitle} {{2022 Snowmass Summer
  Study}}}}\ (\bibinfo {year} {2022})\ \Eprint
  {https://arxiv.org/abs/2207.01851} {arXiv:2207.01851 [hep-ph]} \BibitemShut
  {NoStop}%
\bibitem [{\citenamefont {Bailey}\ \emph {et~al.}(2016)\citenamefont {Bailey}
  \emph {et~al.}}]{Bailey:2015dka}%
  \BibitemOpen
  \bibfield  {author} {\bibinfo {author} {\bibfnamefont {J.~A.}\ \bibnamefont
  {Bailey}} \emph {et~al.} (\bibinfo {collaboration} {Fermilab Lattice,
  MILC}),\ }\bibfield  {title} {\bibinfo {title} {{$B\to Kl^+l^-$} decay form
  factors from three-flavor lattice {QCD}},\ }\href
  {https://doi.org/10.1103/PhysRevD.93.025026} {\bibfield  {journal} {\bibinfo
  {journal} {Phys. Rev. D}\ }\textbf {\bibinfo {volume} {93}},\ \bibinfo
  {pages} {025026} (\bibinfo {year} {2016})},\ \Eprint
  {https://arxiv.org/abs/1509.06235} {arXiv:1509.06235 [hep-lat]} \BibitemShut
  {NoStop}%
\bibitem [{\citenamefont {Du}\ \emph {et~al.}(2016)\citenamefont {Du},
  \citenamefont {El-Khadra}, \citenamefont {Gottlieb}, \citenamefont
  {Kronfeld}, \citenamefont {Laiho}, \citenamefont {Lunghi}, \citenamefont
  {Van~de Water},\ and\ \citenamefont {Zhou}}]{Du:2015tda}%
  \BibitemOpen
  \bibfield  {author} {\bibinfo {author} {\bibfnamefont {D.}~\bibnamefont
  {Du}}, \bibinfo {author} {\bibfnamefont {A.~X.}\ \bibnamefont {El-Khadra}},
  \bibinfo {author} {\bibfnamefont {S.}~\bibnamefont {Gottlieb}}, \bibinfo
  {author} {\bibfnamefont {A.~S.}\ \bibnamefont {Kronfeld}}, \bibinfo {author}
  {\bibfnamefont {J.}~\bibnamefont {Laiho}}, \bibinfo {author} {\bibfnamefont
  {E.}~\bibnamefont {Lunghi}}, \bibinfo {author} {\bibfnamefont {R.~S.}\
  \bibnamefont {Van~de Water}},\ and\ \bibinfo {author} {\bibfnamefont
  {R.}~\bibnamefont {Zhou}} (\bibinfo {collaboration} {Fermilab Lattice}),\
  }\bibfield  {title} {\bibinfo {title} {Phenomenology of semileptonic
  {$B$}-meson decays with form factors from lattice {QCD}},\ }\href
  {https://doi.org/10.1103/PhysRevD.93.034005} {\bibfield  {journal} {\bibinfo
  {journal} {Phys. Rev. D}\ }\textbf {\bibinfo {volume} {93}},\ \bibinfo
  {pages} {034005} (\bibinfo {year} {2016})},\ \Eprint
  {https://arxiv.org/abs/1510.02349} {arXiv:1510.02349 [hep-ph]} \BibitemShut
  {NoStop}%
\bibitem [{\citenamefont {Bouchard}\ \emph {et~al.}(2013)\citenamefont
  {Bouchard}, \citenamefont {Lepage}, \citenamefont {Monahan}, \citenamefont
  {Na},\ and\ \citenamefont {Shigemitsu}}]{Bouchard:2013pna}%
  \BibitemOpen
  \bibfield  {author} {\bibinfo {author} {\bibfnamefont {C.}~\bibnamefont
  {Bouchard}}, \bibinfo {author} {\bibfnamefont {G.~P.}\ \bibnamefont
  {Lepage}}, \bibinfo {author} {\bibfnamefont {C.}~\bibnamefont {Monahan}},
  \bibinfo {author} {\bibfnamefont {H.}~\bibnamefont {Na}},\ and\ \bibinfo
  {author} {\bibfnamefont {J.}~\bibnamefont {Shigemitsu}} (\bibinfo
  {collaboration} {HPQCD}),\ }\bibfield  {title} {\bibinfo {title} {{Rare decay
  $B \to K \ell^+ \ell^-$ form factors from lattice QCD}},\ }\href
  {https://doi.org/10.1103/PhysRevD.88.054509} {\bibfield  {journal} {\bibinfo
  {journal} {Phys. Rev. D}\ }\textbf {\bibinfo {volume} {88}},\ \bibinfo
  {pages} {054509} (\bibinfo {year} {2013})},\ \bibinfo {note} {(E)
  \href{https://doi.org/10.1103/PhysRevD.88.079901}{\textbf{88}, 079901
  (2013)}},\ \Eprint {https://arxiv.org/abs/1306.2384} {arXiv:1306.2384
  [hep-lat]} \BibitemShut {NoStop}%
\bibitem [{\citenamefont {Horgan}\ \emph {et~al.}(2014)\citenamefont {Horgan},
  \citenamefont {Liu}, \citenamefont {Meinel},\ and\ \citenamefont
  {Wingate}}]{Horgan:2013pva}%
  \BibitemOpen
  \bibfield  {author} {\bibinfo {author} {\bibfnamefont {R.~R.}\ \bibnamefont
  {Horgan}}, \bibinfo {author} {\bibfnamefont {Z.}~\bibnamefont {Liu}},
  \bibinfo {author} {\bibfnamefont {S.}~\bibnamefont {Meinel}},\ and\ \bibinfo
  {author} {\bibfnamefont {M.}~\bibnamefont {Wingate}},\ }\bibfield  {title}
  {\bibinfo {title} {{Calculation of $B^0 \to K^{*0} \mu^+ \mu^-$ and $B_s^0
  \to \phi \mu^+ \mu^-$ observables using form factors from lattice QCD}},\
  }\href {https://doi.org/10.1103/PhysRevLett.112.212003} {\bibfield  {journal}
  {\bibinfo  {journal} {Phys. Rev. Lett.}\ }\textbf {\bibinfo {volume} {112}},\
  \bibinfo {pages} {212003} (\bibinfo {year} {2014})},\ \Eprint
  {https://arxiv.org/abs/1310.3887} {arXiv:1310.3887 [hep-ph]} \BibitemShut
  {NoStop}%
\bibitem [{\citenamefont {Aaij}\ \emph {et~al.}(2021)\citenamefont {Aaij} \emph
  {et~al.}}]{LHCb:2021zwz}%
  \BibitemOpen
  \bibfield  {author} {\bibinfo {author} {\bibfnamefont {R.}~\bibnamefont
  {Aaij}} \emph {et~al.} (\bibinfo {collaboration} {LHCb}),\ }\bibfield
  {title} {\bibinfo {title} {Branching fraction measurements of the rare
  {$B^0_s\to\phi\mu^+\mu^-$ and $B^0_s\to f_2'(1525)\mu^+\mu^-$} decays},\
  }\href {https://doi.org/10.1103/PhysRevLett.127.151801} {\bibfield  {journal}
  {\bibinfo  {journal} {Phys. Rev. Lett.}\ }\textbf {\bibinfo {volume} {127}},\
  \bibinfo {pages} {151801} (\bibinfo {year} {2021})},\ \Eprint
  {https://arxiv.org/abs/2105.14007} {arXiv:2105.14007 [hep-ex]} \BibitemShut
  {NoStop}%
\bibitem [{\citenamefont {Aaij}\ \emph {et~al.}(2017)\citenamefont {Aaij} \emph
  {et~al.}}]{LHCb:2017avl}%
  \BibitemOpen
  \bibfield  {author} {\bibinfo {author} {\bibfnamefont {R.}~\bibnamefont
  {Aaij}} \emph {et~al.} (\bibinfo {collaboration} {LHCb}),\ }\bibfield
  {title} {\bibinfo {title} {{Test of lepton universality with $B^{0}
  \rightarrow K^{*0}\ell^{+}\ell^{-}$ decays}},\ }\href
  {https://doi.org/10.1007/JHEP08(2017)055} {\bibfield  {journal} {\bibinfo
  {journal} {JHEP}\ }\textbf {\bibinfo {volume} {08}},\ \bibinfo {pages}
  {055}},\ \Eprint {https://arxiv.org/abs/1705.05802} {arXiv:1705.05802
  [hep-ex]} \BibitemShut {NoStop}%
\bibitem [{\citenamefont {Aaij}\ \emph
  {et~al.}(2022{\natexlab{a}})\citenamefont {Aaij} \emph
  {et~al.}}]{LHCb:2021trn}%
  \BibitemOpen
  \bibfield  {author} {\bibinfo {author} {\bibfnamefont {R.}~\bibnamefont
  {Aaij}} \emph {et~al.} (\bibinfo {collaboration} {LHCb}),\ }\bibfield
  {title} {\bibinfo {title} {{Test of lepton universality in beauty-quark
  decays}},\ }\href {https://doi.org/10.1038/s41567-021-01478-8} {\bibfield
  {journal} {\bibinfo  {journal} {Nature Phys.}\ }\textbf {\bibinfo {volume}
  {18}},\ \bibinfo {pages} {277} (\bibinfo {year} {2022}{\natexlab{a}})},\
  \Eprint {https://arxiv.org/abs/2103.11769} {arXiv:2103.11769 [hep-ex]}
  \BibitemShut {NoStop}%
\bibitem [{\citenamefont {Aaij}\ \emph
  {et~al.}(2022{\natexlab{b}})\citenamefont {Aaij} \emph
  {et~al.}}]{LHCb:2021lvy}%
  \BibitemOpen
  \bibfield  {author} {\bibinfo {author} {\bibfnamefont {R.}~\bibnamefont
  {Aaij}} \emph {et~al.} (\bibinfo {collaboration} {LHCb}),\ }\bibfield
  {title} {\bibinfo {title} {{Tests of lepton universality using $B^0\to K^0_S
  \ell^+ \ell^-$ and $B^+\to K^{*+} \ell^+ \ell^-$ decays}},\ }\href
  {https://doi.org/10.1103/PhysRevLett.128.191802} {\bibfield  {journal}
  {\bibinfo  {journal} {Phys. Rev. Lett.}\ }\textbf {\bibinfo {volume} {128}},\
  \bibinfo {pages} {191802} (\bibinfo {year} {2022}{\natexlab{b}})},\ \Eprint
  {https://arxiv.org/abs/2110.09501} {arXiv:2110.09501 [hep-ex]} \BibitemShut
  {NoStop}%
\bibitem [{\citenamefont
  {\href{https://cds.cern.ch/record/2727207/}{LHCb-CONF-2020-002}}(2020)}]{LHCb:2020zud}%
  \BibitemOpen
  \bibfield  {author} {\bibinfo {author} {\bibnamefont
  {\href{https://cds.cern.ch/record/2727207/}{LHCb-CONF-2020-002}}} (\bibinfo
  {collaboration} {{LHCb, ATLAS, and CMS}}),\ }\bibfield  {title} {\bibinfo
  {title} {{Combination of the ATLAS, CMS, and LHCb results on the
  $B^0_{(s)}\to\mu^+\mu^-$ decays}},\ }\href@noop {} {\  (\bibinfo {year}
  {2020})}\BibitemShut {NoStop}%
\bibitem [{\citenamefont {Aaij}\ \emph
  {et~al.}(2022{\natexlab{c}})\citenamefont {Aaij} \emph
  {et~al.}}]{LHCb:2021awg}%
  \BibitemOpen
  \bibfield  {author} {\bibinfo {author} {\bibfnamefont {R.}~\bibnamefont
  {Aaij}} \emph {et~al.} (\bibinfo {collaboration} {LHCb}),\ }\bibfield
  {title} {\bibinfo {title} {{Measurement of the $B^0_s\to\mu^+\mu^-$ decay
  properties and search for the $B^0\to\mu^+\mu^-$ and
  $B^0_s\to\mu^+\mu^-\gamma$ decays}},\ }\href
  {https://doi.org/10.1103/PhysRevD.105.012010} {\bibfield  {journal} {\bibinfo
   {journal} {Phys. Rev. D}\ }\textbf {\bibinfo {volume} {105}},\ \bibinfo
  {pages} {012010} (\bibinfo {year} {2022}{\natexlab{c}})},\ \Eprint
  {https://arxiv.org/abs/2108.09283} {arXiv:2108.09283 [hep-ex]} \BibitemShut
  {NoStop}%
\bibitem [{\citenamefont
  {\href{https://cds.cern.ch/record/2815334/}{CMS-PAS-BPH-21-006}}(2022)}]{CMS-PAS-BPH-21-006}%
  \BibitemOpen
  \bibfield  {author} {\bibinfo {author} {\bibnamefont
  {\href{https://cds.cern.ch/record/2815334/}{CMS-PAS-BPH-21-006}}} (\bibinfo
  {collaboration} {{CMS}}),\ }\bibfield  {title} {\bibinfo {title} {Measurement
  of $b^0_s\to\mu^+\mu^-$ decay properties and search for the
  $b^0\to\mu^+\mu^-$ decay in proton-proton collisions at $\sqrt{s}=13$~tev},\
  }\href@noop {} {\  (\bibinfo {year} {2022})}\BibitemShut {NoStop}%
\bibitem [{\citenamefont {Detmold}\ and\ \citenamefont
  {Meinel}(2016)}]{Detmold:2016pkz}%
  \BibitemOpen
  \bibfield  {author} {\bibinfo {author} {\bibfnamefont {W.}~\bibnamefont
  {Detmold}}\ and\ \bibinfo {author} {\bibfnamefont {S.}~\bibnamefont
  {Meinel}},\ }\bibfield  {title} {\bibinfo {title} {{$\Lambda_b \to \Lambda
  \ell^+ \ell^-$} form factors, differential branching fraction, and angular
  observables from lattice qcd with relativistic $b$ quarks},\ }\href
  {https://doi.org/10.1103/PhysRevD.93.074501} {\bibfield  {journal} {\bibinfo
  {journal} {Phys. Rev. D}\ }\textbf {\bibinfo {volume} {93}},\ \bibinfo
  {pages} {074501} (\bibinfo {year} {2016})},\ \Eprint
  {https://arxiv.org/abs/1602.01399} {arXiv:1602.01399 [hep-lat]} \BibitemShut
  {NoStop}%
\bibitem [{\citenamefont {Aaij}\ \emph {et~al.}(2015)\citenamefont {Aaij} \emph
  {et~al.}}]{LHCb:2015tgy}%
  \BibitemOpen
  \bibfield  {author} {\bibinfo {author} {\bibfnamefont {R.}~\bibnamefont
  {Aaij}} \emph {et~al.} (\bibinfo {collaboration} {LHCb}),\ }\bibfield
  {title} {\bibinfo {title} {{Differential branching fraction and angular
  analysis of $\Lambda^{0}_{b} \rightarrow \Lambda \mu^+\mu^-$ decays}},\
  }\href {https://doi.org/10.1007/JHEP06(2015)115} {\bibfield  {journal}
  {\bibinfo  {journal} {JHEP}\ }\textbf {\bibinfo {volume} {06}},\ \bibinfo
  {pages} {115}},\ \bibinfo {note} {(E)
  \href{https://doi.org/10.1007/JHEP09(2018)145}{\textbf{09}, 145 (2018)}},\
  \Eprint {https://arxiv.org/abs/1503.07138} {arXiv:1503.07138 [hep-ex]}
  \BibitemShut {NoStop}%
\bibitem [{\citenamefont {Aaij}\ \emph
  {et~al.}(2018{\natexlab{a}})\citenamefont {Aaij} \emph
  {et~al.}}]{LHCb:2018jna}%
  \BibitemOpen
  \bibfield  {author} {\bibinfo {author} {\bibfnamefont {R.}~\bibnamefont
  {Aaij}} \emph {et~al.} (\bibinfo {collaboration} {LHCb}),\ }\bibfield
  {title} {\bibinfo {title} {{Angular moments of the decay $\Lambda_b^0
  \rightarrow \Lambda \mu^{+} \mu^{-}$ at low hadronic recoil}},\ }\href
  {https://doi.org/10.1007/JHEP09(2018)146} {\bibfield  {journal} {\bibinfo
  {journal} {JHEP}\ }\textbf {\bibinfo {volume} {09}},\ \bibinfo {pages}
  {146}},\ \Eprint {https://arxiv.org/abs/1808.00264} {arXiv:1808.00264
  [hep-ex]} \BibitemShut {NoStop}%
\bibitem [{\citenamefont {Blake}\ \emph {et~al.}(2020)\citenamefont {Blake},
  \citenamefont {Meinel},\ and\ \citenamefont {van Dyk}}]{Blake:2019guk}%
  \BibitemOpen
  \bibfield  {author} {\bibinfo {author} {\bibfnamefont {T.}~\bibnamefont
  {Blake}}, \bibinfo {author} {\bibfnamefont {S.}~\bibnamefont {Meinel}},\ and\
  \bibinfo {author} {\bibfnamefont {D.}~\bibnamefont {van Dyk}},\ }\bibfield
  {title} {\bibinfo {title} {{Bayesian} analysis of $b\to s\mu^+\mu^-$ {Wilson}
  coefficients using the full angular distribution of {$\Lambda_b\to
  \Lambda(\to p\, \pi^-)\mu^+\mu^-$} decays},\ }\href
  {https://doi.org/10.1103/PhysRevD.101.035023} {\bibfield  {journal} {\bibinfo
   {journal} {Phys. Rev. D}\ }\textbf {\bibinfo {volume} {101}},\ \bibinfo
  {pages} {035023} (\bibinfo {year} {2020})},\ \Eprint
  {https://arxiv.org/abs/1912.05811} {arXiv:1912.05811 [hep-ph]} \BibitemShut
  {NoStop}%
\bibitem [{\citenamefont {Blake}\ \emph {et~al.}(2022)\citenamefont {Blake},
  \citenamefont {Meinel}, \citenamefont {Rahimi},\ and\ \citenamefont {van
  Dyk}}]{Blake:2022vfl}%
  \BibitemOpen
  \bibfield  {author} {\bibinfo {author} {\bibfnamefont {T.}~\bibnamefont
  {Blake}}, \bibinfo {author} {\bibfnamefont {S.}~\bibnamefont {Meinel}},
  \bibinfo {author} {\bibfnamefont {M.}~\bibnamefont {Rahimi}},\ and\ \bibinfo
  {author} {\bibfnamefont {D.}~\bibnamefont {van Dyk}},\ }\bibfield  {title}
  {\bibinfo {title} {{Dispersive bounds for local form factors in $\Lambda_b
  \to \Lambda$ transitions}},\ }\href@noop {} {\  (\bibinfo {year} {2022})},\
  \Eprint {https://arxiv.org/abs/2205.06041} {arXiv:2205.06041 [hep-ph]}
  \BibitemShut {NoStop}%
\bibitem [{\citenamefont {Briceño}\ \emph {et~al.}(2015)\citenamefont
  {Briceño}, \citenamefont {Hansen},\ and\ \citenamefont
  {Walker-Loud}}]{Briceno:2014uqa}%
  \BibitemOpen
  \bibfield  {author} {\bibinfo {author} {\bibfnamefont {R.~A.}\ \bibnamefont
  {Briceño}}, \bibinfo {author} {\bibfnamefont {M.~T.}\ \bibnamefont
  {Hansen}},\ and\ \bibinfo {author} {\bibfnamefont {A.}~\bibnamefont
  {Walker-Loud}},\ }\bibfield  {title} {\bibinfo {title} {Multichannel $1\to2$
  transition amplitudes in a finite volume},\ }\href
  {https://doi.org/10.1103/PhysRevD.91.034501} {\bibfield  {journal} {\bibinfo
  {journal} {Phys. Rev. D}\ }\textbf {\bibinfo {volume} {91}},\ \bibinfo
  {pages} {034501} (\bibinfo {year} {2015})},\ \Eprint
  {https://arxiv.org/abs/1406.5965} {arXiv:1406.5965 [hep-lat]} \BibitemShut
  {NoStop}%
\bibitem [{\citenamefont {Rendon}\ \emph {et~al.}(2019)\citenamefont {Rendon},
  \citenamefont {Leskovec}, \citenamefont {Meinel}, \citenamefont {Negele},
  \citenamefont {Paul}, \citenamefont {Petschlies}, \citenamefont {Pochinsky},
  \citenamefont {Silvi},\ and\ \citenamefont {Syritsyn}}]{Rendon:2018fem}%
  \BibitemOpen
  \bibfield  {author} {\bibinfo {author} {\bibfnamefont {G.}~\bibnamefont
  {Rendon}}, \bibinfo {author} {\bibfnamefont {L.}~\bibnamefont {Leskovec}},
  \bibinfo {author} {\bibfnamefont {S.}~\bibnamefont {Meinel}}, \bibinfo
  {author} {\bibfnamefont {J.}~\bibnamefont {Negele}}, \bibinfo {author}
  {\bibfnamefont {S.}~\bibnamefont {Paul}}, \bibinfo {author} {\bibfnamefont
  {M.}~\bibnamefont {Petschlies}}, \bibinfo {author} {\bibfnamefont
  {A.}~\bibnamefont {Pochinsky}}, \bibinfo {author} {\bibfnamefont
  {G.}~\bibnamefont {Silvi}},\ and\ \bibinfo {author} {\bibfnamefont
  {S.}~\bibnamefont {Syritsyn}},\ }\bibfield  {title} {\bibinfo {title} {{$K
  \pi$ scattering and the $K^*(892)$ resonance in 2+1 flavor QCD}},\ }\href
  {https://doi.org/10.22323/1.334.0073} {\bibfield  {journal} {\bibinfo
  {journal} {PoS}\ }\textbf {\bibinfo {volume} {LATTICE2018}},\ \bibinfo
  {pages} {073} (\bibinfo {year} {2019})},\ \Eprint
  {https://arxiv.org/abs/1811.10750} {arXiv:1811.10750 [hep-lat]} \BibitemShut
  {NoStop}%
\bibitem [{\citenamefont {Bazavov}\ \emph {et~al.}(2016)\citenamefont {Bazavov}
  \emph {et~al.}}]{Bazavov:2016nty}%
  \BibitemOpen
  \bibfield  {author} {\bibinfo {author} {\bibfnamefont {A.}~\bibnamefont
  {Bazavov}} \emph {et~al.} (\bibinfo {collaboration} {Fermilab Lattice,
  MILC}),\ }\bibfield  {title} {\bibinfo {title} {{$B^0_{(s)}$-mixing matrix
  elements from lattice QCD for the Standard Model and beyond}},\ }\href
  {https://doi.org/10.1103/PhysRevD.93.113016} {\bibfield  {journal} {\bibinfo
  {journal} {Phys. Rev. D}\ }\textbf {\bibinfo {volume} {93}},\ \bibinfo
  {pages} {113016} (\bibinfo {year} {2016})},\ \Eprint
  {https://arxiv.org/abs/1602.03560} {arXiv:1602.03560 [hep-lat]} \BibitemShut
  {NoStop}%
\bibitem [{\citenamefont {Dowdall}\ \emph {et~al.}(2019)\citenamefont
  {Dowdall}, \citenamefont {Davies}, \citenamefont {Horgan}, \citenamefont
  {Lepage}, \citenamefont {Monahan}, \citenamefont {Shigemitsu},\ and\
  \citenamefont {Wingate}}]{Dowdall:2019bea}%
  \BibitemOpen
  \bibfield  {author} {\bibinfo {author} {\bibfnamefont {R.~J.}\ \bibnamefont
  {Dowdall}}, \bibinfo {author} {\bibfnamefont {C.~T.~H.}\ \bibnamefont
  {Davies}}, \bibinfo {author} {\bibfnamefont {R.~R.}\ \bibnamefont {Horgan}},
  \bibinfo {author} {\bibfnamefont {G.~P.}\ \bibnamefont {Lepage}}, \bibinfo
  {author} {\bibfnamefont {C.~J.}\ \bibnamefont {Monahan}}, \bibinfo {author}
  {\bibfnamefont {J.}~\bibnamefont {Shigemitsu}},\ and\ \bibinfo {author}
  {\bibfnamefont {M.}~\bibnamefont {Wingate}} (\bibinfo {collaboration}
  {HPQCD}),\ }\bibfield  {title} {\bibinfo {title} {{Neutral $B$-meson mixing
  from full lattice QCD at the physical point}},\ }\href
  {https://doi.org/10.1103/PhysRevD.100.094508} {\bibfield  {journal} {\bibinfo
   {journal} {Phys. Rev. D}\ }\textbf {\bibinfo {volume} {100}},\ \bibinfo
  {pages} {094508} (\bibinfo {year} {2019})},\ \Eprint
  {https://arxiv.org/abs/1907.01025} {arXiv:1907.01025 [hep-lat]} \BibitemShut
  {NoStop}%
\bibitem [{\citenamefont {Amhis}\ \emph {et~al.}(2022)\citenamefont {Amhis}
  \emph {et~al.}}]{Amhis:2022mac}%
  \BibitemOpen
  \bibfield  {author} {\bibinfo {author} {\bibfnamefont {Y.}~\bibnamefont
  {Amhis}} \emph {et~al.},\ }\bibfield  {title} {\bibinfo {title} {{Averages of
  $b$-hadron, $c$-hadron, and $\tau$-lepton properties as of 2021}},\
  }\href@noop {} {\  (\bibinfo {year} {2022})},\ \Eprint
  {https://arxiv.org/abs/2206.07501} {arXiv:2206.07501 [hep-ex]} \BibitemShut
  {NoStop}%
\bibitem [{\citenamefont {Harrison}\ \emph {et~al.}(2020)\citenamefont
  {Harrison}, \citenamefont {Davies},\ and\ \citenamefont
  {Lytle}}]{Harrison:2020nrv}%
  \BibitemOpen
  \bibfield  {author} {\bibinfo {author} {\bibfnamefont {J.}~\bibnamefont
  {Harrison}}, \bibinfo {author} {\bibfnamefont {C.~T.~H.}\ \bibnamefont
  {Davies}},\ and\ \bibinfo {author} {\bibfnamefont {A.}~\bibnamefont {Lytle}}
  (\bibinfo {collaboration} {HPQCD}),\ }\bibfield  {title} {\bibinfo {title}
  {{$R(J/\psi)$ and $B_c^-\to J/\psi \ell^-\bar{\nu}_\ell$} lepton flavor
  universality violating observables from lattice {QCD}},\ }\href
  {https://doi.org/10.1103/PhysRevLett.125.222003} {\bibfield  {journal}
  {\bibinfo  {journal} {Phys. Rev. Lett.}\ }\textbf {\bibinfo {volume} {125}},\
  \bibinfo {pages} {222003} (\bibinfo {year} {2020})},\ \Eprint
  {https://arxiv.org/abs/2007.06956} {arXiv:2007.06956 [hep-lat]} \BibitemShut
  {NoStop}%
\bibitem [{\citenamefont {Aaij}\ \emph
  {et~al.}(2018{\natexlab{b}})\citenamefont {Aaij} \emph
  {et~al.}}]{LHCb:2017vlu}%
  \BibitemOpen
  \bibfield  {author} {\bibinfo {author} {\bibfnamefont {R.}~\bibnamefont
  {Aaij}} \emph {et~al.} (\bibinfo {collaboration} {LHCb}),\ }\bibfield
  {title} {\bibinfo {title} {{Measurement of the ratio of branching fractions
  $\mathcal{B}(B_c^+\to J/\psi\tau^+\nu_\tau)$/$\mathcal{B}(B_c^+\to
  J/\psi\mu^+\nu_\mu)$}},\ }\href
  {https://doi.org/10.1103/PhysRevLett.120.121801} {\bibfield  {journal}
  {\bibinfo  {journal} {Phys. Rev. Lett.}\ }\textbf {\bibinfo {volume} {120}},\
  \bibinfo {pages} {121801} (\bibinfo {year} {2018}{\natexlab{b}})},\ \Eprint
  {https://arxiv.org/abs/1711.05623} {arXiv:1711.05623 [hep-ex]} \BibitemShut
  {NoStop}%
\bibitem [{\citenamefont {Aaij}\ \emph
  {et~al.}(2022{\natexlab{d}})\citenamefont {Aaij} \emph
  {et~al.}}]{LHCb:2022piu}%
  \BibitemOpen
  \bibfield  {author} {\bibinfo {author} {\bibfnamefont {R.}~\bibnamefont
  {Aaij}} \emph {et~al.} (\bibinfo {collaboration} {LHCb}),\ }\bibfield
  {title} {\bibinfo {title} {Observation of the decay {$\Lambda_b^0\to
  \Lambda_c^+\tau^-\bar{\nu}_{\tau}$}},\ }\href
  {https://doi.org/10.1103/PhysRevLett.128.191803} {\bibfield  {journal}
  {\bibinfo  {journal} {Phys. Rev. Lett.}\ }\textbf {\bibinfo {volume} {128}},\
  \bibinfo {pages} {191803} (\bibinfo {year} {2022}{\natexlab{d}})},\ \Eprint
  {https://arxiv.org/abs/2201.03497} {arXiv:2201.03497 [hep-ex]} \BibitemShut
  {NoStop}%
\bibitem [{\citenamefont {Gambino}\ \emph {et~al.}(2020)\citenamefont {Gambino}
  \emph {et~al.}}]{Gambino:2020jvv}%
  \BibitemOpen
  \bibfield  {author} {\bibinfo {author} {\bibfnamefont {P.}~\bibnamefont
  {Gambino}} \emph {et~al.},\ }\bibfield  {title} {\bibinfo {title}
  {{Challenges in semileptonic $B$ decays}},\ }\href
  {https://doi.org/10.1140/epjc/s10052-020-08490-x} {\bibfield  {journal}
  {\bibinfo  {journal} {Eur. Phys. J. C}\ }\textbf {\bibinfo {volume} {80}},\
  \bibinfo {pages} {966} (\bibinfo {year} {2020})},\ \Eprint
  {https://arxiv.org/abs/2006.07287} {arXiv:2006.07287 [hep-ph]} \BibitemShut
  {NoStop}%
\bibitem [{\citenamefont {Gimenez}\ \emph {et~al.}(1997)\citenamefont
  {Gimenez}, \citenamefont {Martinelli},\ and\ \citenamefont
  {Sachrajda}}]{Gimenez:1996av}%
  \BibitemOpen
  \bibfield  {author} {\bibinfo {author} {\bibfnamefont {V.}~\bibnamefont
  {Gimenez}}, \bibinfo {author} {\bibfnamefont {G.}~\bibnamefont
  {Martinelli}},\ and\ \bibinfo {author} {\bibfnamefont {C.~T.}\ \bibnamefont
  {Sachrajda}},\ }\bibfield  {title} {\bibinfo {title} {A high statistics
  lattice calculation of $\lambda_1$ and $\lambda_2$ in the {$B$} meson},\
  }\href {https://doi.org/10.1016/S0550-3213(96)00696-7} {\bibfield  {journal}
  {\bibinfo  {journal} {Nucl. Phys. B}\ }\textbf {\bibinfo {volume} {486}},\
  \bibinfo {pages} {227} (\bibinfo {year} {1997})},\ \Eprint
  {https://arxiv.org/abs/hep-lat/9607055} {arXiv:hep-lat/9607055} \BibitemShut
  {NoStop}%
\bibitem [{\citenamefont {Kronfeld}\ and\ \citenamefont
  {Simone}(2000)}]{Kronfeld:2000gk}%
  \BibitemOpen
  \bibfield  {author} {\bibinfo {author} {\bibfnamefont {A.~S.}\ \bibnamefont
  {Kronfeld}}\ and\ \bibinfo {author} {\bibfnamefont {J.~N.}\ \bibnamefont
  {Simone}},\ }\bibfield  {title} {\bibinfo {title} {{Computation of
  $\bar\Lambda$ and $\lambda_1$ with lattice QCD}},\ }\href
  {https://doi.org/10.1016/S0370-2693(00)00833-9} {\bibfield  {journal}
  {\bibinfo  {journal} {Phys. Lett. B}\ }\textbf {\bibinfo {volume} {490}},\
  \bibinfo {pages} {228} (\bibinfo {year} {2000})},\ \bibinfo {note} {(E)
  \href{https://doi.org/10.1016/S0370-2693(00)01277-6}{\textbf{495}, 441--442
  (2000)}},\ \Eprint {https://arxiv.org/abs/hep-ph/0006345}
  {arXiv:hep-ph/0006345} \BibitemShut {NoStop}%
\bibitem [{\citenamefont {Gambino}\ \emph {et~al.}(2017)\citenamefont
  {Gambino}, \citenamefont {Melis},\ and\ \citenamefont
  {Simula}}]{Gambino:2017vkx}%
  \BibitemOpen
  \bibfield  {author} {\bibinfo {author} {\bibfnamefont {P.}~\bibnamefont
  {Gambino}}, \bibinfo {author} {\bibfnamefont {A.}~\bibnamefont {Melis}},\
  and\ \bibinfo {author} {\bibfnamefont {S.}~\bibnamefont {Simula}},\
  }\bibfield  {title} {\bibinfo {title} {{Extraction of heavy-quark-expansion
  parameters from unquenched lattice data on pseudoscalar and vector
  heavy-light meson masses}},\ }\href
  {https://doi.org/10.1103/PhysRevD.96.014511} {\bibfield  {journal} {\bibinfo
  {journal} {Phys. Rev. D}\ }\textbf {\bibinfo {volume} {96}},\ \bibinfo
  {pages} {014511} (\bibinfo {year} {2017})},\ \Eprint
  {https://arxiv.org/abs/1704.06105} {arXiv:1704.06105 [hep-lat]} \BibitemShut
  {NoStop}%
\bibitem [{\citenamefont {Bazavov}\ \emph
  {et~al.}(2018{\natexlab{b}})\citenamefont {Bazavov} \emph
  {et~al.}}]{FermilabLattice:2018est}%
  \BibitemOpen
  \bibfield  {author} {\bibinfo {author} {\bibfnamefont {A.}~\bibnamefont
  {Bazavov}} \emph {et~al.} (\bibinfo {collaboration} {Fermilab Lattice, MILC,
  TUMQCD}),\ }\bibfield  {title} {\bibinfo {title} {{Up-, down-, strange-,
  charm-, and bottom-quark masses from four-flavor lattice QCD}},\ }\href
  {https://doi.org/10.1103/PhysRevD.98.054517} {\bibfield  {journal} {\bibinfo
  {journal} {Phys. Rev. D}\ }\textbf {\bibinfo {volume} {98}},\ \bibinfo
  {pages} {054517} (\bibinfo {year} {2018}{\natexlab{b}})},\ \Eprint
  {https://arxiv.org/abs/1802.04248} {arXiv:1802.04248 [hep-lat]} \BibitemShut
  {NoStop}%
\bibitem [{\citenamefont {Hashimoto}(2017)}]{Hashimoto:2017wqo}%
  \BibitemOpen
  \bibfield  {author} {\bibinfo {author} {\bibfnamefont {S.}~\bibnamefont
  {Hashimoto}},\ }\bibfield  {title} {\bibinfo {title} {{Inclusive
  semi-leptonic $B$ meson decay structure functions from lattice QCD}},\ }\href
  {https://doi.org/10.1093/ptep/ptx052} {\bibfield  {journal} {\bibinfo
  {journal} {PTEP}\ }\textbf {\bibinfo {volume} {2017}},\ \bibinfo {pages}
  {053B03} (\bibinfo {year} {2017})},\ \Eprint
  {https://arxiv.org/abs/1703.01881} {arXiv:1703.01881 [hep-lat]} \BibitemShut
  {NoStop}%
\bibitem [{\citenamefont {Hansen}\ \emph {et~al.}(2017)\citenamefont {Hansen},
  \citenamefont {Meyer},\ and\ \citenamefont {Robaina}}]{Hansen:2017mnd}%
  \BibitemOpen
  \bibfield  {author} {\bibinfo {author} {\bibfnamefont {M.~T.}\ \bibnamefont
  {Hansen}}, \bibinfo {author} {\bibfnamefont {H.~B.}\ \bibnamefont {Meyer}},\
  and\ \bibinfo {author} {\bibfnamefont {D.}~\bibnamefont {Robaina}},\
  }\bibfield  {title} {\bibinfo {title} {{From deep inelastic scattering to
  heavy-flavor semileptonic decays: Total rates into multihadron final states
  from lattice QCD}},\ }\href {https://doi.org/10.1103/PhysRevD.96.094513}
  {\bibfield  {journal} {\bibinfo  {journal} {Phys. Rev. D}\ }\textbf {\bibinfo
  {volume} {96}},\ \bibinfo {pages} {094513} (\bibinfo {year} {2017})},\
  \Eprint {https://arxiv.org/abs/1704.08993} {arXiv:1704.08993 [hep-lat]}
  \BibitemShut {NoStop}%
\bibitem [{\citenamefont {Fukaya}\ \emph {et~al.}(2020)\citenamefont {Fukaya},
  \citenamefont {Hashimoto}, \citenamefont {Kaneko},\ and\ \citenamefont
  {Ohki}}]{Fukaya:2020wpp}%
  \BibitemOpen
  \bibfield  {author} {\bibinfo {author} {\bibfnamefont {H.}~\bibnamefont
  {Fukaya}}, \bibinfo {author} {\bibfnamefont {S.}~\bibnamefont {Hashimoto}},
  \bibinfo {author} {\bibfnamefont {T.}~\bibnamefont {Kaneko}},\ and\ \bibinfo
  {author} {\bibfnamefont {H.}~\bibnamefont {Ohki}},\ }\bibfield  {title}
  {\bibinfo {title} {{Towards fully nonperturbative computations of inelastic
  $\ell N$ scattering cross sections from lattice QCD}},\ }\href
  {https://doi.org/10.1103/PhysRevD.102.114516} {\bibfield  {journal} {\bibinfo
   {journal} {Phys. Rev. D}\ }\textbf {\bibinfo {volume} {102}},\ \bibinfo
  {pages} {114516} (\bibinfo {year} {2020})},\ \Eprint
  {https://arxiv.org/abs/2010.01253} {arXiv:2010.01253 [hep-lat]} \BibitemShut
  {NoStop}%
\bibitem [{\citenamefont {Gambino}\ and\ \citenamefont
  {Hashimoto}(2020)}]{Gambino:2020crt}%
  \BibitemOpen
  \bibfield  {author} {\bibinfo {author} {\bibfnamefont {P.}~\bibnamefont
  {Gambino}}\ and\ \bibinfo {author} {\bibfnamefont {S.}~\bibnamefont
  {Hashimoto}},\ }\bibfield  {title} {\bibinfo {title} {Inclusive semileptonic
  decays from lattice {QCD}},\ }\href
  {https://doi.org/10.1103/PhysRevLett.125.032001} {\bibfield  {journal}
  {\bibinfo  {journal} {Phys. Rev. Lett.}\ }\textbf {\bibinfo {volume} {125}},\
  \bibinfo {pages} {032001} (\bibinfo {year} {2020})},\ \Eprint
  {https://arxiv.org/abs/2005.13730} {arXiv:2005.13730 [hep-lat]} \BibitemShut
  {NoStop}%
\bibitem [{\citenamefont {Ishikawa}\ and\ \citenamefont
  {Hashimoto}(2021)}]{Ishikawa:2021txe}%
  \BibitemOpen
  \bibfield  {author} {\bibinfo {author} {\bibfnamefont {T.}~\bibnamefont
  {Ishikawa}}\ and\ \bibinfo {author} {\bibfnamefont {S.}~\bibnamefont
  {Hashimoto}},\ }\bibfield  {title} {\bibinfo {title} {{Spectral sum of
  current correlators from lattice QCD}},\ }\href
  {https://doi.org/10.1103/PhysRevD.104.074521} {\bibfield  {journal} {\bibinfo
   {journal} {Phys. Rev. D}\ }\textbf {\bibinfo {volume} {104}},\ \bibinfo
  {pages} {074521} (\bibinfo {year} {2021})},\ \Eprint
  {https://arxiv.org/abs/2103.06539} {arXiv:2103.06539 [hep-lat]} \BibitemShut
  {NoStop}%
\bibitem [{\citenamefont {DeGrand}(2022)}]{DeGrand:2022lmc}%
  \BibitemOpen
  \bibfield  {author} {\bibinfo {author} {\bibfnamefont {T.}~\bibnamefont
  {DeGrand}},\ }\bibfield  {title} {\bibinfo {title} {{Remarks about weighted
  energy integrals over Minkowski spectral functions from Euclidean lattice
  data}},\ }\href@noop {} {\  (\bibinfo {year} {2022})},\ \Eprint
  {https://arxiv.org/abs/2203.04393} {arXiv:2203.04393 [hep-lat]} \BibitemShut
  {NoStop}%
\bibitem [{\citenamefont {Gambino}\ \emph {et~al.}(2022)\citenamefont
  {Gambino}, \citenamefont {Hashimoto}, \citenamefont {M\"achler},
  \citenamefont {Panero}, \citenamefont {Sanfilippo}, \citenamefont {Simula},
  \citenamefont {Smecca},\ and\ \citenamefont {Tantalo}}]{Gambino:2022dvu}%
  \BibitemOpen
  \bibfield  {author} {\bibinfo {author} {\bibfnamefont {P.}~\bibnamefont
  {Gambino}}, \bibinfo {author} {\bibfnamefont {S.}~\bibnamefont {Hashimoto}},
  \bibinfo {author} {\bibfnamefont {S.}~\bibnamefont {M\"achler}}, \bibinfo
  {author} {\bibfnamefont {M.}~\bibnamefont {Panero}}, \bibinfo {author}
  {\bibfnamefont {F.}~\bibnamefont {Sanfilippo}}, \bibinfo {author}
  {\bibfnamefont {S.}~\bibnamefont {Simula}}, \bibinfo {author} {\bibfnamefont
  {A.}~\bibnamefont {Smecca}},\ and\ \bibinfo {author} {\bibfnamefont
  {N.}~\bibnamefont {Tantalo}},\ }\bibfield  {title} {\bibinfo {title}
  {{Lattice QCD study of inclusive semileptonic decays of heavy mesons}},\
  }\href@noop {} {\  (\bibinfo {year} {2022})},\ \Eprint
  {https://arxiv.org/abs/2203.11762} {arXiv:2203.11762 [hep-lat]} \BibitemShut
  {NoStop}%
\bibitem [{\citenamefont {Chakraborty}\ \emph {et~al.}(2015)\citenamefont
  {Chakraborty}, \citenamefont {Davies}, \citenamefont {Galloway},
  \citenamefont {Knecht}, \citenamefont {Koponen}, \citenamefont {Donald},
  \citenamefont {Dowdall}, \citenamefont {Lepage},\ and\ \citenamefont
  {McNeile}}]{Chakraborty:2014aca}%
  \BibitemOpen
  \bibfield  {author} {\bibinfo {author} {\bibfnamefont {B.}~\bibnamefont
  {Chakraborty}}, \bibinfo {author} {\bibfnamefont {C.~T.~H.}\ \bibnamefont
  {Davies}}, \bibinfo {author} {\bibfnamefont {B.}~\bibnamefont {Galloway}},
  \bibinfo {author} {\bibfnamefont {P.}~\bibnamefont {Knecht}}, \bibinfo
  {author} {\bibfnamefont {J.}~\bibnamefont {Koponen}}, \bibinfo {author}
  {\bibfnamefont {G.~C.}\ \bibnamefont {Donald}}, \bibinfo {author}
  {\bibfnamefont {R.~J.}\ \bibnamefont {Dowdall}}, \bibinfo {author}
  {\bibfnamefont {G.~P.}\ \bibnamefont {Lepage}},\ and\ \bibinfo {author}
  {\bibfnamefont {C.}~\bibnamefont {McNeile}},\ }\bibfield  {title} {\bibinfo
  {title} {{High-precision quark masses and QCD coupling from $n_f=4$ lattice
  QCD}},\ }\href {https://doi.org/10.1103/PhysRevD.91.054508} {\bibfield
  {journal} {\bibinfo  {journal} {Phys. Rev. D}\ }\textbf {\bibinfo {volume}
  {91}},\ \bibinfo {pages} {054508} (\bibinfo {year} {2015})},\ \Eprint
  {https://arxiv.org/abs/1408.4169} {arXiv:1408.4169 [hep-lat]} \BibitemShut
  {NoStop}%
\bibitem [{\citenamefont {Lytle}\ \emph {et~al.}(2018)\citenamefont {Lytle},
  \citenamefont {Davies}, \citenamefont {Hatton}, \citenamefont {Lepage},\ and\
  \citenamefont {Sturm}}]{Lytle:2018evc}%
  \BibitemOpen
  \bibfield  {author} {\bibinfo {author} {\bibfnamefont {A.~T.}\ \bibnamefont
  {Lytle}}, \bibinfo {author} {\bibfnamefont {C.~T.~H.}\ \bibnamefont
  {Davies}}, \bibinfo {author} {\bibfnamefont {D.}~\bibnamefont {Hatton}},
  \bibinfo {author} {\bibfnamefont {G.~P.}\ \bibnamefont {Lepage}},\ and\
  \bibinfo {author} {\bibfnamefont {C.}~\bibnamefont {Sturm}} (\bibinfo
  {collaboration} {HPQCD}),\ }\bibfield  {title} {\bibinfo {title}
  {{Determination of quark masses from $n_f=4$ lattice QCD and the RI-SMOM
  intermediate scheme}},\ }\href {https://doi.org/10.1103/PhysRevD.98.014513}
  {\bibfield  {journal} {\bibinfo  {journal} {Phys. Rev. D}\ }\textbf {\bibinfo
  {volume} {98}},\ \bibinfo {pages} {014513} (\bibinfo {year} {2018})},\
  \Eprint {https://arxiv.org/abs/1805.06225} {arXiv:1805.06225 [hep-lat]}
  \BibitemShut {NoStop}%
\bibitem [{\citenamefont {Hatton}\ \emph {et~al.}(2021)\citenamefont {Hatton},
  \citenamefont {Davies}, \citenamefont {Koponen}, \citenamefont {Lepage},\
  and\ \citenamefont {Lytle}}]{Hatton:2021syc}%
  \BibitemOpen
  \bibfield  {author} {\bibinfo {author} {\bibfnamefont {D.}~\bibnamefont
  {Hatton}}, \bibinfo {author} {\bibfnamefont {C.~T.~H.}\ \bibnamefont
  {Davies}}, \bibinfo {author} {\bibfnamefont {J.}~\bibnamefont {Koponen}},
  \bibinfo {author} {\bibfnamefont {G.~P.}\ \bibnamefont {Lepage}},\ and\
  \bibinfo {author} {\bibfnamefont {A.~T.}\ \bibnamefont {Lytle}},\ }\bibfield
  {title} {\bibinfo {title} {{Determination of $\overline{m}_b/\overline{m}_c$
  and $\overline{m}_b$ from $n_f=4$ lattice QCD+QED}},\ }\href
  {https://doi.org/10.1103/PhysRevD.103.114508} {\bibfield  {journal} {\bibinfo
   {journal} {Phys. Rev. D}\ }\textbf {\bibinfo {volume} {103}},\ \bibinfo
  {pages} {114508} (\bibinfo {year} {2021})},\ \Eprint
  {https://arxiv.org/abs/2102.09609} {arXiv:2102.09609 [hep-lat]} \BibitemShut
  {NoStop}%
\bibitem [{\citenamefont {Jay}\ \emph {et~al.}(2022)\citenamefont {Jay},
  \citenamefont {Lytle}, \citenamefont {DeTar}, \citenamefont {El-Khadra},
  \citenamefont {Gamiz}, \citenamefont {Gelzer}, \citenamefont {Gottlieb},
  \citenamefont {Kronfeld}, \citenamefont {Simone},\ and\ \citenamefont
  {Vaquero}}]{FermilabLattice:2021bxu}%
  \BibitemOpen
  \bibfield  {author} {\bibinfo {author} {\bibfnamefont {W.~I.}\ \bibnamefont
  {Jay}}, \bibinfo {author} {\bibfnamefont {A.}~\bibnamefont {Lytle}}, \bibinfo
  {author} {\bibfnamefont {C.}~\bibnamefont {DeTar}}, \bibinfo {author}
  {\bibfnamefont {A.~X.}\ \bibnamefont {El-Khadra}}, \bibinfo {author}
  {\bibfnamefont {E.}~\bibnamefont {Gamiz}}, \bibinfo {author} {\bibfnamefont
  {Z.}~\bibnamefont {Gelzer}}, \bibinfo {author} {\bibfnamefont
  {S.}~\bibnamefont {Gottlieb}}, \bibinfo {author} {\bibfnamefont
  {A.}~\bibnamefont {Kronfeld}}, \bibinfo {author} {\bibfnamefont
  {J.}~\bibnamefont {Simone}},\ and\ \bibinfo {author} {\bibfnamefont
  {A.}~\bibnamefont {Vaquero}} (\bibinfo {collaboration} {Fermilab Lattice,
  MILC}),\ }\bibfield  {title} {\bibinfo {title} {{$B$- and $D$-meson
  semileptonic decays with highly improved staggered quarks}},\ }\href
  {https://doi.org/10.22323/1.396.0109} {\bibfield  {journal} {\bibinfo
  {journal} {PoS}\ }\textbf {\bibinfo {volume} {LATTICE2021}},\ \bibinfo
  {pages} {109} (\bibinfo {year} {2022})},\ \Eprint
  {https://arxiv.org/abs/2111.05184} {arXiv:2111.05184 [hep-lat]} \BibitemShut
  {NoStop}%
\bibitem [{\citenamefont {Boyle}\ \emph
  {et~al.}(2022{\natexlab{b}})\citenamefont {Boyle} \emph
  {et~al.}}]{Boyle:2022uba}%
  \BibitemOpen
  \bibfield  {author} {\bibinfo {author} {\bibfnamefont {P.~A.}\ \bibnamefont
  {Boyle}} \emph {et~al.},\ }\bibfield  {title} {\bibinfo {title} {A lattice
  {QCD} perspective on weak decays of $b$ and $c$ quarks},\ }in\ \href@noop {}
  {\emph {\bibinfo {booktitle} {{2022 Snowmass Summer Study}}}}\ (\bibinfo
  {year} {2022})\ \Eprint {https://arxiv.org/abs/2205.15373} {arXiv:2205.15373
  [hep-lat]} \BibitemShut {NoStop}%
\bibitem [{\citenamefont {Charles}\ \emph {et~al.}(2020)\citenamefont
  {Charles}, \citenamefont {Descotes-Genon}, \citenamefont {Ligeti},
  \citenamefont {Monteil}, \citenamefont {Papucci}, \citenamefont {Trabelsi},\
  and\ \citenamefont {Vale~Silva}}]{Charles:2020dfl}%
  \BibitemOpen
  \bibfield  {author} {\bibinfo {author} {\bibfnamefont {J.}~\bibnamefont
  {Charles}}, \bibinfo {author} {\bibfnamefont {S.}~\bibnamefont
  {Descotes-Genon}}, \bibinfo {author} {\bibfnamefont {Z.}~\bibnamefont
  {Ligeti}}, \bibinfo {author} {\bibfnamefont {S.}~\bibnamefont {Monteil}},
  \bibinfo {author} {\bibfnamefont {M.}~\bibnamefont {Papucci}}, \bibinfo
  {author} {\bibfnamefont {K.}~\bibnamefont {Trabelsi}},\ and\ \bibinfo
  {author} {\bibfnamefont {L.}~\bibnamefont {Vale~Silva}},\ }\bibfield  {title}
  {\bibinfo {title} {{New physics in $B$ meson mixing: future sensitivity and
  limitations}},\ }\href {https://doi.org/10.1103/PhysRevD.102.056023}
  {\bibfield  {journal} {\bibinfo  {journal} {Phys. Rev. D}\ }\textbf {\bibinfo
  {volume} {102}},\ \bibinfo {pages} {056023} (\bibinfo {year} {2020})},\
  \Eprint {https://arxiv.org/abs/2006.04824} {arXiv:2006.04824 [hep-ph]}
  \BibitemShut {NoStop}%
\bibitem [{\citenamefont {Cheng}\ \emph {et~al.}(2022)\citenamefont {Cheng},
  \citenamefont {Lyu},\ and\ \citenamefont {Xing}}]{Cheng:2022tog}%
  \BibitemOpen
  \bibfield  {author} {\bibinfo {author} {\bibfnamefont {H.-Y.}\ \bibnamefont
  {Cheng}}, \bibinfo {author} {\bibfnamefont {X.-R.}\ \bibnamefont {Lyu}},\
  and\ \bibinfo {author} {\bibfnamefont {Z.-Z.}\ \bibnamefont {Xing}},\
  }\bibfield  {title} {\bibinfo {title} {Charm physics in the high-luminosity
  super $\tau$-charm factory},\ }in\ \href@noop {} {\emph {\bibinfo {booktitle}
  {{2022 Snowmass Summer Study}}}}\ (\bibinfo {year} {2022})\ \Eprint
  {https://arxiv.org/abs/2203.03211} {arXiv:2203.03211 [hep-ph]} \BibitemShut
  {NoStop}%
\bibitem [{\citenamefont
  {\href{https://cds.cern.ch/record/2806113}{LHCb-PUB-2022-012}}(2022)}]{LHCb:2022ine}%
  \BibitemOpen
  \bibfield  {author} {\bibinfo {author} {\bibnamefont
  {\href{https://cds.cern.ch/record/2806113}{LHCb-PUB-2022-012}}} (\bibinfo
  {collaboration} {LHCb}),\ }\bibfield  {title} {\bibinfo {title} {Future
  physics potential of {LHCb}},\ }in\ \href@noop {} {\emph {\bibinfo
  {booktitle} {{2022 Snowmass Summer Study}}}}\ (\bibinfo {year}
  {2022})\BibitemShut {NoStop}%
\bibitem [{\citenamefont {Aggarwal}\ \emph {et~al.}(2022)\citenamefont
  {Aggarwal} \emph {et~al.}}]{BelleII:2022yoy}%
  \BibitemOpen
  \bibfield  {author} {\bibinfo {author} {\bibfnamefont {L.}~\bibnamefont
  {Aggarwal}} \emph {et~al.} (\bibinfo {collaboration} {Belle-II}),\ }\bibfield
   {title} {\bibinfo {title} {{Belle II physics reach and plans for the next
  decade and beyond}},\ }in\ \href@noop {} {\emph {\bibinfo {booktitle} {{2022
  Snowmass Summer Study}}}}\ (\bibinfo {year} {2022})\ \Eprint
  {https://arxiv.org/abs/2207.06307} {arXiv:2207.06307 [hep-ex]} \BibitemShut
  {NoStop}%
\bibitem [{\citenamefont {Li}\ \emph {et~al.}(2022)\citenamefont {Li} \emph
  {et~al.}}]{BESIII:2022mxl}%
  \BibitemOpen
  \bibfield  {author} {\bibinfo {author} {\bibfnamefont {H.~B.}\ \bibnamefont
  {Li}} \emph {et~al.} (\bibinfo {collaboration} {BESIII}),\ }\bibfield
  {title} {\bibinfo {title} {Physics in the $\tau$-charm region at {BESIII}},\
  }in\ \href@noop {} {\emph {\bibinfo {booktitle} {{2022 Snowmass Summer
  Study}}}}\ (\bibinfo {year} {2022})\ \Eprint
  {https://arxiv.org/abs/2204.08943} {arXiv:2204.08943 [hep-ex]} \BibitemShut
  {NoStop}%
\bibitem [{\citenamefont {Bhattacharya}\ \emph {et~al.}(2022)\citenamefont
  {Bhattacharya}, \citenamefont {Browder}, \citenamefont {Campagna},
  \citenamefont {Datta}, \citenamefont {Dubey}, \citenamefont {Mukherjee},\
  and\ \citenamefont {Sibidanov}}]{Bhattacharya:2022cna}%
  \BibitemOpen
  \bibfield  {author} {\bibinfo {author} {\bibfnamefont {B.}~\bibnamefont
  {Bhattacharya}}, \bibinfo {author} {\bibfnamefont {T.}~\bibnamefont
  {Browder}}, \bibinfo {author} {\bibfnamefont {Q.}~\bibnamefont {Campagna}},
  \bibinfo {author} {\bibfnamefont {A.}~\bibnamefont {Datta}}, \bibinfo
  {author} {\bibfnamefont {S.}~\bibnamefont {Dubey}}, \bibinfo {author}
  {\bibfnamefont {L.}~\bibnamefont {Mukherjee}},\ and\ \bibinfo {author}
  {\bibfnamefont {A.}~\bibnamefont {Sibidanov}},\ }\bibfield  {title} {\bibinfo
  {title} {A new tool to search for physics beyond the {Standard Model in
  $\bar{B}\to D^{*+}\ell^-\bar{\nu}$}},\ }in\ \href@noop {} {\emph {\bibinfo
  {booktitle} {{2022 Snowmass Summer Study}}}}\ (\bibinfo {year} {2022})\
  \Eprint {https://arxiv.org/abs/2203.07189} {arXiv:2203.07189 [hep-ph]}
  \BibitemShut {NoStop}%
\bibitem [{\citenamefont {Sibidanov}\ \emph {et~al.}(2022)\citenamefont
  {Sibidanov}, \citenamefont {Browder}, \citenamefont {Dubey}, \citenamefont
  {Kohani}, \citenamefont {Mandal}, \citenamefont {Sandilya}, \citenamefont
  {Sinha},\ and\ \citenamefont {Vahsen}}]{Sibidanov:2022gvb}%
  \BibitemOpen
  \bibfield  {author} {\bibinfo {author} {\bibfnamefont {A.}~\bibnamefont
  {Sibidanov}}, \bibinfo {author} {\bibfnamefont {T.~E.}\ \bibnamefont
  {Browder}}, \bibinfo {author} {\bibfnamefont {S.}~\bibnamefont {Dubey}},
  \bibinfo {author} {\bibfnamefont {S.}~\bibnamefont {Kohani}}, \bibinfo
  {author} {\bibfnamefont {R.}~\bibnamefont {Mandal}}, \bibinfo {author}
  {\bibfnamefont {S.}~\bibnamefont {Sandilya}}, \bibinfo {author}
  {\bibfnamefont {R.}~\bibnamefont {Sinha}},\ and\ \bibinfo {author}
  {\bibfnamefont {S.~E.}\ \bibnamefont {Vahsen}},\ }\bibfield  {title}
  {\bibinfo {title} {A new tool for detecting {BSM} physics in {$B\to
  K^*\ell\ell$} decays},\ }in\ \href@noop {} {\emph {\bibinfo {booktitle}
  {{2022 Snowmass Summer Study}}}}\ (\bibinfo {year} {2022})\ \Eprint
  {https://arxiv.org/abs/2203.06827} {arXiv:2203.06827 [hep-ph]} \BibitemShut
  {NoStop}%
\bibitem [{\citenamefont {Golowich}\ \emph {et~al.}(2007)\citenamefont
  {Golowich}, \citenamefont {Hewett}, \citenamefont {Pakvasa},\ and\
  \citenamefont {Petrov}}]{Golowich:2007ka}%
  \BibitemOpen
  \bibfield  {author} {\bibinfo {author} {\bibfnamefont {E.}~\bibnamefont
  {Golowich}}, \bibinfo {author} {\bibfnamefont {J.}~\bibnamefont {Hewett}},
  \bibinfo {author} {\bibfnamefont {S.}~\bibnamefont {Pakvasa}},\ and\ \bibinfo
  {author} {\bibfnamefont {A.~A.}\ \bibnamefont {Petrov}},\ }\bibfield  {title}
  {\bibinfo {title} {Implications of {$D^0$-$\bar{D}^0$} mixing for new
  physics},\ }\href {https://doi.org/10.1103/PhysRevD.76.095009} {\bibfield
  {journal} {\bibinfo  {journal} {Phys. Rev. D}\ }\textbf {\bibinfo {volume}
  {76}},\ \bibinfo {pages} {095009} (\bibinfo {year} {2007})},\ \Eprint
  {https://arxiv.org/abs/0705.3650} {arXiv:0705.3650 [hep-ph]} \BibitemShut
  {NoStop}%
\bibitem [{\citenamefont {Golowich}\ \emph {et~al.}(2009)\citenamefont
  {Golowich}, \citenamefont {Hewett}, \citenamefont {Pakvasa},\ and\
  \citenamefont {Petrov}}]{Golowich:2009ii}%
  \BibitemOpen
  \bibfield  {author} {\bibinfo {author} {\bibfnamefont {E.}~\bibnamefont
  {Golowich}}, \bibinfo {author} {\bibfnamefont {J.}~\bibnamefont {Hewett}},
  \bibinfo {author} {\bibfnamefont {S.}~\bibnamefont {Pakvasa}},\ and\ \bibinfo
  {author} {\bibfnamefont {A.~A.}\ \bibnamefont {Petrov}},\ }\bibfield  {title}
  {\bibinfo {title} {Relating {$D^0$-$\bar{D}^0$} mixing and {$D^0\to l^+ l^-$}
  with new physics},\ }\href {https://doi.org/10.1103/PhysRevD.79.114030}
  {\bibfield  {journal} {\bibinfo  {journal} {Phys. Rev. D}\ }\textbf {\bibinfo
  {volume} {79}},\ \bibinfo {pages} {114030} (\bibinfo {year} {2009})},\
  \Eprint {https://arxiv.org/abs/0903.2830} {arXiv:0903.2830 [hep-ph]}
  \BibitemShut {NoStop}%
\bibitem [{\citenamefont {Christ}\ \emph {et~al.}(2015)\citenamefont {Christ},
  \citenamefont {Feng}, \citenamefont {Martinelli},\ and\ \citenamefont
  {Sachrajda}}]{Christ:2015pwa}%
  \BibitemOpen
  \bibfield  {author} {\bibinfo {author} {\bibfnamefont {N.~H.}\ \bibnamefont
  {Christ}}, \bibinfo {author} {\bibfnamefont {X.}~\bibnamefont {Feng}},
  \bibinfo {author} {\bibfnamefont {G.}~\bibnamefont {Martinelli}},\ and\
  \bibinfo {author} {\bibfnamefont {C.~T.}\ \bibnamefont {Sachrajda}},\
  }\bibfield  {title} {\bibinfo {title} {{Effects of finite volume on the
  $K_L$-$K_S$ mass difference}},\ }\href
  {https://doi.org/10.1103/PhysRevD.91.114510} {\bibfield  {journal} {\bibinfo
  {journal} {Phys. Rev. D}\ }\textbf {\bibinfo {volume} {91}},\ \bibinfo
  {pages} {114510} (\bibinfo {year} {2015})},\ \Eprint
  {https://arxiv.org/abs/1504.01170} {arXiv:1504.01170 [hep-lat]} \BibitemShut
  {NoStop}%
\bibitem [{\citenamefont {Bai}\ \emph {et~al.}(2018{\natexlab{a}})\citenamefont
  {Bai}, \citenamefont {Christ},\ and\ \citenamefont
  {Sachrajda}}]{Bai:2018mdv}%
  \BibitemOpen
  \bibfield  {author} {\bibinfo {author} {\bibfnamefont {Z.}~\bibnamefont
  {Bai}}, \bibinfo {author} {\bibfnamefont {N.~H.}\ \bibnamefont {Christ}},\
  and\ \bibinfo {author} {\bibfnamefont {C.~T.}\ \bibnamefont {Sachrajda}},\
  }\bibfield  {title} {\bibinfo {title} {The {$K_L$-$K_S$} mass difference},\
  }\href {https://doi.org/10.1051/epjconf/201817513017} {\bibfield  {journal}
  {\bibinfo  {journal} {EPJ Web Conf.}\ }\textbf {\bibinfo {volume} {175}},\
  \bibinfo {pages} {13017} (\bibinfo {year} {2018}{\natexlab{a}})}\BibitemShut
  {NoStop}%
\bibitem [{\citenamefont {Bai}\ \emph {et~al.}(2014)\citenamefont {Bai},
  \citenamefont {Christ}, \citenamefont {Izubuchi}, \citenamefont {Sachrajda},
  \citenamefont {Soni},\ and\ \citenamefont {Yu}}]{Bai:2014cva}%
  \BibitemOpen
  \bibfield  {author} {\bibinfo {author} {\bibfnamefont {Z.}~\bibnamefont
  {Bai}}, \bibinfo {author} {\bibfnamefont {N.~H.}\ \bibnamefont {Christ}},
  \bibinfo {author} {\bibfnamefont {T.}~\bibnamefont {Izubuchi}}, \bibinfo
  {author} {\bibfnamefont {C.~T.}\ \bibnamefont {Sachrajda}}, \bibinfo {author}
  {\bibfnamefont {A.}~\bibnamefont {Soni}},\ and\ \bibinfo {author}
  {\bibfnamefont {J.}~\bibnamefont {Yu}},\ }\bibfield  {title} {\bibinfo
  {title} {{$K_L$-$K_S$} mass difference from lattice {QCD}},\ }\href
  {https://doi.org/10.1103/PhysRevLett.113.112003} {\bibfield  {journal}
  {\bibinfo  {journal} {Phys. Rev. Lett.}\ }\textbf {\bibinfo {volume} {113}},\
  \bibinfo {pages} {112003} (\bibinfo {year} {2014})},\ \Eprint
  {https://arxiv.org/abs/1406.0916} {arXiv:1406.0916 [hep-lat]} \BibitemShut
  {NoStop}%
\bibitem [{\citenamefont {Bai}\ \emph {et~al.}(2015)\citenamefont {Bai} \emph
  {et~al.}}]{Bai:2015nea}%
  \BibitemOpen
  \bibfield  {author} {\bibinfo {author} {\bibfnamefont {Z.}~\bibnamefont
  {Bai}} \emph {et~al.} (\bibinfo {collaboration} {RBC, UKQCD}),\ }\bibfield
  {title} {\bibinfo {title} {{Standard Model} prediction for direct {$CP$}
  violation in {$K\to\pi\pi$} decay},\ }\href
  {https://doi.org/10.1103/PhysRevLett.115.212001} {\bibfield  {journal}
  {\bibinfo  {journal} {Phys. Rev. Lett.}\ }\textbf {\bibinfo {volume} {115}},\
  \bibinfo {pages} {212001} (\bibinfo {year} {2015})},\ \Eprint
  {https://arxiv.org/abs/1505.07863} {arXiv:1505.07863 [hep-lat]} \BibitemShut
  {NoStop}%
\bibitem [{\citenamefont {Kelly}(2017)}]{Kelly:2016mxf}%
  \BibitemOpen
  \bibfield  {author} {\bibinfo {author} {\bibfnamefont {C.}~\bibnamefont
  {Kelly}} (\bibinfo {collaboration} {RBC, UKQCD}),\ }\bibfield  {title}
  {\bibinfo {title} {Calculation of $\epsilon'/\epsilon$ on the lattice},\
  }\href {https://doi.org/10.22323/1.280.0017} {\bibfield  {journal} {\bibinfo
  {journal} {PoS}\ }\textbf {\bibinfo {volume} {FPCP2016}},\ \bibinfo {pages}
  {017} (\bibinfo {year} {2017})}\BibitemShut {NoStop}%
\bibitem [{\citenamefont {Abbott}\ \emph {et~al.}(2020)\citenamefont {Abbott}
  \emph {et~al.}}]{RBC:2020kdj}%
  \BibitemOpen
  \bibfield  {author} {\bibinfo {author} {\bibfnamefont {R.}~\bibnamefont
  {Abbott}} \emph {et~al.} (\bibinfo {collaboration} {RBC, UKQCD}),\ }\bibfield
   {title} {\bibinfo {title} {{Direct {$CP$} violation and the $\Delta I=1/2$
  rule in $K\to\pi\pi$ decay from the standard model}},\ }\href
  {https://doi.org/10.1103/PhysRevD.102.054509} {\bibfield  {journal} {\bibinfo
   {journal} {Phys. Rev. D}\ }\textbf {\bibinfo {volume} {102}},\ \bibinfo
  {pages} {054509} (\bibinfo {year} {2020})},\ \Eprint
  {https://arxiv.org/abs/2004.09440} {arXiv:2004.09440 [hep-lat]} \BibitemShut
  {NoStop}%
\bibitem [{\citenamefont {Alavi-Harati}\ \emph {et~al.}(2003)\citenamefont
  {Alavi-Harati} \emph {et~al.}}]{AlaviHarati:2002ye}%
  \BibitemOpen
  \bibfield  {author} {\bibinfo {author} {\bibfnamefont {A.}~\bibnamefont
  {Alavi-Harati}} \emph {et~al.} (\bibinfo {collaboration} {KTeV}),\ }\bibfield
   {title} {\bibinfo {title} {{Measurements of direct $CP$ violation, $CPT$
  symmetry, and other parameters in the neutral kaon system}},\ }\href
  {https://doi.org/10.1103/PhysRevD.67.012005} {\bibfield  {journal} {\bibinfo
  {journal} {Phys. Rev. D}\ }\textbf {\bibinfo {volume} {67}},\ \bibinfo
  {pages} {012005} (\bibinfo {year} {2003})},\ \bibinfo {note} {(E)
  \href{https://doi.org/10.1103/PhysRevD.70.079904}{\textbf{70}, 079904
  (2004)}},\ \Eprint {https://arxiv.org/abs/hep-ex/0208007}
  {arXiv:hep-ex/0208007 [hep-ex]} \BibitemShut {NoStop}%
\bibitem [{\citenamefont {Batley}\ \emph {et~al.}(2002)\citenamefont {Batley}
  \emph {et~al.}}]{Batley:2002gn}%
  \BibitemOpen
  \bibfield  {author} {\bibinfo {author} {\bibfnamefont {J.~R.}\ \bibnamefont
  {Batley}} \emph {et~al.} (\bibinfo {collaboration} {NA48}),\ }\bibfield
  {title} {\bibinfo {title} {{A Precision measurement of direct $CP$ violation
  in the decay of neutral kaons into two pions}},\ }\href
  {https://doi.org/10.1016/S0370-2693(02)02476-0} {\bibfield  {journal}
  {\bibinfo  {journal} {Phys. Lett. B}\ }\textbf {\bibinfo {volume} {544}},\
  \bibinfo {pages} {97} (\bibinfo {year} {2002})},\ \Eprint
  {https://arxiv.org/abs/hep-ex/0208009} {arXiv:hep-ex/0208009 [hep-ex]}
  \BibitemShut {NoStop}%
\bibitem [{\citenamefont {Seng}\ \emph {et~al.}(2018)\citenamefont {Seng},
  \citenamefont {Gorchtein}, \citenamefont {Patel},\ and\ \citenamefont
  {Ramsey-Musolf}}]{Seng:2018yzq}%
  \BibitemOpen
  \bibfield  {author} {\bibinfo {author} {\bibfnamefont {C.-Y.}\ \bibnamefont
  {Seng}}, \bibinfo {author} {\bibfnamefont {M.}~\bibnamefont {Gorchtein}},
  \bibinfo {author} {\bibfnamefont {H.~H.}\ \bibnamefont {Patel}},\ and\
  \bibinfo {author} {\bibfnamefont {M.~J.}\ \bibnamefont {Ramsey-Musolf}},\
  }\bibfield  {title} {\bibinfo {title} {Reduced hadronic uncertainty in the
  determination of {$V_{ud}$}},\ }\href
  {https://doi.org/10.1103/PhysRevLett.121.241804} {\bibfield  {journal}
  {\bibinfo  {journal} {Phys. Rev. Lett.}\ }\textbf {\bibinfo {volume} {121}},\
  \bibinfo {pages} {241804} (\bibinfo {year} {2018})},\ \Eprint
  {https://arxiv.org/abs/1807.10197} {arXiv:1807.10197 [hep-ph]} \BibitemShut
  {NoStop}%
\bibitem [{\citenamefont {Seng}\ \emph
  {et~al.}(2022{\natexlab{a}})\citenamefont {Seng}, \citenamefont {Galviz},
  \citenamefont {Gorchtein},\ and\ \citenamefont {Mei\ss{}ner}}]{Seng:2022wcw}%
  \BibitemOpen
  \bibfield  {author} {\bibinfo {author} {\bibfnamefont {C.-Y.}\ \bibnamefont
  {Seng}}, \bibinfo {author} {\bibfnamefont {D.}~\bibnamefont {Galviz}},
  \bibinfo {author} {\bibfnamefont {M.}~\bibnamefont {Gorchtein}},\ and\
  \bibinfo {author} {\bibfnamefont {U.-G.}\ \bibnamefont {Mei\ss{}ner}},\
  }\bibfield  {title} {\bibinfo {title} {Full {$K_{\ell 3}$} radiative
  corrections and implications on {$|V_{us}|$}},\ }\href@noop {} {\  (\bibinfo
  {year} {2022}{\natexlab{a}})},\ \Eprint {https://arxiv.org/abs/2203.05217}
  {arXiv:2203.05217 [hep-ph]} \BibitemShut {NoStop}%
\bibitem [{\citenamefont {Seng}\ \emph
  {et~al.}(2022{\natexlab{b}})\citenamefont {Seng}, \citenamefont {Marciano},\
  and\ \citenamefont {Mei\ss{}ner}}]{Seng:2022tjh}%
  \BibitemOpen
  \bibfield  {author} {\bibinfo {author} {\bibfnamefont {C.-Y.}\ \bibnamefont
  {Seng}}, \bibinfo {author} {\bibfnamefont {W.~J.}\ \bibnamefont {Marciano}},\
  and\ \bibinfo {author} {\bibfnamefont {U.-G.}\ \bibnamefont {Mei\ss{}ner}},\
  }\bibfield  {title} {\bibinfo {title} {Electron mass singularities in
  semileptonic kaon decays},\ }\href@noop {} {\  (\bibinfo {year}
  {2022}{\natexlab{b}})},\ \Eprint {https://arxiv.org/abs/2206.01513}
  {arXiv:2206.01513 [hep-ph]} \BibitemShut {NoStop}%
\bibitem [{\citenamefont {Bai}\ \emph {et~al.}(2018{\natexlab{b}})\citenamefont
  {Bai}, \citenamefont {Christ}, \citenamefont {Feng}, \citenamefont {Lawson},
  \citenamefont {Portelli},\ and\ \citenamefont {Sachrajda}}]{Bai:2018hqu}%
  \BibitemOpen
  \bibfield  {author} {\bibinfo {author} {\bibfnamefont {Z.}~\bibnamefont
  {Bai}}, \bibinfo {author} {\bibfnamefont {N.~H.}\ \bibnamefont {Christ}},
  \bibinfo {author} {\bibfnamefont {X.}~\bibnamefont {Feng}}, \bibinfo {author}
  {\bibfnamefont {A.}~\bibnamefont {Lawson}}, \bibinfo {author} {\bibfnamefont
  {A.}~\bibnamefont {Portelli}},\ and\ \bibinfo {author} {\bibfnamefont
  {C.~T.}\ \bibnamefont {Sachrajda}},\ }\bibfield  {title} {\bibinfo {title}
  {{$K^+\to\pi^+\nu\bar{\nu}$ decay amplitude from lattice QCD}},\ }\href
  {https://doi.org/10.1103/PhysRevD.98.074509} {\bibfield  {journal} {\bibinfo
  {journal} {Phys. Rev. D}\ }\textbf {\bibinfo {volume} {98}},\ \bibinfo
  {pages} {074509} (\bibinfo {year} {2018}{\natexlab{b}})},\ \Eprint
  {https://arxiv.org/abs/1806.11520} {arXiv:1806.11520 [hep-lat]} \BibitemShut
  {NoStop}%
\bibitem [{\citenamefont {Blum}\ \emph {et~al.}(2022)\citenamefont {Blum} \emph
  {et~al.}}]{Blum:2022wsz}%
  \BibitemOpen
  \bibfield  {author} {\bibinfo {author} {\bibfnamefont {T.}~\bibnamefont
  {Blum}} \emph {et~al.},\ }\bibfield  {title} {\bibinfo {title} {{Discovering
  new physics in rare kaon decays}},\ }in\ \href@noop {} {\emph {\bibinfo
  {booktitle} {{2022 Snowmass Summer Study}}}}\ (\bibinfo {year} {2022})\
  \Eprint {https://arxiv.org/abs/2203.10998} {arXiv:2203.10998 [hep-lat]}
  \BibitemShut {NoStop}%
\bibitem [{\citenamefont {Aebischer}\ \emph {et~al.}(2022)\citenamefont
  {Aebischer}, \citenamefont {Buras},\ and\ \citenamefont
  {Kumar}}]{Aebischer:2022vky}%
  \BibitemOpen
  \bibfield  {author} {\bibinfo {author} {\bibfnamefont {J.}~\bibnamefont
  {Aebischer}}, \bibinfo {author} {\bibfnamefont {A.~J.}\ \bibnamefont
  {Buras}},\ and\ \bibinfo {author} {\bibfnamefont {J.}~\bibnamefont {Kumar}},\
  }\bibfield  {title} {\bibinfo {title} {On the importance of rare kaon
  decays},\ }in\ \href@noop {} {\emph {\bibinfo {booktitle} {{2022 Snowmass
  Summer Study}}}}\ (\bibinfo {year} {2022})\ \Eprint
  {https://arxiv.org/abs/2203.09524} {arXiv:2203.09524 [hep-ph]} \BibitemShut
  {NoStop}%
\bibitem [{\citenamefont {Goudzovski}\ \emph {et~al.}(2022)\citenamefont
  {Goudzovski} \emph {et~al.}}]{Goudzovski:2022vbt}%
  \BibitemOpen
  \bibfield  {author} {\bibinfo {author} {\bibfnamefont {E.}~\bibnamefont
  {Goudzovski}} \emph {et~al.},\ }\bibfield  {title} {\bibinfo {title} {New
  physics searches at kaon and hyperon factories},\ }in\ \href@noop {} {\emph
  {\bibinfo {booktitle} {{2022 Snowmass Summer Study}}}}\ (\bibinfo {year}
  {2022})\ \Eprint {https://arxiv.org/abs/2201.07805} {arXiv:2201.07805
  [hep-ph]} \BibitemShut {NoStop}%
\bibitem [{\citenamefont {Moulson}\ \emph {et~al.}(2022)\citenamefont {Moulson}
  \emph {et~al.}}]{NA62KLEVER:2022nea}%
  \BibitemOpen
  \bibfield  {author} {\bibinfo {author} {\bibfnamefont {M.}~\bibnamefont
  {Moulson}} \emph {et~al.} (\bibinfo {collaboration} {KOTO, LHCb, NA62/KLEVER,
  and the US Kaon Interest Group}),\ }\bibfield  {title} {\bibinfo {title}
  {Searches for new physics with high-intensity kaon beams},\ }in\ \href@noop
  {} {\emph {\bibinfo {booktitle} {{2022 Snowmass Summer Study}}}}\ (\bibinfo
  {year} {2022})\ \Eprint {https://arxiv.org/abs/2204.13394} {arXiv:2204.13394
  [hep-ex]} \BibitemShut {NoStop}%
\bibitem [{\citenamefont {Christ}\ \emph {et~al.}(2010)\citenamefont {Christ},
  \citenamefont {Dawson}, \citenamefont {Izubuchi}, \citenamefont {Jung},
  \citenamefont {Liu}, \citenamefont {Mawhinney}, \citenamefont {Sachrajda},
  \citenamefont {Soni},\ and\ \citenamefont {Zhou}}]{Christ:2010dd}%
  \BibitemOpen
  \bibfield  {author} {\bibinfo {author} {\bibfnamefont {N.~H.}\ \bibnamefont
  {Christ}}, \bibinfo {author} {\bibfnamefont {C.}~\bibnamefont {Dawson}},
  \bibinfo {author} {\bibfnamefont {T.}~\bibnamefont {Izubuchi}}, \bibinfo
  {author} {\bibfnamefont {C.}~\bibnamefont {Jung}}, \bibinfo {author}
  {\bibfnamefont {Q.}~\bibnamefont {Liu}}, \bibinfo {author} {\bibfnamefont
  {R.~D.}\ \bibnamefont {Mawhinney}}, \bibinfo {author} {\bibfnamefont {C.~T.}\
  \bibnamefont {Sachrajda}}, \bibinfo {author} {\bibfnamefont {A.}~\bibnamefont
  {Soni}},\ and\ \bibinfo {author} {\bibfnamefont {R.}~\bibnamefont {Zhou}},\
  }\bibfield  {title} {\bibinfo {title} {The $\eta$ and $\eta'$ mesons from
  lattice {QCD}},\ }\href {https://doi.org/10.1103/PhysRevLett.105.241601}
  {\bibfield  {journal} {\bibinfo  {journal} {Phys. Rev. Lett.}\ }\textbf
  {\bibinfo {volume} {105}},\ \bibinfo {pages} {241601} (\bibinfo {year}
  {2010})},\ \Eprint {https://arxiv.org/abs/1002.2999} {arXiv:1002.2999
  [hep-lat]} \BibitemShut {NoStop}%
\bibitem [{\citenamefont {Michael}\ \emph {et~al.}(2013)\citenamefont
  {Michael}, \citenamefont {Ottnad},\ and\ \citenamefont
  {Urbach}}]{Michael:2013gka}%
  \BibitemOpen
  \bibfield  {author} {\bibinfo {author} {\bibfnamefont {C.}~\bibnamefont
  {Michael}}, \bibinfo {author} {\bibfnamefont {K.}~\bibnamefont {Ottnad}},\
  and\ \bibinfo {author} {\bibfnamefont {C.}~\bibnamefont {Urbach}} (\bibinfo
  {collaboration} {ETM}),\ }\bibfield  {title} {\bibinfo {title} {$\eta$ and
  $\eta'$ mixing from lattice {QCD}},\ }\href
  {https://doi.org/10.1103/PhysRevLett.111.181602} {\bibfield  {journal}
  {\bibinfo  {journal} {Phys. Rev. Lett.}\ }\textbf {\bibinfo {volume} {111}},\
  \bibinfo {pages} {181602} (\bibinfo {year} {2013})},\ \Eprint
  {https://arxiv.org/abs/1310.1207} {arXiv:1310.1207 [hep-lat]} \BibitemShut
  {NoStop}%
\bibitem [{\citenamefont {Fukaya}\ \emph {et~al.}(2015)\citenamefont {Fukaya},
  \citenamefont {Aoki}, \citenamefont {Cossu}, \citenamefont {Hashimoto},
  \citenamefont {Kaneko},\ and\ \citenamefont {Noaki}}]{Fukaya:2015ara}%
  \BibitemOpen
  \bibfield  {author} {\bibinfo {author} {\bibfnamefont {H.}~\bibnamefont
  {Fukaya}}, \bibinfo {author} {\bibfnamefont {S.}~\bibnamefont {Aoki}},
  \bibinfo {author} {\bibfnamefont {G.}~\bibnamefont {Cossu}}, \bibinfo
  {author} {\bibfnamefont {S.}~\bibnamefont {Hashimoto}}, \bibinfo {author}
  {\bibfnamefont {T.}~\bibnamefont {Kaneko}},\ and\ \bibinfo {author}
  {\bibfnamefont {J.}~\bibnamefont {Noaki}} (\bibinfo {collaboration}
  {JLQCD}),\ }\bibfield  {title} {\bibinfo {title} {{$\eta^\prime$ meson mass
  from topological charge density correlator in QCD}},\ }\href
  {https://doi.org/10.1103/PhysRevD.92.111501} {\bibfield  {journal} {\bibinfo
  {journal} {Phys. Rev. D}\ }\textbf {\bibinfo {volume} {92}},\ \bibinfo
  {pages} {111501} (\bibinfo {year} {2015})},\ \Eprint
  {https://arxiv.org/abs/1509.00944} {arXiv:1509.00944 [hep-lat]} \BibitemShut
  {NoStop}%
\bibitem [{\citenamefont {Bali}\ \emph {et~al.}(2021)\citenamefont {Bali},
  \citenamefont {Braun}, \citenamefont {Collins}, \citenamefont {Sch\"afer},\
  and\ \citenamefont {Simeth}}]{Bali:2021qem}%
  \BibitemOpen
  \bibfield  {author} {\bibinfo {author} {\bibfnamefont {G.~S.}\ \bibnamefont
  {Bali}}, \bibinfo {author} {\bibfnamefont {V.}~\bibnamefont {Braun}},
  \bibinfo {author} {\bibfnamefont {S.}~\bibnamefont {Collins}}, \bibinfo
  {author} {\bibfnamefont {A.}~\bibnamefont {Sch\"afer}},\ and\ \bibinfo
  {author} {\bibfnamefont {J.}~\bibnamefont {Simeth}} (\bibinfo {collaboration}
  {RQCD}),\ }\bibfield  {title} {\bibinfo {title} {{Masses and decay constants
  of the $\eta$ and $\eta'$ mesons from lattice QCD}},\ }\href
  {https://doi.org/10.1007/JHEP08(2021)137} {\bibfield  {journal} {\bibinfo
  {journal} {JHEP}\ }\textbf {\bibinfo {volume} {08}},\ \bibinfo {pages}
  {137}},\ \Eprint {https://arxiv.org/abs/2106.05398} {arXiv:2106.05398
  [hep-lat]} \BibitemShut {NoStop}%
\bibitem [{\citenamefont {Kordov}\ \emph {et~al.}(2021)\citenamefont {Kordov}
  \emph {et~al.}}]{CSSMQCDSFUKQCD:2021rvs}%
  \BibitemOpen
  \bibfield  {author} {\bibinfo {author} {\bibfnamefont {Z.~R.}\ \bibnamefont
  {Kordov}} \emph {et~al.} (\bibinfo {collaboration} {CSSM, QCDSF, UKQCD}),\
  }\bibfield  {title} {\bibinfo {title} {{State mixing and masses of the
  $\pi^0$, $\eta$, and $\eta'$ mesons from $n_f=1+1+1$ lattice QCD+QED}},\
  }\href {https://doi.org/10.1103/PhysRevD.104.114514} {\bibfield  {journal}
  {\bibinfo  {journal} {Phys. Rev. D}\ }\textbf {\bibinfo {volume} {104}},\
  \bibinfo {pages} {114514} (\bibinfo {year} {2021})},\ \Eprint
  {https://arxiv.org/abs/2110.11533} {arXiv:2110.11533 [hep-lat]} \BibitemShut
  {NoStop}%
\bibitem [{\citenamefont {Elam}\ \emph {et~al.}(2022)\citenamefont {Elam} \emph
  {et~al.}}]{REDTOP:2022slw}%
  \BibitemOpen
  \bibfield  {author} {\bibinfo {author} {\bibfnamefont {J.}~\bibnamefont
  {Elam}} \emph {et~al.} (\bibinfo {collaboration} {REDTOP}),\ }\bibfield
  {title} {\bibinfo {title} {The {REDTOP} experiment: Rare $\eta/\eta^{\prime}$
  decays to probe new physics},\ }in\ \href@noop {} {\emph {\bibinfo
  {booktitle} {{2022 Snowmass Summer Study}}}}\ (\bibinfo {year} {2022})\
  \Eprint {https://arxiv.org/abs/2203.07651} {arXiv:2203.07651 [hep-ex]}
  \BibitemShut {NoStop}%
\bibitem [{\citenamefont {Lautrup}\ and\ \citenamefont
  {de~Rafael}(1974)}]{Lautrup:1974ic}%
  \BibitemOpen
  \bibfield  {author} {\bibinfo {author} {\bibfnamefont {B.~E.}\ \bibnamefont
  {Lautrup}}\ and\ \bibinfo {author} {\bibfnamefont {E.}~\bibnamefont
  {de~Rafael}},\ }\bibfield  {title} {\bibinfo {title} {{The anomalous magnetic
  moment of the muon and short-distance behaviour of quantum
  electrodynamics}},\ }\href {https://doi.org/10.1016/0550-3213(74)90481-7}
  {\bibfield  {journal} {\bibinfo  {journal} {Nucl. Phys. B}\ }\textbf
  {\bibinfo {volume} {70}},\ \bibinfo {pages} {317} (\bibinfo {year} {1974})},\
  \bibinfo {note} {(E)
  \href{https://doi.org/10.1016/0550-3213(74)90600-2}{\textbf{78}, 576
  (1974)}}\BibitemShut {NoStop}%
\bibitem [{\citenamefont {Blum}(2003)}]{Blum:2002ii}%
  \BibitemOpen
  \bibfield  {author} {\bibinfo {author} {\bibfnamefont {T.}~\bibnamefont
  {Blum}},\ }\bibfield  {title} {\bibinfo {title} {Lattice calculation of the
  lowest order hadronic contribution to the muon anomalous magnetic moment},\
  }\href {https://doi.org/10.1103/PhysRevLett.91.052001} {\bibfield  {journal}
  {\bibinfo  {journal} {Phys. Rev. Lett.}\ }\textbf {\bibinfo {volume} {91}},\
  \bibinfo {pages} {052001} (\bibinfo {year} {2003})},\ \Eprint
  {https://arxiv.org/abs/hep-lat/0212018} {arXiv:hep-lat/0212018 [hep-lat]}
  \BibitemShut {NoStop}%
\bibitem [{\citenamefont {Burger}\ \emph {et~al.}(2014)\citenamefont {Burger},
  \citenamefont {Feng}, \citenamefont {Hotzel}, \citenamefont {Jansen},
  \citenamefont {Petschlies},\ and\ \citenamefont {Renner}}]{Burger:2013jya}%
  \BibitemOpen
  \bibfield  {author} {\bibinfo {author} {\bibfnamefont {F.}~\bibnamefont
  {Burger}}, \bibinfo {author} {\bibfnamefont {X.}~\bibnamefont {Feng}},
  \bibinfo {author} {\bibfnamefont {G.}~\bibnamefont {Hotzel}}, \bibinfo
  {author} {\bibfnamefont {K.}~\bibnamefont {Jansen}}, \bibinfo {author}
  {\bibfnamefont {M.}~\bibnamefont {Petschlies}},\ and\ \bibinfo {author}
  {\bibfnamefont {D.~B.}\ \bibnamefont {Renner}} (\bibinfo {collaboration}
  {ETM}),\ }\bibfield  {title} {\bibinfo {title} {Four-flavour leading-order
  hadronic contribution to the muon anomalous magnetic moment},\ }\href
  {https://doi.org/10.1007/JHEP02(2014)099} {\bibfield  {journal} {\bibinfo
  {journal} {JHEP}\ }\textbf {\bibinfo {volume} {02}},\ \bibinfo {pages}
  {099}},\ \Eprint {https://arxiv.org/abs/1308.4327} {arXiv:1308.4327
  [hep-lat]} \BibitemShut {NoStop}%
\bibitem [{\citenamefont {Blum}\ \emph
  {et~al.}(2016{\natexlab{a}})\citenamefont {Blum} \emph
  {et~al.}}]{Blum:2015you}%
  \BibitemOpen
  \bibfield  {author} {\bibinfo {author} {\bibfnamefont {T.}~\bibnamefont
  {Blum}} \emph {et~al.} (\bibinfo {collaboration} {RBC, UKQCD}),\ }\bibfield
  {title} {\bibinfo {title} {{Calculation of the hadronic vacuum polarization
  disconnected contribution to the muon anomalous magnetic moment}},\ }\href
  {https://doi.org/10.1103/PhysRevLett.116.232002} {\bibfield  {journal}
  {\bibinfo  {journal} {Phys. Rev. Lett.}\ }\textbf {\bibinfo {volume} {116}},\
  \bibinfo {pages} {232002} (\bibinfo {year} {2016}{\natexlab{a}})},\ \Eprint
  {https://arxiv.org/abs/1512.09054} {arXiv:1512.09054 [hep-lat]} \BibitemShut
  {NoStop}%
\bibitem [{\citenamefont {Blum}\ \emph {et~al.}(2018)\citenamefont {Blum} \emph
  {et~al.}}]{Blum:2018mom}%
  \BibitemOpen
  \bibfield  {author} {\bibinfo {author} {\bibfnamefont {T.}~\bibnamefont
  {Blum}} \emph {et~al.} (\bibinfo {collaboration} {RBC, UKQCD}),\ }\bibfield
  {title} {\bibinfo {title} {Calculation of the hadronic vacuum polarization
  contribution to the muon anomalous magnetic moment},\ }\href
  {https://doi.org/10.1103/PhysRevLett.121.022003} {\bibfield  {journal}
  {\bibinfo  {journal} {Phys. Rev. Lett.}\ }\textbf {\bibinfo {volume} {121}},\
  \bibinfo {pages} {022003} (\bibinfo {year} {2018})},\ \Eprint
  {https://arxiv.org/abs/1801.07224} {arXiv:1801.07224 [hep-lat]} \BibitemShut
  {NoStop}%
\bibitem [{\citenamefont {Chakraborty}\ \emph {et~al.}(2017)\citenamefont
  {Chakraborty}, \citenamefont {Davies}, \citenamefont {de~Oliviera},
  \citenamefont {Koponen}, \citenamefont {Lepage},\ and\ \citenamefont {Van~de
  Water}}]{Chakraborty:2016mwy}%
  \BibitemOpen
  \bibfield  {author} {\bibinfo {author} {\bibfnamefont {B.}~\bibnamefont
  {Chakraborty}}, \bibinfo {author} {\bibfnamefont {C.~T.~H.}\ \bibnamefont
  {Davies}}, \bibinfo {author} {\bibfnamefont {P.~G.}\ \bibnamefont
  {de~Oliviera}}, \bibinfo {author} {\bibfnamefont {J.}~\bibnamefont
  {Koponen}}, \bibinfo {author} {\bibfnamefont {G.~P.}\ \bibnamefont
  {Lepage}},\ and\ \bibinfo {author} {\bibfnamefont {R.~S.}\ \bibnamefont
  {Van~de Water}} (\bibinfo {collaboration} {HPQCD}),\ }\bibfield  {title}
  {\bibinfo {title} {The hadronic vacuum polarization contribution to $a_{\mu}$
  from full lattice {QCD}},\ }\href
  {https://doi.org/10.1103/PhysRevD.96.034516} {\bibfield  {journal} {\bibinfo
  {journal} {Phys. Rev. D}\ }\textbf {\bibinfo {volume} {96}},\ \bibinfo
  {pages} {034516} (\bibinfo {year} {2017})},\ \Eprint
  {https://arxiv.org/abs/1601.03071} {arXiv:1601.03071 [hep-lat]} \BibitemShut
  {NoStop}%
\bibitem [{\citenamefont {Borsányi}\ \emph {et~al.}(2018)\citenamefont
  {Borsányi} \emph {et~al.}}]{Borsanyi:2017zdw}%
  \BibitemOpen
  \bibfield  {author} {\bibinfo {author} {\bibfnamefont {S.}~\bibnamefont
  {Borsányi}} \emph {et~al.} (\bibinfo {collaboration}
  {Budapest-Marseille-Wuppertal}),\ }\bibfield  {title} {\bibinfo {title}
  {Hadronic vacuum polarization contribution to the anomalous magnetic moments
  of leptons from first principles},\ }\href
  {https://doi.org/10.1103/PhysRevLett.121.022002} {\bibfield  {journal}
  {\bibinfo  {journal} {Phys. Rev. Lett.}\ }\textbf {\bibinfo {volume} {121}},\
  \bibinfo {pages} {022002} (\bibinfo {year} {2018})},\ \Eprint
  {https://arxiv.org/abs/1711.04980} {arXiv:1711.04980 [hep-lat]} \BibitemShut
  {NoStop}%
\bibitem [{\citenamefont {Chakraborty}\ \emph
  {et~al.}(2018{\natexlab{a}})\citenamefont {Chakraborty} \emph
  {et~al.}}]{Chakraborty:2017tqp}%
  \BibitemOpen
  \bibfield  {author} {\bibinfo {author} {\bibfnamefont {B.}~\bibnamefont
  {Chakraborty}} \emph {et~al.} (\bibinfo {collaboration} {Fermilab Lattice,
  HPQCD, MILC}),\ }\bibfield  {title} {\bibinfo {title}
  {Strong-isospin-breaking correction to the muon anomalous magnetic moment
  from lattice {QCD} at the physical point},\ }\href
  {https://doi.org/10.1103/PhysRevLett.120.152001} {\bibfield  {journal}
  {\bibinfo  {journal} {Phys. Rev. Lett.}\ }\textbf {\bibinfo {volume} {120}},\
  \bibinfo {pages} {152001} (\bibinfo {year} {2018}{\natexlab{a}})},\ \Eprint
  {https://arxiv.org/abs/1710.11212} {arXiv:1710.11212 [hep-lat]} \BibitemShut
  {NoStop}%
\bibitem [{\citenamefont {Chakraborty}\ \emph
  {et~al.}(2018{\natexlab{b}})\citenamefont {Chakraborty} \emph
  {et~al.}}]{Chakraborty:2018iyb}%
  \BibitemOpen
  \bibfield  {author} {\bibinfo {author} {\bibfnamefont {B.}~\bibnamefont
  {Chakraborty}} \emph {et~al.} (\bibinfo {collaboration} {Fermilab Lattice,
  HPQCD, MILC}),\ }\bibfield  {title} {\bibinfo {title} {Higher-order
  hadronic-vacuum-polarization contribution to the muon $g-2$ from lattice
  {QCD}},\ }\href {https://doi.org/10.1103/PhysRevD.98.094503} {\bibfield
  {journal} {\bibinfo  {journal} {Phys. Rev. D}\ }\textbf {\bibinfo {volume}
  {98}},\ \bibinfo {pages} {094503} (\bibinfo {year} {2018}{\natexlab{b}})},\
  \Eprint {https://arxiv.org/abs/1806.08190} {arXiv:1806.08190 [hep-lat]}
  \BibitemShut {NoStop}%
\bibitem [{\citenamefont {Davies}\ \emph {et~al.}(2020)\citenamefont {Davies}
  \emph {et~al.}}]{Davies:2019efs}%
  \BibitemOpen
  \bibfield  {author} {\bibinfo {author} {\bibfnamefont {C.~T.~H.}\
  \bibnamefont {Davies}} \emph {et~al.} (\bibinfo {collaboration} {Fermilab
  Lattice, HPQCD, MILC}),\ }\bibfield  {title} {\bibinfo {title}
  {Hadronic-vacuum-polarization contribution to the muon's anomalous magnetic
  moment from four-flavor lattice {QCD}},\ }\href
  {https://doi.org/10.1103/PhysRevD.101.034512} {\bibfield  {journal} {\bibinfo
   {journal} {Phys. Rev. D}\ }\textbf {\bibinfo {volume} {101}},\ \bibinfo
  {pages} {034512} (\bibinfo {year} {2020})},\ \Eprint
  {https://arxiv.org/abs/1902.04223} {arXiv:1902.04223 [hep-lat]} \BibitemShut
  {NoStop}%
\bibitem [{\citenamefont {Davies}\ \emph {et~al.}(2022)\citenamefont {Davies}
  \emph {et~al.}}]{Davies:2022epg}%
  \BibitemOpen
  \bibfield  {author} {\bibinfo {author} {\bibfnamefont {C.~T.~H.}\
  \bibnamefont {Davies}} \emph {et~al.} (\bibinfo {collaboration} {Fermilab
  Lattice, HPQCD, and MILC}),\ }\bibfield  {title} {\bibinfo {title} {{Windows
  on the hadronic vacuum polarization contribution to the muon anomalous
  magnetic moment}},\ }\href@noop {} {\  (\bibinfo {year} {2022})},\ \Eprint
  {https://arxiv.org/abs/2207.04765} {arXiv:2207.04765 [hep-lat]} \BibitemShut
  {NoStop}%
\bibitem [{\citenamefont {Della~Morte}\ \emph {et~al.}(2017)\citenamefont
  {Della~Morte}, \citenamefont {Francis}, \citenamefont {Gülpers},
  \citenamefont {Herdoíza}, \citenamefont {von Hippel}, \citenamefont {Horch},
  \citenamefont {Jäger}, \citenamefont {Meyer}, \citenamefont {Nyffeler},\
  and\ \citenamefont {Wittig}}]{DellaMorte:2017dyu}%
  \BibitemOpen
  \bibfield  {author} {\bibinfo {author} {\bibfnamefont {M.}~\bibnamefont
  {Della~Morte}}, \bibinfo {author} {\bibfnamefont {A.}~\bibnamefont
  {Francis}}, \bibinfo {author} {\bibfnamefont {V.}~\bibnamefont {Gülpers}},
  \bibinfo {author} {\bibfnamefont {G.}~\bibnamefont {Herdoíza}}, \bibinfo
  {author} {\bibfnamefont {G.}~\bibnamefont {von Hippel}}, \bibinfo {author}
  {\bibfnamefont {H.}~\bibnamefont {Horch}}, \bibinfo {author} {\bibfnamefont
  {B.}~\bibnamefont {Jäger}}, \bibinfo {author} {\bibfnamefont {H.~B.}\
  \bibnamefont {Meyer}}, \bibinfo {author} {\bibfnamefont {A.}~\bibnamefont
  {Nyffeler}},\ and\ \bibinfo {author} {\bibfnamefont {H.}~\bibnamefont
  {Wittig}},\ }\bibfield  {title} {\bibinfo {title} {The hadronic vacuum
  polarization contribution to the muon $g-2$ from lattice {QCD}},\ }\href
  {https://doi.org/10.1007/JHEP10(2017)020} {\bibfield  {journal} {\bibinfo
  {journal} {JHEP}\ }\textbf {\bibinfo {volume} {10}},\ \bibinfo {pages}
  {020}},\ \Eprint {https://arxiv.org/abs/1705.01775} {arXiv:1705.01775
  [hep-lat]} \BibitemShut {NoStop}%
\bibitem [{\citenamefont {Gérardin}\ \emph {et~al.}(2019)\citenamefont
  {Gérardin}, \citenamefont {Cè}, \citenamefont {von Hippel}, \citenamefont
  {Hörz}, \citenamefont {Meyer}, \citenamefont {Mohler}, \citenamefont
  {Ottnad}, \citenamefont {Wilhelm},\ and\ \citenamefont
  {Wittig}}]{Gerardin:2019rua}%
  \BibitemOpen
  \bibfield  {author} {\bibinfo {author} {\bibfnamefont {A.}~\bibnamefont
  {Gérardin}}, \bibinfo {author} {\bibfnamefont {M.}~\bibnamefont {Cè}},
  \bibinfo {author} {\bibfnamefont {G.}~\bibnamefont {von Hippel}}, \bibinfo
  {author} {\bibfnamefont {B.}~\bibnamefont {Hörz}}, \bibinfo {author}
  {\bibfnamefont {H.~B.}\ \bibnamefont {Meyer}}, \bibinfo {author}
  {\bibfnamefont {D.}~\bibnamefont {Mohler}}, \bibinfo {author} {\bibfnamefont
  {K.}~\bibnamefont {Ottnad}}, \bibinfo {author} {\bibfnamefont
  {J.}~\bibnamefont {Wilhelm}},\ and\ \bibinfo {author} {\bibfnamefont
  {H.}~\bibnamefont {Wittig}},\ }\bibfield  {title} {\bibinfo {title} {{The
  leading hadronic contribution to $(g-2)_\mu$ from lattice QCD with $N_f=2+1$
  flavors of $\mathrm{O}(a)$ improved Wilson quarks}},\ }\href
  {https://doi.org/10.1103/PhysRevD.100.014510} {\bibfield  {journal} {\bibinfo
   {journal} {Phys. Rev. D}\ }\textbf {\bibinfo {volume} {100}},\ \bibinfo
  {pages} {014510} (\bibinfo {year} {2019})},\ \Eprint
  {https://arxiv.org/abs/1904.03120} {arXiv:1904.03120 [hep-lat]} \BibitemShut
  {NoStop}%
\bibitem [{\citenamefont {C\`e}\ \emph {et~al.}(2022)\citenamefont {C\`e},
  \citenamefont {Gérardin}, \citenamefont {von Hippel}, \citenamefont
  {Hudspith}, \citenamefont {Kuberski}, \citenamefont {Meyer}, \citenamefont
  {Miura}, \citenamefont {Mohler}, \citenamefont {Ottnad}, \citenamefont
  {Paul}, \citenamefont {Risch}, \citenamefont {San~José},\ and\ \citenamefont
  {Wittig}}]{Ce:2022kxy}%
  \BibitemOpen
  \bibfield  {author} {\bibinfo {author} {\bibfnamefont {M.}~\bibnamefont
  {C\`e}}, \bibinfo {author} {\bibfnamefont {A.}~\bibnamefont {Gérardin}},
  \bibinfo {author} {\bibfnamefont {G.}~\bibnamefont {von Hippel}}, \bibinfo
  {author} {\bibfnamefont {R.~J.}\ \bibnamefont {Hudspith}}, \bibinfo {author}
  {\bibfnamefont {S.}~\bibnamefont {Kuberski}}, \bibinfo {author}
  {\bibfnamefont {H.~B.}\ \bibnamefont {Meyer}}, \bibinfo {author}
  {\bibfnamefont {K.}~\bibnamefont {Miura}}, \bibinfo {author} {\bibfnamefont
  {D.}~\bibnamefont {Mohler}}, \bibinfo {author} {\bibfnamefont
  {K.}~\bibnamefont {Ottnad}}, \bibinfo {author} {\bibfnamefont
  {S.}~\bibnamefont {Paul}}, \bibinfo {author} {\bibfnamefont {A.}~\bibnamefont
  {Risch}}, \bibinfo {author} {\bibfnamefont {T.}~\bibnamefont {San~José}},\
  and\ \bibinfo {author} {\bibfnamefont {T.}~\bibnamefont {Wittig}},\
  }\bibfield  {title} {\bibinfo {title} {{Window observable for the hadronic
  vacuum polarization contribution to the muon $g-2$ from lattice QCD}},\
  }\href@noop {} {\  (\bibinfo {year} {2022})},\ \Eprint
  {https://arxiv.org/abs/2206.06582} {arXiv:2206.06582 [hep-lat]} \BibitemShut
  {NoStop}%
\bibitem [{\citenamefont {Giusti}\ \emph {et~al.}(2018)\citenamefont {Giusti},
  \citenamefont {Sanfilippo},\ and\ \citenamefont {Simula}}]{Giusti:2018mdh}%
  \BibitemOpen
  \bibfield  {author} {\bibinfo {author} {\bibfnamefont {D.}~\bibnamefont
  {Giusti}}, \bibinfo {author} {\bibfnamefont {F.}~\bibnamefont {Sanfilippo}},\
  and\ \bibinfo {author} {\bibfnamefont {S.}~\bibnamefont {Simula}} (\bibinfo
  {collaboration} {ETM}),\ }\bibfield  {title} {\bibinfo {title} {Light-quark
  contribution to the leading hadronic vacuum polarization term of the muon
  $g-2$ from twisted-mass fermions},\ }\href
  {https://doi.org/10.1103/PhysRevD.98.114504} {\bibfield  {journal} {\bibinfo
  {journal} {Phys. Rev. D}\ }\textbf {\bibinfo {volume} {98}},\ \bibinfo
  {pages} {114504} (\bibinfo {year} {2018})},\ \Eprint
  {https://arxiv.org/abs/1808.00887} {arXiv:1808.00887 [hep-lat]} \BibitemShut
  {NoStop}%
\bibitem [{\citenamefont {Giusti}\ \emph {et~al.}(2019)\citenamefont {Giusti},
  \citenamefont {Lubicz}, \citenamefont {Martinelli}, \citenamefont
  {Sanfilippo},\ and\ \citenamefont {Simula}}]{Giusti:2019xct}%
  \BibitemOpen
  \bibfield  {author} {\bibinfo {author} {\bibfnamefont {D.}~\bibnamefont
  {Giusti}}, \bibinfo {author} {\bibfnamefont {V.}~\bibnamefont {Lubicz}},
  \bibinfo {author} {\bibfnamefont {G.}~\bibnamefont {Martinelli}}, \bibinfo
  {author} {\bibfnamefont {F.}~\bibnamefont {Sanfilippo}},\ and\ \bibinfo
  {author} {\bibfnamefont {S.}~\bibnamefont {Simula}} (\bibinfo {collaboration}
  {ETM}),\ }\bibfield  {title} {\bibinfo {title} {Electromagnetic and strong
  isospin-breaking corrections to the muon $g-2$ from lattice {QCD+QED}},\
  }\href {https://doi.org/10.1103/PhysRevD.99.114502} {\bibfield  {journal}
  {\bibinfo  {journal} {Phys. Rev. D}\ }\textbf {\bibinfo {volume} {99}},\
  \bibinfo {pages} {114502} (\bibinfo {year} {2019})},\ \Eprint
  {https://arxiv.org/abs/1901.10462} {arXiv:1901.10462 [hep-lat]} \BibitemShut
  {NoStop}%
\bibitem [{\citenamefont {Aubin}\ \emph {et~al.}(2020)\citenamefont {Aubin},
  \citenamefont {Blum}, \citenamefont {Tu}, \citenamefont {Golterman},
  \citenamefont {Jung},\ and\ \citenamefont {Peris}}]{Aubin:2019usy}%
  \BibitemOpen
  \bibfield  {author} {\bibinfo {author} {\bibfnamefont {C.}~\bibnamefont
  {Aubin}}, \bibinfo {author} {\bibfnamefont {T.}~\bibnamefont {Blum}},
  \bibinfo {author} {\bibfnamefont {C.}~\bibnamefont {Tu}}, \bibinfo {author}
  {\bibfnamefont {M.}~\bibnamefont {Golterman}}, \bibinfo {author}
  {\bibfnamefont {C.}~\bibnamefont {Jung}},\ and\ \bibinfo {author}
  {\bibfnamefont {S.}~\bibnamefont {Peris}},\ }\bibfield  {title} {\bibinfo
  {title} {{Light quark vacuum polarization at the physical point and
  contribution to the muon $g-2$}},\ }\href
  {https://doi.org/10.1103/PhysRevD.101.014503} {\bibfield  {journal} {\bibinfo
   {journal} {Phys. Rev. D}\ }\textbf {\bibinfo {volume} {101}},\ \bibinfo
  {pages} {014503} (\bibinfo {year} {2020})},\ \Eprint
  {https://arxiv.org/abs/1905.09307} {arXiv:1905.09307 [hep-lat]} \BibitemShut
  {NoStop}%
\bibitem [{\citenamefont {Aubin}\ \emph {et~al.}(2022)\citenamefont {Aubin},
  \citenamefont {Blum}, \citenamefont {Golterman},\ and\ \citenamefont
  {Peris}}]{Aubin:2022hgm}%
  \BibitemOpen
  \bibfield  {author} {\bibinfo {author} {\bibfnamefont {C.}~\bibnamefont
  {Aubin}}, \bibinfo {author} {\bibfnamefont {T.}~\bibnamefont {Blum}},
  \bibinfo {author} {\bibfnamefont {M.}~\bibnamefont {Golterman}},\ and\
  \bibinfo {author} {\bibfnamefont {S.}~\bibnamefont {Peris}},\ }\bibfield
  {title} {\bibinfo {title} {{The muon anomalous magnetic moment with staggered
  fermions: is the lattice spacing small enough?}},\ }\href@noop {} {\
  (\bibinfo {year} {2022})},\ \Eprint {https://arxiv.org/abs/2204.12256}
  {arXiv:2204.12256 [hep-lat]} \BibitemShut {NoStop}%
\bibitem [{\citenamefont {Shintani}\ and\ \citenamefont
  {Kuramashi}(2019)}]{Shintani:2019wai}%
  \BibitemOpen
  \bibfield  {author} {\bibinfo {author} {\bibfnamefont {E.}~\bibnamefont
  {Shintani}}\ and\ \bibinfo {author} {\bibfnamefont {Y.}~\bibnamefont
  {Kuramashi}} (\bibinfo {collaboration} {PACS}),\ }\bibfield  {title}
  {\bibinfo {title} {Study of systematic uncertainties in hadronic vacuum
  polarization contribution to muon $g-2$ with $2+1$-flavor lattice {QCD}},\
  }\href {https://doi.org/10.1103/PhysRevD.100.034517} {\bibfield  {journal}
  {\bibinfo  {journal} {Phys. Rev. D}\ }\textbf {\bibinfo {volume} {100}},\
  \bibinfo {pages} {034517} (\bibinfo {year} {2019})},\ \Eprint
  {https://arxiv.org/abs/1902.00885} {arXiv:1902.00885 [hep-lat]} \BibitemShut
  {NoStop}%
\bibitem [{\citenamefont {Borsányi}\ \emph {et~al.}(2021)\citenamefont
  {Borsányi} \emph {et~al.}}]{Borsanyi:2020mff}%
  \BibitemOpen
  \bibfield  {author} {\bibinfo {author} {\bibfnamefont {S.}~\bibnamefont
  {Borsányi}} \emph {et~al.} (\bibinfo {collaboration}
  {Budapest-Marseille-Wuppertal}),\ }\bibfield  {title} {\bibinfo {title}
  {Leading hadronic contribution to the muon magnetic moment from lattice
  {QCD}},\ }\href {https://doi.org/10.1038/s41586-021-03418-1} {\bibfield
  {journal} {\bibinfo  {journal} {Nature}\ }\textbf {\bibinfo {volume} {593}},\
  \bibinfo {pages} {51} (\bibinfo {year} {2021})},\ \Eprint
  {https://arxiv.org/abs/2002.12347} {arXiv:2002.12347 [hep-lat]} \BibitemShut
  {NoStop}%
\bibitem [{\citenamefont {Lehner}\ and\ \citenamefont
  {Meyer}(2020)}]{Lehner:2020crt}%
  \BibitemOpen
  \bibfield  {author} {\bibinfo {author} {\bibfnamefont {C.}~\bibnamefont
  {Lehner}}\ and\ \bibinfo {author} {\bibfnamefont {A.~S.}\ \bibnamefont
  {Meyer}},\ }\bibfield  {title} {\bibinfo {title} {{Consistency of hadronic
  vacuum polarization between lattice QCD and the $R$~ratio}},\ }\href
  {https://doi.org/10.1103/PhysRevD.101.074515} {\bibfield  {journal} {\bibinfo
   {journal} {Phys. Rev. D}\ }\textbf {\bibinfo {volume} {101}},\ \bibinfo
  {pages} {074515} (\bibinfo {year} {2020})},\ \Eprint
  {https://arxiv.org/abs/2003.04177} {arXiv:2003.04177 [hep-lat]} \BibitemShut
  {NoStop}%
\bibitem [{\citenamefont {Wang}\ \emph {et~al.}(2022)\citenamefont {Wang},
  \citenamefont {Draper}, \citenamefont {Liu},\ and\ \citenamefont
  {Yang}}]{Wang:2022lkq}%
  \BibitemOpen
  \bibfield  {author} {\bibinfo {author} {\bibfnamefont {G.}~\bibnamefont
  {Wang}}, \bibinfo {author} {\bibfnamefont {T.}~\bibnamefont {Draper}},
  \bibinfo {author} {\bibfnamefont {K.-F.}\ \bibnamefont {Liu}},\ and\ \bibinfo
  {author} {\bibfnamefont {Y.-B.}\ \bibnamefont {Yang}} (\bibinfo
  {collaboration} {\ensuremath{\chi}QCD}),\ }\bibfield  {title} {\bibinfo
  {title} {{Muon $g-2$ with overlap valence fermion}},\ }\href@noop {} {\
  (\bibinfo {year} {2022})},\ \Eprint {https://arxiv.org/abs/2204.01280}
  {arXiv:2204.01280 [hep-lat]} \BibitemShut {NoStop}%
\bibitem [{\citenamefont {Alexandrou}\ \emph {et~al.}(2022)\citenamefont
  {Alexandrou} \emph {et~al.}}]{Alexandrou:2022amy}%
  \BibitemOpen
  \bibfield  {author} {\bibinfo {author} {\bibfnamefont {C.}~\bibnamefont
  {Alexandrou}} \emph {et~al.} (\bibinfo {collaboration} {ETM}),\ }\bibfield
  {title} {\bibinfo {title} {{Lattice calculation of the short and intermediate
  time-distance hadronic vacuum polarization contributions to the muon magnetic
  moment using twisted-mass fermions}},\ }\href@noop {} {\  (\bibinfo {year}
  {2022})},\ \Eprint {https://arxiv.org/abs/2206.15084} {arXiv:2206.15084
  [hep-lat]} \BibitemShut {NoStop}%
\bibitem [{\citenamefont {Abi}\ \emph {et~al.}(2021)\citenamefont {Abi} \emph
  {et~al.}}]{Muong-2:2021ojo}%
  \BibitemOpen
  \bibfield  {author} {\bibinfo {author} {\bibfnamefont {B.}~\bibnamefont
  {Abi}} \emph {et~al.} (\bibinfo {collaboration} {Muon $g-2$}),\ }\bibfield
  {title} {\bibinfo {title} {Measurement of the positive muon anomalous
  magnetic moment to 0.46~ppm},\ }\href
  {https://doi.org/10.1103/PhysRevLett.126.141801} {\bibfield  {journal}
  {\bibinfo  {journal} {Phys. Rev. Lett.}\ }\textbf {\bibinfo {volume} {126}},\
  \bibinfo {pages} {141801} (\bibinfo {year} {2021})},\ \Eprint
  {https://arxiv.org/abs/2104.03281} {arXiv:2104.03281 [hep-ex]} \BibitemShut
  {NoStop}%
\bibitem [{\citenamefont {Albahri}\ \emph {et~al.}(2021)\citenamefont {Albahri}
  \emph {et~al.}}]{Muong-2:2021vma}%
  \BibitemOpen
  \bibfield  {author} {\bibinfo {author} {\bibfnamefont {T.}~\bibnamefont
  {Albahri}} \emph {et~al.} (\bibinfo {collaboration} {Muon $g-2$}),\
  }\bibfield  {title} {\bibinfo {title} {{Measurement of the anomalous
  precession frequency of the muon in the Fermilab Muon $g-2$ Experiment}},\
  }\href {https://doi.org/10.1103/PhysRevD.103.072002} {\bibfield  {journal}
  {\bibinfo  {journal} {Phys. Rev. D}\ }\textbf {\bibinfo {volume} {103}},\
  \bibinfo {pages} {072002} (\bibinfo {year} {2021})},\ \Eprint
  {https://arxiv.org/abs/2104.03247} {arXiv:2104.03247 [hep-ex]} \BibitemShut
  {NoStop}%
\bibitem [{\citenamefont {Bennett}\ \emph {et~al.}(2006)\citenamefont {Bennett}
  \emph {et~al.}}]{Muong-2:2006rrc}%
  \BibitemOpen
  \bibfield  {author} {\bibinfo {author} {\bibfnamefont {G.~W.}\ \bibnamefont
  {Bennett}} \emph {et~al.} (\bibinfo {collaboration} {Muon $g-2$}),\
  }\bibfield  {title} {\bibinfo {title} {Final report of the muon {E821}
  anomalous magnetic moment measurement at {BNL}},\ }\href
  {https://doi.org/10.1103/PhysRevD.73.072003} {\bibfield  {journal} {\bibinfo
  {journal} {Phys. Rev. D}\ }\textbf {\bibinfo {volume} {73}},\ \bibinfo
  {pages} {072003} (\bibinfo {year} {2006})},\ \Eprint
  {https://arxiv.org/abs/hep-ex/0602035} {arXiv:hep-ex/0602035} \BibitemShut
  {NoStop}%
\bibitem [{muo(2022)}]{muon:gm2:TI}%
  \BibitemOpen
  \href@noop {} {\bibinfo {title}
  {\href{https://muon-gm2-theory.illinois.edu/}{Muon $g-2$ Theory Initiative}}}
  (\bibinfo {year} {2017--2022})\BibitemShut {NoStop}%
\bibitem [{\citenamefont {Aoyama}\ \emph {et~al.}(2020)\citenamefont {Aoyama}
  \emph {et~al.}}]{Aoyama:2020ynm}%
  \BibitemOpen
  \bibfield  {author} {\bibinfo {author} {\bibfnamefont {T.}~\bibnamefont
  {Aoyama}} \emph {et~al.} (\bibinfo {collaboration} {Muon $g-2$ Theory
  Initiative}),\ }\bibfield  {title} {\bibinfo {title} {The anomalous magnetic
  moment of the muon in the {Standard Model}},\ }\href
  {https://doi.org/10.1016/j.physrep.2020.07.006} {\bibfield  {journal}
  {\bibinfo  {journal} {Phys. Rept.}\ }\textbf {\bibinfo {volume} {887}},\
  \bibinfo {pages} {1} (\bibinfo {year} {2020})},\ \Eprint
  {https://arxiv.org/abs/2006.04822} {arXiv:2006.04822 [hep-ph]} \BibitemShut
  {NoStop}%
\bibitem [{\citenamefont {Colangelo}\ \emph {et~al.}(2022)\citenamefont
  {Colangelo} \emph {et~al.}}]{Colangelo:2022jxc}%
  \BibitemOpen
  \bibfield  {author} {\bibinfo {author} {\bibfnamefont {G.}~\bibnamefont
  {Colangelo}} \emph {et~al.} (\bibinfo {collaboration} {Muon $g-2$ Theory
  Initiative}),\ }\bibfield  {title} {\bibinfo {title} {Prospects for precise
  predictions of $a_\mu$ in the {Standard Model}},\ }in\ \href@noop {} {\emph
  {\bibinfo {booktitle} {{2022 Snowmass Summer Study}}}}\ (\bibinfo {year}
  {2022})\ \Eprint {https://arxiv.org/abs/2203.15810} {arXiv:2203.15810
  [hep-ph]} \BibitemShut {NoStop}%
\bibitem [{\citenamefont {Davier}\ \emph {et~al.}(2020)\citenamefont {Davier},
  \citenamefont {Hoecker}, \citenamefont {Malaescu},\ and\ \citenamefont
  {Zhang}}]{Davier:2019can}%
  \BibitemOpen
  \bibfield  {author} {\bibinfo {author} {\bibfnamefont {M.}~\bibnamefont
  {Davier}}, \bibinfo {author} {\bibfnamefont {A.}~\bibnamefont {Hoecker}},
  \bibinfo {author} {\bibfnamefont {B.}~\bibnamefont {Malaescu}},\ and\
  \bibinfo {author} {\bibfnamefont {Z.}~\bibnamefont {Zhang}},\ }\bibfield
  {title} {\bibinfo {title} {{A new evaluation of the hadronic vacuum
  polarization contributions to the muon anomalous magnetic moment and to
  $\alpha(m_Z^2)$}},\ }\href {https://doi.org/10.1140/epjc/s10052-020-7792-2}
  {\bibfield  {journal} {\bibinfo  {journal} {Eur. Phys. J. C}\ }\textbf
  {\bibinfo {volume} {80}},\ \bibinfo {pages} {241} (\bibinfo {year} {2020})},\
  \bibinfo {note} {(E)
  \href{https://doi.org/10.1140/epjc/s10052-020-7857-2}{\textbf{80}, 410
  (2020)}},\ \Eprint {https://arxiv.org/abs/1908.00921} {arXiv:1908.00921
  [hep-ph]} \BibitemShut {NoStop}%
\bibitem [{\citenamefont {Keshavarzi}\ \emph
  {et~al.}(2020{\natexlab{a}})\citenamefont {Keshavarzi}, \citenamefont
  {Nomura},\ and\ \citenamefont {Teubner}}]{Keshavarzi:2019abf}%
  \BibitemOpen
  \bibfield  {author} {\bibinfo {author} {\bibfnamefont {A.}~\bibnamefont
  {Keshavarzi}}, \bibinfo {author} {\bibfnamefont {D.}~\bibnamefont {Nomura}},\
  and\ \bibinfo {author} {\bibfnamefont {T.}~\bibnamefont {Teubner}},\
  }\bibfield  {title} {\bibinfo {title} {{$g-2$ of charged leptons, $\alpha
  (M^2_Z)$, and the hyperfine splitting of muonium}},\ }\href
  {https://doi.org/10.1103/PhysRevD.101.014029} {\bibfield  {journal} {\bibinfo
   {journal} {Phys. Rev. D}\ }\textbf {\bibinfo {volume} {101}},\ \bibinfo
  {pages} {014029} (\bibinfo {year} {2020}{\natexlab{a}})},\ \Eprint
  {https://arxiv.org/abs/1911.00367} {arXiv:1911.00367 [hep-ph]} \BibitemShut
  {NoStop}%
\bibitem [{\citenamefont {Athron}\ \emph {et~al.}(2021)\citenamefont {Athron},
  \citenamefont {Bal\'azs}, \citenamefont {Jacob}, \citenamefont {Kotlarski},
  \citenamefont {St\"ockinger},\ and\ \citenamefont
  {St\"ockinger-Kim}}]{Athron:2021iuf}%
  \BibitemOpen
  \bibfield  {author} {\bibinfo {author} {\bibfnamefont {P.}~\bibnamefont
  {Athron}}, \bibinfo {author} {\bibfnamefont {C.}~\bibnamefont {Bal\'azs}},
  \bibinfo {author} {\bibfnamefont {D.~H.~J.}\ \bibnamefont {Jacob}}, \bibinfo
  {author} {\bibfnamefont {W.}~\bibnamefont {Kotlarski}}, \bibinfo {author}
  {\bibfnamefont {D.}~\bibnamefont {St\"ockinger}},\ and\ \bibinfo {author}
  {\bibfnamefont {H.}~\bibnamefont {St\"ockinger-Kim}},\ }\bibfield  {title}
  {\bibinfo {title} {{New physics explanations of $a_\mu$ in light of the FNAL
  Muon $g-2$ measurement}},\ }\href {https://doi.org/10.1007/JHEP09(2021)080}
  {\bibfield  {journal} {\bibinfo  {journal} {JHEP}\ }\textbf {\bibinfo
  {volume} {09}},\ \bibinfo {pages} {080}},\ \Eprint
  {https://arxiv.org/abs/2104.03691} {arXiv:2104.03691 [hep-ph]} \BibitemShut
  {NoStop}%
\bibitem [{\citenamefont {Passera}\ \emph {et~al.}(2008)\citenamefont
  {Passera}, \citenamefont {Marciano},\ and\ \citenamefont
  {Sirlin}}]{Passera:2008jk}%
  \BibitemOpen
  \bibfield  {author} {\bibinfo {author} {\bibfnamefont {M.}~\bibnamefont
  {Passera}}, \bibinfo {author} {\bibfnamefont {W.~J.}\ \bibnamefont
  {Marciano}},\ and\ \bibinfo {author} {\bibfnamefont {A.}~\bibnamefont
  {Sirlin}},\ }\bibfield  {title} {\bibinfo {title} {{The Muon g-2 and the
  bounds on the Higgs boson mass}},\ }\href
  {https://doi.org/10.1103/PhysRevD.78.013009} {\bibfield  {journal} {\bibinfo
  {journal} {Phys. Rev. D}\ }\textbf {\bibinfo {volume} {78}},\ \bibinfo
  {pages} {013009} (\bibinfo {year} {2008})},\ \Eprint
  {https://arxiv.org/abs/0804.1142} {arXiv:0804.1142 [hep-ph]} \BibitemShut
  {NoStop}%
\bibitem [{\citenamefont {Keshavarzi}\ \emph
  {et~al.}(2020{\natexlab{b}})\citenamefont {Keshavarzi}, \citenamefont
  {Marciano}, \citenamefont {Passera},\ and\ \citenamefont
  {Sirlin}}]{Keshavarzi:2020bfy}%
  \BibitemOpen
  \bibfield  {author} {\bibinfo {author} {\bibfnamefont {A.}~\bibnamefont
  {Keshavarzi}}, \bibinfo {author} {\bibfnamefont {W.~J.}\ \bibnamefont
  {Marciano}}, \bibinfo {author} {\bibfnamefont {M.}~\bibnamefont {Passera}},\
  and\ \bibinfo {author} {\bibfnamefont {A.}~\bibnamefont {Sirlin}},\
  }\bibfield  {title} {\bibinfo {title} {{Muon $g-2$ and $\Delta \alpha$
  connection}},\ }\href {https://doi.org/10.1103/PhysRevD.102.033002}
  {\bibfield  {journal} {\bibinfo  {journal} {Phys. Rev. D}\ }\textbf {\bibinfo
  {volume} {102}},\ \bibinfo {pages} {033002} (\bibinfo {year}
  {2020}{\natexlab{b}})},\ \Eprint {https://arxiv.org/abs/2006.12666}
  {arXiv:2006.12666 [hep-ph]} \BibitemShut {NoStop}%
\bibitem [{\citenamefont {Crivellin}\ \emph {et~al.}(2020)\citenamefont
  {Crivellin}, \citenamefont {Hoferichter}, \citenamefont {Manzari},\ and\
  \citenamefont {Montull}}]{Crivellin:2020zul}%
  \BibitemOpen
  \bibfield  {author} {\bibinfo {author} {\bibfnamefont {A.}~\bibnamefont
  {Crivellin}}, \bibinfo {author} {\bibfnamefont {M.}~\bibnamefont
  {Hoferichter}}, \bibinfo {author} {\bibfnamefont {C.~A.}\ \bibnamefont
  {Manzari}},\ and\ \bibinfo {author} {\bibfnamefont {M.}~\bibnamefont
  {Montull}},\ }\bibfield  {title} {\bibinfo {title} {Hadronic vacuum
  polarization: $(g-2)_\mu$ versus global electroweak fits},\ }\href
  {https://doi.org/10.1103/PhysRevLett.125.091801} {\bibfield  {journal}
  {\bibinfo  {journal} {Phys. Rev. Lett.}\ }\textbf {\bibinfo {volume} {125}},\
  \bibinfo {pages} {091801} (\bibinfo {year} {2020})},\ \Eprint
  {https://arxiv.org/abs/2003.04886} {arXiv:2003.04886 [hep-ph]} \BibitemShut
  {NoStop}%
\bibitem [{\citenamefont {Miura}\ \emph {et~al.}(2022)\citenamefont {Miura},
  \citenamefont {C\`e}, \citenamefont {G\'erardin}, \citenamefont {von Hippel},
  \citenamefont {Meyer}, \citenamefont {Ottnad}, \citenamefont {Risch},
  \citenamefont {San~Jose}, \citenamefont {Wilhelm},\ and\ \citenamefont
  {Wittig}}]{Miura:2022trp}%
  \BibitemOpen
  \bibfield  {author} {\bibinfo {author} {\bibfnamefont {K.}~\bibnamefont
  {Miura}}, \bibinfo {author} {\bibfnamefont {M.}~\bibnamefont {C\`e}},
  \bibinfo {author} {\bibfnamefont {A.}~\bibnamefont {G\'erardin}}, \bibinfo
  {author} {\bibfnamefont {G.}~\bibnamefont {von Hippel}}, \bibinfo {author}
  {\bibfnamefont {H.~B.}\ \bibnamefont {Meyer}}, \bibinfo {author}
  {\bibfnamefont {K.}~\bibnamefont {Ottnad}}, \bibinfo {author} {\bibfnamefont
  {A.}~\bibnamefont {Risch}}, \bibinfo {author} {\bibfnamefont
  {T.}~\bibnamefont {San~Jose}}, \bibinfo {author} {\bibfnamefont
  {J.}~\bibnamefont {Wilhelm}},\ and\ \bibinfo {author} {\bibfnamefont
  {H.}~\bibnamefont {Wittig}},\ }\bibfield  {title} {\bibinfo {title} {{HVP}
  contribution to running coupling and electroweak precision science},\ }\href
  {https://doi.org/10.22323/1.396.0342} {\bibfield  {journal} {\bibinfo
  {journal} {PoS}\ }\textbf {\bibinfo {volume} {LATTICE2021}},\ \bibinfo
  {pages} {342} (\bibinfo {year} {2022})}\BibitemShut {NoStop}%
\bibitem [{\citenamefont {Prades}\ \emph {et~al.}(2009)\citenamefont {Prades},
  \citenamefont {de~Rafael},\ and\ \citenamefont {Vainshtein}}]{Prades:2009tw}%
  \BibitemOpen
  \bibfield  {author} {\bibinfo {author} {\bibfnamefont {J.}~\bibnamefont
  {Prades}}, \bibinfo {author} {\bibfnamefont {E.}~\bibnamefont {de~Rafael}},\
  and\ \bibinfo {author} {\bibfnamefont {A.}~\bibnamefont {Vainshtein}},\
  }\bibfield  {title} {\bibinfo {title} {The hadronic light-by-light scattering
  contribution to the muon and electron anomalous magnetic moments},\ }\href
  {https://doi.org/10.1142/9789814271844_0009} {\bibfield  {journal} {\bibinfo
  {journal} {Adv. Ser. Direct. High Energy Phys.}\ }\textbf {\bibinfo {volume}
  {20}},\ \bibinfo {pages} {303} (\bibinfo {year} {2009})},\ \Eprint
  {https://arxiv.org/abs/0901.0306} {arXiv:0901.0306 [hep-ph]} \BibitemShut
  {NoStop}%
\bibitem [{\citenamefont {Nyffeler}(2009)}]{Nyffeler:2009tw}%
  \BibitemOpen
  \bibfield  {author} {\bibinfo {author} {\bibfnamefont {A.}~\bibnamefont
  {Nyffeler}},\ }\bibfield  {title} {\bibinfo {title} {{Hadronic light-by-light
  scattering in the muon $g-2$: A new short-distance constraint on
  pion-exchange}},\ }\href {https://doi.org/10.1103/PhysRevD.79.073012}
  {\bibfield  {journal} {\bibinfo  {journal} {Phys. Rev. D}\ }\textbf {\bibinfo
  {volume} {79}},\ \bibinfo {pages} {073012} (\bibinfo {year} {2009})},\
  \Eprint {https://arxiv.org/abs/0901.1172} {arXiv:0901.1172 [hep-ph]}
  \BibitemShut {NoStop}%
\bibitem [{\citenamefont {Jegerlehner}\ and\ \citenamefont
  {Nyffeler}(2009)}]{Jegerlehner:2009ry}%
  \BibitemOpen
  \bibfield  {author} {\bibinfo {author} {\bibfnamefont {F.}~\bibnamefont
  {Jegerlehner}}\ and\ \bibinfo {author} {\bibfnamefont {A.}~\bibnamefont
  {Nyffeler}},\ }\bibfield  {title} {\bibinfo {title} {The muon $g-2$},\ }\href
  {https://doi.org/10.1016/j.physrep.2009.04.003} {\bibfield  {journal}
  {\bibinfo  {journal} {Phys. Rept.}\ }\textbf {\bibinfo {volume} {477}},\
  \bibinfo {pages} {1} (\bibinfo {year} {2009})},\ \Eprint
  {https://arxiv.org/abs/0902.3360} {arXiv:0902.3360 [hep-ph]} \BibitemShut
  {NoStop}%
\bibitem [{\citenamefont {Jegerlehner}(2017)}]{Jegerlehner:2017gek}%
  \BibitemOpen
  \bibfield  {author} {\bibinfo {author} {\bibfnamefont {F.}~\bibnamefont
  {Jegerlehner}},\ }\href {https://doi.org/10.1007/978-3-319-63577-4} {\emph
  {\bibinfo {title} {{The Anomalous Magnetic Moment of the Muon}}}}\ (\bibinfo
  {publisher} {Springer},\ \bibinfo {address} {Cham},\ \bibinfo {year}
  {2017})\BibitemShut {NoStop}%
\bibitem [{\citenamefont {Colangelo}\ \emph {et~al.}(2015)\citenamefont
  {Colangelo}, \citenamefont {Hoferichter}, \citenamefont {Procura},\ and\
  \citenamefont {Stoffer}}]{Colangelo:2015ama}%
  \BibitemOpen
  \bibfield  {author} {\bibinfo {author} {\bibfnamefont {G.}~\bibnamefont
  {Colangelo}}, \bibinfo {author} {\bibfnamefont {M.}~\bibnamefont
  {Hoferichter}}, \bibinfo {author} {\bibfnamefont {M.}~\bibnamefont
  {Procura}},\ and\ \bibinfo {author} {\bibfnamefont {P.}~\bibnamefont
  {Stoffer}},\ }\bibfield  {title} {\bibinfo {title} {{Dispersion relation for
  hadronic light-by-light scattering: theoretical foundations}},\ }\href
  {https://doi.org/10.1007/JHEP09(2015)074} {\bibfield  {journal} {\bibinfo
  {journal} {JHEP}\ }\textbf {\bibinfo {volume} {09}},\ \bibinfo {pages}
  {074}},\ \Eprint {https://arxiv.org/abs/1506.01386} {arXiv:1506.01386
  [hep-ph]} \BibitemShut {NoStop}%
\bibitem [{\citenamefont {Blum}\ \emph {et~al.}(2015)\citenamefont {Blum},
  \citenamefont {Chowdhury}, \citenamefont {Hayakawa},\ and\ \citenamefont
  {Izubuchi}}]{Blum:2014oka}%
  \BibitemOpen
  \bibfield  {author} {\bibinfo {author} {\bibfnamefont {T.}~\bibnamefont
  {Blum}}, \bibinfo {author} {\bibfnamefont {S.}~\bibnamefont {Chowdhury}},
  \bibinfo {author} {\bibfnamefont {M.}~\bibnamefont {Hayakawa}},\ and\
  \bibinfo {author} {\bibfnamefont {T.}~\bibnamefont {Izubuchi}},\ }\bibfield
  {title} {\bibinfo {title} {{Hadronic light-by-light scattering contribution
  to the muon anomalous magnetic moment from lattice QCD}},\ }\href
  {https://doi.org/10.1103/PhysRevLett.114.012001} {\bibfield  {journal}
  {\bibinfo  {journal} {Phys. Rev. Lett.}\ }\textbf {\bibinfo {volume} {114}},\
  \bibinfo {pages} {012001} (\bibinfo {year} {2015})},\ \Eprint
  {https://arxiv.org/abs/1407.2923} {arXiv:1407.2923 [hep-lat]} \BibitemShut
  {NoStop}%
\bibitem [{\citenamefont {Blum}\ \emph
  {et~al.}(2016{\natexlab{b}})\citenamefont {Blum}, \citenamefont {Christ},
  \citenamefont {Hayakawa}, \citenamefont {Izubuchi}, \citenamefont {Jin},\
  and\ \citenamefont {Lehner}}]{Blum:2015gfa}%
  \BibitemOpen
  \bibfield  {author} {\bibinfo {author} {\bibfnamefont {T.}~\bibnamefont
  {Blum}}, \bibinfo {author} {\bibfnamefont {N.}~\bibnamefont {Christ}},
  \bibinfo {author} {\bibfnamefont {M.}~\bibnamefont {Hayakawa}}, \bibinfo
  {author} {\bibfnamefont {T.}~\bibnamefont {Izubuchi}}, \bibinfo {author}
  {\bibfnamefont {L.}~\bibnamefont {Jin}},\ and\ \bibinfo {author}
  {\bibfnamefont {C.}~\bibnamefont {Lehner}},\ }\bibfield  {title} {\bibinfo
  {title} {Lattice calculation of hadronic light-by-light contribution to the
  muon anomalous magnetic moment},\ }\href
  {https://doi.org/10.1103/PhysRevD.93.014503} {\bibfield  {journal} {\bibinfo
  {journal} {Phys. Rev. D}\ }\textbf {\bibinfo {volume} {93}},\ \bibinfo
  {pages} {014503} (\bibinfo {year} {2016}{\natexlab{b}})},\ \Eprint
  {https://arxiv.org/abs/1510.07100} {arXiv:1510.07100 [hep-lat]} \BibitemShut
  {NoStop}%
\bibitem [{\citenamefont {Blum}\ \emph
  {et~al.}(2017{\natexlab{a}})\citenamefont {Blum}, \citenamefont {Christ},
  \citenamefont {Hayakawa}, \citenamefont {Izubuchi}, \citenamefont {Jin},
  \citenamefont {Jung},\ and\ \citenamefont {Lehner}}]{Blum:2016lnc}%
  \BibitemOpen
  \bibfield  {author} {\bibinfo {author} {\bibfnamefont {T.}~\bibnamefont
  {Blum}}, \bibinfo {author} {\bibfnamefont {N.}~\bibnamefont {Christ}},
  \bibinfo {author} {\bibfnamefont {M.}~\bibnamefont {Hayakawa}}, \bibinfo
  {author} {\bibfnamefont {T.}~\bibnamefont {Izubuchi}}, \bibinfo {author}
  {\bibfnamefont {L.}~\bibnamefont {Jin}}, \bibinfo {author} {\bibfnamefont
  {C.}~\bibnamefont {Jung}},\ and\ \bibinfo {author} {\bibfnamefont
  {C.}~\bibnamefont {Lehner}},\ }\bibfield  {title} {\bibinfo {title}
  {Connected and leading disconnected hadronic light-by-light contribution to
  the muon anomalous magnetic moment with a physical pion mass},\ }\href
  {https://doi.org/10.1103/PhysRevLett.118.022005} {\bibfield  {journal}
  {\bibinfo  {journal} {Phys. Rev. Lett.}\ }\textbf {\bibinfo {volume} {118}},\
  \bibinfo {pages} {022005} (\bibinfo {year} {2017}{\natexlab{a}})},\ \Eprint
  {https://arxiv.org/abs/1610.04603} {arXiv:1610.04603 [hep-lat]} \BibitemShut
  {NoStop}%
\bibitem [{\citenamefont {Blum}\ \emph
  {et~al.}(2017{\natexlab{b}})\citenamefont {Blum}, \citenamefont {Christ},
  \citenamefont {Hayakawa}, \citenamefont {Izubuchi}, \citenamefont {Jin},
  \citenamefont {Jung},\ and\ \citenamefont {Lehner}}]{Blum:2017cer}%
  \BibitemOpen
  \bibfield  {author} {\bibinfo {author} {\bibfnamefont {T.}~\bibnamefont
  {Blum}}, \bibinfo {author} {\bibfnamefont {N.}~\bibnamefont {Christ}},
  \bibinfo {author} {\bibfnamefont {M.}~\bibnamefont {Hayakawa}}, \bibinfo
  {author} {\bibfnamefont {T.}~\bibnamefont {Izubuchi}}, \bibinfo {author}
  {\bibfnamefont {L.}~\bibnamefont {Jin}}, \bibinfo {author} {\bibfnamefont
  {C.}~\bibnamefont {Jung}},\ and\ \bibinfo {author} {\bibfnamefont
  {C.}~\bibnamefont {Lehner}},\ }\bibfield  {title} {\bibinfo {title} {Using
  infinite volume, continuum {QED} and lattice {QCD} for the hadronic
  light-by-light contribution to the muon anomalous magnetic moment},\ }\href
  {https://doi.org/10.1103/PhysRevD.96.034515} {\bibfield  {journal} {\bibinfo
  {journal} {Phys. Rev. D}\ }\textbf {\bibinfo {volume} {96}},\ \bibinfo
  {pages} {034515} (\bibinfo {year} {2017}{\natexlab{b}})},\ \Eprint
  {https://arxiv.org/abs/1705.01067} {arXiv:1705.01067 [hep-lat]} \BibitemShut
  {NoStop}%
\bibitem [{\citenamefont {Asmussen}\ \emph {et~al.}(2016)\citenamefont
  {Asmussen}, \citenamefont {Green}, \citenamefont {Meyer},\ and\ \citenamefont
  {Nyffeler}}]{Asmussen:2016lse}%
  \BibitemOpen
  \bibfield  {author} {\bibinfo {author} {\bibfnamefont {N.}~\bibnamefont
  {Asmussen}}, \bibinfo {author} {\bibfnamefont {J.}~\bibnamefont {Green}},
  \bibinfo {author} {\bibfnamefont {H.~B.}\ \bibnamefont {Meyer}},\ and\
  \bibinfo {author} {\bibfnamefont {A.}~\bibnamefont {Nyffeler}},\ }\bibfield
  {title} {\bibinfo {title} {{Position-space approach to hadronic
  light-by-light scattering in the muon $g-2$ on the lattice}},\ }\href
  {https://doi.org/10.22323/1.256.0164} {\bibfield  {journal} {\bibinfo
  {journal} {PoS}\ }\textbf {\bibinfo {volume} {LATTICE2016}},\ \bibinfo
  {pages} {164} (\bibinfo {year} {2016})},\ \Eprint
  {https://arxiv.org/abs/1609.08454} {arXiv:1609.08454 [hep-lat]} \BibitemShut
  {NoStop}%
\bibitem [{\citenamefont {Blum}\ \emph {et~al.}(2020)\citenamefont {Blum},
  \citenamefont {Christ}, \citenamefont {Hayakawa}, \citenamefont {Izubuchi},
  \citenamefont {Jin}, \citenamefont {Jung},\ and\ \citenamefont
  {Lehner}}]{Blum:2019ugy}%
  \BibitemOpen
  \bibfield  {author} {\bibinfo {author} {\bibfnamefont {T.}~\bibnamefont
  {Blum}}, \bibinfo {author} {\bibfnamefont {N.}~\bibnamefont {Christ}},
  \bibinfo {author} {\bibfnamefont {M.}~\bibnamefont {Hayakawa}}, \bibinfo
  {author} {\bibfnamefont {T.}~\bibnamefont {Izubuchi}}, \bibinfo {author}
  {\bibfnamefont {L.}~\bibnamefont {Jin}}, \bibinfo {author} {\bibfnamefont
  {C.}~\bibnamefont {Jung}},\ and\ \bibinfo {author} {\bibfnamefont
  {C.}~\bibnamefont {Lehner}},\ }\bibfield  {title} {\bibinfo {title} {Hadronic
  light-by-light scattering contribution to the muon anomalous magnetic moment
  from lattice {QCD}},\ }\href {https://doi.org/10.1103/PhysRevLett.124.132002}
  {\bibfield  {journal} {\bibinfo  {journal} {Phys. Rev. Lett.}\ }\textbf
  {\bibinfo {volume} {124}},\ \bibinfo {pages} {132002} (\bibinfo {year}
  {2020})},\ \Eprint {https://arxiv.org/abs/1911.08123} {arXiv:1911.08123
  [hep-lat]} \BibitemShut {NoStop}%
\bibitem [{\citenamefont {Chao}\ \emph {et~al.}(2021)\citenamefont {Chao},
  \citenamefont {Hudspith}, \citenamefont {G\'erardin}, \citenamefont {Green},
  \citenamefont {Meyer},\ and\ \citenamefont {Ottnad}}]{Chao:2021tvp}%
  \BibitemOpen
  \bibfield  {author} {\bibinfo {author} {\bibfnamefont {E.-H.}\ \bibnamefont
  {Chao}}, \bibinfo {author} {\bibfnamefont {R.~J.}\ \bibnamefont {Hudspith}},
  \bibinfo {author} {\bibfnamefont {A.}~\bibnamefont {G\'erardin}}, \bibinfo
  {author} {\bibfnamefont {J.~R.}\ \bibnamefont {Green}}, \bibinfo {author}
  {\bibfnamefont {H.~B.}\ \bibnamefont {Meyer}},\ and\ \bibinfo {author}
  {\bibfnamefont {K.}~\bibnamefont {Ottnad}},\ }\bibfield  {title} {\bibinfo
  {title} {{Hadronic light-by-light contribution to $(g-2)_\mu$ from lattice
  QCD: a complete calculation}},\ }\href
  {https://doi.org/10.1140/epjc/s10052-021-09455-4} {\bibfield  {journal}
  {\bibinfo  {journal} {Eur. Phys. J. C}\ }\textbf {\bibinfo {volume} {81}},\
  \bibinfo {pages} {651} (\bibinfo {year} {2021})},\ \Eprint
  {https://arxiv.org/abs/2104.02632} {arXiv:2104.02632 [hep-lat]} \BibitemShut
  {NoStop}%
\bibitem [{\citenamefont {Chao}\ \emph {et~al.}(2022)\citenamefont {Chao},
  \citenamefont {Hudspith}, \citenamefont {G\'erardin}, \citenamefont {Green},\
  and\ \citenamefont {Meyer}}]{Chao:2022xzg}%
  \BibitemOpen
  \bibfield  {author} {\bibinfo {author} {\bibfnamefont {E.-H.}\ \bibnamefont
  {Chao}}, \bibinfo {author} {\bibfnamefont {R.~J.}\ \bibnamefont {Hudspith}},
  \bibinfo {author} {\bibfnamefont {A.}~\bibnamefont {G\'erardin}}, \bibinfo
  {author} {\bibfnamefont {J.~R.}\ \bibnamefont {Green}},\ and\ \bibinfo
  {author} {\bibfnamefont {H.~B.}\ \bibnamefont {Meyer}},\ }\bibfield  {title}
  {\bibinfo {title} {{The charm-quark contribution to light-by-light scattering
  in the muon $(g-2)$ from lattice QCD}},\ }\href@noop {} {\  (\bibinfo {year}
  {2022})},\ \Eprint {https://arxiv.org/abs/2204.08844} {arXiv:2204.08844
  [hep-lat]} \BibitemShut {NoStop}%
\bibitem [{\citenamefont {Pospelov}\ and\ \citenamefont
  {Ritz}(2005)}]{Pospelov:2005pr}%
  \BibitemOpen
  \bibfield  {author} {\bibinfo {author} {\bibfnamefont {M.}~\bibnamefont
  {Pospelov}}\ and\ \bibinfo {author} {\bibfnamefont {A.}~\bibnamefont
  {Ritz}},\ }\bibfield  {title} {\bibinfo {title} {{Electric dipole moments as
  probes of new physics}},\ }\href {https://doi.org/10.1016/j.aop.2005.04.002}
  {\bibfield  {journal} {\bibinfo  {journal} {Annals Phys.}\ }\textbf {\bibinfo
  {volume} {318}},\ \bibinfo {pages} {119} (\bibinfo {year} {2005})},\ \Eprint
  {https://arxiv.org/abs/hep-ph/0504231} {arXiv:hep-ph/0504231} \BibitemShut
  {NoStop}%
\bibitem [{\citenamefont {Blinov}\ \emph {et~al.}(2022)\citenamefont {Blinov},
  \citenamefont {Craig}, \citenamefont {Dolan}, \citenamefont {de~Vries},
  \citenamefont {Draper}, \citenamefont {Garcia~Garcia}, \citenamefont
  {Lillard},\ and\ \citenamefont {Shelton}}]{Blinov:2022tfy}%
  \BibitemOpen
  \bibfield  {author} {\bibinfo {author} {\bibfnamefont {N.}~\bibnamefont
  {Blinov}}, \bibinfo {author} {\bibfnamefont {N.}~\bibnamefont {Craig}},
  \bibinfo {author} {\bibfnamefont {M.~J.}\ \bibnamefont {Dolan}}, \bibinfo
  {author} {\bibfnamefont {J.}~\bibnamefont {de~Vries}}, \bibinfo {author}
  {\bibfnamefont {P.}~\bibnamefont {Draper}}, \bibinfo {author} {\bibfnamefont
  {I.}~\bibnamefont {Garcia~Garcia}}, \bibinfo {author} {\bibfnamefont
  {B.}~\bibnamefont {Lillard}},\ and\ \bibinfo {author} {\bibfnamefont
  {J.}~\bibnamefont {Shelton}},\ }\bibfield  {title} {\bibinfo {title} {Strong
  {$CP$} beyond axion direct detection},\ }in\ \href@noop {} {\emph {\bibinfo
  {booktitle} {{2022 Snowmass Summer Study}}}}\ (\bibinfo {year} {2022})\
  \Eprint {https://arxiv.org/abs/2203.07218} {arXiv:2203.07218 [hep-ph]}
  \BibitemShut {NoStop}%
\bibitem [{\citenamefont {Bhattacharya}\ \emph {et~al.}(2021)\citenamefont
  {Bhattacharya}, \citenamefont {Cirigliano}, \citenamefont {Gupta},
  \citenamefont {Mereghetti},\ and\ \citenamefont
  {Yoon}}]{Bhattacharya:2021lol}%
  \BibitemOpen
  \bibfield  {author} {\bibinfo {author} {\bibfnamefont {T.}~\bibnamefont
  {Bhattacharya}}, \bibinfo {author} {\bibfnamefont {V.}~\bibnamefont
  {Cirigliano}}, \bibinfo {author} {\bibfnamefont {R.}~\bibnamefont {Gupta}},
  \bibinfo {author} {\bibfnamefont {E.}~\bibnamefont {Mereghetti}},\ and\
  \bibinfo {author} {\bibfnamefont {B.}~\bibnamefont {Yoon}},\ }\bibfield
  {title} {\bibinfo {title} {{Contribution of the QCD $\Theta$-term to the
  nucleon electric dipole moment}},\ }\href
  {https://doi.org/10.1103/PhysRevD.103.114507} {\bibfield  {journal} {\bibinfo
   {journal} {Phys. Rev. D}\ }\textbf {\bibinfo {volume} {103}},\ \bibinfo
  {pages} {114507} (\bibinfo {year} {2021})},\ \Eprint
  {https://arxiv.org/abs/2101.07230} {arXiv:2101.07230 [hep-lat]} \BibitemShut
  {NoStop}%
\bibitem [{\citenamefont {Shindler}(2021)}]{Shindler:2021bcx}%
  \BibitemOpen
  \bibfield  {author} {\bibinfo {author} {\bibfnamefont {A.}~\bibnamefont
  {Shindler}},\ }\bibfield  {title} {\bibinfo {title} {{Flavor-diagonal $CP$
  violation: The electric dipole moment}},\ }\href
  {https://doi.org/10.1140/epja/s10050-021-00421-y} {\bibfield  {journal}
  {\bibinfo  {journal} {Eur. Phys. J. A}\ }\textbf {\bibinfo {volume} {57}},\
  \bibinfo {pages} {128} (\bibinfo {year} {2021})}\BibitemShut {NoStop}%
\bibitem [{\citenamefont {Shintani}\ \emph {et~al.}(2005)\citenamefont
  {Shintani}, \citenamefont {Aoki}, \citenamefont {Ishizuka}, \citenamefont
  {Kanaya}, \citenamefont {Kikukawa}, \citenamefont {Kuramashi}, \citenamefont
  {Okawa}, \citenamefont {Taniguchi}, \citenamefont {Ukawa},\ and\
  \citenamefont {Yoshi\'e}}]{Shintani:2005xg}%
  \BibitemOpen
  \bibfield  {author} {\bibinfo {author} {\bibfnamefont {E.}~\bibnamefont
  {Shintani}}, \bibinfo {author} {\bibfnamefont {S.}~\bibnamefont {Aoki}},
  \bibinfo {author} {\bibfnamefont {N.}~\bibnamefont {Ishizuka}}, \bibinfo
  {author} {\bibfnamefont {K.}~\bibnamefont {Kanaya}}, \bibinfo {author}
  {\bibfnamefont {Y.}~\bibnamefont {Kikukawa}}, \bibinfo {author}
  {\bibfnamefont {Y.}~\bibnamefont {Kuramashi}}, \bibinfo {author}
  {\bibfnamefont {M.}~\bibnamefont {Okawa}}, \bibinfo {author} {\bibfnamefont
  {Y.}~\bibnamefont {Taniguchi}}, \bibinfo {author} {\bibfnamefont
  {A.}~\bibnamefont {Ukawa}},\ and\ \bibinfo {author} {\bibfnamefont
  {T.}~\bibnamefont {Yoshi\'e}} (\bibinfo {collaboration} {CP-PACS}),\
  }\bibfield  {title} {\bibinfo {title} {{Neutron electric dipole moment from
  lattice QCD}},\ }\href {https://doi.org/10.1103/PhysRevD.72.014504}
  {\bibfield  {journal} {\bibinfo  {journal} {Phys. Rev. D}\ }\textbf {\bibinfo
  {volume} {72}},\ \bibinfo {pages} {014504} (\bibinfo {year} {2005})},\
  \Eprint {https://arxiv.org/abs/hep-lat/0505022} {arXiv:hep-lat/0505022}
  \BibitemShut {NoStop}%
\bibitem [{\citenamefont {Fodor}\ \emph
  {et~al.}(2016{\natexlab{a}})\citenamefont {Fodor}, \citenamefont {Hoelbling},
  \citenamefont {Krieg}, \citenamefont {Lellouch}, \citenamefont {Lippert},
  \citenamefont {Portelli}, \citenamefont {Sastre}, \citenamefont {Szabo},\
  and\ \citenamefont {Varnhorst}}]{Fodor:2016bgu}%
  \BibitemOpen
  \bibfield  {author} {\bibinfo {author} {\bibfnamefont {Z.}~\bibnamefont
  {Fodor}}, \bibinfo {author} {\bibfnamefont {C.}~\bibnamefont {Hoelbling}},
  \bibinfo {author} {\bibfnamefont {S.}~\bibnamefont {Krieg}}, \bibinfo
  {author} {\bibfnamefont {L.}~\bibnamefont {Lellouch}}, \bibinfo {author}
  {\bibfnamefont {T.}~\bibnamefont {Lippert}}, \bibinfo {author} {\bibfnamefont
  {A.}~\bibnamefont {Portelli}}, \bibinfo {author} {\bibfnamefont
  {A.}~\bibnamefont {Sastre}}, \bibinfo {author} {\bibfnamefont {K.~K.}\
  \bibnamefont {Szabo}},\ and\ \bibinfo {author} {\bibfnamefont
  {L.}~\bibnamefont {Varnhorst}} (\bibinfo {collaboration}
  {Budapest-Marseille-Wuppertal}),\ }\bibfield  {title} {\bibinfo {title} {{Up
  and down quark masses and corrections to Dashen's theorem from lattice QCD
  and quenched QED}},\ }\href {https://doi.org/10.1103/PhysRevLett.117.082001}
  {\bibfield  {journal} {\bibinfo  {journal} {Phys. Rev. Lett.}\ }\textbf
  {\bibinfo {volume} {117}},\ \bibinfo {pages} {082001} (\bibinfo {year}
  {2016}{\natexlab{a}})},\ \Eprint {https://arxiv.org/abs/1604.07112}
  {arXiv:1604.07112 [hep-lat]} \BibitemShut {NoStop}%
\bibitem [{\citenamefont {Giusti}\ \emph {et~al.}(2017)\citenamefont {Giusti},
  \citenamefont {Lubicz}, \citenamefont {Tarantino}, \citenamefont
  {Martinelli}, \citenamefont {Sanfilippo}, \citenamefont {Simula},\ and\
  \citenamefont {Tantalo}}]{Giusti:2017dmp}%
  \BibitemOpen
  \bibfield  {author} {\bibinfo {author} {\bibfnamefont {D.}~\bibnamefont
  {Giusti}}, \bibinfo {author} {\bibfnamefont {V.}~\bibnamefont {Lubicz}},
  \bibinfo {author} {\bibfnamefont {C.}~\bibnamefont {Tarantino}}, \bibinfo
  {author} {\bibfnamefont {G.}~\bibnamefont {Martinelli}}, \bibinfo {author}
  {\bibfnamefont {F.}~\bibnamefont {Sanfilippo}}, \bibinfo {author}
  {\bibfnamefont {S.}~\bibnamefont {Simula}},\ and\ \bibinfo {author}
  {\bibfnamefont {N.}~\bibnamefont {Tantalo}} (\bibinfo {collaboration}
  {RM123}),\ }\bibfield  {title} {\bibinfo {title} {{Leading isospin-breaking
  corrections to pion, kaon and charmed-meson masses with Twisted-Mass
  fermions}},\ }\href {https://doi.org/10.1103/PhysRevD.95.114504} {\bibfield
  {journal} {\bibinfo  {journal} {Phys. Rev. D}\ }\textbf {\bibinfo {volume}
  {95}},\ \bibinfo {pages} {114504} (\bibinfo {year} {2017})},\ \Eprint
  {https://arxiv.org/abs/1704.06561} {arXiv:1704.06561 [hep-lat]} \BibitemShut
  {NoStop}%
\bibitem [{\citenamefont {Alexandrou}\ \emph
  {et~al.}(2020{\natexlab{a}})\citenamefont {Alexandrou}, \citenamefont
  {Finkenrath}, \citenamefont {Funcke}, \citenamefont {Jansen}, \citenamefont
  {Kostrzewa}, \citenamefont {Pittler},\ and\ \citenamefont
  {Urbach}}]{Alexandrou:2020bkd}%
  \BibitemOpen
  \bibfield  {author} {\bibinfo {author} {\bibfnamefont {C.}~\bibnamefont
  {Alexandrou}}, \bibinfo {author} {\bibfnamefont {J.}~\bibnamefont
  {Finkenrath}}, \bibinfo {author} {\bibfnamefont {L.}~\bibnamefont {Funcke}},
  \bibinfo {author} {\bibfnamefont {K.}~\bibnamefont {Jansen}}, \bibinfo
  {author} {\bibfnamefont {B.}~\bibnamefont {Kostrzewa}}, \bibinfo {author}
  {\bibfnamefont {F.}~\bibnamefont {Pittler}},\ and\ \bibinfo {author}
  {\bibfnamefont {C.}~\bibnamefont {Urbach}},\ }\bibfield  {title} {\bibinfo
  {title} {Ruling out the massless up-quark solution to the strong {$CP$}
  problem by computing the topological mass contribution with lattice {QCD}},\
  }\href {https://doi.org/10.1103/PhysRevLett.125.232001} {\bibfield  {journal}
  {\bibinfo  {journal} {Phys. Rev. Lett.}\ }\textbf {\bibinfo {volume} {125}},\
  \bibinfo {pages} {232001} (\bibinfo {year} {2020}{\natexlab{a}})},\ \Eprint
  {https://arxiv.org/abs/2002.07802} {arXiv:2002.07802 [hep-lat]} \BibitemShut
  {NoStop}%
\bibitem [{\citenamefont {Chupp}\ \emph {et~al.}(2019)\citenamefont {Chupp},
  \citenamefont {Fierlinger}, \citenamefont {Ramsey-Musolf},\ and\
  \citenamefont {Singh}}]{Chupp:2017rkp}%
  \BibitemOpen
  \bibfield  {author} {\bibinfo {author} {\bibfnamefont {T.}~\bibnamefont
  {Chupp}}, \bibinfo {author} {\bibfnamefont {P.}~\bibnamefont {Fierlinger}},
  \bibinfo {author} {\bibfnamefont {M.}~\bibnamefont {Ramsey-Musolf}},\ and\
  \bibinfo {author} {\bibfnamefont {J.}~\bibnamefont {Singh}},\ }\bibfield
  {title} {\bibinfo {title} {{Electric dipole moments of atoms, molecules,
  nuclei, and particles}},\ }\href
  {https://doi.org/10.1103/RevModPhys.91.015001} {\bibfield  {journal}
  {\bibinfo  {journal} {Rev. Mod. Phys.}\ }\textbf {\bibinfo {volume} {91}},\
  \bibinfo {pages} {015001} (\bibinfo {year} {2019})},\ \Eprint
  {https://arxiv.org/abs/1710.02504} {arXiv:1710.02504 [physics.atom-ph]}
  \BibitemShut {NoStop}%
\bibitem [{\citenamefont {Alarcon}\ \emph {et~al.}(2022)\citenamefont {Alarcon}
  \emph {et~al.}}]{Alarcon:2022ero}%
  \BibitemOpen
  \bibfield  {author} {\bibinfo {author} {\bibfnamefont {R.}~\bibnamefont
  {Alarcon}} \emph {et~al.},\ }\bibfield  {title} {\bibinfo {title} {{Electric
  dipole moments and the search for new physics}},\ }in\ \href@noop {} {\emph
  {\bibinfo {booktitle} {{2022 Snowmass Summer Study}}}}\ (\bibinfo {year}
  {2022})\ \Eprint {https://arxiv.org/abs/2203.08103} {arXiv:2203.08103
  [hep-ph]} \BibitemShut {NoStop}%
\bibitem [{\citenamefont {Anastassopoulos}\ \emph {et~al.}(2016)\citenamefont
  {Anastassopoulos} \emph {et~al.}}]{Anastassopoulos:2015ura}%
  \BibitemOpen
  \bibfield  {author} {\bibinfo {author} {\bibfnamefont {V.}~\bibnamefont
  {Anastassopoulos}} \emph {et~al.},\ }\bibfield  {title} {\bibinfo {title} {A
  storage ring experiment to detect a proton electric dipole moment},\ }\href
  {https://doi.org/10.1063/1.4967465} {\bibfield  {journal} {\bibinfo
  {journal} {Rev. Sci. Instrum.}\ }\textbf {\bibinfo {volume} {87}},\ \bibinfo
  {pages} {115116} (\bibinfo {year} {2016})},\ \Eprint
  {https://arxiv.org/abs/1502.04317} {arXiv:1502.04317 [physics.acc-ph]}
  \BibitemShut {NoStop}%
\bibitem [{\citenamefont {Omarov}\ \emph {et~al.}(2022)\citenamefont {Omarov},
  \citenamefont {Davoudiasl}, \citenamefont {Haciomeroglu}, \citenamefont
  {Lebedev}, \citenamefont {Morse}, \citenamefont {Semertzidis}, \citenamefont
  {Silenko}, \citenamefont {Stephenson},\ and\ \citenamefont
  {Suleiman}}]{Omarov:2020kws}%
  \BibitemOpen
  \bibfield  {author} {\bibinfo {author} {\bibfnamefont {Z.}~\bibnamefont
  {Omarov}}, \bibinfo {author} {\bibfnamefont {H.}~\bibnamefont {Davoudiasl}},
  \bibinfo {author} {\bibfnamefont {S.}~\bibnamefont {Haciomeroglu}}, \bibinfo
  {author} {\bibfnamefont {V.}~\bibnamefont {Lebedev}}, \bibinfo {author}
  {\bibfnamefont {W.~M.}\ \bibnamefont {Morse}}, \bibinfo {author}
  {\bibfnamefont {Y.~K.}\ \bibnamefont {Semertzidis}}, \bibinfo {author}
  {\bibfnamefont {A.~J.}\ \bibnamefont {Silenko}}, \bibinfo {author}
  {\bibfnamefont {E.~J.}\ \bibnamefont {Stephenson}},\ and\ \bibinfo {author}
  {\bibfnamefont {R.}~\bibnamefont {Suleiman}},\ }\bibfield  {title} {\bibinfo
  {title} {{Comprehensive symmetric-hybrid ring design for a proton EDM
  experiment at below $10^{-29} e\cdot\text{cm}$}},\ }\href
  {https://doi.org/10.1103/PhysRevD.105.032001} {\bibfield  {journal} {\bibinfo
   {journal} {Phys. Rev. D}\ }\textbf {\bibinfo {volume} {105}},\ \bibinfo
  {pages} {032001} (\bibinfo {year} {2022})},\ \Eprint
  {https://arxiv.org/abs/2007.10332} {arXiv:2007.10332 [physics.acc-ph]}
  \BibitemShut {NoStop}%
\bibitem [{\citenamefont {Alexander}\ \emph {et~al.}(2022)\citenamefont
  {Alexander} \emph {et~al.}}]{Alexander:2022rmq}%
  \BibitemOpen
  \bibfield  {author} {\bibinfo {author} {\bibfnamefont {J.}~\bibnamefont
  {Alexander}} \emph {et~al.},\ }\bibfield  {title} {\bibinfo {title} {{The
  storage ring proton EDM experiment}},\ }in\ \href@noop {} {\emph {\bibinfo
  {booktitle} {{2022 Snowmass Summer Study}}}}\ (\bibinfo {year} {2022})\
  \Eprint {https://arxiv.org/abs/2205.00830} {arXiv:2205.00830 [hep-ph]}
  \BibitemShut {NoStop}%
\bibitem [{\citenamefont {Guo}\ \emph {et~al.}(2015)\citenamefont {Guo},
  \citenamefont {Horsley}, \citenamefont {Mei{\ss}ner}, \citenamefont
  {Nakamura}, \citenamefont {Perlt}, \citenamefont {Rakow}, \citenamefont
  {Schierholz}, \citenamefont {Schiller},\ and\ \citenamefont
  {Zanotti}}]{Guo:2015tla}%
  \BibitemOpen
  \bibfield  {author} {\bibinfo {author} {\bibfnamefont {F.~K.}\ \bibnamefont
  {Guo}}, \bibinfo {author} {\bibfnamefont {R.}~\bibnamefont {Horsley}},
  \bibinfo {author} {\bibfnamefont {U.~G.}\ \bibnamefont {Mei{\ss}ner}},
  \bibinfo {author} {\bibfnamefont {Y.}~\bibnamefont {Nakamura}}, \bibinfo
  {author} {\bibfnamefont {H.}~\bibnamefont {Perlt}}, \bibinfo {author}
  {\bibfnamefont {P.~E.~L.}\ \bibnamefont {Rakow}}, \bibinfo {author}
  {\bibfnamefont {G.}~\bibnamefont {Schierholz}}, \bibinfo {author}
  {\bibfnamefont {A.}~\bibnamefont {Schiller}},\ and\ \bibinfo {author}
  {\bibfnamefont {J.~M.}\ \bibnamefont {Zanotti}},\ }\bibfield  {title}
  {\bibinfo {title} {{The electric dipole moment of the neutron from 2+1 flavor
  lattice QCD}},\ }\href {https://doi.org/10.1103/PhysRevLett.115.062001}
  {\bibfield  {journal} {\bibinfo  {journal} {Phys. Rev. Lett.}\ }\textbf
  {\bibinfo {volume} {115}},\ \bibinfo {pages} {062001} (\bibinfo {year}
  {2015})},\ \Eprint {https://arxiv.org/abs/1502.02295} {arXiv:1502.02295
  [hep-lat]} \BibitemShut {NoStop}%
\bibitem [{\citenamefont {Shintani}\ \emph {et~al.}(2016)\citenamefont
  {Shintani}, \citenamefont {Blum}, \citenamefont {Izubuchi},\ and\
  \citenamefont {Soni}}]{Shintani:2015vsx}%
  \BibitemOpen
  \bibfield  {author} {\bibinfo {author} {\bibfnamefont {E.}~\bibnamefont
  {Shintani}}, \bibinfo {author} {\bibfnamefont {T.}~\bibnamefont {Blum}},
  \bibinfo {author} {\bibfnamefont {T.}~\bibnamefont {Izubuchi}},\ and\
  \bibinfo {author} {\bibfnamefont {A.}~\bibnamefont {Soni}} (\bibinfo
  {collaboration} {RBC, UKQCD}),\ }\bibfield  {title} {\bibinfo {title}
  {{Neutron and proton electric dipole moments from $N_f=2+1$ domain-wall
  fermion lattice QCD}},\ }\href {https://doi.org/10.1103/PhysRevD.93.094503}
  {\bibfield  {journal} {\bibinfo  {journal} {Phys. Rev. D}\ }\textbf {\bibinfo
  {volume} {93}},\ \bibinfo {pages} {094503} (\bibinfo {year} {2016})},\
  \Eprint {https://arxiv.org/abs/1512.00566} {arXiv:1512.00566 [hep-lat]}
  \BibitemShut {NoStop}%
\bibitem [{\citenamefont {Abramczyk}\ \emph {et~al.}(2017)\citenamefont
  {Abramczyk}, \citenamefont {Aoki}, \citenamefont {Blum}, \citenamefont
  {Izubuchi}, \citenamefont {Ohki},\ and\ \citenamefont
  {Syritsyn}}]{Abramczyk:2017oxr}%
  \BibitemOpen
  \bibfield  {author} {\bibinfo {author} {\bibfnamefont {M.}~\bibnamefont
  {Abramczyk}}, \bibinfo {author} {\bibfnamefont {S.}~\bibnamefont {Aoki}},
  \bibinfo {author} {\bibfnamefont {T.}~\bibnamefont {Blum}}, \bibinfo {author}
  {\bibfnamefont {T.}~\bibnamefont {Izubuchi}}, \bibinfo {author}
  {\bibfnamefont {H.}~\bibnamefont {Ohki}},\ and\ \bibinfo {author}
  {\bibfnamefont {S.}~\bibnamefont {Syritsyn}},\ }\bibfield  {title} {\bibinfo
  {title} {{Lattice calculation of electric dipole moments and form factors of
  the nucleon}},\ }\href {https://doi.org/10.1103/PhysRevD.96.014501}
  {\bibfield  {journal} {\bibinfo  {journal} {Phys. Rev. D}\ }\textbf {\bibinfo
  {volume} {96}},\ \bibinfo {pages} {014501} (\bibinfo {year} {2017})},\
  \Eprint {https://arxiv.org/abs/1701.07792} {arXiv:1701.07792 [hep-lat]}
  \BibitemShut {NoStop}%
\bibitem [{\citenamefont {Syritsyn}\ \emph {et~al.}(2018)\citenamefont
  {Syritsyn}, \citenamefont {Izubuchi},\ and\ \citenamefont
  {Ohki}}]{Syritsyn:2018mon}%
  \BibitemOpen
  \bibfield  {author} {\bibinfo {author} {\bibfnamefont {S.}~\bibnamefont
  {Syritsyn}}, \bibinfo {author} {\bibfnamefont {T.}~\bibnamefont {Izubuchi}},\
  and\ \bibinfo {author} {\bibfnamefont {H.}~\bibnamefont {Ohki}},\ }\bibfield
  {title} {\bibinfo {title} {Progress in the nucleon electric dipole moment
  calculations in lattice {QCD}},\ }in\ \href@noop {} {\emph {\bibinfo
  {booktitle} {13$^{\rm th}$ Conference on the Intersections of Particle and
  Nuclear Physics}}},\ \bibinfo {editor} {edited by\ \bibinfo {editor}
  {\bibfnamefont {W.}~\bibnamefont {Haxton}}, \bibinfo {editor} {\bibfnamefont
  {B.}~\bibnamefont {Casey}}, \emph {et~al.}}\ (\bibinfo {year} {2018})\
  \Eprint {https://arxiv.org/abs/1810.03721} {arXiv:1810.03721 [hep-lat]}
  \BibitemShut {NoStop}%
\bibitem [{\citenamefont {Bhattacharya}\ \emph {et~al.}(2018)\citenamefont
  {Bhattacharya}, \citenamefont {Yoon}, \citenamefont {Gupta},\ and\
  \citenamefont {Cirigliano}}]{Bhattacharya:2018qat}%
  \BibitemOpen
  \bibfield  {author} {\bibinfo {author} {\bibfnamefont {T.}~\bibnamefont
  {Bhattacharya}}, \bibinfo {author} {\bibfnamefont {B.}~\bibnamefont {Yoon}},
  \bibinfo {author} {\bibfnamefont {R.}~\bibnamefont {Gupta}},\ and\ \bibinfo
  {author} {\bibfnamefont {V.}~\bibnamefont {Cirigliano}},\ }\bibfield  {title}
  {\bibinfo {title} {Neutron electric dipole moment from beyond the {Standard
  Model}},\ }\href {https://doi.org/10.22323/1.334.0123} {\bibfield  {journal}
  {\bibinfo  {journal} {PoS}\ }\textbf {\bibinfo {volume} {LATTICE2018}},\
  \bibinfo {pages} {123} (\bibinfo {year} {2018})},\ \Eprint
  {https://arxiv.org/abs/1812.06233} {arXiv:1812.06233 [hep-lat]} \BibitemShut
  {NoStop}%
\bibitem [{\citenamefont {Dragos}\ \emph {et~al.}(2021)\citenamefont {Dragos},
  \citenamefont {Luu}, \citenamefont {Shindler}, \citenamefont {de~Vries},\
  and\ \citenamefont {Yousif}}]{Dragos:2019oxn}%
  \BibitemOpen
  \bibfield  {author} {\bibinfo {author} {\bibfnamefont {J.}~\bibnamefont
  {Dragos}}, \bibinfo {author} {\bibfnamefont {T.}~\bibnamefont {Luu}},
  \bibinfo {author} {\bibfnamefont {A.}~\bibnamefont {Shindler}}, \bibinfo
  {author} {\bibfnamefont {J.}~\bibnamefont {de~Vries}},\ and\ \bibinfo
  {author} {\bibfnamefont {A.}~\bibnamefont {Yousif}},\ }\bibfield  {title}
  {\bibinfo {title} {Confirming the existence of the strong {$CP$} problem in
  lattice {QCD} with the gradient flow},\ }\href
  {https://doi.org/10.1103/PhysRevC.103.015202} {\bibfield  {journal} {\bibinfo
   {journal} {Phys. Rev. C}\ }\textbf {\bibinfo {volume} {103}},\ \bibinfo
  {pages} {015202} (\bibinfo {year} {2021})},\ \Eprint
  {https://arxiv.org/abs/1902.03254} {arXiv:1902.03254 [hep-lat]} \BibitemShut
  {NoStop}%
\bibitem [{\citenamefont {Alexandrou}\ \emph
  {et~al.}(2021{\natexlab{a}})\citenamefont {Alexandrou}, \citenamefont
  {Athenodorou}, \citenamefont {Hadjiyiannakou},\ and\ \citenamefont
  {Todaro}}]{Alexandrou:2020mds}%
  \BibitemOpen
  \bibfield  {author} {\bibinfo {author} {\bibfnamefont {C.}~\bibnamefont
  {Alexandrou}}, \bibinfo {author} {\bibfnamefont {A.}~\bibnamefont
  {Athenodorou}}, \bibinfo {author} {\bibfnamefont {K.}~\bibnamefont
  {Hadjiyiannakou}},\ and\ \bibinfo {author} {\bibfnamefont {A.}~\bibnamefont
  {Todaro}},\ }\bibfield  {title} {\bibinfo {title} {{Neutron electric dipole
  moment using lattice QCD simulations at the physical point}},\ }\href
  {https://doi.org/10.1103/PhysRevD.103.054501} {\bibfield  {journal} {\bibinfo
   {journal} {Phys. Rev. D}\ }\textbf {\bibinfo {volume} {103}},\ \bibinfo
  {pages} {054501} (\bibinfo {year} {2021}{\natexlab{a}})},\ \Eprint
  {https://arxiv.org/abs/2011.01084} {arXiv:2011.01084 [hep-lat]} \BibitemShut
  {NoStop}%
\bibitem [{\citenamefont {Gupta}\ \emph
  {et~al.}(2018{\natexlab{a}})\citenamefont {Gupta}, \citenamefont {Yoon},
  \citenamefont {Bhattacharya}, \citenamefont {Cirigliano}, \citenamefont
  {Jang},\ and\ \citenamefont {Lin}}]{Gupta:2018lvp}%
  \BibitemOpen
  \bibfield  {author} {\bibinfo {author} {\bibfnamefont {R.}~\bibnamefont
  {Gupta}}, \bibinfo {author} {\bibfnamefont {B.}~\bibnamefont {Yoon}},
  \bibinfo {author} {\bibfnamefont {T.}~\bibnamefont {Bhattacharya}}, \bibinfo
  {author} {\bibfnamefont {V.}~\bibnamefont {Cirigliano}}, \bibinfo {author}
  {\bibfnamefont {Y.-C.}\ \bibnamefont {Jang}},\ and\ \bibinfo {author}
  {\bibfnamefont {H.-W.}\ \bibnamefont {Lin}} (\bibinfo {collaboration}
  {PNDME}),\ }\bibfield  {title} {\bibinfo {title} {{Flavor-diagonal tensor
  charges of the nucleon from $(2+1+1)$-flavor lattice QCD}},\ }\href
  {https://doi.org/10.1103/PhysRevD.98.091501} {\bibfield  {journal} {\bibinfo
  {journal} {Phys. Rev. D}\ }\textbf {\bibinfo {volume} {98}},\ \bibinfo
  {pages} {091501} (\bibinfo {year} {2018}{\natexlab{a}})},\ \Eprint
  {https://arxiv.org/abs/1808.07597} {arXiv:1808.07597 [hep-lat]} \BibitemShut
  {NoStop}%
\bibitem [{\citenamefont {Aoki}\ \emph {et~al.}(2020)\citenamefont {Aoki} \emph
  {et~al.}}]{Aoki:2019cca}%
  \BibitemOpen
  \bibfield  {author} {\bibinfo {author} {\bibfnamefont {S.}~\bibnamefont
  {Aoki}} \emph {et~al.} (\bibinfo {collaboration} {Flavor Lattice Averaging
  Group}),\ }\bibfield  {title} {\bibinfo {title} {{FLAG} review 2019},\ }\href
  {https://doi.org/10.1140/epjc/s10052-019-7354-7} {\bibfield  {journal}
  {\bibinfo  {journal} {Eur. Phys. J. C}\ }\textbf {\bibinfo {volume} {80}},\
  \bibinfo {pages} {113} (\bibinfo {year} {2020})},\ \Eprint
  {https://arxiv.org/abs/1902.08191} {arXiv:1902.08191 [hep-lat]} \BibitemShut
  {NoStop}%
\bibitem [{\citenamefont {Aoki}\ \emph {et~al.}(2021)\citenamefont {Aoki} \emph
  {et~al.}}]{Aoki:2021kgd}%
  \BibitemOpen
  \bibfield  {author} {\bibinfo {author} {\bibfnamefont {Y.}~\bibnamefont
  {Aoki}} \emph {et~al.} (\bibinfo {collaboration} {Flavor Lattice Averaging
  Group}),\ }\bibfield  {title} {\bibinfo {title} {{FLAG Review 2021}},\
  }\href@noop {} {\  (\bibinfo {year} {2021})},\ \Eprint
  {https://arxiv.org/abs/2111.09849} {arXiv:2111.09849 [hep-lat]} \BibitemShut
  {NoStop}%
\bibitem [{\citenamefont {Bhattacharya}\ \emph {et~al.}(2015)\citenamefont
  {Bhattacharya}, \citenamefont {Cirigliano}, \citenamefont {Gupta},
  \citenamefont {Mereghetti},\ and\ \citenamefont
  {Yoon}}]{Bhattacharya:2015rsa}%
  \BibitemOpen
  \bibfield  {author} {\bibinfo {author} {\bibfnamefont {T.}~\bibnamefont
  {Bhattacharya}}, \bibinfo {author} {\bibfnamefont {V.}~\bibnamefont
  {Cirigliano}}, \bibinfo {author} {\bibfnamefont {R.}~\bibnamefont {Gupta}},
  \bibinfo {author} {\bibfnamefont {E.}~\bibnamefont {Mereghetti}},\ and\
  \bibinfo {author} {\bibfnamefont {B.}~\bibnamefont {Yoon}},\ }\bibfield
  {title} {\bibinfo {title} {{Dimension-5 $CP$-odd operators: QCD mixing and
  renormalization}},\ }\href {https://doi.org/10.1103/PhysRevD.92.114026}
  {\bibfield  {journal} {\bibinfo  {journal} {Phys. Rev. D}\ }\textbf {\bibinfo
  {volume} {92}},\ \bibinfo {pages} {114026} (\bibinfo {year} {2015})},\
  \Eprint {https://arxiv.org/abs/1502.07325} {arXiv:1502.07325 [hep-ph]}
  \BibitemShut {NoStop}%
\bibitem [{\citenamefont {Cirigliano}\ \emph {et~al.}(2020)\citenamefont
  {Cirigliano}, \citenamefont {Mereghetti},\ and\ \citenamefont
  {Stoffer}}]{Cirigliano:2020msr}%
  \BibitemOpen
  \bibfield  {author} {\bibinfo {author} {\bibfnamefont {V.}~\bibnamefont
  {Cirigliano}}, \bibinfo {author} {\bibfnamefont {E.}~\bibnamefont
  {Mereghetti}},\ and\ \bibinfo {author} {\bibfnamefont {P.}~\bibnamefont
  {Stoffer}},\ }\bibfield  {title} {\bibinfo {title} {{Non-perturbative
  renormalization scheme for the $CP$-odd three-gluon operator}},\ }\href
  {https://doi.org/10.1007/JHEP09(2020)094} {\bibfield  {journal} {\bibinfo
  {journal} {JHEP}\ }\textbf {\bibinfo {volume} {09}},\ \bibinfo {pages}
  {094}},\ \Eprint {https://arxiv.org/abs/2004.03576} {arXiv:2004.03576
  [hep-ph]} \BibitemShut {NoStop}%
\bibitem [{\citenamefont {Rizik}\ \emph {et~al.}(2020)\citenamefont {Rizik},
  \citenamefont {Monahan},\ and\ \citenamefont {Shindler}}]{Rizik:2020naq}%
  \BibitemOpen
  \bibfield  {author} {\bibinfo {author} {\bibfnamefont {M.~D.}\ \bibnamefont
  {Rizik}}, \bibinfo {author} {\bibfnamefont {C.~J.}\ \bibnamefont {Monahan}},\
  and\ \bibinfo {author} {\bibfnamefont {A.}~\bibnamefont {Shindler}} (\bibinfo
  {collaboration} {SymLat}),\ }\bibfield  {title} {\bibinfo {title} {{Short
  flow-time coefficients of $CP$-violating operators}},\ }\href
  {https://doi.org/10.1103/PhysRevD.102.034509} {\bibfield  {journal} {\bibinfo
   {journal} {Phys. Rev. D}\ }\textbf {\bibinfo {volume} {102}},\ \bibinfo
  {pages} {034509} (\bibinfo {year} {2020})},\ \Eprint
  {https://arxiv.org/abs/2005.04199} {arXiv:2005.04199 [hep-lat]} \BibitemShut
  {NoStop}%
\bibitem [{\citenamefont {Mereghetti}\ \emph {et~al.}(2022)\citenamefont
  {Mereghetti}, \citenamefont {Monahan}, \citenamefont {Rizik}, \citenamefont
  {Shindler},\ and\ \citenamefont {Stoffer}}]{Mereghetti:2021nkt}%
  \BibitemOpen
  \bibfield  {author} {\bibinfo {author} {\bibfnamefont {E.}~\bibnamefont
  {Mereghetti}}, \bibinfo {author} {\bibfnamefont {C.~J.}\ \bibnamefont
  {Monahan}}, \bibinfo {author} {\bibfnamefont {M.~D.}\ \bibnamefont {Rizik}},
  \bibinfo {author} {\bibfnamefont {A.}~\bibnamefont {Shindler}},\ and\
  \bibinfo {author} {\bibfnamefont {P.}~\bibnamefont {Stoffer}},\ }\bibfield
  {title} {\bibinfo {title} {{One-loop matching for quark dipole operators in a
  gradient-flow scheme}},\ }\href {https://doi.org/10.1007/JHEP04(2022)050}
  {\bibfield  {journal} {\bibinfo  {journal} {JHEP}\ }\textbf {\bibinfo
  {volume} {04}},\ \bibinfo {pages} {050}},\ \Eprint
  {https://arxiv.org/abs/2111.11449} {arXiv:2111.11449 [hep-lat]} \BibitemShut
  {NoStop}%
\bibitem [{\citenamefont {Kudryavtsev}(2016)}]{Kudryavtsev:2016ybl}%
  \BibitemOpen
  \bibfield  {author} {\bibinfo {author} {\bibfnamefont {V.~A.}\ \bibnamefont
  {Kudryavtsev}} (\bibinfo {collaboration} {DUNE}),\ }\bibfield  {title}
  {\bibinfo {title} {Underground physics with {DUNE}},\ }\href
  {https://doi.org/10.1088/1742-6596/718/6/062032} {\bibfield  {journal}
  {\bibinfo  {journal} {J. Phys. Conf. Ser.}\ }\textbf {\bibinfo {volume}
  {718}},\ \bibinfo {pages} {062032} (\bibinfo {year} {2016})},\ \Eprint
  {https://arxiv.org/abs/1601.03496} {arXiv:1601.03496 [physics.ins-det]}
  \BibitemShut {NoStop}%
\bibitem [{\citenamefont {Abe}\ \emph {et~al.}(2018)\citenamefont {Abe} \emph
  {et~al.}}]{Abe:2018uyc}%
  \BibitemOpen
  \bibfield  {author} {\bibinfo {author} {\bibfnamefont {K.}~\bibnamefont
  {Abe}} \emph {et~al.} (\bibinfo {collaboration} {Hyper-Kamiokande}),\
  }\bibfield  {title} {\bibinfo {title} {{Hyper-Kamiokande} design report},\
  }\href@noop {} {\bibfield  {journal} {\bibinfo  {journal} {unpublished}\ }
  (\bibinfo {year} {2018})},\ \Eprint {https://arxiv.org/abs/1805.04163}
  {arXiv:1805.04163 [physics.ins-det]} \BibitemShut {NoStop}%
\bibitem [{\citenamefont {Dev}\ \emph {et~al.}(2022)\citenamefont {Dev} \emph
  {et~al.}}]{Dev:2022jbf}%
  \BibitemOpen
  \bibfield  {author} {\bibinfo {author} {\bibfnamefont {P.~S.~B.}\
  \bibnamefont {Dev}} \emph {et~al.},\ }\bibfield  {title} {\bibinfo {title}
  {Searches for baryon number violation in neutrino experiments},\ }in\
  \href@noop {} {\emph {\bibinfo {booktitle} {{2022 Snowmass Summer Study}}}}\
  (\bibinfo {year} {2022})\ \Eprint {https://arxiv.org/abs/2203.08771}
  {arXiv:2203.08771 [hep-ex]} \BibitemShut {NoStop}%
\bibitem [{\citenamefont {Phillips}\ \emph {et~al.}(2016)\citenamefont
  {Phillips} \emph {et~al.}}]{Phillips:2014fgb}%
  \BibitemOpen
  \bibfield  {author} {\bibinfo {author} {\bibfnamefont {D.~G.}\ \bibnamefont
  {Phillips}} \emph {et~al.},\ }\bibfield  {title} {\bibinfo {title}
  {Neutron-antineutron oscillations: Theoretical status and experimental
  prospects},\ }\href {https://doi.org/10.1016/j.physrep.2015.11.001}
  {\bibfield  {journal} {\bibinfo  {journal} {Phys. Rept.}\ }\textbf {\bibinfo
  {volume} {612}},\ \bibinfo {pages} {1} (\bibinfo {year} {2016})},\ \Eprint
  {https://arxiv.org/abs/1410.1100} {arXiv:1410.1100 [hep-ex]} \BibitemShut
  {NoStop}%
\bibitem [{\citenamefont {Addazi}\ \emph {et~al.}(2021)\citenamefont {Addazi}
  \emph {et~al.}}]{Addazi:2020nlz}%
  \BibitemOpen
  \bibfield  {author} {\bibinfo {author} {\bibfnamefont {A.}~\bibnamefont
  {Addazi}} \emph {et~al.},\ }\bibfield  {title} {\bibinfo {title} {{New
  high-sensitivity searches for neutrons converting into antineutrons and/or
  sterile neutrons at the HIBEAM/NNBAR experiment at the European Spallation
  Source}},\ }\href {https://doi.org/10.1088/1361-6471/abf429} {\bibfield
  {journal} {\bibinfo  {journal} {J. Phys. G}\ }\textbf {\bibinfo {volume}
  {48}},\ \bibinfo {pages} {070501} (\bibinfo {year} {2021})},\ \Eprint
  {https://arxiv.org/abs/2006.04907} {arXiv:2006.04907 [physics.ins-det]}
  \BibitemShut {NoStop}%
\bibitem [{\citenamefont {Oosterhof}\ \emph {et~al.}(2019)\citenamefont
  {Oosterhof}, \citenamefont {Long}, \citenamefont {de~Vries}, \citenamefont
  {Timmermans},\ and\ \citenamefont {van Kolck}}]{Oosterhof:2019dlo}%
  \BibitemOpen
  \bibfield  {author} {\bibinfo {author} {\bibfnamefont {F.}~\bibnamefont
  {Oosterhof}}, \bibinfo {author} {\bibfnamefont {B.}~\bibnamefont {Long}},
  \bibinfo {author} {\bibfnamefont {J.}~\bibnamefont {de~Vries}}, \bibinfo
  {author} {\bibfnamefont {R.~G.~E.}\ \bibnamefont {Timmermans}},\ and\
  \bibinfo {author} {\bibfnamefont {U.}~\bibnamefont {van Kolck}},\ }\bibfield
  {title} {\bibinfo {title} {{Baryon-number violation by two units and the
  deuteron lifetime}},\ }\href {https://doi.org/10.1103/PhysRevLett.122.172501}
  {\bibfield  {journal} {\bibinfo  {journal} {Phys. Rev. Lett.}\ }\textbf
  {\bibinfo {volume} {122}},\ \bibinfo {pages} {172501} (\bibinfo {year}
  {2019})},\ \Eprint {https://arxiv.org/abs/1902.05342} {arXiv:1902.05342
  [hep-ph]} \BibitemShut {NoStop}%
\bibitem [{\citenamefont {Haidenbauer}\ and\ \citenamefont
  {Mei\ss{}ner}(2020)}]{Haidenbauer:2019fyd}%
  \BibitemOpen
  \bibfield  {author} {\bibinfo {author} {\bibfnamefont {J.}~\bibnamefont
  {Haidenbauer}}\ and\ \bibinfo {author} {\bibfnamefont {U.-G.}\ \bibnamefont
  {Mei\ss{}ner}},\ }\bibfield  {title} {\bibinfo {title} {{Neutron-antineutron
  oscillations in the deuteron studied with $NN$ and $\bar NN$ interactions
  based on chiral effective field theory}},\ }\href
  {https://doi.org/10.1088/1674-1137/44/3/033101} {\bibfield  {journal}
  {\bibinfo  {journal} {Chin. Phys. C}\ }\textbf {\bibinfo {volume} {44}},\
  \bibinfo {pages} {033101} (\bibinfo {year} {2020})},\ \Eprint
  {https://arxiv.org/abs/1910.14423} {arXiv:1910.14423 [hep-ph]} \BibitemShut
  {NoStop}%
\bibitem [{\citenamefont {Golubeva}\ \emph {et~al.}(2019)\citenamefont
  {Golubeva}, \citenamefont {Barrow},\ and\ \citenamefont
  {Ladd}}]{Golubeva:2018mrz}%
  \BibitemOpen
  \bibfield  {author} {\bibinfo {author} {\bibfnamefont {E.~S.}\ \bibnamefont
  {Golubeva}}, \bibinfo {author} {\bibfnamefont {J.~L.}\ \bibnamefont
  {Barrow}},\ and\ \bibinfo {author} {\bibfnamefont {C.~G.}\ \bibnamefont
  {Ladd}},\ }\bibfield  {title} {\bibinfo {title} {{Model of $\bar n$
  annihilation in experimental searches for $\bar n$ transformations}},\ }\href
  {https://doi.org/10.1103/PhysRevD.99.035002} {\bibfield  {journal} {\bibinfo
  {journal} {Phys. Rev. D}\ }\textbf {\bibinfo {volume} {99}},\ \bibinfo
  {pages} {035002} (\bibinfo {year} {2019})},\ \Eprint
  {https://arxiv.org/abs/1804.10270} {arXiv:1804.10270 [hep-ex]} \BibitemShut
  {NoStop}%
\bibitem [{\citenamefont {Barrow}\ \emph {et~al.}(2020)\citenamefont {Barrow},
  \citenamefont {Golubeva}, \citenamefont {Paryev},\ and\ \citenamefont
  {Richard}}]{Barrow:2019viz}%
  \BibitemOpen
  \bibfield  {author} {\bibinfo {author} {\bibfnamefont {J.~L.}\ \bibnamefont
  {Barrow}}, \bibinfo {author} {\bibfnamefont {E.~S.}\ \bibnamefont
  {Golubeva}}, \bibinfo {author} {\bibfnamefont {E.}~\bibnamefont {Paryev}},\
  and\ \bibinfo {author} {\bibfnamefont {J.-M.}\ \bibnamefont {Richard}},\
  }\bibfield  {title} {\bibinfo {title} {{Progress and simulations for
  intranuclear neutron-antineutron transformations in ${}^{40}_{18}Ar$}},\
  }\href {https://doi.org/10.1103/PhysRevD.101.036008} {\bibfield  {journal}
  {\bibinfo  {journal} {Phys. Rev. D}\ }\textbf {\bibinfo {volume} {101}},\
  \bibinfo {pages} {036008} (\bibinfo {year} {2020})},\ \Eprint
  {https://arxiv.org/abs/1906.02833} {arXiv:1906.02833 [hep-ex]} \BibitemShut
  {NoStop}%
\bibitem [{\citenamefont {Tsutsui}\ \emph {et~al.}(2004)\citenamefont {Tsutsui}
  \emph {et~al.}}]{CP-PACS:2004wqk}%
  \BibitemOpen
  \bibfield  {author} {\bibinfo {author} {\bibfnamefont {N.}~\bibnamefont
  {Tsutsui}} \emph {et~al.} (\bibinfo {collaboration} {CP-PACS, JLQCD}),\
  }\bibfield  {title} {\bibinfo {title} {{Lattice QCD calculation of the proton
  decay matrix element in the continuum limit}},\ }\href
  {https://doi.org/10.1103/PhysRevD.70.111501} {\bibfield  {journal} {\bibinfo
  {journal} {Phys. Rev. D}\ }\textbf {\bibinfo {volume} {70}},\ \bibinfo
  {pages} {111501} (\bibinfo {year} {2004})},\ \Eprint
  {https://arxiv.org/abs/hep-lat/0402026} {arXiv:hep-lat/0402026} \BibitemShut
  {NoStop}%
\bibitem [{\citenamefont {Aoki}\ \emph {et~al.}(2008)\citenamefont {Aoki},
  \citenamefont {Boyle}, \citenamefont {Cooney}, \citenamefont {Del~Debbio},
  \citenamefont {Kenway}, \citenamefont {Maynard}, \citenamefont {Soni},\ and\
  \citenamefont {Tweedie}}]{Aoki:2008ku}%
  \BibitemOpen
  \bibfield  {author} {\bibinfo {author} {\bibfnamefont {Y.}~\bibnamefont
  {Aoki}}, \bibinfo {author} {\bibfnamefont {P.}~\bibnamefont {Boyle}},
  \bibinfo {author} {\bibfnamefont {P.}~\bibnamefont {Cooney}}, \bibinfo
  {author} {\bibfnamefont {L.}~\bibnamefont {Del~Debbio}}, \bibinfo {author}
  {\bibfnamefont {R.}~\bibnamefont {Kenway}}, \bibinfo {author} {\bibfnamefont
  {C.~M.}\ \bibnamefont {Maynard}}, \bibinfo {author} {\bibfnamefont
  {A.}~\bibnamefont {Soni}},\ and\ \bibinfo {author} {\bibfnamefont
  {R.}~\bibnamefont {Tweedie}} (\bibinfo {collaboration} {RBC, UKQCD}),\
  }\bibfield  {title} {\bibinfo {title} {Proton lifetime bounds from chirally
  symmetric lattice {QCD}},\ }\href
  {https://doi.org/10.1103/PhysRevD.78.054505} {\bibfield  {journal} {\bibinfo
  {journal} {Phys. Rev. D}\ }\textbf {\bibinfo {volume} {78}},\ \bibinfo
  {pages} {054505} (\bibinfo {year} {2008})},\ \Eprint
  {https://arxiv.org/abs/0806.1031} {arXiv:0806.1031 [hep-lat]} \BibitemShut
  {NoStop}%
\bibitem [{\citenamefont {Aoki}\ \emph {et~al.}(2014)\citenamefont {Aoki},
  \citenamefont {Shintani},\ and\ \citenamefont {Soni}}]{Aoki:2013yxa}%
  \BibitemOpen
  \bibfield  {author} {\bibinfo {author} {\bibfnamefont {Y.}~\bibnamefont
  {Aoki}}, \bibinfo {author} {\bibfnamefont {E.}~\bibnamefont {Shintani}},\
  and\ \bibinfo {author} {\bibfnamefont {A.}~\bibnamefont {Soni}},\ }\bibfield
  {title} {\bibinfo {title} {{Proton decay matrix elements on the lattice}},\
  }\href {https://doi.org/10.1103/PhysRevD.89.014505} {\bibfield  {journal}
  {\bibinfo  {journal} {Phys. Rev. D}\ }\textbf {\bibinfo {volume} {89}},\
  \bibinfo {pages} {014505} (\bibinfo {year} {2014})},\ \Eprint
  {https://arxiv.org/abs/1304.7424} {arXiv:1304.7424 [hep-lat]} \BibitemShut
  {NoStop}%
\bibitem [{\citenamefont {Aoki}\ \emph {et~al.}(2017)\citenamefont {Aoki},
  \citenamefont {Izubuchi}, \citenamefont {Shintani},\ and\ \citenamefont
  {Soni}}]{Aoki:2017puj}%
  \BibitemOpen
  \bibfield  {author} {\bibinfo {author} {\bibfnamefont {Y.}~\bibnamefont
  {Aoki}}, \bibinfo {author} {\bibfnamefont {T.}~\bibnamefont {Izubuchi}},
  \bibinfo {author} {\bibfnamefont {E.}~\bibnamefont {Shintani}},\ and\
  \bibinfo {author} {\bibfnamefont {A.}~\bibnamefont {Soni}},\ }\bibfield
  {title} {\bibinfo {title} {{Improved lattice computation of proton decay
  matrix elements}},\ }\href {https://doi.org/10.1103/PhysRevD.96.014506}
  {\bibfield  {journal} {\bibinfo  {journal} {Phys. Rev. D}\ }\textbf {\bibinfo
  {volume} {96}},\ \bibinfo {pages} {014506} (\bibinfo {year} {2017})},\
  \Eprint {https://arxiv.org/abs/1705.01338} {arXiv:1705.01338 [hep-lat]}
  \BibitemShut {NoStop}%
\bibitem [{\citenamefont {Yoo}\ \emph {et~al.}(2022)\citenamefont {Yoo},
  \citenamefont {Aoki}, \citenamefont {Boyle}, \citenamefont {Izubuchi},
  \citenamefont {Soni},\ and\ \citenamefont {Syritsyn}}]{Yoo:2021gql}%
  \BibitemOpen
  \bibfield  {author} {\bibinfo {author} {\bibfnamefont {J.-S.}\ \bibnamefont
  {Yoo}}, \bibinfo {author} {\bibfnamefont {Y.}~\bibnamefont {Aoki}}, \bibinfo
  {author} {\bibfnamefont {P.}~\bibnamefont {Boyle}}, \bibinfo {author}
  {\bibfnamefont {T.}~\bibnamefont {Izubuchi}}, \bibinfo {author}
  {\bibfnamefont {A.}~\bibnamefont {Soni}},\ and\ \bibinfo {author}
  {\bibfnamefont {S.}~\bibnamefont {Syritsyn}},\ }\bibfield  {title} {\bibinfo
  {title} {{Proton decay matrix elements on the lattice at physical pion
  mass}},\ }\href {https://doi.org/10.1103/PhysRevD.105.074501} {\bibfield
  {journal} {\bibinfo  {journal} {Phys. Rev. D}\ }\textbf {\bibinfo {volume}
  {105}},\ \bibinfo {pages} {074501} (\bibinfo {year} {2022})},\ \Eprint
  {https://arxiv.org/abs/2111.01608} {arXiv:2111.01608 [hep-lat]} \BibitemShut
  {NoStop}%
\bibitem [{\citenamefont {Buchoff}\ and\ \citenamefont
  {Wagman}(2016)}]{Buchoff:2015qwa}%
  \BibitemOpen
  \bibfield  {author} {\bibinfo {author} {\bibfnamefont {M.~I.}\ \bibnamefont
  {Buchoff}}\ and\ \bibinfo {author} {\bibfnamefont {M.}~\bibnamefont
  {Wagman}},\ }\bibfield  {title} {\bibinfo {title} {Perturbative
  renormalization of neutron-antineutron operators},\ }\href
  {https://doi.org/10.1103/PhysRevD.93.016005} {\bibfield  {journal} {\bibinfo
  {journal} {Phys. Rev. D}\ }\textbf {\bibinfo {volume} {93}},\ \bibinfo
  {pages} {016005} (\bibinfo {year} {2016})},\ \bibinfo {note} {(E)
  \href{http://doi.org/10.1103/PhysRevD.98.079901}{\textbf{98}, 079901
  (2018)}},\ \Eprint {https://arxiv.org/abs/1506.00647} {arXiv:1506.00647
  [hep-ph]} \BibitemShut {NoStop}%
\bibitem [{\citenamefont {Rinaldi}\ \emph
  {et~al.}(2019{\natexlab{a}})\citenamefont {Rinaldi}, \citenamefont
  {Syritsyn}, \citenamefont {Wagman}, \citenamefont {Buchoff}, \citenamefont
  {Schroeder},\ and\ \citenamefont {Wasem}}]{Rinaldi:2018osy}%
  \BibitemOpen
  \bibfield  {author} {\bibinfo {author} {\bibfnamefont {E.}~\bibnamefont
  {Rinaldi}}, \bibinfo {author} {\bibfnamefont {S.}~\bibnamefont {Syritsyn}},
  \bibinfo {author} {\bibfnamefont {M.~L.}\ \bibnamefont {Wagman}}, \bibinfo
  {author} {\bibfnamefont {M.~I.}\ \bibnamefont {Buchoff}}, \bibinfo {author}
  {\bibfnamefont {C.}~\bibnamefont {Schroeder}},\ and\ \bibinfo {author}
  {\bibfnamefont {J.}~\bibnamefont {Wasem}},\ }\bibfield  {title} {\bibinfo
  {title} {Neutron-antineutron oscillations from lattice {QCD}},\ }\href
  {https://doi.org/10.1103/PhysRevLett.122.162001} {\bibfield  {journal}
  {\bibinfo  {journal} {Phys. Rev. Lett.}\ }\textbf {\bibinfo {volume} {122}},\
  \bibinfo {pages} {162001} (\bibinfo {year} {2019}{\natexlab{a}})},\ \Eprint
  {https://arxiv.org/abs/1809.00246} {arXiv:1809.00246 [hep-lat]} \BibitemShut
  {NoStop}%
\bibitem [{\citenamefont {Rinaldi}\ \emph
  {et~al.}(2019{\natexlab{b}})\citenamefont {Rinaldi}, \citenamefont
  {Syritsyn}, \citenamefont {Wagman}, \citenamefont {Buchoff}, \citenamefont
  {Schroeder},\ and\ \citenamefont {Wasem}}]{Rinaldi:2019thf}%
  \BibitemOpen
  \bibfield  {author} {\bibinfo {author} {\bibfnamefont {E.}~\bibnamefont
  {Rinaldi}}, \bibinfo {author} {\bibfnamefont {S.}~\bibnamefont {Syritsyn}},
  \bibinfo {author} {\bibfnamefont {M.~L.}\ \bibnamefont {Wagman}}, \bibinfo
  {author} {\bibfnamefont {M.~I.}\ \bibnamefont {Buchoff}}, \bibinfo {author}
  {\bibfnamefont {C.}~\bibnamefont {Schroeder}},\ and\ \bibinfo {author}
  {\bibfnamefont {J.}~\bibnamefont {Wasem}},\ }\bibfield  {title} {\bibinfo
  {title} {Lattice {QCD} determination of neutron-antineutron matrix elements
  with physical quark masses},\ }\href
  {https://doi.org/10.1103/PhysRevD.99.074510} {\bibfield  {journal} {\bibinfo
  {journal} {Phys. Rev. D}\ }\textbf {\bibinfo {volume} {99}},\ \bibinfo
  {pages} {074510} (\bibinfo {year} {2019}{\natexlab{b}})},\ \Eprint
  {https://arxiv.org/abs/1901.07519} {arXiv:1901.07519 [hep-lat]} \BibitemShut
  {NoStop}%
\bibitem [{\citenamefont {Dolinski}\ \emph {et~al.}(2019)\citenamefont
  {Dolinski}, \citenamefont {Poon},\ and\ \citenamefont
  {Rodejohann}}]{Dolinski:2019nrj}%
  \BibitemOpen
  \bibfield  {author} {\bibinfo {author} {\bibfnamefont {M.~J.}\ \bibnamefont
  {Dolinski}}, \bibinfo {author} {\bibfnamefont {A.~W.~P.}\ \bibnamefont
  {Poon}},\ and\ \bibinfo {author} {\bibfnamefont {W.}~\bibnamefont
  {Rodejohann}},\ }\bibfield  {title} {\bibinfo {title} {Neutrinoless
  double-beta decay: Status and prospects},\ }\href
  {https://doi.org/10.1146/annurev-nucl-101918-023407} {\bibfield  {journal}
  {\bibinfo  {journal} {Ann. Rev. Nucl. Part. Sci.}\ }\textbf {\bibinfo
  {volume} {69}},\ \bibinfo {pages} {219} (\bibinfo {year} {2019})},\ \Eprint
  {https://arxiv.org/abs/1902.04097} {arXiv:1902.04097 [nucl-ex]} \BibitemShut
  {NoStop}%
\bibitem [{\citenamefont {Cirigliano}\ \emph {et~al.}(2018)\citenamefont
  {Cirigliano}, \citenamefont {Dekens}, \citenamefont {de~Vries}, \citenamefont
  {Graesser}, \citenamefont {Mereghetti}, \citenamefont {Pastore},\ and\
  \citenamefont {van Kolck}}]{Cirigliano:2018hja}%
  \BibitemOpen
  \bibfield  {author} {\bibinfo {author} {\bibfnamefont {V.}~\bibnamefont
  {Cirigliano}}, \bibinfo {author} {\bibfnamefont {W.}~\bibnamefont {Dekens}},
  \bibinfo {author} {\bibfnamefont {J.}~\bibnamefont {de~Vries}}, \bibinfo
  {author} {\bibfnamefont {M.~L.}\ \bibnamefont {Graesser}}, \bibinfo {author}
  {\bibfnamefont {E.}~\bibnamefont {Mereghetti}}, \bibinfo {author}
  {\bibfnamefont {S.}~\bibnamefont {Pastore}},\ and\ \bibinfo {author}
  {\bibfnamefont {U.}~\bibnamefont {van Kolck}},\ }\bibfield  {title} {\bibinfo
  {title} {New leading contribution to neutrinoless double-$\beta$ decay},\
  }\href {https://doi.org/10.1103/PhysRevLett.120.202001} {\bibfield  {journal}
  {\bibinfo  {journal} {Phys. Rev. Lett.}\ }\textbf {\bibinfo {volume} {120}},\
  \bibinfo {pages} {202001} (\bibinfo {year} {2018})},\ \Eprint
  {https://arxiv.org/abs/1802.10097} {arXiv:1802.10097 [hep-ph]} \BibitemShut
  {NoStop}%
\bibitem [{\citenamefont {Cirigliano}\ \emph {et~al.}(2021)\citenamefont
  {Cirigliano}, \citenamefont {Dekens}, \citenamefont {de~Vries}, \citenamefont
  {Hoferichter},\ and\ \citenamefont {Mereghetti}}]{Cirigliano:2021qko}%
  \BibitemOpen
  \bibfield  {author} {\bibinfo {author} {\bibfnamefont {V.}~\bibnamefont
  {Cirigliano}}, \bibinfo {author} {\bibfnamefont {W.}~\bibnamefont {Dekens}},
  \bibinfo {author} {\bibfnamefont {J.}~\bibnamefont {de~Vries}}, \bibinfo
  {author} {\bibfnamefont {M.}~\bibnamefont {Hoferichter}},\ and\ \bibinfo
  {author} {\bibfnamefont {E.}~\bibnamefont {Mereghetti}},\ }\bibfield  {title}
  {\bibinfo {title} {{Determining the leading-order contact term in
  neutrinoless double $\beta$ decay}},\ }\href
  {https://doi.org/10.1007/JHEP05(2021)289} {\bibfield  {journal} {\bibinfo
  {journal} {JHEP}\ }\textbf {\bibinfo {volume} {05}},\ \bibinfo {pages}
  {289}},\ \Eprint {https://arxiv.org/abs/2102.03371} {arXiv:2102.03371
  [nucl-th]} \BibitemShut {NoStop}%
\bibitem [{\citenamefont {Wirth}\ \emph {et~al.}(2021)\citenamefont {Wirth},
  \citenamefont {Yao},\ and\ \citenamefont {Hergert}}]{Wirth:2021pij}%
  \BibitemOpen
  \bibfield  {author} {\bibinfo {author} {\bibfnamefont {R.}~\bibnamefont
  {Wirth}}, \bibinfo {author} {\bibfnamefont {J.~M.}\ \bibnamefont {Yao}},\
  and\ \bibinfo {author} {\bibfnamefont {H.}~\bibnamefont {Hergert}},\
  }\bibfield  {title} {\bibinfo {title} {\emph{Ab~Initio} calculation of the
  contact operator contribution in the standard mechanism for neutrinoless
  double beta decay},\ }\href {https://doi.org/10.1103/PhysRevLett.127.242502}
  {\bibfield  {journal} {\bibinfo  {journal} {Phys. Rev. Lett.}\ }\textbf
  {\bibinfo {volume} {127}},\ \bibinfo {pages} {242502} (\bibinfo {year}
  {2021})},\ \Eprint {https://arxiv.org/abs/2105.05415} {arXiv:2105.05415
  [nucl-th]} \BibitemShut {NoStop}%
\bibitem [{\citenamefont {Weiss}\ \emph {et~al.}(2021)\citenamefont {Weiss},
  \citenamefont {Soriano}, \citenamefont {Lovato}, \citenamefont {Menendez},\
  and\ \citenamefont {Wiringa}}]{Weiss:2021rig}%
  \BibitemOpen
  \bibfield  {author} {\bibinfo {author} {\bibfnamefont {R.}~\bibnamefont
  {Weiss}}, \bibinfo {author} {\bibfnamefont {P.}~\bibnamefont {Soriano}},
  \bibinfo {author} {\bibfnamefont {A.}~\bibnamefont {Lovato}}, \bibinfo
  {author} {\bibfnamefont {J.}~\bibnamefont {Menendez}},\ and\ \bibinfo
  {author} {\bibfnamefont {R.~B.}\ \bibnamefont {Wiringa}},\ }\bibfield
  {title} {\bibinfo {title} {{Neutrinoless double-beta decay: combining quantum
  Monte Carlo and the nuclear shell model with the generalized contact
  formalism}},\ }\href@noop {} {\  (\bibinfo {year} {2021})},\ \Eprint
  {https://arxiv.org/abs/2112.08146} {arXiv:2112.08146 [nucl-th]} \BibitemShut
  {NoStop}%
\bibitem [{\citenamefont {Jokiniemi}\ \emph {et~al.}(2021)\citenamefont
  {Jokiniemi}, \citenamefont {Soriano},\ and\ \citenamefont
  {Men\'endez}}]{Jokiniemi:2021qqv}%
  \BibitemOpen
  \bibfield  {author} {\bibinfo {author} {\bibfnamefont {L.}~\bibnamefont
  {Jokiniemi}}, \bibinfo {author} {\bibfnamefont {P.}~\bibnamefont {Soriano}},\
  and\ \bibinfo {author} {\bibfnamefont {J.}~\bibnamefont {Men\'endez}},\
  }\bibfield  {title} {\bibinfo {title} {{Impact of the leading-order
  short-range nuclear matrix element on the neutrinoless double-beta decay of
  medium-mass and heavy nuclei}},\ }\href
  {https://doi.org/10.1016/j.physletb.2021.136720} {\bibfield  {journal}
  {\bibinfo  {journal} {Phys. Lett. B}\ }\textbf {\bibinfo {volume} {823}},\
  \bibinfo {pages} {136720} (\bibinfo {year} {2021})},\ \Eprint
  {https://arxiv.org/abs/2107.13354} {arXiv:2107.13354 [nucl-th]} \BibitemShut
  {NoStop}%
\bibitem [{\citenamefont {Shanahan}\ \emph {et~al.}(2017)\citenamefont
  {Shanahan}, \citenamefont {Tiburzi}, \citenamefont {Wagman}, \citenamefont
  {Winter}, \citenamefont {Chang}, \citenamefont {Davoudi}, \citenamefont
  {Detmold}, \citenamefont {Orginos},\ and\ \citenamefont
  {Savage}}]{Shanahan:2017bgi}%
  \BibitemOpen
  \bibfield  {author} {\bibinfo {author} {\bibfnamefont {P.~E.}\ \bibnamefont
  {Shanahan}}, \bibinfo {author} {\bibfnamefont {B.~C.}\ \bibnamefont
  {Tiburzi}}, \bibinfo {author} {\bibfnamefont {M.~L.}\ \bibnamefont {Wagman}},
  \bibinfo {author} {\bibfnamefont {F.}~\bibnamefont {Winter}}, \bibinfo
  {author} {\bibfnamefont {E.}~\bibnamefont {Chang}}, \bibinfo {author}
  {\bibfnamefont {Z.}~\bibnamefont {Davoudi}}, \bibinfo {author} {\bibfnamefont
  {W.}~\bibnamefont {Detmold}}, \bibinfo {author} {\bibfnamefont
  {K.}~\bibnamefont {Orginos}},\ and\ \bibinfo {author} {\bibfnamefont {M.~J.}\
  \bibnamefont {Savage}} (\bibinfo {collaboration} {NPLQCD}),\ }\bibfield
  {title} {\bibinfo {title} {Isotensor axial polarizability and lattice {QCD}
  input for nuclear double-$\beta$ decay phenomenology},\ }\href
  {https://doi.org/10.1103/PhysRevLett.119.062003} {\bibfield  {journal}
  {\bibinfo  {journal} {Phys. Rev. Lett.}\ }\textbf {\bibinfo {volume} {119}},\
  \bibinfo {pages} {062003} (\bibinfo {year} {2017})},\ \Eprint
  {https://arxiv.org/abs/1701.03456} {arXiv:1701.03456 [hep-lat]} \BibitemShut
  {NoStop}%
\bibitem [{\citenamefont {Tiburzi}\ \emph {et~al.}(2017)\citenamefont
  {Tiburzi}, \citenamefont {Wagman}, \citenamefont {Winter}, \citenamefont
  {Chang}, \citenamefont {Davoudi}, \citenamefont {Detmold}, \citenamefont
  {Orginos}, \citenamefont {Savage},\ and\ \citenamefont
  {Shanahan}}]{Tiburzi:2017iux}%
  \BibitemOpen
  \bibfield  {author} {\bibinfo {author} {\bibfnamefont {B.~C.}\ \bibnamefont
  {Tiburzi}}, \bibinfo {author} {\bibfnamefont {M.~L.}\ \bibnamefont {Wagman}},
  \bibinfo {author} {\bibfnamefont {F.}~\bibnamefont {Winter}}, \bibinfo
  {author} {\bibfnamefont {E.}~\bibnamefont {Chang}}, \bibinfo {author}
  {\bibfnamefont {Z.}~\bibnamefont {Davoudi}}, \bibinfo {author} {\bibfnamefont
  {W.}~\bibnamefont {Detmold}}, \bibinfo {author} {\bibfnamefont
  {K.}~\bibnamefont {Orginos}}, \bibinfo {author} {\bibfnamefont {M.~J.}\
  \bibnamefont {Savage}},\ and\ \bibinfo {author} {\bibfnamefont {P.~E.}\
  \bibnamefont {Shanahan}} (\bibinfo {collaboration} {NPLQCD}),\ }\bibfield
  {title} {\bibinfo {title} {Double-$\beta$ decay matrix elements from lattice
  quantum chromodynamics},\ }\href {https://doi.org/10.1103/PhysRevD.96.054505}
  {\bibfield  {journal} {\bibinfo  {journal} {Phys. Rev. D}\ }\textbf {\bibinfo
  {volume} {96}},\ \bibinfo {pages} {054505} (\bibinfo {year} {2017})},\
  \Eprint {https://arxiv.org/abs/1702.02929} {arXiv:1702.02929 [hep-lat]}
  \BibitemShut {NoStop}%
\bibitem [{\citenamefont {Feng}\ \emph {et~al.}(2019)\citenamefont {Feng},
  \citenamefont {Jin}, \citenamefont {Tuo},\ and\ \citenamefont
  {Xia}}]{Feng:2018pdq}%
  \BibitemOpen
  \bibfield  {author} {\bibinfo {author} {\bibfnamefont {X.}~\bibnamefont
  {Feng}}, \bibinfo {author} {\bibfnamefont {L.-C.}\ \bibnamefont {Jin}},
  \bibinfo {author} {\bibfnamefont {X.-Y.}\ \bibnamefont {Tuo}},\ and\ \bibinfo
  {author} {\bibfnamefont {S.-C.}\ \bibnamefont {Xia}},\ }\bibfield  {title}
  {\bibinfo {title} {Light-neutrino exchange and long-distance contributions to
  $0\nu2\beta$ decays: An exploratory study on $\pi\pi\to ee$},\ }\href
  {https://doi.org/10.1103/PhysRevLett.122.022001} {\bibfield  {journal}
  {\bibinfo  {journal} {Phys. Rev. Lett.}\ }\textbf {\bibinfo {volume} {122}},\
  \bibinfo {pages} {022001} (\bibinfo {year} {2019})},\ \Eprint
  {https://arxiv.org/abs/1809.10511} {arXiv:1809.10511 [hep-lat]} \BibitemShut
  {NoStop}%
\bibitem [{\citenamefont {Detmold}\ and\ \citenamefont
  {Murphy}(2020)}]{Detmold:2020jqv}%
  \BibitemOpen
  \bibfield  {author} {\bibinfo {author} {\bibfnamefont {W.}~\bibnamefont
  {Detmold}}\ and\ \bibinfo {author} {\bibfnamefont {D.~J.}\ \bibnamefont
  {Murphy}} (\bibinfo {collaboration} {NPLQCD}),\ }\bibfield  {title} {\bibinfo
  {title} {Neutrinoless double beta decay from lattice {QCD}: The long-distance
  $\pi^-\to\pi^+e^-e^-$ amplitude},\ }\href@noop {} {\  (\bibinfo {year}
  {2020})},\ \Eprint {https://arxiv.org/abs/2004.07404} {arXiv:2004.07404
  [hep-lat]} \BibitemShut {NoStop}%
\bibitem [{\citenamefont {Tuo}\ \emph {et~al.}(2019)\citenamefont {Tuo},
  \citenamefont {Feng},\ and\ \citenamefont {Jin}}]{Tuo:2019bue}%
  \BibitemOpen
  \bibfield  {author} {\bibinfo {author} {\bibfnamefont {X.-Y.}\ \bibnamefont
  {Tuo}}, \bibinfo {author} {\bibfnamefont {X.}~\bibnamefont {Feng}},\ and\
  \bibinfo {author} {\bibfnamefont {L.-C.}\ \bibnamefont {Jin}},\ }\bibfield
  {title} {\bibinfo {title} {{Long-distance contributions to neutrinoless
  double beta decay $\pi^- \to\pi^+ e e$}},\ }\href
  {https://doi.org/10.1103/PhysRevD.100.094511} {\bibfield  {journal} {\bibinfo
   {journal} {Phys. Rev. D}\ }\textbf {\bibinfo {volume} {100}},\ \bibinfo
  {pages} {094511} (\bibinfo {year} {2019})},\ \Eprint
  {https://arxiv.org/abs/1909.13525} {arXiv:1909.13525 [hep-lat]} \BibitemShut
  {NoStop}%
\bibitem [{\citenamefont {Nicholson}\ \emph {et~al.}(2018)\citenamefont
  {Nicholson} \emph {et~al.}}]{Nicholson:2018mwc}%
  \BibitemOpen
  \bibfield  {author} {\bibinfo {author} {\bibfnamefont {A.}~\bibnamefont
  {Nicholson}} \emph {et~al.},\ }\bibfield  {title} {\bibinfo {title} {Heavy
  physics contributions to neutrinoless double beta decay from {QCD}},\ }\href
  {https://doi.org/10.1103/PhysRevLett.121.172501} {\bibfield  {journal}
  {\bibinfo  {journal} {Phys. Rev. Lett.}\ }\textbf {\bibinfo {volume} {121}},\
  \bibinfo {pages} {172501} (\bibinfo {year} {2018})},\ \Eprint
  {https://arxiv.org/abs/1805.02634} {arXiv:1805.02634 [nucl-th]} \BibitemShut
  {NoStop}%
\bibitem [{\citenamefont {Cirigliano}\ \emph {et~al.}(2022)\citenamefont
  {Cirigliano} \emph {et~al.}}]{Cirigliano:2022oqy}%
  \BibitemOpen
  \bibfield  {author} {\bibinfo {author} {\bibfnamefont {V.}~\bibnamefont
  {Cirigliano}} \emph {et~al.},\ }\bibfield  {title} {\bibinfo {title}
  {Neutrinoless double-beta decay: A roadmap for matching theory to
  experiment},\ }in\ \href@noop {} {\emph {\bibinfo {booktitle} {{2022 Snowmass
  Summer Study}}}}\ (\bibinfo {year} {2022})\ \Eprint
  {https://arxiv.org/abs/2203.12169} {arXiv:2203.12169 [hep-ph]} \BibitemShut
  {NoStop}%
\bibitem [{\citenamefont {Kosmas}\ and\ \citenamefont
  {Vergados}(1996)}]{Kosmas:1994pt}%
  \BibitemOpen
  \bibfield  {author} {\bibinfo {author} {\bibfnamefont {T.~S.}\ \bibnamefont
  {Kosmas}}\ and\ \bibinfo {author} {\bibfnamefont {J.~D.}\ \bibnamefont
  {Vergados}},\ }\bibfield  {title} {\bibinfo {title} {$(\mu^-, e^-)$
  conversion: A symbiosis of particle and nuclear physics},\ }\href
  {https://doi.org/10.1016/0370-1573(95)00041-0} {\bibfield  {journal}
  {\bibinfo  {journal} {Phys. Rept.}\ }\textbf {\bibinfo {volume} {264}},\
  \bibinfo {pages} {251} (\bibinfo {year} {1996})},\ \Eprint
  {https://arxiv.org/abs/nucl-th/9408011} {arXiv:nucl-th/9408011} \BibitemShut
  {NoStop}%
\bibitem [{\citenamefont {Cirigliano}\ \emph {et~al.}(2009)\citenamefont
  {Cirigliano}, \citenamefont {Kitano}, \citenamefont {Okada},\ and\
  \citenamefont {Tuzon}}]{Cirigliano:2009bz}%
  \BibitemOpen
  \bibfield  {author} {\bibinfo {author} {\bibfnamefont {V.}~\bibnamefont
  {Cirigliano}}, \bibinfo {author} {\bibfnamefont {R.}~\bibnamefont {Kitano}},
  \bibinfo {author} {\bibfnamefont {Y.}~\bibnamefont {Okada}},\ and\ \bibinfo
  {author} {\bibfnamefont {P.}~\bibnamefont {Tuzon}},\ }\bibfield  {title}
  {\bibinfo {title} {On the model discriminating power of $\mu\to e$ conversion
  in nuclei},\ }\href {https://doi.org/10.1103/PhysRevD.80.013002} {\bibfield
  {journal} {\bibinfo  {journal} {Phys. Rev. D}\ }\textbf {\bibinfo {volume}
  {80}},\ \bibinfo {pages} {013002} (\bibinfo {year} {2009})},\ \Eprint
  {https://arxiv.org/abs/0904.0957} {arXiv:0904.0957 [hep-ph]} \BibitemShut
  {NoStop}%
\bibitem [{\citenamefont {Freeman}\ and\ \citenamefont
  {Toussaint}(2013)}]{Freeman:2012ry}%
  \BibitemOpen
  \bibfield  {author} {\bibinfo {author} {\bibfnamefont {W.}~\bibnamefont
  {Freeman}}\ and\ \bibinfo {author} {\bibfnamefont {D.}~\bibnamefont
  {Toussaint}} (\bibinfo {collaboration} {MILC}),\ }\bibfield  {title}
  {\bibinfo {title} {{Intrinsic strangeness and charm of the nucleon using
  improved staggered fermions}},\ }\href
  {https://doi.org/10.1103/PhysRevD.88.054503} {\bibfield  {journal} {\bibinfo
  {journal} {Phys. Rev. D}\ }\textbf {\bibinfo {volume} {88}},\ \bibinfo
  {pages} {054503} (\bibinfo {year} {2013})},\ \Eprint
  {https://arxiv.org/abs/1204.3866} {arXiv:1204.3866 [hep-lat]} \BibitemShut
  {NoStop}%
\bibitem [{\citenamefont {Junnarkar}\ and\ \citenamefont
  {Walker-Loud}(2013)}]{Junnarkar:2013ac}%
  \BibitemOpen
  \bibfield  {author} {\bibinfo {author} {\bibfnamefont {P.}~\bibnamefont
  {Junnarkar}}\ and\ \bibinfo {author} {\bibfnamefont {A.}~\bibnamefont
  {Walker-Loud}},\ }\bibfield  {title} {\bibinfo {title} {{Scalar strange
  content of the nucleon from lattice QCD}},\ }\href
  {https://doi.org/10.1103/PhysRevD.87.114510} {\bibfield  {journal} {\bibinfo
  {journal} {Phys. Rev. D}\ }\textbf {\bibinfo {volume} {87}},\ \bibinfo
  {pages} {114510} (\bibinfo {year} {2013})},\ \Eprint
  {https://arxiv.org/abs/1301.1114} {arXiv:1301.1114 [hep-lat]} \BibitemShut
  {NoStop}%
\bibitem [{\citenamefont {Alexandrou}\ \emph {et~al.}(2014)\citenamefont
  {Alexandrou}, \citenamefont {Drach}, \citenamefont {Jansen}, \citenamefont
  {Kallidonis},\ and\ \citenamefont {Koutsou}}]{Alexandrou:2014sha}%
  \BibitemOpen
  \bibfield  {author} {\bibinfo {author} {\bibfnamefont {C.}~\bibnamefont
  {Alexandrou}}, \bibinfo {author} {\bibfnamefont {V.}~\bibnamefont {Drach}},
  \bibinfo {author} {\bibfnamefont {K.}~\bibnamefont {Jansen}}, \bibinfo
  {author} {\bibfnamefont {C.}~\bibnamefont {Kallidonis}},\ and\ \bibinfo
  {author} {\bibfnamefont {G.}~\bibnamefont {Koutsou}},\ }\bibfield  {title}
  {\bibinfo {title} {{Baryon spectrum with $N_f=2+1+1$ twisted mass
  fermions}},\ }\href {https://doi.org/10.1103/PhysRevD.90.074501} {\bibfield
  {journal} {\bibinfo  {journal} {Phys. Rev. D}\ }\textbf {\bibinfo {volume}
  {90}},\ \bibinfo {pages} {074501} (\bibinfo {year} {2014})},\ \Eprint
  {https://arxiv.org/abs/1406.4310} {arXiv:1406.4310 [hep-lat]} \BibitemShut
  {NoStop}%
\bibitem [{\citenamefont {Yang}\ \emph {et~al.}(2016)\citenamefont {Yang},
  \citenamefont {Alexandru}, \citenamefont {Draper}, \citenamefont {Liang},\
  and\ \citenamefont {Liu}}]{Yang:2015uis}%
  \BibitemOpen
  \bibfield  {author} {\bibinfo {author} {\bibfnamefont {Y.-B.}\ \bibnamefont
  {Yang}}, \bibinfo {author} {\bibfnamefont {A.}~\bibnamefont {Alexandru}},
  \bibinfo {author} {\bibfnamefont {T.}~\bibnamefont {Draper}}, \bibinfo
  {author} {\bibfnamefont {J.}~\bibnamefont {Liang}},\ and\ \bibinfo {author}
  {\bibfnamefont {K.-F.}\ \bibnamefont {Liu}} (\bibinfo {collaboration}
  {$\chi$QCD}),\ }\bibfield  {title} {\bibinfo {title} {{$\pi N$ and
  strangeness sigma terms at the physical point with chiral fermions}},\ }\href
  {https://doi.org/10.1103/PhysRevD.94.054503} {\bibfield  {journal} {\bibinfo
  {journal} {Phys. Rev. D}\ }\textbf {\bibinfo {volume} {94}},\ \bibinfo
  {pages} {054503} (\bibinfo {year} {2016})},\ \Eprint
  {https://arxiv.org/abs/1511.09089} {arXiv:1511.09089 [hep-lat]} \BibitemShut
  {NoStop}%
\bibitem [{\citenamefont {D\"urr}\ \emph {et~al.}(2012)\citenamefont {D\"urr}
  \emph {et~al.}}]{Durr:2011mp}%
  \BibitemOpen
  \bibfield  {author} {\bibinfo {author} {\bibfnamefont {S.}~\bibnamefont
  {D\"urr}} \emph {et~al.},\ }\bibfield  {title} {\bibinfo {title} {{Sigma term
  and strangeness content of octet baryons}},\ }\href
  {https://doi.org/10.1103/PhysRevD.85.014509} {\bibfield  {journal} {\bibinfo
  {journal} {Phys. Rev. D}\ }\textbf {\bibinfo {volume} {85}},\ \bibinfo
  {pages} {014509} (\bibinfo {year} {2012})},\ \bibinfo {note} {(E)
  \href{https://doi.org/10.1103/PhysRevD.93.039905}{\textbf{93}, 039905
  (2016)}},\ \Eprint {https://arxiv.org/abs/1109.4265} {arXiv:1109.4265
  [hep-lat]} \BibitemShut {NoStop}%
\bibitem [{\citenamefont {D\"urr}\ \emph {et~al.}(2016)\citenamefont {D\"urr}
  \emph {et~al.}}]{Durr:2015dna}%
  \BibitemOpen
  \bibfield  {author} {\bibinfo {author} {\bibfnamefont {S.}~\bibnamefont
  {D\"urr}} \emph {et~al.},\ }\bibfield  {title} {\bibinfo {title} {{Lattice
  computation of the nucleon scalar quark contents at the physical point}},\
  }\href {https://doi.org/10.1103/PhysRevLett.116.172001} {\bibfield  {journal}
  {\bibinfo  {journal} {Phys. Rev. Lett.}\ }\textbf {\bibinfo {volume} {116}},\
  \bibinfo {pages} {172001} (\bibinfo {year} {2016})},\ \Eprint
  {https://arxiv.org/abs/1510.08013} {arXiv:1510.08013 [hep-lat]} \BibitemShut
  {NoStop}%
\bibitem [{\citenamefont {Engelhardt}(2012)}]{Engelhardt:2012gd}%
  \BibitemOpen
  \bibfield  {author} {\bibinfo {author} {\bibfnamefont {M.}~\bibnamefont
  {Engelhardt}},\ }\bibfield  {title} {\bibinfo {title} {Strange quark
  contributions to nucleon mass and spin from lattice {QCD}},\ }\href
  {https://doi.org/10.1103/PhysRevD.86.114510} {\bibfield  {journal} {\bibinfo
  {journal} {Phys. Rev. D}\ }\textbf {\bibinfo {volume} {86}},\ \bibinfo
  {pages} {114510} (\bibinfo {year} {2012})},\ \Eprint
  {https://arxiv.org/abs/1210.0025} {arXiv:1210.0025 [hep-lat]} \BibitemShut
  {NoStop}%
\bibitem [{\citenamefont {Hoferichter}\ \emph {et~al.}(2015)\citenamefont
  {Hoferichter}, \citenamefont {Ruiz~de Elvira}, \citenamefont {Kubis},\ and\
  \citenamefont {Mei\ss{}ner}}]{Hoferichter:2015dsa}%
  \BibitemOpen
  \bibfield  {author} {\bibinfo {author} {\bibfnamefont {M.}~\bibnamefont
  {Hoferichter}}, \bibinfo {author} {\bibfnamefont {J.}~\bibnamefont {Ruiz~de
  Elvira}}, \bibinfo {author} {\bibfnamefont {B.}~\bibnamefont {Kubis}},\ and\
  \bibinfo {author} {\bibfnamefont {U.-G.}\ \bibnamefont {Mei\ss{}ner}},\
  }\bibfield  {title} {\bibinfo {title} {{High-Precision Determination of the
  Pion-Nucleon \ensuremath{\sigma} Term from Roy-Steiner Equations}},\ }\href
  {https://doi.org/10.1103/PhysRevLett.115.092301} {\bibfield  {journal}
  {\bibinfo  {journal} {Phys. Rev. Lett.}\ }\textbf {\bibinfo {volume} {115}},\
  \bibinfo {pages} {092301} (\bibinfo {year} {2015})},\ \Eprint
  {https://arxiv.org/abs/1506.04142} {arXiv:1506.04142 [hep-ph]} \BibitemShut
  {NoStop}%
\bibitem [{\citenamefont {Ruiz~de Elvira}\ \emph {et~al.}(2018)\citenamefont
  {Ruiz~de Elvira}, \citenamefont {Hoferichter}, \citenamefont {Kubis},\ and\
  \citenamefont {Mei\ss{}ner}}]{RuizdeElvira:2017stg}%
  \BibitemOpen
  \bibfield  {author} {\bibinfo {author} {\bibfnamefont {J.}~\bibnamefont
  {Ruiz~de Elvira}}, \bibinfo {author} {\bibfnamefont {M.}~\bibnamefont
  {Hoferichter}}, \bibinfo {author} {\bibfnamefont {B.}~\bibnamefont {Kubis}},\
  and\ \bibinfo {author} {\bibfnamefont {U.-G.}\ \bibnamefont {Mei\ss{}ner}},\
  }\bibfield  {title} {\bibinfo {title} {{Extracting the $\sigma$-term from
  low-energy pion-nucleon scattering}},\ }\href
  {https://doi.org/10.1088/1361-6471/aa9422} {\bibfield  {journal} {\bibinfo
  {journal} {J. Phys. G}\ }\textbf {\bibinfo {volume} {45}},\ \bibinfo {pages}
  {024001} (\bibinfo {year} {2018})},\ \Eprint
  {https://arxiv.org/abs/1706.01465} {arXiv:1706.01465 [hep-ph]} \BibitemShut
  {NoStop}%
\bibitem [{\citenamefont {Gupta}\ \emph {et~al.}(2021)\citenamefont {Gupta},
  \citenamefont {Park}, \citenamefont {Hoferichter}, \citenamefont
  {Mereghetti}, \citenamefont {Yoon},\ and\ \citenamefont
  {Bhattacharya}}]{Gupta:2021ahb}%
  \BibitemOpen
  \bibfield  {author} {\bibinfo {author} {\bibfnamefont {R.}~\bibnamefont
  {Gupta}}, \bibinfo {author} {\bibfnamefont {S.}~\bibnamefont {Park}},
  \bibinfo {author} {\bibfnamefont {M.}~\bibnamefont {Hoferichter}}, \bibinfo
  {author} {\bibfnamefont {E.}~\bibnamefont {Mereghetti}}, \bibinfo {author}
  {\bibfnamefont {B.}~\bibnamefont {Yoon}},\ and\ \bibinfo {author}
  {\bibfnamefont {T.}~\bibnamefont {Bhattacharya}},\ }\bibfield  {title}
  {\bibinfo {title} {{Pion\textendash{}Nucleon Sigma Term from Lattice QCD}},\
  }\href {https://doi.org/10.1103/PhysRevLett.127.242002} {\bibfield  {journal}
  {\bibinfo  {journal} {Phys. Rev. Lett.}\ }\textbf {\bibinfo {volume} {127}},\
  \bibinfo {pages} {242002} (\bibinfo {year} {2021})},\ \Eprint
  {https://arxiv.org/abs/2105.12095} {arXiv:2105.12095 [hep-lat]} \BibitemShut
  {NoStop}%
\bibitem [{\citenamefont {Shifman}\ \emph {et~al.}(1978)\citenamefont
  {Shifman}, \citenamefont {Vainshtein},\ and\ \citenamefont
  {Zakharov}}]{Shifman:1978zn}%
  \BibitemOpen
  \bibfield  {author} {\bibinfo {author} {\bibfnamefont {M.~A.}\ \bibnamefont
  {Shifman}}, \bibinfo {author} {\bibfnamefont {A.~I.}\ \bibnamefont
  {Vainshtein}},\ and\ \bibinfo {author} {\bibfnamefont {V.~I.}\ \bibnamefont
  {Zakharov}},\ }\bibfield  {title} {\bibinfo {title} {Remarks on {Higgs} boson
  interactions with nucleons},\ }\href
  {https://doi.org/10.1016/0370-2693(78)90481-1} {\bibfield  {journal}
  {\bibinfo  {journal} {Phys. Lett. B}\ }\textbf {\bibinfo {volume} {78}},\
  \bibinfo {pages} {443} (\bibinfo {year} {1978})}\BibitemShut {NoStop}%
\bibitem [{\citenamefont {Gong}\ \emph {et~al.}(2013)\citenamefont {Gong} \emph
  {et~al.}}]{XQCD:2013odc}%
  \BibitemOpen
  \bibfield  {author} {\bibinfo {author} {\bibfnamefont {M.}~\bibnamefont
  {Gong}} \emph {et~al.} (\bibinfo {collaboration} {$\chi$QCD}),\ }\bibfield
  {title} {\bibinfo {title} {{Strangeness and charmness content of the nucleon
  from overlap fermions on $2+1$-flavor domain-wall fermion configurations}},\
  }\href {https://doi.org/10.1103/PhysRevD.88.014503} {\bibfield  {journal}
  {\bibinfo  {journal} {Phys. Rev. D}\ }\textbf {\bibinfo {volume} {88}},\
  \bibinfo {pages} {014503} (\bibinfo {year} {2013})},\ \Eprint
  {https://arxiv.org/abs/1304.1194} {arXiv:1304.1194 [hep-ph]} \BibitemShut
  {NoStop}%
\bibitem [{\citenamefont {Alexandrou}\ \emph
  {et~al.}(2020{\natexlab{b}})\citenamefont {Alexandrou}, \citenamefont
  {Bacchio}, \citenamefont {Constantinou}, \citenamefont {Finkenrath},
  \citenamefont {Hadjiyiannakou}, \citenamefont {Jansen}, \citenamefont
  {Koutsou},\ and\ \citenamefont {Vaquero Aviles-Casco}}]{Alexandrou:2019brg}%
  \BibitemOpen
  \bibfield  {author} {\bibinfo {author} {\bibfnamefont {C.}~\bibnamefont
  {Alexandrou}}, \bibinfo {author} {\bibfnamefont {S.}~\bibnamefont {Bacchio}},
  \bibinfo {author} {\bibfnamefont {M.}~\bibnamefont {Constantinou}}, \bibinfo
  {author} {\bibfnamefont {J.}~\bibnamefont {Finkenrath}}, \bibinfo {author}
  {\bibfnamefont {K.}~\bibnamefont {Hadjiyiannakou}}, \bibinfo {author}
  {\bibfnamefont {K.}~\bibnamefont {Jansen}}, \bibinfo {author} {\bibfnamefont
  {G.}~\bibnamefont {Koutsou}},\ and\ \bibinfo {author} {\bibfnamefont
  {A.}~\bibnamefont {Vaquero Aviles-Casco}},\ }\bibfield  {title} {\bibinfo
  {title} {{Nucleon axial, tensor, and scalar charges and $\sigma$-terms in
  lattice QCD}},\ }\href {https://doi.org/10.1103/PhysRevD.102.054517}
  {\bibfield  {journal} {\bibinfo  {journal} {Phys. Rev. D}\ }\textbf {\bibinfo
  {volume} {102}},\ \bibinfo {pages} {054517} (\bibinfo {year}
  {2020}{\natexlab{b}})},\ \Eprint {https://arxiv.org/abs/1909.00485}
  {arXiv:1909.00485 [hep-lat]} \BibitemShut {NoStop}%
\bibitem [{\citenamefont {Cirigliano}\ \emph {et~al.}(2017)\citenamefont
  {Cirigliano}, \citenamefont {Davidson},\ and\ \citenamefont
  {Kuno}}]{Cirigliano:2017azj}%
  \BibitemOpen
  \bibfield  {author} {\bibinfo {author} {\bibfnamefont {V.}~\bibnamefont
  {Cirigliano}}, \bibinfo {author} {\bibfnamefont {S.}~\bibnamefont
  {Davidson}},\ and\ \bibinfo {author} {\bibfnamefont {Y.}~\bibnamefont
  {Kuno}},\ }\bibfield  {title} {\bibinfo {title} {Spin-dependent $\mu \to e$
  conversion},\ }\href {https://doi.org/10.1016/j.physletb.2017.05.053}
  {\bibfield  {journal} {\bibinfo  {journal} {Phys. Lett. B}\ }\textbf
  {\bibinfo {volume} {771}},\ \bibinfo {pages} {242} (\bibinfo {year}
  {2017})},\ \Eprint {https://arxiv.org/abs/1703.02057} {arXiv:1703.02057
  [hep-ph]} \BibitemShut {NoStop}%
\bibitem [{\citenamefont {Chang}\ \emph
  {et~al.}(2018{\natexlab{a}})\citenamefont {Chang}, \citenamefont {Davoudi},
  \citenamefont {Detmold}, \citenamefont {Gambhir}, \citenamefont {Orginos},
  \citenamefont {Savage}, \citenamefont {Shanahan}, \citenamefont {Wagman},\
  and\ \citenamefont {Winter}}]{Chang:2017eiq}%
  \BibitemOpen
  \bibfield  {author} {\bibinfo {author} {\bibfnamefont {E.}~\bibnamefont
  {Chang}}, \bibinfo {author} {\bibfnamefont {Z.}~\bibnamefont {Davoudi}},
  \bibinfo {author} {\bibfnamefont {W.}~\bibnamefont {Detmold}}, \bibinfo
  {author} {\bibfnamefont {A.~S.}\ \bibnamefont {Gambhir}}, \bibinfo {author}
  {\bibfnamefont {K.}~\bibnamefont {Orginos}}, \bibinfo {author} {\bibfnamefont
  {M.~J.}\ \bibnamefont {Savage}}, \bibinfo {author} {\bibfnamefont {P.~E.}\
  \bibnamefont {Shanahan}}, \bibinfo {author} {\bibfnamefont {M.~L.}\
  \bibnamefont {Wagman}},\ and\ \bibinfo {author} {\bibfnamefont
  {F.}~\bibnamefont {Winter}} (\bibinfo {collaboration} {NPLQCD}),\ }\bibfield
  {title} {\bibinfo {title} {Scalar, axial, and tensor interactions of light
  nuclei from lattice {QCD}},\ }\href
  {https://doi.org/10.1103/PhysRevLett.120.152002} {\bibfield  {journal}
  {\bibinfo  {journal} {Phys. Rev. Lett.}\ }\textbf {\bibinfo {volume} {120}},\
  \bibinfo {pages} {152002} (\bibinfo {year} {2018}{\natexlab{a}})},\ \Eprint
  {https://arxiv.org/abs/1712.03221} {arXiv:1712.03221 [hep-lat]} \BibitemShut
  {NoStop}%
\bibitem [{\citenamefont {Workman}\ \emph {et~al.}(2022)\citenamefont {Workman}
  \emph {et~al.}}]{ParticleDataGroup:2022ssz}%
  \BibitemOpen
  \bibfield  {author} {\bibinfo {author} {\bibfnamefont {R.~L.}\ \bibnamefont
  {Workman}} \emph {et~al.} (\bibinfo {collaboration} {Particle Data Group}),\
  }\bibfield  {title} {\bibinfo {title} {Review of particle physics},\
  }\href@noop {} {\bibfield  {journal} {\bibinfo  {journal} {PTEP}\ }\textbf
  {\bibinfo {volume} {2022}},\ \bibinfo {pages} {083C01} (\bibinfo {year}
  {2022})},\ \bibinfo {note} {and updates at \href{http://pdg.lbl.gov}{\tt
  http://pdg.lbl.gov/}}\BibitemShut {NoStop}%
\bibitem [{\citenamefont {Czarnecki}\ \emph {et~al.}(2018)\citenamefont
  {Czarnecki}, \citenamefont {Marciano},\ and\ \citenamefont
  {Sirlin}}]{Czarnecki:2018okw}%
  \BibitemOpen
  \bibfield  {author} {\bibinfo {author} {\bibfnamefont {A.}~\bibnamefont
  {Czarnecki}}, \bibinfo {author} {\bibfnamefont {W.~J.}\ \bibnamefont
  {Marciano}},\ and\ \bibinfo {author} {\bibfnamefont {A.}~\bibnamefont
  {Sirlin}},\ }\bibfield  {title} {\bibinfo {title} {Neutron lifetime and axial
  coupling connection},\ }\href
  {https://doi.org/10.1103/PhysRevLett.120.202002} {\bibfield  {journal}
  {\bibinfo  {journal} {Phys. Rev. Lett.}\ }\textbf {\bibinfo {volume} {120}},\
  \bibinfo {pages} {202002} (\bibinfo {year} {2018})},\ \Eprint
  {https://arxiv.org/abs/1802.01804} {arXiv:1802.01804 [hep-ph]} \BibitemShut
  {NoStop}%
\bibitem [{\citenamefont {Chang}\ \emph
  {et~al.}(2018{\natexlab{b}})\citenamefont {Chang} \emph
  {et~al.}}]{Chang:2018uxx}%
  \BibitemOpen
  \bibfield  {author} {\bibinfo {author} {\bibfnamefont {C.~C.}\ \bibnamefont
  {Chang}} \emph {et~al.},\ }\bibfield  {title} {\bibinfo {title} {A
  percent-level determination of the nucleon axial coupling from quantum
  chromodynamics},\ }\href {https://doi.org/10.1038/s41586-018-0161-8}
  {\bibfield  {journal} {\bibinfo  {journal} {Nature}\ }\textbf {\bibinfo
  {volume} {558}},\ \bibinfo {pages} {91} (\bibinfo {year}
  {2018}{\natexlab{b}})},\ \Eprint {https://arxiv.org/abs/1805.12130}
  {arXiv:1805.12130 [hep-lat]} \BibitemShut {NoStop}%
\bibitem [{\citenamefont {Berkowitz}\ \emph {et~al.}(2017)\citenamefont
  {Berkowitz} \emph {et~al.}}]{Berkowitz:2017gql}%
  \BibitemOpen
  \bibfield  {author} {\bibinfo {author} {\bibfnamefont {E.}~\bibnamefont
  {Berkowitz}} \emph {et~al.},\ }\bibfield  {title} {\bibinfo {title} {An
  accurate calculation of the nucleon axial charge with lattice {QCD}},\
  }\href@noop {} {\bibfield  {journal} {\bibinfo  {journal} {unpublished}\ }
  (\bibinfo {year} {2017})},\ \Eprint {https://arxiv.org/abs/1704.01114}
  {arXiv:1704.01114 [hep-lat]} \BibitemShut {NoStop}%
\bibitem [{\citenamefont {Gupta}\ \emph
  {et~al.}(2018{\natexlab{b}})\citenamefont {Gupta}, \citenamefont {Jang},
  \citenamefont {Yoon}, \citenamefont {Lin}, \citenamefont {Cirigliano},\ and\
  \citenamefont {Bhattacharya}}]{Gupta:2018qil}%
  \BibitemOpen
  \bibfield  {author} {\bibinfo {author} {\bibfnamefont {R.}~\bibnamefont
  {Gupta}}, \bibinfo {author} {\bibfnamefont {Y.-C.}\ \bibnamefont {Jang}},
  \bibinfo {author} {\bibfnamefont {B.}~\bibnamefont {Yoon}}, \bibinfo {author}
  {\bibfnamefont {H.-W.}\ \bibnamefont {Lin}}, \bibinfo {author} {\bibfnamefont
  {V.}~\bibnamefont {Cirigliano}},\ and\ \bibinfo {author} {\bibfnamefont
  {T.}~\bibnamefont {Bhattacharya}} (\bibinfo {collaboration} {PNDME}),\
  }\bibfield  {title} {\bibinfo {title} {Isovector charges of the nucleon from
  2+1+1-flavor lattice {QCD}},\ }\href
  {https://doi.org/10.1103/PhysRevD.98.034503} {\bibfield  {journal} {\bibinfo
  {journal} {Phys. Rev. D}\ }\textbf {\bibinfo {volume} {98}},\ \bibinfo
  {pages} {034503} (\bibinfo {year} {2018}{\natexlab{b}})},\ \Eprint
  {https://arxiv.org/abs/1806.09006} {arXiv:1806.09006 [hep-lat]} \BibitemShut
  {NoStop}%
\bibitem [{\citenamefont {Bhattacharya}\ \emph {et~al.}(2016)\citenamefont
  {Bhattacharya}, \citenamefont {Cirigliano}, \citenamefont {Cohen},
  \citenamefont {Gupta}, \citenamefont {Lin},\ and\ \citenamefont
  {Yoon}}]{Bhattacharya:2016zcn}%
  \BibitemOpen
  \bibfield  {author} {\bibinfo {author} {\bibfnamefont {T.}~\bibnamefont
  {Bhattacharya}}, \bibinfo {author} {\bibfnamefont {V.}~\bibnamefont
  {Cirigliano}}, \bibinfo {author} {\bibfnamefont {S.}~\bibnamefont {Cohen}},
  \bibinfo {author} {\bibfnamefont {R.}~\bibnamefont {Gupta}}, \bibinfo
  {author} {\bibfnamefont {H.-W.}\ \bibnamefont {Lin}},\ and\ \bibinfo {author}
  {\bibfnamefont {B.}~\bibnamefont {Yoon}},\ }\bibfield  {title} {\bibinfo
  {title} {Axial, scalar and tensor charges of the nucleon from 2+1+1-flavor
  lattice {QCD}},\ }\href {https://doi.org/10.1103/PhysRevD.94.054508}
  {\bibfield  {journal} {\bibinfo  {journal} {Phys. Rev. D}\ }\textbf {\bibinfo
  {volume} {94}},\ \bibinfo {pages} {054508} (\bibinfo {year} {2016})},\
  \Eprint {https://arxiv.org/abs/1606.07049} {arXiv:1606.07049 [hep-lat]}
  \BibitemShut {NoStop}%
\bibitem [{\citenamefont {Horkel}\ \emph {et~al.}(2020)\citenamefont {Horkel},
  \citenamefont {Bi}, \citenamefont {Constantinou}, \citenamefont {Draper},
  \citenamefont {Liang}, \citenamefont {Liu}, \citenamefont {Liu},\ and\
  \citenamefont {Yang}}]{Horkel:2020hpi}%
  \BibitemOpen
  \bibfield  {author} {\bibinfo {author} {\bibfnamefont {D.}~\bibnamefont
  {Horkel}}, \bibinfo {author} {\bibfnamefont {Y.}~\bibnamefont {Bi}}, \bibinfo
  {author} {\bibfnamefont {M.}~\bibnamefont {Constantinou}}, \bibinfo {author}
  {\bibfnamefont {T.}~\bibnamefont {Draper}}, \bibinfo {author} {\bibfnamefont
  {J.}~\bibnamefont {Liang}}, \bibinfo {author} {\bibfnamefont {K.-F.}\
  \bibnamefont {Liu}}, \bibinfo {author} {\bibfnamefont {Z.}~\bibnamefont
  {Liu}},\ and\ \bibinfo {author} {\bibfnamefont {Y.-B.}\ \bibnamefont {Yang}}
  (\bibinfo {collaboration} {\ensuremath{\chi}QCD}),\ }\bibfield  {title}
  {\bibinfo {title} {{Nucleon isovector tensor charge from lattice QCD using
  chiral fermions}},\ }\href {https://doi.org/10.1103/PhysRevD.101.094501}
  {\bibfield  {journal} {\bibinfo  {journal} {Phys. Rev. D}\ }\textbf {\bibinfo
  {volume} {101}},\ \bibinfo {pages} {094501} (\bibinfo {year} {2020})},\
  \Eprint {https://arxiv.org/abs/2002.06699} {arXiv:2002.06699 [hep-lat]}
  \BibitemShut {NoStop}%
\bibitem [{\citenamefont {Liu}\ \emph {et~al.}(2021)\citenamefont {Liu},
  \citenamefont {Chen}, \citenamefont {Draper}, \citenamefont {Liang},
  \citenamefont {Liu}, \citenamefont {Wang},\ and\ \citenamefont
  {Yang}}]{Liu:2021irg}%
  \BibitemOpen
  \bibfield  {author} {\bibinfo {author} {\bibfnamefont {L.}~\bibnamefont
  {Liu}}, \bibinfo {author} {\bibfnamefont {T.}~\bibnamefont {Chen}}, \bibinfo
  {author} {\bibfnamefont {T.}~\bibnamefont {Draper}}, \bibinfo {author}
  {\bibfnamefont {J.}~\bibnamefont {Liang}}, \bibinfo {author} {\bibfnamefont
  {K.-F.}\ \bibnamefont {Liu}}, \bibinfo {author} {\bibfnamefont
  {G.}~\bibnamefont {Wang}},\ and\ \bibinfo {author} {\bibfnamefont {Y.-B.}\
  \bibnamefont {Yang}} (\bibinfo {collaboration} {\ensuremath{\chi}QCD}),\
  }\bibfield  {title} {\bibinfo {title} {{Nucleon isovector scalar charge from
  overlap fermions}},\ }\href {https://doi.org/10.1103/PhysRevD.104.094503}
  {\bibfield  {journal} {\bibinfo  {journal} {Phys. Rev. D}\ }\textbf {\bibinfo
  {volume} {104}},\ \bibinfo {pages} {094503} (\bibinfo {year} {2021})},\
  \Eprint {https://arxiv.org/abs/2103.12933} {arXiv:2103.12933 [hep-lat]}
  \BibitemShut {NoStop}%
\bibitem [{\citenamefont {Bhattacharya}\ \emph {et~al.}(2012)\citenamefont
  {Bhattacharya}, \citenamefont {Cirigliano}, \citenamefont {Cohen},
  \citenamefont {Filipuzzi}, \citenamefont {Gonzalez-Alonso}, \citenamefont
  {Graesser}, \citenamefont {Gupta},\ and\ \citenamefont
  {Lin}}]{Bhattacharya:2011qm}%
  \BibitemOpen
  \bibfield  {author} {\bibinfo {author} {\bibfnamefont {T.}~\bibnamefont
  {Bhattacharya}}, \bibinfo {author} {\bibfnamefont {V.}~\bibnamefont
  {Cirigliano}}, \bibinfo {author} {\bibfnamefont {S.~D.}\ \bibnamefont
  {Cohen}}, \bibinfo {author} {\bibfnamefont {A.}~\bibnamefont {Filipuzzi}},
  \bibinfo {author} {\bibfnamefont {M.}~\bibnamefont {Gonzalez-Alonso}},
  \bibinfo {author} {\bibfnamefont {M.~L.}\ \bibnamefont {Graesser}}, \bibinfo
  {author} {\bibfnamefont {R.}~\bibnamefont {Gupta}},\ and\ \bibinfo {author}
  {\bibfnamefont {H.-W.}\ \bibnamefont {Lin}},\ }\bibfield  {title} {\bibinfo
  {title} {Probing novel scalar and tensor interactions from (ultra)cold
  neutrons to the {LHC}},\ }\href {https://doi.org/10.1103/PhysRevD.85.054512}
  {\bibfield  {journal} {\bibinfo  {journal} {Phys. Rev. D}\ }\textbf {\bibinfo
  {volume} {85}},\ \bibinfo {pages} {054512} (\bibinfo {year} {2012})},\
  \Eprint {https://arxiv.org/abs/1110.6448} {arXiv:1110.6448 [hep-ph]}
  \BibitemShut {NoStop}%
\bibitem [{\citenamefont {Gonzalez-Alonso}\ \emph {et~al.}(2019)\citenamefont
  {Gonzalez-Alonso}, \citenamefont {Naviliat-Cuncic},\ and\ \citenamefont
  {Severijns}}]{Gonzalez-Alonso:2018omy}%
  \BibitemOpen
  \bibfield  {author} {\bibinfo {author} {\bibfnamefont {M.}~\bibnamefont
  {Gonzalez-Alonso}}, \bibinfo {author} {\bibfnamefont {O.}~\bibnamefont
  {Naviliat-Cuncic}},\ and\ \bibinfo {author} {\bibfnamefont {N.}~\bibnamefont
  {Severijns}},\ }\bibfield  {title} {\bibinfo {title} {{New physics searches
  in nuclear and neutron $\beta$ decay}},\ }\href
  {https://doi.org/10.1016/j.ppnp.2018.08.002} {\bibfield  {journal} {\bibinfo
  {journal} {Prog. Part. Nucl. Phys.}\ }\textbf {\bibinfo {volume} {104}},\
  \bibinfo {pages} {165} (\bibinfo {year} {2019})},\ \Eprint
  {https://arxiv.org/abs/1803.08732} {arXiv:1803.08732 [hep-ph]} \BibitemShut
  {NoStop}%
\bibitem [{\citenamefont {Kronfeld}(2012)}]{Kronfeld:2012uk}%
  \BibitemOpen
  \bibfield  {author} {\bibinfo {author} {\bibfnamefont {A.~S.}\ \bibnamefont
  {Kronfeld}},\ }\bibfield  {title} {\bibinfo {title} {Twenty-first century
  lattice gauge theory: Results from the {QCD Lagrangian}},\ }\href
  {https://doi.org/10.1146/annurev-nucl-102711-094942} {\bibfield  {journal}
  {\bibinfo  {journal} {Ann. Rev. Nucl. Part. Sci.}\ }\textbf {\bibinfo
  {volume} {62}},\ \bibinfo {pages} {265} (\bibinfo {year} {2012})},\ \Eprint
  {https://arxiv.org/abs/1203.1204} {arXiv:1203.1204 [hep-lat]} \BibitemShut
  {NoStop}%
\bibitem [{\citenamefont {Brambilla}\ \emph {et~al.}(2020)\citenamefont
  {Brambilla}, \citenamefont {Eidelman}, \citenamefont {Hanhart}, \citenamefont
  {Nefediev}, \citenamefont {Shen}, \citenamefont {Thomas}, \citenamefont
  {Vairo},\ and\ \citenamefont {Yuan}}]{Brambilla:2019esw}%
  \BibitemOpen
  \bibfield  {author} {\bibinfo {author} {\bibfnamefont {N.}~\bibnamefont
  {Brambilla}}, \bibinfo {author} {\bibfnamefont {S.}~\bibnamefont {Eidelman}},
  \bibinfo {author} {\bibfnamefont {C.}~\bibnamefont {Hanhart}}, \bibinfo
  {author} {\bibfnamefont {A.}~\bibnamefont {Nefediev}}, \bibinfo {author}
  {\bibfnamefont {C.-P.}\ \bibnamefont {Shen}}, \bibinfo {author}
  {\bibfnamefont {C.~E.}\ \bibnamefont {Thomas}}, \bibinfo {author}
  {\bibfnamefont {A.}~\bibnamefont {Vairo}},\ and\ \bibinfo {author}
  {\bibfnamefont {C.-Z.}\ \bibnamefont {Yuan}},\ }\bibfield  {title} {\bibinfo
  {title} {{The $XYZ$ states: experimental and theoretical status and
  perspectives}},\ }\href {https://doi.org/10.1016/j.physrep.2020.05.001}
  {\bibfield  {journal} {\bibinfo  {journal} {Phys. Rept.}\ }\textbf {\bibinfo
  {volume} {873}},\ \bibinfo {pages} {1} (\bibinfo {year} {2020})},\ \Eprint
  {https://arxiv.org/abs/1907.07583} {arXiv:1907.07583 [hep-ex]} \BibitemShut
  {NoStop}%
\bibitem [{\citenamefont {Bulava}\ \emph {et~al.}(2022)\citenamefont {Bulava}
  \emph {et~al.}}]{Bulava:2022ovd}%
  \BibitemOpen
  \bibfield  {author} {\bibinfo {author} {\bibfnamefont {J.}~\bibnamefont
  {Bulava}} \emph {et~al.},\ }\bibfield  {title} {\bibinfo {title} {Hadron
  spectroscopy with lattice {QCD}},\ }in\ \href@noop {} {\emph {\bibinfo
  {booktitle} {{2022 Snowmass Summer Study}}}}\ (\bibinfo {year} {2022})\
  \Eprint {https://arxiv.org/abs/2203.03230} {arXiv:2203.03230 [hep-lat]}
  \BibitemShut {NoStop}%
\bibitem [{\citenamefont {Brambilla}\ \emph {et~al.}(2022)\citenamefont
  {Brambilla} \emph {et~al.}}]{Brambilla:2022ura}%
  \BibitemOpen
  \bibfield  {author} {\bibinfo {author} {\bibfnamefont {N.}~\bibnamefont
  {Brambilla}} \emph {et~al.},\ }\bibfield  {title} {\bibinfo {title}
  {Substructure of multiquark hadrons},\ }in\ \href@noop {} {\emph {\bibinfo
  {booktitle} {{2022 Snowmass Summer Study}}}}\ (\bibinfo {year} {2022})\
  \Eprint {https://arxiv.org/abs/2203.16583} {arXiv:2203.16583 [hep-ph]}
  \BibitemShut {NoStop}%
\bibitem [{\citenamefont {Beane}\ \emph {et~al.}(2012)\citenamefont {Beane},
  \citenamefont {Chang}, \citenamefont {Detmold}, \citenamefont {Lin},
  \citenamefont {Luu}, \citenamefont {Orginos}, \citenamefont {Parreno},
  \citenamefont {Savage}, \citenamefont {Torok},\ and\ \citenamefont
  {Walker-Loud}}]{NPLQCD:2011naw}%
  \BibitemOpen
  \bibfield  {author} {\bibinfo {author} {\bibfnamefont {S.~R.}\ \bibnamefont
  {Beane}}, \bibinfo {author} {\bibfnamefont {E.}~\bibnamefont {Chang}},
  \bibinfo {author} {\bibfnamefont {W.}~\bibnamefont {Detmold}}, \bibinfo
  {author} {\bibfnamefont {H.~W.}\ \bibnamefont {Lin}}, \bibinfo {author}
  {\bibfnamefont {T.~C.}\ \bibnamefont {Luu}}, \bibinfo {author} {\bibfnamefont
  {K.}~\bibnamefont {Orginos}}, \bibinfo {author} {\bibfnamefont
  {A.}~\bibnamefont {Parreno}}, \bibinfo {author} {\bibfnamefont {M.~J.}\
  \bibnamefont {Savage}}, \bibinfo {author} {\bibfnamefont {A.}~\bibnamefont
  {Torok}},\ and\ \bibinfo {author} {\bibfnamefont {A.}~\bibnamefont
  {Walker-Loud}} (\bibinfo {collaboration} {NPLQCD}),\ }\bibfield  {title}
  {\bibinfo {title} {The deuteron and exotic two-body bound states from lattice
  {QCD}},\ }\href {https://doi.org/10.1103/PhysRevD.85.054511} {\bibfield
  {journal} {\bibinfo  {journal} {Phys. Rev. D}\ }\textbf {\bibinfo {volume}
  {85}},\ \bibinfo {pages} {054511} (\bibinfo {year} {2012})},\ \Eprint
  {https://arxiv.org/abs/1109.2889} {arXiv:1109.2889 [hep-lat]} \BibitemShut
  {NoStop}%
\bibitem [{\citenamefont {Prelovsek}\ and\ \citenamefont
  {Leskovec}(2013)}]{Prelovsek:2013cra}%
  \BibitemOpen
  \bibfield  {author} {\bibinfo {author} {\bibfnamefont {S.}~\bibnamefont
  {Prelovsek}}\ and\ \bibinfo {author} {\bibfnamefont {L.}~\bibnamefont
  {Leskovec}},\ }\bibfield  {title} {\bibinfo {title} {{Evidence for $X(3872)$
  from $DD^*$ scattering on the lattice}},\ }\href
  {https://doi.org/10.1103/PhysRevLett.111.192001} {\bibfield  {journal}
  {\bibinfo  {journal} {Phys. Rev. Lett.}\ }\textbf {\bibinfo {volume} {111}},\
  \bibinfo {pages} {192001} (\bibinfo {year} {2013})},\ \Eprint
  {https://arxiv.org/abs/1307.5172} {arXiv:1307.5172 [hep-lat]} \BibitemShut
  {NoStop}%
\bibitem [{\citenamefont {Francis}\ \emph {et~al.}(2017)\citenamefont
  {Francis}, \citenamefont {Hudspith}, \citenamefont {Lewis},\ and\
  \citenamefont {Maltman}}]{Francis:2016hui}%
  \BibitemOpen
  \bibfield  {author} {\bibinfo {author} {\bibfnamefont {A.}~\bibnamefont
  {Francis}}, \bibinfo {author} {\bibfnamefont {R.~J.}\ \bibnamefont
  {Hudspith}}, \bibinfo {author} {\bibfnamefont {R.}~\bibnamefont {Lewis}},\
  and\ \bibinfo {author} {\bibfnamefont {K.}~\bibnamefont {Maltman}},\
  }\bibfield  {title} {\bibinfo {title} {Lattice prediction for deeply bound
  doubly heavy tetraquarks},\ }\href
  {https://doi.org/10.1103/PhysRevLett.118.142001} {\bibfield  {journal}
  {\bibinfo  {journal} {Phys. Rev. Lett.}\ }\textbf {\bibinfo {volume} {118}},\
  \bibinfo {pages} {142001} (\bibinfo {year} {2017})},\ \Eprint
  {https://arxiv.org/abs/1607.05214} {arXiv:1607.05214 [hep-lat]} \BibitemShut
  {NoStop}%
\bibitem [{\citenamefont {Francis}\ \emph
  {et~al.}(2019{\natexlab{a}})\citenamefont {Francis}, \citenamefont {Green},
  \citenamefont {Junnarkar}, \citenamefont {Miao}, \citenamefont {Rae},\ and\
  \citenamefont {Wittig}}]{Francis:2018qch}%
  \BibitemOpen
  \bibfield  {author} {\bibinfo {author} {\bibfnamefont {A.}~\bibnamefont
  {Francis}}, \bibinfo {author} {\bibfnamefont {J.~R.}\ \bibnamefont {Green}},
  \bibinfo {author} {\bibfnamefont {P.~M.}\ \bibnamefont {Junnarkar}}, \bibinfo
  {author} {\bibfnamefont {C.}~\bibnamefont {Miao}}, \bibinfo {author}
  {\bibfnamefont {T.~D.}\ \bibnamefont {Rae}},\ and\ \bibinfo {author}
  {\bibfnamefont {H.}~\bibnamefont {Wittig}},\ }\bibfield  {title} {\bibinfo
  {title} {{Lattice QCD study of the $H$ dibaryon using hexaquark and
  two-baryon interpolators}},\ }\href
  {https://doi.org/10.1103/PhysRevD.99.074505} {\bibfield  {journal} {\bibinfo
  {journal} {Phys. Rev. D}\ }\textbf {\bibinfo {volume} {99}},\ \bibinfo
  {pages} {074505} (\bibinfo {year} {2019}{\natexlab{a}})},\ \Eprint
  {https://arxiv.org/abs/1805.03966} {arXiv:1805.03966 [hep-lat]} \BibitemShut
  {NoStop}%
\bibitem [{\citenamefont {Francis}\ \emph
  {et~al.}(2019{\natexlab{b}})\citenamefont {Francis}, \citenamefont
  {Hudspith}, \citenamefont {Lewis},\ and\ \citenamefont
  {Maltman}}]{Francis:2018jyb}%
  \BibitemOpen
  \bibfield  {author} {\bibinfo {author} {\bibfnamefont {A.}~\bibnamefont
  {Francis}}, \bibinfo {author} {\bibfnamefont {R.~J.}\ \bibnamefont
  {Hudspith}}, \bibinfo {author} {\bibfnamefont {R.}~\bibnamefont {Lewis}},\
  and\ \bibinfo {author} {\bibfnamefont {K.}~\bibnamefont {Maltman}},\
  }\bibfield  {title} {\bibinfo {title} {{Evidence for charm-bottom tetraquarks
  and the mass dependence of heavy-light tetraquark states from lattice QCD}},\
  }\href {https://doi.org/10.1103/PhysRevD.99.054505} {\bibfield  {journal}
  {\bibinfo  {journal} {Phys. Rev. D}\ }\textbf {\bibinfo {volume} {99}},\
  \bibinfo {pages} {054505} (\bibinfo {year} {2019}{\natexlab{b}})},\ \Eprint
  {https://arxiv.org/abs/1810.10550} {arXiv:1810.10550 [hep-lat]} \BibitemShut
  {NoStop}%
\bibitem [{\citenamefont {Junnarkar}\ \emph {et~al.}(2019)\citenamefont
  {Junnarkar}, \citenamefont {Mathur},\ and\ \citenamefont
  {Padmanath}}]{Junnarkar:2018twb}%
  \BibitemOpen
  \bibfield  {author} {\bibinfo {author} {\bibfnamefont {P.}~\bibnamefont
  {Junnarkar}}, \bibinfo {author} {\bibfnamefont {N.}~\bibnamefont {Mathur}},\
  and\ \bibinfo {author} {\bibfnamefont {M.}~\bibnamefont {Padmanath}},\
  }\bibfield  {title} {\bibinfo {title} {Study of doubly heavy tetraquarks in
  lattice {QCD}},\ }\href {https://doi.org/10.1103/PhysRevD.99.034507}
  {\bibfield  {journal} {\bibinfo  {journal} {Phys. Rev. D}\ }\textbf {\bibinfo
  {volume} {99}},\ \bibinfo {pages} {034507} (\bibinfo {year} {2019})},\
  \Eprint {https://arxiv.org/abs/1810.12285} {arXiv:1810.12285 [hep-lat]}
  \BibitemShut {NoStop}%
\bibitem [{\citenamefont {Padmanath}\ and\ \citenamefont
  {Mathur}(2022)}]{Padmanath:2021qje}%
  \BibitemOpen
  \bibfield  {author} {\bibinfo {author} {\bibfnamefont {M.}~\bibnamefont
  {Padmanath}}\ and\ \bibinfo {author} {\bibfnamefont {N.}~\bibnamefont
  {Mathur}},\ }\bibfield  {title} {\bibinfo {title} {$\bar{b}\bar{c}\,q_1q_2$
  four-quark states from lattice {QCD}},\ }\href
  {https://doi.org/10.22323/1.396.0443} {\bibfield  {journal} {\bibinfo
  {journal} {PoS}\ }\textbf {\bibinfo {volume} {LATTICE2021}},\ \bibinfo
  {pages} {443} (\bibinfo {year} {2022})},\ \Eprint
  {https://arxiv.org/abs/2111.01147} {arXiv:2111.01147 [hep-lat]} \BibitemShut
  {NoStop}%
\bibitem [{\citenamefont {Junnarkar}\ and\ \citenamefont
  {Mathur}(2019)}]{Junnarkar:2019equ}%
  \BibitemOpen
  \bibfield  {author} {\bibinfo {author} {\bibfnamefont {P.}~\bibnamefont
  {Junnarkar}}\ and\ \bibinfo {author} {\bibfnamefont {N.}~\bibnamefont
  {Mathur}},\ }\bibfield  {title} {\bibinfo {title} {Deuteronlike heavy
  dibaryons from lattice quantum chromodynamics},\ }\href
  {https://doi.org/10.1103/PhysRevLett.123.162003} {\bibfield  {journal}
  {\bibinfo  {journal} {Phys. Rev. Lett.}\ }\textbf {\bibinfo {volume} {123}},\
  \bibinfo {pages} {162003} (\bibinfo {year} {2019})},\ \Eprint
  {https://arxiv.org/abs/1906.06054} {arXiv:1906.06054 [hep-lat]} \BibitemShut
  {NoStop}%
\bibitem [{\citenamefont {Hudspith}\ \emph {et~al.}(2020)\citenamefont
  {Hudspith}, \citenamefont {Colquhoun}, \citenamefont {Francis}, \citenamefont
  {Lewis},\ and\ \citenamefont {Maltman}}]{Hudspith:2020tdf}%
  \BibitemOpen
  \bibfield  {author} {\bibinfo {author} {\bibfnamefont {R.~J.}\ \bibnamefont
  {Hudspith}}, \bibinfo {author} {\bibfnamefont {B.}~\bibnamefont {Colquhoun}},
  \bibinfo {author} {\bibfnamefont {A.}~\bibnamefont {Francis}}, \bibinfo
  {author} {\bibfnamefont {R.}~\bibnamefont {Lewis}},\ and\ \bibinfo {author}
  {\bibfnamefont {K.}~\bibnamefont {Maltman}},\ }\bibfield  {title} {\bibinfo
  {title} {{A lattice investigation of exotic tetraquark channels}},\ }\href
  {https://doi.org/10.1103/PhysRevD.102.114506} {\bibfield  {journal} {\bibinfo
   {journal} {Phys. Rev. D}\ }\textbf {\bibinfo {volume} {102}},\ \bibinfo
  {pages} {114506} (\bibinfo {year} {2020})},\ \Eprint
  {https://arxiv.org/abs/2006.14294} {arXiv:2006.14294 [hep-lat]} \BibitemShut
  {NoStop}%
\bibitem [{\citenamefont {Prelovsek}\ \emph {et~al.}(2021)\citenamefont
  {Prelovsek}, \citenamefont {Collins}, \citenamefont {Mohler}, \citenamefont
  {Padmanath},\ and\ \citenamefont {Piemonte}}]{Prelovsek:2020eiw}%
  \BibitemOpen
  \bibfield  {author} {\bibinfo {author} {\bibfnamefont {S.}~\bibnamefont
  {Prelovsek}}, \bibinfo {author} {\bibfnamefont {S.}~\bibnamefont {Collins}},
  \bibinfo {author} {\bibfnamefont {D.}~\bibnamefont {Mohler}}, \bibinfo
  {author} {\bibfnamefont {M.}~\bibnamefont {Padmanath}},\ and\ \bibinfo
  {author} {\bibfnamefont {S.}~\bibnamefont {Piemonte}},\ }\bibfield  {title}
  {\bibinfo {title} {{Charmonium-like resonances with $J^{PC}=0^{++}$, $2^{++}$
  in coupled $D\bar{D}$, $D_s\bar{D}_s$ scattering on the lattice}},\ }\href
  {https://doi.org/10.1007/JHEP06(2021)035} {\bibfield  {journal} {\bibinfo
  {journal} {JHEP}\ }\textbf {\bibinfo {volume} {06}},\ \bibinfo {pages}
  {035}},\ \Eprint {https://arxiv.org/abs/2011.02542} {arXiv:2011.02542
  [hep-lat]} \BibitemShut {NoStop}%
\bibitem [{\citenamefont {Green}\ \emph {et~al.}(2021)\citenamefont {Green},
  \citenamefont {Hanlon}, \citenamefont {Junnarkar},\ and\ \citenamefont
  {Wittig}}]{Green:2021qol}%
  \BibitemOpen
  \bibfield  {author} {\bibinfo {author} {\bibfnamefont {J.~R.}\ \bibnamefont
  {Green}}, \bibinfo {author} {\bibfnamefont {A.~D.}\ \bibnamefont {Hanlon}},
  \bibinfo {author} {\bibfnamefont {P.~M.}\ \bibnamefont {Junnarkar}},\ and\
  \bibinfo {author} {\bibfnamefont {H.}~\bibnamefont {Wittig}},\ }\bibfield
  {title} {\bibinfo {title} {Weakly bound {$H$} dibaryon from
  {SU}(3)-flavor-symmetric {QCD}},\ }\href
  {https://doi.org/10.1103/PhysRevLett.127.242003} {\bibfield  {journal}
  {\bibinfo  {journal} {Phys. Rev. Lett.}\ }\textbf {\bibinfo {volume} {127}},\
  \bibinfo {pages} {242003} (\bibinfo {year} {2021})},\ \Eprint
  {https://arxiv.org/abs/2103.01054} {arXiv:2103.01054 [hep-lat]} \BibitemShut
  {NoStop}%
\bibitem [{\citenamefont {Francis}\ \emph {et~al.}(2022)\citenamefont
  {Francis}, \citenamefont {de~Forcrand}, \citenamefont {Lewis},\ and\
  \citenamefont {Maltman}}]{Francis:2021vrr}%
  \BibitemOpen
  \bibfield  {author} {\bibinfo {author} {\bibfnamefont {A.}~\bibnamefont
  {Francis}}, \bibinfo {author} {\bibfnamefont {P.}~\bibnamefont
  {de~Forcrand}}, \bibinfo {author} {\bibfnamefont {R.}~\bibnamefont {Lewis}},\
  and\ \bibinfo {author} {\bibfnamefont {K.}~\bibnamefont {Maltman}},\
  }\bibfield  {title} {\bibinfo {title} {{Diquark properties from full QCD
  lattice simulations}},\ }\href {https://doi.org/10.1007/JHEP05(2022)062}
  {\bibfield  {journal} {\bibinfo  {journal} {JHEP}\ }\textbf {\bibinfo
  {volume} {05}},\ \bibinfo {pages} {062}},\ \Eprint
  {https://arxiv.org/abs/2106.09080} {arXiv:2106.09080 [hep-lat]} \BibitemShut
  {NoStop}%
\bibitem [{\citenamefont {Padmanath}\ and\ \citenamefont
  {Prelovsek}(2022)}]{Padmanath:2022cvl}%
  \BibitemOpen
  \bibfield  {author} {\bibinfo {author} {\bibfnamefont {M.}~\bibnamefont
  {Padmanath}}\ and\ \bibinfo {author} {\bibfnamefont {S.}~\bibnamefont
  {Prelovsek}},\ }\bibfield  {title} {\bibinfo {title} {{Evidence for a doubly
  charm tetraquark pole in $DD^*$ scattering on the lattice}},\ }\href@noop {}
  {\  (\bibinfo {year} {2022})},\ \Eprint {https://arxiv.org/abs/2202.10110}
  {arXiv:2202.10110 [hep-lat]} \BibitemShut {NoStop}%
\bibitem [{\citenamefont {{L\"uscher}}(1986)}]{Luscher:1986pf}%
  \BibitemOpen
  \bibfield  {author} {\bibinfo {author} {\bibfnamefont {M.}~\bibnamefont
  {{L\"uscher}}},\ }\bibfield  {title} {\bibinfo {title} {Volume dependence of
  the energy spectrum in massive quantum field theories~2: Scattering states},\
  }\href {https://doi.org/10.1007/BF01211097} {\bibfield  {journal} {\bibinfo
  {journal} {Commun. Math. Phys.}\ }\textbf {\bibinfo {volume} {105}},\
  \bibinfo {pages} {153} (\bibinfo {year} {1986})}\BibitemShut {NoStop}%
\bibitem [{\citenamefont {{L\"uscher}}(1991)}]{Luscher:1990ux}%
  \BibitemOpen
  \bibfield  {author} {\bibinfo {author} {\bibfnamefont {M.}~\bibnamefont
  {{L\"uscher}}},\ }\bibfield  {title} {\bibinfo {title} {{Two-particle states
  on a torus and their relation to the scattering matrix}},\ }\href
  {https://doi.org/10.1016/0550-3213(91)90366-6} {\bibfield  {journal}
  {\bibinfo  {journal} {Nucl. Phys. B}\ }\textbf {\bibinfo {volume} {354}},\
  \bibinfo {pages} {531} (\bibinfo {year} {1991})}\BibitemShut {NoStop}%
\bibitem [{\citenamefont {Rummukainen}\ and\ \citenamefont
  {Gottlieb}(1995)}]{Rummukainen:1995vs}%
  \BibitemOpen
  \bibfield  {author} {\bibinfo {author} {\bibfnamefont {K.}~\bibnamefont
  {Rummukainen}}\ and\ \bibinfo {author} {\bibfnamefont {S.~A.}\ \bibnamefont
  {Gottlieb}},\ }\bibfield  {title} {\bibinfo {title} {{Resonance scattering
  phase shifts on a nonrest frame lattice}},\ }\href
  {https://doi.org/10.1016/0550-3213(95)00313-H} {\bibfield  {journal}
  {\bibinfo  {journal} {Nucl. Phys. B}\ }\textbf {\bibinfo {volume} {450}},\
  \bibinfo {pages} {397} (\bibinfo {year} {1995})},\ \Eprint
  {https://arxiv.org/abs/hep-lat/9503028} {arXiv:hep-lat/9503028} \BibitemShut
  {NoStop}%
\bibitem [{\citenamefont {Lellouch}\ and\ \citenamefont
  {{L\"uscher}}(2001)}]{Lellouch:2000pv}%
  \BibitemOpen
  \bibfield  {author} {\bibinfo {author} {\bibfnamefont {L.}~\bibnamefont
  {Lellouch}}\ and\ \bibinfo {author} {\bibfnamefont {M.}~\bibnamefont
  {{L\"uscher}}},\ }\bibfield  {title} {\bibinfo {title} {{Weak transition
  matrix elements from finite volume correlation functions}},\ }\href
  {https://doi.org/10.1007/s002200100410} {\bibfield  {journal} {\bibinfo
  {journal} {Commun. Math. Phys.}\ }\textbf {\bibinfo {volume} {219}},\
  \bibinfo {pages} {31} (\bibinfo {year} {2001})},\ \Eprint
  {https://arxiv.org/abs/hep-lat/0003023} {arXiv:hep-lat/0003023} \BibitemShut
  {NoStop}%
\bibitem [{\citenamefont {Kim}\ \emph {et~al.}(2005)\citenamefont {Kim},
  \citenamefont {Sachrajda},\ and\ \citenamefont {Sharpe}}]{Kim:2005gf}%
  \BibitemOpen
  \bibfield  {author} {\bibinfo {author} {\bibfnamefont {C.~h.}\ \bibnamefont
  {Kim}}, \bibinfo {author} {\bibfnamefont {C.~T.}\ \bibnamefont {Sachrajda}},\
  and\ \bibinfo {author} {\bibfnamefont {S.~R.}\ \bibnamefont {Sharpe}},\
  }\bibfield  {title} {\bibinfo {title} {{Finite-volume effects for two-hadron
  states in moving frames}},\ }\href
  {https://doi.org/10.1016/j.nuclphysb.2005.08.029} {\bibfield  {journal}
  {\bibinfo  {journal} {Nucl. Phys. B}\ }\textbf {\bibinfo {volume} {727}},\
  \bibinfo {pages} {218} (\bibinfo {year} {2005})},\ \Eprint
  {https://arxiv.org/abs/hep-lat/0507006} {arXiv:hep-lat/0507006} \BibitemShut
  {NoStop}%
\bibitem [{\citenamefont {Lage}\ \emph {et~al.}(2009)\citenamefont {Lage},
  \citenamefont {Meißner},\ and\ \citenamefont {Rusetsky}}]{Lage:2009zv}%
  \BibitemOpen
  \bibfield  {author} {\bibinfo {author} {\bibfnamefont {M.}~\bibnamefont
  {Lage}}, \bibinfo {author} {\bibfnamefont {U.-G.}\ \bibnamefont {Meißner}},\
  and\ \bibinfo {author} {\bibfnamefont {A.}~\bibnamefont {Rusetsky}},\
  }\bibfield  {title} {\bibinfo {title} {A method to measure the
  antikaon-nucleon scattering length in lattice {QCD}},\ }\href
  {https://doi.org/10.1016/j.physletb.2009.10.055} {\bibfield  {journal}
  {\bibinfo  {journal} {Phys. Lett. B}\ }\textbf {\bibinfo {volume} {681}},\
  \bibinfo {pages} {439} (\bibinfo {year} {2009})},\ \Eprint
  {https://arxiv.org/abs/0905.0069} {arXiv:0905.0069 [hep-lat]} \BibitemShut
  {NoStop}%
\bibitem [{\citenamefont {Hansen}\ and\ \citenamefont
  {Sharpe}(2012)}]{Hansen:2012tf}%
  \BibitemOpen
  \bibfield  {author} {\bibinfo {author} {\bibfnamefont {M.~T.}\ \bibnamefont
  {Hansen}}\ and\ \bibinfo {author} {\bibfnamefont {S.~R.}\ \bibnamefont
  {Sharpe}},\ }\bibfield  {title} {\bibinfo {title} {{Multiple-channel
  generalization of Lellouch-L\"uscher formula}},\ }\href
  {https://doi.org/10.1103/PhysRevD.86.016007} {\bibfield  {journal} {\bibinfo
  {journal} {Phys. Rev. D}\ }\textbf {\bibinfo {volume} {86}},\ \bibinfo
  {pages} {016007} (\bibinfo {year} {2012})},\ \Eprint
  {https://arxiv.org/abs/1204.0826} {arXiv:1204.0826 [hep-lat]} \BibitemShut
  {NoStop}%
\bibitem [{\citenamefont {G\"ockeler}\ \emph {et~al.}(2012)\citenamefont
  {G\"ockeler}, \citenamefont {Horsley}, \citenamefont {Lage}, \citenamefont
  {Meißner}, \citenamefont {Rakow}, \citenamefont {Rusetsky}, \citenamefont
  {Schierholz},\ and\ \citenamefont {Zanotti}}]{Gockeler:2012yj}%
  \BibitemOpen
  \bibfield  {author} {\bibinfo {author} {\bibfnamefont {M.}~\bibnamefont
  {G\"ockeler}}, \bibinfo {author} {\bibfnamefont {R.}~\bibnamefont {Horsley}},
  \bibinfo {author} {\bibfnamefont {M.}~\bibnamefont {Lage}}, \bibinfo {author}
  {\bibfnamefont {U.~G.}\ \bibnamefont {Meißner}}, \bibinfo {author}
  {\bibfnamefont {P.~E.~L.}\ \bibnamefont {Rakow}}, \bibinfo {author}
  {\bibfnamefont {A.}~\bibnamefont {Rusetsky}}, \bibinfo {author}
  {\bibfnamefont {G.}~\bibnamefont {Schierholz}},\ and\ \bibinfo {author}
  {\bibfnamefont {J.~M.}\ \bibnamefont {Zanotti}},\ }\bibfield  {title}
  {\bibinfo {title} {{Scattering phases for meson and baryon resonances on
  general moving-frame lattices}},\ }\href
  {https://doi.org/10.1103/PhysRevD.86.094513} {\bibfield  {journal} {\bibinfo
  {journal} {Phys. Rev. D}\ }\textbf {\bibinfo {volume} {86}},\ \bibinfo
  {pages} {094513} (\bibinfo {year} {2012})},\ \Eprint
  {https://arxiv.org/abs/1206.4141} {arXiv:1206.4141 [hep-lat]} \BibitemShut
  {NoStop}%
\bibitem [{\citenamefont {{Brice\~no}}(2014)}]{Briceno:2014oea}%
  \BibitemOpen
  \bibfield  {author} {\bibinfo {author} {\bibfnamefont {R.~A.}\ \bibnamefont
  {{Brice\~no}}},\ }\bibfield  {title} {\bibinfo {title} {{Two-particle
  multichannel systems in a finite volume with arbitrary spin}},\ }\href
  {https://doi.org/10.1103/PhysRevD.89.074507} {\bibfield  {journal} {\bibinfo
  {journal} {Phys. Rev. D}\ }\textbf {\bibinfo {volume} {89}},\ \bibinfo
  {pages} {074507} (\bibinfo {year} {2014})},\ \Eprint
  {https://arxiv.org/abs/1401.3312} {arXiv:1401.3312 [hep-lat]} \BibitemShut
  {NoStop}%
\bibitem [{\citenamefont {{Brice\~no}}\ and\ \citenamefont
  {Hansen}(2015)}]{Briceno:2015csa}%
  \BibitemOpen
  \bibfield  {author} {\bibinfo {author} {\bibfnamefont {R.~A.}\ \bibnamefont
  {{Brice\~no}}}\ and\ \bibinfo {author} {\bibfnamefont {M.~T.}\ \bibnamefont
  {Hansen}},\ }\bibfield  {title} {\bibinfo {title} {{Multichannel 0 $\to$ 2
  and 1 $\to$ 2 transition amplitudes for arbitrary spin particles in a finite
  volume}},\ }\href {https://doi.org/10.1103/PhysRevD.92.074509} {\bibfield
  {journal} {\bibinfo  {journal} {Phys. Rev. D}\ }\textbf {\bibinfo {volume}
  {92}},\ \bibinfo {pages} {074509} (\bibinfo {year} {2015})},\ \Eprint
  {https://arxiv.org/abs/1502.04314} {arXiv:1502.04314 [hep-lat]} \BibitemShut
  {NoStop}%
\bibitem [{\citenamefont {{Brice\~no}}\ \emph {et~al.}(2017)\citenamefont
  {{Brice\~no}}, \citenamefont {Hansen},\ and\ \citenamefont
  {Sharpe}}]{Briceno:2017tce}%
  \BibitemOpen
  \bibfield  {author} {\bibinfo {author} {\bibfnamefont {R.~A.}\ \bibnamefont
  {{Brice\~no}}}, \bibinfo {author} {\bibfnamefont {M.~T.}\ \bibnamefont
  {Hansen}},\ and\ \bibinfo {author} {\bibfnamefont {S.~R.}\ \bibnamefont
  {Sharpe}},\ }\bibfield  {title} {\bibinfo {title} {{Relating the
  finite-volume spectrum and the two-and-three-particle $S$ matrix for
  relativistic systems of identical scalar particles}},\ }\href
  {https://doi.org/10.1103/PhysRevD.95.074510} {\bibfield  {journal} {\bibinfo
  {journal} {Phys. Rev. D}\ }\textbf {\bibinfo {volume} {95}},\ \bibinfo
  {pages} {074510} (\bibinfo {year} {2017})},\ \Eprint
  {https://arxiv.org/abs/1701.07465} {arXiv:1701.07465 [hep-lat]} \BibitemShut
  {NoStop}%
\bibitem [{\citenamefont {{Brice\~no}}\ \emph {et~al.}(2018)\citenamefont
  {{Brice\~no}}, \citenamefont {Dudek},\ and\ \citenamefont
  {Young}}]{Briceno:2017max}%
  \BibitemOpen
  \bibfield  {author} {\bibinfo {author} {\bibfnamefont {R.~A.}\ \bibnamefont
  {{Brice\~no}}}, \bibinfo {author} {\bibfnamefont {J.~J.}\ \bibnamefont
  {Dudek}},\ and\ \bibinfo {author} {\bibfnamefont {R.~D.}\ \bibnamefont
  {Young}},\ }\bibfield  {title} {\bibinfo {title} {{Scattering processes and
  resonances from lattice QCD}},\ }\href
  {https://doi.org/10.1103/RevModPhys.90.025001} {\bibfield  {journal}
  {\bibinfo  {journal} {Rev. Mod. Phys.}\ }\textbf {\bibinfo {volume} {90}},\
  \bibinfo {pages} {025001} (\bibinfo {year} {2018})},\ \Eprint
  {https://arxiv.org/abs/1706.06223} {arXiv:1706.06223 [hep-lat]} \BibitemShut
  {NoStop}%
\bibitem [{\citenamefont {Brice\~no}\ \emph {et~al.}(2015)\citenamefont
  {Brice\~no}, \citenamefont {Dudek}, \citenamefont {Edwards}, \citenamefont
  {Shultz}, \citenamefont {Thomas},\ and\ \citenamefont
  {Wilson}}]{Briceno:2015dca}%
  \BibitemOpen
  \bibfield  {author} {\bibinfo {author} {\bibfnamefont {R.~A.}\ \bibnamefont
  {Brice\~no}}, \bibinfo {author} {\bibfnamefont {J.~J.}\ \bibnamefont
  {Dudek}}, \bibinfo {author} {\bibfnamefont {R.~G.}\ \bibnamefont {Edwards}},
  \bibinfo {author} {\bibfnamefont {C.~J.}\ \bibnamefont {Shultz}}, \bibinfo
  {author} {\bibfnamefont {C.~E.}\ \bibnamefont {Thomas}},\ and\ \bibinfo
  {author} {\bibfnamefont {D.~J.}\ \bibnamefont {Wilson}} (\bibinfo
  {collaboration} {Hadron Spectrum}),\ }\bibfield  {title} {\bibinfo {title}
  {{The resonant $\pi^+\gamma\to\pi^+\pi^0$ amplitude from quantum
  chromodynamics}},\ }\href {https://doi.org/10.1103/PhysRevLett.115.242001}
  {\bibfield  {journal} {\bibinfo  {journal} {Phys. Rev. Lett.}\ }\textbf
  {\bibinfo {volume} {115}},\ \bibinfo {pages} {242001} (\bibinfo {year}
  {2015})},\ \Eprint {https://arxiv.org/abs/1507.06622} {arXiv:1507.06622
  [hep-ph]} \BibitemShut {NoStop}%
\bibitem [{\citenamefont {Alexandrou}\ \emph
  {et~al.}(2018{\natexlab{a}})\citenamefont {Alexandrou}, \citenamefont
  {Leskovec}, \citenamefont {Meinel}, \citenamefont {Negele}, \citenamefont
  {Paul}, \citenamefont {Petschlies}, \citenamefont {Pochinsky}, \citenamefont
  {Rendon},\ and\ \citenamefont {Syritsyn}}]{Alexandrou:2018jbt}%
  \BibitemOpen
  \bibfield  {author} {\bibinfo {author} {\bibfnamefont {C.}~\bibnamefont
  {Alexandrou}}, \bibinfo {author} {\bibfnamefont {L.}~\bibnamefont
  {Leskovec}}, \bibinfo {author} {\bibfnamefont {S.}~\bibnamefont {Meinel}},
  \bibinfo {author} {\bibfnamefont {J.}~\bibnamefont {Negele}}, \bibinfo
  {author} {\bibfnamefont {S.}~\bibnamefont {Paul}}, \bibinfo {author}
  {\bibfnamefont {M.}~\bibnamefont {Petschlies}}, \bibinfo {author}
  {\bibfnamefont {A.}~\bibnamefont {Pochinsky}}, \bibinfo {author}
  {\bibfnamefont {G.}~\bibnamefont {Rendon}},\ and\ \bibinfo {author}
  {\bibfnamefont {S.}~\bibnamefont {Syritsyn}},\ }\bibfield  {title} {\bibinfo
  {title} {{$\pi\gamma \to \pi\pi$ transition and the $\rho$ radiative decay
  width from lattice QCD}},\ }\href
  {https://doi.org/10.1103/PhysRevD.98.074502} {\bibfield  {journal} {\bibinfo
  {journal} {Phys. Rev. D}\ }\textbf {\bibinfo {volume} {98}},\ \bibinfo
  {pages} {074502} (\bibinfo {year} {2018}{\natexlab{a}})},\ \Eprint
  {https://arxiv.org/abs/1807.08357} {arXiv:1807.08357 [hep-lat]} \BibitemShut
  {NoStop}%
\bibitem [{\citenamefont {Moir}\ \emph {et~al.}(2016)\citenamefont {Moir},
  \citenamefont {Peardon}, \citenamefont {Ryan}, \citenamefont {Thomas},\ and\
  \citenamefont {Wilson}}]{Moir:2016srx}%
  \BibitemOpen
  \bibfield  {author} {\bibinfo {author} {\bibfnamefont {G.}~\bibnamefont
  {Moir}}, \bibinfo {author} {\bibfnamefont {M.}~\bibnamefont {Peardon}},
  \bibinfo {author} {\bibfnamefont {S.~M.}\ \bibnamefont {Ryan}}, \bibinfo
  {author} {\bibfnamefont {C.~E.}\ \bibnamefont {Thomas}},\ and\ \bibinfo
  {author} {\bibfnamefont {D.~J.}\ \bibnamefont {Wilson}},\ }\bibfield  {title}
  {\bibinfo {title} {Coupled-channel {$D\pi$}, {$D\eta$}, and {$D_{s}\bar{K}$}
  scattering from lattice {QCD}},\ }\href
  {https://doi.org/10.1007/JHEP10(2016)011} {\bibfield  {journal} {\bibinfo
  {journal} {JHEP}\ }\textbf {\bibinfo {volume} {10}},\ \bibinfo {pages}
  {011}},\ \Eprint {https://arxiv.org/abs/1607.07093} {arXiv:1607.07093
  [hep-lat]} \BibitemShut {NoStop}%
\bibitem [{\citenamefont {Albaladejo}\ \emph {et~al.}(2017)\citenamefont
  {Albaladejo}, \citenamefont {Fernandez-Soler}, \citenamefont {Guo},\ and\
  \citenamefont {Nieves}}]{Albaladejo:2016lbb}%
  \BibitemOpen
  \bibfield  {author} {\bibinfo {author} {\bibfnamefont {M.}~\bibnamefont
  {Albaladejo}}, \bibinfo {author} {\bibfnamefont {P.}~\bibnamefont
  {Fernandez-Soler}}, \bibinfo {author} {\bibfnamefont {F.-K.}\ \bibnamefont
  {Guo}},\ and\ \bibinfo {author} {\bibfnamefont {J.}~\bibnamefont {Nieves}},\
  }\bibfield  {title} {\bibinfo {title} {{Two-pole structure of the
  $D^\ast_0(2400)$}},\ }\href {https://doi.org/10.1016/j.physletb.2017.02.036}
  {\bibfield  {journal} {\bibinfo  {journal} {Phys. Lett. B}\ }\textbf
  {\bibinfo {volume} {767}},\ \bibinfo {pages} {465} (\bibinfo {year}
  {2017})},\ \Eprint {https://arxiv.org/abs/1610.06727} {arXiv:1610.06727
  [hep-ph]} \BibitemShut {NoStop}%
\bibitem [{\citenamefont {Lang}\ and\ \citenamefont
  {Wilson}(2022)}]{Lang:2022elg}%
  \BibitemOpen
  \bibfield  {author} {\bibinfo {author} {\bibfnamefont {N.}~\bibnamefont
  {Lang}}\ and\ \bibinfo {author} {\bibfnamefont {D.~J.}\ \bibnamefont
  {Wilson}} (\bibinfo {collaboration} {Hadron Spectrum}),\ }\bibfield  {title}
  {\bibinfo {title} {{Axial-vector $D_1$ hadrons in $D^\ast\pi$ scattering from
  QCD}},\ }\href@noop {} {\  (\bibinfo {year} {2022})},\ \Eprint
  {https://arxiv.org/abs/2205.05026} {arXiv:2205.05026 [hep-ph]} \BibitemShut
  {NoStop}%
\bibitem [{\citenamefont {Dowdall}\ \emph
  {et~al.}(2012{\natexlab{a}})\citenamefont {Dowdall} \emph
  {et~al.}}]{HPQCD:2011qwj}%
  \BibitemOpen
  \bibfield  {author} {\bibinfo {author} {\bibfnamefont {R.~J.}\ \bibnamefont
  {Dowdall}} \emph {et~al.} (\bibinfo {collaboration} {HPQCD}),\ }\bibfield
  {title} {\bibinfo {title} {{The Upsilon spectrum and the determination of the
  lattice spacing from lattice QCD including charm quarks in the sea}},\ }\href
  {https://doi.org/10.1103/PhysRevD.85.054509} {\bibfield  {journal} {\bibinfo
  {journal} {Phys. Rev. D}\ }\textbf {\bibinfo {volume} {85}},\ \bibinfo
  {pages} {054509} (\bibinfo {year} {2012}{\natexlab{a}})},\ \Eprint
  {https://arxiv.org/abs/1110.6887} {arXiv:1110.6887 [hep-lat]} \BibitemShut
  {NoStop}%
\bibitem [{\citenamefont {DeTar}\ \emph {et~al.}(2019)\citenamefont {DeTar},
  \citenamefont {Kronfeld}, \citenamefont {Lee}, \citenamefont {Mohler},\ and\
  \citenamefont {Simone}}]{DeTar:2018uko}%
  \BibitemOpen
  \bibfield  {author} {\bibinfo {author} {\bibfnamefont {C.}~\bibnamefont
  {DeTar}}, \bibinfo {author} {\bibfnamefont {A.~S.}\ \bibnamefont {Kronfeld}},
  \bibinfo {author} {\bibfnamefont {S.-h.}\ \bibnamefont {Lee}}, \bibinfo
  {author} {\bibfnamefont {D.}~\bibnamefont {Mohler}},\ and\ \bibinfo {author}
  {\bibfnamefont {J.~N.}\ \bibnamefont {Simone}} (\bibinfo {collaboration}
  {Fermilab Lattice, MILC}),\ }\bibfield  {title} {\bibinfo {title}
  {{Splittings of low-lying charmonium masses at the physical point}},\ }\href
  {https://doi.org/10.1103/PhysRevD.99.034509} {\bibfield  {journal} {\bibinfo
  {journal} {Phys. Rev. D}\ }\textbf {\bibinfo {volume} {99}},\ \bibinfo
  {pages} {034509} (\bibinfo {year} {2019})},\ \Eprint
  {https://arxiv.org/abs/1810.09983} {arXiv:1810.09983 [hep-lat]} \BibitemShut
  {NoStop}%
\bibitem [{\citenamefont {Liu}\ \emph {et~al.}(2012)\citenamefont {Liu},
  \citenamefont {Moir}, \citenamefont {Peardon}, \citenamefont {Ryan},
  \citenamefont {Thomas}, \citenamefont {Vilaseca}, \citenamefont {Dudek},
  \citenamefont {Edwards}, \citenamefont {Joo},\ and\ \citenamefont
  {Richards}}]{HadronSpectrum:2012gic}%
  \BibitemOpen
  \bibfield  {author} {\bibinfo {author} {\bibfnamefont {L.}~\bibnamefont
  {Liu}}, \bibinfo {author} {\bibfnamefont {G.}~\bibnamefont {Moir}}, \bibinfo
  {author} {\bibfnamefont {M.}~\bibnamefont {Peardon}}, \bibinfo {author}
  {\bibfnamefont {S.~M.}\ \bibnamefont {Ryan}}, \bibinfo {author}
  {\bibfnamefont {C.~E.}\ \bibnamefont {Thomas}}, \bibinfo {author}
  {\bibfnamefont {P.}~\bibnamefont {Vilaseca}}, \bibinfo {author}
  {\bibfnamefont {J.~J.}\ \bibnamefont {Dudek}}, \bibinfo {author}
  {\bibfnamefont {R.~G.}\ \bibnamefont {Edwards}}, \bibinfo {author}
  {\bibfnamefont {B.}~\bibnamefont {Joo}},\ and\ \bibinfo {author}
  {\bibfnamefont {D.~G.}\ \bibnamefont {Richards}} (\bibinfo {collaboration}
  {Hadron Spectrum}),\ }\bibfield  {title} {\bibinfo {title} {{Excited and
  exotic charmonium spectroscopy from lattice QCD}},\ }\href
  {https://doi.org/10.1007/JHEP07(2012)126} {\bibfield  {journal} {\bibinfo
  {journal} {JHEP}\ }\textbf {\bibinfo {volume} {07}},\ \bibinfo {pages}
  {126}},\ \Eprint {https://arxiv.org/abs/1204.5425} {arXiv:1204.5425 [hep-ph]}
  \BibitemShut {NoStop}%
\bibitem [{\citenamefont {Ryan}\ and\ \citenamefont
  {Wilson}(2021)}]{Ryan:2020iog}%
  \BibitemOpen
  \bibfield  {author} {\bibinfo {author} {\bibfnamefont {S.~M.}\ \bibnamefont
  {Ryan}}\ and\ \bibinfo {author} {\bibfnamefont {D.~J.}\ \bibnamefont
  {Wilson}} (\bibinfo {collaboration} {Hadron Spectrum}),\ }\bibfield  {title}
  {\bibinfo {title} {{Excited and exotic bottomonium spectroscopy from lattice
  QCD}},\ }\href {https://doi.org/10.1007/JHEP02(2021)214} {\bibfield
  {journal} {\bibinfo  {journal} {JHEP}\ }\textbf {\bibinfo {volume} {02}},\
  \bibinfo {pages} {214}},\ \Eprint {https://arxiv.org/abs/2008.02656}
  {arXiv:2008.02656 [hep-lat]} \BibitemShut {NoStop}%
\bibitem [{\citenamefont {Kronfeld}(2004)}]{Kronfeld:2003sd}%
  \BibitemOpen
  \bibfield  {author} {\bibinfo {author} {\bibfnamefont {A.~S.}\ \bibnamefont
  {Kronfeld}},\ }\bibfield  {title} {\bibinfo {title} {Heavy quarks and lattice
  {QCD}},\ }\href {https://doi.org/10.1016/S0920-5632(03)02506-4} {\bibfield
  {journal} {\bibinfo  {journal} {Nucl. Phys. B Proc. Suppl.}\ }\textbf
  {\bibinfo {volume} {129}},\ \bibinfo {pages} {46} (\bibinfo {year} {2004})},\
  \Eprint {https://arxiv.org/abs/hep-lat/0310063} {arXiv:hep-lat/0310063}
  \BibitemShut {NoStop}%
\bibitem [{\citenamefont {Caswell}\ and\ \citenamefont
  {Lepage}(1986)}]{Caswell:1985ui}%
  \BibitemOpen
  \bibfield  {author} {\bibinfo {author} {\bibfnamefont {W.~E.}\ \bibnamefont
  {Caswell}}\ and\ \bibinfo {author} {\bibfnamefont {G.~P.}\ \bibnamefont
  {Lepage}},\ }\bibfield  {title} {\bibinfo {title} {{Effective Lagrangians for
  bound state problems in QED, QCD, and other field theories}},\ }\href
  {https://doi.org/10.1016/0370-2693(86)91297-9} {\bibfield  {journal}
  {\bibinfo  {journal} {Phys. Lett. B}\ }\textbf {\bibinfo {volume} {167}},\
  \bibinfo {pages} {437} (\bibinfo {year} {1986})}\BibitemShut {NoStop}%
\bibitem [{\citenamefont {Lepage}\ and\ \citenamefont
  {Thacker}(1988)}]{Lepage:1987gg}%
  \BibitemOpen
  \bibfield  {author} {\bibinfo {author} {\bibfnamefont {G.~P.}\ \bibnamefont
  {Lepage}}\ and\ \bibinfo {author} {\bibfnamefont {B.~A.}\ \bibnamefont
  {Thacker}},\ }\bibfield  {title} {\bibinfo {title} {Effective lagrangians for
  simulating heavy quark systems},\ }\href
  {https://doi.org/10.1016/0920-5632(88)90102-8} {\bibfield  {journal}
  {\bibinfo  {journal} {Nucl. Phys. B Proc. Suppl.}\ }\textbf {\bibinfo
  {volume} {4}},\ \bibinfo {pages} {199} (\bibinfo {year} {1988})}\BibitemShut
  {NoStop}%
\bibitem [{\citenamefont {Thacker}\ and\ \citenamefont
  {Lepage}(1991)}]{Thacker:1990bm}%
  \BibitemOpen
  \bibfield  {author} {\bibinfo {author} {\bibfnamefont {B.~A.}\ \bibnamefont
  {Thacker}}\ and\ \bibinfo {author} {\bibfnamefont {G.~P.}\ \bibnamefont
  {Lepage}},\ }\bibfield  {title} {\bibinfo {title} {Heavy quark bound states
  in lattice {QCD}},\ }\href {https://doi.org/10.1103/PhysRevD.43.196}
  {\bibfield  {journal} {\bibinfo  {journal} {Phys. Rev. D}\ }\textbf {\bibinfo
  {volume} {43}},\ \bibinfo {pages} {196} (\bibinfo {year} {1991})}\BibitemShut
  {NoStop}%
\bibitem [{\citenamefont {Lepage}\ \emph {et~al.}(1992)\citenamefont {Lepage},
  \citenamefont {Magnea}, \citenamefont {Nakhleh}, \citenamefont {Magnea},\
  and\ \citenamefont {Hornbostel}}]{Lepage:1992tx}%
  \BibitemOpen
  \bibfield  {author} {\bibinfo {author} {\bibfnamefont {G.~P.}\ \bibnamefont
  {Lepage}}, \bibinfo {author} {\bibfnamefont {L.}~\bibnamefont {Magnea}},
  \bibinfo {author} {\bibfnamefont {C.}~\bibnamefont {Nakhleh}}, \bibinfo
  {author} {\bibfnamefont {U.}~\bibnamefont {Magnea}},\ and\ \bibinfo {author}
  {\bibfnamefont {K.}~\bibnamefont {Hornbostel}},\ }\bibfield  {title}
  {\bibinfo {title} {Improved nonrelativistic {QCD} for heavy quark physics},\
  }\href {https://doi.org/10.1103/PhysRevD.46.4052} {\bibfield  {journal}
  {\bibinfo  {journal} {Phys. Rev. D}\ }\textbf {\bibinfo {volume} {46}},\
  \bibinfo {pages} {4052} (\bibinfo {year} {1992})},\ \Eprint
  {https://arxiv.org/abs/hep-lat/9205007} {arXiv:hep-lat/9205007} \BibitemShut
  {NoStop}%
\bibitem [{\citenamefont {Eichten}(1988)}]{Eichten:1987xu}%
  \BibitemOpen
  \bibfield  {author} {\bibinfo {author} {\bibfnamefont {E.}~\bibnamefont
  {Eichten}},\ }\bibfield  {title} {\bibinfo {title} {Heavy quarks on the
  lattice},\ }\href {https://doi.org/10.1016/0920-5632(88)90097-7} {\bibfield
  {journal} {\bibinfo  {journal} {Nucl. Phys. B Proc. Suppl.}\ }\textbf
  {\bibinfo {volume} {4}},\ \bibinfo {pages} {170} (\bibinfo {year}
  {1988})}\BibitemShut {NoStop}%
\bibitem [{\citenamefont {Eichten}\ and\ \citenamefont
  {Hill}(1990{\natexlab{a}})}]{Eichten:1989zv}%
  \BibitemOpen
  \bibfield  {author} {\bibinfo {author} {\bibfnamefont {E.}~\bibnamefont
  {Eichten}}\ and\ \bibinfo {author} {\bibfnamefont {B.~R.}\ \bibnamefont
  {Hill}},\ }\bibfield  {title} {\bibinfo {title} {An effective field theory
  for the calculation of matrix elements involving heavy quarks},\ }\href
  {https://doi.org/10.1016/0370-2693(90)92049-O} {\bibfield  {journal}
  {\bibinfo  {journal} {Phys. Lett. B}\ }\textbf {\bibinfo {volume} {234}},\
  \bibinfo {pages} {511} (\bibinfo {year} {1990}{\natexlab{a}})}\BibitemShut
  {NoStop}%
\bibitem [{\citenamefont {Eichten}\ and\ \citenamefont
  {Hill}(1990{\natexlab{b}})}]{Eichten:1990vp}%
  \BibitemOpen
  \bibfield  {author} {\bibinfo {author} {\bibfnamefont {E.}~\bibnamefont
  {Eichten}}\ and\ \bibinfo {author} {\bibfnamefont {B.~R.}\ \bibnamefont
  {Hill}},\ }\bibfield  {title} {\bibinfo {title} {Static effective field
  theory: $1/m$ corrections},\ }\href
  {https://doi.org/10.1016/0370-2693(90)91408-4} {\bibfield  {journal}
  {\bibinfo  {journal} {Phys. Lett. B}\ }\textbf {\bibinfo {volume} {243}},\
  \bibinfo {pages} {427} (\bibinfo {year} {1990}{\natexlab{b}})}\BibitemShut
  {NoStop}%
\bibitem [{\citenamefont {Allison}\ \emph {et~al.}(2005)\citenamefont
  {Allison}, \citenamefont {Davies}, \citenamefont {Gray}, \citenamefont
  {Kronfeld}, \citenamefont {Mackenzie},\ and\ \citenamefont
  {Simone}}]{Allison:2004be}%
  \BibitemOpen
  \bibfield  {author} {\bibinfo {author} {\bibfnamefont {I.~F.}\ \bibnamefont
  {Allison}}, \bibinfo {author} {\bibfnamefont {C.~T.~H.}\ \bibnamefont
  {Davies}}, \bibinfo {author} {\bibfnamefont {A.}~\bibnamefont {Gray}},
  \bibinfo {author} {\bibfnamefont {A.~S.}\ \bibnamefont {Kronfeld}}, \bibinfo
  {author} {\bibfnamefont {P.~B.}\ \bibnamefont {Mackenzie}},\ and\ \bibinfo
  {author} {\bibfnamefont {J.~N.}\ \bibnamefont {Simone}} (\bibinfo
  {collaboration} {HPQCD, Fermilab Lattice}),\ }\bibfield  {title} {\bibinfo
  {title} {{Mass of the $B_c$ meson in three-flavor lattice QCD}},\ }\href
  {https://doi.org/10.1103/PhysRevLett.94.172001} {\bibfield  {journal}
  {\bibinfo  {journal} {Phys. Rev. Lett.}\ }\textbf {\bibinfo {volume} {94}},\
  \bibinfo {pages} {172001} (\bibinfo {year} {2005})},\ \Eprint
  {https://arxiv.org/abs/hep-lat/0411027} {arXiv:hep-lat/0411027} \BibitemShut
  {NoStop}%
\bibitem [{\citenamefont {Gregory}\ \emph {et~al.}(2011)\citenamefont {Gregory}
  \emph {et~al.}}]{Gregory:2010gm}%
  \BibitemOpen
  \bibfield  {author} {\bibinfo {author} {\bibfnamefont {E.~B.}\ \bibnamefont
  {Gregory}} \emph {et~al.} (\bibinfo {collaboration} {HPQCD}),\ }\bibfield
  {title} {\bibinfo {title} {{Precise $B, B_s$ and $B_c$ meson spectroscopy
  from full lattice QCD}},\ }\href {https://doi.org/10.1103/PhysRevD.83.014506}
  {\bibfield  {journal} {\bibinfo  {journal} {Phys. Rev. D}\ }\textbf {\bibinfo
  {volume} {83}},\ \bibinfo {pages} {014506} (\bibinfo {year} {2011})},\
  \Eprint {https://arxiv.org/abs/1010.3848} {arXiv:1010.3848 [hep-lat]}
  \BibitemShut {NoStop}%
\bibitem [{\citenamefont {Abulencia}\ \emph {et~al.}(2006)\citenamefont
  {Abulencia} \emph {et~al.}}]{CDF:2005yjh}%
  \BibitemOpen
  \bibfield  {author} {\bibinfo {author} {\bibfnamefont {A.}~\bibnamefont
  {Abulencia}} \emph {et~al.} (\bibinfo {collaboration} {CDF}),\ }\bibfield
  {title} {\bibinfo {title} {{Evidence for the exclusive decay $B_c^\pm \to
  J/\psi \pi^\pm$ and measurement of the mass of the $B_c$ meson}},\ }\href
  {https://doi.org/10.1103/PhysRevLett.96.082002} {\bibfield  {journal}
  {\bibinfo  {journal} {Phys. Rev. Lett.}\ }\textbf {\bibinfo {volume} {96}},\
  \bibinfo {pages} {082002} (\bibinfo {year} {2006})},\ \Eprint
  {https://arxiv.org/abs/hep-ex/0505076} {arXiv:hep-ex/0505076} \BibitemShut
  {NoStop}%
\bibitem [{\citenamefont {Gregory}\ \emph {et~al.}(2010)\citenamefont
  {Gregory}, \citenamefont {Davies}, \citenamefont {Follana}, \citenamefont
  {Gámiz}, \citenamefont {Kendall}, \citenamefont {Lepage}, \citenamefont
  {Na}, \citenamefont {Shigemitsu},\ and\ \citenamefont
  {Wong}}]{Gregory:2009hq}%
  \BibitemOpen
  \bibfield  {author} {\bibinfo {author} {\bibfnamefont {E.~B.}\ \bibnamefont
  {Gregory}}, \bibinfo {author} {\bibfnamefont {C.~T.~H.}\ \bibnamefont
  {Davies}}, \bibinfo {author} {\bibfnamefont {E.}~\bibnamefont {Follana}},
  \bibinfo {author} {\bibfnamefont {E.}~\bibnamefont {Gámiz}}, \bibinfo
  {author} {\bibfnamefont {I.~D.}\ \bibnamefont {Kendall}}, \bibinfo {author}
  {\bibfnamefont {G.~P.}\ \bibnamefont {Lepage}}, \bibinfo {author}
  {\bibfnamefont {H.}~\bibnamefont {Na}}, \bibinfo {author} {\bibfnamefont
  {J.}~\bibnamefont {Shigemitsu}},\ and\ \bibinfo {author} {\bibfnamefont
  {K.~Y.}\ \bibnamefont {Wong}} (\bibinfo {collaboration} {HPQCD}),\ }\bibfield
   {title} {\bibinfo {title} {{Prediction of the $B^*_c$ mass in full lattice
  QCD}},\ }\href {https://doi.org/10.1103/PhysRevLett.104.022001} {\bibfield
  {journal} {\bibinfo  {journal} {Phys. Rev. Lett.}\ }\textbf {\bibinfo
  {volume} {104}},\ \bibinfo {pages} {022001} (\bibinfo {year} {2010})},\
  \Eprint {https://arxiv.org/abs/0909.4462} {arXiv:0909.4462 [hep-lat]}
  \BibitemShut {NoStop}%
\bibitem [{\citenamefont {Dowdall}\ \emph
  {et~al.}(2012{\natexlab{b}})\citenamefont {Dowdall}, \citenamefont {Davies},
  \citenamefont {Hammant},\ and\ \citenamefont {Horgan}}]{Dowdall:2012ab}%
  \BibitemOpen
  \bibfield  {author} {\bibinfo {author} {\bibfnamefont {R.~J.}\ \bibnamefont
  {Dowdall}}, \bibinfo {author} {\bibfnamefont {C.~T.~H.}\ \bibnamefont
  {Davies}}, \bibinfo {author} {\bibfnamefont {T.~C.}\ \bibnamefont
  {Hammant}},\ and\ \bibinfo {author} {\bibfnamefont {R.~R.}\ \bibnamefont
  {Horgan}} (\bibinfo {collaboration} {HPQCD}),\ }\bibfield  {title} {\bibinfo
  {title} {{Precise heavy-light meson masses and hyperfine splittings from
  lattice QCD including charm quarks in the sea}},\ }\href
  {https://doi.org/10.1103/PhysRevD.86.094510} {\bibfield  {journal} {\bibinfo
  {journal} {Phys. Rev. D}\ }\textbf {\bibinfo {volume} {86}},\ \bibinfo
  {pages} {094510} (\bibinfo {year} {2012}{\natexlab{b}})},\ \Eprint
  {https://arxiv.org/abs/1207.5149} {arXiv:1207.5149 [hep-lat]} \BibitemShut
  {NoStop}%
\bibitem [{\citenamefont {Aad}\ \emph {et~al.}(2014)\citenamefont {Aad} \emph
  {et~al.}}]{ATLAS:2014lga}%
  \BibitemOpen
  \bibfield  {author} {\bibinfo {author} {\bibfnamefont {G.}~\bibnamefont
  {Aad}} \emph {et~al.} (\bibinfo {collaboration} {ATLAS}),\ }\bibfield
  {title} {\bibinfo {title} {Observation of an excited {$B_c^\pm$} meson state
  with the {ATLAS} detector},\ }\href
  {https://doi.org/10.1103/PhysRevLett.113.212004} {\bibfield  {journal}
  {\bibinfo  {journal} {Phys. Rev. Lett.}\ }\textbf {\bibinfo {volume} {113}},\
  \bibinfo {pages} {212004} (\bibinfo {year} {2014})},\ \Eprint
  {https://arxiv.org/abs/1407.1032} {arXiv:1407.1032 [hep-ex]} \BibitemShut
  {NoStop}%
\bibitem [{\citenamefont {Pontecorvo}(1957)}]{Pontecorvo:1957cp}%
  \BibitemOpen
  \bibfield  {author} {\bibinfo {author} {\bibfnamefont {B.}~\bibnamefont
  {Pontecorvo}},\ }\bibfield  {title} {\bibinfo {title} {Mesonium and
  antimesonium},\ }\href@noop {} {\bibfield  {journal} {\bibinfo  {journal}
  {Sov. Phys. JETP}\ }\textbf {\bibinfo {volume} {6}},\ \bibinfo {pages} {429}
  (\bibinfo {year} {1957})},\ \bibinfo {note}
  {[\href{http://www.jetp.ac.ru/cgi-bin/e/index/e/6/2/p429?a=list}{Zh. Eksp.
  Teor. Fiz.~\textbf{33}, 549 (1957)}]}\BibitemShut {NoStop}%
\bibitem [{\citenamefont {Pontecorvo}(1968)}]{Pontecorvo:1967fh}%
  \BibitemOpen
  \bibfield  {author} {\bibinfo {author} {\bibfnamefont {B.}~\bibnamefont
  {Pontecorvo}},\ }\bibfield  {title} {\bibinfo {title} {Neutrino experiments
  and the problem of conservation of leptonic charge},\ }\href@noop {}
  {\bibfield  {journal} {\bibinfo  {journal} {Sov. Phys. JETP}\ }\textbf
  {\bibinfo {volume} {26}},\ \bibinfo {pages} {984} (\bibinfo {year} {1968})},\
  \bibinfo {note}
  {[\href{http://www.jetp.ac.ru/cgi-bin/e/index/e/26/5/p984?a=list}{Zh. Eksp.
  Teor. Fiz.~\textbf{53}, 1717 (1967)}]}\BibitemShut {NoStop}%
\bibitem [{\citenamefont {Maki}\ \emph {et~al.}(1962)\citenamefont {Maki},
  \citenamefont {Nakagawa},\ and\ \citenamefont {Sakata}}]{Maki:1962mu}%
  \BibitemOpen
  \bibfield  {author} {\bibinfo {author} {\bibfnamefont {Z.}~\bibnamefont
  {Maki}}, \bibinfo {author} {\bibfnamefont {M.}~\bibnamefont {Nakagawa}},\
  and\ \bibinfo {author} {\bibfnamefont {S.}~\bibnamefont {Sakata}},\
  }\bibfield  {title} {\bibinfo {title} {{Remarks on the unified model of
  elementary particles}},\ }\href {https://doi.org/10.1143/PTP.28.870}
  {\bibfield  {journal} {\bibinfo  {journal} {Prog. Theor. Phys.}\ }\textbf
  {\bibinfo {volume} {28}},\ \bibinfo {pages} {870} (\bibinfo {year}
  {1962})}\BibitemShut {NoStop}%
\bibitem [{\citenamefont {Huber}(2017)}]{Huber:2016mki}%
  \BibitemOpen
  \bibfield  {author} {\bibinfo {author} {\bibfnamefont {P.}~\bibnamefont
  {Huber}},\ }\bibfield  {title} {\bibinfo {title} {{Prospects for neutrino
  oscillation parameters}},\ }\href {https://doi.org/10.22323/1.283.0025}
  {\bibfield  {journal} {\bibinfo  {journal} {PoS}\ }\textbf {\bibinfo {volume}
  {NOW2016}},\ \bibinfo {pages} {025} (\bibinfo {year} {2017})},\ \Eprint
  {https://arxiv.org/abs/1612.04843} {arXiv:1612.04843 [hep-ph]} \BibitemShut
  {NoStop}%
\bibitem [{\citenamefont {Alvarez-Ruso}\ \emph {et~al.}(2018)\citenamefont
  {Alvarez-Ruso} \emph {et~al.}}]{Alvarez-Ruso:2017oui}%
  \BibitemOpen
  \bibfield  {author} {\bibinfo {author} {\bibfnamefont {L.}~\bibnamefont
  {Alvarez-Ruso}} \emph {et~al.} (\bibinfo {collaboration} {NuSTEC}),\
  }\bibfield  {title} {\bibinfo {title} {Status and challenges of
  neutrino–nucleus scattering},\ }\href
  {https://doi.org/10.1016/j.ppnp.2018.01.006} {\bibfield  {journal} {\bibinfo
  {journal} {Prog. Part. Nucl. Phys.}\ }\textbf {\bibinfo {volume} {100}},\
  \bibinfo {pages} {1} (\bibinfo {year} {2018})},\ \Eprint
  {https://arxiv.org/abs/1706.03621} {arXiv:1706.03621 [hep-ph]} \BibitemShut
  {NoStop}%
\bibitem [{\citenamefont {Coloma}\ \emph {et~al.}(2013)\citenamefont {Coloma},
  \citenamefont {Huber}, \citenamefont {Kopp},\ and\ \citenamefont
  {Winter}}]{Coloma:2012ji}%
  \BibitemOpen
  \bibfield  {author} {\bibinfo {author} {\bibfnamefont {P.}~\bibnamefont
  {Coloma}}, \bibinfo {author} {\bibfnamefont {P.}~\bibnamefont {Huber}},
  \bibinfo {author} {\bibfnamefont {J.}~\bibnamefont {Kopp}},\ and\ \bibinfo
  {author} {\bibfnamefont {W.}~\bibnamefont {Winter}},\ }\bibfield  {title}
  {\bibinfo {title} {Systematic uncertainties in long-baseline neutrino
  oscillations for large $\theta_{13}$},\ }\href
  {https://doi.org/10.1103/PhysRevD.87.033004} {\bibfield  {journal} {\bibinfo
  {journal} {Phys. Rev. D}\ }\textbf {\bibinfo {volume} {87}},\ \bibinfo
  {pages} {033004} (\bibinfo {year} {2013})},\ \Eprint
  {https://arxiv.org/abs/1209.5973} {arXiv:1209.5973 [hep-ph]} \BibitemShut
  {NoStop}%
\bibitem [{\citenamefont {Coloma}\ \emph {et~al.}(2014)\citenamefont {Coloma},
  \citenamefont {Huber}, \citenamefont {Jen},\ and\ \citenamefont
  {Mariani}}]{Coloma:2013tba}%
  \BibitemOpen
  \bibfield  {author} {\bibinfo {author} {\bibfnamefont {P.}~\bibnamefont
  {Coloma}}, \bibinfo {author} {\bibfnamefont {P.}~\bibnamefont {Huber}},
  \bibinfo {author} {\bibfnamefont {C.-M.}\ \bibnamefont {Jen}},\ and\ \bibinfo
  {author} {\bibfnamefont {C.}~\bibnamefont {Mariani}},\ }\bibfield  {title}
  {\bibinfo {title} {{Neutrino-nucleus interaction models and their impact on
  oscillation analyses}},\ }\href {https://doi.org/10.1103/PhysRevD.89.073015}
  {\bibfield  {journal} {\bibinfo  {journal} {Phys. Rev. D}\ }\textbf {\bibinfo
  {volume} {89}},\ \bibinfo {pages} {073015} (\bibinfo {year} {2014})},\
  \Eprint {https://arxiv.org/abs/1311.4506} {arXiv:1311.4506 [hep-ph]}
  \BibitemShut {NoStop}%
\bibitem [{\citenamefont {Yamazaki}\ \emph {et~al.}(2009)\citenamefont
  {Yamazaki}, \citenamefont {Aoki}, \citenamefont {Blum}, \citenamefont {Lin},
  \citenamefont {Ohta}, \citenamefont {Sasaki}, \citenamefont {Tweedie},\ and\
  \citenamefont {Zanotti}}]{Yamazaki:2009zq}%
  \BibitemOpen
  \bibfield  {author} {\bibinfo {author} {\bibfnamefont {T.}~\bibnamefont
  {Yamazaki}}, \bibinfo {author} {\bibfnamefont {Y.}~\bibnamefont {Aoki}},
  \bibinfo {author} {\bibfnamefont {T.}~\bibnamefont {Blum}}, \bibinfo {author}
  {\bibfnamefont {H.-W.}\ \bibnamefont {Lin}}, \bibinfo {author} {\bibfnamefont
  {S.}~\bibnamefont {Ohta}}, \bibinfo {author} {\bibfnamefont {S.}~\bibnamefont
  {Sasaki}}, \bibinfo {author} {\bibfnamefont {R.}~\bibnamefont {Tweedie}},\
  and\ \bibinfo {author} {\bibfnamefont {J.}~\bibnamefont {Zanotti}},\
  }\bibfield  {title} {\bibinfo {title} {Nucleon form factors with $2+1$ flavor
  dynamical domain-wall fermions},\ }\href
  {https://doi.org/10.1103/PhysRevD.79.114505} {\bibfield  {journal} {\bibinfo
  {journal} {Phys. Rev. D}\ }\textbf {\bibinfo {volume} {79}},\ \bibinfo
  {pages} {114505} (\bibinfo {year} {2009})},\ \Eprint
  {https://arxiv.org/abs/0904.2039} {arXiv:0904.2039 [hep-lat]} \BibitemShut
  {NoStop}%
\bibitem [{\citenamefont {Babich}\ \emph {et~al.}(2012)\citenamefont {Babich},
  \citenamefont {Brower}, \citenamefont {Clark}, \citenamefont {Fleming},
  \citenamefont {Osborn}, \citenamefont {Rebbi},\ and\ \citenamefont
  {Schaich}}]{Babich:2010at}%
  \BibitemOpen
  \bibfield  {author} {\bibinfo {author} {\bibfnamefont {R.}~\bibnamefont
  {Babich}}, \bibinfo {author} {\bibfnamefont {R.~C.}\ \bibnamefont {Brower}},
  \bibinfo {author} {\bibfnamefont {M.~A.}\ \bibnamefont {Clark}}, \bibinfo
  {author} {\bibfnamefont {G.~T.}\ \bibnamefont {Fleming}}, \bibinfo {author}
  {\bibfnamefont {J.~C.}\ \bibnamefont {Osborn}}, \bibinfo {author}
  {\bibfnamefont {C.}~\bibnamefont {Rebbi}},\ and\ \bibinfo {author}
  {\bibfnamefont {D.}~\bibnamefont {Schaich}},\ }\bibfield  {title} {\bibinfo
  {title} {Exploring strange nucleon form factors on the lattice},\ }\href
  {https://doi.org/10.1103/PhysRevD.85.054510} {\bibfield  {journal} {\bibinfo
  {journal} {Phys. Rev. D}\ }\textbf {\bibinfo {volume} {85}},\ \bibinfo
  {pages} {054510} (\bibinfo {year} {2012})},\ \Eprint
  {https://arxiv.org/abs/1012.0562} {arXiv:1012.0562 [hep-lat]} \BibitemShut
  {NoStop}%
\bibitem [{\citenamefont {Green}\ \emph {et~al.}(2017)\citenamefont {Green},
  \citenamefont {Hasan}, \citenamefont {Meinel}, \citenamefont {Engelhardt},
  \citenamefont {Krieg}, \citenamefont {Laeuchli}, \citenamefont {Negele},
  \citenamefont {Orginos}, \citenamefont {Pochinsky},\ and\ \citenamefont
  {Syritsyn}}]{Green:2017keo}%
  \BibitemOpen
  \bibfield  {author} {\bibinfo {author} {\bibfnamefont {J.}~\bibnamefont
  {Green}}, \bibinfo {author} {\bibfnamefont {N.}~\bibnamefont {Hasan}},
  \bibinfo {author} {\bibfnamefont {S.}~\bibnamefont {Meinel}}, \bibinfo
  {author} {\bibfnamefont {M.}~\bibnamefont {Engelhardt}}, \bibinfo {author}
  {\bibfnamefont {S.}~\bibnamefont {Krieg}}, \bibinfo {author} {\bibfnamefont
  {J.}~\bibnamefont {Laeuchli}}, \bibinfo {author} {\bibfnamefont
  {J.}~\bibnamefont {Negele}}, \bibinfo {author} {\bibfnamefont
  {K.}~\bibnamefont {Orginos}}, \bibinfo {author} {\bibfnamefont
  {A.}~\bibnamefont {Pochinsky}},\ and\ \bibinfo {author} {\bibfnamefont
  {S.}~\bibnamefont {Syritsyn}},\ }\bibfield  {title} {\bibinfo {title} {Up,
  down, and strange nucleon axial form factors from lattice {QCD}},\ }\href
  {https://doi.org/10.1103/PhysRevD.95.114502} {\bibfield  {journal} {\bibinfo
  {journal} {Phys. Rev. D}\ }\textbf {\bibinfo {volume} {95}},\ \bibinfo
  {pages} {114502} (\bibinfo {year} {2017})},\ \Eprint
  {https://arxiv.org/abs/1703.06703} {arXiv:1703.06703 [hep-lat]} \BibitemShut
  {NoStop}%
\bibitem [{\citenamefont {Hasan}\ \emph {et~al.}(2018)\citenamefont {Hasan},
  \citenamefont {Green}, \citenamefont {Meinel}, \citenamefont {Engelhardt},
  \citenamefont {Krieg}, \citenamefont {Negele}, \citenamefont {Pochinsky},\
  and\ \citenamefont {Syritsyn}}]{Hasan:2017wwt}%
  \BibitemOpen
  \bibfield  {author} {\bibinfo {author} {\bibfnamefont {N.}~\bibnamefont
  {Hasan}}, \bibinfo {author} {\bibfnamefont {J.}~\bibnamefont {Green}},
  \bibinfo {author} {\bibfnamefont {S.}~\bibnamefont {Meinel}}, \bibinfo
  {author} {\bibfnamefont {M.}~\bibnamefont {Engelhardt}}, \bibinfo {author}
  {\bibfnamefont {S.}~\bibnamefont {Krieg}}, \bibinfo {author} {\bibfnamefont
  {J.}~\bibnamefont {Negele}}, \bibinfo {author} {\bibfnamefont
  {A.}~\bibnamefont {Pochinsky}},\ and\ \bibinfo {author} {\bibfnamefont
  {S.}~\bibnamefont {Syritsyn}},\ }\bibfield  {title} {\bibinfo {title}
  {{Computing the nucleon charge and axial radii directly at $Q^2=0$ in lattice
  QCD}},\ }\href {https://doi.org/10.1103/PhysRevD.97.034504} {\bibfield
  {journal} {\bibinfo  {journal} {Phys. Rev. D}\ }\textbf {\bibinfo {volume}
  {97}},\ \bibinfo {pages} {034504} (\bibinfo {year} {2018})},\ \Eprint
  {https://arxiv.org/abs/1711.11385} {arXiv:1711.11385 [hep-lat]} \BibitemShut
  {NoStop}%
\bibitem [{\citenamefont {Gupta}\ \emph {et~al.}(2017)\citenamefont {Gupta},
  \citenamefont {Jang}, \citenamefont {Lin}, \citenamefont {Yoon},\ and\
  \citenamefont {Bhattacharya}}]{Rajan:2017lxk}%
  \BibitemOpen
  \bibfield  {author} {\bibinfo {author} {\bibfnamefont {R.}~\bibnamefont
  {Gupta}}, \bibinfo {author} {\bibfnamefont {Y.-C.}\ \bibnamefont {Jang}},
  \bibinfo {author} {\bibfnamefont {H.-W.}\ \bibnamefont {Lin}}, \bibinfo
  {author} {\bibfnamefont {B.}~\bibnamefont {Yoon}},\ and\ \bibinfo {author}
  {\bibfnamefont {T.}~\bibnamefont {Bhattacharya}} (\bibinfo {collaboration}
  {PNDME}),\ }\bibfield  {title} {\bibinfo {title} {Axial vector form factors
  of the nucleon from lattice {QCD}},\ }\href
  {https://doi.org/10.1103/PhysRevD.96.114503} {\bibfield  {journal} {\bibinfo
  {journal} {Phys. Rev. D}\ }\textbf {\bibinfo {volume} {96}},\ \bibinfo
  {pages} {114503} (\bibinfo {year} {2017})},\ \Eprint
  {https://arxiv.org/abs/1705.06834} {arXiv:1705.06834 [hep-lat]} \BibitemShut
  {NoStop}%
\bibitem [{\citenamefont {Jang}\ \emph {et~al.}(2020)\citenamefont {Jang},
  \citenamefont {Gupta}, \citenamefont {Yoon},\ and\ \citenamefont
  {Bhattacharya}}]{Jang:2019vkm}%
  \BibitemOpen
  \bibfield  {author} {\bibinfo {author} {\bibfnamefont {Y.-C.}\ \bibnamefont
  {Jang}}, \bibinfo {author} {\bibfnamefont {R.}~\bibnamefont {Gupta}},
  \bibinfo {author} {\bibfnamefont {B.}~\bibnamefont {Yoon}},\ and\ \bibinfo
  {author} {\bibfnamefont {T.}~\bibnamefont {Bhattacharya}},\ }\bibfield
  {title} {\bibinfo {title} {Axial vector form factors from lattice {QCD} that
  satisfy the {PCAC} relation},\ }\href
  {https://doi.org/10.1103/PhysRevLett.124.072002} {\bibfield  {journal}
  {\bibinfo  {journal} {Phys. Rev. Lett.}\ }\textbf {\bibinfo {volume} {124}},\
  \bibinfo {pages} {072002} (\bibinfo {year} {2020})},\ \Eprint
  {https://arxiv.org/abs/1905.06470} {arXiv:1905.06470 [hep-lat]} \BibitemShut
  {NoStop}%
\bibitem [{\citenamefont {Ishikawa}\ \emph {et~al.}(2018)\citenamefont
  {Ishikawa}, \citenamefont {Kuramashi}, \citenamefont {Sasaki}, \citenamefont
  {Tsukamoto}, \citenamefont {Ukawa},\ and\ \citenamefont
  {Yamazaki}}]{Ishikawa:2018rew}%
  \BibitemOpen
  \bibfield  {author} {\bibinfo {author} {\bibfnamefont {K.-I.}\ \bibnamefont
  {Ishikawa}}, \bibinfo {author} {\bibfnamefont {Y.}~\bibnamefont {Kuramashi}},
  \bibinfo {author} {\bibfnamefont {S.}~\bibnamefont {Sasaki}}, \bibinfo
  {author} {\bibfnamefont {N.}~\bibnamefont {Tsukamoto}}, \bibinfo {author}
  {\bibfnamefont {A.}~\bibnamefont {Ukawa}},\ and\ \bibinfo {author}
  {\bibfnamefont {T.}~\bibnamefont {Yamazaki}} (\bibinfo {collaboration}
  {PACS}),\ }\bibfield  {title} {\bibinfo {title} {Nucleon form factors on a
  large volume lattice near the physical point in $2+1$ flavor {QCD}},\ }\href
  {https://doi.org/10.1103/PhysRevD.98.074510} {\bibfield  {journal} {\bibinfo
  {journal} {Phys. Rev. D}\ }\textbf {\bibinfo {volume} {98}},\ \bibinfo
  {pages} {074510} (\bibinfo {year} {2018})},\ \Eprint
  {https://arxiv.org/abs/1807.03974} {arXiv:1807.03974 [hep-lat]} \BibitemShut
  {NoStop}%
\bibitem [{\citenamefont {Shintani}\ \emph {et~al.}(2019)\citenamefont
  {Shintani}, \citenamefont {Ishikawa}, \citenamefont {Kuramashi},
  \citenamefont {Sasaki},\ and\ \citenamefont {Yamazaki}}]{Shintani:2018ozy}%
  \BibitemOpen
  \bibfield  {author} {\bibinfo {author} {\bibfnamefont {E.}~\bibnamefont
  {Shintani}}, \bibinfo {author} {\bibfnamefont {K.-I.}\ \bibnamefont
  {Ishikawa}}, \bibinfo {author} {\bibfnamefont {Y.}~\bibnamefont {Kuramashi}},
  \bibinfo {author} {\bibfnamefont {S.}~\bibnamefont {Sasaki}},\ and\ \bibinfo
  {author} {\bibfnamefont {T.}~\bibnamefont {Yamazaki}},\ }\bibfield  {title}
  {\bibinfo {title} {{Nucleon form factors and root-mean-square radii on a
  (10.8~fm)$^4$ lattice at the physical point}},\ }\href
  {https://doi.org/10.1103/PhysRevD.99.014510} {\bibfield  {journal} {\bibinfo
  {journal} {Phys. Rev. D}\ }\textbf {\bibinfo {volume} {99}},\ \bibinfo
  {pages} {014510} (\bibinfo {year} {2019})},\ \bibinfo {note} {(E)
  \href{http://doi.org/10.1103/PhysRevD.102.019902}{\textbf{102}, 019902
  (2020)}},\ \Eprint {https://arxiv.org/abs/1811.07292} {arXiv:1811.07292
  [hep-lat]} \BibitemShut {NoStop}%
\bibitem [{\citenamefont {Ishikawa}\ \emph {et~al.}(2021)\citenamefont
  {Ishikawa}, \citenamefont {Kuramashi}, \citenamefont {Sasaki}, \citenamefont
  {Shintani},\ and\ \citenamefont {Yamazaki}}]{Ishikawa:2021eut}%
  \BibitemOpen
  \bibfield  {author} {\bibinfo {author} {\bibfnamefont {K.-I.}\ \bibnamefont
  {Ishikawa}}, \bibinfo {author} {\bibfnamefont {Y.}~\bibnamefont {Kuramashi}},
  \bibinfo {author} {\bibfnamefont {S.}~\bibnamefont {Sasaki}}, \bibinfo
  {author} {\bibfnamefont {E.}~\bibnamefont {Shintani}},\ and\ \bibinfo
  {author} {\bibfnamefont {T.}~\bibnamefont {Yamazaki}} (\bibinfo
  {collaboration} {PACS}),\ }\bibfield  {title} {\bibinfo {title} {{Calculation
  of the derivative of nucleon form factors in $N_f=2+1$ lattice QCD at
  $M_\pi=138$~MeV on a $(5.5~\text{fm})^3$ volume}},\ }\href
  {https://doi.org/10.1103/PhysRevD.104.074514} {\bibfield  {journal} {\bibinfo
   {journal} {Phys. Rev. D}\ }\textbf {\bibinfo {volume} {104}},\ \bibinfo
  {pages} {074514} (\bibinfo {year} {2021})},\ \Eprint
  {https://arxiv.org/abs/2107.07085} {arXiv:2107.07085 [hep-lat]} \BibitemShut
  {NoStop}%
\bibitem [{\citenamefont {Bali}\ \emph {et~al.}(2020)\citenamefont {Bali},
  \citenamefont {Barca}, \citenamefont {Collins}, \citenamefont {Gruber},
  \citenamefont {L\"offler}, \citenamefont {Sch\"afer}, \citenamefont
  {S\"oldner}, \citenamefont {Wein}, \citenamefont {Weish\"aupl},\ and\
  \citenamefont {Wurm}}]{RQCD:2019jai}%
  \BibitemOpen
  \bibfield  {author} {\bibinfo {author} {\bibfnamefont {G.~S.}\ \bibnamefont
  {Bali}}, \bibinfo {author} {\bibfnamefont {L.}~\bibnamefont {Barca}},
  \bibinfo {author} {\bibfnamefont {S.}~\bibnamefont {Collins}}, \bibinfo
  {author} {\bibfnamefont {M.}~\bibnamefont {Gruber}}, \bibinfo {author}
  {\bibfnamefont {M.}~\bibnamefont {L\"offler}}, \bibinfo {author}
  {\bibfnamefont {A.}~\bibnamefont {Sch\"afer}}, \bibinfo {author}
  {\bibfnamefont {W.}~\bibnamefont {S\"oldner}}, \bibinfo {author}
  {\bibfnamefont {P.}~\bibnamefont {Wein}}, \bibinfo {author} {\bibfnamefont
  {S.}~\bibnamefont {Weish\"aupl}},\ and\ \bibinfo {author} {\bibfnamefont
  {T.}~\bibnamefont {Wurm}} (\bibinfo {collaboration} {RQCD}),\ }\bibfield
  {title} {\bibinfo {title} {{Nucleon axial structure from lattice QCD}},\
  }\href {https://doi.org/10.1007/JHEP05(2020)126} {\bibfield  {journal}
  {\bibinfo  {journal} {JHEP}\ }\textbf {\bibinfo {volume} {05}},\ \bibinfo
  {pages} {126}},\ \Eprint {https://arxiv.org/abs/1911.13150} {arXiv:1911.13150
  [hep-lat]} \BibitemShut {NoStop}%
\bibitem [{\citenamefont {Alexandrou}\ \emph
  {et~al.}(2021{\natexlab{b}})\citenamefont {Alexandrou} \emph
  {et~al.}}]{Alexandrou:2020okk}%
  \BibitemOpen
  \bibfield  {author} {\bibinfo {author} {\bibfnamefont {C.}~\bibnamefont
  {Alexandrou}} \emph {et~al.},\ }\bibfield  {title} {\bibinfo {title}
  {{Nucleon axial and pseudoscalar form factors from lattice QCD at the
  physical point}},\ }\href {https://doi.org/10.1103/PhysRevD.103.034509}
  {\bibfield  {journal} {\bibinfo  {journal} {Phys. Rev. D}\ }\textbf {\bibinfo
  {volume} {103}},\ \bibinfo {pages} {034509} (\bibinfo {year}
  {2021}{\natexlab{b}})},\ \Eprint {https://arxiv.org/abs/2011.13342}
  {arXiv:2011.13342 [hep-lat]} \BibitemShut {NoStop}%
\bibitem [{\citenamefont {Park}\ \emph {et~al.}(2022)\citenamefont {Park},
  \citenamefont {Gupta}, \citenamefont {Yoon}, \citenamefont {Mondal},
  \citenamefont {Bhattacharya}, \citenamefont {Jang}, \citenamefont {Jo\'o},\
  and\ \citenamefont {Winter}}]{Park:2021ypf}%
  \BibitemOpen
  \bibfield  {author} {\bibinfo {author} {\bibfnamefont {S.}~\bibnamefont
  {Park}}, \bibinfo {author} {\bibfnamefont {R.}~\bibnamefont {Gupta}},
  \bibinfo {author} {\bibfnamefont {B.}~\bibnamefont {Yoon}}, \bibinfo {author}
  {\bibfnamefont {S.}~\bibnamefont {Mondal}}, \bibinfo {author} {\bibfnamefont
  {T.}~\bibnamefont {Bhattacharya}}, \bibinfo {author} {\bibfnamefont {Y.-C.}\
  \bibnamefont {Jang}}, \bibinfo {author} {\bibfnamefont {B.}~\bibnamefont
  {Jo\'o}},\ and\ \bibinfo {author} {\bibfnamefont {F.}~\bibnamefont {Winter}}
  (\bibinfo {collaboration} {NME}),\ }\bibfield  {title} {\bibinfo {title}
  {Precision nucleon charges and form factors using (2+1)-flavor lattice
  {QCD}},\ }\href {https://doi.org/10.1103/PhysRevD.105.054505} {\bibfield
  {journal} {\bibinfo  {journal} {Phys. Rev. D}\ }\textbf {\bibinfo {volume}
  {105}},\ \bibinfo {pages} {054505} (\bibinfo {year} {2022})},\ \Eprint
  {https://arxiv.org/abs/2103.05599} {arXiv:2103.05599 [hep-lat]} \BibitemShut
  {NoStop}%
\bibitem [{\citenamefont {Ankowski}\ \emph {et~al.}(2022)\citenamefont
  {Ankowski} \emph {et~al.}}]{Ankowski:2022thw}%
  \BibitemOpen
  \bibfield  {author} {\bibinfo {author} {\bibfnamefont {A.~M.}\ \bibnamefont
  {Ankowski}} \emph {et~al.},\ }\bibfield  {title} {\bibinfo {title} {Electron
  scattering and neutrino physics},\ }in\ \href@noop {} {\emph {\bibinfo
  {booktitle} {{2022 Snowmass Summer Study}}}}\ (\bibinfo {year} {2022})\
  \Eprint {https://arxiv.org/abs/2203.06853} {arXiv:2203.06853 [hep-ex]}
  \BibitemShut {NoStop}%
\bibitem [{\citenamefont {Kelly}(2004)}]{Kelly:2004hm}%
  \BibitemOpen
  \bibfield  {author} {\bibinfo {author} {\bibfnamefont {J.~J.}\ \bibnamefont
  {Kelly}},\ }\bibfield  {title} {\bibinfo {title} {{Simple parametrization of
  nucleon form factors}},\ }\href {https://doi.org/10.1103/PhysRevC.70.068202}
  {\bibfield  {journal} {\bibinfo  {journal} {Phys. Rev. C}\ }\textbf {\bibinfo
  {volume} {70}},\ \bibinfo {pages} {068202} (\bibinfo {year}
  {2004})}\BibitemShut {NoStop}%
\bibitem [{\citenamefont {Liang}\ \emph {et~al.}(2018)\citenamefont {Liang},
  \citenamefont {Yang}, \citenamefont {Draper}, \citenamefont {Gong},\ and\
  \citenamefont {Liu}}]{Liang:2018pis}%
  \BibitemOpen
  \bibfield  {author} {\bibinfo {author} {\bibfnamefont {J.}~\bibnamefont
  {Liang}}, \bibinfo {author} {\bibfnamefont {Y.-B.}\ \bibnamefont {Yang}},
  \bibinfo {author} {\bibfnamefont {T.}~\bibnamefont {Draper}}, \bibinfo
  {author} {\bibfnamefont {M.}~\bibnamefont {Gong}},\ and\ \bibinfo {author}
  {\bibfnamefont {K.-F.}\ \bibnamefont {Liu}},\ }\bibfield  {title} {\bibinfo
  {title} {Quark spins and anomalous {Ward} identity},\ }\href
  {https://doi.org/10.1103/PhysRevD.98.074505} {\bibfield  {journal} {\bibinfo
  {journal} {Phys. Rev. D}\ }\textbf {\bibinfo {volume} {98}},\ \bibinfo
  {pages} {074505} (\bibinfo {year} {2018})},\ \Eprint
  {https://arxiv.org/abs/1806.08366} {arXiv:1806.08366 [hep-ph]} \BibitemShut
  {NoStop}%
\bibitem [{\citenamefont {Harris}\ \emph {et~al.}(2019)\citenamefont {Harris},
  \citenamefont {von Hippel}, \citenamefont {Junnarkar}, \citenamefont {Meyer},
  \citenamefont {Ottnad}, \citenamefont {Wilhelm}, \citenamefont {Wittig},\
  and\ \citenamefont {Wrang}}]{Harris:2019bih}%
  \BibitemOpen
  \bibfield  {author} {\bibinfo {author} {\bibfnamefont {T.}~\bibnamefont
  {Harris}}, \bibinfo {author} {\bibfnamefont {G.}~\bibnamefont {von Hippel}},
  \bibinfo {author} {\bibfnamefont {P.}~\bibnamefont {Junnarkar}}, \bibinfo
  {author} {\bibfnamefont {H.~B.}\ \bibnamefont {Meyer}}, \bibinfo {author}
  {\bibfnamefont {K.}~\bibnamefont {Ottnad}}, \bibinfo {author} {\bibfnamefont
  {J.}~\bibnamefont {Wilhelm}}, \bibinfo {author} {\bibfnamefont
  {H.}~\bibnamefont {Wittig}},\ and\ \bibinfo {author} {\bibfnamefont
  {L.}~\bibnamefont {Wrang}},\ }\bibfield  {title} {\bibinfo {title} {{Nucleon
  isovector charges and twist-2 matrix elements with $N_f=2+1$ dynamical Wilson
  quarks}},\ }\href {https://doi.org/10.1103/PhysRevD.100.034513} {\bibfield
  {journal} {\bibinfo  {journal} {Phys. Rev. D}\ }\textbf {\bibinfo {volume}
  {100}},\ \bibinfo {pages} {034513} (\bibinfo {year} {2019})},\ \Eprint
  {https://arxiv.org/abs/1905.01291} {arXiv:1905.01291 [hep-lat]} \BibitemShut
  {NoStop}%
\bibitem [{\citenamefont {Meyer}\ \emph
  {et~al.}(2022{\natexlab{a}})\citenamefont {Meyer}, \citenamefont
  {Walker-Loud},\ and\ \citenamefont {Wilkinson}}]{Meyer:2022mix}%
  \BibitemOpen
  \bibfield  {author} {\bibinfo {author} {\bibfnamefont {A.~S.}\ \bibnamefont
  {Meyer}}, \bibinfo {author} {\bibfnamefont {A.}~\bibnamefont {Walker-Loud}},\
  and\ \bibinfo {author} {\bibfnamefont {C.}~\bibnamefont {Wilkinson}},\
  }\bibfield  {title} {\bibinfo {title} {Status of lattice-{QCD} determination
  of nucleon form factors and their relevance for the few-{GeV} neutrino
  program},\ }\href@noop {} {\bibfield  {journal} {\bibinfo  {journal} {Ann.
  Rev. Nucl. Part. Sci.}\ } (\bibinfo {year} {2022}{\natexlab{a}})},\ \Eprint
  {https://arxiv.org/abs/2201.01839} {arXiv:2201.01839 [hep-lat]} \BibitemShut
  {NoStop}%
\bibitem [{\citenamefont {Meyer}\ \emph {et~al.}(2016)\citenamefont {Meyer},
  \citenamefont {Betancourt}, \citenamefont {Gran},\ and\ \citenamefont
  {Hill}}]{Meyer:2016oeg}%
  \BibitemOpen
  \bibfield  {author} {\bibinfo {author} {\bibfnamefont {A.~S.}\ \bibnamefont
  {Meyer}}, \bibinfo {author} {\bibfnamefont {M.}~\bibnamefont {Betancourt}},
  \bibinfo {author} {\bibfnamefont {R.}~\bibnamefont {Gran}},\ and\ \bibinfo
  {author} {\bibfnamefont {R.~J.}\ \bibnamefont {Hill}},\ }\bibfield  {title}
  {\bibinfo {title} {{Deuterium target data for precision neutrino-nucleus
  cross sections}},\ }\href {https://doi.org/10.1103/PhysRevD.93.113015}
  {\bibfield  {journal} {\bibinfo  {journal} {Phys. Rev. D}\ }\textbf {\bibinfo
  {volume} {93}},\ \bibinfo {pages} {113015} (\bibinfo {year} {2016})},\
  \Eprint {https://arxiv.org/abs/1603.03048} {arXiv:1603.03048 [hep-ph]}
  \BibitemShut {NoStop}%
\bibitem [{\citenamefont {Meyer}\ \emph
  {et~al.}(2022{\natexlab{b}})\citenamefont {Meyer} \emph
  {et~al.}}]{Meyer:2021vfq}%
  \BibitemOpen
  \bibfield  {author} {\bibinfo {author} {\bibfnamefont {A.~S.}\ \bibnamefont
  {Meyer}} \emph {et~al.},\ }\bibfield  {title} {\bibinfo {title} {Nucleon
  axial form factor from domain wall on {HISQ}},\ }\href
  {https://doi.org/10.22323/1.396.0081} {\bibfield  {journal} {\bibinfo
  {journal} {PoS}\ }\textbf {\bibinfo {volume} {LATTICE2021}},\ \bibinfo
  {pages} {081} (\bibinfo {year} {2022}{\natexlab{b}})},\ \Eprint
  {https://arxiv.org/abs/2111.06333} {arXiv:2111.06333 [hep-lat]} \BibitemShut
  {NoStop}%
\bibitem [{\citenamefont {Djukanovic}\ \emph {et~al.}(2022)\citenamefont
  {Djukanovic}, \citenamefont {von Hippel}, \citenamefont {Koponen},
  \citenamefont {Meyer}, \citenamefont {Ottnad}, \citenamefont {Schulz},\ and\
  \citenamefont {Wittig}}]{Schulz:2021kwz}%
  \BibitemOpen
  \bibfield  {author} {\bibinfo {author} {\bibfnamefont {D.}~\bibnamefont
  {Djukanovic}}, \bibinfo {author} {\bibfnamefont {G.}~\bibnamefont {von
  Hippel}}, \bibinfo {author} {\bibfnamefont {J.}~\bibnamefont {Koponen}},
  \bibinfo {author} {\bibfnamefont {H.~B.}\ \bibnamefont {Meyer}}, \bibinfo
  {author} {\bibfnamefont {K.}~\bibnamefont {Ottnad}}, \bibinfo {author}
  {\bibfnamefont {T.}~\bibnamefont {Schulz}},\ and\ \bibinfo {author}
  {\bibfnamefont {H.}~\bibnamefont {Wittig}},\ }\bibfield  {title} {\bibinfo
  {title} {Isovector axial vector form factors of the nucleon from lattice
  {QCD} with $n_f=2+1$ $\text{O}(a)$-improved {Wilson} fermions},\ }\href
  {https://doi.org/10.22323/1.396.0577} {\bibfield  {journal} {\bibinfo
  {journal} {PoS}\ }\textbf {\bibinfo {volume} {LATTICE2021}},\ \bibinfo
  {pages} {577} (\bibinfo {year} {2022})},\ \Eprint
  {https://arxiv.org/abs/2112.00127} {arXiv:2112.00127 [hep-lat]} \BibitemShut
  {NoStop}%
\bibitem [{\citenamefont {Bernard}\ \emph {et~al.}(2002)\citenamefont
  {Bernard}, \citenamefont {Elouadrhiri},\ and\ \citenamefont
  {Mei\ss{}ner}}]{Bernard:2001rs}%
  \BibitemOpen
  \bibfield  {author} {\bibinfo {author} {\bibfnamefont {V.}~\bibnamefont
  {Bernard}}, \bibinfo {author} {\bibfnamefont {L.}~\bibnamefont
  {Elouadrhiri}},\ and\ \bibinfo {author} {\bibfnamefont {U.-G.}\ \bibnamefont
  {Mei\ss{}ner}},\ }\bibfield  {title} {\bibinfo {title} {{Axial structure of
  the nucleon}},\ }\href {https://doi.org/10.1088/0954-3899/28/1/201}
  {\bibfield  {journal} {\bibinfo  {journal} {J. Phys. G}\ }\textbf {\bibinfo
  {volume} {28}},\ \bibinfo {pages} {R1} (\bibinfo {year} {2002})},\ \Eprint
  {https://arxiv.org/abs/hep-ph/0107088} {arXiv:hep-ph/0107088} \BibitemShut
  {NoStop}%
\bibitem [{\citenamefont {Bodek}\ \emph {et~al.}(2008)\citenamefont {Bodek},
  \citenamefont {Avvakumov}, \citenamefont {Bradford},\ and\ \citenamefont
  {Budd}}]{Bodek:2007ym}%
  \BibitemOpen
  \bibfield  {author} {\bibinfo {author} {\bibfnamefont {A.}~\bibnamefont
  {Bodek}}, \bibinfo {author} {\bibfnamefont {S.}~\bibnamefont {Avvakumov}},
  \bibinfo {author} {\bibfnamefont {R.}~\bibnamefont {Bradford}},\ and\
  \bibinfo {author} {\bibfnamefont {H.~S.}\ \bibnamefont {Budd}},\ }\bibfield
  {title} {\bibinfo {title} {Vector and axial nucleon form factors: A duality
  constrained parameterization},\ }\href
  {https://doi.org/10.1140/epjc/s10052-007-0491-4} {\bibfield  {journal}
  {\bibinfo  {journal} {Eur. Phys. J. C}\ }\textbf {\bibinfo {volume} {53}},\
  \bibinfo {pages} {349} (\bibinfo {year} {2008})},\ \Eprint
  {https://arxiv.org/abs/0708.1946} {arXiv:0708.1946 [hep-ex]} \BibitemShut
  {NoStop}%
\bibitem [{\citenamefont {Hill}\ \emph {et~al.}(2018)\citenamefont {Hill},
  \citenamefont {Kammel}, \citenamefont {Marciano},\ and\ \citenamefont
  {Sirlin}}]{Hill:2017wgb}%
  \BibitemOpen
  \bibfield  {author} {\bibinfo {author} {\bibfnamefont {R.~J.}\ \bibnamefont
  {Hill}}, \bibinfo {author} {\bibfnamefont {P.}~\bibnamefont {Kammel}},
  \bibinfo {author} {\bibfnamefont {W.~J.}\ \bibnamefont {Marciano}},\ and\
  \bibinfo {author} {\bibfnamefont {A.}~\bibnamefont {Sirlin}},\ }\bibfield
  {title} {\bibinfo {title} {Nucleon axial radius and muonic hydrogen: A new
  analysis and review},\ }\href {https://doi.org/10.1088/1361-6633/aac190}
  {\bibfield  {journal} {\bibinfo  {journal} {Rept. Prog. Phys.}\ }\textbf
  {\bibinfo {volume} {81}},\ \bibinfo {pages} {096301} (\bibinfo {year}
  {2018})},\ \Eprint {https://arxiv.org/abs/1708.08462} {arXiv:1708.08462
  [hep-ph]} \BibitemShut {NoStop}%
\bibitem [{\citenamefont {Liang}\ \emph {et~al.}(2020)\citenamefont {Liang},
  \citenamefont {Draper}, \citenamefont {Liu}, \citenamefont {Rothkopf},\ and\
  \citenamefont {Yang}}]{Liang:2019frk}%
  \BibitemOpen
  \bibfield  {author} {\bibinfo {author} {\bibfnamefont {J.}~\bibnamefont
  {Liang}}, \bibinfo {author} {\bibfnamefont {T.}~\bibnamefont {Draper}},
  \bibinfo {author} {\bibfnamefont {K.-F.}\ \bibnamefont {Liu}}, \bibinfo
  {author} {\bibfnamefont {A.}~\bibnamefont {Rothkopf}},\ and\ \bibinfo
  {author} {\bibfnamefont {Y.-B.}\ \bibnamefont {Yang}} (\bibinfo
  {collaboration} {$\chi$QCD}),\ }\bibfield  {title} {\bibinfo {title} {Towards
  the nucleon hadronic tensor from lattice {QCD}},\ }\href
  {https://doi.org/10.1103/PhysRevD.101.114503} {\bibfield  {journal} {\bibinfo
   {journal} {Phys. Rev. D}\ }\textbf {\bibinfo {volume} {101}},\ \bibinfo
  {pages} {114503} (\bibinfo {year} {2020})},\ \Eprint
  {https://arxiv.org/abs/1906.05312} {arXiv:1906.05312 [hep-ph]} \BibitemShut
  {NoStop}%
\bibitem [{\citenamefont {Savage}\ \emph {et~al.}(2017)\citenamefont {Savage},
  \citenamefont {Shanahan}, \citenamefont {Tiburzi}, \citenamefont {Wagman},
  \citenamefont {Winter}, \citenamefont {Beane}, \citenamefont {Chang},
  \citenamefont {Davoudi}, \citenamefont {Detmold},\ and\ \citenamefont
  {Orginos}}]{Savage:2016kon}%
  \BibitemOpen
  \bibfield  {author} {\bibinfo {author} {\bibfnamefont {M.~J.}\ \bibnamefont
  {Savage}}, \bibinfo {author} {\bibfnamefont {P.~E.}\ \bibnamefont
  {Shanahan}}, \bibinfo {author} {\bibfnamefont {B.~C.}\ \bibnamefont
  {Tiburzi}}, \bibinfo {author} {\bibfnamefont {M.~L.}\ \bibnamefont {Wagman}},
  \bibinfo {author} {\bibfnamefont {F.}~\bibnamefont {Winter}}, \bibinfo
  {author} {\bibfnamefont {S.~R.}\ \bibnamefont {Beane}}, \bibinfo {author}
  {\bibfnamefont {E.}~\bibnamefont {Chang}}, \bibinfo {author} {\bibfnamefont
  {Z.}~\bibnamefont {Davoudi}}, \bibinfo {author} {\bibfnamefont
  {W.}~\bibnamefont {Detmold}},\ and\ \bibinfo {author} {\bibfnamefont
  {K.}~\bibnamefont {Orginos}},\ }\bibfield  {title} {\bibinfo {title}
  {Proton-proton fusion and tritium $\beta$ decay from lattice quantum
  chromodynamics},\ }\href {https://doi.org/10.1103/PhysRevLett.119.062002}
  {\bibfield  {journal} {\bibinfo  {journal} {Phys. Rev. Lett.}\ }\textbf
  {\bibinfo {volume} {119}},\ \bibinfo {pages} {062002} (\bibinfo {year}
  {2017})},\ \Eprint {https://arxiv.org/abs/1610.04545} {arXiv:1610.04545
  [hep-lat]} \BibitemShut {NoStop}%
\bibitem [{\citenamefont {Parre\~no}\ \emph {et~al.}(2021)\citenamefont
  {Parre\~no}, \citenamefont {Shanahan}, \citenamefont {Wagman}, \citenamefont
  {Winter}, \citenamefont {Chang}, \citenamefont {Detmold},\ and\ \citenamefont
  {Illa}}]{Parreno:2021ovq}%
  \BibitemOpen
  \bibfield  {author} {\bibinfo {author} {\bibfnamefont {A.}~\bibnamefont
  {Parre\~no}}, \bibinfo {author} {\bibfnamefont {P.~E.}\ \bibnamefont
  {Shanahan}}, \bibinfo {author} {\bibfnamefont {M.~L.}\ \bibnamefont
  {Wagman}}, \bibinfo {author} {\bibfnamefont {F.}~\bibnamefont {Winter}},
  \bibinfo {author} {\bibfnamefont {E.}~\bibnamefont {Chang}}, \bibinfo
  {author} {\bibfnamefont {W.}~\bibnamefont {Detmold}},\ and\ \bibinfo {author}
  {\bibfnamefont {M.}~\bibnamefont {Illa}} (\bibinfo {collaboration}
  {NPLQCD}),\ }\bibfield  {title} {\bibinfo {title} {{Axial charge of the
  triton from lattice QCD}},\ }\href
  {https://doi.org/10.1103/PhysRevD.103.074511} {\bibfield  {journal} {\bibinfo
   {journal} {Phys. Rev. D}\ }\textbf {\bibinfo {volume} {103}},\ \bibinfo
  {pages} {074511} (\bibinfo {year} {2021})},\ \Eprint
  {https://arxiv.org/abs/2102.03805} {arXiv:2102.03805 [hep-lat]} \BibitemShut
  {NoStop}%
\bibitem [{\citenamefont {Winter}\ \emph {et~al.}(2017)\citenamefont {Winter},
  \citenamefont {Detmold}, \citenamefont {Gambhir}, \citenamefont {Orginos},
  \citenamefont {Savage}, \citenamefont {Shanahan},\ and\ \citenamefont
  {Wagman}}]{Winter:2017bfs}%
  \BibitemOpen
  \bibfield  {author} {\bibinfo {author} {\bibfnamefont {F.}~\bibnamefont
  {Winter}}, \bibinfo {author} {\bibfnamefont {W.}~\bibnamefont {Detmold}},
  \bibinfo {author} {\bibfnamefont {A.~S.}\ \bibnamefont {Gambhir}}, \bibinfo
  {author} {\bibfnamefont {K.}~\bibnamefont {Orginos}}, \bibinfo {author}
  {\bibfnamefont {M.~J.}\ \bibnamefont {Savage}}, \bibinfo {author}
  {\bibfnamefont {P.~E.}\ \bibnamefont {Shanahan}},\ and\ \bibinfo {author}
  {\bibfnamefont {M.~L.}\ \bibnamefont {Wagman}} (\bibinfo {collaboration}
  {NPLQCD}),\ }\bibfield  {title} {\bibinfo {title} {{First lattice QCD study
  of the gluonic structure of light nuclei}},\ }\href
  {https://doi.org/10.1103/PhysRevD.96.094512} {\bibfield  {journal} {\bibinfo
  {journal} {Phys. Rev. D}\ }\textbf {\bibinfo {volume} {96}},\ \bibinfo
  {pages} {094512} (\bibinfo {year} {2017})},\ \Eprint
  {https://arxiv.org/abs/1709.00395} {arXiv:1709.00395 [hep-lat]} \BibitemShut
  {NoStop}%
\bibitem [{\citenamefont {Detmold}\ \emph {et~al.}(2021)\citenamefont
  {Detmold}, \citenamefont {Illa}, \citenamefont {Murphy}, \citenamefont
  {Oare}, \citenamefont {Orginos}, \citenamefont {Shanahan}, \citenamefont
  {Wagman},\ and\ \citenamefont {Winter}}]{Detmold:2020snb}%
  \BibitemOpen
  \bibfield  {author} {\bibinfo {author} {\bibfnamefont {W.}~\bibnamefont
  {Detmold}}, \bibinfo {author} {\bibfnamefont {M.}~\bibnamefont {Illa}},
  \bibinfo {author} {\bibfnamefont {D.~J.}\ \bibnamefont {Murphy}}, \bibinfo
  {author} {\bibfnamefont {P.}~\bibnamefont {Oare}}, \bibinfo {author}
  {\bibfnamefont {K.}~\bibnamefont {Orginos}}, \bibinfo {author} {\bibfnamefont
  {P.~E.}\ \bibnamefont {Shanahan}}, \bibinfo {author} {\bibfnamefont {M.~L.}\
  \bibnamefont {Wagman}},\ and\ \bibinfo {author} {\bibfnamefont
  {F.}~\bibnamefont {Winter}} (\bibinfo {collaboration} {NPLQCD}),\ }\bibfield
  {title} {\bibinfo {title} {Lattice {QCD} constraints on the parton
  distribution functions of {$^3$He}},\ }\href
  {https://doi.org/10.1103/PhysRevLett.126.202001} {\bibfield  {journal}
  {\bibinfo  {journal} {Phys. Rev. Lett.}\ }\textbf {\bibinfo {volume} {126}},\
  \bibinfo {pages} {202001} (\bibinfo {year} {2021})},\ \Eprint
  {https://arxiv.org/abs/2009.05522} {arXiv:2009.05522 [hep-lat]} \BibitemShut
  {NoStop}%
\bibitem [{\citenamefont {Davoudi}\ \emph {et~al.}(2021)\citenamefont
  {Davoudi}, \citenamefont {Detmold}, \citenamefont {Orginos}, \citenamefont
  {Parre\~no}, \citenamefont {Savage}, \citenamefont {Shanahan},\ and\
  \citenamefont {Wagman}}]{Davoudi:2020ngi}%
  \BibitemOpen
  \bibfield  {author} {\bibinfo {author} {\bibfnamefont {Z.}~\bibnamefont
  {Davoudi}}, \bibinfo {author} {\bibfnamefont {W.}~\bibnamefont {Detmold}},
  \bibinfo {author} {\bibfnamefont {K.}~\bibnamefont {Orginos}}, \bibinfo
  {author} {\bibfnamefont {A.}~\bibnamefont {Parre\~no}}, \bibinfo {author}
  {\bibfnamefont {M.~J.}\ \bibnamefont {Savage}}, \bibinfo {author}
  {\bibfnamefont {P.}~\bibnamefont {Shanahan}},\ and\ \bibinfo {author}
  {\bibfnamefont {M.~L.}\ \bibnamefont {Wagman}},\ }\bibfield  {title}
  {\bibinfo {title} {{Nuclear matrix elements from lattice QCD for electroweak
  and beyond-Standard-Model processes}},\ }\href
  {https://doi.org/10.1016/j.physrep.2020.10.004} {\bibfield  {journal}
  {\bibinfo  {journal} {Phys. Rept.}\ }\textbf {\bibinfo {volume} {900}},\
  \bibinfo {pages} {1} (\bibinfo {year} {2021})},\ \Eprint
  {https://arxiv.org/abs/2008.11160} {arXiv:2008.11160 [hep-lat]} \BibitemShut
  {NoStop}%
\bibitem [{\citenamefont {Aad}\ \emph {et~al.}(2012)\citenamefont {Aad} \emph
  {et~al.}}]{Aad:2012tfa}%
  \BibitemOpen
  \bibfield  {author} {\bibinfo {author} {\bibfnamefont {G.}~\bibnamefont
  {Aad}} \emph {et~al.} (\bibinfo {collaboration} {ATLAS}),\ }\bibfield
  {title} {\bibinfo {title} {Observation of a new particle in the search for
  the {Standard Model Higgs} boson with the {ATLAS} detector at the {LHC}},\
  }\href {https://doi.org/10.1016/j.physletb.2012.08.020} {\bibfield  {journal}
  {\bibinfo  {journal} {Phys. Lett. B}\ }\textbf {\bibinfo {volume} {716}},\
  \bibinfo {pages} {1} (\bibinfo {year} {2012})},\ \Eprint
  {https://arxiv.org/abs/1207.7214} {arXiv:1207.7214 [hep-ex]} \BibitemShut
  {NoStop}%
\bibitem [{\citenamefont {Chatrchyan}\ \emph {et~al.}(2012)\citenamefont
  {Chatrchyan} \emph {et~al.}}]{Chatrchyan:2012xdj}%
  \BibitemOpen
  \bibfield  {author} {\bibinfo {author} {\bibfnamefont {S.}~\bibnamefont
  {Chatrchyan}} \emph {et~al.} (\bibinfo {collaboration} {CMS}),\ }\bibfield
  {title} {\bibinfo {title} {Observation of a new boson at a mass of {125~GeV}
  with the {CMS} experiment at the {LHC}},\ }\href
  {https://doi.org/10.1016/j.physletb.2012.08.021} {\bibfield  {journal}
  {\bibinfo  {journal} {Phys. Lett. B}\ }\textbf {\bibinfo {volume} {716}},\
  \bibinfo {pages} {30} (\bibinfo {year} {2012})},\ \Eprint
  {https://arxiv.org/abs/1207.7235} {arXiv:1207.7235 [hep-ex]} \BibitemShut
  {NoStop}%
\bibitem [{\citenamefont {Maltman}\ \emph {et~al.}(2008)\citenamefont
  {Maltman}, \citenamefont {Leinweber}, \citenamefont {Moran},\ and\
  \citenamefont {Sternbeck}}]{Maltman:2008bx}%
  \BibitemOpen
  \bibfield  {author} {\bibinfo {author} {\bibfnamefont {K.}~\bibnamefont
  {Maltman}}, \bibinfo {author} {\bibfnamefont {D.}~\bibnamefont {Leinweber}},
  \bibinfo {author} {\bibfnamefont {P.}~\bibnamefont {Moran}},\ and\ \bibinfo
  {author} {\bibfnamefont {A.}~\bibnamefont {Sternbeck}},\ }\bibfield  {title}
  {\bibinfo {title} {The realistic lattice determination of {$\alpha_s(m_Z)$}
  revisited},\ }\href {https://doi.org/10.1103/PhysRevD.78.114504} {\bibfield
  {journal} {\bibinfo  {journal} {Phys. Rev. D}\ }\textbf {\bibinfo {volume}
  {78}},\ \bibinfo {pages} {114504} (\bibinfo {year} {2008})},\ \Eprint
  {https://arxiv.org/abs/0807.2020} {arXiv:0807.2020 [hep-lat]} \BibitemShut
  {NoStop}%
\bibitem [{\citenamefont {Aoki}\ \emph
  {et~al.}(2009{\natexlab{a}})\citenamefont {Aoki} \emph
  {et~al.}}]{PACS-CS:2009zxm}%
  \BibitemOpen
  \bibfield  {author} {\bibinfo {author} {\bibfnamefont {S.}~\bibnamefont
  {Aoki}} \emph {et~al.} (\bibinfo {collaboration} {PACS-CS}),\ }\bibfield
  {title} {\bibinfo {title} {{Precise determination of the strong coupling
  constant in $N_f=2+1$ lattice QCD with the Schr\"odinger functional
  scheme}},\ }\href {https://doi.org/10.1088/1126-6708/2009/10/053} {\bibfield
  {journal} {\bibinfo  {journal} {JHEP}\ }\textbf {\bibinfo {volume} {10}},\
  \bibinfo {pages} {053}},\ \Eprint {https://arxiv.org/abs/0906.3906}
  {arXiv:0906.3906 [hep-lat]} \BibitemShut {NoStop}%
\bibitem [{\citenamefont {McNeile}\ \emph {et~al.}(2010)\citenamefont
  {McNeile}, \citenamefont {Davies}, \citenamefont {Follana}, \citenamefont
  {Hornbostel},\ and\ \citenamefont {Lepage}}]{McNeile:2010ji}%
  \BibitemOpen
  \bibfield  {author} {\bibinfo {author} {\bibfnamefont {C.}~\bibnamefont
  {McNeile}}, \bibinfo {author} {\bibfnamefont {C.~T.~H.}\ \bibnamefont
  {Davies}}, \bibinfo {author} {\bibfnamefont {E.}~\bibnamefont {Follana}},
  \bibinfo {author} {\bibfnamefont {K.}~\bibnamefont {Hornbostel}},\ and\
  \bibinfo {author} {\bibfnamefont {G.~P.}\ \bibnamefont {Lepage}},\ }\bibfield
   {title} {\bibinfo {title} {High-precision $c$ and $b$ masses, and {QCD}
  coupling from current-current correlators in lattice and continuum {QCD}},\
  }\href {https://doi.org/10.1103/PhysRevD.82.034512} {\bibfield  {journal}
  {\bibinfo  {journal} {Phys. Rev. D}\ }\textbf {\bibinfo {volume} {82}},\
  \bibinfo {pages} {034512} (\bibinfo {year} {2010})},\ \Eprint
  {https://arxiv.org/abs/1004.4285} {arXiv:1004.4285 [hep-lat]} \BibitemShut
  {NoStop}%
\bibitem [{\citenamefont {Maezawa}\ and\ \citenamefont
  {Petreczky}(2016)}]{Maezawa:2016vgv}%
  \BibitemOpen
  \bibfield  {author} {\bibinfo {author} {\bibfnamefont {Y.}~\bibnamefont
  {Maezawa}}\ and\ \bibinfo {author} {\bibfnamefont {P.}~\bibnamefont
  {Petreczky}},\ }\bibfield  {title} {\bibinfo {title} {Quark masses and
  strong-coupling constant in 2+1 flavor {QCD}},\ }\href
  {https://doi.org/10.1103/PhysRevD.94.034507} {\bibfield  {journal} {\bibinfo
  {journal} {Phys. Rev. D}\ }\textbf {\bibinfo {volume} {94}},\ \bibinfo
  {pages} {034507} (\bibinfo {year} {2016})},\ \Eprint
  {https://arxiv.org/abs/1606.08798} {arXiv:1606.08798 [hep-lat]} \BibitemShut
  {NoStop}%
\bibitem [{\citenamefont {Bruno}\ \emph
  {et~al.}(2017{\natexlab{a}})\citenamefont {Bruno}, \citenamefont
  {Dalla~Brida}, \citenamefont {Fritzsch}, \citenamefont {Korzec},
  \citenamefont {Ramos}, \citenamefont {Schaefer}, \citenamefont {Simma},
  \citenamefont {Sint},\ and\ \citenamefont {Sommer}}]{Bruno:2017gxd}%
  \BibitemOpen
  \bibfield  {author} {\bibinfo {author} {\bibfnamefont {M.}~\bibnamefont
  {Bruno}}, \bibinfo {author} {\bibfnamefont {M.}~\bibnamefont {Dalla~Brida}},
  \bibinfo {author} {\bibfnamefont {P.}~\bibnamefont {Fritzsch}}, \bibinfo
  {author} {\bibfnamefont {T.}~\bibnamefont {Korzec}}, \bibinfo {author}
  {\bibfnamefont {A.}~\bibnamefont {Ramos}}, \bibinfo {author} {\bibfnamefont
  {S.}~\bibnamefont {Schaefer}}, \bibinfo {author} {\bibfnamefont
  {H.}~\bibnamefont {Simma}}, \bibinfo {author} {\bibfnamefont
  {S.}~\bibnamefont {Sint}},\ and\ \bibinfo {author} {\bibfnamefont
  {R.}~\bibnamefont {Sommer}} (\bibinfo {collaboration} {ALPHA}),\ }\bibfield
  {title} {\bibinfo {title} {{QCD} coupling from a nonperturbative
  determination of the three-flavor {$\Lambda$} parameter},\ }\href
  {https://doi.org/10.1103/PhysRevLett.119.102001} {\bibfield  {journal}
  {\bibinfo  {journal} {Phys. Rev. Lett.}\ }\textbf {\bibinfo {volume} {119}},\
  \bibinfo {pages} {102001} (\bibinfo {year} {2017}{\natexlab{a}})},\ \Eprint
  {https://arxiv.org/abs/1706.03821} {arXiv:1706.03821 [hep-lat]} \BibitemShut
  {NoStop}%
\bibitem [{\citenamefont {Petreczky}\ and\ \citenamefont
  {Weber}(2019)}]{Petreczky:2019ozv}%
  \BibitemOpen
  \bibfield  {author} {\bibinfo {author} {\bibfnamefont {P.}~\bibnamefont
  {Petreczky}}\ and\ \bibinfo {author} {\bibfnamefont {J.~H.}\ \bibnamefont
  {Weber}},\ }\bibfield  {title} {\bibinfo {title} {{Strong coupling constant
  and heavy quark masses in $(2+1)$-flavor QCD}},\ }\href
  {https://doi.org/10.1103/PhysRevD.100.034519} {\bibfield  {journal} {\bibinfo
   {journal} {Phys. Rev. D}\ }\textbf {\bibinfo {volume} {100}},\ \bibinfo
  {pages} {034519} (\bibinfo {year} {2019})},\ \Eprint
  {https://arxiv.org/abs/1901.06424} {arXiv:1901.06424 [hep-lat]} \BibitemShut
  {NoStop}%
\bibitem [{\citenamefont {Bazavov}\ \emph
  {et~al.}(2019{\natexlab{d}})\citenamefont {Bazavov}, \citenamefont
  {Brambilla}, \citenamefont {Garcia~i Tormo}, \citenamefont {Petreczky},
  \citenamefont {Soto}, \citenamefont {Vairo},\ and\ \citenamefont
  {Weber}}]{Bazavov:2019qoo}%
  \BibitemOpen
  \bibfield  {author} {\bibinfo {author} {\bibfnamefont {A.}~\bibnamefont
  {Bazavov}}, \bibinfo {author} {\bibfnamefont {N.}~\bibnamefont {Brambilla}},
  \bibinfo {author} {\bibfnamefont {X.}~\bibnamefont {Garcia~i Tormo}},
  \bibinfo {author} {\bibfnamefont {P.}~\bibnamefont {Petreczky}}, \bibinfo
  {author} {\bibfnamefont {J.}~\bibnamefont {Soto}}, \bibinfo {author}
  {\bibfnamefont {A.}~\bibnamefont {Vairo}},\ and\ \bibinfo {author}
  {\bibfnamefont {J.~H.}\ \bibnamefont {Weber}} (\bibinfo {collaboration}
  {TUMQCD}),\ }\bibfield  {title} {\bibinfo {title} {{Determination of the QCD
  coupling from the static energy and the free energy}},\ }\href
  {https://doi.org/10.1103/PhysRevD.100.114511} {\bibfield  {journal} {\bibinfo
   {journal} {Phys. Rev. D}\ }\textbf {\bibinfo {volume} {100}},\ \bibinfo
  {pages} {114511} (\bibinfo {year} {2019}{\natexlab{d}})},\ \Eprint
  {https://arxiv.org/abs/1907.11747} {arXiv:1907.11747 [hep-lat]} \BibitemShut
  {NoStop}%
\bibitem [{\citenamefont {Cali}\ \emph {et~al.}(2020)\citenamefont {Cali},
  \citenamefont {Cichy}, \citenamefont {Korcyl},\ and\ \citenamefont
  {Simeth}}]{Cali:2020hrj}%
  \BibitemOpen
  \bibfield  {author} {\bibinfo {author} {\bibfnamefont {S.}~\bibnamefont
  {Cali}}, \bibinfo {author} {\bibfnamefont {K.}~\bibnamefont {Cichy}},
  \bibinfo {author} {\bibfnamefont {P.}~\bibnamefont {Korcyl}},\ and\ \bibinfo
  {author} {\bibfnamefont {J.}~\bibnamefont {Simeth}},\ }\bibfield  {title}
  {\bibinfo {title} {{Running coupling constant from position-space
  current-current correlation functions in three-flavor lattice QCD}},\ }\href
  {https://doi.org/10.1103/PhysRevLett.125.242002} {\bibfield  {journal}
  {\bibinfo  {journal} {Phys. Rev. Lett.}\ }\textbf {\bibinfo {volume} {125}},\
  \bibinfo {pages} {242002} (\bibinfo {year} {2020})},\ \Eprint
  {https://arxiv.org/abs/2003.05781} {arXiv:2003.05781 [hep-lat]} \BibitemShut
  {NoStop}%
\bibitem [{\citenamefont {Ayala}\ \emph {et~al.}(2020)\citenamefont {Ayala},
  \citenamefont {Lobregat},\ and\ \citenamefont {Pineda}}]{Ayala:2020odx}%
  \BibitemOpen
  \bibfield  {author} {\bibinfo {author} {\bibfnamefont {C.}~\bibnamefont
  {Ayala}}, \bibinfo {author} {\bibfnamefont {X.}~\bibnamefont {Lobregat}},\
  and\ \bibinfo {author} {\bibfnamefont {A.}~\bibnamefont {Pineda}},\
  }\bibfield  {title} {\bibinfo {title} {{Determination of $\alpha(m_Z)$ from
  an hyperasymptotic approximation to the energy of a static quark-antiquark
  pair}},\ }\href {https://doi.org/10.1007/JHEP09(2020)016} {\bibfield
  {journal} {\bibinfo  {journal} {JHEP}\ }\textbf {\bibinfo {volume} {09}},\
  \bibinfo {pages} {016}},\ \Eprint {https://arxiv.org/abs/2005.12301}
  {arXiv:2005.12301 [hep-ph]} \BibitemShut {NoStop}%
\bibitem [{\citenamefont {Petreczky}\ and\ \citenamefont
  {Weber}(2022)}]{Petreczky:2020tky}%
  \BibitemOpen
  \bibfield  {author} {\bibinfo {author} {\bibfnamefont {P.}~\bibnamefont
  {Petreczky}}\ and\ \bibinfo {author} {\bibfnamefont {J.~H.}\ \bibnamefont
  {Weber}},\ }\bibfield  {title} {\bibinfo {title} {{Strong coupling constant
  from moments of quarkonium correlators revisited}},\ }\href
  {https://doi.org/10.1140/epjc/s10052-022-09998-0} {\bibfield  {journal}
  {\bibinfo  {journal} {Eur. Phys. J. C}\ }\textbf {\bibinfo {volume} {82}},\
  \bibinfo {pages} {64} (\bibinfo {year} {2022})},\ \Eprint
  {https://arxiv.org/abs/2012.06193} {arXiv:2012.06193 [hep-lat]} \BibitemShut
  {NoStop}%
\bibitem [{\citenamefont {Colquhoun}\ \emph {et~al.}(2015)\citenamefont
  {Colquhoun}, \citenamefont {Dowdall}, \citenamefont {Davies}, \citenamefont
  {Hornbostel},\ and\ \citenamefont {Lepage}}]{Colquhoun:2014ica}%
  \BibitemOpen
  \bibfield  {author} {\bibinfo {author} {\bibfnamefont {B.}~\bibnamefont
  {Colquhoun}}, \bibinfo {author} {\bibfnamefont {R.~J.}\ \bibnamefont
  {Dowdall}}, \bibinfo {author} {\bibfnamefont {C.~T.~H.}\ \bibnamefont
  {Davies}}, \bibinfo {author} {\bibfnamefont {K.}~\bibnamefont {Hornbostel}},\
  and\ \bibinfo {author} {\bibfnamefont {G.~P.}\ \bibnamefont {Lepage}},\
  }\bibfield  {title} {\bibinfo {title} {{$\Upsilon$ and $\Upsilon^{\prime}$
  Leptonic Widths, $a_{\mu}^b$ and $m_b$ from full lattice QCD}},\ }\href
  {https://doi.org/10.1103/PhysRevD.91.074514} {\bibfield  {journal} {\bibinfo
  {journal} {Phys. Rev. D}\ }\textbf {\bibinfo {volume} {91}},\ \bibinfo
  {pages} {074514} (\bibinfo {year} {2015})},\ \Eprint
  {https://arxiv.org/abs/1408.5768} {arXiv:1408.5768 [hep-lat]} \BibitemShut
  {NoStop}%
\bibitem [{\citenamefont {Yang}\ \emph {et~al.}(2015)\citenamefont {Yang} \emph
  {et~al.}}]{Yang:2014sea}%
  \BibitemOpen
  \bibfield  {author} {\bibinfo {author} {\bibfnamefont {Y.-B.}\ \bibnamefont
  {Yang}} \emph {et~al.} (\bibinfo {collaboration} {$\chi$QCD}),\ }\bibfield
  {title} {\bibinfo {title} {Charm and strange quark masses and {$f_{D_s}$}
  from overlap fermions},\ }\href {https://doi.org/10.1103/PhysRevD.92.034517}
  {\bibfield  {journal} {\bibinfo  {journal} {Phys. Rev. D}\ }\textbf {\bibinfo
  {volume} {92}},\ \bibinfo {pages} {034517} (\bibinfo {year} {2015})},\
  \Eprint {https://arxiv.org/abs/1410.3343} {arXiv:1410.3343 [hep-lat]}
  \BibitemShut {NoStop}%
\bibitem [{\citenamefont {Bussone}\ \emph {et~al.}(2016)\citenamefont {Bussone}
  \emph {et~al.}}]{ETM:2016nbo}%
  \BibitemOpen
  \bibfield  {author} {\bibinfo {author} {\bibfnamefont {A.}~\bibnamefont
  {Bussone}} \emph {et~al.} (\bibinfo {collaboration} {ETM}),\ }\bibfield
  {title} {\bibinfo {title} {{Mass of the $b$ quark and $B$-meson decay
  constants from $N_f=2+1+1$ twisted-mass lattice QCD}},\ }\href
  {https://doi.org/10.1103/PhysRevD.93.114505} {\bibfield  {journal} {\bibinfo
  {journal} {Phys. Rev. D}\ }\textbf {\bibinfo {volume} {93}},\ \bibinfo
  {pages} {114505} (\bibinfo {year} {2016})},\ \Eprint
  {https://arxiv.org/abs/1603.04306} {arXiv:1603.04306 [hep-lat]} \BibitemShut
  {NoStop}%
\bibitem [{\citenamefont {Lepage}\ \emph {et~al.}(2014)\citenamefont {Lepage},
  \citenamefont {Mackenzie},\ and\ \citenamefont {Peskin}}]{Lepage:2014fla}%
  \BibitemOpen
  \bibfield  {author} {\bibinfo {author} {\bibfnamefont {G.~P.}\ \bibnamefont
  {Lepage}}, \bibinfo {author} {\bibfnamefont {P.~B.}\ \bibnamefont
  {Mackenzie}},\ and\ \bibinfo {author} {\bibfnamefont {M.~E.}\ \bibnamefont
  {Peskin}},\ }\bibfield  {title} {\bibinfo {title} {Expected precision of
  {Higgs} boson partial widths within the {Standard Model}},\ }\href@noop {}
  {\bibfield  {journal} {\bibinfo  {journal} {unpublished}\ } (\bibinfo {year}
  {2014})},\ \Eprint {https://arxiv.org/abs/1404.0319} {arXiv:1404.0319
  [hep-ph]} \BibitemShut {NoStop}%
\bibitem [{\citenamefont {Bazavov}\ \emph {et~al.}(2010)\citenamefont {Bazavov}
  \emph {et~al.}}]{MILC:2010pul}%
  \BibitemOpen
  \bibfield  {author} {\bibinfo {author} {\bibfnamefont {A.}~\bibnamefont
  {Bazavov}} \emph {et~al.} (\bibinfo {collaboration} {MILC}),\ }\bibfield
  {title} {\bibinfo {title} {Scaling studies of {QCD} with the dynamical {HISQ}
  action},\ }\href {https://doi.org/10.1103/PhysRevD.82.074501} {\bibfield
  {journal} {\bibinfo  {journal} {Phys. Rev. D}\ }\textbf {\bibinfo {volume}
  {82}},\ \bibinfo {pages} {074501} (\bibinfo {year} {2010})},\ \Eprint
  {https://arxiv.org/abs/1004.0342} {arXiv:1004.0342 [hep-lat]} \BibitemShut
  {NoStop}%
\bibitem [{\citenamefont {Bazavov}\ \emph
  {et~al.}(2013{\natexlab{b}})\citenamefont {Bazavov} \emph
  {et~al.}}]{MILC:2012znn}%
  \BibitemOpen
  \bibfield  {author} {\bibinfo {author} {\bibfnamefont {A.}~\bibnamefont
  {Bazavov}} \emph {et~al.} (\bibinfo {collaboration} {MILC}),\ }\bibfield
  {title} {\bibinfo {title} {Lattice {QCD} ensembles with four flavors of
  highly improved staggered quarks},\ }\href
  {https://doi.org/10.1103/PhysRevD.87.054505} {\bibfield  {journal} {\bibinfo
  {journal} {Phys. Rev. D}\ }\textbf {\bibinfo {volume} {87}},\ \bibinfo
  {pages} {054505} (\bibinfo {year} {2013}{\natexlab{b}})},\ \Eprint
  {https://arxiv.org/abs/1212.4768} {arXiv:1212.4768 [hep-lat]} \BibitemShut
  {NoStop}%
\bibitem [{\citenamefont {Alexandrou}\ \emph
  {et~al.}(2021{\natexlab{c}})\citenamefont {Alexandrou} \emph
  {et~al.}}]{ExtendedTwistedMass:2021gbo}%
  \BibitemOpen
  \bibfield  {author} {\bibinfo {author} {\bibfnamefont {C.}~\bibnamefont
  {Alexandrou}} \emph {et~al.} (\bibinfo {collaboration} {Extended Twisted
  Mass}),\ }\bibfield  {title} {\bibinfo {title} {{Quark masses using
  twisted-mass fermion gauge ensembles}},\ }\href
  {https://doi.org/10.1103/PhysRevD.104.074515} {\bibfield  {journal} {\bibinfo
   {journal} {Phys. Rev. D}\ }\textbf {\bibinfo {volume} {104}},\ \bibinfo
  {pages} {074515} (\bibinfo {year} {2021}{\natexlab{c}})},\ \Eprint
  {https://arxiv.org/abs/2104.13408} {arXiv:2104.13408 [hep-lat]} \BibitemShut
  {NoStop}%
\bibitem [{\citenamefont {d'Enterria}\ \emph {et~al.}(2022)\citenamefont
  {d'Enterria} \emph {et~al.}}]{dEnterria:2022hzv}%
  \BibitemOpen
  \bibfield  {author} {\bibinfo {author} {\bibfnamefont {D.}~\bibnamefont
  {d'Enterria}} \emph {et~al.},\ }\bibfield  {title} {\bibinfo {title} {{The
  strong coupling constant: State of the art and the decade ahead}},\ }in\
  \href@noop {} {\emph {\bibinfo {booktitle} {{2022 Snowmass Summer Study}}}}\
  (\bibinfo {year} {2022})\ \Eprint {https://arxiv.org/abs/2203.08271}
  {arXiv:2203.08271 [hep-ph]} \BibitemShut {NoStop}%
\bibitem [{\citenamefont {Mondal}\ \emph
  {et~al.}(2020{\natexlab{a}})\citenamefont {Mondal}, \citenamefont {Gupta},
  \citenamefont {Park}, \citenamefont {Yoon}, \citenamefont {Bhattacharya},\
  and\ \citenamefont {Lin}}]{Mondal:2020cmt}%
  \BibitemOpen
  \bibfield  {author} {\bibinfo {author} {\bibfnamefont {S.}~\bibnamefont
  {Mondal}}, \bibinfo {author} {\bibfnamefont {R.}~\bibnamefont {Gupta}},
  \bibinfo {author} {\bibfnamefont {S.}~\bibnamefont {Park}}, \bibinfo {author}
  {\bibfnamefont {B.}~\bibnamefont {Yoon}}, \bibinfo {author} {\bibfnamefont
  {T.}~\bibnamefont {Bhattacharya}},\ and\ \bibinfo {author} {\bibfnamefont
  {H.-W.}\ \bibnamefont {Lin}},\ }\bibfield  {title} {\bibinfo {title}
  {{Moments of nucleon isovector structure functions in $2+1+1$-flavor QCD}},\
  }\href {https://doi.org/10.1103/PhysRevD.102.054512} {\bibfield  {journal}
  {\bibinfo  {journal} {Phys. Rev. D}\ }\textbf {\bibinfo {volume} {102}},\
  \bibinfo {pages} {054512} (\bibinfo {year} {2020}{\natexlab{a}})},\ \Eprint
  {https://arxiv.org/abs/2005.13779} {arXiv:2005.13779 [hep-lat]} \BibitemShut
  {NoStop}%
\bibitem [{\citenamefont {Mondal}\ \emph
  {et~al.}(2020{\natexlab{b}})\citenamefont {Mondal}, \citenamefont {Gupta},
  \citenamefont {Park}, \citenamefont {Yoon}, \citenamefont {Bhattacharya},
  \citenamefont {Jo\'o},\ and\ \citenamefont {Winter}}]{Mondal:2020ela}%
  \BibitemOpen
  \bibfield  {author} {\bibinfo {author} {\bibfnamefont {S.}~\bibnamefont
  {Mondal}}, \bibinfo {author} {\bibfnamefont {R.}~\bibnamefont {Gupta}},
  \bibinfo {author} {\bibfnamefont {S.}~\bibnamefont {Park}}, \bibinfo {author}
  {\bibfnamefont {B.}~\bibnamefont {Yoon}}, \bibinfo {author} {\bibfnamefont
  {T.}~\bibnamefont {Bhattacharya}}, \bibinfo {author} {\bibfnamefont
  {B.}~\bibnamefont {Jo\'o}},\ and\ \bibinfo {author} {\bibfnamefont
  {F.}~\bibnamefont {Winter}} (\bibinfo {collaboration} {NME}),\ }\bibfield
  {title} {\bibinfo {title} {{Nucleon momentum fraction, helicity and
  transversity from 2+1-flavor lattice QCD}},\ }\href
  {https://doi.org/10.1007/JHEP04(2021)044} {\bibfield  {journal} {\bibinfo
  {journal} {JHEP}\ }\textbf {\bibinfo {volume} {21}},\ \bibinfo {pages}
  {004}},\ \Eprint {https://arxiv.org/abs/2011.12787} {arXiv:2011.12787
  [hep-lat]} \BibitemShut {NoStop}%
\bibitem [{\citenamefont {Davoudi}\ and\ \citenamefont
  {Savage}(2012)}]{Davoudi:2012ya}%
  \BibitemOpen
  \bibfield  {author} {\bibinfo {author} {\bibfnamefont {Z.}~\bibnamefont
  {Davoudi}}\ and\ \bibinfo {author} {\bibfnamefont {M.~J.}\ \bibnamefont
  {Savage}},\ }\bibfield  {title} {\bibinfo {title} {Restoration of rotational
  symmetry in the continuum limit of lattice field theories},\ }\href
  {https://doi.org/10.1103/PhysRevD.86.054505} {\bibfield  {journal} {\bibinfo
  {journal} {Phys. Rev. D}\ }\textbf {\bibinfo {volume} {86}},\ \bibinfo
  {pages} {054505} (\bibinfo {year} {2012})},\ \Eprint
  {https://arxiv.org/abs/1204.4146} {arXiv:1204.4146 [hep-lat]} \BibitemShut
  {NoStop}%
\bibitem [{\citenamefont {Monahan}\ and\ \citenamefont
  {Orginos}(2015)}]{Monahan:2015lha}%
  \BibitemOpen
  \bibfield  {author} {\bibinfo {author} {\bibfnamefont {C.}~\bibnamefont
  {Monahan}}\ and\ \bibinfo {author} {\bibfnamefont {K.}~\bibnamefont
  {Orginos}},\ }\bibfield  {title} {\bibinfo {title} {{Locally smeared operator
  product expansions in scalar field theory}},\ }\href
  {https://doi.org/10.1103/PhysRevD.91.074513} {\bibfield  {journal} {\bibinfo
  {journal} {Phys. Rev. D}\ }\textbf {\bibinfo {volume} {91}},\ \bibinfo
  {pages} {074513} (\bibinfo {year} {2015})},\ \Eprint
  {https://arxiv.org/abs/1501.05348} {arXiv:1501.05348 [hep-lat]} \BibitemShut
  {NoStop}%
\bibitem [{\citenamefont {Detmold}\ and\ \citenamefont
  {Lin}(2006)}]{Detmold:2005gg}%
  \BibitemOpen
  \bibfield  {author} {\bibinfo {author} {\bibfnamefont {W.}~\bibnamefont
  {Detmold}}\ and\ \bibinfo {author} {\bibfnamefont {C.~J.~D.}\ \bibnamefont
  {Lin}},\ }\bibfield  {title} {\bibinfo {title} {{Deep-inelastic scattering
  and the operator product expansion in lattice QCD}},\ }\href
  {https://doi.org/10.1103/PhysRevD.73.014501} {\bibfield  {journal} {\bibinfo
  {journal} {Phys. Rev. D}\ }\textbf {\bibinfo {volume} {73}},\ \bibinfo
  {pages} {014501} (\bibinfo {year} {2006})},\ \Eprint
  {https://arxiv.org/abs/hep-lat/0507007} {arXiv:hep-lat/0507007} \BibitemShut
  {NoStop}%
\bibitem [{\citenamefont {Braun}\ and\ \citenamefont
  {M\"uller}(2008)}]{Braun:2007wv}%
  \BibitemOpen
  \bibfield  {author} {\bibinfo {author} {\bibfnamefont {V.}~\bibnamefont
  {Braun}}\ and\ \bibinfo {author} {\bibfnamefont {D.}~\bibnamefont
  {M\"uller}},\ }\bibfield  {title} {\bibinfo {title} {{Exclusive processes in
  position space and the pion distribution amplitude}},\ }\href
  {https://doi.org/10.1140/epjc/s10052-008-0608-4} {\bibfield  {journal}
  {\bibinfo  {journal} {Eur. Phys. J. C}\ }\textbf {\bibinfo {volume} {55}},\
  \bibinfo {pages} {349} (\bibinfo {year} {2008})},\ \Eprint
  {https://arxiv.org/abs/0709.1348} {arXiv:0709.1348 [hep-ph]} \BibitemShut
  {NoStop}%
\bibitem [{\citenamefont {Chambers}\ \emph {et~al.}(2017)\citenamefont
  {Chambers}, \citenamefont {Horsley}, \citenamefont {Nakamura}, \citenamefont
  {Perlt}, \citenamefont {Rakow}, \citenamefont {Schierholz}, \citenamefont
  {Schiller}, \citenamefont {Somfleth}, \citenamefont {Young},\ and\
  \citenamefont {Zanotti}}]{Chambers:2017dov}%
  \BibitemOpen
  \bibfield  {author} {\bibinfo {author} {\bibfnamefont {A.~J.}\ \bibnamefont
  {Chambers}}, \bibinfo {author} {\bibfnamefont {R.}~\bibnamefont {Horsley}},
  \bibinfo {author} {\bibfnamefont {Y.}~\bibnamefont {Nakamura}}, \bibinfo
  {author} {\bibfnamefont {H.}~\bibnamefont {Perlt}}, \bibinfo {author}
  {\bibfnamefont {P.~E.~L.}\ \bibnamefont {Rakow}}, \bibinfo {author}
  {\bibfnamefont {G.}~\bibnamefont {Schierholz}}, \bibinfo {author}
  {\bibfnamefont {A.}~\bibnamefont {Schiller}}, \bibinfo {author}
  {\bibfnamefont {K.}~\bibnamefont {Somfleth}}, \bibinfo {author}
  {\bibfnamefont {R.~D.}\ \bibnamefont {Young}},\ and\ \bibinfo {author}
  {\bibfnamefont {J.~M.}\ \bibnamefont {Zanotti}},\ }\bibfield  {title}
  {\bibinfo {title} {Nucleon structure functions from operator product
  expansion on the lattice},\ }\href
  {https://doi.org/10.1103/PhysRevLett.118.242001} {\bibfield  {journal}
  {\bibinfo  {journal} {Phys. Rev. Lett.}\ }\textbf {\bibinfo {volume} {118}},\
  \bibinfo {pages} {242001} (\bibinfo {year} {2017})},\ \Eprint
  {https://arxiv.org/abs/1703.01153} {arXiv:1703.01153 [hep-lat]} \BibitemShut
  {NoStop}%
\bibitem [{\citenamefont {Liu}\ and\ \citenamefont {Dong}(1994)}]{Liu:1993cv}%
  \BibitemOpen
  \bibfield  {author} {\bibinfo {author} {\bibfnamefont {K.-F.}\ \bibnamefont
  {Liu}}\ and\ \bibinfo {author} {\bibfnamefont {S.-J.}\ \bibnamefont {Dong}},\
  }\bibfield  {title} {\bibinfo {title} {{Origin of difference between
  $\bar{d}$ and $\bar{u}$ partons in the nucleon}},\ }\href
  {https://doi.org/10.1103/PhysRevLett.72.1790} {\bibfield  {journal} {\bibinfo
   {journal} {Phys. Rev. Lett.}\ }\textbf {\bibinfo {volume} {72}},\ \bibinfo
  {pages} {1790} (\bibinfo {year} {1994})},\ \Eprint
  {https://arxiv.org/abs/hep-ph/9306299} {arXiv:hep-ph/9306299} \BibitemShut
  {NoStop}%
\bibitem [{\citenamefont {Ma}\ and\ \citenamefont
  {Qiu}(2018{\natexlab{a}})}]{Ma:2014jla}%
  \BibitemOpen
  \bibfield  {author} {\bibinfo {author} {\bibfnamefont {Y.-Q.}\ \bibnamefont
  {Ma}}\ and\ \bibinfo {author} {\bibfnamefont {J.-W.}\ \bibnamefont {Qiu}},\
  }\bibfield  {title} {\bibinfo {title} {Extracting parton distribution
  functions from lattice {QCD} calculations},\ }\href
  {https://doi.org/10.1103/PhysRevD.98.074021} {\bibfield  {journal} {\bibinfo
  {journal} {Phys. Rev. D}\ }\textbf {\bibinfo {volume} {98}},\ \bibinfo
  {pages} {074021} (\bibinfo {year} {2018}{\natexlab{a}})},\ \Eprint
  {https://arxiv.org/abs/1404.6860} {arXiv:1404.6860 [hep-ph]} \BibitemShut
  {NoStop}%
\bibitem [{\citenamefont {Ma}\ and\ \citenamefont
  {Qiu}(2018{\natexlab{b}})}]{Ma:2017pxb}%
  \BibitemOpen
  \bibfield  {author} {\bibinfo {author} {\bibfnamefont {Y.-Q.}\ \bibnamefont
  {Ma}}\ and\ \bibinfo {author} {\bibfnamefont {J.-W.}\ \bibnamefont {Qiu}},\
  }\bibfield  {title} {\bibinfo {title} {Exploring partonic structure of
  hadrons using \emph{ab initio} lattice {QCD} calculations},\ }\href
  {https://doi.org/10.1103/PhysRevLett.120.022003} {\bibfield  {journal}
  {\bibinfo  {journal} {Phys. Rev. Lett.}\ }\textbf {\bibinfo {volume} {120}},\
  \bibinfo {pages} {022003} (\bibinfo {year} {2018}{\natexlab{b}})},\ \Eprint
  {https://arxiv.org/abs/1709.03018} {arXiv:1709.03018 [hep-ph]} \BibitemShut
  {NoStop}%
\bibitem [{\citenamefont {Ji}(2013)}]{Ji:2013dva}%
  \BibitemOpen
  \bibfield  {author} {\bibinfo {author} {\bibfnamefont {X.}~\bibnamefont
  {Ji}},\ }\bibfield  {title} {\bibinfo {title} {Parton physics on a
  {Euclidean} lattice},\ }\href
  {https://doi.org/10.1103/PhysRevLett.110.262002} {\bibfield  {journal}
  {\bibinfo  {journal} {Phys. Rev. Lett.}\ }\textbf {\bibinfo {volume} {110}},\
  \bibinfo {pages} {262002} (\bibinfo {year} {2013})},\ \Eprint
  {https://arxiv.org/abs/1305.1539} {arXiv:1305.1539 [hep-ph]} \BibitemShut
  {NoStop}%
\bibitem [{\citenamefont {Xiong}\ \emph {et~al.}(2014)\citenamefont {Xiong},
  \citenamefont {Ji}, \citenamefont {Zhang},\ and\ \citenamefont
  {Zhao}}]{Xiong:2013bka}%
  \BibitemOpen
  \bibfield  {author} {\bibinfo {author} {\bibfnamefont {X.}~\bibnamefont
  {Xiong}}, \bibinfo {author} {\bibfnamefont {X.}~\bibnamefont {Ji}}, \bibinfo
  {author} {\bibfnamefont {J.-H.}\ \bibnamefont {Zhang}},\ and\ \bibinfo
  {author} {\bibfnamefont {Y.}~\bibnamefont {Zhao}},\ }\bibfield  {title}
  {\bibinfo {title} {{One-loop matching for parton distributions: Nonsinglet
  case}},\ }\href {https://doi.org/10.1103/PhysRevD.90.014051} {\bibfield
  {journal} {\bibinfo  {journal} {Phys. Rev. D}\ }\textbf {\bibinfo {volume}
  {90}},\ \bibinfo {pages} {014051} (\bibinfo {year} {2014})},\ \Eprint
  {https://arxiv.org/abs/1310.7471} {arXiv:1310.7471 [hep-ph]} \BibitemShut
  {NoStop}%
\bibitem [{\citenamefont {Ji}(2014)}]{Ji:2014gla}%
  \BibitemOpen
  \bibfield  {author} {\bibinfo {author} {\bibfnamefont {X.}~\bibnamefont
  {Ji}},\ }\bibfield  {title} {\bibinfo {title} {Parton physics from
  large-momentum effective field theory},\ }\href
  {https://doi.org/10.1007/s11433-014-5492-3} {\bibfield  {journal} {\bibinfo
  {journal} {Sci. China Phys. Mech. Astron.}\ }\textbf {\bibinfo {volume}
  {57}},\ \bibinfo {pages} {1407} (\bibinfo {year} {2014})},\ \Eprint
  {https://arxiv.org/abs/1404.6680} {arXiv:1404.6680 [hep-ph]} \BibitemShut
  {NoStop}%
\bibitem [{\citenamefont {Ji}\ \emph {et~al.}(2021{\natexlab{a}})\citenamefont
  {Ji}, \citenamefont {Liu}, \citenamefont {Liu}, \citenamefont {Zhang},\ and\
  \citenamefont {Zhao}}]{Ji:2020ect}%
  \BibitemOpen
  \bibfield  {author} {\bibinfo {author} {\bibfnamefont {X.}~\bibnamefont
  {Ji}}, \bibinfo {author} {\bibfnamefont {Y.-S.}\ \bibnamefont {Liu}},
  \bibinfo {author} {\bibfnamefont {Y.}~\bibnamefont {Liu}}, \bibinfo {author}
  {\bibfnamefont {J.-H.}\ \bibnamefont {Zhang}},\ and\ \bibinfo {author}
  {\bibfnamefont {Y.}~\bibnamefont {Zhao}},\ }\bibfield  {title} {\bibinfo
  {title} {Large-momentum effective theory},\ }\href
  {https://doi.org/10.1103/RevModPhys.93.035005} {\bibfield  {journal}
  {\bibinfo  {journal} {Rev. Mod. Phys.}\ }\textbf {\bibinfo {volume} {93}},\
  \bibinfo {pages} {035005} (\bibinfo {year} {2021}{\natexlab{a}})},\ \Eprint
  {https://arxiv.org/abs/2004.03543} {arXiv:2004.03543 [hep-ph]} \BibitemShut
  {NoStop}%
\bibitem [{\citenamefont {Lin}\ \emph {et~al.}(2015)\citenamefont {Lin},
  \citenamefont {Chen}, \citenamefont {Cohen},\ and\ \citenamefont
  {Ji}}]{Lin:2014zya}%
  \BibitemOpen
  \bibfield  {author} {\bibinfo {author} {\bibfnamefont {H.-W.}\ \bibnamefont
  {Lin}}, \bibinfo {author} {\bibfnamefont {J.-W.}\ \bibnamefont {Chen}},
  \bibinfo {author} {\bibfnamefont {S.~D.}\ \bibnamefont {Cohen}},\ and\
  \bibinfo {author} {\bibfnamefont {X.}~\bibnamefont {Ji}},\ }\bibfield
  {title} {\bibinfo {title} {Flavor structure of the nucleon sea from lattice
  {QCD}},\ }\href {https://doi.org/10.1103/PhysRevD.91.054510} {\bibfield
  {journal} {\bibinfo  {journal} {Phys. Rev. D}\ }\textbf {\bibinfo {volume}
  {91}},\ \bibinfo {pages} {054510} (\bibinfo {year} {2015})},\ \Eprint
  {https://arxiv.org/abs/1402.1462} {arXiv:1402.1462 [hep-ph]} \BibitemShut
  {NoStop}%
\bibitem [{\citenamefont {Alexandrou}\ \emph {et~al.}(2015)\citenamefont
  {Alexandrou}, \citenamefont {Cichy}, \citenamefont {Drach}, \citenamefont
  {Garcia-Ramos}, \citenamefont {Hadjiyiannakou}, \citenamefont {Jansen},
  \citenamefont {Steffens},\ and\ \citenamefont {Wiese}}]{Alexandrou:2015rja}%
  \BibitemOpen
  \bibfield  {author} {\bibinfo {author} {\bibfnamefont {C.}~\bibnamefont
  {Alexandrou}}, \bibinfo {author} {\bibfnamefont {K.}~\bibnamefont {Cichy}},
  \bibinfo {author} {\bibfnamefont {V.}~\bibnamefont {Drach}}, \bibinfo
  {author} {\bibfnamefont {E.}~\bibnamefont {Garcia-Ramos}}, \bibinfo {author}
  {\bibfnamefont {K.}~\bibnamefont {Hadjiyiannakou}}, \bibinfo {author}
  {\bibfnamefont {K.}~\bibnamefont {Jansen}}, \bibinfo {author} {\bibfnamefont
  {F.}~\bibnamefont {Steffens}},\ and\ \bibinfo {author} {\bibfnamefont
  {C.}~\bibnamefont {Wiese}},\ }\bibfield  {title} {\bibinfo {title} {{Lattice
  calculation of parton distributions}},\ }\href
  {https://doi.org/10.1103/PhysRevD.92.014502} {\bibfield  {journal} {\bibinfo
  {journal} {Phys. Rev. D}\ }\textbf {\bibinfo {volume} {92}},\ \bibinfo
  {pages} {014502} (\bibinfo {year} {2015})},\ \Eprint
  {https://arxiv.org/abs/1504.07455} {arXiv:1504.07455 [hep-lat]} \BibitemShut
  {NoStop}%
\bibitem [{\citenamefont {Ji}\ and\ \citenamefont {Zhang}(2015)}]{Ji:2015jwa}%
  \BibitemOpen
  \bibfield  {author} {\bibinfo {author} {\bibfnamefont {X.}~\bibnamefont
  {Ji}}\ and\ \bibinfo {author} {\bibfnamefont {J.-H.}\ \bibnamefont {Zhang}},\
  }\bibfield  {title} {\bibinfo {title} {{Renormalization of quasiparton
  distribution}},\ }\href {https://doi.org/10.1103/PhysRevD.92.034006}
  {\bibfield  {journal} {\bibinfo  {journal} {Phys. Rev. D}\ }\textbf {\bibinfo
  {volume} {92}},\ \bibinfo {pages} {034006} (\bibinfo {year} {2015})},\
  \Eprint {https://arxiv.org/abs/1505.07699} {arXiv:1505.07699 [hep-ph]}
  \BibitemShut {NoStop}%
\bibitem [{\citenamefont {Monahan}\ and\ \citenamefont
  {Orginos}(2017)}]{Monahan:2016bvm}%
  \BibitemOpen
  \bibfield  {author} {\bibinfo {author} {\bibfnamefont {C.}~\bibnamefont
  {Monahan}}\ and\ \bibinfo {author} {\bibfnamefont {K.}~\bibnamefont
  {Orginos}},\ }\bibfield  {title} {\bibinfo {title} {{Quasi parton
  distributions and the gradient flow}},\ }\href
  {https://doi.org/10.1007/JHEP03(2017)116} {\bibfield  {journal} {\bibinfo
  {journal} {JHEP}\ }\textbf {\bibinfo {volume} {03}},\ \bibinfo {pages}
  {116}},\ \Eprint {https://arxiv.org/abs/1612.01584} {arXiv:1612.01584
  [hep-lat]} \BibitemShut {NoStop}%
\bibitem [{\citenamefont
  {Radyushkin}(2017{\natexlab{a}})}]{Radyushkin:2016hsy}%
  \BibitemOpen
  \bibfield  {author} {\bibinfo {author} {\bibfnamefont {A.~V.}\ \bibnamefont
  {Radyushkin}},\ }\bibfield  {title} {\bibinfo {title} {Nonperturbative
  evolution of parton quasi-distributions},\ }\href
  {https://doi.org/10.1016/j.physletb.2017.02.019} {\bibfield  {journal}
  {\bibinfo  {journal} {Phys. Lett. B}\ }\textbf {\bibinfo {volume} {767}},\
  \bibinfo {pages} {314} (\bibinfo {year} {2017}{\natexlab{a}})},\ \Eprint
  {https://arxiv.org/abs/1612.05170} {arXiv:1612.05170 [hep-ph]} \BibitemShut
  {NoStop}%
\bibitem [{\citenamefont
  {Radyushkin}(2017{\natexlab{b}})}]{Radyushkin:2017cyf}%
  \BibitemOpen
  \bibfield  {author} {\bibinfo {author} {\bibfnamefont {A.~V.}\ \bibnamefont
  {Radyushkin}},\ }\bibfield  {title} {\bibinfo {title} {{Quasi-parton
  distribution functions, momentum distributions, and pseudo-parton
  distribution functions}},\ }\href
  {https://doi.org/10.1103/PhysRevD.96.034025} {\bibfield  {journal} {\bibinfo
  {journal} {Phys. Rev. D}\ }\textbf {\bibinfo {volume} {96}},\ \bibinfo
  {pages} {034025} (\bibinfo {year} {2017}{\natexlab{b}})},\ \Eprint
  {https://arxiv.org/abs/1705.01488} {arXiv:1705.01488 [hep-ph]} \BibitemShut
  {NoStop}%
\bibitem [{\citenamefont {Radyushkin}(2019)}]{Radyushkin:2018nbf}%
  \BibitemOpen
  \bibfield  {author} {\bibinfo {author} {\bibfnamefont {A.~V.}\ \bibnamefont
  {Radyushkin}},\ }\bibfield  {title} {\bibinfo {title} {{Structure of parton
  quasi-distributions and their moments}},\ }\href
  {https://doi.org/10.1016/j.physletb.2018.11.047} {\bibfield  {journal}
  {\bibinfo  {journal} {Phys. Lett. B}\ }\textbf {\bibinfo {volume} {788}},\
  \bibinfo {pages} {380} (\bibinfo {year} {2019})},\ \Eprint
  {https://arxiv.org/abs/1807.07509} {arXiv:1807.07509 [hep-ph]} \BibitemShut
  {NoStop}%
\bibitem [{\citenamefont {Constantinou}\ and\ \citenamefont
  {Panagopoulos}(2017)}]{Constantinou:2017sej}%
  \BibitemOpen
  \bibfield  {author} {\bibinfo {author} {\bibfnamefont {M.}~\bibnamefont
  {Constantinou}}\ and\ \bibinfo {author} {\bibfnamefont {H.}~\bibnamefont
  {Panagopoulos}},\ }\bibfield  {title} {\bibinfo {title} {Perturbative
  renormalization of quasi-parton distribution functions},\ }\href
  {https://doi.org/10.1103/PhysRevD.96.054506} {\bibfield  {journal} {\bibinfo
  {journal} {Phys. Rev. D}\ }\textbf {\bibinfo {volume} {96}},\ \bibinfo
  {pages} {054506} (\bibinfo {year} {2017})},\ \Eprint
  {https://arxiv.org/abs/1705.11193} {arXiv:1705.11193 [hep-lat]} \BibitemShut
  {NoStop}%
\bibitem [{\citenamefont {Alexandrou}\ \emph
  {et~al.}(2017{\natexlab{a}})\citenamefont {Alexandrou}, \citenamefont
  {Cichy}, \citenamefont {Constantinou}, \citenamefont {Hadjiyiannakou},
  \citenamefont {Jansen}, \citenamefont {Panagopoulos},\ and\ \citenamefont
  {Steffens}}]{Alexandrou:2017huk}%
  \BibitemOpen
  \bibfield  {author} {\bibinfo {author} {\bibfnamefont {C.}~\bibnamefont
  {Alexandrou}}, \bibinfo {author} {\bibfnamefont {K.}~\bibnamefont {Cichy}},
  \bibinfo {author} {\bibfnamefont {M.}~\bibnamefont {Constantinou}}, \bibinfo
  {author} {\bibfnamefont {K.}~\bibnamefont {Hadjiyiannakou}}, \bibinfo
  {author} {\bibfnamefont {K.}~\bibnamefont {Jansen}}, \bibinfo {author}
  {\bibfnamefont {H.}~\bibnamefont {Panagopoulos}},\ and\ \bibinfo {author}
  {\bibfnamefont {F.}~\bibnamefont {Steffens}},\ }\bibfield  {title} {\bibinfo
  {title} {A complete non-perturbative renormalization prescription for
  quasi-{PDF}s},\ }\href {https://doi.org/10.1016/j.nuclphysb.2017.08.012}
  {\bibfield  {journal} {\bibinfo  {journal} {Nucl. Phys. B}\ }\textbf
  {\bibinfo {volume} {923}},\ \bibinfo {pages} {394} (\bibinfo {year}
  {2017}{\natexlab{a}})},\ \Eprint {https://arxiv.org/abs/1706.00265}
  {arXiv:1706.00265 [hep-lat]} \BibitemShut {NoStop}%
\bibitem [{\citenamefont {Chen}\ \emph {et~al.}(2018)\citenamefont {Chen} \emph
  {et~al.}}]{Chen:2017mzz}%
  \BibitemOpen
  \bibfield  {author} {\bibinfo {author} {\bibfnamefont {J.-W.}\ \bibnamefont
  {Chen}} \emph {et~al.} (\bibinfo {collaboration} {LP$^3$}),\ }\bibfield
  {title} {\bibinfo {title} {Parton distribution function with nonperturbative
  renormalization from lattice {QCD}},\ }\href
  {https://doi.org/10.1103/PhysRevD.97.014505} {\bibfield  {journal} {\bibinfo
  {journal} {Phys. Rev. D}\ }\textbf {\bibinfo {volume} {97}},\ \bibinfo
  {pages} {014505} (\bibinfo {year} {2018})},\ \Eprint
  {https://arxiv.org/abs/1706.01295} {arXiv:1706.01295 [hep-lat]} \BibitemShut
  {NoStop}%
\bibitem [{\citenamefont {Ji}\ \emph {et~al.}(2018)\citenamefont {Ji},
  \citenamefont {Zhang},\ and\ \citenamefont {Zhao}}]{Ji:2017oey}%
  \BibitemOpen
  \bibfield  {author} {\bibinfo {author} {\bibfnamefont {X.}~\bibnamefont
  {Ji}}, \bibinfo {author} {\bibfnamefont {J.-H.}\ \bibnamefont {Zhang}},\ and\
  \bibinfo {author} {\bibfnamefont {Y.}~\bibnamefont {Zhao}},\ }\bibfield
  {title} {\bibinfo {title} {Renormalization in large momentum effective theory
  of parton physics},\ }\href {https://doi.org/10.1103/PhysRevLett.120.112001}
  {\bibfield  {journal} {\bibinfo  {journal} {Phys. Rev. Lett.}\ }\textbf
  {\bibinfo {volume} {120}},\ \bibinfo {pages} {112001} (\bibinfo {year}
  {2018})},\ \Eprint {https://arxiv.org/abs/1706.08962} {arXiv:1706.08962
  [hep-ph]} \BibitemShut {NoStop}%
\bibitem [{\citenamefont {Ishikawa}\ \emph {et~al.}(2017)\citenamefont
  {Ishikawa}, \citenamefont {Ma}, \citenamefont {Qiu},\ and\ \citenamefont
  {Yoshida}}]{Ishikawa:2017faj}%
  \BibitemOpen
  \bibfield  {author} {\bibinfo {author} {\bibfnamefont {T.}~\bibnamefont
  {Ishikawa}}, \bibinfo {author} {\bibfnamefont {Y.-Q.}\ \bibnamefont {Ma}},
  \bibinfo {author} {\bibfnamefont {J.-W.}\ \bibnamefont {Qiu}},\ and\ \bibinfo
  {author} {\bibfnamefont {S.}~\bibnamefont {Yoshida}},\ }\bibfield  {title}
  {\bibinfo {title} {{Renormalizability of quasiparton distribution
  functions}},\ }\href {https://doi.org/10.1103/PhysRevD.96.094019} {\bibfield
  {journal} {\bibinfo  {journal} {Phys. Rev. D}\ }\textbf {\bibinfo {volume}
  {96}},\ \bibinfo {pages} {094019} (\bibinfo {year} {2017})},\ \Eprint
  {https://arxiv.org/abs/1707.03107} {arXiv:1707.03107 [hep-ph]} \BibitemShut
  {NoStop}%
\bibitem [{\citenamefont {Green}\ \emph {et~al.}(2018)\citenamefont {Green},
  \citenamefont {Jansen},\ and\ \citenamefont {Steffens}}]{Green:2017xeu}%
  \BibitemOpen
  \bibfield  {author} {\bibinfo {author} {\bibfnamefont {J.}~\bibnamefont
  {Green}}, \bibinfo {author} {\bibfnamefont {K.}~\bibnamefont {Jansen}},\ and\
  \bibinfo {author} {\bibfnamefont {F.}~\bibnamefont {Steffens}},\ }\bibfield
  {title} {\bibinfo {title} {Nonperturbative renormalization of nonlocal quark
  bilinears for parton quasidistribution functions on the lattice using an
  auxiliary field},\ }\href {https://doi.org/10.1103/PhysRevLett.121.022004}
  {\bibfield  {journal} {\bibinfo  {journal} {Phys. Rev. Lett.}\ }\textbf
  {\bibinfo {volume} {121}},\ \bibinfo {pages} {022004} (\bibinfo {year}
  {2018})},\ \Eprint {https://arxiv.org/abs/1707.07152} {arXiv:1707.07152
  [hep-lat]} \BibitemShut {NoStop}%
\bibitem [{\citenamefont {Stewart}\ and\ \citenamefont
  {Zhao}(2018)}]{Stewart:2017tvs}%
  \BibitemOpen
  \bibfield  {author} {\bibinfo {author} {\bibfnamefont {I.~W.}\ \bibnamefont
  {Stewart}}\ and\ \bibinfo {author} {\bibfnamefont {Y.}~\bibnamefont {Zhao}},\
  }\bibfield  {title} {\bibinfo {title} {{Matching the quasiparton distribution
  in a momentum subtraction scheme}},\ }\href
  {https://doi.org/10.1103/PhysRevD.97.054512} {\bibfield  {journal} {\bibinfo
  {journal} {Phys. Rev. D}\ }\textbf {\bibinfo {volume} {97}},\ \bibinfo
  {pages} {054512} (\bibinfo {year} {2018})},\ \Eprint
  {https://arxiv.org/abs/1709.04933} {arXiv:1709.04933 [hep-ph]} \BibitemShut
  {NoStop}%
\bibitem [{\citenamefont {Izubuchi}\ \emph {et~al.}(2018)\citenamefont
  {Izubuchi}, \citenamefont {Ji}, \citenamefont {Jin}, \citenamefont
  {Stewart},\ and\ \citenamefont {Zhao}}]{Izubuchi:2018srq}%
  \BibitemOpen
  \bibfield  {author} {\bibinfo {author} {\bibfnamefont {T.}~\bibnamefont
  {Izubuchi}}, \bibinfo {author} {\bibfnamefont {X.}~\bibnamefont {Ji}},
  \bibinfo {author} {\bibfnamefont {L.}~\bibnamefont {Jin}}, \bibinfo {author}
  {\bibfnamefont {I.~W.}\ \bibnamefont {Stewart}},\ and\ \bibinfo {author}
  {\bibfnamefont {Y.}~\bibnamefont {Zhao}},\ }\bibfield  {title} {\bibinfo
  {title} {Factorization theorem relating {Euclidean} and light-cone parton
  distributions},\ }\href {https://doi.org/10.1103/PhysRevD.98.056004}
  {\bibfield  {journal} {\bibinfo  {journal} {Phys. Rev. D}\ }\textbf {\bibinfo
  {volume} {98}},\ \bibinfo {pages} {056004} (\bibinfo {year} {2018})},\
  \Eprint {https://arxiv.org/abs/1801.03917} {arXiv:1801.03917 [hep-ph]}
  \BibitemShut {NoStop}%
\bibitem [{\citenamefont {Li}\ \emph {et~al.}(2021)\citenamefont {Li},
  \citenamefont {Ma},\ and\ \citenamefont {Qiu}}]{Li:2020xml}%
  \BibitemOpen
  \bibfield  {author} {\bibinfo {author} {\bibfnamefont {Z.-Y.}\ \bibnamefont
  {Li}}, \bibinfo {author} {\bibfnamefont {Y.-Q.}\ \bibnamefont {Ma}},\ and\
  \bibinfo {author} {\bibfnamefont {J.-W.}\ \bibnamefont {Qiu}},\ }\bibfield
  {title} {\bibinfo {title} {Extraction of next-to-next-to-leading-order parton
  distribution functions from lattice-{QCD} calculations},\ }\href
  {https://doi.org/10.1103/PhysRevLett.126.072001} {\bibfield  {journal}
  {\bibinfo  {journal} {Phys. Rev. Lett.}\ }\textbf {\bibinfo {volume} {126}},\
  \bibinfo {pages} {072001} (\bibinfo {year} {2021})},\ \Eprint
  {https://arxiv.org/abs/2006.12370} {arXiv:2006.12370 [hep-ph]} \BibitemShut
  {NoStop}%
\bibitem [{\citenamefont {Chen}\ \emph {et~al.}(2021)\citenamefont {Chen},
  \citenamefont {Wang},\ and\ \citenamefont {Zhu}}]{Chen:2020ody}%
  \BibitemOpen
  \bibfield  {author} {\bibinfo {author} {\bibfnamefont {L.-B.}\ \bibnamefont
  {Chen}}, \bibinfo {author} {\bibfnamefont {W.}~\bibnamefont {Wang}},\ and\
  \bibinfo {author} {\bibfnamefont {R.}~\bibnamefont {Zhu}},\ }\bibfield
  {title} {\bibinfo {title} {Next-to-next-to-leading order calculation of
  quasiparton distribution functions},\ }\href
  {https://doi.org/10.1103/PhysRevLett.126.072002} {\bibfield  {journal}
  {\bibinfo  {journal} {Phys. Rev. Lett.}\ }\textbf {\bibinfo {volume} {126}},\
  \bibinfo {pages} {072002} (\bibinfo {year} {2021})},\ \Eprint
  {https://arxiv.org/abs/2006.14825} {arXiv:2006.14825 [hep-ph]} \BibitemShut
  {NoStop}%
\bibitem [{\citenamefont {Ji}\ \emph {et~al.}(2021{\natexlab{b}})\citenamefont
  {Ji}, \citenamefont {Liu}, \citenamefont {Sch\"afer}, \citenamefont {Wang},
  \citenamefont {Yang}, \citenamefont {Zhang},\ and\ \citenamefont
  {Zhao}}]{Ji:2020brr}%
  \BibitemOpen
  \bibfield  {author} {\bibinfo {author} {\bibfnamefont {X.}~\bibnamefont
  {Ji}}, \bibinfo {author} {\bibfnamefont {Y.}~\bibnamefont {Liu}}, \bibinfo
  {author} {\bibfnamefont {A.}~\bibnamefont {Sch\"afer}}, \bibinfo {author}
  {\bibfnamefont {W.}~\bibnamefont {Wang}}, \bibinfo {author} {\bibfnamefont
  {Y.-B.}\ \bibnamefont {Yang}}, \bibinfo {author} {\bibfnamefont {J.-H.}\
  \bibnamefont {Zhang}},\ and\ \bibinfo {author} {\bibfnamefont
  {Y.}~\bibnamefont {Zhao}},\ }\bibfield  {title} {\bibinfo {title} {A hybrid
  renormalization scheme for quasi light-front correlations in large-momentum
  effective theory},\ }\href {https://doi.org/10.1016/j.nuclphysb.2021.115311}
  {\bibfield  {journal} {\bibinfo  {journal} {Nucl. Phys. B}\ }\textbf
  {\bibinfo {volume} {964}},\ \bibinfo {pages} {115311} (\bibinfo {year}
  {2021}{\natexlab{b}})},\ \Eprint {https://arxiv.org/abs/2008.03886}
  {arXiv:2008.03886 [hep-ph]} \BibitemShut {NoStop}%
\bibitem [{\citenamefont {Gao}\ \emph {et~al.}(2021)\citenamefont {Gao},
  \citenamefont {Lee}, \citenamefont {Mukherjee}, \citenamefont {Shugert},\
  and\ \citenamefont {Zhao}}]{Gao:2021hxl}%
  \BibitemOpen
  \bibfield  {author} {\bibinfo {author} {\bibfnamefont {X.}~\bibnamefont
  {Gao}}, \bibinfo {author} {\bibfnamefont {K.}~\bibnamefont {Lee}}, \bibinfo
  {author} {\bibfnamefont {S.}~\bibnamefont {Mukherjee}}, \bibinfo {author}
  {\bibfnamefont {C.}~\bibnamefont {Shugert}},\ and\ \bibinfo {author}
  {\bibfnamefont {Y.}~\bibnamefont {Zhao}},\ }\bibfield  {title} {\bibinfo
  {title} {{Origin and resummation of threshold logarithms in the lattice QCD
  calculations of PDFs}},\ }\href {https://doi.org/10.1103/PhysRevD.103.094504}
  {\bibfield  {journal} {\bibinfo  {journal} {Phys. Rev. D}\ }\textbf {\bibinfo
  {volume} {103}},\ \bibinfo {pages} {094504} (\bibinfo {year} {2021})},\
  \Eprint {https://arxiv.org/abs/2102.01101} {arXiv:2102.01101 [hep-ph]}
  \BibitemShut {NoStop}%
\bibitem [{\citenamefont {Huo}\ \emph {et~al.}(2021)\citenamefont {Huo} \emph
  {et~al.}}]{LatticePartonCollaborationLPC:2021xdx}%
  \BibitemOpen
  \bibfield  {author} {\bibinfo {author} {\bibfnamefont {Y.-K.}\ \bibnamefont
  {Huo}} \emph {et~al.} (\bibinfo {collaboration} {Lattice Parton (LPC)}),\
  }\bibfield  {title} {\bibinfo {title} {{Self-renormalization of
  quasi-light-front correlators on the lattice}},\ }\href
  {https://doi.org/10.1016/j.nuclphysb.2021.115443} {\bibfield  {journal}
  {\bibinfo  {journal} {Nucl. Phys. B}\ }\textbf {\bibinfo {volume} {969}},\
  \bibinfo {pages} {115443} (\bibinfo {year} {2021})},\ \Eprint
  {https://arxiv.org/abs/2103.02965} {arXiv:2103.02965 [hep-lat]} \BibitemShut
  {NoStop}%
\bibitem [{\citenamefont {Monahan}(2018)}]{Monahan:2018euv}%
  \BibitemOpen
  \bibfield  {author} {\bibinfo {author} {\bibfnamefont {C.}~\bibnamefont
  {Monahan}},\ }\bibfield  {title} {\bibinfo {title} {Recent developments in
  $x$-dependent structure calculations},\ }\href
  {https://doi.org/10.22323/1.334.0018} {\bibfield  {journal} {\bibinfo
  {journal} {PoS}\ }\textbf {\bibinfo {volume} {LATTICE2018}},\ \bibinfo
  {pages} {018} (\bibinfo {year} {2018})},\ \Eprint
  {https://arxiv.org/abs/1811.00678} {arXiv:1811.00678 [hep-lat]} \BibitemShut
  {NoStop}%
\bibitem [{\citenamefont {Lin}\ \emph {et~al.}(2018{\natexlab{a}})\citenamefont
  {Lin}, \citenamefont {Chen}, \citenamefont {Ji}, \citenamefont {Jin},
  \citenamefont {Li}, \citenamefont {Liu}, \citenamefont {Yang}, \citenamefont
  {Zhang},\ and\ \citenamefont {Zhao}}]{Lin:2018pvv}%
  \BibitemOpen
  \bibfield  {author} {\bibinfo {author} {\bibfnamefont {H.-W.}\ \bibnamefont
  {Lin}}, \bibinfo {author} {\bibfnamefont {J.-W.}\ \bibnamefont {Chen}},
  \bibinfo {author} {\bibfnamefont {X.}~\bibnamefont {Ji}}, \bibinfo {author}
  {\bibfnamefont {L.}~\bibnamefont {Jin}}, \bibinfo {author} {\bibfnamefont
  {R.}~\bibnamefont {Li}}, \bibinfo {author} {\bibfnamefont {Y.-S.}\
  \bibnamefont {Liu}}, \bibinfo {author} {\bibfnamefont {Y.-B.}\ \bibnamefont
  {Yang}}, \bibinfo {author} {\bibfnamefont {J.-H.}\ \bibnamefont {Zhang}},\
  and\ \bibinfo {author} {\bibfnamefont {Y.}~\bibnamefont {Zhao}},\ }\bibfield
  {title} {\bibinfo {title} {Proton isovector helicity distribution on the
  lattice at physical pion mass},\ }\href
  {https://doi.org/10.1103/PhysRevLett.121.242003} {\bibfield  {journal}
  {\bibinfo  {journal} {Phys. Rev. Lett.}\ }\textbf {\bibinfo {volume} {121}},\
  \bibinfo {pages} {242003} (\bibinfo {year} {2018}{\natexlab{a}})},\ \Eprint
  {https://arxiv.org/abs/1807.07431} {arXiv:1807.07431 [hep-lat]} \BibitemShut
  {NoStop}%
\bibitem [{\citenamefont {Alexandrou}\ \emph
  {et~al.}(2017{\natexlab{b}})\citenamefont {Alexandrou}, \citenamefont
  {Cichy}, \citenamefont {Constantinou}, \citenamefont {Hadjiyiannakou},
  \citenamefont {Jansen}, \citenamefont {Steffens},\ and\ \citenamefont
  {Wiese}}]{Alexandrou:2016jqi}%
  \BibitemOpen
  \bibfield  {author} {\bibinfo {author} {\bibfnamefont {C.}~\bibnamefont
  {Alexandrou}}, \bibinfo {author} {\bibfnamefont {K.}~\bibnamefont {Cichy}},
  \bibinfo {author} {\bibfnamefont {M.}~\bibnamefont {Constantinou}}, \bibinfo
  {author} {\bibfnamefont {K.}~\bibnamefont {Hadjiyiannakou}}, \bibinfo
  {author} {\bibfnamefont {K.}~\bibnamefont {Jansen}}, \bibinfo {author}
  {\bibfnamefont {F.}~\bibnamefont {Steffens}},\ and\ \bibinfo {author}
  {\bibfnamefont {C.}~\bibnamefont {Wiese}},\ }\bibfield  {title} {\bibinfo
  {title} {Updated lattice results for parton distributions},\ }\href
  {https://doi.org/10.1103/PhysRevD.96.014513} {\bibfield  {journal} {\bibinfo
  {journal} {Phys. Rev. D}\ }\textbf {\bibinfo {volume} {96}},\ \bibinfo
  {pages} {014513} (\bibinfo {year} {2017}{\natexlab{b}})},\ \Eprint
  {https://arxiv.org/abs/1610.03689} {arXiv:1610.03689 [hep-lat]} \BibitemShut
  {NoStop}%
\bibitem [{\citenamefont {Alexandrou}\ \emph
  {et~al.}(2018{\natexlab{b}})\citenamefont {Alexandrou} \emph
  {et~al.}}]{Alexandrou:2018pbm}%
  \BibitemOpen
  \bibfield  {author} {\bibinfo {author} {\bibfnamefont {C.}~\bibnamefont
  {Alexandrou}} \emph {et~al.},\ }\bibfield  {title} {\bibinfo {title}
  {Light-cone parton distribution functions from lattice {QCD}},\ }\href
  {https://doi.org/10.1103/PhysRevLett.121.112001} {\bibfield  {journal}
  {\bibinfo  {journal} {Phys. Rev. Lett.}\ }\textbf {\bibinfo {volume} {121}},\
  \bibinfo {pages} {112001} (\bibinfo {year} {2018}{\natexlab{b}})},\ \Eprint
  {https://arxiv.org/abs/1803.02685} {arXiv:1803.02685 [hep-lat]} \BibitemShut
  {NoStop}%
\bibitem [{\citenamefont {Alexandrou}\ \emph {et~al.}(2019)\citenamefont
  {Alexandrou}, \citenamefont {Cichy}, \citenamefont {Constantinou},
  \citenamefont {Hadjiyiannakou}, \citenamefont {Jansen}, \citenamefont
  {Scapellato},\ and\ \citenamefont {Steffens}}]{Alexandrou:2019lfo}%
  \BibitemOpen
  \bibfield  {author} {\bibinfo {author} {\bibfnamefont {C.}~\bibnamefont
  {Alexandrou}}, \bibinfo {author} {\bibfnamefont {K.}~\bibnamefont {Cichy}},
  \bibinfo {author} {\bibfnamefont {M.}~\bibnamefont {Constantinou}}, \bibinfo
  {author} {\bibfnamefont {K.}~\bibnamefont {Hadjiyiannakou}}, \bibinfo
  {author} {\bibfnamefont {K.}~\bibnamefont {Jansen}}, \bibinfo {author}
  {\bibfnamefont {A.}~\bibnamefont {Scapellato}},\ and\ \bibinfo {author}
  {\bibfnamefont {F.}~\bibnamefont {Steffens}},\ }\bibfield  {title} {\bibinfo
  {title} {Systematic uncertainties in parton distribution functions from
  lattice {QCD} simulations at the physical point},\ }\href
  {https://doi.org/10.1103/PhysRevD.99.114504} {\bibfield  {journal} {\bibinfo
  {journal} {Phys. Rev. D}\ }\textbf {\bibinfo {volume} {99}},\ \bibinfo
  {pages} {114504} (\bibinfo {year} {2019})},\ \Eprint
  {https://arxiv.org/abs/1902.00587} {arXiv:1902.00587 [hep-lat]} \BibitemShut
  {NoStop}%
\bibitem [{\citenamefont {Alexandrou}\ \emph
  {et~al.}(2021{\natexlab{d}})\citenamefont {Alexandrou}, \citenamefont
  {Constantinou}, \citenamefont {Hadjiyiannakou}, \citenamefont {Jansen},\ and\
  \citenamefont {Manigrasso}}]{Alexandrou:2020uyt}%
  \BibitemOpen
  \bibfield  {author} {\bibinfo {author} {\bibfnamefont {C.}~\bibnamefont
  {Alexandrou}}, \bibinfo {author} {\bibfnamefont {M.}~\bibnamefont
  {Constantinou}}, \bibinfo {author} {\bibfnamefont {K.}~\bibnamefont
  {Hadjiyiannakou}}, \bibinfo {author} {\bibfnamefont {K.}~\bibnamefont
  {Jansen}},\ and\ \bibinfo {author} {\bibfnamefont {F.}~\bibnamefont
  {Manigrasso}} (\bibinfo {collaboration} {ETM}),\ }\bibfield  {title}
  {\bibinfo {title} {{Flavor decomposition for the proton helicity parton
  distribution functions}},\ }\href
  {https://doi.org/10.1103/PhysRevLett.126.102003} {\bibfield  {journal}
  {\bibinfo  {journal} {Phys. Rev. Lett.}\ }\textbf {\bibinfo {volume} {126}},\
  \bibinfo {pages} {102003} (\bibinfo {year} {2021}{\natexlab{d}})},\ \Eprint
  {https://arxiv.org/abs/2009.13061} {arXiv:2009.13061 [hep-lat]} \BibitemShut
  {NoStop}%
\bibitem [{\citenamefont {Alexandrou}\ \emph
  {et~al.}(2021{\natexlab{e}})\citenamefont {Alexandrou}, \citenamefont
  {Cichy}, \citenamefont {Constantinou}, \citenamefont {Green}, \citenamefont
  {Hadjiyiannakou}, \citenamefont {Jansen}, \citenamefont {Manigrasso},
  \citenamefont {Scapellato},\ and\ \citenamefont
  {Steffens}}]{Alexandrou:2020qtt}%
  \BibitemOpen
  \bibfield  {author} {\bibinfo {author} {\bibfnamefont {C.}~\bibnamefont
  {Alexandrou}}, \bibinfo {author} {\bibfnamefont {K.}~\bibnamefont {Cichy}},
  \bibinfo {author} {\bibfnamefont {M.}~\bibnamefont {Constantinou}}, \bibinfo
  {author} {\bibfnamefont {J.~R.}\ \bibnamefont {Green}}, \bibinfo {author}
  {\bibfnamefont {K.}~\bibnamefont {Hadjiyiannakou}}, \bibinfo {author}
  {\bibfnamefont {K.}~\bibnamefont {Jansen}}, \bibinfo {author} {\bibfnamefont
  {F.}~\bibnamefont {Manigrasso}}, \bibinfo {author} {\bibfnamefont
  {A.}~\bibnamefont {Scapellato}},\ and\ \bibinfo {author} {\bibfnamefont
  {F.}~\bibnamefont {Steffens}},\ }\bibfield  {title} {\bibinfo {title}
  {{Lattice continuum-limit study of nucleon quasi-PDFs}},\ }\href
  {https://doi.org/10.1103/PhysRevD.103.094512} {\bibfield  {journal} {\bibinfo
   {journal} {Phys. Rev. D}\ }\textbf {\bibinfo {volume} {103}},\ \bibinfo
  {pages} {094512} (\bibinfo {year} {2021}{\natexlab{e}})},\ \Eprint
  {https://arxiv.org/abs/2011.00964} {arXiv:2011.00964 [hep-lat]} \BibitemShut
  {NoStop}%
\bibitem [{\citenamefont {Gao}\ \emph {et~al.}(2022)\citenamefont {Gao},
  \citenamefont {Hanlon}, \citenamefont {Mukherjee}, \citenamefont {Petreczky},
  \citenamefont {Scior}, \citenamefont {Syritsyn},\ and\ \citenamefont
  {Zhao}}]{Gao:2021dbh}%
  \BibitemOpen
  \bibfield  {author} {\bibinfo {author} {\bibfnamefont {X.}~\bibnamefont
  {Gao}}, \bibinfo {author} {\bibfnamefont {A.~D.}\ \bibnamefont {Hanlon}},
  \bibinfo {author} {\bibfnamefont {S.}~\bibnamefont {Mukherjee}}, \bibinfo
  {author} {\bibfnamefont {P.}~\bibnamefont {Petreczky}}, \bibinfo {author}
  {\bibfnamefont {P.}~\bibnamefont {Scior}}, \bibinfo {author} {\bibfnamefont
  {S.}~\bibnamefont {Syritsyn}},\ and\ \bibinfo {author} {\bibfnamefont
  {Y.}~\bibnamefont {Zhao}},\ }\bibfield  {title} {\bibinfo {title} {Lattice
  {QCD} determination of the {Bjorken}-$x$ dependence of parton distribution
  functions at next-to-next-to-leading order},\ }\href
  {https://doi.org/10.1103/PhysRevLett.128.142003} {\bibfield  {journal}
  {\bibinfo  {journal} {Phys. Rev. Lett.}\ }\textbf {\bibinfo {volume} {128}},\
  \bibinfo {pages} {142003} (\bibinfo {year} {2022})},\ \Eprint
  {https://arxiv.org/abs/2112.02208} {arXiv:2112.02208 [hep-lat]} \BibitemShut
  {NoStop}%
\bibitem [{\citenamefont {Orginos}\ \emph {et~al.}(2017)\citenamefont
  {Orginos}, \citenamefont {Radyushkin}, \citenamefont {Karpie},\ and\
  \citenamefont {Zafeiropoulos}}]{Orginos:2017kos}%
  \BibitemOpen
  \bibfield  {author} {\bibinfo {author} {\bibfnamefont {K.}~\bibnamefont
  {Orginos}}, \bibinfo {author} {\bibfnamefont {A.}~\bibnamefont {Radyushkin}},
  \bibinfo {author} {\bibfnamefont {J.}~\bibnamefont {Karpie}},\ and\ \bibinfo
  {author} {\bibfnamefont {S.}~\bibnamefont {Zafeiropoulos}},\ }\bibfield
  {title} {\bibinfo {title} {Lattice {QCD} exploration of parton
  pseudo-distribution functions},\ }\href
  {https://doi.org/10.1103/PhysRevD.96.094503} {\bibfield  {journal} {\bibinfo
  {journal} {Phys. Rev. D}\ }\textbf {\bibinfo {volume} {96}},\ \bibinfo
  {pages} {094503} (\bibinfo {year} {2017})},\ \Eprint
  {https://arxiv.org/abs/1706.05373} {arXiv:1706.05373 [hep-ph]} \BibitemShut
  {NoStop}%
\bibitem [{\citenamefont {Karpie}\ \emph {et~al.}(2018)\citenamefont {Karpie},
  \citenamefont {Orginos},\ and\ \citenamefont
  {Zafeiropoulos}}]{Karpie:2018zaz}%
  \BibitemOpen
  \bibfield  {author} {\bibinfo {author} {\bibfnamefont {J.}~\bibnamefont
  {Karpie}}, \bibinfo {author} {\bibfnamefont {K.}~\bibnamefont {Orginos}},\
  and\ \bibinfo {author} {\bibfnamefont {S.}~\bibnamefont {Zafeiropoulos}},\
  }\bibfield  {title} {\bibinfo {title} {Moments of {Ioffe} time parton
  distribution functions from non-local matrix elements},\ }\href
  {https://doi.org/10.1007/JHEP11(2018)178} {\bibfield  {journal} {\bibinfo
  {journal} {JHEP}\ }\textbf {\bibinfo {volume} {11}},\ \bibinfo {pages}
  {178}},\ \Eprint {https://arxiv.org/abs/1807.10933} {arXiv:1807.10933
  [hep-lat]} \BibitemShut {NoStop}%
\bibitem [{\citenamefont {Karpie}\ \emph {et~al.}(2019)\citenamefont {Karpie},
  \citenamefont {Orginos}, \citenamefont {Rothkopf},\ and\ \citenamefont
  {Zafeiropoulos}}]{Karpie:2019eiq}%
  \BibitemOpen
  \bibfield  {author} {\bibinfo {author} {\bibfnamefont {J.}~\bibnamefont
  {Karpie}}, \bibinfo {author} {\bibfnamefont {K.}~\bibnamefont {Orginos}},
  \bibinfo {author} {\bibfnamefont {A.}~\bibnamefont {Rothkopf}},\ and\
  \bibinfo {author} {\bibfnamefont {S.}~\bibnamefont {Zafeiropoulos}},\
  }\bibfield  {title} {\bibinfo {title} {Reconstructing parton distribution
  functions from {Ioffe} time data: From {Bayesian} methods to neural
  networks},\ }\href {https://doi.org/10.1007/JHEP04(2019)057} {\bibfield
  {journal} {\bibinfo  {journal} {JHEP}\ }\textbf {\bibinfo {volume} {04}},\
  \bibinfo {pages} {057}},\ \Eprint {https://arxiv.org/abs/1901.05408}
  {arXiv:1901.05408 [hep-lat]} \BibitemShut {NoStop}%
\bibitem [{\citenamefont {Jo\'o}\ \emph {et~al.}(2019)\citenamefont {Jo\'o},
  \citenamefont {Karpie}, \citenamefont {Orginos}, \citenamefont {Radyushkin},
  \citenamefont {Richards},\ and\ \citenamefont {Zafeiropoulos}}]{Joo:2019jct}%
  \BibitemOpen
  \bibfield  {author} {\bibinfo {author} {\bibfnamefont {B.}~\bibnamefont
  {Jo\'o}}, \bibinfo {author} {\bibfnamefont {J.}~\bibnamefont {Karpie}},
  \bibinfo {author} {\bibfnamefont {K.}~\bibnamefont {Orginos}}, \bibinfo
  {author} {\bibfnamefont {A.}~\bibnamefont {Radyushkin}}, \bibinfo {author}
  {\bibfnamefont {D.}~\bibnamefont {Richards}},\ and\ \bibinfo {author}
  {\bibfnamefont {S.}~\bibnamefont {Zafeiropoulos}},\ }\bibfield  {title}
  {\bibinfo {title} {Parton distribution functions from {Ioffe} time
  pseudo-distributions},\ }\href {https://doi.org/10.1007/JHEP12(2019)081}
  {\bibfield  {journal} {\bibinfo  {journal} {JHEP}\ }\textbf {\bibinfo
  {volume} {12}},\ \bibinfo {pages} {081}},\ \Eprint
  {https://arxiv.org/abs/1908.09771} {arXiv:1908.09771 [hep-lat]} \BibitemShut
  {NoStop}%
\bibitem [{\citenamefont {Jo\'o}\ \emph {et~al.}(2020)\citenamefont {Jo\'o},
  \citenamefont {Karpie}, \citenamefont {Orginos}, \citenamefont {Radyushkin},
  \citenamefont {Richards},\ and\ \citenamefont {Zafeiropoulos}}]{Joo:2020spy}%
  \BibitemOpen
  \bibfield  {author} {\bibinfo {author} {\bibfnamefont {B.}~\bibnamefont
  {Jo\'o}}, \bibinfo {author} {\bibfnamefont {J.}~\bibnamefont {Karpie}},
  \bibinfo {author} {\bibfnamefont {K.}~\bibnamefont {Orginos}}, \bibinfo
  {author} {\bibfnamefont {A.~V.}\ \bibnamefont {Radyushkin}}, \bibinfo
  {author} {\bibfnamefont {D.~G.}\ \bibnamefont {Richards}},\ and\ \bibinfo
  {author} {\bibfnamefont {S.}~\bibnamefont {Zafeiropoulos}},\ }\bibfield
  {title} {\bibinfo {title} {Parton distribution functions from {Ioffe} time
  pseudodistributions from lattice calculations: Approaching the physical
  point},\ }\href {https://doi.org/10.1103/PhysRevLett.125.232003} {\bibfield
  {journal} {\bibinfo  {journal} {Phys. Rev. Lett.}\ }\textbf {\bibinfo
  {volume} {125}},\ \bibinfo {pages} {232003} (\bibinfo {year} {2020})},\
  \Eprint {https://arxiv.org/abs/2004.01687} {arXiv:2004.01687 [hep-lat]}
  \BibitemShut {NoStop}%
\bibitem [{\citenamefont {Karpie}\ \emph {et~al.}(2021)\citenamefont {Karpie},
  \citenamefont {Orginos}, \citenamefont {Radyushkin},\ and\ \citenamefont
  {Zafeiropoulos}}]{Karpie:2021pap}%
  \BibitemOpen
  \bibfield  {author} {\bibinfo {author} {\bibfnamefont {J.}~\bibnamefont
  {Karpie}}, \bibinfo {author} {\bibfnamefont {K.}~\bibnamefont {Orginos}},
  \bibinfo {author} {\bibfnamefont {A.}~\bibnamefont {Radyushkin}},\ and\
  \bibinfo {author} {\bibfnamefont {S.}~\bibnamefont {Zafeiropoulos}} (\bibinfo
  {collaboration} {HadStruc}),\ }\bibfield  {title} {\bibinfo {title} {{The
  continuum and leading twist limits of parton distribution functions in
  lattice QCD}},\ }\href {https://doi.org/10.1007/JHEP11(2021)024} {\bibfield
  {journal} {\bibinfo  {journal} {JHEP}\ }\textbf {\bibinfo {volume} {11}},\
  \bibinfo {pages} {024}},\ \Eprint {https://arxiv.org/abs/2105.13313}
  {arXiv:2105.13313 [hep-lat]} \BibitemShut {NoStop}%
\bibitem [{\citenamefont {Del~Debbio}\ \emph {et~al.}(2021)\citenamefont
  {Del~Debbio}, \citenamefont {Giani}, \citenamefont {Karpie}, \citenamefont
  {Orginos}, \citenamefont {Radyushkin},\ and\ \citenamefont
  {Zafeiropoulos}}]{DelDebbio:2020rgv}%
  \BibitemOpen
  \bibfield  {author} {\bibinfo {author} {\bibfnamefont {L.}~\bibnamefont
  {Del~Debbio}}, \bibinfo {author} {\bibfnamefont {T.}~\bibnamefont {Giani}},
  \bibinfo {author} {\bibfnamefont {J.}~\bibnamefont {Karpie}}, \bibinfo
  {author} {\bibfnamefont {K.}~\bibnamefont {Orginos}}, \bibinfo {author}
  {\bibfnamefont {A.}~\bibnamefont {Radyushkin}},\ and\ \bibinfo {author}
  {\bibfnamefont {S.}~\bibnamefont {Zafeiropoulos}},\ }\bibfield  {title}
  {\bibinfo {title} {Neural-network analysis of parton distribution functions
  from {Ioffe-time} pseudodistributions},\ }\href
  {https://doi.org/10.1007/JHEP02(2021)138} {\bibfield  {journal} {\bibinfo
  {journal} {JHEP}\ }\textbf {\bibinfo {volume} {02}},\ \bibinfo {pages}
  {138}},\ \Eprint {https://arxiv.org/abs/2010.03996} {arXiv:2010.03996
  [hep-ph]} \BibitemShut {NoStop}%
\bibitem [{\citenamefont {Bhat}\ \emph {et~al.}(2022)\citenamefont {Bhat},
  \citenamefont {Chomicki}, \citenamefont {Cichy}, \citenamefont
  {Constantinou}, \citenamefont {Green},\ and\ \citenamefont
  {Scapellato}}]{Bhat:2022zrw}%
  \BibitemOpen
  \bibfield  {author} {\bibinfo {author} {\bibfnamefont {M.}~\bibnamefont
  {Bhat}}, \bibinfo {author} {\bibfnamefont {W.}~\bibnamefont {Chomicki}},
  \bibinfo {author} {\bibfnamefont {K.}~\bibnamefont {Cichy}}, \bibinfo
  {author} {\bibfnamefont {M.}~\bibnamefont {Constantinou}}, \bibinfo {author}
  {\bibfnamefont {J.~R.}\ \bibnamefont {Green}},\ and\ \bibinfo {author}
  {\bibfnamefont {A.}~\bibnamefont {Scapellato}},\ }\bibfield  {title}
  {\bibinfo {title} {{Continuum limit of parton distribution functions from the
  pseudo-distribution approach on the lattice}},\ }\href@noop {} {\  (\bibinfo
  {year} {2022})},\ \Eprint {https://arxiv.org/abs/2205.07585}
  {arXiv:2205.07585 [hep-lat]} \BibitemShut {NoStop}%
\bibitem [{\citenamefont {Constantinou}\ \emph {et~al.}(2022)\citenamefont
  {Constantinou} \emph {et~al.}}]{Constantinou:2022yye}%
  \BibitemOpen
  \bibfield  {author} {\bibinfo {author} {\bibfnamefont {M.}~\bibnamefont
  {Constantinou}} \emph {et~al.},\ }\bibfield  {title} {\bibinfo {title}
  {Lattice {QCD} calculations of parton physics},\ }in\ \href@noop {} {\emph
  {\bibinfo {booktitle} {{2022 Snowmass Summer Study}}}}\ (\bibinfo {year}
  {2022})\ \Eprint {https://arxiv.org/abs/2202.07193} {arXiv:2202.07193
  [hep-lat]} \BibitemShut {NoStop}%
\bibitem [{\citenamefont {Lin}\ \emph {et~al.}(2018{\natexlab{b}})\citenamefont
  {Lin} \emph {et~al.}}]{Lin:2017snn}%
  \BibitemOpen
  \bibfield  {author} {\bibinfo {author} {\bibfnamefont {H.-W.}\ \bibnamefont
  {Lin}} \emph {et~al.},\ }\bibfield  {title} {\bibinfo {title} {Parton
  distributions and lattice {QCD} calculations: A community white paper},\
  }\href {https://doi.org/10.1016/j.ppnp.2018.01.007} {\bibfield  {journal}
  {\bibinfo  {journal} {Prog. Part. Nucl. Phys.}\ }\textbf {\bibinfo {volume}
  {100}},\ \bibinfo {pages} {107} (\bibinfo {year} {2018}{\natexlab{b}})},\
  \Eprint {https://arxiv.org/abs/1711.07916} {arXiv:1711.07916 [hep-ph]}
  \BibitemShut {NoStop}%
\bibitem [{\citenamefont {Amoroso}\ \emph {et~al.}(2022)\citenamefont {Amoroso}
  \emph {et~al.}}]{Amoroso:2022eow}%
  \BibitemOpen
  \bibfield  {author} {\bibinfo {author} {\bibfnamefont {S.}~\bibnamefont
  {Amoroso}} \emph {et~al.},\ }\bibfield  {title} {\bibinfo {title} {Proton
  structure at the precision frontier},\ }in\ \href@noop {} {\emph {\bibinfo
  {booktitle} {{2022 Snowmass Summer Study}}}}\ (\bibinfo {year} {2022})\
  \Eprint {https://arxiv.org/abs/2203.13923} {arXiv:2203.13923 [hep-ph]}
  \BibitemShut {NoStop}%
\bibitem [{\citenamefont {Hou}\ \emph {et~al.}(2022)\citenamefont {Hou},
  \citenamefont {Lin}, \citenamefont {Yan},\ and\ \citenamefont
  {Yuan}}]{Hou:2022sdf}%
  \BibitemOpen
  \bibfield  {author} {\bibinfo {author} {\bibfnamefont {T.-J.}\ \bibnamefont
  {Hou}}, \bibinfo {author} {\bibfnamefont {H.-W.}\ \bibnamefont {Lin}},
  \bibinfo {author} {\bibfnamefont {M.}~\bibnamefont {Yan}},\ and\ \bibinfo
  {author} {\bibfnamefont {C.~P.}\ \bibnamefont {Yuan}},\ }\bibfield  {title}
  {\bibinfo {title} {{Impact of lattice $s(x)$-$\bar{s}(x)$ data in the
  CTEQ-TEA global analysis}},\ }in\ \href@noop {} {\emph {\bibinfo {booktitle}
  {{2022 Snowmass Summer Study}}}}\ (\bibinfo {year} {2022})\ \Eprint
  {https://arxiv.org/abs/2204.07944} {arXiv:2204.07944 [hep-ph]} \BibitemShut
  {NoStop}%
\bibitem [{\citenamefont {Khalek}\ \emph {et~al.}(2022)\citenamefont {Khalek}
  \emph {et~al.}}]{Khalek:2022bzd}%
  \BibitemOpen
  \bibfield  {author} {\bibinfo {author} {\bibfnamefont {R.~A.}\ \bibnamefont
  {Khalek}} \emph {et~al.},\ }\bibfield  {title} {\bibinfo {title} {Electron
  ion collider for high energy physics},\ }in\ \href@noop {} {\emph {\bibinfo
  {booktitle} {{2022 Snowmass Summer Study}}}}\ (\bibinfo {year} {2022})\
  \Eprint {https://arxiv.org/abs/2203.13199} {arXiv:2203.13199 [hep-ph]}
  \BibitemShut {NoStop}%
\bibitem [{\citenamefont {Müller}(2013)}]{Muller:2013dea}%
  \BibitemOpen
  \bibfield  {author} {\bibinfo {author} {\bibfnamefont {B.}~\bibnamefont
  {Müller}},\ }\bibfield  {title} {\bibinfo {title} {Investigation of hot
  {QCD} matter: Theoretical aspects},\ }\href
  {https://doi.org/10.1088/0031-8949/2013/T158/014004} {\bibfield  {journal}
  {\bibinfo  {journal} {Phys. Scripta}\ }\textbf {\bibinfo {volume} {T158}},\
  \bibinfo {pages} {014004} (\bibinfo {year} {2013})},\ \Eprint
  {https://arxiv.org/abs/1309.7616} {arXiv:1309.7616 [nucl-th]} \BibitemShut
  {NoStop}%
\bibitem [{\citenamefont {Aoki}\ \emph
  {et~al.}(2006{\natexlab{a}})\citenamefont {Aoki}, \citenamefont
  {Endr\H{o}di}, \citenamefont {Fodor}, \citenamefont {Katz},\ and\
  \citenamefont {Szab\'o}}]{Aoki:2006we}%
  \BibitemOpen
  \bibfield  {author} {\bibinfo {author} {\bibfnamefont {Y.}~\bibnamefont
  {Aoki}}, \bibinfo {author} {\bibfnamefont {G.}~\bibnamefont {Endr\H{o}di}},
  \bibinfo {author} {\bibfnamefont {Z.}~\bibnamefont {Fodor}}, \bibinfo
  {author} {\bibfnamefont {S.~D.}\ \bibnamefont {Katz}},\ and\ \bibinfo
  {author} {\bibfnamefont {K.~K.}\ \bibnamefont {Szab\'o}},\ }\bibfield
  {title} {\bibinfo {title} {{The order of the quantum chromodynamics
  transition predicted by the standard model of particle physics}},\ }\href
  {https://doi.org/10.1038/nature05120} {\bibfield  {journal} {\bibinfo
  {journal} {Nature}\ }\textbf {\bibinfo {volume} {443}},\ \bibinfo {pages}
  {675} (\bibinfo {year} {2006}{\natexlab{a}})},\ \Eprint
  {https://arxiv.org/abs/hep-lat/0611014} {arXiv:hep-lat/0611014} \BibitemShut
  {NoStop}%
\bibitem [{\citenamefont {Bhattacharya}\ \emph {et~al.}(2014)\citenamefont
  {Bhattacharya} \emph {et~al.}}]{Bhattacharya:2014ara}%
  \BibitemOpen
  \bibfield  {author} {\bibinfo {author} {\bibfnamefont {T.}~\bibnamefont
  {Bhattacharya}} \emph {et~al.} (\bibinfo {collaboration} {HotQCD}),\
  }\bibfield  {title} {\bibinfo {title} {{QCD} phase transition with chiral
  quarks and physical quark masses},\ }\href
  {https://doi.org/10.1103/PhysRevLett.113.082001} {\bibfield  {journal}
  {\bibinfo  {journal} {Phys. Rev. Lett.}\ }\textbf {\bibinfo {volume} {113}},\
  \bibinfo {pages} {082001} (\bibinfo {year} {2014})},\ \Eprint
  {https://arxiv.org/abs/1402.5175} {arXiv:1402.5175 [hep-lat]} \BibitemShut
  {NoStop}%
\bibitem [{\citenamefont {Aoki}\ \emph
  {et~al.}(2006{\natexlab{b}})\citenamefont {Aoki}, \citenamefont {Fodor},
  \citenamefont {Katz},\ and\ \citenamefont {Szabo}}]{Aoki:2006br}%
  \BibitemOpen
  \bibfield  {author} {\bibinfo {author} {\bibfnamefont {Y.}~\bibnamefont
  {Aoki}}, \bibinfo {author} {\bibfnamefont {Z.}~\bibnamefont {Fodor}},
  \bibinfo {author} {\bibfnamefont {S.~D.}\ \bibnamefont {Katz}},\ and\
  \bibinfo {author} {\bibfnamefont {K.~K.}\ \bibnamefont {Szabo}},\ }\bibfield
  {title} {\bibinfo {title} {{The QCD transition temperature: Results with
  physical masses in the continuum limit}},\ }\href
  {https://doi.org/10.1016/j.physletb.2006.10.021} {\bibfield  {journal}
  {\bibinfo  {journal} {Phys. Lett. B}\ }\textbf {\bibinfo {volume} {643}},\
  \bibinfo {pages} {46} (\bibinfo {year} {2006}{\natexlab{b}})},\ \Eprint
  {https://arxiv.org/abs/hep-lat/0609068} {arXiv:hep-lat/0609068} \BibitemShut
  {NoStop}%
\bibitem [{\citenamefont {Aoki}\ \emph
  {et~al.}(2009{\natexlab{b}})\citenamefont {Aoki}, \citenamefont {Borsanyi},
  \citenamefont {Durr}, \citenamefont {Fodor}, \citenamefont {Katz},
  \citenamefont {Krieg},\ and\ \citenamefont {Szab\'o}}]{Aoki:2009sc}%
  \BibitemOpen
  \bibfield  {author} {\bibinfo {author} {\bibfnamefont {Y.}~\bibnamefont
  {Aoki}}, \bibinfo {author} {\bibfnamefont {S.}~\bibnamefont {Borsanyi}},
  \bibinfo {author} {\bibfnamefont {S.}~\bibnamefont {Durr}}, \bibinfo {author}
  {\bibfnamefont {Z.}~\bibnamefont {Fodor}}, \bibinfo {author} {\bibfnamefont
  {S.~D.}\ \bibnamefont {Katz}}, \bibinfo {author} {\bibfnamefont
  {S.}~\bibnamefont {Krieg}},\ and\ \bibinfo {author} {\bibfnamefont {K.~K.}\
  \bibnamefont {Szab\'o}},\ }\bibfield  {title} {\bibinfo {title} {{The QCD
  transition temperature: results with physical masses in the continuum limit
  II.}},\ }\href {https://doi.org/10.1088/1126-6708/2009/06/088} {\bibfield
  {journal} {\bibinfo  {journal} {JHEP}\ }\textbf {\bibinfo {volume} {06}},\
  \bibinfo {pages} {088}},\ \Eprint {https://arxiv.org/abs/0903.4155}
  {arXiv:0903.4155 [hep-lat]} \BibitemShut {NoStop}%
\bibitem [{\citenamefont {Borsanyi}\ \emph {et~al.}(2010)\citenamefont
  {Borsanyi}, \citenamefont {Fodor}, \citenamefont {Hoelbling}, \citenamefont
  {Katz}, \citenamefont {Krieg}, \citenamefont {Ratti},\ and\ \citenamefont
  {Szab\'o}}]{Borsanyi:2010bp}%
  \BibitemOpen
  \bibfield  {author} {\bibinfo {author} {\bibfnamefont {S.}~\bibnamefont
  {Borsanyi}}, \bibinfo {author} {\bibfnamefont {Z.}~\bibnamefont {Fodor}},
  \bibinfo {author} {\bibfnamefont {C.}~\bibnamefont {Hoelbling}}, \bibinfo
  {author} {\bibfnamefont {S.~D.}\ \bibnamefont {Katz}}, \bibinfo {author}
  {\bibfnamefont {S.}~\bibnamefont {Krieg}}, \bibinfo {author} {\bibfnamefont
  {C.}~\bibnamefont {Ratti}},\ and\ \bibinfo {author} {\bibfnamefont {K.~K.}\
  \bibnamefont {Szab\'o}} (\bibinfo {collaboration} {Wuppertal-Budapest}),\
  }\bibfield  {title} {\bibinfo {title} {{Is there still any $T_c$ mystery in
  lattice QCD? Results with physical masses in the continuum limit III}},\
  }\href {https://doi.org/10.1007/JHEP09(2010)073} {\bibfield  {journal}
  {\bibinfo  {journal} {JHEP}\ }\textbf {\bibinfo {volume} {09}},\ \bibinfo
  {pages} {073}},\ \Eprint {https://arxiv.org/abs/1005.3508} {arXiv:1005.3508
  [hep-lat]} \BibitemShut {NoStop}%
\bibitem [{\citenamefont {Bazavov}\ \emph {et~al.}(2012)\citenamefont {Bazavov}
  \emph {et~al.}}]{Bazavov:2011nk}%
  \BibitemOpen
  \bibfield  {author} {\bibinfo {author} {\bibfnamefont {A.}~\bibnamefont
  {Bazavov}} \emph {et~al.} (\bibinfo {collaboration} {HotQCD}),\ }\bibfield
  {title} {\bibinfo {title} {{The chiral and deconfinement aspects of the QCD
  transition}},\ }\href {https://doi.org/10.1103/PhysRevD.85.054503} {\bibfield
   {journal} {\bibinfo  {journal} {Phys. Rev. D}\ }\textbf {\bibinfo {volume}
  {85}},\ \bibinfo {pages} {054503} (\bibinfo {year} {2012})},\ \Eprint
  {https://arxiv.org/abs/1111.1710} {arXiv:1111.1710 [hep-lat]} \BibitemShut
  {NoStop}%
\bibitem [{\citenamefont {Bazavov}\ \emph
  {et~al.}(2019{\natexlab{e}})\citenamefont {Bazavov} \emph
  {et~al.}}]{HotQCD:2018pds}%
  \BibitemOpen
  \bibfield  {author} {\bibinfo {author} {\bibfnamefont {A.}~\bibnamefont
  {Bazavov}} \emph {et~al.} (\bibinfo {collaboration} {HotQCD}),\ }\bibfield
  {title} {\bibinfo {title} {{Chiral crossover in QCD at zero and non-zero
  chemical potentials}},\ }\href
  {https://doi.org/10.1016/j.physletb.2019.05.013} {\bibfield  {journal}
  {\bibinfo  {journal} {Phys. Lett. B}\ }\textbf {\bibinfo {volume} {795}},\
  \bibinfo {pages} {15} (\bibinfo {year} {2019}{\natexlab{e}})},\ \Eprint
  {https://arxiv.org/abs/1812.08235} {arXiv:1812.08235 [hep-lat]} \BibitemShut
  {NoStop}%
\bibitem [{\citenamefont {Andronic}\ \emph {et~al.}(2018)\citenamefont
  {Andronic}, \citenamefont {Braun-Munzinger}, \citenamefont {Redlich},\ and\
  \citenamefont {Stachel}}]{Andronic:2017pug}%
  \BibitemOpen
  \bibfield  {author} {\bibinfo {author} {\bibfnamefont {A.}~\bibnamefont
  {Andronic}}, \bibinfo {author} {\bibfnamefont {P.}~\bibnamefont
  {Braun-Munzinger}}, \bibinfo {author} {\bibfnamefont {K.}~\bibnamefont
  {Redlich}},\ and\ \bibinfo {author} {\bibfnamefont {J.}~\bibnamefont
  {Stachel}},\ }\bibfield  {title} {\bibinfo {title} {{Decoding the phase
  structure of QCD via particle production at high energy}},\ }\href
  {https://doi.org/10.1038/s41586-018-0491-6} {\bibfield  {journal} {\bibinfo
  {journal} {Nature}\ }\textbf {\bibinfo {volume} {561}},\ \bibinfo {pages}
  {321} (\bibinfo {year} {2018})},\ \Eprint {https://arxiv.org/abs/1710.09425}
  {arXiv:1710.09425 [nucl-th]} \BibitemShut {NoStop}%
\bibitem [{\citenamefont {Borsányi}\ \emph {et~al.}(2010)\citenamefont
  {Borsányi}, \citenamefont {Endr\H{o}di}, \citenamefont {Fodor},
  \citenamefont {Jakovác}, \citenamefont {Katz}, \citenamefont {Krieg},
  \citenamefont {Ratti},\ and\ \citenamefont {Szab\'o}}]{Borsanyi:2010cj}%
  \BibitemOpen
  \bibfield  {author} {\bibinfo {author} {\bibfnamefont {S.}~\bibnamefont
  {Borsányi}}, \bibinfo {author} {\bibfnamefont {G.}~\bibnamefont
  {Endr\H{o}di}}, \bibinfo {author} {\bibfnamefont {Z.}~\bibnamefont {Fodor}},
  \bibinfo {author} {\bibfnamefont {A.}~\bibnamefont {Jakovác}}, \bibinfo
  {author} {\bibfnamefont {S.~D.}\ \bibnamefont {Katz}}, \bibinfo {author}
  {\bibfnamefont {S.}~\bibnamefont {Krieg}}, \bibinfo {author} {\bibfnamefont
  {C.}~\bibnamefont {Ratti}},\ and\ \bibinfo {author} {\bibfnamefont {K.~K.}\
  \bibnamefont {Szab\'o}},\ }\bibfield  {title} {\bibinfo {title} {{The QCD
  equation of state with dynamical quarks}},\ }\href
  {https://doi.org/10.1007/JHEP11(2010)077} {\bibfield  {journal} {\bibinfo
  {journal} {JHEP}\ }\textbf {\bibinfo {volume} {11}},\ \bibinfo {pages}
  {077}},\ \Eprint {https://arxiv.org/abs/1007.2580} {arXiv:1007.2580
  [hep-lat]} \BibitemShut {NoStop}%
\bibitem [{\citenamefont {Borsányi}\ \emph {et~al.}(2014)\citenamefont
  {Borsányi}, \citenamefont {Fodor}, \citenamefont {Hoelbling}, \citenamefont
  {Katz}, \citenamefont {Krieg},\ and\ \citenamefont
  {Szabó}}]{Borsanyi:2013bia}%
  \BibitemOpen
  \bibfield  {author} {\bibinfo {author} {\bibfnamefont {S.}~\bibnamefont
  {Borsányi}}, \bibinfo {author} {\bibfnamefont {Z.}~\bibnamefont {Fodor}},
  \bibinfo {author} {\bibfnamefont {C.}~\bibnamefont {Hoelbling}}, \bibinfo
  {author} {\bibfnamefont {S.~D.}\ \bibnamefont {Katz}}, \bibinfo {author}
  {\bibfnamefont {S.}~\bibnamefont {Krieg}},\ and\ \bibinfo {author}
  {\bibfnamefont {K.~K.}\ \bibnamefont {Szabó}},\ }\bibfield  {title}
  {\bibinfo {title} {{Full result for the QCD equation of state with 2+1
  flavors}},\ }\href {https://doi.org/10.1016/j.physletb.2014.01.007}
  {\bibfield  {journal} {\bibinfo  {journal} {Phys. Lett. B}\ }\textbf
  {\bibinfo {volume} {730}},\ \bibinfo {pages} {99} (\bibinfo {year} {2014})},\
  \Eprint {https://arxiv.org/abs/1309.5258} {arXiv:1309.5258 [hep-lat]}
  \BibitemShut {NoStop}%
\bibitem [{\citenamefont {Bazavov}\ \emph
  {et~al.}(2014{\natexlab{b}})\citenamefont {Bazavov} \emph
  {et~al.}}]{HotQCD:2014kol}%
  \BibitemOpen
  \bibfield  {author} {\bibinfo {author} {\bibfnamefont {A.}~\bibnamefont
  {Bazavov}} \emph {et~al.} (\bibinfo {collaboration} {HotQCD}),\ }\bibfield
  {title} {\bibinfo {title} {{Equation of state in ( 2+1 )-flavor QCD}},\
  }\href {https://doi.org/10.1103/PhysRevD.90.094503} {\bibfield  {journal}
  {\bibinfo  {journal} {Phys. Rev. D}\ }\textbf {\bibinfo {volume} {90}},\
  \bibinfo {pages} {094503} (\bibinfo {year} {2014}{\natexlab{b}})},\ \Eprint
  {https://arxiv.org/abs/1407.6387} {arXiv:1407.6387 [hep-lat]} \BibitemShut
  {NoStop}%
\bibitem [{\citenamefont {Contino}(2011)}]{Contino:2010rs}%
  \BibitemOpen
  \bibfield  {author} {\bibinfo {author} {\bibfnamefont {R.}~\bibnamefont
  {Contino}},\ }\bibfield  {title} {\bibinfo {title} {The {Higgs} as a
  composite {Nambu-Goldstone} boson},\ }in\ \href
  {https://doi.org/10.1142/9789814327183_0005} {\emph {\bibinfo {booktitle}
  {{TASI 2009: Physics of the large and the small}}}},\ \bibinfo {editor}
  {edited by\ \bibinfo {editor} {\bibfnamefont {C.}~\bibnamefont {Csaki}}\ and\
  \bibinfo {editor} {\bibfnamefont {S.}~\bibnamefont {Dodelson}}}\ (\bibinfo
  {publisher} {World Scientific},\ \bibinfo {address} {Singapore},\ \bibinfo
  {year} {2011})\ pp.\ \bibinfo {pages} {235--306},\ \Eprint
  {https://arxiv.org/abs/1005.4269} {arXiv:1005.4269 [hep-ph]} \BibitemShut
  {NoStop}%
\bibitem [{\citenamefont {Panico}\ and\ \citenamefont
  {Wulzer}(2016)}]{Panico:2015jxa}%
  \BibitemOpen
  \bibfield  {author} {\bibinfo {author} {\bibfnamefont {G.}~\bibnamefont
  {Panico}}\ and\ \bibinfo {author} {\bibfnamefont {A.}~\bibnamefont
  {Wulzer}},\ }\bibfield  {title} {\bibinfo {title} {{The composite
  Nambu-Goldstone Higgs}},\ }\href {https://doi.org/10.1007/978-3-319-22617-0}
  {\bibfield  {journal} {\bibinfo  {journal} {Lect. Notes Phys.}\ }\textbf
  {\bibinfo {volume} {913}},\ \bibinfo {pages} {1} (\bibinfo {year} {2016})},\
  \Eprint {https://arxiv.org/abs/1506.01961} {arXiv:1506.01961 [hep-ph]}
  \BibitemShut {NoStop}%
\bibitem [{\citenamefont {Del~Debbio}(2018)}]{DelDebbio:2018szp}%
  \BibitemOpen
  \bibfield  {author} {\bibinfo {author} {\bibfnamefont {L.}~\bibnamefont
  {Del~Debbio}},\ }\bibfield  {title} {\bibinfo {title} {{BSM} physics and the
  lattice},\ }\href {https://doi.org/10.22323/1.330.0022} {\bibfield  {journal}
  {\bibinfo  {journal} {PoS}\ }\textbf {\bibinfo {volume} {ALPS2018}},\
  \bibinfo {pages} {022} (\bibinfo {year} {2018})}\BibitemShut {NoStop}%
\bibitem [{\citenamefont {Banerjee}\ \emph {et~al.}(2022)\citenamefont
  {Banerjee} \emph {et~al.}}]{Banerjee:2022xmu}%
  \BibitemOpen
  \bibfield  {author} {\bibinfo {author} {\bibfnamefont {A.}~\bibnamefont
  {Banerjee}} \emph {et~al.},\ }\bibfield  {title} {\bibinfo {title}
  {Phenomenological aspects of composite {Higgs} scenarios: Exotic scalars and
  vector-like quarks},\ }in\ \href@noop {} {\emph {\bibinfo {booktitle} {{2022
  Snowmass Summer Study}}}}\ (\bibinfo {year} {2022})\ \Eprint
  {https://arxiv.org/abs/2203.07270} {arXiv:2203.07270 [hep-ph]} \BibitemShut
  {NoStop}%
\bibitem [{\citenamefont {Fodor}\ \emph
  {et~al.}(2016{\natexlab{b}})\citenamefont {Fodor}, \citenamefont {Holland},
  \citenamefont {Kuti}, \citenamefont {Mondal}, \citenamefont {Nogradi},\ and\
  \citenamefont {Wong}}]{Fodor:2016wal}%
  \BibitemOpen
  \bibfield  {author} {\bibinfo {author} {\bibfnamefont {Z.}~\bibnamefont
  {Fodor}}, \bibinfo {author} {\bibfnamefont {K.}~\bibnamefont {Holland}},
  \bibinfo {author} {\bibfnamefont {J.}~\bibnamefont {Kuti}}, \bibinfo {author}
  {\bibfnamefont {S.}~\bibnamefont {Mondal}}, \bibinfo {author} {\bibfnamefont
  {D.}~\bibnamefont {Nogradi}},\ and\ \bibinfo {author} {\bibfnamefont {C.~H.}\
  \bibnamefont {Wong}},\ }\bibfield  {title} {\bibinfo {title} {Electroweak
  interactions and dark baryons in the sextet {BSM} model with a composite
  {Higgs} particle},\ }\href {https://doi.org/10.1103/PhysRevD.94.014503}
  {\bibfield  {journal} {\bibinfo  {journal} {Phys. Rev. D}\ }\textbf {\bibinfo
  {volume} {94}},\ \bibinfo {pages} {014503} (\bibinfo {year}
  {2016}{\natexlab{b}})},\ \Eprint {https://arxiv.org/abs/1601.03302}
  {arXiv:1601.03302 [hep-lat]} \BibitemShut {NoStop}%
\bibitem [{\citenamefont {Ayyar}\ \emph
  {et~al.}(2018{\natexlab{a}})\citenamefont {Ayyar} \emph
  {et~al.}}]{Ayyar:2017qdf}%
  \BibitemOpen
  \bibfield  {author} {\bibinfo {author} {\bibfnamefont {V.}~\bibnamefont
  {Ayyar}} \emph {et~al.},\ }\bibfield  {title} {\bibinfo {title}
  {{Spectroscopy of SU(4) composite Higgs theory with two distinct fermion
  representations}},\ }\href {https://doi.org/10.1103/PhysRevD.97.074505}
  {\bibfield  {journal} {\bibinfo  {journal} {Phys. Rev. D}\ }\textbf {\bibinfo
  {volume} {97}},\ \bibinfo {pages} {074505} (\bibinfo {year}
  {2018}{\natexlab{a}})},\ \Eprint {https://arxiv.org/abs/1710.00806}
  {arXiv:1710.00806 [hep-lat]} \BibitemShut {NoStop}%
\bibitem [{\citenamefont {Ayyar}\ \emph
  {et~al.}(2018{\natexlab{b}})\citenamefont {Ayyar} \emph
  {et~al.}}]{Ayyar:2018zuk}%
  \BibitemOpen
  \bibfield  {author} {\bibinfo {author} {\bibfnamefont {V.}~\bibnamefont
  {Ayyar}} \emph {et~al.},\ }\bibfield  {title} {\bibinfo {title} {{Baryon
  spectrum of SU(4) composite Higgs theory with two distinct fermion
  representations}},\ }\href {https://doi.org/10.1103/PhysRevD.97.114505}
  {\bibfield  {journal} {\bibinfo  {journal} {Phys. Rev. D}\ }\textbf {\bibinfo
  {volume} {97}},\ \bibinfo {pages} {114505} (\bibinfo {year}
  {2018}{\natexlab{b}})},\ \Eprint {https://arxiv.org/abs/1801.05809}
  {arXiv:1801.05809 [hep-ph]} \BibitemShut {NoStop}%
\bibitem [{\citenamefont {Ayyar}\ \emph {et~al.}(2019)\citenamefont {Ayyar}
  \emph {et~al.}}]{Ayyar:2018glg}%
  \BibitemOpen
  \bibfield  {author} {\bibinfo {author} {\bibfnamefont {V.}~\bibnamefont
  {Ayyar}} \emph {et~al.},\ }\bibfield  {title} {\bibinfo {title} {{Partial
  compositeness and baryon matrix elements on the lattice}},\ }\href
  {https://doi.org/10.1103/PhysRevD.99.094502} {\bibfield  {journal} {\bibinfo
  {journal} {Phys. Rev. D}\ }\textbf {\bibinfo {volume} {99}},\ \bibinfo
  {pages} {094502} (\bibinfo {year} {2019})},\ \Eprint
  {https://arxiv.org/abs/1812.02727} {arXiv:1812.02727 [hep-ph]} \BibitemShut
  {NoStop}%
\bibitem [{\citenamefont {Brower}\ \emph {et~al.}(2016)\citenamefont {Brower},
  \citenamefont {Hasenfratz}, \citenamefont {Rebbi}, \citenamefont {Weinberg},\
  and\ \citenamefont {Witzel}}]{Brower:2015owo}%
  \BibitemOpen
  \bibfield  {author} {\bibinfo {author} {\bibfnamefont {R.~C.}\ \bibnamefont
  {Brower}}, \bibinfo {author} {\bibfnamefont {A.}~\bibnamefont {Hasenfratz}},
  \bibinfo {author} {\bibfnamefont {C.}~\bibnamefont {Rebbi}}, \bibinfo
  {author} {\bibfnamefont {E.}~\bibnamefont {Weinberg}},\ and\ \bibinfo
  {author} {\bibfnamefont {O.}~\bibnamefont {Witzel}},\ }\bibfield  {title}
  {\bibinfo {title} {{Composite Higgs model at a conformal fixed point}},\
  }\href {https://doi.org/10.1103/PhysRevD.93.075028} {\bibfield  {journal}
  {\bibinfo  {journal} {Phys. Rev. D}\ }\textbf {\bibinfo {volume} {93}},\
  \bibinfo {pages} {075028} (\bibinfo {year} {2016})},\ \Eprint
  {https://arxiv.org/abs/1512.02576} {arXiv:1512.02576 [hep-ph]} \BibitemShut
  {NoStop}%
\bibitem [{\citenamefont {Hasenfratz}\ \emph {et~al.}(2017)\citenamefont
  {Hasenfratz}, \citenamefont {Rebbi},\ and\ \citenamefont
  {Witzel}}]{Hasenfratz:2016gut}%
  \BibitemOpen
  \bibfield  {author} {\bibinfo {author} {\bibfnamefont {A.}~\bibnamefont
  {Hasenfratz}}, \bibinfo {author} {\bibfnamefont {C.}~\bibnamefont {Rebbi}},\
  and\ \bibinfo {author} {\bibfnamefont {O.}~\bibnamefont {Witzel}},\
  }\bibfield  {title} {\bibinfo {title} {{Large scale separation and resonances
  within LHC range from a prototype BSM model}},\ }\href
  {https://doi.org/10.1016/j.physletb.2017.07.058} {\bibfield  {journal}
  {\bibinfo  {journal} {Phys. Lett. B}\ }\textbf {\bibinfo {volume} {773}},\
  \bibinfo {pages} {86} (\bibinfo {year} {2017})},\ \Eprint
  {https://arxiv.org/abs/1609.01401} {arXiv:1609.01401 [hep-ph]} \BibitemShut
  {NoStop}%
\bibitem [{\citenamefont {Appelquist}\ \emph {et~al.}(2016)\citenamefont
  {Appelquist} \emph {et~al.}}]{Appelquist:2016viq}%
  \BibitemOpen
  \bibfield  {author} {\bibinfo {author} {\bibfnamefont {T.}~\bibnamefont
  {Appelquist}} \emph {et~al.} (\bibinfo {collaboration} {Lattice Strong
  Dynamics}),\ }\bibfield  {title} {\bibinfo {title} {{Strongly interacting
  dynamics and the search for new physics at the LHC}},\ }\href
  {https://doi.org/10.1103/PhysRevD.93.114514} {\bibfield  {journal} {\bibinfo
  {journal} {Phys. Rev. D}\ }\textbf {\bibinfo {volume} {93}},\ \bibinfo
  {pages} {114514} (\bibinfo {year} {2016})},\ \Eprint
  {https://arxiv.org/abs/1601.04027} {arXiv:1601.04027 [hep-lat]} \BibitemShut
  {NoStop}%
\bibitem [{\citenamefont {Appelquist}\ \emph {et~al.}(2019)\citenamefont
  {Appelquist} \emph {et~al.}}]{Appelquist:2018yqe}%
  \BibitemOpen
  \bibfield  {author} {\bibinfo {author} {\bibfnamefont {T.}~\bibnamefont
  {Appelquist}} \emph {et~al.} (\bibinfo {collaboration} {Lattice Strong
  Dynamics}),\ }\bibfield  {title} {\bibinfo {title} {{Nonperturbative
  investigations of SU(3) gauge theory with eight dynamical flavors}},\ }\href
  {https://doi.org/10.1103/PhysRevD.99.014509} {\bibfield  {journal} {\bibinfo
  {journal} {Phys. Rev. D}\ }\textbf {\bibinfo {volume} {99}},\ \bibinfo
  {pages} {014509} (\bibinfo {year} {2019})},\ \Eprint
  {https://arxiv.org/abs/1807.08411} {arXiv:1807.08411 [hep-lat]} \BibitemShut
  {NoStop}%
\bibitem [{\citenamefont {Appelquist}\ \emph {et~al.}(2021)\citenamefont
  {Appelquist} \emph {et~al.}}]{LatticeStrongDynamics:2020uwo}%
  \BibitemOpen
  \bibfield  {author} {\bibinfo {author} {\bibfnamefont {T.}~\bibnamefont
  {Appelquist}} \emph {et~al.} (\bibinfo {collaboration} {Lattice Strong
  Dynamics}),\ }\bibfield  {title} {\bibinfo {title} {{Near-conformal dynamics
  in a chirally broken system}},\ }\href
  {https://doi.org/10.1103/PhysRevD.103.014504} {\bibfield  {journal} {\bibinfo
   {journal} {Phys. Rev. D}\ }\textbf {\bibinfo {volume} {103}},\ \bibinfo
  {pages} {014504} (\bibinfo {year} {2021})},\ \Eprint
  {https://arxiv.org/abs/2007.01810} {arXiv:2007.01810 [hep-ph]} \BibitemShut
  {NoStop}%
\bibitem [{\citenamefont {Kilic}\ \emph {et~al.}(2010)\citenamefont {Kilic},
  \citenamefont {Okui},\ and\ \citenamefont {Sundrum}}]{Kilic:2009mi}%
  \BibitemOpen
  \bibfield  {author} {\bibinfo {author} {\bibfnamefont {C.}~\bibnamefont
  {Kilic}}, \bibinfo {author} {\bibfnamefont {T.}~\bibnamefont {Okui}},\ and\
  \bibinfo {author} {\bibfnamefont {R.}~\bibnamefont {Sundrum}},\ }\bibfield
  {title} {\bibinfo {title} {Vectorlike confinement at the {LHC}},\ }\href
  {https://doi.org/10.1007/JHEP02(2010)018} {\bibfield  {journal} {\bibinfo
  {journal} {JHEP}\ }\textbf {\bibinfo {volume} {02}},\ \bibinfo {pages}
  {018}},\ \Eprint {https://arxiv.org/abs/0906.0577} {arXiv:0906.0577 [hep-ph]}
  \BibitemShut {NoStop}%
\bibitem [{\citenamefont {Daci}\ \emph {et~al.}(2015)\citenamefont {Daci},
  \citenamefont {De~Bruyn}, \citenamefont {Lowette}, \citenamefont {Tytgat},\
  and\ \citenamefont {Zaldivar}}]{Daci:2015hca}%
  \BibitemOpen
  \bibfield  {author} {\bibinfo {author} {\bibfnamefont {N.}~\bibnamefont
  {Daci}}, \bibinfo {author} {\bibfnamefont {I.}~\bibnamefont {De~Bruyn}},
  \bibinfo {author} {\bibfnamefont {S.}~\bibnamefont {Lowette}}, \bibinfo
  {author} {\bibfnamefont {M.~H.~G.}\ \bibnamefont {Tytgat}},\ and\ \bibinfo
  {author} {\bibfnamefont {B.}~\bibnamefont {Zaldivar}},\ }\bibfield  {title}
  {\bibinfo {title} {{Simplified SIMPs and the LHC}},\ }\href
  {https://doi.org/10.1007/JHEP11(2015)108} {\bibfield  {journal} {\bibinfo
  {journal} {JHEP}\ }\textbf {\bibinfo {volume} {11}},\ \bibinfo {pages}
  {108}},\ \Eprint {https://arxiv.org/abs/1503.05505} {arXiv:1503.05505
  [hep-ph]} \BibitemShut {NoStop}%
\bibitem [{\citenamefont {Kribs}\ \emph {et~al.}(2019)\citenamefont {Kribs},
  \citenamefont {Martin},\ and\ \citenamefont {Tong}}]{Kribs:2018oad}%
  \BibitemOpen
  \bibfield  {author} {\bibinfo {author} {\bibfnamefont {G.~D.}\ \bibnamefont
  {Kribs}}, \bibinfo {author} {\bibfnamefont {A.}~\bibnamefont {Martin}},\ and\
  \bibinfo {author} {\bibfnamefont {T.}~\bibnamefont {Tong}},\ }\bibfield
  {title} {\bibinfo {title} {Effective theories of dark mesons with custodial
  symmetry},\ }\href@noop {} {\bibfield  {journal} {\bibinfo  {journal} {JHEP}\
  }\textbf {\bibinfo {volume} {08}},\ \bibinfo {pages} {020}},\ \Eprint
  {https://arxiv.org/abs/1809.10183} {arXiv:1809.10183 [hep-ph]} \BibitemShut
  {NoStop}%
\bibitem [{\citenamefont {Kribs}\ and\ \citenamefont
  {Neil}(2016)}]{Kribs:2016cew}%
  \BibitemOpen
  \bibfield  {author} {\bibinfo {author} {\bibfnamefont {G.~D.}\ \bibnamefont
  {Kribs}}\ and\ \bibinfo {author} {\bibfnamefont {E.~T.}\ \bibnamefont
  {Neil}},\ }\bibfield  {title} {\bibinfo {title} {Review of strongly-coupled
  composite dark matter models and lattice simulations},\ }\href
  {https://doi.org/10.1142/S0217751X16430041} {\bibfield  {journal} {\bibinfo
  {journal} {Int. J. Mod. Phys.}\ }\textbf {\bibinfo {volume} {A31}},\ \bibinfo
  {pages} {1643004} (\bibinfo {year} {2016})},\ \Eprint
  {https://arxiv.org/abs/1604.04627} {arXiv:1604.04627 [hep-ph]} \BibitemShut
  {NoStop}%
\bibitem [{\citenamefont {Berkowitz}\ \emph {et~al.}(2015)\citenamefont
  {Berkowitz}, \citenamefont {Buchoff},\ and\ \citenamefont
  {Rinaldi}}]{Berkowitz:2015aua}%
  \BibitemOpen
  \bibfield  {author} {\bibinfo {author} {\bibfnamefont {E.}~\bibnamefont
  {Berkowitz}}, \bibinfo {author} {\bibfnamefont {M.~I.}\ \bibnamefont
  {Buchoff}},\ and\ \bibinfo {author} {\bibfnamefont {E.}~\bibnamefont
  {Rinaldi}},\ }\bibfield  {title} {\bibinfo {title} {{Lattice QCD input for
  axion cosmology}},\ }\href {https://doi.org/10.1103/PhysRevD.92.034507}
  {\bibfield  {journal} {\bibinfo  {journal} {Phys. Rev. D}\ }\textbf {\bibinfo
  {volume} {92}},\ \bibinfo {pages} {034507} (\bibinfo {year} {2015})},\
  \Eprint {https://arxiv.org/abs/1505.07455} {arXiv:1505.07455 [hep-ph]}
  \BibitemShut {NoStop}%
\bibitem [{\citenamefont {Kitano}\ and\ \citenamefont
  {Yamada}(2015)}]{Kitano:2015fla}%
  \BibitemOpen
  \bibfield  {author} {\bibinfo {author} {\bibfnamefont {R.}~\bibnamefont
  {Kitano}}\ and\ \bibinfo {author} {\bibfnamefont {N.}~\bibnamefont
  {Yamada}},\ }\bibfield  {title} {\bibinfo {title} {{Topology in QCD and the
  axion abundance}},\ }\href {https://doi.org/10.1007/JHEP10(2015)136}
  {\bibfield  {journal} {\bibinfo  {journal} {JHEP}\ }\textbf {\bibinfo
  {volume} {10}},\ \bibinfo {pages} {136}},\ \Eprint
  {https://arxiv.org/abs/1506.00370} {arXiv:1506.00370 [hep-ph]} \BibitemShut
  {NoStop}%
\bibitem [{\citenamefont {Borsányi}\ \emph
  {et~al.}(2016{\natexlab{a}})\citenamefont {Borsányi}, \citenamefont
  {Dierigl}, \citenamefont {Fodor}, \citenamefont {Katz}, \citenamefont
  {Mages}, \citenamefont {Nogradi}, \citenamefont {Redondo}, \citenamefont
  {Ringwald},\ and\ \citenamefont {Szabó}}]{Borsanyi:2015cka}%
  \BibitemOpen
  \bibfield  {author} {\bibinfo {author} {\bibfnamefont {S.}~\bibnamefont
  {Borsányi}}, \bibinfo {author} {\bibfnamefont {M.}~\bibnamefont {Dierigl}},
  \bibinfo {author} {\bibfnamefont {Z.}~\bibnamefont {Fodor}}, \bibinfo
  {author} {\bibfnamefont {S.~D.}\ \bibnamefont {Katz}}, \bibinfo {author}
  {\bibfnamefont {S.~W.}\ \bibnamefont {Mages}}, \bibinfo {author}
  {\bibfnamefont {D.}~\bibnamefont {Nogradi}}, \bibinfo {author} {\bibfnamefont
  {J.}~\bibnamefont {Redondo}}, \bibinfo {author} {\bibfnamefont
  {A.}~\bibnamefont {Ringwald}},\ and\ \bibinfo {author} {\bibfnamefont
  {K.~K.}\ \bibnamefont {Szabó}},\ }\bibfield  {title} {\bibinfo {title}
  {{Axion cosmology, lattice QCD and the dilute instanton gas}},\ }\href
  {https://doi.org/10.1016/j.physletb.2015.11.020} {\bibfield  {journal}
  {\bibinfo  {journal} {Phys. Lett. B}\ }\textbf {\bibinfo {volume} {752}},\
  \bibinfo {pages} {175} (\bibinfo {year} {2016}{\natexlab{a}})},\ \Eprint
  {https://arxiv.org/abs/1508.06917} {arXiv:1508.06917 [hep-lat]} \BibitemShut
  {NoStop}%
\bibitem [{\citenamefont {Borsányi}\ \emph
  {et~al.}(2016{\natexlab{b}})\citenamefont {Borsányi} \emph
  {et~al.}}]{Borsanyi:2016ksw}%
  \BibitemOpen
  \bibfield  {author} {\bibinfo {author} {\bibfnamefont {S.}~\bibnamefont
  {Borsányi}} \emph {et~al.},\ }\bibfield  {title} {\bibinfo {title}
  {{Calculation of the axion mass based on high-temperature lattice quantum
  chromodynamics}},\ }\href {https://doi.org/10.1038/nature20115} {\bibfield
  {journal} {\bibinfo  {journal} {Nature}\ }\textbf {\bibinfo {volume} {539}},\
  \bibinfo {pages} {69} (\bibinfo {year} {2016}{\natexlab{b}})},\ \Eprint
  {https://arxiv.org/abs/1606.07494} {arXiv:1606.07494 [hep-lat]} \BibitemShut
  {NoStop}%
\bibitem [{\citenamefont {Bonati}\ \emph {et~al.}(2016)\citenamefont {Bonati},
  \citenamefont {D'Elia}, \citenamefont {Mariti}, \citenamefont {Martinelli},
  \citenamefont {Mesiti}, \citenamefont {Negro}, \citenamefont {Sanfilippo},\
  and\ \citenamefont {Villadoro}}]{Bonati:2015vqz}%
  \BibitemOpen
  \bibfield  {author} {\bibinfo {author} {\bibfnamefont {C.}~\bibnamefont
  {Bonati}}, \bibinfo {author} {\bibfnamefont {M.}~\bibnamefont {D'Elia}},
  \bibinfo {author} {\bibfnamefont {M.}~\bibnamefont {Mariti}}, \bibinfo
  {author} {\bibfnamefont {G.}~\bibnamefont {Martinelli}}, \bibinfo {author}
  {\bibfnamefont {M.}~\bibnamefont {Mesiti}}, \bibinfo {author} {\bibfnamefont
  {F.}~\bibnamefont {Negro}}, \bibinfo {author} {\bibfnamefont
  {F.}~\bibnamefont {Sanfilippo}},\ and\ \bibinfo {author} {\bibfnamefont
  {G.}~\bibnamefont {Villadoro}},\ }\bibfield  {title} {\bibinfo {title}
  {{Axion phenomenology and $\theta$-dependence from $N_f = 2+1$ lattice
  QCD}},\ }\href {https://doi.org/10.1007/JHEP03(2016)155} {\bibfield
  {journal} {\bibinfo  {journal} {JHEP}\ }\textbf {\bibinfo {volume} {03}},\
  \bibinfo {pages} {155}},\ \Eprint {https://arxiv.org/abs/1512.06746}
  {arXiv:1512.06746 [hep-lat]} \BibitemShut {NoStop}%
\bibitem [{\citenamefont {Petreczky}\ \emph {et~al.}(2016)\citenamefont
  {Petreczky}, \citenamefont {Schadler},\ and\ \citenamefont
  {Sharma}}]{Petreczky:2016vrs}%
  \BibitemOpen
  \bibfield  {author} {\bibinfo {author} {\bibfnamefont {P.}~\bibnamefont
  {Petreczky}}, \bibinfo {author} {\bibfnamefont {H.-P.}\ \bibnamefont
  {Schadler}},\ and\ \bibinfo {author} {\bibfnamefont {S.}~\bibnamefont
  {Sharma}},\ }\bibfield  {title} {\bibinfo {title} {{The topological
  susceptibility in finite temperature QCD and axion cosmology}},\ }\href
  {https://doi.org/10.1016/j.physletb.2016.09.063} {\bibfield  {journal}
  {\bibinfo  {journal} {Phys. Lett. B}\ }\textbf {\bibinfo {volume} {762}},\
  \bibinfo {pages} {498} (\bibinfo {year} {2016})},\ \Eprint
  {https://arxiv.org/abs/1606.03145} {arXiv:1606.03145 [hep-lat]} \BibitemShut
  {NoStop}%
\bibitem [{\citenamefont {Azcoiti}(2016)}]{Azcoiti:2016zbi}%
  \BibitemOpen
  \bibfield  {author} {\bibinfo {author} {\bibfnamefont {V.}~\bibnamefont
  {Azcoiti}},\ }\bibfield  {title} {\bibinfo {title} {{Topology in the
  SU($N_f$) chiral symmetry restored phase of unquenched QCD and axion
  cosmology}},\ }\href {https://doi.org/10.1103/PhysRevD.94.094505} {\bibfield
  {journal} {\bibinfo  {journal} {Phys. Rev. D}\ }\textbf {\bibinfo {volume}
  {94}},\ \bibinfo {pages} {094505} (\bibinfo {year} {2016})},\ \Eprint
  {https://arxiv.org/abs/1609.01230} {arXiv:1609.01230 [hep-lat]} \BibitemShut
  {NoStop}%
\bibitem [{\citenamefont {Azcoiti}(2017)}]{Azcoiti:2017jsh}%
  \BibitemOpen
  \bibfield  {author} {\bibinfo {author} {\bibfnamefont {V.}~\bibnamefont
  {Azcoiti}},\ }\bibfield  {title} {\bibinfo {title} {{Topology in the
  SU($N_f$) chiral symmetry restored phase of unquenched QCD and axion
  cosmology~2}},\ }\href {https://doi.org/10.1103/PhysRevD.96.014505}
  {\bibfield  {journal} {\bibinfo  {journal} {Phys. Rev. D}\ }\textbf {\bibinfo
  {volume} {96}},\ \bibinfo {pages} {014505} (\bibinfo {year} {2017})},\
  \Eprint {https://arxiv.org/abs/1704.04906} {arXiv:1704.04906 [hep-lat]}
  \BibitemShut {NoStop}%
\bibitem [{\citenamefont {Jaeckel}\ \emph {et~al.}(2022)\citenamefont
  {Jaeckel}, \citenamefont {Rybka}, \citenamefont {Winslow} \emph
  {et~al.}}]{Adams:2022pbo}%
  \BibitemOpen
  \bibfield  {author} {\bibinfo {author} {\bibfnamefont {J.}~\bibnamefont
  {Jaeckel}}, \bibinfo {author} {\bibfnamefont {G.}~\bibnamefont {Rybka}},
  \bibinfo {author} {\bibfnamefont {L.}~\bibnamefont {Winslow}}, \emph
  {et~al.},\ }\bibfield  {title} {\bibinfo {title} {Axion dark matter},\ }in\
  \href@noop {} {\emph {\bibinfo {booktitle} {{2022 Snowmass Summer Study}}}}\
  (\bibinfo {year} {2022})\ \Eprint {https://arxiv.org/abs/2203.14923}
  {arXiv:2203.14923 [hep-ex]} \BibitemShut {NoStop}%
\bibitem [{\citenamefont {Schwaller}(2015)}]{Schwaller:2015tja}%
  \BibitemOpen
  \bibfield  {author} {\bibinfo {author} {\bibfnamefont {P.}~\bibnamefont
  {Schwaller}},\ }\bibfield  {title} {\bibinfo {title} {Gravitational waves
  from a dark phase transition},\ }\href
  {https://doi.org/10.1103/PhysRevLett.115.181101} {\bibfield  {journal}
  {\bibinfo  {journal} {Phys. Rev. Lett.}\ }\textbf {\bibinfo {volume} {115}},\
  \bibinfo {pages} {181101} (\bibinfo {year} {2015})},\ \Eprint
  {https://arxiv.org/abs/1504.07263} {arXiv:1504.07263 [hep-ph]} \BibitemShut
  {NoStop}%
\bibitem [{\citenamefont {Catterall}\ \emph {et~al.}(2009)\citenamefont
  {Catterall}, \citenamefont {Kaplan},\ and\ \citenamefont
  {Ünsal}}]{Catterall:2009it}%
  \BibitemOpen
  \bibfield  {author} {\bibinfo {author} {\bibfnamefont {S.}~\bibnamefont
  {Catterall}}, \bibinfo {author} {\bibfnamefont {D.~B.}\ \bibnamefont
  {Kaplan}},\ and\ \bibinfo {author} {\bibfnamefont {M.}~\bibnamefont
  {Ünsal}},\ }\bibfield  {title} {\bibinfo {title} {{Exact lattice
  supersymmetry}},\ }\href {https://doi.org/10.1016/j.physrep.2009.09.001}
  {\bibfield  {journal} {\bibinfo  {journal} {Phys. Rept.}\ }\textbf {\bibinfo
  {volume} {484}},\ \bibinfo {pages} {71} (\bibinfo {year} {2009})},\ \Eprint
  {https://arxiv.org/abs/0903.4881} {arXiv:0903.4881 [hep-lat]} \BibitemShut
  {NoStop}%
\bibitem [{\citenamefont {Berkowitz}\ \emph {et~al.}(2016)\citenamefont
  {Berkowitz}, \citenamefont {Rinaldi}, \citenamefont {Hanada}, \citenamefont
  {Ishiki}, \citenamefont {Shimasaki},\ and\ \citenamefont
  {Vranas}}]{Berkowitz:2016jlq}%
  \BibitemOpen
  \bibfield  {author} {\bibinfo {author} {\bibfnamefont {E.}~\bibnamefont
  {Berkowitz}}, \bibinfo {author} {\bibfnamefont {E.}~\bibnamefont {Rinaldi}},
  \bibinfo {author} {\bibfnamefont {M.}~\bibnamefont {Hanada}}, \bibinfo
  {author} {\bibfnamefont {G.}~\bibnamefont {Ishiki}}, \bibinfo {author}
  {\bibfnamefont {S.}~\bibnamefont {Shimasaki}},\ and\ \bibinfo {author}
  {\bibfnamefont {P.}~\bibnamefont {Vranas}} (\bibinfo {collaboration}
  {MCSM}),\ }\bibfield  {title} {\bibinfo {title} {{Precision lattice test of
  the gauge/gravity duality at large-$N$}},\ }\href
  {https://doi.org/10.1103/PhysRevD.94.094501} {\bibfield  {journal} {\bibinfo
  {journal} {Phys. Rev. D}\ }\textbf {\bibinfo {volume} {94}},\ \bibinfo
  {pages} {094501} (\bibinfo {year} {2016})},\ \Eprint
  {https://arxiv.org/abs/1606.04951} {arXiv:1606.04951 [hep-lat]} \BibitemShut
  {NoStop}%
\bibitem [{\citenamefont {Catterall}\ \emph {et~al.}(2010)\citenamefont
  {Catterall}, \citenamefont {Joseph},\ and\ \citenamefont
  {Wiseman}}]{Catterall:2010fx}%
  \BibitemOpen
  \bibfield  {author} {\bibinfo {author} {\bibfnamefont {S.}~\bibnamefont
  {Catterall}}, \bibinfo {author} {\bibfnamefont {A.}~\bibnamefont {Joseph}},\
  and\ \bibinfo {author} {\bibfnamefont {T.}~\bibnamefont {Wiseman}},\
  }\bibfield  {title} {\bibinfo {title} {{Thermal phases of D1-branes on a
  circle from lattice super Yang-Mills}},\ }\href
  {https://doi.org/10.1007/JHEP12(2010)022} {\bibfield  {journal} {\bibinfo
  {journal} {JHEP}\ }\textbf {\bibinfo {volume} {12}},\ \bibinfo {pages}
  {022}},\ \Eprint {https://arxiv.org/abs/1008.4964} {arXiv:1008.4964 [hep-th]}
  \BibitemShut {NoStop}%
\bibitem [{\citenamefont {Catterall}\ \emph {et~al.}(2018)\citenamefont
  {Catterall}, \citenamefont {Jha}, \citenamefont {Schaich},\ and\
  \citenamefont {Wiseman}}]{Catterall:2017lub}%
  \BibitemOpen
  \bibfield  {author} {\bibinfo {author} {\bibfnamefont {S.}~\bibnamefont
  {Catterall}}, \bibinfo {author} {\bibfnamefont {R.~G.}\ \bibnamefont {Jha}},
  \bibinfo {author} {\bibfnamefont {D.}~\bibnamefont {Schaich}},\ and\ \bibinfo
  {author} {\bibfnamefont {T.}~\bibnamefont {Wiseman}},\ }\bibfield  {title}
  {\bibinfo {title} {{Testing holography using lattice super-Yang-Mills theory
  on a 2-torus}},\ }\href {https://doi.org/10.1103/PhysRevD.97.086020}
  {\bibfield  {journal} {\bibinfo  {journal} {Phys. Rev. D}\ }\textbf {\bibinfo
  {volume} {97}},\ \bibinfo {pages} {086020} (\bibinfo {year} {2018})},\
  \Eprint {https://arxiv.org/abs/1709.07025} {arXiv:1709.07025 [hep-th]}
  \BibitemShut {NoStop}%
\bibitem [{\citenamefont {Polley}\ and\ \citenamefont
  {Wiese}(1991)}]{Polley:1990tf}%
  \BibitemOpen
  \bibfield  {author} {\bibinfo {author} {\bibfnamefont {L.}~\bibnamefont
  {Polley}}\ and\ \bibinfo {author} {\bibfnamefont {U.~J.}\ \bibnamefont
  {Wiese}},\ }\bibfield  {title} {\bibinfo {title} {Monopole condensate and
  monopole mass in {U(1)} lattice gauge theory},\ }\href
  {https://doi.org/10.1016/0550-3213(91)90380-G} {\bibfield  {journal}
  {\bibinfo  {journal} {Nucl. Phys. B}\ }\textbf {\bibinfo {volume} {356}},\
  \bibinfo {pages} {629} (\bibinfo {year} {1991})}\BibitemShut {NoStop}%
\bibitem [{\citenamefont {Kronfeld}\ and\ \citenamefont
  {Wiese}(1991)}]{Kronfeld:1990qu}%
  \BibitemOpen
  \bibfield  {author} {\bibinfo {author} {\bibfnamefont {A.~S.}\ \bibnamefont
  {Kronfeld}}\ and\ \bibinfo {author} {\bibfnamefont {U.~J.}\ \bibnamefont
  {Wiese}},\ }\bibfield  {title} {\bibinfo {title} {{SU($N$) gauge theories
  with $C$-periodic boundary conditions~1: Topological structure}},\ }\href
  {https://doi.org/10.1016/0550-3213(91)90479-H} {\bibfield  {journal}
  {\bibinfo  {journal} {Nucl. Phys. B}\ }\textbf {\bibinfo {volume} {357}},\
  \bibinfo {pages} {521} (\bibinfo {year} {1991})}\BibitemShut {NoStop}%
\bibitem [{\citenamefont {Wiese}(1992)}]{Wiese:1991ku}%
  \BibitemOpen
  \bibfield  {author} {\bibinfo {author} {\bibfnamefont {U.~J.}\ \bibnamefont
  {Wiese}},\ }\bibfield  {title} {\bibinfo {title} {{$C$-periodic and
  $G$-periodic QCD at finite temperature}},\ }\href
  {https://doi.org/10.1016/0550-3213(92)90333-7} {\bibfield  {journal}
  {\bibinfo  {journal} {Nucl. Phys. B}\ }\textbf {\bibinfo {volume} {375}},\
  \bibinfo {pages} {45} (\bibinfo {year} {1992})}\BibitemShut {NoStop}%
\bibitem [{\citenamefont {Kanamori}\ \emph {et~al.}(2008)\citenamefont
  {Kanamori}, \citenamefont {Sugino},\ and\ \citenamefont
  {Suzuki}}]{Kanamori:2007yx}%
  \BibitemOpen
  \bibfield  {author} {\bibinfo {author} {\bibfnamefont {I.}~\bibnamefont
  {Kanamori}}, \bibinfo {author} {\bibfnamefont {F.}~\bibnamefont {Sugino}},\
  and\ \bibinfo {author} {\bibfnamefont {H.}~\bibnamefont {Suzuki}},\
  }\bibfield  {title} {\bibinfo {title} {Observing dynamical supersymmetry
  breaking with {Euclidean} lattice simulations},\ }\href
  {https://doi.org/10.1143/PTP.119.797} {\bibfield  {journal} {\bibinfo
  {journal} {Prog. Theor. Phys.}\ }\textbf {\bibinfo {volume} {119}},\ \bibinfo
  {pages} {797} (\bibinfo {year} {2008})},\ \Eprint
  {https://arxiv.org/abs/0711.2132} {arXiv:0711.2132 [hep-lat]} \BibitemShut
  {NoStop}%
\bibitem [{\citenamefont {Catterall}\ and\ \citenamefont
  {Veernala}(2015)}]{Catterall:2015tta}%
  \BibitemOpen
  \bibfield  {author} {\bibinfo {author} {\bibfnamefont {S.}~\bibnamefont
  {Catterall}}\ and\ \bibinfo {author} {\bibfnamefont {A.}~\bibnamefont
  {Veernala}},\ }\bibfield  {title} {\bibinfo {title} {{Spontaneous
  supersymmetry breaking in two dimensional lattice super QCD}},\ }\href
  {https://doi.org/10.1007/JHEP10(2015)013} {\bibfield  {journal} {\bibinfo
  {journal} {JHEP}\ }\textbf {\bibinfo {volume} {10}},\ \bibinfo {pages}
  {013}},\ \Eprint {https://arxiv.org/abs/1505.00467} {arXiv:1505.00467
  [hep-lat]} \BibitemShut {NoStop}%
\bibitem [{\citenamefont {Sugino}(2005)}]{Sugino:2004uv}%
  \BibitemOpen
  \bibfield  {author} {\bibinfo {author} {\bibfnamefont {F.}~\bibnamefont
  {Sugino}},\ }\bibfield  {title} {\bibinfo {title} {{Various super Yang-Mills
  theories with exact supersymmetry on the lattice}},\ }\href
  {https://doi.org/10.1088/1126-6708/2005/01/016} {\bibfield  {journal}
  {\bibinfo  {journal} {JHEP}\ }\textbf {\bibinfo {volume} {01}},\ \bibinfo
  {pages} {016}},\ \Eprint {https://arxiv.org/abs/hep-lat/0410035}
  {arXiv:hep-lat/0410035} \BibitemShut {NoStop}%
\bibitem [{\citenamefont {Hanada}\ \emph {et~al.}(2012)\citenamefont {Hanada},
  \citenamefont {Matsuura},\ and\ \citenamefont {Sugino}}]{Hanada:2011qx}%
  \BibitemOpen
  \bibfield  {author} {\bibinfo {author} {\bibfnamefont {M.}~\bibnamefont
  {Hanada}}, \bibinfo {author} {\bibfnamefont {S.}~\bibnamefont {Matsuura}},\
  and\ \bibinfo {author} {\bibfnamefont {F.}~\bibnamefont {Sugino}},\
  }\bibfield  {title} {\bibinfo {title} {{Non-perturbative construction of 2D
  and 4D supersymmetric Yang-Mills theories with 8 supercharges}},\ }\href
  {https://doi.org/10.1016/j.nuclphysb.2011.12.014} {\bibfield  {journal}
  {\bibinfo  {journal} {Nucl. Phys. B}\ }\textbf {\bibinfo {volume} {857}},\
  \bibinfo {pages} {335} (\bibinfo {year} {2012})},\ \Eprint
  {https://arxiv.org/abs/1109.6807} {arXiv:1109.6807 [hep-lat]} \BibitemShut
  {NoStop}%
\bibitem [{\citenamefont {Matsuura}\ and\ \citenamefont
  {Sugino}(2014)}]{Matsuura:2014pua}%
  \BibitemOpen
  \bibfield  {author} {\bibinfo {author} {\bibfnamefont {S.}~\bibnamefont
  {Matsuura}}\ and\ \bibinfo {author} {\bibfnamefont {F.}~\bibnamefont
  {Sugino}},\ }\bibfield  {title} {\bibinfo {title} {{Lattice formulation for
  2d $\mathcal{N}=(2,2)$, $(4,4)$ super Yang-Mills theories without
  admissibility conditions}},\ }\href {https://doi.org/10.1007/JHEP04(2014)088}
  {\bibfield  {journal} {\bibinfo  {journal} {JHEP}\ }\textbf {\bibinfo
  {volume} {04}},\ \bibinfo {pages} {088}},\ \Eprint
  {https://arxiv.org/abs/1402.0952} {arXiv:1402.0952 [hep-lat]} \BibitemShut
  {NoStop}%
\bibitem [{\citenamefont {Catterall}\ and\ \citenamefont
  {Giedt}(2022)}]{Catterall:2022qzs}%
  \BibitemOpen
  \bibfield  {author} {\bibinfo {author} {\bibfnamefont {S.}~\bibnamefont
  {Catterall}}\ and\ \bibinfo {author} {\bibfnamefont {J.}~\bibnamefont
  {Giedt}},\ }\bibfield  {title} {\bibinfo {title} {Supersymmetric lattice
  theories},\ }in\ \href@noop {} {\emph {\bibinfo {booktitle} {{2022 Snowmass
  Summer Study}}}}\ (\bibinfo {year} {2022})\ \Eprint
  {https://arxiv.org/abs/2202.08154} {arXiv:2202.08154 [hep-lat]} \BibitemShut
  {NoStop}%
\bibitem [{\citenamefont {C\'ordova}\ \emph {et~al.}(2022)\citenamefont
  {C\'ordova}, \citenamefont {Dumitrescu}, \citenamefont {Intriligator},\ and\
  \citenamefont {Shao}}]{Cordova:2022ruw}%
  \BibitemOpen
  \bibfield  {author} {\bibinfo {author} {\bibfnamefont {C.}~\bibnamefont
  {C\'ordova}}, \bibinfo {author} {\bibfnamefont {T.~T.}\ \bibnamefont
  {Dumitrescu}}, \bibinfo {author} {\bibfnamefont {K.}~\bibnamefont
  {Intriligator}},\ and\ \bibinfo {author} {\bibfnamefont {S.-H.}\ \bibnamefont
  {Shao}},\ }\bibfield  {title} {\bibinfo {title} {Generalized symmetries in
  quantum field theory and beyond},\ }in\ \href@noop {} {\emph {\bibinfo
  {booktitle} {{2022 Snowmass Summer Study}}}}\ (\bibinfo {year} {2022})\
  \Eprint {https://arxiv.org/abs/2205.09545} {arXiv:2205.09545 [hep-th]}
  \BibitemShut {NoStop}%
\bibitem [{\citenamefont {Schaich}\ and\ \citenamefont
  {DeGrand}(2015)}]{Schaich:2014pda}%
  \BibitemOpen
  \bibfield  {author} {\bibinfo {author} {\bibfnamefont {D.}~\bibnamefont
  {Schaich}}\ and\ \bibinfo {author} {\bibfnamefont {T.}~\bibnamefont
  {DeGrand}},\ }\bibfield  {title} {\bibinfo {title} {{Parallel software for
  lattice $\mathcal{N}=4$ supersymmetric Yang-Mills theory}},\ }\href
  {https://doi.org/10.1016/j.cpc.2014.12.025} {\bibfield  {journal} {\bibinfo
  {journal} {Comput. Phys. Commun.}\ }\textbf {\bibinfo {volume} {190}},\
  \bibinfo {pages} {200} (\bibinfo {year} {2015})},\ \Eprint
  {https://arxiv.org/abs/1410.6971} {arXiv:1410.6971 [hep-lat]} \BibitemShut
  {NoStop}%
\bibitem [{\citenamefont {Kronfeld}(2002)}]{Kronfeld:2002pi}%
  \BibitemOpen
  \bibfield  {author} {\bibinfo {author} {\bibfnamefont {A.~S.}\ \bibnamefont
  {Kronfeld}},\ }\bibinfo {title} {Uses of effective field theory in lattice
  {QCD}},\ in\ \href {https://doi.org/10.1142/9789812777270_0004} {\emph
  {\bibinfo {booktitle} {At the Frontiers of Particle Physics: Handbook of
  {QCD}}}},\ Vol.~\bibinfo {volume} {4},\ \bibinfo {editor} {edited by\
  \bibinfo {editor} {\bibfnamefont {M.}~\bibnamefont {Shifman}}}\ (\bibinfo
  {publisher} {World Scientific},\ \bibinfo {address} {Singapore},\ \bibinfo
  {year} {2002})\ Chap.~\bibinfo {chapter} {39}, pp.\ \bibinfo {pages}
  {2411--2477},\ \Eprint {https://arxiv.org/abs/hep-lat/0205021}
  {arXiv:hep-lat/0205021} \BibitemShut {NoStop}%
\bibitem [{\citenamefont {Symanzik}(1980)}]{Symanzik:1979ph}%
  \BibitemOpen
  \bibfield  {author} {\bibinfo {author} {\bibfnamefont {K.}~\bibnamefont
  {Symanzik}},\ }\bibinfo {title} {Cutoff dependence in lattice $\phi^4$ in
  four-dimensions theory},\ in\ \href
  {https://doi.org/10.1007/978-1-4684-7571-5_18} {\emph {\bibinfo {booktitle}
  {Recent Developments in Gauge Theories}}},\ \bibinfo {editor} {edited by\
  \bibinfo {editor} {\bibfnamefont {G.}~\bibnamefont {'t~Hooft}}, \bibinfo
  {editor} {\bibfnamefont {C.}~\bibnamefont {Itzykson}}, \bibinfo {editor}
  {\bibfnamefont {A.}~\bibnamefont {Jaffe}}, \bibinfo {editor} {\bibfnamefont
  {H.}~\bibnamefont {Lehmann}}, \bibinfo {editor} {\bibfnamefont {P.~K.}\
  \bibnamefont {Mitter}}, \bibinfo {editor} {\bibfnamefont {I.~M.}\
  \bibnamefont {Singer}},\ and\ \bibinfo {editor} {\bibfnamefont
  {R.}~\bibnamefont {Stora}}}\ (\bibinfo  {publisher} {Plenum},\ \bibinfo
  {address} {New York},\ \bibinfo {year} {1980})\ pp.\ \bibinfo {pages}
  {313--330}\BibitemShut {NoStop}%
\bibitem [{\citenamefont {Symanzik}(1983{\natexlab{a}})}]{Symanzik:1983dc}%
  \BibitemOpen
  \bibfield  {author} {\bibinfo {author} {\bibfnamefont {K.}~\bibnamefont
  {Symanzik}},\ }\bibfield  {title} {\bibinfo {title} {Continuum limit and
  improved action in lattice theories~1: Principles and $\phi^4$ theory},\
  }\href {https://doi.org/10.1016/0550-3213(83)90468-6} {\bibfield  {journal}
  {\bibinfo  {journal} {Nucl. Phys. B}\ }\textbf {\bibinfo {volume} {226}},\
  \bibinfo {pages} {187} (\bibinfo {year} {1983}{\natexlab{a}})}\BibitemShut
  {NoStop}%
\bibitem [{\citenamefont {Symanzik}(1983{\natexlab{b}})}]{Symanzik:1983gh}%
  \BibitemOpen
  \bibfield  {author} {\bibinfo {author} {\bibfnamefont {K.}~\bibnamefont
  {Symanzik}},\ }\bibfield  {title} {\bibinfo {title} {Continuum limit and
  improved action in lattice theories~2: {$\text{O}(N)$} nonlinear sigma model
  in perturbation theory},\ }\href
  {https://doi.org/10.1016/0550-3213(83)90469-8} {\bibfield  {journal}
  {\bibinfo  {journal} {Nucl. Phys. B}\ }\textbf {\bibinfo {volume} {226}},\
  \bibinfo {pages} {205} (\bibinfo {year} {1983}{\natexlab{b}})}\BibitemShut
  {NoStop}%
\bibitem [{\citenamefont {Symanzik}(1970)}]{Symanzik:1970rt}%
  \BibitemOpen
  \bibfield  {author} {\bibinfo {author} {\bibfnamefont {K.}~\bibnamefont
  {Symanzik}},\ }\bibfield  {title} {\bibinfo {title} {{Small distance behavior
  in field theory and power counting}},\ }\href
  {https://doi.org/10.1007/BF01649434} {\bibfield  {journal} {\bibinfo
  {journal} {Commun. Math. Phys.}\ }\textbf {\bibinfo {volume} {18}},\ \bibinfo
  {pages} {227} (\bibinfo {year} {1970})}\BibitemShut {NoStop}%
\bibitem [{\citenamefont {Weisz}(1983)}]{Weisz:1982zw}%
  \BibitemOpen
  \bibfield  {author} {\bibinfo {author} {\bibfnamefont {P.}~\bibnamefont
  {Weisz}},\ }\bibfield  {title} {\bibinfo {title} {Continuum limit improved
  lattice action for pure {Yang-Mills} theory~1},\ }\href
  {https://doi.org/10.1016/0550-3213(83)90595-3} {\bibfield  {journal}
  {\bibinfo  {journal} {Nucl. Phys. B}\ }\textbf {\bibinfo {volume} {212}},\
  \bibinfo {pages} {1} (\bibinfo {year} {1983})}\BibitemShut {NoStop}%
\bibitem [{\citenamefont {Weisz}\ and\ \citenamefont
  {Wohlert}(1984)}]{Weisz:1983bn}%
  \BibitemOpen
  \bibfield  {author} {\bibinfo {author} {\bibfnamefont {P.}~\bibnamefont
  {Weisz}}\ and\ \bibinfo {author} {\bibfnamefont {R.}~\bibnamefont
  {Wohlert}},\ }\bibfield  {title} {\bibinfo {title} {Continuum limit improved
  lattice action for pure {Yang-Mills} theory~2},\ }\href
  {https://doi.org/10.1016/0550-3213(84)90563-7} {\bibfield  {journal}
  {\bibinfo  {journal} {Nucl. Phys. B}\ }\textbf {\bibinfo {volume} {236}},\
  \bibinfo {pages} {397} (\bibinfo {year} {1984})},\ \bibinfo {note} {(E)
  \href{https://doi.org/10.1016/0550-3213(84)90543-1}{\textbf{247}, 544
  (1984)}}\BibitemShut {NoStop}%
\bibitem [{\citenamefont {Curci}\ \emph {et~al.}(1983)\citenamefont {Curci},
  \citenamefont {Menotti},\ and\ \citenamefont {Paffuti}}]{Curci:1983an}%
  \BibitemOpen
  \bibfield  {author} {\bibinfo {author} {\bibfnamefont {G.}~\bibnamefont
  {Curci}}, \bibinfo {author} {\bibfnamefont {P.}~\bibnamefont {Menotti}},\
  and\ \bibinfo {author} {\bibfnamefont {G.}~\bibnamefont {Paffuti}},\
  }\bibfield  {title} {\bibinfo {title} {{Symanzik's} improved lagrangian for
  lattice gauge theory},\ }\href {https://doi.org/10.1016/0370-2693(83)91043-2}
  {\bibfield  {journal} {\bibinfo  {journal} {Phys. Lett. B}\ }\textbf
  {\bibinfo {volume} {130}},\ \bibinfo {pages} {205} (\bibinfo {year}
  {1983})},\ \bibinfo {note} {(E)
  \href{https://doi.org/10.1016/0370-2693(84)90326-5}{\textbf{135}, 516
  (1984)}}\BibitemShut {NoStop}%
\bibitem [{\citenamefont {Lüscher}\ and\ \citenamefont
  {Weisz}(1985{\natexlab{a}})}]{Luscher:1984xn}%
  \BibitemOpen
  \bibfield  {author} {\bibinfo {author} {\bibfnamefont {M.}~\bibnamefont
  {Lüscher}}\ and\ \bibinfo {author} {\bibfnamefont {P.}~\bibnamefont
  {Weisz}},\ }\bibfield  {title} {\bibinfo {title} {On-shell improved lattice
  gauge theories},\ }\href {https://doi.org/10.1007/BF01206178} {\bibfield
  {journal} {\bibinfo  {journal} {Commun. Math. Phys.}\ }\textbf {\bibinfo
  {volume} {97}},\ \bibinfo {pages} {59} (\bibinfo {year}
  {1985}{\natexlab{a}})},\ \bibinfo {note} {(E)
  \href{https://doi.org/10.1007/BF01205792}{\textbf{98}, 433
  (1985)}}\BibitemShut {NoStop}%
\bibitem [{\citenamefont {Sheikholeslami}\ and\ \citenamefont
  {Wohlert}(1985)}]{Sheikholeslami:1985ij}%
  \BibitemOpen
  \bibfield  {author} {\bibinfo {author} {\bibfnamefont {B.}~\bibnamefont
  {Sheikholeslami}}\ and\ \bibinfo {author} {\bibfnamefont {R.}~\bibnamefont
  {Wohlert}},\ }\bibfield  {title} {\bibinfo {title} {Improved continuum limit
  lattice action for {QCD} with {Wilson} fermions},\ }\href
  {https://doi.org/10.1016/0550-3213(85)90002-1} {\bibfield  {journal}
  {\bibinfo  {journal} {Nucl. Phys. B}\ }\textbf {\bibinfo {volume} {259}},\
  \bibinfo {pages} {572} (\bibinfo {year} {1985})}\BibitemShut {NoStop}%
\bibitem [{\citenamefont {Naik}(1989)}]{Naik:1986bn}%
  \BibitemOpen
  \bibfield  {author} {\bibinfo {author} {\bibfnamefont {S.}~\bibnamefont
  {Naik}},\ }\bibfield  {title} {\bibinfo {title} {On-shell improved lattice
  action for {QCD} with {Susskind} fermions and asymptotic freedom scale},\
  }\href {https://doi.org/10.1016/0550-3213(89)90394-5} {\bibfield  {journal}
  {\bibinfo  {journal} {Nucl. Phys. B}\ }\textbf {\bibinfo {volume} {316}},\
  \bibinfo {pages} {238} (\bibinfo {year} {1989})}\BibitemShut {NoStop}%
\bibitem [{\citenamefont {Lüscher}\ and\ \citenamefont
  {Weisz}(1985{\natexlab{b}})}]{Luscher:1985zq}%
  \BibitemOpen
  \bibfield  {author} {\bibinfo {author} {\bibfnamefont {M.}~\bibnamefont
  {Lüscher}}\ and\ \bibinfo {author} {\bibfnamefont {P.}~\bibnamefont
  {Weisz}},\ }\bibfield  {title} {\bibinfo {title} {Computation of the action
  for on-shell improved lattice gauge theories at weak coupling},\ }\href
  {https://doi.org/10.1016/0370-2693(85)90966-9} {\bibfield  {journal}
  {\bibinfo  {journal} {Phys. Lett. B}\ }\textbf {\bibinfo {volume} {158}},\
  \bibinfo {pages} {250} (\bibinfo {year} {1985}{\natexlab{b}})}\BibitemShut
  {NoStop}%
\bibitem [{\citenamefont {Lüscher}\ and\ \citenamefont
  {Weisz}(1986)}]{Luscher:1985wf}%
  \BibitemOpen
  \bibfield  {author} {\bibinfo {author} {\bibfnamefont {M.}~\bibnamefont
  {Lüscher}}\ and\ \bibinfo {author} {\bibfnamefont {P.}~\bibnamefont
  {Weisz}},\ }\bibfield  {title} {\bibinfo {title} {Efficient numerical
  techniques for perturbative lattice gauge theory computations},\ }\href
  {https://doi.org/10.1016/0550-3213(86)90094-5} {\bibfield  {journal}
  {\bibinfo  {journal} {Nucl. Phys. B}\ }\textbf {\bibinfo {volume} {266}},\
  \bibinfo {pages} {309} (\bibinfo {year} {1986})}\BibitemShut {NoStop}%
\bibitem [{\citenamefont {Hart}\ \emph {et~al.}(2009)\citenamefont {Hart},
  \citenamefont {von Hippel},\ and\ \citenamefont {Horgan}}]{Hart:2008sq}%
  \BibitemOpen
  \bibfield  {author} {\bibinfo {author} {\bibfnamefont {A.}~\bibnamefont
  {Hart}}, \bibinfo {author} {\bibfnamefont {G.~M.}\ \bibnamefont {von
  Hippel}},\ and\ \bibinfo {author} {\bibfnamefont {R.~R.}\ \bibnamefont
  {Horgan}} (\bibinfo {collaboration} {HPQCD}),\ }\bibfield  {title} {\bibinfo
  {title} {Radiative corrections to the lattice gluon action for {HISQ}
  improved staggered quarks and the effect of such corrections on the static
  potential},\ }\href {https://doi.org/10.1103/PhysRevD.79.074008} {\bibfield
  {journal} {\bibinfo  {journal} {Phys. Rev. D}\ }\textbf {\bibinfo {volume}
  {79}},\ \bibinfo {pages} {074008} (\bibinfo {year} {2009})},\ \Eprint
  {https://arxiv.org/abs/0812.0503} {arXiv:0812.0503 [hep-lat]} \BibitemShut
  {NoStop}%
\bibitem [{\citenamefont {Jansen}\ \emph {et~al.}(1996)\citenamefont {Jansen},
  \citenamefont {Liu}, \citenamefont {Luscher}, \citenamefont {Simma},
  \citenamefont {Sint}, \citenamefont {Sommer}, \citenamefont {Weisz},\ and\
  \citenamefont {Wolff}}]{Jansen:1995ck}%
  \BibitemOpen
  \bibfield  {author} {\bibinfo {author} {\bibfnamefont {K.}~\bibnamefont
  {Jansen}}, \bibinfo {author} {\bibfnamefont {C.}~\bibnamefont {Liu}},
  \bibinfo {author} {\bibfnamefont {M.}~\bibnamefont {Luscher}}, \bibinfo
  {author} {\bibfnamefont {H.}~\bibnamefont {Simma}}, \bibinfo {author}
  {\bibfnamefont {S.}~\bibnamefont {Sint}}, \bibinfo {author} {\bibfnamefont
  {R.}~\bibnamefont {Sommer}}, \bibinfo {author} {\bibfnamefont
  {P.}~\bibnamefont {Weisz}},\ and\ \bibinfo {author} {\bibfnamefont
  {U.}~\bibnamefont {Wolff}},\ }\bibfield  {title} {\bibinfo {title}
  {{Nonperturbative renormalization of lattice QCD at all scales}},\ }\href
  {https://doi.org/10.1016/0370-2693(96)00075-5} {\bibfield  {journal}
  {\bibinfo  {journal} {Phys. Lett. B}\ }\textbf {\bibinfo {volume} {372}},\
  \bibinfo {pages} {275} (\bibinfo {year} {1996})},\ \Eprint
  {https://arxiv.org/abs/hep-lat/9512009} {arXiv:hep-lat/9512009} \BibitemShut
  {NoStop}%
\bibitem [{\citenamefont {Weinberg}(1979)}]{Weinberg:1978kz}%
  \BibitemOpen
  \bibfield  {author} {\bibinfo {author} {\bibfnamefont {S.}~\bibnamefont
  {Weinberg}},\ }\bibfield  {title} {\bibinfo {title} {{Phenomenological
  Lagrangians}},\ }\href {https://doi.org/10.1016/0378-4371(79)90223-1}
  {\bibfield  {journal} {\bibinfo  {journal} {Physica A}\ }\textbf {\bibinfo
  {volume} {96}},\ \bibinfo {pages} {327} (\bibinfo {year} {1979})}\BibitemShut
  {NoStop}%
\bibitem [{\citenamefont {Leutwyler}(1994)}]{Leutwyler:1993iq}%
  \BibitemOpen
  \bibfield  {author} {\bibinfo {author} {\bibfnamefont {H.}~\bibnamefont
  {Leutwyler}},\ }\bibfield  {title} {\bibinfo {title} {{On the foundations of
  chiral perturbation theory}},\ }\href
  {https://doi.org/10.1006/aphy.1994.1094} {\bibfield  {journal} {\bibinfo
  {journal} {Annals Phys.}\ }\textbf {\bibinfo {volume} {235}},\ \bibinfo
  {pages} {165} (\bibinfo {year} {1994})},\ \Eprint
  {https://arxiv.org/abs/hep-ph/9311274} {arXiv:hep-ph/9311274} \BibitemShut
  {NoStop}%
\bibitem [{\citenamefont {Bernard}\ and\ \citenamefont
  {Golterman}(1992)}]{Bernard:1992mk}%
  \BibitemOpen
  \bibfield  {author} {\bibinfo {author} {\bibfnamefont {C.~W.}\ \bibnamefont
  {Bernard}}\ and\ \bibinfo {author} {\bibfnamefont {M.~F.~L.}\ \bibnamefont
  {Golterman}},\ }\bibfield  {title} {\bibinfo {title} {{Chiral perturbation
  theory for the quenched approximation of QCD}},\ }\href
  {https://doi.org/10.1103/PhysRevD.46.853} {\bibfield  {journal} {\bibinfo
  {journal} {Phys. Rev. D}\ }\textbf {\bibinfo {volume} {46}},\ \bibinfo
  {pages} {853} (\bibinfo {year} {1992})},\ \Eprint
  {https://arxiv.org/abs/hep-lat/9204007} {arXiv:hep-lat/9204007} \BibitemShut
  {NoStop}%
\bibitem [{\citenamefont {Bernard}\ and\ \citenamefont
  {Golterman}(1994)}]{Bernard:1993sv}%
  \BibitemOpen
  \bibfield  {author} {\bibinfo {author} {\bibfnamefont {C.~W.}\ \bibnamefont
  {Bernard}}\ and\ \bibinfo {author} {\bibfnamefont {M.~F.~L.}\ \bibnamefont
  {Golterman}},\ }\bibfield  {title} {\bibinfo {title} {{Partially quenched
  gauge theories and an application to staggered fermions}},\ }\href
  {https://doi.org/10.1103/PhysRevD.49.486} {\bibfield  {journal} {\bibinfo
  {journal} {Phys. Rev. D}\ }\textbf {\bibinfo {volume} {49}},\ \bibinfo
  {pages} {486} (\bibinfo {year} {1994})},\ \Eprint
  {https://arxiv.org/abs/hep-lat/9306005} {arXiv:hep-lat/9306005} \BibitemShut
  {NoStop}%
\bibitem [{\citenamefont {Sharpe}(1997)}]{Sharpe:1997by}%
  \BibitemOpen
  \bibfield  {author} {\bibinfo {author} {\bibfnamefont {S.~R.}\ \bibnamefont
  {Sharpe}},\ }\bibfield  {title} {\bibinfo {title} {{Enhanced chiral
  logarithms in partially quenched QCD}},\ }\href
  {https://doi.org/10.1103/PhysRevD.62.099901} {\bibfield  {journal} {\bibinfo
  {journal} {Phys. Rev. D}\ }\textbf {\bibinfo {volume} {56}},\ \bibinfo
  {pages} {7052} (\bibinfo {year} {1997})},\ \bibinfo {note} {(E)
  \href{https://doi.org/10.1103/PhysRevD.62.099901}{\textbf{62}, 099901
  (2000)}},\ \Eprint {https://arxiv.org/abs/hep-lat/9707018}
  {arXiv:hep-lat/9707018} \BibitemShut {NoStop}%
\bibitem [{\citenamefont {Sharpe}\ and\ \citenamefont
  {Shoresh}(2000)}]{Sharpe:2000bc}%
  \BibitemOpen
  \bibfield  {author} {\bibinfo {author} {\bibfnamefont {S.~R.}\ \bibnamefont
  {Sharpe}}\ and\ \bibinfo {author} {\bibfnamefont {N.}~\bibnamefont
  {Shoresh}},\ }\bibfield  {title} {\bibinfo {title} {{Physical results from
  unphysical simulations}},\ }\href
  {https://doi.org/10.1103/PhysRevD.62.094503} {\bibfield  {journal} {\bibinfo
  {journal} {Phys. Rev. D}\ }\textbf {\bibinfo {volume} {62}},\ \bibinfo
  {pages} {094503} (\bibinfo {year} {2000})},\ \Eprint
  {https://arxiv.org/abs/hep-lat/0006017} {arXiv:hep-lat/0006017} \BibitemShut
  {NoStop}%
\bibitem [{\citenamefont {Bernard}\ and\ \citenamefont
  {Golterman}(2013)}]{Bernard:2013kwa}%
  \BibitemOpen
  \bibfield  {author} {\bibinfo {author} {\bibfnamefont {C.}~\bibnamefont
  {Bernard}}\ and\ \bibinfo {author} {\bibfnamefont {M.}~\bibnamefont
  {Golterman}},\ }\bibfield  {title} {\bibinfo {title} {{On the foundations of
  partially quenched chiral perturbation theory}},\ }\href
  {https://doi.org/10.1103/PhysRevD.88.014004} {\bibfield  {journal} {\bibinfo
  {journal} {Phys. Rev. D}\ }\textbf {\bibinfo {volume} {88}},\ \bibinfo
  {pages} {014004} (\bibinfo {year} {2013})},\ \Eprint
  {https://arxiv.org/abs/1304.1948} {arXiv:1304.1948 [hep-lat]} \BibitemShut
  {NoStop}%
\bibitem [{\citenamefont {Bijnens}(2015)}]{Bijnens:2014gsa}%
  \BibitemOpen
  \bibfield  {author} {\bibinfo {author} {\bibfnamefont {J.}~\bibnamefont
  {Bijnens}},\ }\bibfield  {title} {\bibinfo {title} {{CHIRON: a package for
  ChPT numerical results at two loops}},\ }\href
  {https://doi.org/10.1140/epjc/s10052-014-3249-9} {\bibfield  {journal}
  {\bibinfo  {journal} {Eur. Phys. J. C}\ }\textbf {\bibinfo {volume} {75}},\
  \bibinfo {pages} {27} (\bibinfo {year} {2015})},\ \Eprint
  {https://arxiv.org/abs/1412.0887} {arXiv:1412.0887 [hep-ph]} \BibitemShut
  {NoStop}%
\bibitem [{\citenamefont {Lüscher}(1986)}]{Luscher:1985dn}%
  \BibitemOpen
  \bibfield  {author} {\bibinfo {author} {\bibfnamefont {M.}~\bibnamefont
  {Lüscher}},\ }\bibfield  {title} {\bibinfo {title} {Volume dependence of the
  energy spectrum in massive quantum field theories~1: Stable particle
  states},\ }\href {https://doi.org/10.1007/BF01211589} {\bibfield  {journal}
  {\bibinfo  {journal} {Commun. Math. Phys.}\ }\textbf {\bibinfo {volume}
  {104}},\ \bibinfo {pages} {177} (\bibinfo {year} {1986})}\BibitemShut
  {NoStop}%
\bibitem [{\citenamefont {Gasser}\ and\ \citenamefont
  {Leutwyler}(1988)}]{Gasser:1987zq}%
  \BibitemOpen
  \bibfield  {author} {\bibinfo {author} {\bibfnamefont {J.}~\bibnamefont
  {Gasser}}\ and\ \bibinfo {author} {\bibfnamefont {H.}~\bibnamefont
  {Leutwyler}},\ }\bibfield  {title} {\bibinfo {title} {Spontaneously broken
  symmetries: Effective {Lagrangians} at finite volume},\ }\href
  {https://doi.org/10.1016/0550-3213(88)90107-1} {\bibfield  {journal}
  {\bibinfo  {journal} {Nucl. Phys. B}\ }\textbf {\bibinfo {volume} {307}},\
  \bibinfo {pages} {763} (\bibinfo {year} {1988})}\BibitemShut {NoStop}%
\bibitem [{\citenamefont {Colangelo}\ \emph {et~al.}(2005)\citenamefont
  {Colangelo}, \citenamefont {Dürr},\ and\ \citenamefont
  {Haefeli}}]{Colangelo:2005gd}%
  \BibitemOpen
  \bibfield  {author} {\bibinfo {author} {\bibfnamefont {G.}~\bibnamefont
  {Colangelo}}, \bibinfo {author} {\bibfnamefont {S.}~\bibnamefont {Dürr}},\
  and\ \bibinfo {author} {\bibfnamefont {C.}~\bibnamefont {Haefeli}},\
  }\bibfield  {title} {\bibinfo {title} {{Finite volume effects for meson
  masses and decay constants}},\ }\href
  {https://doi.org/10.1016/j.nuclphysb.2005.05.015} {\bibfield  {journal}
  {\bibinfo  {journal} {Nucl. Phys. B}\ }\textbf {\bibinfo {volume} {721}},\
  \bibinfo {pages} {136} (\bibinfo {year} {2005})},\ \Eprint
  {https://arxiv.org/abs/hep-lat/0503014} {arXiv:hep-lat/0503014} \BibitemShut
  {NoStop}%
\bibitem [{\citenamefont {Leutwyler}\ and\ \citenamefont
  {Smilga}(1992)}]{Leutwyler:1992yt}%
  \BibitemOpen
  \bibfield  {author} {\bibinfo {author} {\bibfnamefont {H.}~\bibnamefont
  {Leutwyler}}\ and\ \bibinfo {author} {\bibfnamefont {A.~V.}\ \bibnamefont
  {Smilga}},\ }\bibfield  {title} {\bibinfo {title} {{Spectrum of Dirac
  operator and role of winding number in QCD}},\ }\href
  {https://doi.org/10.1103/PhysRevD.46.5607} {\bibfield  {journal} {\bibinfo
  {journal} {Phys. Rev. D}\ }\textbf {\bibinfo {volume} {46}},\ \bibinfo
  {pages} {5607} (\bibinfo {year} {1992})}\BibitemShut {NoStop}%
\bibitem [{\citenamefont {Brower}\ \emph {et~al.}(2003)\citenamefont {Brower},
  \citenamefont {Chandrasekharan}, \citenamefont {Negele},\ and\ \citenamefont
  {Wiese}}]{Brower:2003yx}%
  \BibitemOpen
  \bibfield  {author} {\bibinfo {author} {\bibfnamefont {R.}~\bibnamefont
  {Brower}}, \bibinfo {author} {\bibfnamefont {S.}~\bibnamefont
  {Chandrasekharan}}, \bibinfo {author} {\bibfnamefont {J.~W.}\ \bibnamefont
  {Negele}},\ and\ \bibinfo {author} {\bibfnamefont {U.~J.}\ \bibnamefont
  {Wiese}},\ }\bibfield  {title} {\bibinfo {title} {{QCD at fixed topology}},\
  }\href {https://doi.org/10.1016/S0370-2693(03)00369-1} {\bibfield  {journal}
  {\bibinfo  {journal} {Phys. Lett. B}\ }\textbf {\bibinfo {volume} {560}},\
  \bibinfo {pages} {64} (\bibinfo {year} {2003})},\ \Eprint
  {https://arxiv.org/abs/hep-lat/0302005} {arXiv:hep-lat/0302005} \BibitemShut
  {NoStop}%
\bibitem [{\citenamefont {Aoki}\ and\ \citenamefont
  {Fukaya}(2010)}]{Aoki:2009mx}%
  \BibitemOpen
  \bibfield  {author} {\bibinfo {author} {\bibfnamefont {S.}~\bibnamefont
  {Aoki}}\ and\ \bibinfo {author} {\bibfnamefont {H.}~\bibnamefont {Fukaya}},\
  }\bibfield  {title} {\bibinfo {title} {{Chiral perturbation theory in a
  $\theta$ vacuum}},\ }\href {https://doi.org/10.1103/PhysRevD.81.034022}
  {\bibfield  {journal} {\bibinfo  {journal} {Phys. Rev. D}\ }\textbf {\bibinfo
  {volume} {81}},\ \bibinfo {pages} {034022} (\bibinfo {year} {2010})},\
  \Eprint {https://arxiv.org/abs/0906.4852} {arXiv:0906.4852 [hep-lat]}
  \BibitemShut {NoStop}%
\bibitem [{\citenamefont {Bernard}\ and\ \citenamefont
  {Toussaint}(2018)}]{Bernard:2017npd}%
  \BibitemOpen
  \bibfield  {author} {\bibinfo {author} {\bibfnamefont {C.}~\bibnamefont
  {Bernard}}\ and\ \bibinfo {author} {\bibfnamefont {D.}~\bibnamefont
  {Toussaint}} (\bibinfo {collaboration} {MILC}),\ }\bibfield  {title}
  {\bibinfo {title} {{Effects of nonequilibrated topological charge
  distributions on pseudoscalar meson masses and decay constants}},\ }\href
  {https://doi.org/10.1103/PhysRevD.97.074502} {\bibfield  {journal} {\bibinfo
  {journal} {Phys. Rev. D}\ }\textbf {\bibinfo {volume} {97}},\ \bibinfo
  {pages} {074502} (\bibinfo {year} {2018})},\ \Eprint
  {https://arxiv.org/abs/1707.05430} {arXiv:1707.05430 [hep-lat]} \BibitemShut
  {NoStop}%
\bibitem [{\citenamefont {Duncan}\ \emph {et~al.}(1996)\citenamefont {Duncan},
  \citenamefont {Eichten},\ and\ \citenamefont {Thacker}}]{Duncan:1996xy}%
  \BibitemOpen
  \bibfield  {author} {\bibinfo {author} {\bibfnamefont {A.}~\bibnamefont
  {Duncan}}, \bibinfo {author} {\bibfnamefont {E.}~\bibnamefont {Eichten}},\
  and\ \bibinfo {author} {\bibfnamefont {H.}~\bibnamefont {Thacker}},\
  }\bibfield  {title} {\bibinfo {title} {{Electromagnetic splittings and light
  quark masses in lattice QCD}},\ }\href
  {https://doi.org/10.1103/PhysRevLett.76.3894} {\bibfield  {journal} {\bibinfo
   {journal} {Phys. Rev. Lett.}\ }\textbf {\bibinfo {volume} {76}},\ \bibinfo
  {pages} {3894} (\bibinfo {year} {1996})},\ \Eprint
  {https://arxiv.org/abs/hep-lat/9602005} {arXiv:hep-lat/9602005} \BibitemShut
  {NoStop}%
\bibitem [{\citenamefont {Davoudi}\ \emph {et~al.}(2019)\citenamefont
  {Davoudi}, \citenamefont {Harrison}, \citenamefont {J\"uttner}, \citenamefont
  {Portelli},\ and\ \citenamefont {Savage}}]{Davoudi:2018qpl}%
  \BibitemOpen
  \bibfield  {author} {\bibinfo {author} {\bibfnamefont {Z.}~\bibnamefont
  {Davoudi}}, \bibinfo {author} {\bibfnamefont {J.}~\bibnamefont {Harrison}},
  \bibinfo {author} {\bibfnamefont {A.}~\bibnamefont {J\"uttner}}, \bibinfo
  {author} {\bibfnamefont {A.}~\bibnamefont {Portelli}},\ and\ \bibinfo
  {author} {\bibfnamefont {M.~J.}\ \bibnamefont {Savage}},\ }\bibfield  {title}
  {\bibinfo {title} {{Theoretical aspects of quantum electrodynamics in a
  finite volume with periodic boundary conditions}},\ }\href
  {https://doi.org/10.1103/PhysRevD.99.034510} {\bibfield  {journal} {\bibinfo
  {journal} {Phys. Rev. D}\ }\textbf {\bibinfo {volume} {99}},\ \bibinfo
  {pages} {034510} (\bibinfo {year} {2019})},\ \Eprint
  {https://arxiv.org/abs/1810.05923} {arXiv:1810.05923 [hep-lat]} \BibitemShut
  {NoStop}%
\bibitem [{\citenamefont {Feng}\ and\ \citenamefont
  {Jin}(2019)}]{Feng:2018qpx}%
  \BibitemOpen
  \bibfield  {author} {\bibinfo {author} {\bibfnamefont {X.}~\bibnamefont
  {Feng}}\ and\ \bibinfo {author} {\bibfnamefont {L.}~\bibnamefont {Jin}},\
  }\bibfield  {title} {\bibinfo {title} {{QED self energies from lattice QCD
  without power-law finite-volume errors}},\ }\href
  {https://doi.org/10.1103/PhysRevD.100.094509} {\bibfield  {journal} {\bibinfo
   {journal} {Phys. Rev. D}\ }\textbf {\bibinfo {volume} {100}},\ \bibinfo
  {pages} {094509} (\bibinfo {year} {2019})},\ \Eprint
  {https://arxiv.org/abs/1812.09817} {arXiv:1812.09817 [hep-lat]} \BibitemShut
  {NoStop}%
\bibitem [{\citenamefont {Patella}(2017)}]{Patella:2017fgk}%
  \BibitemOpen
  \bibfield  {author} {\bibinfo {author} {\bibfnamefont {A.}~\bibnamefont
  {Patella}},\ }\bibfield  {title} {\bibinfo {title} {{QED} corrections to
  hadronic observables},\ }\href {https://doi.org/10.22323/1.256.0020}
  {\bibfield  {journal} {\bibinfo  {journal} {PoS}\ }\textbf {\bibinfo {volume}
  {LATTICE2016}},\ \bibinfo {pages} {020} (\bibinfo {year} {2017})},\ \Eprint
  {https://arxiv.org/abs/1702.03857} {arXiv:1702.03857 [hep-lat]} \BibitemShut
  {NoStop}%
\bibitem [{\citenamefont {Hernandez}\ and\ \citenamefont
  {Hill}(1990)}]{Hernandez:1989ch}%
  \BibitemOpen
  \bibfield  {author} {\bibinfo {author} {\bibfnamefont {O.~F.}\ \bibnamefont
  {Hernandez}}\ and\ \bibinfo {author} {\bibfnamefont {B.~R.}\ \bibnamefont
  {Hill}},\ }\bibfield  {title} {\bibinfo {title} {The static approximation,
  staggered fermions and {$f_B$}},\ }\href
  {https://doi.org/10.1016/0370-2693(90)90469-M} {\bibfield  {journal}
  {\bibinfo  {journal} {Phys. Lett. B}\ }\textbf {\bibinfo {volume} {237}},\
  \bibinfo {pages} {95} (\bibinfo {year} {1990})}\BibitemShut {NoStop}%
\bibitem [{\citenamefont {Heitger}\ and\ \citenamefont
  {Sommer}(2004)}]{Heitger:2003nj}%
  \BibitemOpen
  \bibfield  {author} {\bibinfo {author} {\bibfnamefont {J.}~\bibnamefont
  {Heitger}}\ and\ \bibinfo {author} {\bibfnamefont {R.}~\bibnamefont {Sommer}}
  (\bibinfo {collaboration} {ALPHA}),\ }\bibfield  {title} {\bibinfo {title}
  {{Nonperturbative heavy quark effective theory}},\ }\href
  {https://doi.org/10.1088/1126-6708/2004/02/022} {\bibfield  {journal}
  {\bibinfo  {journal} {JHEP}\ }\textbf {\bibinfo {volume} {02}},\ \bibinfo
  {pages} {022}},\ \Eprint {https://arxiv.org/abs/hep-lat/0310035}
  {arXiv:hep-lat/0310035} \BibitemShut {NoStop}%
\bibitem [{\citenamefont {El-Khadra}\ \emph {et~al.}(1997)\citenamefont
  {El-Khadra}, \citenamefont {Kronfeld},\ and\ \citenamefont
  {Mackenzie}}]{El-Khadra:1996wdx}%
  \BibitemOpen
  \bibfield  {author} {\bibinfo {author} {\bibfnamefont {A.~X.}\ \bibnamefont
  {El-Khadra}}, \bibinfo {author} {\bibfnamefont {A.~S.}\ \bibnamefont
  {Kronfeld}},\ and\ \bibinfo {author} {\bibfnamefont {P.~B.}\ \bibnamefont
  {Mackenzie}},\ }\bibfield  {title} {\bibinfo {title} {{Massive fermions in
  lattice gauge theory}},\ }\href {https://doi.org/10.1103/PhysRevD.55.3933}
  {\bibfield  {journal} {\bibinfo  {journal} {Phys. Rev. D}\ }\textbf {\bibinfo
  {volume} {55}},\ \bibinfo {pages} {3933} (\bibinfo {year} {1997})},\ \Eprint
  {https://arxiv.org/abs/hep-lat/9604004} {arXiv:hep-lat/9604004} \BibitemShut
  {NoStop}%
\bibitem [{\citenamefont {Christ}\ \emph {et~al.}(2007)\citenamefont {Christ},
  \citenamefont {Li},\ and\ \citenamefont {Lin}}]{Christ:2006us}%
  \BibitemOpen
  \bibfield  {author} {\bibinfo {author} {\bibfnamefont {N.~H.}\ \bibnamefont
  {Christ}}, \bibinfo {author} {\bibfnamefont {M.}~\bibnamefont {Li}},\ and\
  \bibinfo {author} {\bibfnamefont {H.-W.}\ \bibnamefont {Lin}},\ }\bibfield
  {title} {\bibinfo {title} {Relativistic heavy quark effective action},\
  }\href {https://doi.org/10.1103/PhysRevD.76.074505} {\bibfield  {journal}
  {\bibinfo  {journal} {Phys. Rev. D}\ }\textbf {\bibinfo {volume} {76}},\
  \bibinfo {pages} {074505} (\bibinfo {year} {2007})},\ \Eprint
  {https://arxiv.org/abs/hep-lat/0608006} {arXiv:hep-lat/0608006} \BibitemShut
  {NoStop}%
\bibitem [{\citenamefont {Kronfeld}(2000)}]{Kronfeld:2000ck}%
  \BibitemOpen
  \bibfield  {author} {\bibinfo {author} {\bibfnamefont {A.~S.}\ \bibnamefont
  {Kronfeld}},\ }\bibfield  {title} {\bibinfo {title} {{Application of heavy
  quark effective theory to lattice QCD~1: Power corrections}},\ }\href
  {https://doi.org/10.1103/PhysRevD.62.014505} {\bibfield  {journal} {\bibinfo
  {journal} {Phys. Rev. D}\ }\textbf {\bibinfo {volume} {62}},\ \bibinfo
  {pages} {014505} (\bibinfo {year} {2000})},\ \Eprint
  {https://arxiv.org/abs/hep-lat/0002008} {arXiv:hep-lat/0002008} \BibitemShut
  {NoStop}%
\bibitem [{\citenamefont {Harada}\ \emph
  {et~al.}(2002{\natexlab{a}})\citenamefont {Harada}, \citenamefont
  {Hashimoto}, \citenamefont {Ishikawa}, \citenamefont {Kronfeld},
  \citenamefont {Onogi},\ and\ \citenamefont {Yamada}}]{Harada:2001fi}%
  \BibitemOpen
  \bibfield  {author} {\bibinfo {author} {\bibfnamefont {J.}~\bibnamefont
  {Harada}}, \bibinfo {author} {\bibfnamefont {S.}~\bibnamefont {Hashimoto}},
  \bibinfo {author} {\bibfnamefont {K.-I.}\ \bibnamefont {Ishikawa}}, \bibinfo
  {author} {\bibfnamefont {A.~S.}\ \bibnamefont {Kronfeld}}, \bibinfo {author}
  {\bibfnamefont {T.}~\bibnamefont {Onogi}},\ and\ \bibinfo {author}
  {\bibfnamefont {N.}~\bibnamefont {Yamada}},\ }\bibfield  {title} {\bibinfo
  {title} {{Application of heavy quark effective theory to lattice QCD~2:
  Radiative corrections to heavy light currents}},\ }\href
  {https://doi.org/10.1103/PhysRevD.71.019903} {\bibfield  {journal} {\bibinfo
  {journal} {Phys. Rev. D}\ }\textbf {\bibinfo {volume} {65}},\ \bibinfo
  {pages} {094513} (\bibinfo {year} {2002}{\natexlab{a}})},\ \bibinfo {note}
  {(E) \href{https://doi.org/10.1103/PhysRevD.71.019903}{\textbf{71}, 019903
  (2005)}},\ \Eprint {https://arxiv.org/abs/hep-lat/0112044}
  {arXiv:hep-lat/0112044} \BibitemShut {NoStop}%
\bibitem [{\citenamefont {Harada}\ \emph
  {et~al.}(2002{\natexlab{b}})\citenamefont {Harada}, \citenamefont
  {Hashimoto}, \citenamefont {Kronfeld},\ and\ \citenamefont
  {Onogi}}]{Harada:2001fj}%
  \BibitemOpen
  \bibfield  {author} {\bibinfo {author} {\bibfnamefont {J.}~\bibnamefont
  {Harada}}, \bibinfo {author} {\bibfnamefont {S.}~\bibnamefont {Hashimoto}},
  \bibinfo {author} {\bibfnamefont {A.~S.}\ \bibnamefont {Kronfeld}},\ and\
  \bibinfo {author} {\bibfnamefont {T.}~\bibnamefont {Onogi}},\ }\bibfield
  {title} {\bibinfo {title} {{Application of heavy quark effective theory to
  lattice QCD~3: Radiative corrections to heavy-heavy currents}},\ }\href
  {https://doi.org/10.1103/PhysRevD.65.094514} {\bibfield  {journal} {\bibinfo
  {journal} {Phys. Rev. D}\ }\textbf {\bibinfo {volume} {65}},\ \bibinfo
  {pages} {094514} (\bibinfo {year} {2002}{\natexlab{b}})},\ \Eprint
  {https://arxiv.org/abs/hep-lat/0112045} {arXiv:hep-lat/0112045} \BibitemShut
  {NoStop}%
\bibitem [{\citenamefont {Oktay}\ and\ \citenamefont
  {Kronfeld}(2008)}]{Oktay:2008ex}%
  \BibitemOpen
  \bibfield  {author} {\bibinfo {author} {\bibfnamefont {M.~B.}\ \bibnamefont
  {Oktay}}\ and\ \bibinfo {author} {\bibfnamefont {A.~S.}\ \bibnamefont
  {Kronfeld}},\ }\bibfield  {title} {\bibinfo {title} {{New lattice action for
  heavy quarks}},\ }\href {https://doi.org/10.1103/PhysRevD.78.014504}
  {\bibfield  {journal} {\bibinfo  {journal} {Phys. Rev. D}\ }\textbf {\bibinfo
  {volume} {78}},\ \bibinfo {pages} {014504} (\bibinfo {year} {2008})},\
  \Eprint {https://arxiv.org/abs/0803.0523} {arXiv:0803.0523 [hep-lat]}
  \BibitemShut {NoStop}%
\bibitem [{\citenamefont {Hammer}\ \emph {et~al.}(2020)\citenamefont {Hammer},
  \citenamefont {K\"onig},\ and\ \citenamefont {van Kolck}}]{Hammer:2019poc}%
  \BibitemOpen
  \bibfield  {author} {\bibinfo {author} {\bibfnamefont {H.~W.}\ \bibnamefont
  {Hammer}}, \bibinfo {author} {\bibfnamefont {S.}~\bibnamefont {K\"onig}},\
  and\ \bibinfo {author} {\bibfnamefont {U.}~\bibnamefont {van Kolck}},\
  }\bibfield  {title} {\bibinfo {title} {{Nuclear effective field theory:
  Status and perspectives}},\ }\href
  {https://doi.org/10.1103/RevModPhys.92.025004} {\bibfield  {journal}
  {\bibinfo  {journal} {Rev. Mod. Phys.}\ }\textbf {\bibinfo {volume} {92}},\
  \bibinfo {pages} {025004} (\bibinfo {year} {2020})},\ \Eprint
  {https://arxiv.org/abs/1906.12122} {arXiv:1906.12122 [nucl-th]} \BibitemShut
  {NoStop}%
\bibitem [{\citenamefont {Campostrini}\ \emph {et~al.}(2006)\citenamefont
  {Campostrini}, \citenamefont {Hasenbusch}, \citenamefont {Pelissetto},\ and\
  \citenamefont {Vicari}}]{Campostrini:2006ms}%
  \BibitemOpen
  \bibfield  {author} {\bibinfo {author} {\bibfnamefont {M.}~\bibnamefont
  {Campostrini}}, \bibinfo {author} {\bibfnamefont {M.}~\bibnamefont
  {Hasenbusch}}, \bibinfo {author} {\bibfnamefont {A.}~\bibnamefont
  {Pelissetto}},\ and\ \bibinfo {author} {\bibfnamefont {E.}~\bibnamefont
  {Vicari}},\ }\bibfield  {title} {\bibinfo {title} {Theoretical estimates of
  the critical exponents of the superfluid transition in {${}^4$He} by lattice
  methods},\ }\href {https://doi.org/10.1103/PhysRevB.74.144506} {\bibfield
  {journal} {\bibinfo  {journal} {Phys. Rev. B}\ }\textbf {\bibinfo {volume}
  {74}},\ \bibinfo {pages} {144506} (\bibinfo {year} {2006})},\ \Eprint
  {https://arxiv.org/abs/cond-mat/0605083} {arXiv:cond-mat/0605083}
  \BibitemShut {NoStop}%
\bibitem [{\citenamefont {Hasenbusch}(2019)}]{Hasenbusch:2019jkj}%
  \BibitemOpen
  \bibfield  {author} {\bibinfo {author} {\bibfnamefont {M.}~\bibnamefont
  {Hasenbusch}},\ }\bibfield  {title} {\bibinfo {title} {{Monte Carlo} study of
  an improved clock model in three dimensions},\ }\href
  {https://doi.org/10.1103/PhysRevB.100.224517} {\bibfield  {journal} {\bibinfo
   {journal} {Phys. Rev. B}\ }\textbf {\bibinfo {volume} {100}},\ \bibinfo
  {pages} {224517} (\bibinfo {year} {2019})},\ \Eprint
  {https://arxiv.org/abs/1910.05916} {arXiv:1910.05916 [cond-mat.stat-mech]}
  \BibitemShut {NoStop}%
\bibitem [{\citenamefont {Lipa}\ \emph {et~al.}(2000)\citenamefont {Lipa},
  \citenamefont {Swanson}, \citenamefont {Nissen}, \citenamefont {Geng},
  \citenamefont {Williamson}, \citenamefont {Stricker}, \citenamefont {Chui},
  \citenamefont {Israelsson},\ and\ \citenamefont {Larson}}]{Lipa:2000zz}%
  \BibitemOpen
  \bibfield  {author} {\bibinfo {author} {\bibfnamefont {J.~A.}\ \bibnamefont
  {Lipa}}, \bibinfo {author} {\bibfnamefont {D.~R.}\ \bibnamefont {Swanson}},
  \bibinfo {author} {\bibfnamefont {J.~A.}\ \bibnamefont {Nissen}}, \bibinfo
  {author} {\bibfnamefont {Z.~K.}\ \bibnamefont {Geng}}, \bibinfo {author}
  {\bibfnamefont {P.~R.}\ \bibnamefont {Williamson}}, \bibinfo {author}
  {\bibfnamefont {D.~A.}\ \bibnamefont {Stricker}}, \bibinfo {author}
  {\bibfnamefont {T.~C.~P.}\ \bibnamefont {Chui}}, \bibinfo {author}
  {\bibfnamefont {U.~E.}\ \bibnamefont {Israelsson}},\ and\ \bibinfo {author}
  {\bibfnamefont {M.}~\bibnamefont {Larson}},\ }\bibfield  {title} {\bibinfo
  {title} {Specific heat of helium confined to a 57-$\mu$m planar geometry near
  the lambda point},\ }\href {https://doi.org/10.1103/PhysRevLett.84.4894}
  {\bibfield  {journal} {\bibinfo  {journal} {Phys. Rev. Lett.}\ }\textbf
  {\bibinfo {volume} {84}},\ \bibinfo {pages} {4894} (\bibinfo {year}
  {2000})}\BibitemShut {NoStop}%
\bibitem [{\citenamefont {Chester}\ \emph {et~al.}(2020)\citenamefont
  {Chester}, \citenamefont {Landry}, \citenamefont {Liu}, \citenamefont
  {Poland}, \citenamefont {Simmons-Duffin}, \citenamefont {Su},\ and\
  \citenamefont {Vichi}}]{Chester:2019ifh}%
  \BibitemOpen
  \bibfield  {author} {\bibinfo {author} {\bibfnamefont {S.~M.}\ \bibnamefont
  {Chester}}, \bibinfo {author} {\bibfnamefont {W.}~\bibnamefont {Landry}},
  \bibinfo {author} {\bibfnamefont {J.}~\bibnamefont {Liu}}, \bibinfo {author}
  {\bibfnamefont {D.}~\bibnamefont {Poland}}, \bibinfo {author} {\bibfnamefont
  {D.}~\bibnamefont {Simmons-Duffin}}, \bibinfo {author} {\bibfnamefont
  {N.}~\bibnamefont {Su}},\ and\ \bibinfo {author} {\bibfnamefont
  {A.}~\bibnamefont {Vichi}},\ }\bibfield  {title} {\bibinfo {title} {{Carving
  out OPE space and precise $\text{O}(2)$ model critical exponents}},\ }\href
  {https://doi.org/10.1007/JHEP06(2020)142} {\bibfield  {journal} {\bibinfo
  {journal} {JHEP}\ }\textbf {\bibinfo {volume} {06}},\ \bibinfo {pages}
  {142}},\ \Eprint {https://arxiv.org/abs/1912.03324} {arXiv:1912.03324
  [hep-th]} \BibitemShut {NoStop}%
\bibitem [{\citenamefont {Pelissetto}\ and\ \citenamefont
  {Vicari}(2002)}]{Pelissetto:2000ek}%
  \BibitemOpen
  \bibfield  {author} {\bibinfo {author} {\bibfnamefont {A.}~\bibnamefont
  {Pelissetto}}\ and\ \bibinfo {author} {\bibfnamefont {E.}~\bibnamefont
  {Vicari}},\ }\bibfield  {title} {\bibinfo {title} {{Critical phenomena and
  renormalization group theory}},\ }\href
  {https://doi.org/10.1016/S0370-1573(02)00219-3} {\bibfield  {journal}
  {\bibinfo  {journal} {Phys. Rept.}\ }\textbf {\bibinfo {volume} {368}},\
  \bibinfo {pages} {549} (\bibinfo {year} {2002})},\ \Eprint
  {https://arxiv.org/abs/cond-mat/0012164} {arXiv:cond-mat/0012164}
  \BibitemShut {NoStop}%
\bibitem [{\citenamefont {Fodor}\ \emph {et~al.}(2018)\citenamefont {Fodor},
  \citenamefont {Holland}, \citenamefont {Kuti}, \citenamefont {Nogradi},\ and\
  \citenamefont {Wong}}]{Fodor:2017die}%
  \BibitemOpen
  \bibfield  {author} {\bibinfo {author} {\bibfnamefont {Z.}~\bibnamefont
  {Fodor}}, \bibinfo {author} {\bibfnamefont {K.}~\bibnamefont {Holland}},
  \bibinfo {author} {\bibfnamefont {J.}~\bibnamefont {Kuti}}, \bibinfo {author}
  {\bibfnamefont {D.}~\bibnamefont {Nogradi}},\ and\ \bibinfo {author}
  {\bibfnamefont {C.~H.}\ \bibnamefont {Wong}},\ }\bibfield  {title} {\bibinfo
  {title} {{A new method for the beta function in the chiral symmetry broken
  phase}},\ }\href {https://doi.org/10.1051/epjconf/201817508027} {\bibfield
  {journal} {\bibinfo  {journal} {EPJ Web Conf.}\ }\textbf {\bibinfo {volume}
  {175}},\ \bibinfo {pages} {08027} (\bibinfo {year} {2018})},\ \Eprint
  {https://arxiv.org/abs/1711.04833} {arXiv:1711.04833 [hep-lat]} \BibitemShut
  {NoStop}%
\bibitem [{\citenamefont {Kuti}\ \emph {et~al.}(2022)\citenamefont {Kuti},
  \citenamefont {Fodor}, \citenamefont {Holland},\ and\ \citenamefont
  {Wong}}]{Kuti:2022ldb}%
  \BibitemOpen
  \bibfield  {author} {\bibinfo {author} {\bibfnamefont {J.}~\bibnamefont
  {Kuti}}, \bibinfo {author} {\bibfnamefont {Z.}~\bibnamefont {Fodor}},
  \bibinfo {author} {\bibfnamefont {K.}~\bibnamefont {Holland}},\ and\ \bibinfo
  {author} {\bibfnamefont {C.~H.}\ \bibnamefont {Wong}},\ }\bibfield  {title}
  {\bibinfo {title} {{From ten-flavor tests of the $\beta$-function to
  $\alpha_s$ at the $Z$~pole}},\ }\href {https://doi.org/10.22323/1.396.0321}
  {\bibfield  {journal} {\bibinfo  {journal} {PoS}\ }\textbf {\bibinfo {volume}
  {LATTICE2021}},\ \bibinfo {pages} {321} (\bibinfo {year} {2022})},\ \Eprint
  {https://arxiv.org/abs/2203.15847} {arXiv:2203.15847 [hep-lat]} \BibitemShut
  {NoStop}%
\bibitem [{\citenamefont {Carosso}\ \emph {et~al.}(2018)\citenamefont
  {Carosso}, \citenamefont {Hasenfratz},\ and\ \citenamefont
  {Neil}}]{Carosso:2018bmz}%
  \BibitemOpen
  \bibfield  {author} {\bibinfo {author} {\bibfnamefont {A.}~\bibnamefont
  {Carosso}}, \bibinfo {author} {\bibfnamefont {A.}~\bibnamefont
  {Hasenfratz}},\ and\ \bibinfo {author} {\bibfnamefont {E.~T.}\ \bibnamefont
  {Neil}},\ }\bibfield  {title} {\bibinfo {title} {Nonperturbative
  renormalization of operators in near-conformal systems using gradient
  flows},\ }\href {https://doi.org/10.1103/PhysRevLett.121.201601} {\bibfield
  {journal} {\bibinfo  {journal} {Phys. Rev. Lett.}\ }\textbf {\bibinfo
  {volume} {121}},\ \bibinfo {pages} {201601} (\bibinfo {year} {2018})},\
  \Eprint {https://arxiv.org/abs/1806.01385} {arXiv:1806.01385 [hep-lat]}
  \BibitemShut {NoStop}%
\bibitem [{\citenamefont {Hasenfratz}\ and\ \citenamefont
  {Witzel}(2020)}]{Hasenfratz:2019hpg}%
  \BibitemOpen
  \bibfield  {author} {\bibinfo {author} {\bibfnamefont {A.}~\bibnamefont
  {Hasenfratz}}\ and\ \bibinfo {author} {\bibfnamefont {O.}~\bibnamefont
  {Witzel}},\ }\bibfield  {title} {\bibinfo {title} {{Continuous
  renormalization group $\beta$ function from lattice simulations}},\ }\href
  {https://doi.org/10.1103/PhysRevD.101.034514} {\bibfield  {journal} {\bibinfo
   {journal} {Phys. Rev. D}\ }\textbf {\bibinfo {volume} {101}},\ \bibinfo
  {pages} {034514} (\bibinfo {year} {2020})},\ \Eprint
  {https://arxiv.org/abs/1910.06408} {arXiv:1910.06408 [hep-lat]} \BibitemShut
  {NoStop}%
\bibitem [{\citenamefont {Peterson}\ \emph {et~al.}(2022)\citenamefont
  {Peterson}, \citenamefont {Hasenfratz}, \citenamefont {van Sickle},\ and\
  \citenamefont {Witzel}}]{Peterson:2021lvb}%
  \BibitemOpen
  \bibfield  {author} {\bibinfo {author} {\bibfnamefont {C.~T.}\ \bibnamefont
  {Peterson}}, \bibinfo {author} {\bibfnamefont {A.}~\bibnamefont
  {Hasenfratz}}, \bibinfo {author} {\bibfnamefont {J.}~\bibnamefont {van
  Sickle}},\ and\ \bibinfo {author} {\bibfnamefont {O.}~\bibnamefont
  {Witzel}},\ }\bibfield  {title} {\bibinfo {title} {{Determination of the
  continuous $\beta$ function of SU(3) Yang-Mills theory}},\ }\href
  {https://doi.org/10.22323/1.396.0174} {\bibfield  {journal} {\bibinfo
  {journal} {PoS}\ }\textbf {\bibinfo {volume} {LATTICE2021}},\ \bibinfo
  {pages} {174} (\bibinfo {year} {2022})},\ \Eprint
  {https://arxiv.org/abs/2109.09720} {arXiv:2109.09720 [hep-lat]} \BibitemShut
  {NoStop}%
\bibitem [{\citenamefont {Brower}\ \emph {et~al.}(2013)\citenamefont {Brower},
  \citenamefont {Fleming},\ and\ \citenamefont {Neuberger}}]{Brower:2012vg}%
  \BibitemOpen
  \bibfield  {author} {\bibinfo {author} {\bibfnamefont {R.~C.}\ \bibnamefont
  {Brower}}, \bibinfo {author} {\bibfnamefont {G.~T.}\ \bibnamefont
  {Fleming}},\ and\ \bibinfo {author} {\bibfnamefont {H.}~\bibnamefont
  {Neuberger}},\ }\bibfield  {title} {\bibinfo {title} {Lattice radial
  quantization: {3D Ising}},\ }\href
  {https://doi.org/10.1016/j.physletb.2013.03.009} {\bibfield  {journal}
  {\bibinfo  {journal} {Phys. Lett. B}\ }\textbf {\bibinfo {volume} {721}},\
  \bibinfo {pages} {299} (\bibinfo {year} {2013})},\ \Eprint
  {https://arxiv.org/abs/1212.6190} {arXiv:1212.6190 [hep-lat]} \BibitemShut
  {NoStop}%
\bibitem [{\citenamefont {Brower}\ \emph {et~al.}(2018)\citenamefont {Brower},
  \citenamefont {Cheng}, \citenamefont {Weinberg}, \citenamefont {Fleming},
  \citenamefont {Gasbarro}, \citenamefont {Raben},\ and\ \citenamefont
  {Tan}}]{Brower:2018szu}%
  \BibitemOpen
  \bibfield  {author} {\bibinfo {author} {\bibfnamefont {R.~C.}\ \bibnamefont
  {Brower}}, \bibinfo {author} {\bibfnamefont {M.}~\bibnamefont {Cheng}},
  \bibinfo {author} {\bibfnamefont {E.~S.}\ \bibnamefont {Weinberg}}, \bibinfo
  {author} {\bibfnamefont {G.~T.}\ \bibnamefont {Fleming}}, \bibinfo {author}
  {\bibfnamefont {A.~D.}\ \bibnamefont {Gasbarro}}, \bibinfo {author}
  {\bibfnamefont {T.~G.}\ \bibnamefont {Raben}},\ and\ \bibinfo {author}
  {\bibfnamefont {C.-I.}\ \bibnamefont {Tan}},\ }\bibfield  {title} {\bibinfo
  {title} {{Lattice $\phi^4$ field theory on Riemann manifolds: Numerical tests
  for the 2-d Ising CFT on $\mathbb{S}^2$}},\ }\href
  {https://doi.org/10.1103/PhysRevD.98.014502} {\bibfield  {journal} {\bibinfo
  {journal} {Phys. Rev. D}\ }\textbf {\bibinfo {volume} {98}},\ \bibinfo
  {pages} {014502} (\bibinfo {year} {2018})},\ \Eprint
  {https://arxiv.org/abs/1803.08512} {arXiv:1803.08512 [hep-lat]} \BibitemShut
  {NoStop}%
\bibitem [{\citenamefont {Brower}\ \emph {et~al.}(2021)\citenamefont {Brower},
  \citenamefont {Fleming}, \citenamefont {Gasbarro}, \citenamefont {Howarth},
  \citenamefont {Raben}, \citenamefont {Tan},\ and\ \citenamefont
  {Weinberg}}]{Brower:2020jqj}%
  \BibitemOpen
  \bibfield  {author} {\bibinfo {author} {\bibfnamefont {R.~C.}\ \bibnamefont
  {Brower}}, \bibinfo {author} {\bibfnamefont {G.~T.}\ \bibnamefont {Fleming}},
  \bibinfo {author} {\bibfnamefont {A.~D.}\ \bibnamefont {Gasbarro}}, \bibinfo
  {author} {\bibfnamefont {D.}~\bibnamefont {Howarth}}, \bibinfo {author}
  {\bibfnamefont {T.~G.}\ \bibnamefont {Raben}}, \bibinfo {author}
  {\bibfnamefont {C.-I.}\ \bibnamefont {Tan}},\ and\ \bibinfo {author}
  {\bibfnamefont {E.~S.}\ \bibnamefont {Weinberg}},\ }\bibfield  {title}
  {\bibinfo {title} {{Radial lattice quantization of 3D $\phi^4$ field
  theory}},\ }\href {https://doi.org/10.1103/PhysRevD.104.094502} {\bibfield
  {journal} {\bibinfo  {journal} {Phys. Rev. D}\ }\textbf {\bibinfo {volume}
  {104}},\ \bibinfo {pages} {094502} (\bibinfo {year} {2021})},\ \Eprint
  {https://arxiv.org/abs/2006.15636} {arXiv:2006.15636 [hep-lat]} \BibitemShut
  {NoStop}%
\bibitem [{\citenamefont {Neuberger}(2014)}]{Neuberger:2014pya}%
  \BibitemOpen
  \bibfield  {author} {\bibinfo {author} {\bibfnamefont {H.}~\bibnamefont
  {Neuberger}},\ }\bibfield  {title} {\bibinfo {title} {{Lattice radial
  quantization by cubature}},\ }\href
  {https://doi.org/10.1103/PhysRevD.90.114501} {\bibfield  {journal} {\bibinfo
  {journal} {Phys. Rev. D}\ }\textbf {\bibinfo {volume} {90}},\ \bibinfo
  {pages} {114501} (\bibinfo {year} {2014})},\ \Eprint
  {https://arxiv.org/abs/1410.2820} {arXiv:1410.2820 [hep-lat]} \BibitemShut
  {NoStop}%
\bibitem [{\citenamefont {Poland}\ and\ \citenamefont
  {Simmons-Duffin}(2022)}]{Poland:2022qrs}%
  \BibitemOpen
  \bibfield  {author} {\bibinfo {author} {\bibfnamefont {D.}~\bibnamefont
  {Poland}}\ and\ \bibinfo {author} {\bibfnamefont {D.}~\bibnamefont
  {Simmons-Duffin}},\ }\bibfield  {title} {\bibinfo {title} {The numerical
  conformal bootstrap},\ }in\ \href@noop {} {\emph {\bibinfo {booktitle} {{2022
  Snowmass Summer Study}}}}\ (\bibinfo {year} {2022})\ \Eprint
  {https://arxiv.org/abs/2203.08117} {arXiv:2203.08117 [hep-th]} \BibitemShut
  {NoStop}%
\bibitem [{\citenamefont {Baron}\ \emph {et~al.}(2010)\citenamefont {Baron}
  \emph {et~al.}}]{Baron:2010bv}%
  \BibitemOpen
  \bibfield  {author} {\bibinfo {author} {\bibfnamefont {R.}~\bibnamefont
  {Baron}} \emph {et~al.} (\bibinfo {collaboration} {ETM}),\ }\bibfield
  {title} {\bibinfo {title} {{Light hadrons from lattice QCD with light
  $(u,d)$, strange and charm dynamical quarks}},\ }\href
  {https://doi.org/10.1007/JHEP06(2010)111} {\bibfield  {journal} {\bibinfo
  {journal} {JHEP}\ }\textbf {\bibinfo {volume} {06}},\ \bibinfo {pages}
  {111}},\ \Eprint {https://arxiv.org/abs/1004.5284} {arXiv:1004.5284
  [hep-lat]} \BibitemShut {NoStop}%
\bibitem [{\citenamefont {Alexandrou}\ \emph
  {et~al.}(2018{\natexlab{c}})\citenamefont {Alexandrou} \emph
  {et~al.}}]{Alexandrou:2018egz}%
  \BibitemOpen
  \bibfield  {author} {\bibinfo {author} {\bibfnamefont {C.}~\bibnamefont
  {Alexandrou}} \emph {et~al.} (\bibinfo {collaboration} {ETM}),\ }\bibfield
  {title} {\bibinfo {title} {{Simulating twisted mass fermions at physical
  light, strange and charm quark masses}},\ }\href
  {https://doi.org/10.1103/PhysRevD.98.054518} {\bibfield  {journal} {\bibinfo
  {journal} {Phys. Rev. D}\ }\textbf {\bibinfo {volume} {98}},\ \bibinfo
  {pages} {054518} (\bibinfo {year} {2018}{\natexlab{c}})},\ \Eprint
  {https://arxiv.org/abs/1807.00495} {arXiv:1807.00495 [hep-lat]} \BibitemShut
  {NoStop}%
\bibitem [{\citenamefont {Aoki}\ \emph {et~al.}(2011)\citenamefont {Aoki} \emph
  {et~al.}}]{Aoki:2010dy}%
  \BibitemOpen
  \bibfield  {author} {\bibinfo {author} {\bibfnamefont {Y.}~\bibnamefont
  {Aoki}} \emph {et~al.} (\bibinfo {collaboration} {RBC, UKQCD}),\ }\bibfield
  {title} {\bibinfo {title} {Continuum limit physics from 2+1 flavor domain
  wall {QCD}},\ }\href {https://doi.org/10.1103/PhysRevD.83.074508} {\bibfield
  {journal} {\bibinfo  {journal} {Phys. Rev. D}\ }\textbf {\bibinfo {volume}
  {83}},\ \bibinfo {pages} {074508} (\bibinfo {year} {2011})},\ \Eprint
  {https://arxiv.org/abs/1011.0892} {arXiv:1011.0892 [hep-lat]} \BibitemShut
  {NoStop}%
\bibitem [{\citenamefont {Blum}\ \emph
  {et~al.}(2016{\natexlab{c}})\citenamefont {Blum} \emph
  {et~al.}}]{RBC:2014ntl}%
  \BibitemOpen
  \bibfield  {author} {\bibinfo {author} {\bibfnamefont {T.}~\bibnamefont
  {Blum}} \emph {et~al.} (\bibinfo {collaboration} {RBC, UKQCD}),\ }\bibfield
  {title} {\bibinfo {title} {{Domain wall QCD with physical quark masses}},\
  }\href {https://doi.org/10.1103/PhysRevD.93.074505} {\bibfield  {journal}
  {\bibinfo  {journal} {Phys. Rev. D}\ }\textbf {\bibinfo {volume} {93}},\
  \bibinfo {pages} {074505} (\bibinfo {year} {2016}{\natexlab{c}})},\ \Eprint
  {https://arxiv.org/abs/1411.7017} {arXiv:1411.7017 [hep-lat]} \BibitemShut
  {NoStop}%
\bibitem [{\citenamefont {Boyle}\ \emph {et~al.}(2017)\citenamefont {Boyle},
  \citenamefont {Del~Debbio}, \citenamefont {J\"uttner}, \citenamefont
  {Khamseh}, \citenamefont {Sanfilippo},\ and\ \citenamefont
  {Tsang}}]{Boyle:2017jwu}%
  \BibitemOpen
  \bibfield  {author} {\bibinfo {author} {\bibfnamefont {P.~A.}\ \bibnamefont
  {Boyle}}, \bibinfo {author} {\bibfnamefont {L.}~\bibnamefont {Del~Debbio}},
  \bibinfo {author} {\bibfnamefont {A.}~\bibnamefont {J\"uttner}}, \bibinfo
  {author} {\bibfnamefont {A.}~\bibnamefont {Khamseh}}, \bibinfo {author}
  {\bibfnamefont {F.}~\bibnamefont {Sanfilippo}},\ and\ \bibinfo {author}
  {\bibfnamefont {J.~T.}\ \bibnamefont {Tsang}} (\bibinfo {collaboration} {RBC,
  UKQCD}),\ }\bibfield  {title} {\bibinfo {title} {The decay constants ${f_D}$
  and ${f_{D_{s}}}$ in the continuum limit of ${N_f=2+1}$ domain wall lattice
  {QCD}},\ }\href {https://doi.org/10.1007/JHEP12(2017)008} {\bibfield
  {journal} {\bibinfo  {journal} {JHEP}\ }\textbf {\bibinfo {volume} {12}},\
  \bibinfo {pages} {008}},\ \Eprint {https://arxiv.org/abs/1701.02644}
  {arXiv:1701.02644 [hep-lat]} \BibitemShut {NoStop}%
\bibitem [{\citenamefont {Bruno}\ \emph {et~al.}(2015)\citenamefont {Bruno}
  \emph {et~al.}}]{Bruno:2014jqa}%
  \BibitemOpen
  \bibfield  {author} {\bibinfo {author} {\bibfnamefont {M.}~\bibnamefont
  {Bruno}} \emph {et~al.},\ }\bibfield  {title} {\bibinfo {title} {{Simulation
  of QCD with $_N{f}=2+1$ flavors of non-perturbatively improved Wilson
  fermions}},\ }\href {https://doi.org/10.1007/JHEP02(2015)043} {\bibfield
  {journal} {\bibinfo  {journal} {JHEP}\ }\textbf {\bibinfo {volume} {02}},\
  \bibinfo {pages} {043}},\ \Eprint {https://arxiv.org/abs/1411.3982}
  {arXiv:1411.3982 [hep-lat]} \BibitemShut {NoStop}%
\bibitem [{\citenamefont {Bruno}\ \emph
  {et~al.}(2017{\natexlab{b}})\citenamefont {Bruno}, \citenamefont {Korzec},\
  and\ \citenamefont {Schaefer}}]{Bruno:2016plf}%
  \BibitemOpen
  \bibfield  {author} {\bibinfo {author} {\bibfnamefont {M.}~\bibnamefont
  {Bruno}}, \bibinfo {author} {\bibfnamefont {T.}~\bibnamefont {Korzec}},\ and\
  \bibinfo {author} {\bibfnamefont {S.}~\bibnamefont {Schaefer}},\ }\bibfield
  {title} {\bibinfo {title} {{Setting the scale for the CLS $2 + 1$ flavor
  ensembles}},\ }\href {https://doi.org/10.1103/PhysRevD.95.074504} {\bibfield
  {journal} {\bibinfo  {journal} {Phys. Rev. D}\ }\textbf {\bibinfo {volume}
  {95}},\ \bibinfo {pages} {074504} (\bibinfo {year} {2017}{\natexlab{b}})},\
  \Eprint {https://arxiv.org/abs/1608.08900} {arXiv:1608.08900 [hep-lat]}
  \BibitemShut {NoStop}%
\bibitem [{\citenamefont {Edwards}\ \emph {et~al.}(2016)\citenamefont
  {Edwards}, \citenamefont {Gupta}, \citenamefont {Joó}, \citenamefont
  {Orginos}, \citenamefont {Richards}, \citenamefont {Winter},\ and\
  \citenamefont {Yoon}}]{Clover:2016csw}%
  \BibitemOpen
  \bibfield  {author} {\bibinfo {author} {\bibfnamefont {R.}~\bibnamefont
  {Edwards}}, \bibinfo {author} {\bibfnamefont {R.}~\bibnamefont {Gupta}},
  \bibinfo {author} {\bibfnamefont {B.}~\bibnamefont {Joó}}, \bibinfo {author}
  {\bibfnamefont {K.}~\bibnamefont {Orginos}}, \bibinfo {author} {\bibfnamefont
  {D.}~\bibnamefont {Richards}}, \bibinfo {author} {\bibfnamefont
  {F.}~\bibnamefont {Winter}},\ and\ \bibinfo {author} {\bibfnamefont
  {B.}~\bibnamefont {Yoon}} (\bibinfo {collaboration} {JLab/W\&M/LANL/MIT}),\
  }\href@noop {} {\bibinfo {title} {{U.S.}~$2+1$ flavor clover lattice
  generation program}} (\bibinfo {year} {2016}),\ \bibinfo {note}
  {unpublished}\BibitemShut {NoStop}%
\bibitem [{\citenamefont {Aaltonen}\ \emph {et~al.}(2022)\citenamefont
  {Aaltonen} \emph {et~al.}}]{CDF:2022hxs}%
  \BibitemOpen
  \bibfield  {author} {\bibinfo {author} {\bibfnamefont {T.}~\bibnamefont
  {Aaltonen}} \emph {et~al.} (\bibinfo {collaboration} {CDF}),\ }\bibfield
  {title} {\bibinfo {title} {{High-precision measurement of the $W$ boson mass
  with the CDF~II detector}},\ }\href {https://doi.org/10.1126/science.abk1781}
  {\bibfield  {journal} {\bibinfo  {journal} {Science}\ }\textbf {\bibinfo
  {volume} {376}},\ \bibinfo {pages} {170} (\bibinfo {year}
  {2022})}\BibitemShut {NoStop}%
\bibitem [{usq(2022)}]{usqcd}%
  \BibitemOpen
  \href@noop {} {\bibinfo {title} {{USQCD Collaboration} web site,
  \href{https://www.usqcd.org/}{\tt https://www.usqcd.org/}}} (\bibinfo {year}
  {updated July 2022})\BibitemShut {NoStop}%
\bibitem [{\citenamefont {Aubin}\ \emph {et~al.}(2019)\citenamefont {Aubin},
  \citenamefont {Bali}, \citenamefont {Del~Debbio}, \citenamefont {Detmold},
  \citenamefont {G\"ulpers}, \citenamefont {Hollitt}, \citenamefont {Lin},
  \citenamefont {Liu},\ and\ \citenamefont {Ryan}}]{Aubin:2019rdf}%
  \BibitemOpen
  \bibfield  {author} {\bibinfo {author} {\bibfnamefont {C.}~\bibnamefont
  {Aubin}}, \bibinfo {author} {\bibfnamefont {G.}~\bibnamefont {Bali}},
  \bibinfo {author} {\bibfnamefont {L.}~\bibnamefont {Del~Debbio}}, \bibinfo
  {author} {\bibfnamefont {W.}~\bibnamefont {Detmold}}, \bibinfo {author}
  {\bibfnamefont {V.}~\bibnamefont {G\"ulpers}}, \bibinfo {author}
  {\bibfnamefont {S.}~\bibnamefont {Hollitt}}, \bibinfo {author} {\bibfnamefont
  {H.-W.}\ \bibnamefont {Lin}}, \bibinfo {author} {\bibfnamefont
  {L.}~\bibnamefont {Liu}},\ and\ \bibinfo {author} {\bibfnamefont {S.~M.}\
  \bibnamefont {Ryan}},\ }\bibfield  {title} {\bibinfo {title} {Report on the
  2019 lattice diversity and inclusivity survey},\ }\href
  {https://doi.org/10.22323/1.363.0295} {\bibfield  {journal} {\bibinfo
  {journal} {PoS}\ }\textbf {\bibinfo {volume} {LATTICE2019}},\ \bibinfo
  {pages} {295} (\bibinfo {year} {2019})},\ \Eprint
  {https://arxiv.org/abs/1910.06800} {arXiv:1910.06800 [hep-lat]} \BibitemShut
  {NoStop}%
\end{thebibliography}%
